\documentclass[graybox,envcountchap,sectrefs]{svmono}
\usepackage{chapterbib}

\usepackage{mathptmx}
\usepackage{helvet}
\usepackage{courier}
\usepackage{type1cm}

\usepackage{makeidx}         
\usepackage{graphicx}        
\usepackage{multicol}        
\usepackage[bottom]{footmisc}

\usepackage{amscd,amsmath,amssymb,verbatim}

\usepackage{multirow}

\usepackage{subfigure}

\usepackage{hyperref}
\hypersetup{
     colorlinks   = true,
     citecolor    = blue
}

\newcommand\ket[1]{\ensuremath{|#1\rangle}}
\newcommand\bra[1]{\ensuremath{\langle#1|}}

\newcommand\Tr{\mathop{\rm Tr}\nolimits}

\newcommand\argmax{\mathop{\rm argmax}\nolimits}

\newcommand{\Z}{\mathbb{Z}}

\renewcommand{\v}[1]{\boldsymbol{#1}}
\renewcommand{\t}[1]{\tilde{#1}}
\newcommand{\ii}{\hspace{1pt}\mathrm{i}\hspace{1pt}}

\newcommand{\eq}[1]{(\ref{#1})}
\newcommand{\eqn}[1]{eqn.~(\ref{#1})}
\newcommand{\Eqn}[1]{Eqn.~(\ref{#1})}

\newcommand{\<}{\langle}
\renewcommand{\>}{\rangle}

\newcommand{\prt}{\partial}

\newcommand{\up}{\uparrow}
\newcommand{\down}{\downarrow}

\newcommand{\ie}{{\it i.e.~}}

\newcommand{\etc}{{\it etc~}}

\newcommand{\al}{\alpha}
\newcommand{\bt}{\beta}

\newcommand{\Del}{\Delta}
\newcommand{\eps}{\epsilon}
\newcommand{\veps}{\varepsilon}
\newcommand{\ga}{\gamma}

\newcommand{\la}{\lambda}

\newcommand{\om}{\omega}

\renewcommand{\th}{\theta}

\newcommand{\si}{\sigma}

\newcommand{\cT}{ {\cal T} }

\newcommand{\enu}[1]{ \noindent
 \begin{enumerate}
 #1
 \end{enumerate}
}

\newcommand{\bpm}{\begin{pmatrix}}
\newcommand{\epm}{\end{pmatrix}}
\newcommand{\bmm}{\begin{matrix}}
\newcommand{\emm}{\end{matrix}}

\newcommand{\be}{\begin{equation}}
\newcommand{\ee}{\end{equation}}

\newcommand{\braket}[2]{\langle #1|#2\rangle}

\newcommand{\mbet}{many-body entanglement\ }

\newcommand{\dg}{\dagger}

\newcommand{\e}{\hspace{1pt}\mathrm{e}}
\newcommand{\imth}{\hspace{1pt}\mathrm{i}\hspace{1pt}}

\makeindex             

\begin{document}

\author{Bei Zeng, Xie Chen, Duan-Lu Zhou, Xiao-Gang Wen}
\title{Quantum Information Meets Quantum Matter}
\subtitle{From Quantum Entanglement to Topological Phase in Many-Body Systems}
\date{\today}
\maketitle

\frontmatter

%
%

\preface

After decades of development, quantum information science and technology has
now come to its golden age. It is not only widely believed that quantum
information processing offers the secure and high rate information
transmission, fast computational solution of certain important problems, which
are at the heart of the modern information technology. But also, it provides
new angles, tools and methods which help in understanding other fields of
science, among which one important area is the link to modern condensed matter
physics.

For a long time, people believe that all phases of matter are described by
Landau's symmetry-breaking theory, and the transitions between those phases are
described by the change of those symmetry-breaking orders. However, after the discovery of
fractional quantum Hall effect, it was realized in 1989 that the fractional
quantum Hall states contain a new type of order (named topological order) which is beyond
Landau symmetry breaking theory. Traditional many-body theory for condensed
matter systems is mostly based on various correlation functions, which suite
Landau symmetry breaking theory very well. But this kind of approaches is
totally inadequate for topological orders, since all different
topological orders have the similar short-range correlations. 

The traditional condensed matter theory mostly only consider two kinds of
many-body states: product states (such as in various mean-field theories) and
states obtained by filling orbitals (such as in Fermi liquid theory).  Those
two types of states fail to include the more general topologically ordered
states. So the big question is, can we understand what is missing in the above two types of states, so
that they fail to capture the topological order?

What quantum information science brings is the information-theoretic
understanding of correlation, and a new concept called `entanglement', which is a
pure quantum correlation that has no classical counterpart. Such input from
quantum information science led to a recent realization that the new
topological order in some strongly correlated systems is nothing but the
pattern of many-body entanglement. The study of topological order and the
related new quantum phases is actually a study of patterns of entanglement. The
non-trivial patterns of entanglement is the root of many highly novel phenomena
in topologically ordered phases (such as fractional quantum Hall states and
spin liquid states), which include fractional charge, fractional statistics,
protected gapless boundary excitations, emergence of gauge theory and Fermi
statistics from purely bosonic systems, etc.

The connection between  quantum information science and condensed matter
physics is not accidental, but has a very deep root.  Quantum theory has
explained and unified many microscopic phenomena, ranging from discrete
spectrum of Hydrogen atom, black-body radiation, to interference of electron
beam, \etc.  However, what quantum theory really unifies is information and
matter.  We know that a change or frequency is a property of information.  But
according to quantum theory, frequency corresponds to energy. According to the theory of 
relativity, energy correspond to mass. Energy and mass are properties of matter.
In this sense frequency leads to mass and information becomes matter.  

But do we believe that matter (and the elementary particles that form the
matter) all come from qubits? Is it possible that qubits are the building
blocks of all the elementary particles?  If matter were formed by simple spin-0
bosonic elementary particles, then it was quite possible that the spin-0
bosonic elementary particles, and the matter that they form, all came from
qubits.  We can simply view the space as a collection of qubits and the 0-state of
qubits as the vacuum. Then the 1-state of qubit will correspond to a  spin-0 bosonic
elementary particle in space.  But our world is much more complicated.  The
matter in our world is formed by particles that have two really strange
properties: Fermi statistics and fractional angular momentum (spin-1/2).  Our
world also have light, which correspond to spin-1 particles that strangely only
have two components. Such spin-1 particles are called gauge bosons.  

Can space formed by simple qubits produce spin-1/2 fermions and spin-1 gauge
bosons?  In the last 20 years (and as explained in this book), we start to realize
that although qubits are very simple, their organization -- their quantum
entanglement -- can be extremely rich and complex.  The long-range quantum
entanglement of qubits make it possible to use simple qubits to produce
spin-1/2 fermions and spin-1 gauge bosons, as well as the matter formed by
those elementary particles.

Thousands of research papers studying the properties of quantum entanglement has been published in the past two decades. Notable progress
includes, but not limited to, extensive study of correlation and entanglement
properties in various strongly-correlated systems, development of concepts of
entanglement area law which results in a new tool called tensor network method,
the role of entanglement play in quantum phase transitions, the concept of long
range entanglement and its use in the study of topological phase of matter. Also, extensive attentions have been attracted on the new states of quantum matter and the emergence
of fractional quantum numbers and fractional/Fermi statistics, with many published papers during the last decades along these directions. 

It is not possible to include all these exciting developments in a single book.
The scope of this book is rather, to introduce some general concepts and basic
ideas and methods that the viewpoints of quantum information scientists have on
condensed matter physics. The style of quantum information theorists treating
physics problem is typically more mathematical than usual condensed matter
physicists. One may understand this as traditional mathematical physics with
tools added from quantum information science. Typical models are studied, but
more general perspectives are also emphasized. For instance, one important
problem widely studied is the so-called `local Hamiltonian problem', which is
based on the real physical situations where Hamiltonians involve only local
interactions with respect to certain lattice geometry. General theory regarding
this problem is developed, which provides powerful tools in understanding the
common properties of these physical systems.

This book aims to introduce the quantum information science viewpoints on condensed matter physics to graduate students in physics (or interested researchers). We keep the writing in a self-consistent way, requiring minimum background in quantum information science. Basic knowledge in undergraduate
quantum physics and condensed matter physics is assumed. We start slowly from the basic ideas in quantum information theory, but wish to eventually bring the readers to the frontiers of research in condensed matter physics, including
topological phases of matter, tensor networks, and symmetry-protected topological phases.

\vspace{5mm}
\noindent{\Large{\bf Structure of the Book}}
\vspace{5mm}

The book has five parts, each includes several chapters. We start from Part I
for introducing the basic concepts in quantum information that will be later
used in the book. Quantum information science is a very large field and many
new ideas and concepts are developed. For a full reference one may turn to
other classical sources such as `Quantum Computation and Quantum Information' by
Nielsen \& Chuang and Preskill's lecture notes for the course of `Quantum Computation' at Caltech.
The goal of this part is to introduce minimum knowledge that will quickly bring
the readers into the more exiting topics of application of quantum information
science to condensed matter physics. 

Three main topics are discussed: 
Chapter 1 summarizes useful tools in the theory of correlation and
entanglement. It introduces the basic idea of correlation from information-theoretic viewpoint, and the basic idea of entanglement and how to
quantify it. Chapter 2 discusses quantum information viewpoint of quantum
evolution and introduces the idea of quantum circuits, and 
the important concept of circuit depth. Chapter
3 summarizes useful tools in the theory of quantum error correction, and the toric
code is introduced for the first time.

Then Part II starts from Chapter 4, discussing a general viewpoint of the local Hamiltonian problem, which
is at the heart of the link between quantum information science and condensed
matter physics. A local Hamiltonian involves only geometrically local few-body interactions. We discuss the ways of determining the ground-state energy of local Hamiltonians, and their hardness. Theories have been developed in quantum information science to show that even with the existence of a quantum computer, there is no efficient way of finding the ground-state energy for a local Hamiltonian in general. However, for practical cases, special structures may lead to simpler method, such as
Hartree's mean-field theory. 
A special kind of local Hamiltonians,
called the frustration-free Hamiltonian, where the ground state of the
Hamiltonian also minimizes the energy of each local term of the Hamiltonian, is
also introduced. These Hamiltonians play important role in later chapters of
the book. 

In Chapter 5, we start to focus our attention on systems of infinite size (i.e. the thermodynamic limit),
which are the central subject of study in condensed matter physics. We introduce important notions for 
the discussion of such quantum many-body systems, like locality, correlation, gap, etc.
In particular, we discuss in depth the notion of many-body entanglement, which is one 
of the most important distinction between quantum and classical many-body systems, and 
is the key to the existence of topological order, a subject which we study in detail in this book. We discuss the important concepts of entanglement area law,
and the topological entanglement entropy. We study the topological entanglement entropy from an information-theoretic viewpoint, which leads to generalizations of topological entanglement entropy that can also be used to study systems without topological order. The corresponding information-theoretic quantity, called the quantum conditional mutual information, provides a universal detector of non-trivial
entanglement in many-body systems.

Entanglement is especially important for the description and understanding of
systems with a special type of order -- topological order. Topological order
has emerged as an exciting research topic in condensed matter physics for
several decades. People have approached the problem using various methods but
many important issues still remain widely open. Recent developments show that
quantum information ideas can contribute greatly to the study of topological
order, the topological entanglement entropy discussed in Chapter 5 is such an example. Part III will further discuss the entanglement properties of topological order in detail. 

In Chapter 6, we give a full review of the basic ideas of topological
order  from the perspective of modern condensed matter theory.
Through this part, we hope to give readers a general idea of what topological
order is, why physicists are interested in it, and what the important issues
are to be solved. This chapter is devoted to the basic concepts and the
characteristic properties of topological order.  After setting the stage up on
both the quantum information and condensed matter physics side, we are then
ready to show that how the combination of these two leads to new discoveries.  

In
Chapter 7, we are going to show how quantum information ideas can be used to
reformulate and characterize topological order and what we have learned from
this new perspective, which leads to a microscopic theory of 
topological order. A new formulation of the basic notion of phase and phase
transition in terms of quantum information concepts is given, based on the concept of local unitary equivalence between systems in the
same gapped phase. We are going to introduce the concept of gapped quantum liquids, and show that topological order corresponds to stable gapped quantum liquids. We also show that symmetry-breaking orders correspond to unstable quantum liquids. This allows us to study both symmetry-breaking and topological order in a same general framework. We also discuss the concept of long-range entanglement, and show that topological orders are patterns of long-range entanglement. 

After that, in part IV, we study gapped phases in one and two dimension (1D and 2D) using the tensor network formalism. First, we focus on one dimensional systems in Chapter 8. It turns out the matrix product state -- the one dimensional version of the tensor network representation -- provides a complete and precise characterization of 1D gapped systems so that we can actually classify all gapped phases in 1D. In particular, we show, after a careful introduction to the matrix product formalism, that there is no topological order in 1D and all gapped states in 1D belong to the same phase (if no symmetry is required). In Chapter 9, we move on to two dimensions, where things become much more complicated and also more interesting. The tensor product state is introduced, whose similarity and difference with matrix product states is emphasized. Apart from the short range entangled phases like symmetry breaking phases, the tensor network states can also represent topological phases in 2D. We discuss examples of such tensor product states and how the topological order is encoded in the local tensors.
In Chapter 10, global symmetry is introduced into the system. It was realized that short range entangled states can be in different phases even when they have the same symmetry. Examples of such `Symmetry Protected Topological (SPT) Phases' are introduced both in 1D and 2D. Moreover, we show that 1D SPT can be fully classified using the matrix product state formalism and a systematic construction exists for SPT states in 2D and higher dimensions in interacting bosonic systems.

The last part (Chapter 11) is devoted to an overview of physics and an outlook
how many-body entanglement may influence how we view our world.  We outline
the developement of our world views in the last a few hundreds of years: from all
matter being fromed by particles to the discovery of wave-like matter
(electromanetics waves and gravitational waves), and to the unification of
particle-like matter and wave-like matter by quantum theory. We feel that we
are in the process of a new revolution where quantum information, matter,
interactions, and even space itself will be all unified.  To make sush a point,
we discuss some simple examples of more general highly entangled quantum states
of matter, which can be gapless.  This leads to a unification of light and
electrons (or all elementary particles) by qubits that form the space.  Those
examples demonstrate a unification of information and matter, the central
theme of this book.

The unified theme of quantum information and quantum matter represents a
totally new world in physics.  This book tries to introduce this new world to
the reader. However, we can only scratch the surface of this new world at this
stage. A lot of new developments are needed to truly reveal this exciting new
world. Even a new mathematical language is needed for such a unified
understanding of information and matter. A comprehensive theory of highly
entangled quantum states of matter requires such a  mathematical theory which
is yet to be developed.

\vspace{\baselineskip}
\begin{flushright}\noindent
February 2018\hfill {\it Bei Zeng}\\
\hfill {\it Xie Chen}\\
\hfill {\it Duan-Lu Zhou}\\
\hfill {\it Xiao-Gang Wen}\\

%
%
%

\end{flushright}

%
%

\extrachap{Acknowledgements}


This is an incomplete list of people that we owe thanks to. Updated list will be included in the published version of Springer.

B.Z. X.-G.W. X.C. would like to thank Institute for Advanced Study at Tsinghua University (IASTU), Beijing, for hospitality. Part of the book has been written during our visit to IASTU for the past five years.

We are grateful to Jianxin Chen, Runyao Duan, David Gosset, Zheng-Cheng Gu, Jame Howard, Zhengfeng Ji, Joel Klassen, Chi-Kwong Li, Yiu Tung Poon, Yi Shen, Changpu Sun, Zhaohui Wei, Zhan Xu, and Nengkun Yu for valuable discussions during writing the first draft of the book.

We appreciate the comments received for Version 1 and Version 2 of the book draft, from Oliver Buerschaper, Abdulah Fawaz, Nicole Yunger Halpern, Junichi Iwasaki, Zeyang Li, David Meyer, Mikio Nakahara, Tomotoshi Nishino,Fernando Pastawski, Mehdi Soleimanifar, Dawson Wang, and Youngliang Zhang.

We appreciate the comments received for Version 3 of the book draft, from Stephen Kwaku Amponsah, Zhi-An Jia, Oleg Kabermik, David Meyer, Mikio Nakahara, Hal Tasaki, Julien Vidal, Mark Wilde, and Mingli Yuan.

We are grateful to Haijing Song for making part of the figures in Chapters 7 and 10, and Zheng An for making part of the figures in Chapters 8 and 9.

We acknowledge volunteers from Swarma Club for making part of the figures in Chapter 11. They are: Song Cheng (Figures 11.9, 11.11(b)),Yanping Dai (Figures 11.4(a,b,c), 11.26(c)),Lei Dong (Figures 11.1, 11.2(a,c)), Yueyuan Hou (Figure 11.6(c),11.25(c)), Weiyi Qiu (Figures 11.7(a),11.26(a), 11.14(a,b,c,d)), Jiannan Wang (Figure 11.3?11.2(b), 11.3, 11.5(a)), Yizhuang You (Figures 11.5(d), 11.10, 11.11(a), 11.25(a,b), 11.26(c)) , Yanbo Zhang (Figures11.5(b), 11.6(a) 11.8(c), 11.11(a), 11.13, 11.26(b)), Yongjie Zhang (Figure 11.25(g)), and Bin Zhao (Figure 11.11(a)).

More comments are welcome.

\tableofcontents

\mainmatter
%
%
%

\begin{partbacktext}
\part{Basic Concepts in Quantum Information Theory}

\end{partbacktext}

%
%
%

\chapter{Correlation and Entanglement}
\label{cp:1} 

\abstract{In this chapter we discuss correlation and entanglement in
  many-body systems. We start from introducing the concepts ofone
  independence and correlation in probability theory, which leads to
  some understanding of the concepts of entropy and mutual
  information, which are of vital importance in modern information
  theory. This builds a framework that allows us to look at the theory
  of a new concept, called quantum entanglement, which serves as a
  fundamental object that we use to develop new theories for
  topological phase of matter later in this book.}

\section{Introduction}
\label{sec:cp1sec1}

The concept of correlation is used ubiquitously in almost every branch
of sciences. Intuitively, correlations describe the dependence of
certain properties for different parts of a composite object. If these
properties of different parts are independent of each other, then we
say that there do not exist correlations between (or among) them. If
they are correlated, then how to characterize the correlation, both
qualitatively and quantitatively, becomes an essential task.

Different branches of science usually have their own way of
characterizing correlation, in particular related to things that
scientists in different fields do care about. For instance, in
many-body physics, people usually characterize correlations in terms
of correlation functions
$\langle O_i O_j\rangle-\langle O_i \rangle\langle O_j \rangle$, where
$O_i$ is some observable on the site $i$, and $\langle \cdot \rangle$
denotes the expectation value with respect to the quantum state of the
system. The behavior of these correlation functions gives lots of
useful information such as the correlation length.

In this chapter we would like to treat correlation in a more formal
way. It will later become clearer that doing so does help with a
better understanding of many-body physics. In other words, there is
something beyond just correlation function to look at, which turns out
to provide new information and characterization of some rather
interesting new physical phenomena, such as the topological phase of
matter.

We start looking at correlation in terms of elementary probability
theory. First of all it is the formal mathematical language of
characterizing the concept of independence and correlation. This
formal language will then be further linked to the concept of entropy
and mutual information, which are key concepts in information theory.
Physicists are indeed familiar with the concept of entropy, which is
in some sense a measure of how chaotic a system is, or how lack of
knowledge we are regarding the system. By looking at it slight
differently, it is then a measure of how much information the system
carries -- in other words, because the lack of knowledge, the system
carries some `information' to tell.

What might be quite surprising to physicists is that the concept of
`entropy' lays the foundation to modern information theory, which
eventually guarantees the correct output of our computers that we rely
on for our everyday research, and fast communication via cell phone or
Internet that we rely on to exchange opinions with our colleagues.
Sitting in this information age, we are proud to know that the basic
concept in physics helps making all this possible. On the other hand,
it is also of vital importance to know `how'. One simple reason is
that we physicists are always curious, which is the essential inert
driving force of our research. But most importantly, one can borrow
the ideas back from information theory to add new ingredient to our
theory of fundamental physics.

One important success in quantum information is the development of the
theory of entanglement. `Entanglement' is widely heard nowadays but
what we would like to emphasize here is that there is nothing
mysterious, in a sense that almost all quantum many-body systems are
entangled. Perhaps you are still quite happy with mean-field theory,
which is valid in most cases, where no entanglement needs to be
considered. This does not mean that the system is not entangled, but
just perhaps not strongly entangled. On the other hand, you may also
be aware of the headache in the theory for strongly-correlated
systems, where the systems turn out to be highly entangled.

We would also like to introduce the theory of entanglement in a more
formal manner, which naturally follows the information theoretic point
of view. One good thing is that this will explain the difference
between `classical correlation' and `quantum entanglement'. More
importantly, it builds on a framework of `tensor product structure' of
Hilbert space for many-body systems, which is natural but not
emphasized in the traditional framework of many-body theory. It will
later become clearer that this `tensor product structure' will indeed
bring new concepts for understanding many-body physics.

We will start our discussion from the simplest case, where we only
consider two objects and their independence/correlation. We look at
the classical correlation case first, and then move to the case of
quantum systems, where the concept of entanglement can be naturally
introduced. Following up all that, we move into looking at the theory
for many-body systems, in terms of both classical correlation and
quantum entanglement.

\section{Correlations in classical probability theory}
\label{sec:cp2sec1}

In this section, we introduce the concepts of independence and
correlation in probability theory, and further link it to vital
concepts in modern information theory, such as entropy and mutual
information.

\subsection{Joint probability without correlations}

We start from looking at the simplest case: two independent objects
$A$ and $B$. Due to conventions of information theory, instead we
usually discuss two people, Alice and Bob, performing some joint
experiments. In this case, assume that Alice has total $d_A$ possible
outcomes, and let us denote the set of these possible outcomes by
$\Omega=\left\{ \omega_{i},i=0,1,\cdots,d_{A}-1\right\} $. For
example, the simplest case is that Alice has only two possible
outcomes, where $\Omega=\{\omega_0,\omega_1\}$.

Similarly, assume that Bob has total $d_B$ possible outcomes, and
denote the set of these possible outcomes for Bob by
$\Lambda=\left\{ \lambda_{m},m=0,1,\cdots,d_{B}-1\right\} $. Again the
simplest case is that Bob has only two possible outcomes, i.e.
$\Lambda=\{\lambda_0,\lambda_1\}$.

A joint possible outcome for Alice and Bob is
$\left(\omega_{i},\lambda_{m}\right)$. All such joint possible
outcomes form a set that we denote by $\Omega\times\Lambda$, which is
the Cartesian product of two sets $\Omega$ and $\Lambda$.
For instance, when $\Omega=\{\omega_0,\omega_1\}$ and
$\Lambda=\{\lambda_0,\lambda_1\}$, we have
$\Omega\times\Lambda=\{(\omega_0,\lambda_0),(\omega_0,\lambda_1),(\omega_1,\lambda_0),(\omega_1,\lambda_1)\}$.
In general, the set $\Omega\times\Lambda$ contains total $d_Ad_B$
elements.

The joint probability distribution
$p_{AB}\left(\omega_{i},\lambda_{m}\right)$ for the joint experiment
Alice and Bob perform needs to satisfy the following conditions.
\begin{eqnarray}
  p_{AB}\left(\omega_{i},\lambda_{m}\right) & \ge & 0,\\
  \sum_{i=0}^{d_{A}-1}\sum_{m=0}^{d_{B}-1}p_{AB}\left(\omega_{i},\lambda_{m}\right) & = & 1.
\end{eqnarray}

The probability for Alice to get the outcome $\omega_{i}\in\Omega$ is
then
\begin{equation}
  p_{A}\left(\omega_{i}\right)=\sum_{m=0}^{d_{B}-1}p_{AB}\left(\omega_{i},\lambda_{m}\right).
\end{equation}
Similarly, the probability for Bob to get the outcome
$\lambda_{m}\in\Lambda$ is
\begin{equation}
  p_{B}\left(\lambda_{m}\right)=\sum_{i=0}^{d_{A}-1}p_{AB}\left(\omega_{i},\lambda_{m}\right).
\end{equation}

As an example, let us again consider the simplest case where
$\Omega=\{\omega_0,\omega_1\}$ and $\Lambda=\{\lambda_0,\lambda_1\}$,
so
$\Omega\times\Lambda=\{(\omega_0,\lambda_0),(\omega_0,\lambda_1),(\omega_1,\lambda_0),(\omega_1,\lambda_1)\}$.
One possible choice of the joint probability distribution could be
\begin{equation}
  \label{eq:jointeg}
  p_{AB}(\omega_0,\lambda_0)=\frac{1}{12},\ p_{AB}(\omega_0,\lambda_1)=\frac{1}{4},
  \ p_{AB}(\omega_1,\lambda_0)=\frac{1}{6},\ p_{AB}(\omega_1,\lambda_1)=\frac{1}{2}.
\end{equation}
It is easy to check that
$\sum_{i=0}^{1}\sum_{m=0}^{1}p_{AB}\left(\omega_{i},\lambda_{m}\right)=1$,
and for Alice,
\begin{eqnarray}
  p_A(\omega_0)&=&\sum_{m=0}^{1}p_{AB}(\omega_0,\lambda_m)=\frac{1}{12}+\frac{1}{4}=\frac{1}{3},\nonumber\\
  p_A(\omega_1)&=&\sum_{m=0}^{1}p_{AB}(\omega_1,\lambda_m)=\frac{1}{6}+\frac{1}{2}=\frac{2}{3}.
\end{eqnarray}
For Bob,
\begin{eqnarray}
  p_B(\lambda_0)&=&\sum_{i=0}^{1}p_{AB}(\omega_i,\lambda_0)=\frac{1}{12}+\frac{1}{6}=\frac{1}{4},\nonumber\\
  p_B(\lambda_1)&=&\sum_{i=0}^{1}p_{AB}(\omega_i,\lambda_1)=\frac{1}{4}+\frac{1}{2}=\frac{3}{4}.
\end{eqnarray}

Now let us try to examine under which circumstances a joint
probability distribution $p_{AB}(\omega_i,\lambda_m)$ has some
correlation between the outcomes of Alice's and Bob's or not. Note
that when Bob gets the outcome $\lambda_{m}$, the probability for
Alice to get the outcome $\omega_{i}$ is then
\begin{equation}
  \label{eq:cond}
  p_{A\vert B}\left(\omega_{i},\lambda_{m}\right)=\frac{p_{AB}\left(\omega_{i},\lambda_{m}\right)}{p_{B}\left(\lambda_{m}\right)}.
\end{equation}
Here $p_{A\vert B}$ is called the conditional probability distribution
for $A$, conditionally on the outcome of $B$. Similarly one can write
down the conditional probability distribution $p_{B\vert A}$ for $B$,
conditionally on the outcome of $A$. That is, when Alice gets the
outcome $\omega_{i}$, the conditional probability for Bob to get the
outcome $\lambda_{m}$ is
\begin{equation}
  p_{B\vert A}\left(\lambda_{m},\omega_{i}\right)=\frac{p_{AB}\left(\omega_{i},\lambda_{m}\right)}{p_{A}\left(\omega_{i}\right)}.\label{def:bca}
\end{equation}

Now suppose that the joint distribution $p_{AB}(\omega_i,\lambda_m)$
has no correlation at all, then from Alice's point of view, her
outcome is independent of Bob's outcome. In other words, whatever
Bob's outcome is, the probability distribution of Alice's outcome
should be just the same. This means that the conditional probability
$p_{A\vert B}\left(\omega_{i},\lambda_{m}\right)$ should not depend on
$\lambda_m$, i.e.
\begin{equation}
  p_{A\vert B}\left(\omega_{i},\lambda_{m}\right)=p_{A\vert B}\left(\omega_{i},\lambda_{n}\right),\forall i,m,n.\label{eq:c0c2}
\end{equation}

Similarly, from Bob's point of view, one should have
\begin{equation}
  p_{B\vert A}\left(\lambda_{m},\omega_{i}\right)=p_{B\vert A}\left(\lambda_{m},\omega_{j}\right),\forall i,j,m.
  \label{eq:c0c1}
\end{equation}

We will show that the condition of (\ref{eq:c0c2}) and (\ref{eq:c0c1}) implies that the
joint probability distribution equals the product of the probability
distributions of each party, i.e.
\begin{equation}
  p_{AB}\left(\omega_{i},\lambda_{m}\right)=p_{A}\left(\omega_{i}\right)p_{B}\left(\lambda_{m}\right),\forall i,m,\label{eq:c0c3}
\end{equation}
and vice versa. In other words, the conditions (\ref{eq:c0c1}) and
(\ref{eq:c0c3}) are just equivalent.

To see this, we first show how to go from (\ref{eq:c0c1}) to
(\ref{eq:c0c3}). For $\forall m,i,$ we have for $\forall j$,
\begin{equation}
  p_{B\vert A}\left(\omega_{i},\lambda_{m}\right)=p_{B\vert A}\left(\omega_{j},\lambda_{m}\right)=\frac{p_{AB}\left(\omega_{j},\lambda_{m}\right)}{p_{A}\left(\omega_{j}\right)}.
\end{equation}
Then
\begin{equation}
  p_{B\vert A}\left(\omega_{i},\lambda_{m}\right)=\frac{\sum_{j=0}^{d_{A}-1}p_{AB}\left(\omega_{j},\lambda_{m}\right)}{\sum_{j=0}^{d_{A}-1}p_{A}\left(\omega_{j}\right)}=p_{B}\left(\lambda_{m}\right).\label{eq:pbca}
\end{equation}
Inserting (\ref{eq:pbca}) into (\ref{def:bca}), we will obtain
(\ref{eq:c0c3}). To show the converse of going from (\ref{eq:c0c3}) to
(\ref{eq:c0c1}): Inserting (\ref{eq:c0c3}) to (\ref{def:bca}), we get
$p_{B\vert
  A}\left(\omega_{i},\lambda_{m}\right)=p_{B}\left(\lambda_{m}\right)$,
which is independent of the index $i$, i.e., we have (\ref{eq:c0c1}).

Because the indices of $A$ and $B$ can be reversed, we also obtain the
equivalence between (\ref{eq:c0c2}) and (\ref{eq:c0c3}). Therefore,
the three conditions (\ref{eq:c0c2}), (\ref{eq:c0c1}) and
(\ref{eq:c0c3}) are essentially equivalent. In other words, Alice's
outcome is independent of Bob's indicates that Bob's outcome is
independent of Alice's and vice versa, and both imply that the joint
probability distribution equals the product of the probability
distributions of each party. We summarize these results in the box
below, which will be our starting point for talking about independent
probability distributions.
\begin{svgraybox}
  \begin{center}
    \textbf{Box 1.1 Independent probability distribution}
  \end{center}

  The following statements are equivalent:
  \begin{enumerate}
  \item There is no correlation in the joint probability distribution
    $p_{AB}(\omega_i,\lambda_m)$.
  \item The probability for Bob's outcome is independent of Alice's
    outcome:
    $$
    p_{B\vert A}\left(\omega_{i},\lambda_{m}\right)=p_{B\vert
      A}\left(\omega_{j},\lambda_{m}\right),\forall i,j,m.
    $$
  \item The probability for Alice's outcome is independent of Bob's
    outcome:
    $$
    p_{A\vert B}\left(\omega_{i},\lambda_{m}\right)=p_{A\vert
      B}\left(\omega_{i},\lambda_{n}\right),\forall i,m,n.
    $$
  \item The joint probability equals the product of probabilities for
    the two parties:
    $$
    p_{AB}\left(\omega_{i},\lambda_{m}\right)=p_{A}\left(\omega_{i}\right)p_{B}\left(\lambda_{m}\right),\forall
    i,m.
    $$
  \end{enumerate}
\end{svgraybox}

As an example, one can show that the joint probability distribution
given in Eq.~(\ref{eq:jointeg}) has no correlation. One can also show
that the joint probability distribution given below in
Eq.~(\ref{eq:jointeg2}) does have some correlation.
\begin{equation}
  \label{eq:jointeg2}
  p_{AB}(\omega_0,\lambda_0)=\frac{1}{6},\ p_{AB}(\omega_0,\lambda_1)=\frac{1}{3},
  \ p_{AB}(\omega_1,\lambda_0)=\frac{1}{4},\ p_{AB}(\omega_1,\lambda_1)=\frac{1}{4}.
\end{equation}

\subsection{Correlation functions}

When the condition given in Eq.(\ref{eq:c0c3}) does not hold, then
there must be correlation between the outcomes of Alice and Bob. We
would like to examine this condition further by relating it to
correlation functions. We first introduce a random variable
$X\left(\Omega\right)$, which is a real function whose domain is the
set of all possible outcomes of Alice. The average value of this
random variable can then be given by
\begin{equation}
  \label{eq:EX}
  E\left(X\right)  =  \sum_{i=0}^{d_{A}-1}p_{A}\left(\omega_{i}\right)X\left(\omega_{i}\right).
\end{equation}
Sometimes for simplicity one will write Eq.~(\ref{eq:EX}) as
\begin{equation}
  E\left(X\right)  =  \sum_{x\in X} p(x) x,
\end{equation}
where the sum runs over all possible values in $X$, and indeed
$p(x)=p(X(\omega_i)=x)=p_A(\omega_i)$. Here we assume that the
correspondence between $\omega_i$ and $x$ is one-to-one.

Similarly, a random variable $Y\left(\Lambda\right)$, a real function
whose domain is the set of all possible outcomes of Bob, has the
average value
\begin{equation}
  E\left(Y\right) = \sum_{m=0}^{d_{B}-1}p_{B}\left(\lambda_{m}\right)Y\left(\lambda_{m}\right),
\end{equation}
and for simplicity we can write
\begin{equation}
  \label{eq:EY}
  E\left(Y\right)  =  \sum_{y\in Y} p(y) y,
\end{equation}
where the sum runs over all possible values in $Y$, and indeed
$p(y)=p(Y(\lambda_i)=y)=p_B(\lambda_i)$.

Note that the direct product of random variables $X$ and $Y$ is a
random variable defined on $\Omega\times\Lambda$. Let us write the
joint probability distribution
\begin{equation}
  \label{eq:pxy}
  p(x,y)=p(X(\omega_{i})=x,Y(\lambda_{m})=y)=p_{AB}\left(\omega_{i},\lambda_{m}\right),
\end{equation} 
then the average value of the random variable $X\times Y$ is (denoted
by $E(X,Y)$)
\begin{equation}
  E\left(X,Y\right)=\sum_{y\in Y}\sum_{x\in X}p(x,y)xy.
  \label{eq:EXY}
\end{equation}

As an example, again consider that $\Omega=\{\omega_0,\omega_1\}$.
Choose a random variable variable $X(\omega_0)=0$ and $X(\omega_1)=1$.
A variable like this is called a `bit', i.e.

\begin{svgraybox}
  \begin{center}
    \textbf{Box 1.2 Bit}

    A bit is a random variable with only two possible values $0$ or
    $1$.
  \end{center}
\end{svgraybox}

Bit is an important concept in information theory -- for instance we
know that the capacity of our hard drive is measured in terms of
`Gigabytes', which is $10^9$ bytes, and $1$ byte is actually $8$ bits.
We will soon be clear what `$8$ bits' mean. Let us continue the
discussion of the above example and further consider
$\Lambda=\{\lambda_0,\lambda_1\}$, and another random variable $Y$
which is also a `bit', i.e. $Y(\lambda_0)=0$ and $Y(\lambda_1)=1$.
Then the possible values of $X\times Y$ will be
$\{(0,0),(0,1),(1,0),(1,1)\}$. And when no confusion arises, we can
simply write it as $\{00,01,10,11\}$. Those are all the possible
values of $2$ bits. In general, if we have $N$ bits, then we have the
following:
\begin{svgraybox}
  \begin{center}
    \textbf{Box 1.3 $N$ Bits}

    A possible value of $N$ bits is a binary string of length $N$,
    i.e. $x_Nx_{N-1}\ldots x_1$, where each $x_i$ is a bit, i.e.
    $x_i\in\{0,1\}$. There are total $2^N$ possible values.
  \end{center}
\end{svgraybox}

Now let us come back to the discussion of the correlation between two
random variables $X$ and $Y$. It is naturally captured by the
`correlation function', which is given by
\begin{equation}
  \label{eq:corrfunction}
  C\left(X,Y\right)=E\left(X, Y\right)-E\left(X\right)E\left(Y\right).
\end{equation}
One direct observation is that if the joint distribution
$p_{AB}\left(\omega_{i},\lambda_{m}\right)$ given by
Eq.~\eqref{eq:pxy} has no correlation, then we should have $C(X,Y)=0$.
To see this, we start from (\ref{eq:c0c3}). Inserting (\ref{eq:c0c3})
into (\ref{eq:EXY}), we get $E(X\times Y)=E(X)E(Y)$, i.e. $C(X,Y)=0$.

Indeed, the converse is also true. To show this, note that
$\forall j,n$, we take $X\left(\omega_{i}\right)=\delta_{ij}$ and
$Y\left(\lambda_{m}\right)=\delta_{mn}$. Then
$E\left(X,Y\right)=p_{AB}\left(\omega_{j},\lambda_{n}\right)$,
$E\left(X\right)=p_{A}\left(\omega_{j}\right)$, and
$E\left(Y\right)=p_{B}\left(\lambda_{n}\right)$. Hence
$C\left(X,Y\right)=0$ implies that
$p_{AB}\left(\omega_{j},\lambda_{n}\right)=p_{A}\left(\omega_{j}\right)p_{B}\left(\lambda_{n}\right)$,
i.e. the joint probability distribution $p_{AB}$ does not have
correlation.

We summarize these observations as below.
\begin{svgraybox}
  \begin{center}
    \textbf{Box 1.4 Correlated joint probability distribution}

    A joint probability distribution $p_{AB}$ does not have
    correlation if and only if $C\left(X,Y\right)=0,\ \forall X,Y.$
  \end{center}
\end{svgraybox}

In other words, Eq.(\ref{eq:c0c3}) holds for the joint probability
distribution $p_{AB}$ if and only if for any random variables
$X\left(\Omega\right)$ and $Y\left(\Lambda\right)$, the correlation
function between them vanishes.

This fact clarifies the role of correlation functions in
characterizing and quantifying of correlations. That is, if all the
correlation functions vanish, then indeed no correlation exists.
However, if one correlation function does not vanish, then the joint
probability distribution cannot have the form of Eq.(\ref{eq:c0c3})
hence there must exist some correlation between the outcomes of Alice
and Bob.

As an example, one can take both the distributions given in
Eq.~\eqref{eq:jointeg} and given in Eq.~\eqref{eq:jointeg2}. Choose
the random variables as the bits discussed above,
$X(\omega_0)=0,X(\omega_1)=1$ and $Y(\lambda_0)=0, Y(\lambda_1)=1$,
compute the correlations functions using Eq.~\eqref{eq:corrfunction}.
Then one will get the values $0$ (with no correlation) and $-1/24$
(with correlation), respectively.

\subsection{Mutual information}

We have seen that correlation functions can indeed give some
information of correlation in the system consisting of two subsystems
-- one is the system of Alice and the other is of Bob. However, we
know that a single correlation function, associated with two given
random variables $X\left(\Omega\right)$ and $Y\left(\Lambda\right)$,
is not sufficient to characterize the correlation in the system. One
indeed has to look at all the correlation functions in some sense, or
has to determine which observables are essentially related to the
physics phenomena one cares about.

Interestingly, in the context of information theory established by
Shannon, there is a concept which nicely quantifies the degree of
correlation with operational meaning for information transmission task
between the two subsystems. The concept is called mutual information,
which is defined for two random variables $X\left(\Omega\right)$ and
$Y\left(\Lambda\right)$, given by
\begin{equation}
  \label{eq:mutual}
  I(X{:}Y)=\sum_{y\in Y}\sum_{x\in X}p(x,y)
  \log\left(\frac{p(x,y)}{p(x)p(y)}\right).
\end{equation}

Note that similar to the correlation functions, mutual information is
defined on two random variables $X\left(\Omega\right)$ and
$Y\left(\Lambda\right)$, however unlike the correlation functions, it
does not depend on the choice of $X\left(\Omega\right)$ and
$Y\left(\Lambda\right)$. In other words, what only matters is the
joint probability distribution of $X\times Y$ but not the values of
the variables $X\left(\Omega\right)$ and $Y\left(\Lambda\right)$.
Therefore, for any two random variables $X\left(\Omega\right)$ and
$Y\left(\Lambda\right)$, Eq.(\ref{eq:mutual}) returns a single value.
In this sense, one can also say that the mutual information is
essentially just for the joint probability distribution.

Intuitively, mutual information measures the information that $X$ and
$Y$ share. Or in other words, how correlated they are in a sense that
how much knowing one of these two variables reduces the uncertainty
about knowing the other. For instance, if $X$ and $Y$ are independent,
then knowing $X$ does not give any information about $Y$ and vice
versa, so their mutual information should be zero. This can be seen
from Eq.(\ref{eq:mutual}) given that Eq.(\ref{eq:c0c3}) now holds. On
the other extreme, if $X$ and $Y$ are identical, which is a case of
`perfect correlation', then all information conveyed by $X$ is shared
with $Y$. Or in other words, knowing $X$ determines the value of $Y$
and vice versa. In this case, the mutual information should be the
same as the uncertainty contained in $Y$ (or $X$) alone.

We will need to clarify what it means by `uncertainty contained in $X$
or $Y$.' We know in physics uncertainty is quantified by entropy.
Information theory does borrow the same concept. For any variable
$X\left(\Omega\right)$, Shannon's entropy is given by
\begin{equation}
  H\left(X\right)=-\sum_{x\in X}p(x)\log p(x),
  \label{eq:shen}
\end{equation}
where, just for convenience, $2$ is taken as the base of the log
function. Again, this quantity of entropy only depends on the
probability distribution of $X$, but not the very values of $X$, so
this is essentially the entropy (or uncertainty) of the probability
distribution.

Now back to the case of `perfect correlation,' which means
mathematically
\begin{equation}
  \label{eq:solution}
  p(x,y) = \left\{ 
    \begin{array}{ll}
      0 & \textrm{if $x\neq y$}\\
      p(x)=p(y) & \textrm{otherwise}
    \end{array} 
  \right.
\end{equation}
Eq.~(\ref{eq:mutual}) then becomes
\begin{equation}
  \label{eq:mutX}
  I(X{:}Y)=\sum_{x\in X}p(x)
  \log\left(\frac{p(x)}{p(x)p(x)}\right)=H(X).
\end{equation}
That is, the mutual information should be the same as the uncertainty
contained in $X$ (or $Y$) alone.

\begin{figure}[htbp]
  \centerline{
  \includegraphics[scale=1.00]{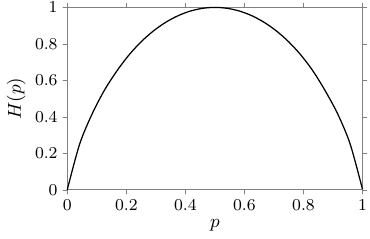}
  }
%
%
  \caption{Binary entropy function $H(p)$}
  \label{fig:binaryentropy} 
\end{figure}

We would like to look at a simple example of Shannon's entropy in case
the random variable $X$ is a bit and the probability distribution is
given by
\begin{equation}
  p(0)=p,\quad p(1)=1-p.
\end{equation}
This gives
\begin{equation}
  H(p)=-p\log p-(1-p)\log (1-p).
\end{equation}
The function $H(p)$ is called the `binary entropy function.' A figure
of this function is shown in Fig.~\ref{fig:binaryentropy}. It vanishes
only for $p=0$ and $p=1$, and reaches the maximum value only at
$p=\frac{1}{2}$ where one has the most `uncertainty': the probability
of getting values $0$ and $1$ is just half and half.

%

In the language of Shannon entropy, for the joint probability
distribution $p(x,y)$ of two random variables, the entropy will be
(denoted by $H(X,Y)$)
\begin{equation}
  H(X,Y)=-\sum_{x\in X}\sum_{y\in Y} p(x,y)\log p(x,y).
\end{equation}

The quantity $H(X|Y=y)$ will then be the entropy of $X$ conditional on
the variable of $Y$ taking the value $y$, i.e.
\begin{equation}
  H(X|Y=y)=-\sum_{x\in X}p(x|y)\log p(x|y),
\end{equation}
where $p(x|y)=p_{A|B}(\omega_i,\lambda_m)$, as given in
Eq.~\eqref{eq:cond}. The conditional entropy $H(X|Y)$ is then given by
\begin{eqnarray}
  H(X|Y)&=&\sum_{y\in Y}p(y)H(X|Y=y)=-\sum_{y\in Y}\sum_{x\in X}p(y)p(x|y)\log p(x|y)\nonumber\\
        &=&-\sum_{x\in X}\sum_{y\in Y} p(x,y)\log\frac{p(x,y)}{p(y)}.
\end{eqnarray}

\begin{figure}[htbp]
  \centerline{
  \includegraphics[width=5cm]{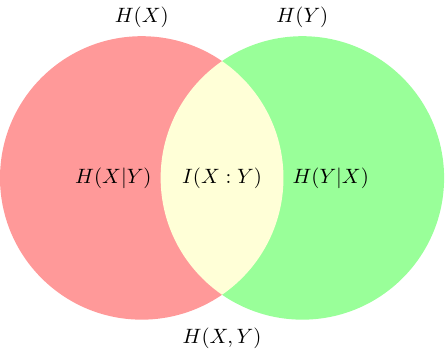}
  }
%
%
  \caption{Mutual information: $H(X)$ and $H(Y)$ are plotted as the
    regions inside two circles, and the mutual information $I(X{:}Y)$
    is just their overlap. The quantities $H(X,Y)$, $H(X|Y)$ and
    $H(Y|X)$ are also illustrated.}
  \label{fig:mutualinformation} 
\end{figure}

In terms of all these quantities, the mutual information can then be
written as
\begin{eqnarray}
  \label{eq:mutual2}
  I(X{:}Y)&=&H(X)+H(Y)-H(X,Y)\nonumber\\
          &=& H(X)-H(X|Y)\nonumber\\
          &=& H(Y)-H(Y|X)\nonumber\\
          &=& H(X,Y)-H(X|Y)-H(Y|X).
              \label{eq:mutent}
\end{eqnarray}
These relationship can be viewed as in
Fig.~\ref{fig:mutualinformation}, which nicely gives intuitively the
meaning of all these quantities.

%

Finally, we summarize the meaning of mutual information below.
\begin{svgraybox}
  \begin{center}
    \textbf{Box 1.5 Mutual information}

    The mutual information $I(X{:}Y)$ given by Eqs.~(\ref{eq:mutual})
    and~(\ref{eq:mutual2}) quantifies the correlation of the joint
    distribution $p(x,y)$.
  \end{center}
\end{svgraybox}

\section{Quantum entanglement}

In this section, we move our discussion of correlation into the
quantum realm. It will soon become clear that there is much more to
expect in the quantum case, due to the superposition principle. Our
discussion will eventually lead to a formal study of the concept of
entanglement.

\subsection{Pure and mixed quantum states}

In quantum mechanics, the state of a quantum system $S$ is represented
by a normalized vector $\vert\psi\rangle$ in the Hilbert space
$\mathcal{H}$. Hence if $\vert\psi_{1}\rangle$ and
$\vert\psi_{2}\rangle$ are two orthogonal quantum states, then any coherent
superposition of the two states
\[
  c_{1}\vert\psi_{1}\rangle+c_{2}\vert\psi_{2}\rangle,
\]
where $c_{1}$ and $c_{2}$ are two complex number satisfying
$\left|c_{1}\right|^{2}+\left|c_{2}\right|^{2}=1$, is also a quantum
state. This obvious property for vectors in a Hilbert space is called
the superposition principle of quantum states in quantum mechanics,
which is a fundamental feature distinguished from classical mechanics.

Let us take a look at the simplest quantum system -- a two-level
system, which could be a spin-$1/2$ particle (here we only care about
the internal states instead of the spatial wavefunction), or a
two-level atom (where all the higher excited states are ignored if
they never enter into the dynamics we care about). The Hilbert space
of the system is then only two-dimensional, with two orthonormal basis
states that we denote as $\ket{0}$ and $\ket{1}$ (which could
represent, for instance, spin up and spin down for the spin-$1/2$
particle, or ground state and the excited state for the two-level
atom).

Any quantum state in this two-dimensional Hilbert space is called
`quantum bit,' or in short `qubit.'
\begin{svgraybox}
  \begin{center}
    \textbf{Box 1.6 Qubit}

    A qubit is a quantum state in a two-dimensional Hilbert space with
    orthonormal basis states $\ket{0}$ and $\ket{1}$, which has the
    form $\ket{\psi}=\alpha\ket{0}+\beta\ket{1}$, where
    $|\alpha|^2+|\beta|^2=1$.
  \end{center}
\end{svgraybox}

Unlikely `bit,' which has only two possible values $0$ and $1$, a
qubit could be in any kind of superposition of the basis states
$\ket{0}$ and $\ket{1}$. This is a direct consequence of the quantum
superposition principle.

\begin{figure}[htbp]
  \centerline{
  \includegraphics[scale=1.05]{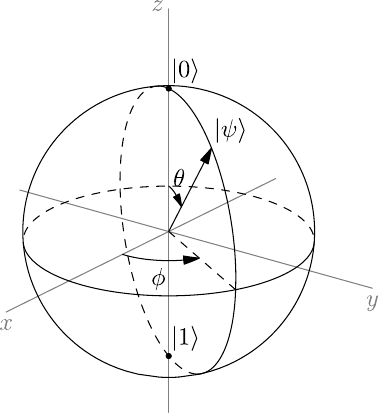}
  }
%
%
  \caption{Bloch sphere.}
  \label{fig:blochsphere} 
\end{figure}

Since $|\alpha|^2+|\beta|^2=1$, we may write $\ket{\psi}$ as
\begin{equation}
  \ket{\psi}=e^{i\gamma}\left(\cos\frac{\theta}{2}\ket{0}+e^{i\phi}\sin\frac{\theta}{2}\ket{1}\right),
\end{equation}
where $\gamma,\theta,\phi$ are real. And by ignoring the overall phase
$e^{i\gamma}$ we can simply write
\begin{equation}
  \ket{\psi}=\cos\frac{\theta}{2}\ket{0}+e^{i\phi}\sin\frac{\theta}{2}\ket{1}.
\end{equation}
This means that $\ket{\psi}$ corresponds to a point on the unit
three-dimensional sphere defined by $\theta$ and $\varphi$, called the
Bloch sphere, as shown in Fig.~\ref{fig:blochsphere}.

%

To understand further about the quantum superposition principle, and
how a qubit could be different from a bit in terms of probability
distribution, let us look at the consequence of quantum measurement.
When a quantum measurement of an observable (i.e. a Hermitian
operator) $M$ is made on the system $S$, we will get one of the
eigenvalues of the operator $M$. We know that $M$ can be written as
\begin{equation}
  M=\sum_i c_i\ket{\phi_i}\bra{\phi_i},
\end{equation}
where each $c_i$ is an eigenvalue of $M$ and $\ket{\phi_i}$ is the
corresponding eigenvector. We know that since $M$ is Hermitian,
$\{\ket{\phi_i}\}$ can always be chosen as an orthonormal basis of the
Hilbert space, that is,
\begin{equation}
  \bra{\phi_i}\phi_j\rangle=\delta_{ij}
\end{equation}
and
\begin{equation}
  \label{eq:identity}
  \sum_i \ket{\phi_i}\bra{\phi_i}=I.
\end{equation}

The probability of getting the value $c_i$ is then
\begin{equation}
  p_i= \bra{\psi}\phi_i\rangle\bra{\phi_i}\psi\rangle,
\end{equation}
and the identity of Eq.\eqref{eq:identity} directly gives
$\sum_{i}p_{i}=1$. That is to say, when a measurement is involved, a
quantum state is associated with a classical probability distribution,
and the correlations discussed in probability theory naturally
generalize to the quantum domain.

Let us look at an example of the qubit case, where
$\ket{\psi}=\alpha\ket{0}+\beta\ket{1}$ is a qubit state. Suppose we
measure an operator whose eigenvectors are $\ket{0}$ and $\ket{1}$,
with eigenvalues $1,-1$, respectively, i.e.
$\ket{0}\bra{0}-\ket{1}\bra{1}$, which is nothing but the Pauli
operator $\sigma_z$. For simplicity we will write it as $Z$ and its
matrix form in the basis $\{\ket{0},\ket{1}\}$ is
\begin{equation}
  \label{eq:PauliZ}
  Z=\sigma_z=\begin{pmatrix} 1 & 0 \\0 & -1 \end{pmatrix}.
\end{equation}
When measuring $Z$, the probabilities $p_0$ of getting $\ket{0}$ and
$p_1$ of getting $\ket{1}$ are
\begin{equation}
  p_0=|\alpha|^2=p,\ p_1=|\beta|^2=1-p,
\end{equation}
respectively.

However, a qubit is indeed different from a bit. To see this, let $W$
be a bit with probabilities of
$p(W=0)=|\alpha|^2, p(W=1)=1-|\alpha|^2$. Let us consider an example
where $\alpha=\beta=\frac{1}{\sqrt{2}}$, so
$p(W=0)=p(W=1)=\frac{1}{2}$. For a corresponding qubit state
$\ket{\psi}=\frac{1}{\sqrt{2}}(\ket{0}+\ket{1})$, measuring the Pauli
operator will return, $\ket{0}$ or $\ket{1}$ with probability
$\frac{1}{2}$. In this sense, the qubit state $\ket{\psi}$ is similar
to the bit $W$.

However, this is more to do for the qubit. Let us write the Pauli
operator $\sigma_x$ as $X$ and $\sigma_y$ as $Y$, i.e.
\begin{equation}
  \label{eq:PauliXY}
  X=\sigma_x=\begin{pmatrix} 0 & 1 \\1 & 0 \end{pmatrix},
  \quad \text{and} \quad
  Y=\sigma_y=\begin{pmatrix} 0 & -i \\i & 0 \end{pmatrix}.
\end{equation}
It is then straightforward to observe that
$\ket{\psi}=\frac{1}{\sqrt{2}}(\ket{0}+\ket{1})$ is an eigenvector of
$X$ with eigenvalue $1$, therefore if we measure $X$, we will get a
definite value $1$. However, when measuring $Y$, we will again get
each eigenvalue of $Y$ of probability half and half.

This example also shows that the probability distribution of a pure
quantum state must be associated with a chosen measurement. In this
sense the chosen measurement is an analog of a random variable in the
classical case. However, it is different from the classical case,
where all the random variables share a single probability
distribution. In the quantum case, if the state happens to be the
eigenstate of the measurement, then the measurement returns a definite
value (i.e. no uncertainty); while if not, there exists some amount of
uncertainty. So it is not consistent to assign a certain value of
uncertainty to a pure quantum state unless the measurement is
specified.

In general, one can further put some probability distribution `on top
of' quantum states, that is, a quantum system may be in the state
$\vert\psi_{i}\rangle$ with probability $p_{i}$, which is represented
by a density operator
\begin{equation}
  \rho=\sum_{i}p_{i}\vert\psi_{i}\rangle\langle\psi_{i}\vert,
  \label{eq:mixstat}
\end{equation}
where $p_i\geq 0$ and $\sum_i p_i=1$.

When the system definitely stays in a state $\vert\psi\rangle$, then
the state is a pure state. Otherwise the state is a mixed state. Note
that any state $\rho$ will satisfy
\begin{enumerate}
\item $\rho$ is Hermitian.
\item $\Tr\rho=\sum_i p_i=1$.
\item $\rho$ is positive (may be written as $\rho\geq 0$), i.e. for
  any $\ket{\psi}$,
  $\bra{\psi}\rho\ket{\psi}=\sum_ip_i|\bra{\psi_i}\psi\rangle|^2\geq
  0$, or all the eigenvalues of $\rho$ are positive. Consequently,
  $\rho$ has a spectral decomposition
  $\rho=\sum_{k}\alpha_{k}\vert\phi_{k}\rangle\langle\phi_{k}\vert$,
  where $\alpha_k\geq 0$ are eigenvalues of $\rho$ and
  $\ket{\phi_{k}}$s are the corresponding eigenvectors which form an
  orthonormal basis.
\item $\Tr\rho^{2}\leq 1$, where the equality is satisfied if and only if
  the state is a pure state.
\end{enumerate}

For a two-dimensional Hilbert space, note that all the three Pauli
operators $X,Y,Z$, together with the identity operator
\begin{equation}
  \label{eq:id}
  I=\begin{pmatrix} 1 & 0 \\0 & 1\end{pmatrix}
\end{equation}
form a basis for $2\times 2$ matrices. Denote
\begin{equation}
  \vec{\sigma}=(\sigma_x,\sigma_y,\sigma_z)=(X,Y,Z),
\end{equation}
then a general quantum state $\rho$ of a qubit can be written as
\begin{equation}
  \rho=\frac{I+\vec{r}\cdot\vec{\sigma}}{2},
\end{equation}  
where $\vec{r}=(r_x,r_y,r_z)$ with $r_x^2+r_y^2+r_z^2\leq 1$.

Now we introduce a measure of uncertainty for a state $\rho$, i.e. the
von Neumman entropy
\begin{equation}
  S\left(\rho\right)=-\Tr\left(\rho\log\rho\right),
  \label{def:vonNeuEnt}
\end{equation}
which is a generalization of the Shannon entropy. This is in a sense
that when writing in its spectral decomposition
$\rho=\sum_{k}\alpha_{k}\vert\phi_{k}\rangle\langle\phi_{k}\vert$, we
have $S\left(\rho\right)=H(\alpha_{k})=-\sum_k\alpha_k\log\alpha_k$.

\subsection{Composite quantum systems, tensor product structure}

Now we consider the case of composite quantum systems. Assume we have
two quantum systems, one for Alice and the other for Bob. We
denote Alice's Hilbert space by $\mathcal{H}_A$, whose dimension is
$d_A$ with an orthonormal basis $\{\ket{i_A}\}: i=0,1,\ldots d_A-1$.
Similarly, we denote Bob's Hilbert space by $\mathcal{H}_B$, whose
dimension is $d_B$ with an orthonormal basis
$\{\ket{m_B}\}: m=0,1,\ldots d_B-1$.

In this case, the basis for the total Hilbert space of both Alice and
Bob will be the Cartesian product of $\{\ket{i_A}\}$ and
$\{\ket{m_B}\}$, i.e. $\{\ket{i_A}\}\times\{\ket{m_B}\}$, which is of
dimension $d_Ad_B$. The corresponding Hilbert space is denoted by
$\mathcal{H}_A\otimes\mathcal{H}_B$, where $\otimes$ is called the
tensor product of two spaces. Therefore, any pure state
$\ket{\psi_{AB}}\in\mathcal{H}_A\otimes\mathcal{H}_B$ can be written
as
\begin{equation}
  \label{eq:2qsystem}
  \ket{\psi_{AB}}=\sum_{im}c_{im}\ket{i_A}\ket{m_B},
\end{equation}
where each term $\ket{i_A}\ket{m_B}$ is sometimes written as
$\ket{i_A}\otimes\ket{m_B}$ to emphasize the tensor product structure
of $\mathcal{H}_A\otimes\mathcal{H}_B$, or sometimes just for
simplicity written as $\ket{i_Am_B}$ or even just $\ket{im}$ if no
confusion arises.

Compared to the case of classical joint probability of two systems,
the difference is that the quantum case deals with a linear space but
the classical case only deals with a certain basis. This is again a
natural consequence of quantum superposition principle. Therefore, any
composite quantum system always has tensor product structure of its
Hilbert space, i.e.
\begin{svgraybox}
  \begin{center}
    \textbf{Box 1.7 Composite quantum system}

    The Hilbert space of a composite quantum system is a
    \textit{tensor product} of the Hilbert spaces of all its
    subsystems.
  \end{center}
\end{svgraybox}

As an example, let us consider the simplest case that both
$\mathcal{H}_A$ and $\mathcal{H}_B$ are two-dimensional, with
orthonormal basis $\{\ket{0_A},\ket{1_A}\}$ and
$\{\ket{0_B},\ket{1_B}\}$, respectively. The basis for the Hilbert
space $\mathcal{H}_A\otimes\mathcal{H}_B$ is then given by
\begin{equation}
  \{\ket{00},\ \ket{01},\ \ket{10},\ \ket{11}\},
\end{equation}
i.e. the basis for $2$ qubits. That is, any two qubit state
$\ket{\psi_{AB}}$ can be written in the form
\begin{equation}
  \label{eq:2qubit}
  \ket{\psi_{AB}}=c_{00}\ket{00}+c_{01}\ket{01}+c_{10}\ket{10}+c_{11}\ket{11}.
\end{equation}
Similarly, we can write a basis for any $n$-qubit state.

\begin{svgraybox}
  \begin{center}
    \textbf{Box 1.8 Computational basis for an $N$-qubit state}

    A basis for an $N$-qubit state are all the $2^N$ binary strings of
    length $n$, i.e. $\ket{x_Nx_{N-1}\ldots x_1}$, where each $x_i$ is
    a bit, i.e. $x_i\in\{0,1\}$. This basis is called the
    `computational basis.'
  \end{center}
\end{svgraybox}

The way to find the quantum state $\rho_B$ for the system $B$ from the
state \ket{\psi_{AB}} given the equation \eqref{eq:2qsystem} is to
`ignore' the subsystem $A$, i.e. to trace (or integrate) over the
subsystem $A$. That is,
\begin{equation}
  \rho_B=\Tr_A \ket{\psi_{AB}}\bra{\psi_{AB}}=\sum_i\langle{i_A}\ket{\psi_{AB}}\bra{\psi_{AB}}{i_A}\rangle
  =\sum_i c_{im}c^{*}_{in}\ket{m_B}\bra{n_B},
\end{equation}
where $c^{*}_{in}$ is the complex conjugate of $c_{in}$. And the
density matrix $\rho_B$ is called the reduced density matrix for the
system $B$.

On the other hand, any density matrix $\rho_{B}$ of the subsystem $B$
can be regarded as reduced state from a pure composite state, i.e. the
system $B$ plus an auxiliary system $A$. That is, for any state
$\rho_{B}$ with spectral decomposition
$\rho_{B}=\sum_i{p_i}\ket{{\phi_{i_B}}}\bra{{\phi_{i_B}}}$, we can
construct a pure state
\begin{equation}
  \vert\psi_{AB}\rangle=\sum_{i}\sqrt{p_{i}}\vert{i_A}\rangle\otimes\vert{\phi_{i_B}}\rangle,\label{eq:puri}
\end{equation}
with $\langle{i_A}\vert{j_A}\rangle=\delta_{ij}$, such that
$\rho_{B}=\Tr\nolimits
_{A}\left(\vert\psi_{AB}\rangle\langle\psi_{AB}\vert\right)$. This
process is called quantum state purification.

Notice that for another orthonormal basis $\{\ket{\varphi_{i_A}}\}$ of
$\mathcal{H}_A$, we can rewrite
\begin{eqnarray*}
  \vert\psi_{AB}\rangle &=&\sum_i\sum_j\ket{\varphi_{j_A}}\bra{\varphi_{j_A}}\sqrt{p_{i}}\vert{i_A}\rangle\otimes\vert{\phi_{i_B}}\rangle\nonumber\\
                        & = & \sum_{j}\sum_{i}\sqrt{p_{i}}\langle{\varphi_{j_A}}\vert{i_A}\rangle\vert{\varphi_{j_A}}\rangle\otimes\vert{\phi_{i_B}}\rangle\\
                        & = & \sum_{j}\sqrt{q_{j}}\vert{\varphi_{j_A}}\rangle\otimes\vert{\xi_{j_B}}\rangle,
\end{eqnarray*}
where
\begin{equation}
  \sum_{i}\langle{\varphi_{j_A}}\vert{i_A}\rangle\sqrt{p_{i}}\vert{\phi_{i_B}}\rangle=\sqrt{q_{j}}\vert{\xi_{j_B}}\rangle.\label{eq:condmix}
\end{equation}
This implies that the state
\[
  \rho_{B}=\Tr\nolimits
  _{A}\left(\vert\psi_{AB}\rangle\langle\psi_{AB}\vert\right)=\sum_{j}q_{j}\ket{\xi_{j_B}}\bra{\xi_{j_B}}.
\]
Therefore any mixed state $\rho_{B}$ can be regarded as the reduced
state of the pure state $\vert\psi_{AB}\rangle$. Different
realizations of the ensemble $\rho_{B}$ correspond to different
measurement bases on the auxiliary system $A$. However, these
different realizations can not be distinguished by any measurements on
the system, in this sense the mixed state $\rho_{B}$ is uniquely
defined.

\subsection{Pure bipartite state, Schmidt decomposition}
\label{sec:Sch}

For a pure bipartite state
\begin{equation}
  \vert\psi_{AB}\rangle=\sum_{i=0}^{d_{A}-1}\sum_{m=0}^{d_{B}-1}c_{im}\vert {i_A}\rangle\vert {m_B}\rangle,
\end{equation}
where $\left\{ \vert i_{A}\rangle\right\} $ and
$\left\{ \vert m_{B}\rangle\right\}$ are orthonormal bases of
$\mathcal{H}_{A}$ and $\mathcal{H}_{B}$ respectively, by choosing
carefully the basis of subsystems $A$ and $B$, one can write the state
$\vert\psi_{AB}\rangle$ in an important standard form, namely, the
Schmidt decomposition.

\begin{svgraybox}
  \begin{center}
    \textbf{Box 1.9 Schmidt decomposition}

    The state $\vert\psi_{AB}\rangle$ can be written in the form
    \begin{equation}
      \vert\psi_{AB}\rangle=\sum_{i=1}^{n_{s}}\sqrt{\lambda_i}\vert{\varphi_{i_A}}\rangle\vert{\phi_{i_B}}\rangle,
    \end{equation}
    where $\lambda_{i}>0$, $\sum_{i}\lambda_i=1$,
    $n_{s}\le\min\left\{ d_{A},d_{B}\right\}$, and
    $\langle{\varphi_{i_A}}\vert{\varphi_{j_A}}\rangle=\langle{\phi_{j_B}}\vert{\phi_{i_B}}\rangle=\delta_{ij}$.
  \end{center}
\end{svgraybox}

Let us show why this works. For the state $\ket{\psi_{AB}}$, we get
the reduced density matrix $\rho_{A}$ of particle $A$, and assume its
spectral decomposition is
$\rho_{A}=\sum_{i=1}^{n_{s}}\lambda_{i}\vert{\varphi_{i_A}}\rangle\langle{\varphi_{i_A}}\vert$
with $\lambda_{i}>0$, $\sum_{i}\lambda_{i}=1$, and
$\langle{\varphi_{i_A}}\vert{\varphi_{j_A}}\rangle=\delta_{ij}$. Now
the state $\ket{\psi_{AB}}$ can be written as
$\vert\psi_{AB}\rangle=\sum_{i=1}^{n_{s}}c_{i}\vert{\varphi_{i_A}}\rangle\vert{\phi_{i_B}}\rangle$,
where $\ket{\phi_{i_{B}}}$ is a normalized vector, and $c_{i}$ is the
corresponding coefficient. Note that we can always take $c_{i}\ge 0$
by choosing the phase factor of $\ket{\phi_{i_{B}}}$. The reduced
state for particle $A$ is then
\begin{equation}
  \rho_{A}=\sum_{ij}\vert{\varphi_{i_A}}\rangle c_{i} c_{j}^{\ast}
  \langle{\phi_{j_B}}\vert{\phi_{i_B}}\rangle\langle{\varphi_{j_A}}\vert. 
\end{equation}
Comparing the above equation with the spectral decomposition of
$\rho_{A}$, we get
$\langle{\phi_{j_B}}\vert{\phi_{i_B}}\rangle=\delta_{ij}$ and
$c_{i}=\sqrt{\lambda_{i}}$. Obviously, $n_{s}\le\min\{d_{A},d_{B}\}$.
This completes our proof.

The Schmidt decomposition plays a key role in characterization of
correlations in a pure bipartite quantum state. The coefficients
$\left\{ \lambda_{i}\right\} $ are called Schmidt coefficients, and
the basis $\ket{\varphi_{i_A}}$ and $\ket{\phi_{i_B}}$ are called
Schmidt basis.

When a joint projective measurement
$\left\{ P_{i_Am_B}=\vert \varphi_{i_{A}}\rangle\langle
  \varphi_{i_{A}}\vert\otimes\vert \phi_{m_{B}}\rangle\langle
  \phi_{m_{B}}\vert\right\} $ is performed, then we get a joint
probability distribution
\begin{equation}
  p_{AB}\left(i,m\right)=\langle\psi_{AB}\vert P_{i_Am_B}\vert\psi_{AB}\rangle.
\end{equation}
Because there are many different choices of projective measurements, a
single bipartite state $\vert\psi_{AB}\rangle$ corresponds to infinite
many numbers of joint probability distributions. And there is no
correlation in the state $\vert\psi_{AB}\rangle$ if none of the joint
probability distributions has any correlation. We summarize this
observation below.
\begin{svgraybox}
  \begin{center}
    \textbf{Box 1.10 Pure state correlation and projective
      measurement}

    A state $\vert\psi_{AB}\rangle$ has no correlation if $\forall$
    projective measurement $P_{i_Am_B}$, the joint probability
    distribution $p_{AB}\left(i,m\right)$ has no correlation.
  \end{center}
\end{svgraybox}
In other words, no correlation can be retrieved from the joint system
by any kind of projective measurement.

Recall that each Hermitian operator corresponds to a random variable
in classical probability theory. Then for two observables $O_{A}$
acting locally on the subsystem $A$, and $O_{B}$ acting locally on
subsystem $B$, the correlation function is given by
\begin{equation}
  C\left(O_{A},O_{B}\right)=\langle O_{A}\otimes O_{B}\rangle-\langle O_{A}\otimes I_{B}\rangle\langle I_{A}\otimes O_{B}\rangle,\label{def:corfun}
\end{equation}
where
$\langle\cdot\rangle=\langle\psi_{AB}\vert\cdot\vert\psi_{AB}\rangle$
is the average value of some observable.

We now ready to state the conditions under which a bipartite pure
state is without correlation.
\begin{svgraybox}
  \begin{center}
    \textbf{Box 1.11 Bipartite pure state without correlation}

    A state $\vert\psi_{AB}\rangle$ has no correlations if and only if

    1.
    $\vert\psi_{AB}\rangle=\vert\psi_{A}\rangle\otimes\vert\psi_{B}\rangle$,
    or

    2. $\forall O_{A},O_{B}$, $C\left(O_{A},O_{B}\right)=0$.
  \end{center}
\end{svgraybox}

Note that a pure state of the form
$\vert\psi_{A}\rangle\otimes\vert\psi_{B}\rangle$ is called a product
state. We now show the necessary and sufficient condition $1$. For the
`if' part: if
$\vert\psi_{AB}\rangle=\vert\psi_{A}\rangle\otimes\vert\psi_{B}\rangle$,
then for $\forall\vert i_{A}\rangle,\vert m_{B}\rangle$,
$p_{AB}\left(i,m\right)=\left|\langle\psi_{A}\vert
  i_{A}\rangle\right|^{2}\left|\langle\psi_{B}\vert
  m_{B}\rangle\right|^{2}=p_{A}\left(i\right)p_{B}\left(m\right)$. For
the `only if' part: if the basis of the projective measurement is
chosen as the Schmidt basis of $\vert\psi_{AB}\rangle$, then we get
$p_{AB}\left(i,j\right)=\lambda_{i}\delta_{ij}$. Hence the condition
for $\vert\psi_{AB}\rangle$ to have no correlations is $n_{s}=1$,
i.e., it is a product state.

We then further show the equivalence of the conditions $1$ and $2$.
The part from $1$ to $2$ is straightforward. For the part from $2$ to
$1$: we start from the Schmidt decomposition
$\vert\psi_{AB}\rangle=\sum_{i}\sqrt{\lambda_{i}}\vert{\varphi_{i_A}}\rangle\otimes\vert\phi_{i_B}\rangle$.
We take $M_{i_A}=\vert\varphi_{i_A}\rangle\langle\varphi_{i_A}\vert$
and $N_{m_B}=\vert\phi_{m_B}\rangle\langle\phi_{m_B}\vert$. Then
$C\left(M_{i_A},N_{m_B}\right)=\delta_{im}\lambda_{i}-\lambda_{i}\lambda_{m}=0$,
which implies that $\lambda_{m}=\delta_{im}$. Therefore there exists
some $m$ such that $\lambda_{m}=1$, so $\vert\psi_{AB}\rangle$ is a
product state.

\subsection{Mixed bipartite state}

We consider the correlations in a mixed bipartite state $\rho_{AB}$.
Now the average value of observable $O_{AB}$ is defined as
$\langle O_{AB}\rangle=\Tr_{AB}\left(O_{AB}\rho_{AB}\right)$. For
example,
\[
  p_{AB}\left(i,m\right)=\Tr\nolimits
  _{AB}\left(P_{i_Am_B}\rho_{AB}\right).
\]
Similar to the pure state case, we have the following observation for
a bipartite state to have no correlation.
\begin{svgraybox}
  \begin{center}
    \textbf{Box 1.12 Correlation and projective measurement}

    A state $\rho_{AB}$ has no correlation if $\forall$ projective
    measurement $P_{i_Am_B}$, the joint probability distribution
    $p_{AB}\left(i,m\right)$ has no correlations.
  \end{center}
\end{svgraybox}

Again, similar as the pure state case, the conditions under which a
bipartite state $\rho_{AB}$ has no correlation can be given by

\begin{svgraybox}
  \begin{center}
    \textbf{Box 1.13 Bipartite state without correlation}

    A state $\rho_{AB}$ has no correlations if and only if

    1. $\rho_{AB}=\rho_{A}\otimes\rho_{B}$, or

    2. $\forall O_{A},O_{B}$, $C\left(O_{A},O_{B}\right)=0$
  \end{center}
\end{svgraybox}

Let us first show the condition $2$. For the `if' part:
$\forall\vert \varphi_{i_{A}}\rangle,\vert \phi_{m_{B}}\rangle$, we
take two types of operators:
$O_{A}=\sum_{i}x_{i}\vert \varphi_{i_{A}}\rangle\langle
\varphi_{i_{A}}\vert$ and
$O_{B}=\sum_{m}y_{m}\vert \phi_{m_{B}}\rangle\langle
\phi_{m_{B}}\vert$. Then we have
\[
  C\left(O_{A},O_{B}\right)=\sum_{im}x_{i}y_{m}\left(p_{AB}\left(i,m\right)-p_{A}\left(i\right)p_{B}\left(m\right)\right)=0.
\]
Since $x_{i}$ and $y_{m}$ can take arbitrary values, we obtain
$p_{AB}\left(i,m\right)=p_{A}\left(i\right)p_{B}\left(m\right)$. The
`only if' part can be shown in a similar way.

The equivalence of conditions $1$ and $2$ can be shown by noticing
that
\[
  \Tr\nolimits _{AB}\left(O_{A}\otimes
    O_{B}\left(\rho_{AB}-\rho_{A}\otimes\rho_{B}\right)\right)=C\left(O_{A},O_{B}\right).
\]

Similar as the case of classical joint probability, we have the
concept of quantum mutual information which measures the total amount
of correlation between $A$ and $B$.
\begin{svgraybox}
  \begin{center}
    \textbf{Box 1.14 Quantum mutual information}

    The correlation in a bipartite state $\rho_{AB}$ is measured by
    the quantum mutual information:
    \[
      I(A{:}B)=S_A+S_B-S_{AB},
    \]
  \end{center}
\end{svgraybox}
Here for simplicity we write $S_A$ for $S\left(\rho_{A}\right)$, $S_B$
for $S\left(\rho_{B}\right)$, and $S_{AB}$ for
$S\left(\rho_{AB}\right)$.

If $\rho_{AB}$ is a pure state, then $S_{AB}=0$ and
$S_A=S_B=H\left(\left\{ \lambda_{i}\right\} \right)$, where
$\lambda_{i}$ is the Schmidt coefficients of the state. Hence
$I(A{:}B)=2H\left(\left\{ \lambda_{i}\right\} \right).$

\subsection{Bell's inequalities}

When correlation exists in a bipartite system, there must be two local
measurements on the two parties respectively, with dependent
measurement results. This is the case for both classical and quantum
bipartite systems. This similarity in classical and quantum
correlations naturally raises the following question: is there any
feature of correlation in a quantum state that is distinct from that
in a classical probability distribution? The Bell's inequalities give
an affirmative answer to this question. Here we look at one of those
inequalities, called the CHSH (Clauser-Horne-Shimony-Holt) inequality.

Let us consider a bipartite system with subsystems $A$ and $B$. Let
$a$, $c$ be local dichotomic variables of $A$, and $b$, $d$ be local
dichotomic variables of $B$. Here a dichotomic variable is a random
variable that takes one of the two possible values $\pm 1$.

Note that
\begin{equation}
  v(a)v(b)+v(a)v(d)+v(c)v(b)-v(c)v(d)=\pm 2,
\end{equation}
where $v(x)$ is the value of the dichotomic variable $x$, which could
be $\pm 1$.

We now obtain the CHSH inequality
\begin{equation}
  \label{eq:CHSHc}
  \vert \langle a b\rangle + \langle a d\rangle +\langle c b\rangle
  -\langle c d\rangle \vert \le 2.
\end{equation} 
 
This CHSH inequality is indeed built on the hidden variable
assumption. That is, if some hidden variable $\lambda$ is given, the
values of the dichotomic variables $a$, $b$, $c$, and $d$ are
specified.

In the quantum situation, a system with dichotomic variables may
correspond to a qubit, and the above bipartite system can correspond
to a two-qubit system. A dichotomic variable will map to the component
of the Pauli operator along a space direction, e.g., $a$ corresponds
to $\vec{\sigma}\cdot \vec{n}_{a}$. Here $\vec{n}_{a}$ is a unit
vector on the Bloch sphere.

Thus the quantum version of the CHSH inequality becomes
\begin{eqnarray}
  \label{eq:CHSH}
  &&\left\vert \langle \vec{\sigma}_{A} \cdot \vec{n}_{a} \vec{\sigma}_{B} \cdot \vec{n}_{b} \rangle +
     \langle \vec{\sigma}_{A} \cdot \vec{n}_{a} \vec{\sigma}_{B} \cdot \vec{n}_{d} \rangle + \langle
     \vec{\sigma}_{A} \cdot \vec{n}_{c} \vec{\sigma}_{B} \cdot
     \vec{n}_{b} \rangle - \langle \vec{\sigma}_{A} \cdot \vec{n}_{c}
     \vec{\sigma}_{B} \cdot \vec{n}_{d} \rangle \right\vert\nonumber\\
  &\le&\left\vert \langle \vec{\sigma}_{A} \cdot \vec{n}_{a}
        \vec{\sigma}_{B} \cdot (\vec{n}_{b} + \vec{n}_{d}) \rangle
        \right\vert + \left\vert \langle \vec{\sigma}_{A} \cdot \vec{n}_{c}
        \vec{\sigma}_{B} \cdot (\vec{n}_{b} - \vec{n}_{d}) \rangle
        \right\vert\nonumber\\
  &\le& \left\vert \vec{n}_{b} + \vec{n}_{d}\right\vert + \left\vert
        \vec{n}_{b} - \vec{n}_{d} \right\vert\nonumber\\
  &\le& \sqrt{2(\left\vert \vec{n}_{b} + \vec{n}_{d}\right\vert^2 + \left\vert
        \vec{n}_{b} - \vec{n}_{d} \right\vert^2)}=2\sqrt{2}.
\end{eqnarray}   
In fact, the maximum value $2\sqrt{2}$ can be reached for a singlet
state
\begin{equation}
  \ket{\psi_{AB}}=\frac{1}{\sqrt{2}}(\ket{01}-\ket{10}),
\end{equation} 
for some $\vec{n}_{a}$, $\vec{n}_{b}$, $\vec{n}_{c}$, and
$\vec{n}_{d}$ which are in the same plane with
$\vec{n}_{a} \perp \vec{n}_{b}$, $\vec{n}_{c} \perp \vec{n}_{d}$, and
$\vec{n}_{a} \perp (\vec{n}_{b} -\vec{n}_{d})$. For instance, one
choice could be that $\vec{n}_{a}$, $\vec{n}_{b}$, $\vec{n}_{c}$, and
$\vec{n}_{d}$ are in the $x-z$ plane of the Bloch sphere with angles
to the $z$ axis as $\theta_a=0$, $\theta_b=\pi/2$, $\theta_c=\pi/4$
and $\theta_d=-\pi/4$.

Comparing the CHSH inequality Eq.~\eqref{eq:CHSHc} with its quantum
version Eq.~\eqref{eq:CHSH}, we then conclude that the quantum
correlation is stronger than its classical counterpart.

\subsection{Entanglement}
\label{sec:entanglement}

Bell inequality implies that a pure bipartite quantum state can have
correlations beyond its classical counterpart. This special type of
correlation is called entanglement. Entanglement is originated from
the superposition principle of quantum states. It is the key resource
for quantum information processing.

For a pure bipartite state $\ket{\psi_{AB}}$, it is natural to state
the following.

\begin{svgraybox}
  \begin{center}
    \textbf{Box 1.15 Bipartite product state}

    A pure bipartite state $\ket{\psi_{AB}}$ is a product state if it
    can be written as $\ket{\psi_{A}}\otimes\ket{\psi_{B}}$ for some
    $\ket{\psi_{A}}\in\mathcal{H}_A$ and
    $\ket{\psi_{B}}\in\mathcal{H}_B$, otherwise it is entangled.
  \end{center}
\end{svgraybox}

Traditionally, entanglement for a pure bipartite state
$\ket{\psi_{AB}}$ is measured by the von Neumann entropy of its the
subsystem.

\begin{svgraybox}
  \begin{center}
    \textbf{Box 1.16 von Neumann entropy as entanglement measure}

    The entanglement for a pure bipartite state $\ket{\psi_{AB}}$ is
    given by the von Neumann entropy of the subsystem state $\rho_A$
    or $\rho_B$:
$$
S(\ket{\psi_{AB}})=-\Tr\rho_A\log\rho_A=-\Tr\rho_B\log\rho_B.
$$
\end{center}
\end{svgraybox}

In terms of the Schmidt coefficients $\{\lambda_i\}$ of the state
$\ket{\psi_{AB}}$, we have $E(\ket{\psi_{AB}})=H(\{\lambda_i\})$.
Therefore, for any bipartite pure state
$\rho_{AB}=\ket{\psi_{AB}}\bra{\psi_{AB}}$, its mutual information
$I(A{:}B)=2H(\{\lambda_i\})$ is twice its entanglement. Since the
mutual information measures the total correlation, this in some sense
means that for $\rho_{AB}$, half of the correlation is `quantum' and
the other half is `classical.'

To see what this might possibly mean, let us write the mutual
information for any bipartite state $\rho_{AB}$ as the following.
\begin{eqnarray}
  \label{eq:qmutual}
  I(A{:}B)&=&S_A+S_B-S_{AB}\nonumber\\
          &=&S_A-S_{A|B}\nonumber\\
          &=&S_B-S_{B|A}\nonumber\\
          &=&S_{AB}-S_{A|B}-S_{B|A}.
\end{eqnarray}
Here the quantum conditional entropy is given by
\begin{eqnarray}
  S_{A|B}&=&S_{AB}-S_B\nonumber\\
  S_{B|A}&=&S_{AB}-S_A\nonumber.
\end{eqnarray}

This looks very similar as Eq.~\eqref{eq:mutual2} for the classical
case. Or in other words, the picture given by
Fig.~\ref{fig:mutualinformation} is in some sense still valid for the
quantum case. We need to emphasize that there is an essential
difference though: in the classical case, the conditional entropy can
never be negative, but the quantum conditional entropy could be.

Any entangled pure state $\ket{\psi_{AB}}$ is an example, where
$S_{AB}=0$ and $S_B=S_A>0$, therefore $S_{A|B}=S_{B|A}<0$. This sounds
strange that how such a `partial information' could be negative, which
could mean that the more you know then the less you know. This puzzle
was solved by quantum information scientists to associate this
negative quantity with future potential to transmit quantum
information.

It is beyond the scope of the book to go into detail of this
operational meaning for quantum mutual information, which would
explain that a pure state contains both classical and quantum
correlation when used for quantum information transmission. Because
there is only a constant factor `2' between its total correlation and
entanglement, in most cases it is both qualitatively and quantitatively
fine that we simply say `all the correlation' in a bipartite pure
state is just `quantum,' i.e. entanglement, at least for the
discussion of this book. Therefore in the rest of the book, we will
simply use the word `entanglement' when talking about correlation in
bipartite pure state.

Indeed `almost all' bipartite pure states are entangled, in a sense
that a very small amount of states can be written into the form of
$\ket{\psi_{AB}}=\ket{\psi_A}\otimes\ket{\psi_B}$. Take the two-qubit
case as an example. A general state can be written as
\begin{equation}
  \ket{\psi}=c_{00}\ket{00}+c_{01}\ket{01}+c_{10}\ket{10}+c_{11}\ket{11},
\end{equation}
where $|c_{00}|^2+|c_{01}|^2+|c_{00}|^2+|c_{01}|^2=1$.

If $\ket{\psi}$ can be written as $\ket{\psi_A}\otimes\ket{\psi_B}$,
then we will have $\ket{\psi_A}=a_0\ket{0}+a_1\ket{1}$ and
$\ket{\psi_B}=b_0\ket{0}+b_1\ket{1}$, then
\begin{equation}
  \ket{\psi}=\ket{\psi_A}\otimes\ket{\psi_B}=a_{0}b_{0}\ket{00}+a_{0}b_1\ket{01}+a_1b_0\ket{10}+a_1b_1\ket{11},
\end{equation}
which means one must have
\begin{equation}
  c_{00}=a_0b_0,\ c_{01}=a_0b_1,\ c_{10}=a_1b_0,\ c_{11}=a_1b_1. 
\end{equation}
However, this is cannot be true in general for $c_{ij}$ satisfying
only $|c_{00}|^2+|c_{01}|^2+|c_{10}|^2+|c_{11}|^2=1$.

One interesting consequence of entanglement is that an unknown quantum
state cannot be `copied'. In other words, there does not exist an
apparatus $\mathcal{A}$ which realizes the following mapping:
\begin{equation}
  \mathcal{A}: \left(\alpha\ket{0}+\beta\ket{1}\right)\otimes \ket{0}\rightarrow \ (\alpha\ket{0}+\beta\ket{1})\otimes(\alpha\ket{0}+\beta\ket{1})
\end{equation}
for any $\alpha,\beta$. To see why this is case, we know that quantum
mechanics is linear. Therefore, if $\mathcal{A}$ can copy the basis
states $\ket{0}$ and $\ket{1}$, i.e.
\begin{equation}
  \mathcal{A}: \ket{00}\rightarrow \ \ket{00},\ \ket{10}\rightarrow \ \ket{11}, 
\end{equation}
then it must map $\left(\alpha\ket{0}+\beta\ket{1}\right)\otimes \ket{0}$ to
$\alpha\ket{00}+\beta\ket{11}$, however we know that
$\alpha\ket{00}+\beta\ket{11}$ is entangled and
\begin{equation}
  \alpha\ket{00}+\beta\ket{11}\neq (\alpha\ket{0}+\beta\ket{1})\otimes(\alpha\ket{0}+\beta\ket{1})
\end{equation}
in general.

This then leads to an important fact which is called the `no-cloning
theorem'.

\begin{svgraybox}
  \begin{center}
    \textbf{Box 1.17 The no-cloning theorem}

    An unknown quantum state cannot be cloned.
  \end{center}
\end{svgraybox}

Having said that `all the correlation in a bipartite pure state is just
quantum', a mixed bipartite state $\rho_{AB}$, however, should contain
both classical correlation and quantum correlation (entanglement).
Here classical correlation means the correlation with origin of
classical probability, i.e. from mixing pure bipartite states. If
initially there is no entanglement in all these pure bipartite states,
then a mixing of them should only result in classical correlation, but
no quantum entanglement. Therefore, a widely-used definition of
entanglement for a general bipartite state is then given as follows.

\begin{svgraybox}
  \begin{center}
    \textbf{Box 1.18 Separable states}

    A state $\rho_{AB}$ is separate if and only if it can be written
    in the form:
$$
\rho_{AB}=\sum_{i}p_{i}\vert\varphi_{i_A}\rangle\langle\varphi_{i_A}\vert\otimes\vert\phi_{i_B}\rangle\langle\phi_{i_B}\vert
$$
Otherwise, it is called entangled.
\end{center}
\end{svgraybox}

The degree of entanglement is a more subtle problem, which has
different definitions in different contexts, and often is very hard to
calculate. As an example, the entanglement of formation is defined as
\[
  E_{F}\left(\rho_{AB}\right)=\min_{\sum_{i}p_{i}\vert\psi_{i_{AB}}\rangle\langle\psi_{i_{AB}}\vert=\rho_{AB}}
  \sum_{i}p_{i}E\left(\vert\psi_{i_{AB}}\rangle\right).
\]
According to the definition, it is obvious that the entanglement of
formation for a state $\rho_{AB}$ is zero if and only if it is
separable.

\section{Correlation and entanglement in many-body quantum systems}

Now let us move on to discuss many-body quantum systems. We start from
a simplest case where there are only three systems $A,B,C$, i.e. the
Hilbert space is now the tensor product of the three systems,
$\mathcal{H}=\mathcal{H}_A\otimes\mathcal{H}_B\otimes\mathcal{H}_C$.
We first start to examine a paradox which shows many-body quantum
correlation is stronger than the classical correlation.

\subsection{The GHZ paradox}

\label{sec:ghz-state-bell}

To further demonstrate the essential differences between classical
correlation and quantum correlation, Greenberg, Horne, and Zeilinger
find that there exists remarkable correlations in the following state
\begin{equation}
  \label{eq:1}
  \vert GHZ \rangle = \frac {1} {\sqrt{2}} \left( \vert 000 \rangle + \vert
    111 \rangle \right).
\end{equation}
Hence the above state is called the GHZ state.

The correlation in the GHZ state can be described as follows. First,
let us observe that it is the unique common eigenstate with
eigenvalues being $1$ of the following observables:
\begin{equation}
  \label{eq:2}
  \{Z_A\otimes Z_B\otimes I_C, I_A \otimes Z_B\otimes Z_C, X_A\otimes X_B\otimes X_C\}.
\end{equation}

Then we use the above set of observables as the generator to generate
the following group:
\begin{eqnarray}
  \label{eq:3}
  \{ I_A\otimes I_B\otimes I_C, Z_A\otimes Z_B\otimes I_C, I_A \otimes Z_B\otimes Z_C, Z_A\otimes I_B\otimes Z_C,\nonumber\\
  X_A\otimes X_B\otimes X_C, - Y_A\otimes Y_B\otimes X_C, - Y_A\otimes
  X_B\otimes Y_C, - X_A\otimes Y_B\otimes Y_C\}.
\end{eqnarray}
Obviously, the GHZ state is also the eigenstate with eigenvalue being
$1$ for all the observables in the group.

Now let us return to the classical world. If we take a measurement of
a Pauli operator $\Lambda$ with $\Lambda\in\{X,Y,Z\}$, we always get
its value, $1$ or $-1$. Then the value of a Pauli matrix $v(\Lambda)$
can take a value $1$ or $-1$. The quantum theory, in the viewpoint of
classical world, implies that
\begin{eqnarray}
  \label{eq:5}
  v(X_A) v(X_B) v(X_C) & = & 1,\\
  -v(Y_A) v(Y_B) v(X_C) & = & 1,\\
  -v(Y_A) v(X_B) v(Y_C) & = & 1,\\
  -v(X_A) v(Y_B) v(Y_C) & = & 1.
\end{eqnarray}
However, this is impossible because the product of the above four
equations leads to $-1=1$. This reflects that it is no longer true in
the quantum world that there always exists a value for a local
observable, as in the classical world. Therefore, the correlation in
the GHZ state can not be simulated by any classical theory.

This GHZ paradox can be viewed as a many-body analogy of the Bell's
inequalities, which show quantum correlation is stronger than
classical correlation. Note that GHZ paradox has an even simpler form
than the Bell's inequalities. This indicates that in the many-body
case, quantum system will more easily behave in a nonclassical manner.

\subsection{Many-body correlation}
\label{sec:irr}

Similar to the bipartite case, we can discuss the correlation for a
tripartite state $\rho_{ABC}$

\begin{svgraybox}
  \begin{center}
    \textbf{Box 1.19 Tripartite states without correlation}

    A state $\rho_{ABC}$ acting on $\mathcal{H}$ has no correlations
    if and only if it can be written as
    $\rho_{ABC}=\rho_{A}\otimes\rho_{B}\otimes\rho_{C}$.
  \end{center}
\end{svgraybox}

Naturally, the degree of the total correlation in a state $\rho_{ABC}$
equals the generalized mutual information of the state, i.e.,
\begin{equation}
  C_{T}(\rho_{ABC})=\mathcal{I}(\rho_{ABC})=S_A+S_B+S_C-S_{ABC}.
\end{equation}

\begin{figure}[htb]
  \centerline{
  \includegraphics[width=6cm]{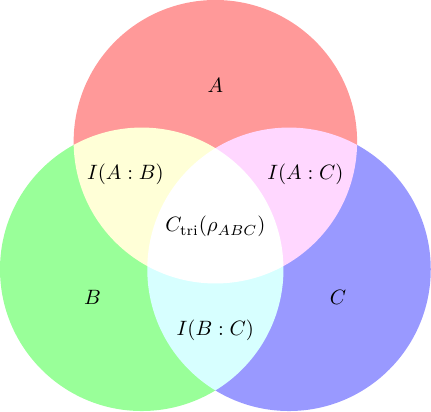}
  }
%
%
  \caption{An inituitive distribution of correlations in a tripartite
    quantum state $\rho_{ABC}$. The true three-body correlation is the
    overlap of $A,B,C$, and the two-body correlations between $A$ and
    $B$, between $A$ and $C$, between $B$ and $C$ are represented by
    the mutual information $I(A:B)$, $I(A:C)$, $I(B:C)$ respectively.}
  \label{fig:three} 
\end{figure}

In general, this total correlation $C_{T}(\rho_{(ABC})$ must contain
both bipartite correlation and tripartite correlation. And we also
know that the bipartite correlations are given by the quantum mutual
information $I(A{:}B),\ I(B{:}C),\ I(A{:}C)$ respectively. So one
simple guess will be that the true tripartite correlation
$C_{tri}(\rho_{ABC})$ should be given by
\begin{eqnarray}
  \label{eq:tri}
  C_{tri}(\rho_{ABC})&=&C_{T}(\rho_{ABC})-(I(A{:}B)+I(B{:}C)+I(A{:}C))\nonumber\\
                     &=&S_{AB}+S_{AC}+S_{BC}-S_{A}-S_{B}-S_{C}-S_{ABC},
\end{eqnarray}
which can be viewed in the graphical manner as illustrated in
Fig.~\ref{fig:three}.

%

Unfortunately, this does not work as $C_{tri}(\rho_{ABC})$ could
sometimes be negative. Indeed, this is even the case of classical
joint probability distribution. As an example, consider a three-qubit
system with the quantum state
\begin{equation}
  \rho^c_{ABC}=\frac{1}{2}(\vert{000}\rangle\langle{000}\vert+\vert{111}\rangle\langle{111}\vert).
\end{equation}
We note that $C_{T}(\rho^c_{ABC})=2$, but
$I(A{:}B)=I(B{:}C)=I(A{:}C)=1$.

To solve the above paradox, for the tripartite state $\rho_{ABC}$, let
us write its one-particle reduced density matrices ($1$-RDMs) as
$\{\rho_A,\rho_B,\rho_C\}$, and two particle reduced density
matrices ($2$-RDMs) as $\{\rho_{AB},\rho_{BC},\rho_{AC}\}$. Now define
\begin{eqnarray}
  \mathbb{L}_{1}& = & \{\sigma_{ABC}|\sigma_{A}=\rho_{A},\sigma_{B}=\rho_{B},\sigma_{C}=\rho_{C}\},\\
  \mathbb{L}_{2}& = &
                      \{\sigma_{ABC}|\sigma_{AB}=\rho_{AB},\sigma_{BC}=\rho_{BC},\sigma_{CA}=\rho_{CA}\},\\
  \mathbb{L}_{3}& = & \{\sigma_{ABC}|\sigma_{ABC}=\rho_{ABC}\}.
\end{eqnarray}
That is, $\mathbb{L}_{k}$ is the set of all tripartite states
$\sigma_{ABC}$ that has the same $k$-RDMs as those of $\rho_{ABC}$.

The idea of $\mathbb{L}_{k}$ can naturally be generalized to the case
of the $N$-particle case. That is, given an $n$-particle state $\rho$,
we have
\begin{eqnarray}
  \mathbb{L}_{k}=\{\sigma|\sigma\ \text{has the same}\ k\text{-RDMs}\ \text{as those of}\ \rho\},
\end{eqnarray}
for $k=1,\ldots,N$.

Now the question is which state in $\mathbb{L}_{k}$ is the best
inference of the global state $\rho$ given only the information of
$k$-RDMs. This is given by the principle of maximum entropy.

\begin{svgraybox}
  \begin{center}
    \textbf{Box 1.20 The principle of maximum entropy}

    For a given set of $k$-RDMs, the best inference of the
    $N$-particle state is the one with the maximum von Neumann entropy
    in $\mathbb{L}_{k}$ .
  \end{center}
\end{svgraybox}

In fact, this state with maximum entropy is unique and let us denote
it by $\rho_k^*$. What the principle of maximum entropy says is that
$\rho_k^*$ contains all the information that is contained in the
$k$-RDMs, but no more. In this sense, the more information we get by
knowing $\rho$ compared to knowing only its $k$-RDMs is given by the
decrease of the uncertainty of our knowledge of the state $\rho$,
i.e.,
\begin{equation}
  \Delta S=S({\rho}_k^*)-S(\rho).
\end{equation}
Similarly the more information we get by knowing the $k$-RDMs compared
to knowing only its $(k-1)$-RDMs is given by
\begin{equation}
  C_k=S(\rho^*_{k-1})-S({\rho}_k^*),
\end{equation}
where $C_k$ measures the degree of $k$-particle correlation that
cannot be learned from the information in $(k-1)$-RDMs. In this sense,
we will call $C_k$ the irreducible $k$-particle correlation.

Now let us come back to the problem of decomposing the correlations in
a tripartite quantum state into bipartite correlation and tripartite
correlation. We can now solve our paradox raised in Eq.~\eqref{eq:tri}
by using the concept of the irreducible $k$-particle correlations.
That is, for a three-particle state $\rho_{ABC}$, the state with
maximum entropy in $\mathbb{L}_{k}$ for $k=1,2,3$ are
\begin{eqnarray}
  {\rho}^*_{1}& = & \argmax\{S(\sigma): \sigma \in \mathbb{L}_{1}(\rho_{ABC})\},\\
  {\rho}^*_{2} & = &
                     \argmax\{S(\sigma):\sigma\in\mathbb{L}_{2}(\rho_{ABC})\},\\
  {\rho}^*_{3}& = &
                    \argmax\{S(\sigma):\sigma\in\mathbb{L}_{3}(\rho_{ABC})\}.
\end{eqnarray}
In fact, it is easy to prove that
${\rho}^*_{1}=\rho_{A}\otimes\rho_{B}\otimes\rho_{C}$ and
${\rho}^*_{3}=\rho_{ABC}$. The total correlation in the tripartite
state $\rho_{ABC}$ is
\begin{eqnarray}
  C_T(\rho_{ABC}) & = & S({\rho}^*_{1}) -
                        S({\rho}^*_{3}),\nonumber\\
                  & = & S_A + S_B + S_C - S_{ABC}. 
\end{eqnarray}
The total correlation can be further decomposed into the irreducible
bipartite correlation and irreducible tripartite correlation. The
degrees of irreducible bipartite correlation and tripartite
correlation are
\begin{eqnarray}
  C_2(\rho_{ABC}) & = & S({\rho}^*_1) -
                        S({\rho}^*_2),\\
  C_3(\rho_{ABC}) & = & S({\rho}^*_2) -
                        S({\rho}^*_3).
\end{eqnarray}

According to the above definitions, $C_T(\rho_{ABC})$,
$C_2(\rho_{ABC})$ and $C_3(\rho_{ABC})$ are nonnegative, and
\begin{equation}
  C_T(\rho_{ABC})=C_2(\rho_{ABC})+C_3(\rho_{ABC}).
\end{equation} 
Therefore the paradox discussed above is resolved. For the state
$\rho_{c}$, $C_T(\rho_{c})=2$, $C_2(\rho_{c})=2$, and
$C_3(\rho_{c})=0$.
  
It is interesting to see how the irreducible tripartite correlation
arise in a tripartite quantum state. The simplest example with
irreducible tripartite correlation is the $GHZ$ state:
\begin{equation}
  \vert{GHZ}\rangle=\frac{1}{\sqrt{2}}(\vert{000}\rangle+\vert{111}\rangle).
\end{equation} 
According to a simple calculation, $C_3(\vert{GHZ}\rangle)=1$. In
fact, it is impossible to specify the relative phase between the
components $\vert{000}\rangle$ and $\vert{111}\rangle$ from only
biparticle correlation, which is the physical reason for the
irreducible tripartite correlation in the $GHZ$ state.

Now let us look at the states where $C_3(\rho_{ABC})=0$, i.e. states
with no irreducible tripartite correlation, or in other words states
contain only irreducible bipartite correlation. This will mean that
\begin{equation}
  \label{eq:2corr}
  \rho_{ABC}={\rho}^*_{2}.
\end{equation}
Given that ${\rho}^*_{2}$ is unique, this means that the state
$\rho_{ABC}$ is uniquely determined by its $2$-RDMs. This does not
mean that there are no other states with the same $2$-RDMs as
$\rho_{ABC}$, but there are no other state with the same $2$-RDMs and a
larger entropy than that of $\rho_{ABC}$. We call these states
$2$-correlated, meaning that they only contain $2$-particle
irreducible correlations, but no more.

If the state satisfying Eq.\eqref{eq:2corr} is a pure state
$\ket{\psi_{ABC}}$, then it really means the state is uniquely
determined by its $2$-RDMs, in a sense that there is no other state,
pure or mixed, which has the same $2$-RDMs as $\ket{\psi_{ABC}}$.
Surprisingly, it is shown that almost all tripartite pure states are
uniquely determined by their $2$-RDMs (i.e. $2$-correlated). It turns
out that the only three-qubit state with non-zero irreducible
tripartite correlation are those equivalent to
\begin{equation}
  \alpha\vert{000}\rangle+\beta\vert{111}\rangle,
\end{equation} 
with $\alpha,\beta\neq 0$.

Obviously the above discussion can be generalized to the case of
many-body systems with $n$-particles. Notice that we always have
$C_k\geq 0$ and $\sum_{i=1}^k C_k=S(\rho)-S(\rho^*_1)$, which is the
total correlation given by the generalized mutual information. In this
sense, $C_k$ gives a hierarchy of correlations contained in the
state $\rho$.

\subsection{Many-body entanglement}

For a pure tripartite state $\ket{\psi_{ABC}}$, it is natural to
state the following.

\begin{svgraybox}
  \begin{center}
    \textbf{Box 1.21 Tripartite product state}

    A pure tripartite state $\ket{\psi_{ABC}}$ is a product state if
    it can be written as
    $\ket{\psi_{A}}\otimes\ket{\psi_{B}}\otimes\ket{\psi_{C}}$ for
    some $\ket{\psi_{A}}\in\mathcal{H}_A$,
    $\ket{\psi_{B}}\in\mathcal{H}_B$ and
    $\ket{\psi_{C}}\in\mathcal{H}_C$, otherwise it is entangled.
  \end{center}
\end{svgraybox}

However, tripartite situation is more complicated than the bipartite
case. For instance, $\ket{\psi_{ABC}}$ may be written as
\begin{equation}
  \ket{\psi_{ABC}}=\ket{\psi_A}\otimes\ket{\psi_{BC}},
\end{equation}
where $\ket{\psi_{BC}}$ is an entangled bipartite state in
$\mathcal{H}_B\otimes\mathcal{H}_C$. In this case, there is indeed no
entanglement between the subsystem $A$ and the subsystems $BC$.

In case we are only interested in tripartite states that contain `genuine'
entanglement, we will need the following statement.

\begin{svgraybox}
  \begin{center}
    \textbf{Box 1.22 Genuine entangled state}

    A pure tripartite state $\ket{\psi_{ABC}}$ is genuinely entangled,
    if it cannot be written as a product state with respect to any
    bipartition of the system.
  \end{center}
\end{svgraybox}

To quantify the entanglement in a pure tripartite state
$\ket{\psi_{ABC}}$, one idea is that we can quantify its `bipartite'
entanglement with respect to any bipartition, using von Neumann
entropy. As we will see in later chapters, in many practical cases,
this provides important information, such as the entanglement area
law. There are also various entanglement measures used in different
scenarios for quantifying pure state entanglement. Here we discuss one
with geometric meaning, namely the geometric measure of entanglement.

\begin{svgraybox}
  \begin{center}
    \textbf{Box 1.23 Geometric measure of entanglement}

    For a pure tripartite state $\ket{\psi_{ABC}}$, consider a
    tripartite product state
    $
    \ket{\alpha}=\ket{\alpha_A}\otimes\ket{\alpha_B}\otimes\ket{\alpha_C}.
    $ The geometric measure of entanglement $E_G(\ket{\psi_{ABC}})$ is
    then revealed by the maximal overlap
    \begin{equation}
      \Lambda_{\max}(\ket{\psi_{ABC}})=\max_{\ket{\alpha}}|\langle\alpha\ket{\psi_{ABC}}|,
    \end{equation}
    and is given by
    \begin{equation}
      E_G(\ket{\psi_{ABC}})=-\log\Lambda^2_{\max}(\ket{\psi_{ABC}}).
    \end{equation}
  \end{center}
\end{svgraybox}

Geometrically, $E_G(\ket{\psi_{ABC}})$ measures how far
$\ket{\psi_{ABC}}$ is from the set of product states $\{\ket{\alpha}\}$. And
$E_G(\ket{\psi_{ABC}})=0$ if and only if $\ket{\psi_{ABC}}$ itself is
a product state. As an example, for the GHZ state
\begin{equation}
  \ket{GHZ}=\frac{1}{\sqrt{2}}(\ket{000}+\ket{111}),
\end{equation}
the maximal overlap
\begin{equation}
  \Lambda^2_{\max}(\ket{GHZ})=\frac{1}{\sqrt{2}},
\end{equation}
with the maximum at either $\alpha=\ket{000}$ or $\alpha=\ket{111}$,
hence the geometric measure of entanglement is
\begin{equation}
  E_G(\ket{GHZ})=1.
\end{equation}

And $E_G$ has a natural generalization to a system with more than
three particles, which is similarly given by the maximal overlap with
a product state.

For mixed many-body states, one can also similarly discuss
entanglement with respect to any bipartition. However, similar to the
bipartite case, in most practical cases, we are more concerned with
`total' correlation rather than just `quantum' correlation for mixed
states, so with respect to any bipartition, we quantify correlation
using mutual information. And when we talk about correlation beyond
just the bipartite ones, we look at the `irreducible' tripartite
correlations as discussed in Sec.~\ref{sec:irr}. The case for
many-body systems with more than three particles can be dealt with
similarly.

\section{Summary and further reading}

In this chapter we have discussed the basic concepts of correlation
and entanglement for many-body quantum systems. We start from
introducing concepts of independence and correlation in probability
theory, which lead to some understanding of the concepts of entropy
and mutual information, which are vital in modern information theory.
Historically, these concepts are introduced by Shannon, who is
considered as the founding father of electronic communications age, in
his 1948 paper `a mathematical theory of
communication'~\cite{shannon-1948}, which builds the foundation for
information theory.

We then continue to examine the correlation in quantum systems. It
turns out that quantum systems possess `somewhat more' correlation beyond
the classical one, which is then called quantum entanglement.
Historically, this issue was first raised by Einstein, Podolsky and
Rosen in 1935~\cite{EPR35}, where they discussed the so called `EPR
paradox', which is a thought experiment revealing what they believed
to be incompleteness of quantum mechanics, that is, quantum mechanics
cannot be reproduced from some hidden variables. The word
`entanglement' was first mentioned by Schr{\"o}dinger in 1935
~\cite{schrodinger:cat}, where he described a famous cat that is
unfortunately both alive and dead due to quantum entanglement, which
later adopts the name `Schr{\"o}dinger's cat'.

A more serious study of quantum entanglement beyond just thought
experiment starts from the study of Bell's inequalities. It was first
proposed by Bell in 1964~\cite{Bel64}. It comes in a form of `Bell's
theorem', which states that no hidden variable theory can reproduce
all of the predictions of quantum mechanics, or in other words,
quantum correlation is beyond classical correlation. There are many
subsequent inequalities following Bell's work, and the CHSH inequality
presented in this chapter is discussed in ~\cite{CHS69}.

The operational meaning of partial quantum information (negative
quantum conditional entropy) was given in~\cite{Horodecki:2005wa}. The
no-cloning theorem is proved in ~\cite{WZ82}. The GHZ paradox has a
spirit similar to the Bell's theorem, but looking at more than two
particles such that inequalities are no longer necessary, is
originally proposed in~\cite{GHZ89}.

The principle of maximum entropy is advocated by Jaynes in the study
on the foundation of statistical mechanics~\cite{Jay57}. The
irreducible correlation is first proposed in~\cite{LPW02} for full
rank states, and then later generalized to non-full rank states
in~\cite{Zho08}. Their work provides a quantum analog of the
information hierarchy idea as studied~\cite{Ama01,SSB+03}.

The idea of separable state is originally from~\cite{Wer89}. The
entanglement of formation was proposed in~\cite{BDS+96}. There are
many aspects of quantum entanglement that have not been mentioned is
this chapter. As already mentioned, we have only chosen those very
basic facts and those will be used to study many-body physics later in
this book. In fact, entanglement theory is an active bunch of study in
the frontier of quantum information and quantum foundation. For a
general review of quantum entanglement emphasizing on mathematical
aspects, we direct the reader to~\cite{HHH09}. For readers interested
in more on entanglement theory in many-body systems, we refer
to~\cite{AFOV08}.

Furthermore, it is also not the goal of this part of the book (i.e.
Chap. 1,2 and 3) to introduce the general theory of quantum
information and computation. Again, we will only introduce those very
basic facts and those will be used to study many-body physics in later
parts of this book. For readers interested in quantum information and
computation in general, there are many good references, such as the
book by Nielsen and Chuang~\cite{nielsenchuang}. There also various
good resources available online, for instance the lecture notes by
Preskill at Caltech~\cite{preskill}.

%
%
\bibliographystyle{plain}
\bibliography{Chap1}


%
%
%
\chapter{Evolution of Quantum Systems}
\label{cp:2} 

\abstract{In this chapter we discuss in general a quantum system with
  Hilbert space $\mathcal{H}_{S}$, whose quantum state is described by
  a density matrix $\rho_S$, under time evolution. In the ideal case,
  the evolution of the wave function
  $\ket{\psi_{S}}\in\mathcal{H}_{S}$ is unitary. This unitary
  evolution gives rise to the circuit model of quantum computation,
  where the computational procedure is to `apply' unitary operations
  to the quantum state carrying information of the computation.
  However, as discussed in Chapter~\ref{cp:1}, in the general case,
  the system $S$ interacts with the environment. Then the quantum
  state $\rho_S$ of the system comes from part of a larger system
  $\mathcal{H}_{S}\otimes \mathcal{H}_{E}$, which is composed of both
  the system and its environment.}
  
\section{Introduction}

The time evolution of a wave function is governed by the
Schr\"{o}dinger's equation and hence is unitary. One may just feel
that in principle this is the end of the story as there is nothing
more than just unitary to talk about. This is indeed, the ideal case.
However, in our real world, there are many factors one has to take
into account when discussing unitary evolution of quantum states. This
will be the topic of this chapter where we introduce those viewpoints
of quantum information science, which turn out to be relevant to real
life.

The first concern is for a many-body system of $N$ particles, in
general the time evolution for a quantum state $\ket{\psi}$ should be
given by an $N$-particle unitary $U$, i.e.
$\ket{\psi(t)}=U\ket{\psi(t_0)}$. However, not all unitaries can be
realized by real-world Hamiltonian $t$ as $U=e^{-iH(t-t_0)}$. That is
because a natural arising many-body Hamiltonian involves only few-body
interactions, i.e. $H=\sum_i H_i$, where each $H_i$ acting
non-trivially on only a few number of particles. By a simple counting
of parameters we know that this kind of natural Hamiltonians cannot
result in all the unitary evolutions for the $N$-particle space.

The idea from quantum information science is to build any $N$-particle
unitaries from those small-particle-number ones. In particular, it is
known that two-qubit unitaries acting on any pair of particles suffice
to produce any $N$-qubit unitary, and further more any single qubit
unitary plus some fixed single two-qubit unitary suffice to produce
any two-qubit unitary. This gives rise to the so called circuit model
of quantum computing, where a diagram is introduced to illustrate how
an $N$-qubit unitary is realized by single and two-particle unitaries.

However, one needs to be aware of that in general, the construction to
realize an $N$-qubit unitary by single and two-particle unitaries is
`not efficient.' This means in general, exponentially many single and
two-particle quantum unitaries are needed (in terms of number of
qubits $N$). Nevertheless, those $N$-particle unitaries from naturally
arising Hamiltonians can be realized with only polynomial number of
single and two-particle quantum unitaries, which is the central idea
of quantum simulation. This is also consistent with the previous
discussion of parameter counting. When realizing an $N$-qubit unitary,
the number of single and two-particle unitaries needed for a quantum
circuit is called its circuit size, and polynomial size circuits (in
terms of the system size $N$) are hence called efficient.

In practice, some of these single and two-particle unitaries on an
$N$-particle system can be implemented in parallel. Therefore, the
real time needed to realize a quantum circuit is the layer of
unitaries where each layer contains parallel realizable single and
two-particle unitaries. The number of layers is hence called the depth
of the circuit. In general, a polynomial size circuit also needs to be
realized by polynomial depth. In special cases it may be realized by a
constant depth circuit, where the depth does not depend on the system
size $N$ (the number of particles). We will see in later chapters of
this book that these constant depth circuits play an important role in
characterizing gapped quantum phases.

Another issue for time evolution of a quantum system is due to
decoherence. As already discussed in Chapter~\ref{cp:1}, the quantum
state $\rho_S$ of the system comes from a lager system
$\mathcal{H}_{S}{\otimes}\mathcal{H}_{E}$, which is composed of both
the system and its environment. The evolution of the total system,
including both the system and its environment, is governed by the
Schr\"{o}dinger's equation and hence is unitary. However, when one
only has access to the system but not the environment, the dynamics of
the system only is in general non-unitary.

The question then becomes what the form of the general dynamics of the
system could be. It should be of course a linear map as quantum
mechanics is linear, but could this be enough? Quantum information
theory developed a method of characterizing the general non-unitary
dynamics of an open quantum system, called the completely positive
trace-preserving map (TPCP). These maps can be characterized by a set
of Kraus operators. The second half of this chapter will introduce
this theory.

Physicists are likely more comfortable with dynamics governed by
differential equations, and indeed theory of the differential equation
for open quantum systems are substantially developed in quantum
optics, which is the so called master equation. It is in general not
possible to have such a differential equation, unless the evolution of
the quantum system is `Markovian,' in a sense that quantum state of a
later time $\rho(t+dt)$ is completely determined by the quantum state
of the previous time $\rho(t)$. Nevertheless, in many cases, the
Markovian description is a very good approximation. We will discuss
the theory of master equation and use it to derive time evolution for
a single qubit system corresponding to some general quantum noise,
such as amplitude damping (i.e. spontaneous emission of a two-level
atom), phase damping (dephasing), and depolarizing.

\section{Unitary evolution}
\label{sec:cp2sec1}

In the ideal case, the evolution of the wave function
$\ket{\psi_{S}}\in\mathcal{H}_{S}$ of the system $S$ is governed by
the Schr\"{o}dinger's equation
\begin{equation}
  \label{eq:shr}
  i\frac{\partial \ket{\psi_{S}(t)}}{\partial t}=H_{S} \ket{\psi_{S}(t)},
\end{equation}
where $H_{S}$ is the Hamiltonian of the system $\mathcal{H}_{S}$, and
we take $\hbar=1$ for simplicity.

The solution of Eq. (\ref{eq:shr}) is given by some unitary operator
$U_S(t,t_0)$, that is,
\begin{equation}
  \label{eq:unitary}
  \ket{\psi_{S}(t)}=U_S(t,t_0) \ket{\psi_{S}(t_0)},
\end{equation}
depending on the initial value of $\ket{\psi_{S}(t_0)}$. In case that
the Hamiltonian $H_S$ is time-independent, one has
$U_S(t,t_0)=\exp[-iH_S(t-t_0)]$.

\subsection{Single qubit unitary}

Let us consider the unitary evolution of the simplest system - a
two-level system (a qubit). Recall that the basis for a two-level
quantum system is typically denoted by $\ket{0}$ and $\ket{1}$, and
its two-dimensional Hilbert space is denoted by $\mathbb{C}_2$. Any
quantum state $\alpha\ket{0}+\beta\ket{1}\in\mathbb{C}_2$ is called a
qubit. Quantum evolution of a qubit is a $2\times 2$ unitary matrix.
The three Pauli matrices $X,Y,Z$ are all unitary matrices, which
together with $I$ form a basis of $2\times 2$ matrices.

Note that $X\ket{0}=\ket{1}$ and $X\ket{1}=\ket{0}$, i.e. the Pauli
$X$ operator flips the qubit basis states
$\ket{0}\leftrightarrow\ket{1}$, therefore the Pauli $X$ operator is
also called `bit flip'. The eigenvalues of $X$ is $\pm 1$ and the
eigenvectors are
\begin{equation}
  \ket{\pm}=\frac{1}{\sqrt{2}}(\ket{0}\pm\ket{1}).
\end{equation}
We call the basis $\{\ket{\pm}\}$ the `$X$ basis'.

Also $Z\ket{0}=\ket{0}$ and $Z\ket{1}=-\ket{1}$, the Pauli $Z$
operator flips the phase of the qubit basis state $\ket{1}$, therefore
the Pauli $Z$ is also called the `phase flip'. Hence the eigenvalues
of $Z$ is $\pm 1$ and the eigenvectors are $\ket{0},\ket{1}$
respectively. We call the basis $\{\ket{0},\ket{1}\}$ the `$Z$ basis'
(or computational basis as discussed in Chapter~\ref{cp:1}).

The Hadamard operation $R$ is given by
\begin{equation}
  \label{eq:Hadamard}
  R=\frac{1}{\sqrt{2}}\begin{pmatrix} 1 & 1 \\1 & -1\end{pmatrix},
\end{equation}
which is unitary. Note that $R^{\dag}=R,\,R^2=I$, and $RXR=Z,\,RZR=X$.
That is, $R$ is the transformation between the `$Z$ basis' and the
`$X$ basis'.

And other important single-qubit unitaries are the $X,Y,Z$ rotations
given by
\begin{equation}
  \label{eq:PauliX}
  X_{\theta}=\exp(-i\theta X/2)=\cos\frac{\theta}{2}I-i\sin\frac{\theta}{2}X
  =\begin{pmatrix} \cos\frac{\theta}{2} & -i\sin\frac{\theta}{2}\\-i\sin\frac{\theta}{2} & \cos\frac{\theta}{2}
  \end{pmatrix},
\end{equation}
and
\begin{equation}
  \label{eq:PauliY}
  Y_{\theta}=\exp(-i\theta Y/2)=\cos\frac{\theta}{2}I-i\sin\frac{\theta}{2}Y
  =\begin{pmatrix} \cos\frac{\theta}{2} & -\sin\frac{\theta}{2}\\
    \sin\frac{\theta}{2} & \cos\frac{\theta}{2}
  \end{pmatrix},
\end{equation}
and
\begin{equation}
  \label{eq:PauliZ}
  Z_{\theta}=\exp(-i\theta Z/2)=\cos\frac{\theta}{2}I-i\sin\frac{\theta}{2}Z
  =\begin{pmatrix} e^{- i\frac{\theta}{2}} & 0 \\
    0 & e^{i\frac{\theta}{2}}
  \end{pmatrix}.
\end{equation}
Note that $X_{\theta},Y_{\theta},Z_{\theta}$ can be realized by the
evolution of the Hamiltonian of the form $X,Y,Z$ respectively, and the
Hadamard operation $R$ is actually $Z_{\pi}Y_{\pi/4}Z_{\pi}$. Actually
we will show that $Y,Z$ rotations together are enough to realize any
single qubit unitary.

\begin{svgraybox}
  \begin{center}
    \textbf{Box 2.1 Single-qubit unitary}

    For any unitary operation on a single qubit, there exist real
    numbers $\alpha,\beta,\gamma,\delta$ such that
    $U=e^{i\alpha}Z_{\beta}Y_{\gamma}Z_{\delta}$.
  \end{center}
\end{svgraybox}

To show why this is the case, note that for any $2\times 2$ unitary
matrix $U$, the rows and columns of $U$ are orthogonal plus that each
row or column is a normalized vector. This then follows that there
exist real numbers $\alpha,\beta,\gamma,\delta$ such that
\begin{equation}
  U=\begin{pmatrix} 
    e^{i(\alpha-\beta/2-\delta/2)} \cos\frac{\gamma}{2} &
    -e^{i(\alpha-\beta/2+\delta/2)} \sin\frac{\gamma}{2}\\
    e^{i(\alpha+\beta/2-\delta/2)} \sin\frac{\gamma}{2} &
    e^{i(\alpha+\beta/2+\delta/2)} \cos\frac{\gamma}{2}
  \end{pmatrix}.
\end{equation}
$U=e^{i\alpha}Z_{\beta}Y_{\gamma}Z_{\delta}$ then follows from
Eq.~\eqref{eq:PauliZ} and Eq.~\eqref{eq:PauliY}.

\subsection{Two-qubit unitary}

Recall that a basis for an $N$-qubit system is chosen as the tensor
products of $\ket{0}$s and $\ket{1}$s. For instance, for $N=2$, the
four basis states are $\{\ket{00},\ \ket{01},\ \ket{10},\ \ket{11}\}$.
As an example, here we discuss a two-qubit unitary operation which is
the most-commonly used in quantum computing, called the controlled-NOT
operation. It takes $\ket{x}\otimes\ket{y}$ to
$\ket{x}\otimes\ket{y\oplus x}$, where $x,y\in\{0,1\}$ and $\oplus$ is
the addition $\mod\ 2$. Here the first qubit is called the control
qubit, which remains unchanged, and the second qubit is called the
target qubit, which is flipped if the control qubit is $1$. In the
basis of $\{\ket{00},\ \ket{01},\ \ket{10},\ \ket{11}\}$ the matrix of
a controlled-NOT gate is then given by
\begin{equation}
  \label{eq:CNOT}
  \begin{pmatrix} 
    1 & 0 & 0 & 0 \\0 & 1 & 0 & 0 \\
    0 & 0 & 0 & 1\\ 0 & 0 & 1 & 0
  \end{pmatrix}.
\end{equation}
Similarly, a controlled-NOT gate with the second qubit as the control
qubit takes $\ket{x}\otimes\ket{y}$ to
$\ket{x\oplus y}\otimes\ket{y}$.

Another important two-qubit unitary is called controlled-$Z$, which
transforms the basis in the following way:
\begin{equation}
  \label{eq:cZ}
  \ket{00}\rightarrow\ket{00},\ \ket{01}\rightarrow\ket{01},\
  \ket{10}\rightarrow\ket{10},\ \ket{11}\rightarrow -\ket{11}. 
\end{equation}
Given that the controlled-$Z$ operation is symmetric between the two
qubits, it is not necessary to specify which one is the control qubit
and which one is the target qubit.

Now let us consider how to realize the controlled-NOT and the
controlled-$Z$ operations using some two-qubit Hamiltonian. Let us
discuss a simple example where the interaction term $H_{in}$ is Ising,
i.e.
\begin{equation}
  H_{in}=-JZ_1\otimes Z_2=-JZ_1Z_2.
\end{equation}
Where $Z_1,Z_2$ are Pauli $Z$ operations acting on the first and
second qubits, respectively. We omit the tensor product symbol
$\otimes$ when no confusion arises. Now observe that
\begin{equation}
  \exp{-i\frac{\pi}{4}(I-Z_1-Z_2+Z_1Z_2)}=e^{-i\frac{\pi}{4}}e^{i\frac{Z_1\pi}{4}}e^{i\frac{Z_2\pi}{4}}e^{-i\frac{Z_1Z_2\pi}{4}}
  =\begin{pmatrix} 1 & 0 & 0 & 0 \\0 & 1 & 0 & 0 \\ 0 & 0 & 1 & 0\\ 0 & 0 & 0 & -1 \end{pmatrix},
\end{equation}
which gives the controlled-$Z$ operation. In other words, the
single-qubit term $Z$ together with the two-qubit Ising interaction
term $H_{in}$ can realize a controlled-$Z$ operation.

For the controlled-NOT operation, note that
\begin{equation}
  \begin{pmatrix} 1 & 0 & 0 & 0 \\0 & 1 & 0 & 0 \\ 0 & 0 & 0 & 1\\ 0 & 0 & 1 & 0 \end{pmatrix}
  =R_2\begin{pmatrix} 1 & 0 & 0 & 0 \\0 & 1 & 0 & 0 \\ 0 & 0 & 1 & 0\\ 0 & 0 & 0 & -1 \end{pmatrix}R_2,
\end{equation}
where $R_2$ is the Hadamard operation acting on the second qubit, i.e.
\begin{equation}
  R_2=I\otimes R=\begin{pmatrix} 1 & 0 \\0 & 1\end{pmatrix}\otimes 
  \frac{1}{\sqrt{2}}\begin{pmatrix} 1 & 1 \\1 & -1\end{pmatrix}=
  \frac{1}{\sqrt{2}}\begin{pmatrix} 1 & 1 & 0 & 0 \\1 & -1 & 0 & 0 \\ 0 & 0 & 1 & 1\\ 0 & 0 & 1 & -1 \end{pmatrix}.
\end{equation}

Therefore, single qubit Hamiltonians of $Y,Z$ terms together with the
two-qubit Ising interaction term $H_{in}$ can realize both the
controlled-$Z$ operation and the controlled-NOT operation.

Now let us look at another kind of two qubit unitary, called
controlled-$U$, denoted by $\Lambda_{12}(U)$, where $U$ is a single
qubit unitary. Here qubit $1$ is the control qubit, and qubit $2$ is
the target qubit. Similar as the controlled-NOT operation,
$\Lambda_{12}(U)$ acts on any computational basis state as
\begin{equation}
  \Lambda_{12} (U)\ket{x}\otimes\ket{y}=\ket{x}\otimes U^x\ket{y},
\end{equation}
where $x,y=0,1$. In this language, controlled-NOT is indeed
$\Lambda_{12}(X)$ and controlled-$Z$ is indeed $\Lambda_{12}(Z)$.

We are now ready to check that the following equation holds.
\begin{equation}
  \Lambda_{12}(U)=(D\otimes A)\Lambda_{12}(X)(I\otimes B)\Lambda_{12}(X)(I\otimes C),
\end{equation}
where
\begin{equation}
  D=\begin{pmatrix} 1 & 0 \\0 & e^{i\alpha}\end{pmatrix},
\end{equation}
and $U,\alpha,A,B,C$ satisfy
\begin{align}
  U&=e^{i\alpha}AXBXC\nonumber\\
  I&=ABC.
\end{align}

To see how this works, note that
\begin{align}
  &(D\otimes A)\Lambda_{12}(X)(I\otimes B)\Lambda_{12}(X)(I\otimes C)\ket{x}\otimes\ket{y}\nonumber\\
  =&(D\otimes A)\Lambda_{12}(X)(I\otimes B)\Lambda_{12}(X)\ket{x}\otimes C\ket{y}\nonumber\\
  =&(D\otimes A)\Lambda_{12}(X)(I\otimes B)\ket{x}\otimes X^{x}C\ket{y}\nonumber\\
  =&(D\otimes A)\Lambda_{12}(X)\ket{x}\otimes BX^{x}C\ket{y}\nonumber\\
  =&(D\otimes A)\ket{x}\otimes X^{x}BX^{x}C\ket{y}\nonumber\\
  =&D\ket{x}\otimes AX^{x}BX^{x}C\ket{y}\nonumber\\
  =&\ket{x}\otimes e^{i\alpha x}AX^{x}BX^{x}C\ket{y}.\nonumber\\
\end{align}
So for $x=0$, the last line reads $\ket{0}\otimes\ket{y}$, and for
$x=1$, the last line reads $\ket{1}\otimes U\ket{y}$.

This then shows that any controlled-$U$ unitary can be written as a
product of single-qubit unitaries and controlled-NOT operations. Or in
terms of Hamiltonians, single-qubit Hamiltonians of $Y,Z$ terms
together with the two-qubit Ising interaction term $H_{in}$ can
realize any two-qubit controlled-$U$. In fact, these are enough to
realize `any' two-qubit unitary, as summarized below.

\begin{svgraybox}
  \begin{center}
    \textbf{Box 2.2 Two-qubit unitary from single-qubit unitary and
      controlled-NOT}

    Any unitary operation on two qubits can be written as a product of
    single-qubit unitaries and controlled-NOT operations.
  \end{center}
\end{svgraybox}

\subsection{$N$-qubit unitary}

Now consider a system of $N$ qubits, whose Hilbert space
$\mathbb{C}_2^{\otimes N}$ is the $N$-fold tensor product of
$\mathbb{C}_2$. An $N$-qubit unitary is now a unitary operator $U$
acting on $N$-qubits. In the general case, the corresponding
Hamiltonian needed to implement this evolution $U$ should contain
interaction terms involving arbitrarily large number of qubits.

However, this is not the case of nature: the natural many-body
Hamiltonians available usually contains only few-body interaction.
That is, usually the Hamiltonian of the system can be written as the
following form
\begin{equation}
  H=\sum_j H_j,
\end{equation}
where each term $H_j$ involves only few-body interactions. Indeed,
most of the time, $H_j$s involve at most two-body interactions. For
examples, the Ising model Hamiltonian $H^{\text{tIsing}}$ in
transverse magnetic field has the form
\begin{equation}
  \label{eq:Ising}
  H^{\text{tIsing}}=-J\sum_{i,j}Z_iZ_j-B\sum_j X_j,
\end{equation}
and the spin-$1/2$ Heisenberg Hamiltonian has the form
\begin{equation}
  \label{eq:Hei}
  H_{\text{Heisenberg}}=-J\sum_{i,j}{S}_i\cdot{S}_j.
\end{equation}

This then raises a question: can we realize an $N$-particle unitary
$U$ using two-body interactions only? This turns out to be possible,
but one needs to pay some price. Before looking into more details, let
us imagine a simpler scenario than those given by the many-body
Hamiltonians (e.g Eq.~\eqref{eq:Ising}~\eqref{eq:Hei}), where one can
indeed engineer the system Hamiltonian such that the two-body
interaction between any of the two particles can be turned on or off
(there are indeed those systems in lab, for instance in certain ion
trap experiments and cavity QED experiments).

This simplified assumption then puts us in the scenario of the
previous section. That is, suppose we have single qubit terms
$Y_i,\,Z_i$ available plus the Ising interaction
$H_{ij}=-J_{ij}Z_iZ_j$, and can turn them on and off freely for any
$i,j$. These then allow us to perform any single qubit unitary
operation and controlled-NOT between any two qubits.

In fact, as already mentioned, these are enough to realize any
$N$-qubit unitary. In order not to get into too much technical
details, we will discuss an example. We show how to implement a
special kind of $3$-qubit unitary, called controlled-controlled-$U$,
denoted by $\Lambda^2_{123}(U)$, where $U$ is a single qubit unitary.
Here the qubits $1,2$ are the controlled qubits, and the qubit $3$ is
the target qubit. Similar as the controlled-$U$ operation,
$\Lambda^2_{123}(U)$ acts on any computational basis state as
\begin{equation}
  \Lambda^2_{123} (U)\ket{x}\otimes\ket{y}\otimes\ket{z}=\ket{x}\otimes\ket{y}\otimes U^{xy}\ket{z},
\end{equation}
where $x,y,z\in\{0,1\}$.

For single qubit unitary $U$ with the matrix form
\begin{equation}
  U=\begin{pmatrix} a & b \\c & d\end{pmatrix},
\end{equation}
$\Lambda^2_{123}(U)$ has the matrix form
\begin{equation}
  \Lambda^2_{123}(U)=\begin{pmatrix} 
    1 & 0 & 0 & 0 & 0 & 0 & 0 & 0\\
    0 & 1 & 0 & 0 & 0 & 0 & 0 & 0\\
    0 & 0 & 1 & 0 & 0 & 0 & 0 & 0\\
    0 & 0 & 0 & 1 & 0 & 0 & 0 & 0\\
    0 & 0 & 0 & 0 & 1 & 0 & 0 & 0\\
    0 & 0 & 0 & 0 & 0 & 1 & 0 & 0\\
    0 & 0 & 0 & 0 & 0 & 0 & a & b\\
    0 & 0 & 0 & 0 & 0 & 0 & c & d\\
  \end{pmatrix}.
\end{equation}

We are now ready to check the following equation holds.
\begin{equation}
  \Lambda^2_{123}(U)=(\Lambda_{13}(V)\otimes I_2)(\Lambda_{12}(X)\otimes I_3)
  (I_1\otimes\Lambda_{23}(V^{\dag}))(\Lambda_{12}(X)\otimes I_3)(I_1\otimes\Lambda_{23}(V)),
\end{equation}
where $V^2=U$.

To see why this is the case, recall that
\begin{align}
  \Lambda_{13}(V)\ket{x}\otimes\ket{y}\otimes\ket{z}&=\ket{x}\otimes\ket{y}\otimes V^{x}\ket{z},\nonumber\\
  \Lambda_{12}(X)\ket{x}\otimes\ket{y}\otimes\ket{z}&=\ket{x}\otimes X^{x}\ket{y}\otimes \ket{z},\nonumber\\
  \Lambda_{23}(V)\ket{x}\otimes\ket{y}\otimes\ket{z}&=\ket{x}\otimes \ket{y}\otimes V^{y}\ket{z},
\end{align}
hence
\begin{align}
  &(\Lambda_{13}(V)\otimes I_2)(\Lambda_{12}(X)\otimes I_3)
     (I_1\otimes\Lambda_{23}(V^{\dag}))(\Lambda_{12}(X)\otimes I_3)(I_1\otimes\Lambda_{23}(V))
     \ket{x}\otimes\ket{y}\otimes \ket{z}\nonumber\\
  =& (\Lambda_{13}(V)\otimes I_2)(\Lambda_{12}(X)\otimes I_3)
      (I_1\otimes\Lambda_{23}(V^{\dag}))(\Lambda_{12}(X)\otimes I_3) \ket{x}\otimes\ket{y}\otimes V^{y}\ket{z},\nonumber\\
  =& (\Lambda_{13}(V)\otimes I_2)(\Lambda_{12}(X)\otimes I_3)
      (I_1\otimes\Lambda_{23}(V^{\dag}))\ket{x}\otimes X^{x}\ket{y}\otimes V^{y}\ket{z},\nonumber\\
  =& (\Lambda_{13}(V)\otimes I_2)(\Lambda_{12}(X)\otimes I_3)
      (I_1\otimes\Lambda_{23}(V^{\dag}))\ket{x}\otimes \ket{y\oplus x}\otimes V^{y}\ket{z},\nonumber\\
  =& (\Lambda_{13}(V)\otimes I_2)(\Lambda_{12}(X)\otimes I_3)
      \ket{x}\otimes \ket{y\oplus x}\otimes (V^{\dag})^{y\oplus x}V^{y}\ket{z},\nonumber\\
  =& (\Lambda_{13}(V)\otimes I_2)
      \ket{x}\otimes X^{x}\ket{y\oplus x}\otimes (V^{\dag})^{y\oplus x}V^{y}\ket{z},\nonumber\\
  =& (\Lambda_{13}(V)\otimes I_2)
      \ket{x}\otimes \ket{y}\otimes (V^{\dag})^{y\oplus x}V^{y}\ket{z},\nonumber\\
  =& 
      \ket{x}\otimes\ket{y}\otimes V^{x}(V^{\dag})^{y\oplus x}V^{y}\ket{z}.
\end{align}
Therefore, only when $x=y=1$, the last line
$V^{x}(V^{\dag})^{y\oplus x}V^{y}=V^2=U$, otherwise
$V^{x}(V^{\dag})^{y\oplus x}V^{y}=I$.

Note that although $\Lambda^2_{123}(U)$ acts on an $8$-dimentional
space, it is effectively a `two-level unitary'. That is, it is a
unitary on the subspace spanned by $\ket{110},\ket{111}$. The
observation is that two-level unitaries are enough to realize any
$N$-qubit unitary, if one can implement two-level unitaries on any
two-level of the system (i.e. any two-dimensional subspace of the
$2^N$-dimensional Hilbert space).

To illustrate the idea, consider an $3\times 3$ unitary $U$. The claim
is that it can be realized as $U=U_1U_2U_3$, where $U_1,U_2,U_3$ are
of the form
\begin{align}
  U_1=\begin{pmatrix}
    a_1 & b_1 & 0\\
    c_1 & d_1& 0\\
    0 & 0 & 1
  \end{pmatrix}, \quad U_2=\begin{pmatrix}
    a_2 & 0 & b_2\\
    0 & 1 & 0\\
    c_2 & 0 & d_2
  \end{pmatrix}, \quad U_3=\begin{pmatrix}
    1 & 0 & 0\\
    0 & a_3 & b_3\\
    0 & c_3 & d_3
  \end{pmatrix}.
\end{align}
The idea of decomposing an $N$-qubit unitary in terms of two-level
unitaries is just similar.

Of course one still needs to show that single qubit unitaries and
controlled-NOT between any two qubits can produce any two-level
unitary. This is indeed possible and we omit the details. We then come
to the result that is summarized below.

\begin{svgraybox}
  \begin{center}
    \textbf{Box 2.2 $N$-qubit unitary from single-qubit unitary and
      controlled-NOT}

    Any unitary operation on $N$ qubits can be written as a product of
    single-qubit unitaries and controlled-NOT operations.
  \end{center}
\end{svgraybox}

In terms of Hamiltonians, we have shown that single-qubit Hamiltonians
of $Y,Z$ terms together with the two-qubit Ising interaction term
$H_{in}$ can realize any $N$-qubit unitary. In fact, there is nothing
special about the Ising interaction $H_{ij}=-J_{ij}Z_iZ_j$. Any
non-trivial two-qubit interaction, in a sense that it is able to
produce entanglement when acting on some input pure state without
entanglement (i.e. product state), is enough to realize any $N$-qubit
unitary. We summarize this observation below.

\newpage
\begin{svgraybox}
  \begin{center}
    \textbf{Box 2.4 $N$-qubit unitary evolutions from single- and
      two-qubit ones}

    Single qubit terms and any non-trivial two-qubit interaction can
    generate an arbitrary $N$-qubit unitary evolution.
  \end{center}
\end{svgraybox}

However, we need to emphasize the that the efficiency of this
realization is in general poor. According to the steps we result in
this realization, an arbitrary $N$-qubit unitary may be written as
$\sim 4^N$ two-level unitary operations, and implementing a two-level
operation needs $\sim N^2$ single particle and controlled-$U$
operations, which gives $\sim N^24^N$ single particle and
controlled-$U$ operations to realize an arbitrary $N$-qubit unitary.
We summarize this observation below.

\begin{svgraybox}
  \begin{center}
    \textbf{Box 2.5 Inefficiency in realizing $N$-qubit unitaries from
      single- and two-qubit ones}

    In general, exponentially many single and two-qubit unitaries are
    needed for generating an $N$-qubit unitary evolution.
  \end{center}
\end{svgraybox}

\section{Quantum Circuits}

In the previous section we have mentioned the name of `quantum
computing', but we do not even tell what a `quantum computer' is. It
is not our goal here to discuss the theory of computation, rather, we
would like to tell that at least one model of quantum computing,
called the circuit model, is based on the unitary evolution discussed
in the previous section.

In this model, the initial $N$-qubit state $\ket{\psi_i}$ is usually
chosen as the all $\ket{0}$ state
$\ket{0}\otimes\ket{0}\cdots\otimes\ket{0}$, which is in short written
as $\ket{00\cdots 0}$ or $\ket{0}^{\otimes N}$. Then a sequence of
single- and two-qubit quantum unitaries are applied on $\ket{\psi_i}$
to result in a final state $\ket{\psi_f}$. And finally single-qubit
measurements are performed on each qubit, usually in the
$\{\ket{0},\ket{1}\}$ basis, to obtain the result of the computation.
And we know that single and two-qubit quantum unitaries are enough to
implement any $N$-qubit unitary, regardless it might be in an
inefficient manner in general.

The sequence of single- and two-qubit unitaries then gives rise to a
diagram called `quantum circuit', and this model is then called the
`circuit model' of quantum computing. We discuss an example of the
circuit diagram. Here each vertical line represents a qubit, and each
box putting on a single line or across two lines are single and
two-qubit uintaries, respectively. Time goes from bottom to top.

%
\begin{figure}[h!]
  \centerline{
  \includegraphics{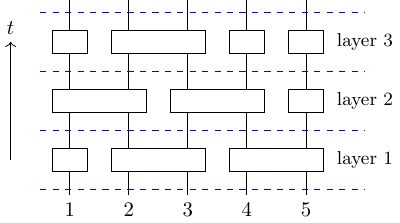}
  }
%
%
  \caption{Circuit diagram with the number of qubits $N=5$ and the
    number of layers $M=3$.}
  \label{fig:circuit} 
\end{figure}

For a given circuit diagram, there are some important parameters to
characterize its efficiency property.

\begin{svgraybox}
  \begin{center}
    \textbf{Box 2.6 Circuit size}

    The number of boxes in a circuit diagram is called its size. The
    circuit is efficient if when the number $N$ of qubits grow, the
    circuit size does not grow exponentially with $N$, in other words,
    the size of the circuit is just some polynomial of $N$.
  \end{center}
\end{svgraybox}

Also, as a box is only cross one or two lines, some of them can be
implemented in one layer, i.e., in parallel. So the evolution time of
the entire circuit will be the number of layers.

\begin{svgraybox}
  \begin{center}
    \textbf{Box 2.7 Circuit depth}

    The number of layers in a circuit diagram is called its depth. A
    constant depth circuit is a circuit with depth which does not
    increase with $N$.
  \end{center}
\end{svgraybox}

As we will discuss later in Chapter~\ref{chap7}, constant depth
circuits play an important role studying gapped quantum phases. Note
that in general, an efficient circuit of size polynomial in $N$ does
not allow a constant depth circuit, so requiring `constant depth' is a
much stronger constraint than requiring `efficiency'.

We now discuss an application of the quantum circuit model. We know
that there are some $N$-qubit unitary evolutions which are hard to
implement. However, we know that the natural occurring Hamiltonians
usually involve only few-body interactions, such as the Ising
Hamiltonian and the Heisenberg Hamiltonian. The good news is that the
evolution of these natural Hamiltonians can be simulated efficiently
by a quantum circuit model, meaning that it is possible to reproduce
the evolution to certain precision using only $~p(N)$ number of single
and two-bit unitary operations, where $p(N)$ is any polynomial in $N$.
In terms of Hamiltonians, we summarize this observation as following.

\newpage
\begin{svgraybox}
  \begin{center}
    \textbf{Box 2.8 Quantum simulation}

    The evolution of few-body Hamiltonians can be simulated
    efficiently by single qubit $Y,Z$ terms and any non-trivial
    two-qubit interaction.
  \end{center}
\end{svgraybox}

To see how this could be possible, recall that the solution to a
Schr\"{o}dinger's equation
$i\frac{\partial\ket{\psi(t)}}{\partial{t}}=H\ket{\psi(t)}$ with a
time independent Hamiltonian $H$ is given by
$\ket{\psi(t)}=\exp[-iH(t-t_0)]$. Now our task is to build an
efficient quantum circuit with only polynomial number of single and
two-qubit unitary operations, which reproduces the unitary evolution
$\exp{-iH(t-t_0)}$ to certain precision. For simplicity we take
$t_0=0$.

For a few body Hamiltonian $H$, we can write
\begin{equation}
  H=\sum_{j=1}^L H_j,
\end{equation}
where each $H_i$ acting nontrivially only on a few number of
particles, and $L$ is a polynomial function of $N$.

In the simplest case, if $[H_j,H_k]=0$ for all $j,k$, i.e. all the
terms $H_j$ commute, then the evolution $\exp{-iHt}$ is given by
\begin{equation}
  \exp[-iHt]=\exp[-it\sum_{j=1}^L H_j]=\prod_{j=1}^L\exp[-iH_jt].
\end{equation}
This directly gives an efficient quantum circuit, as each
$\exp[-iH_jt]$ is a unitary acting on only a few number of particles
(which is independent of the number of particles $N$), hence can be
realized by a constant number (i.e. independent of $N$) of single and
two-particle unitary operations.

The real challenge is when those $H_i$s do not commute. In this case
we need the following Lie product formula.

\begin{svgraybox}
  \begin{center}
    \textbf{Box 2.9 Lie product formula}

    $\lim\limits_{s\rightarrow\infty}(e^{iAt/s}e^{iBt/s})^s=e^{i(A+B)t}$.
  \end{center}
\end{svgraybox}

To prove this formula, note that the Taylor expansion for $e^{iAt/s}$
is given by
\begin{equation}
  e^{iAt/s}=I+\frac{1}{s}(iAt)+O(\frac{1}{s^2}).
\end{equation}
Here $O(\frac{1}{s^2})$ means the terms of the order $\frac{1}{s^2}$
or higher. Therefore,
\begin{equation}
  e^{iAt/s}e^{iBt/s}=I+\frac{1}{s}i(A+B)t+O(\frac{1}{s^2}),
\end{equation}
which gives
\begin{equation}
  \left(e^{iAt/s}e^{iBt/s}\right)^s=\left(I+\frac{1}{s}i(A+B)t+O(\frac{1}{s^2})\right)
  =I+\sum_{k=1}^s{s\choose k}\frac{1}{s^k}\left[i(A+B)t\right]^k+O(\frac{1}{s^2}).
\end{equation}
Since
\begin{equation} {s\choose
    k}\frac{1}{s^k}=\frac{1}{k!}\left[1+O(\frac{1}{s})\right],
\end{equation}
taking the limit $s\rightarrow\infty$ gives
\begin{equation}
  \lim\limits_{s\rightarrow\infty} \left(e^{iAt/s}e^{iBt/s}\right)^s
  =\lim\limits_{s\rightarrow\infty}
  \sum_{k=0}^s\frac{[i(A+B)t]^k}{k!}(1+O(\frac{1}{s}))+O(\frac{1}{s^2})
  = e^{i(A+B)t}. 
\end{equation}

The idea for quantum simulation is now to use a similar reasoning for
proving the Lie product formula to approximate $\exp{-iHt}$ to
certain precision. We look at some examples, and we consider a small
time interval $\Delta t=\frac{t}{s}$. First note that
\begin{equation}
  \label{eq:trotter1}
  e^{i(A+B)\Delta t}=e^{iA\Delta t}e^{iB\Delta t}+O(\Delta t^2),
\end{equation}
similarly
\begin{equation}
  \label{eq:trotter2}
  e^{i(A+B)\Delta t}=e^{iA\Delta t/2}e^{iB\Delta t}e^{iA\Delta t/2}+O(\Delta t^3).
\end{equation}

For $H=\sum_{j=1}^L H_j$, one can further show that
\begin{equation}
  \label{eq:trotter3}
  e^{-2iH\Delta t}=\left[e^{-iH_1\Delta t}e^{-iH_2\Delta t}\ldots e^{-iH_L\Delta t}\right]
  \left[e^{-iH_L\Delta t}e^{-iH_{L-1}\Delta t}\ldots e^{-iH_1\Delta t}\right]+O(\Delta t^3),
\end{equation}
Here each $\exp[-iH_j\Delta t]$ is a unitary operation on only a few
number of particles, hence can be realized by a constant number (i.e.
independent of $N$) of single and two-particle unitary operations.

A more detailed analysis will show that in order to achieve the
precision $\epsilon$ for the simulation, in a sense that the output of
the simulation is $\ket{\psi'(t)}$ such that
\begin{equation}
  |\bra{\psi'(t)}e^{-iHt}\ket{\psi(0)}|^2\geq 1-\epsilon,
\end{equation}
then one would need a quantum circuit with
$\text{poly}(\frac{1}{\epsilon})$ (i.e. polynomial in
$\frac{1}{\epsilon}$) number of single and two-particle unitary
operations.

\section{Open Quantum Systems}
\label{sec:OpenQS}

However, in the general case, the system $S$ is coupled with the
environment $E$, which results in non-unitary evolution of the system.
In this case, the evolution of the wave function
$\ket{\psi_{SE}}\in\mathcal{H}_{S}\otimes \mathcal{H}_{E}$ is governed
by the Schr\"{o}dinger's equation
\begin{equation}
  \label{eq:shrc}
  i\frac{\partial \ket{\psi_{SE}(t)}}{\partial t}=H_{SE} \ket{\psi_{SE}(t)},
\end{equation}
where $H_{SE}$ is the Hamiltonian of the total system
$\mathcal{H}_{S}\otimes \mathcal{H}_{E}$, and the solution of
Eq.\eqref{eq:shrc} is given by some unitary operator $U_{SE}(t,t_0)$.
That is,
\begin{equation}
  \label{eq:unitary}
  \ket{\psi_{SE}(t)}=U_{SE}(t,t_0) \ket{\psi_{SE}(t_0)},
\end{equation}
depending on the initial value of $\ket{\psi_{SE}(t_0)}$.

For any density operator, $\rho_{SE}$ acting on
$\mathcal{H}_{S}\otimes \mathcal{H}_{E}$, its time evolution
$\rho_{SE}(t)$ is then given by
\begin{equation}
  \rho_{SE}(t)=U_{SE}(t,t_0) \rho_{SE}(t_0) U^{\dag}_{SE}(t,t_0).
\end{equation}

What we are interested in is the evolution of the system described by
the density operator
\begin{equation}
  \label{eq:rho}
  \rho_{S}(t)=\Tr_{E}\rho_{SE}(t).
\end{equation}

Suppose initially the system is in a product state with the
environment and the environment is in some pure state, say
$\ket{0_E}$, i.e.
$\rho_{SE}(t_0)=\rho_S(t_0)\otimes \ket{0_E}\bra{0_E}$, then
Eq.\eqref{eq:rho} becomes
\begin{align}
  \label{eq:patrho}
  \rho_{S}(t)&=\Tr_{E}\rho_{SE}(t)\nonumber\\
             &=\Tr_{E}U_{SE}(t,t_0) (\rho_S(t_0)\otimes \ket{0_E}\bra{0_E})U^{\dag}_{SE}(t,t_0)\nonumber\\
             &=\sum_{k}\bra{k_E}U_{SE}(t,t_0)\ket{0_E}\rho_S(t_0)\bra{0_E}U^{\dag}_{SE}(t,t_0)\ket{k_E},
\end{align}
where $\{\ket{k_E}\}$ is an orthonormal basis of $H_E$, and
$\bra{k_E}U_{SE}(t,t_0)\ket{0_E}$ is an operator acting on $H_S$, for
each $k$. Let us write
\begin{equation}
  E_k=\bra{k_E}U_{SE}(t,t_0)\ket{0_E},
\end{equation}
then we have
\begin{equation}
  \label{eq:Kraus}
  \mathcal{E}(\rho_S(0))=\rho_{S}(t)=\sum_kE_k\rho_{S}(t_0)E_k^{\dag}.
\end{equation}

Note that
\begin{align}
  \label{eq:Krausid}
  \sum_k E_k^{\dag}E_k&=\sum_k\bra{0_E}U^{\dag}_{SE}(t,t_0)\ket{k_E}\bra{k_E}U_{SE}(t,t_0)\ket{0_E}\nonumber\\
                      &=\bra{0_E}U^{\dag}_{SE}(t,t_0)U_{SE}(t,t_0)\ket{0_E}=I
\end{align}

The map $\mathcal{E}$ defined by Eq.\eqref{eq:Kraus} is a linear map.
One can write $\mathcal{E}=\{E_1,E_2,\ldots\}$, and when the property
of Eq.\eqref{eq:Krausid} is satisfied, the map $\mathcal{E}$ is then
called a superoperator. Eq.\eqref{eq:Kraus} is then called the
operator sum representation of $\mathcal{E}$, or the Kraus
representation where each $E_k$ is a Kraus operator. We summarize this
Kraus representation for non-unitary evolutions as below.

\begin{svgraybox}
  \begin{center}
    \textbf{Box 2.10 Kraus representation for non-Unitary evolution}

    $\mathcal{E}(\rho_S(0))=\rho_{S}(t)=\sum_kE_k\rho_{S}(t_0)E_k^{\dag}$,
    where $\sum_k E_k^{\dag}E_k=I$.
  \end{center}
\end{svgraybox}

Note that for a given superoperator $\mathcal{E}$, the operator sum
representation is not unique. This is because that in performing the
partial trace as in Eq.\eqref{eq:patrho}. Say, if we instead use
$\{\bra{j_E}=\sum_kU_{jk}\bra{k_E}\}$, then we get another
representation
\begin{equation}
  \mathcal{E}(\rho_S(0))=\rho_{S}(t)=\sum_kF_k\rho_{S}(t_0)F_k^{\dag},
\end{equation}
where $F_k=U_{jk}E_k$.

We now discuss some properties of the superoperator $\mathcal{E}$.
From now on, we omit the superscript $S$ for discussing the system
evolution when no confusion arises. The most important property of
$\mathcal{E}$ is that it maps density operators to density operators.
This can be seen from Eq.\eqref{eq:Kraus} and Eq.\eqref{eq:Krausid}:
\begin{enumerate}
\item $\rho(t)$ is Hermitian:
  \begin{equation}
    \rho(t)^{\dag}=\left(\sum_kE_k\rho(t_0)E_k^{\dag}\right)^{\dag}=\sum_kE_k\rho^{\dag}(t_0)E_k^{\dag}=\rho(t)
  \end{equation}
\item $\rho(t)$ is with unit trace:
  \begin{equation}
    \Tr\rho(t)=\Tr\left(\sum_kE_k\rho(t_0)E_k^{\dag}\right)=\Tr\left(\sum_kE_k^{\dag}E_k\rho(t_0)\right)=\Tr(\rho(t_0))=1
  \end{equation}
\item $\rho(t)$ is positive:
  \begin{equation}
    \bra{\psi}\rho(t)\ket{\psi}=\sum_k(\bra{\psi}E_k)\rho(t_0)(E_k^{\dag}\ket{\psi})\geq 0.
  \end{equation}
\end{enumerate}

Finally, we remark that orthogonal measurements can also be
`interpreted' as in terms of the Kraus representation. In this case,
take a set of operators $\{\Pi_k\}$ which are orthogonal projections
in the Hilbert space $\mathcal{H}$, that is,
\begin{equation}
  \Pi_k=\Pi_k^{\dag},\quad \Pi_j\Pi_k=\delta_{jk}\Pi_k,\quad \sum_k \Pi_k=I,
\end{equation}
then the quantum operation $\mathcal{M}$ describing the measurement is
\begin{equation}
  \mathcal{M}(\rho)=\sum_k\Pi_k\rho\Pi_k.
\end{equation}
When $\rho$ is a pure state $\ket{\psi}$, the measurement will take
$\ket{\psi}\bra{\psi}$ to
\begin{equation}
  \frac{\Pi_k\ket{\psi}\bra{\psi}\Pi_k}{\bra{\psi}\Pi_k\ket{\psi}},
\end{equation}
with probability
\begin{equation}
  p_k=\bra{\psi}\Pi_k\ket{\psi}.
\end{equation}

\section{Master Equation}

We know that the evolution of an open quantum system are given by
superoperator on the density matrix of the system. For a closed
system, the evolution is governed by the integrating the
Schr\"{o}dinger's equation Eq~\eqref{eq:shrc}. A natural question is
what is the differential equation governing the dynamics of an open
system. This is the so-called master equation, which is extensively
studied in the field of quantum optics. We will discuss master
equation in this section.

\subsection{The Lindblad Form}

We start from rewriting the Schr\"{o}dinger's equation
Eq~\eqref{eq:shrc} in terms of density matrices.
\begin{equation}
  \frac{d\rho_{SE}}{dt}=-i[H_{SE},\rho_{SE}],
\end{equation}
where $[H,\rho]=H\rho-\rho H$ is the commutator of $H$ and $\rho$.
Tracing out the environment will give us the time evolution of the
density matrix of the system
\begin{equation}
  \label{eq:drhos}
  \frac{d\rho_S}{dt}=\Tr_E(\frac{d\rho_{SE}}{dt})=\Tr_E(-i[H_{SE},\rho_{SE}]).
\end{equation}

Now we consider the time evolution of the system density matrix
$\rho(t)$, where we omit the subscript $S$. We know that in general
Eq.~\eqref{eq:drhos} will give a time evolution governed by a
superoperation in terms of Kraus operators, i.e.
\begin{equation}
  \label{eq:rhot}
  \rho(t)=\mathcal{E}(\rho)=\sum_k E_k(t)\rho(t_0)E_k^{\dag}(t).
\end{equation}

To derive a differential equation for $\rho(t)$, let us consider the
infinitesimal time interval $dt$, and write
\begin{equation}
  \rho(t+dt)=\rho(t)+O(dt).
\end{equation}

Note here an assumption is made: we assume that the evolution of the
quantum system is `Markovian,' in a sense that $\rho(t+dt)$ is
completely determined by $\rho(t)$. This is not generally guaranteed
by Eq.~\eqref{eq:rhot}, as the environment, though inaccessible, may
have some memory of the system. Nevertheless, in many situations, the
Markovian description is a very good approximation.

Based on the Markovian approximation, we now further expand the Kraus
operators in terms of $dt$, where we will have one of the operators
$E_0$ with order one, that we write as
\begin{equation}
  E_0=I+(-iH+M)dt,
\end{equation}
where both $H,M$ are chosen to be Hermitian and are zeroth order in
$dt$. And the other Kraus operators $E_k$ with order $\sqrt{dt}$,
which has the form
\begin{equation}
  E_k=\sqrt{dt}L_k,\ k>0,
\end{equation}
where $L_k$ are zeroth order in $dt$.

The condition $\sum_kE_k^{\dag}E_k=I$ the gives
\begin{equation}
  M=-\frac{1}{2}\sum_{k>0}L_k^{\dag}L_k.
\end{equation}

The first order of $dt$ from Eq.~\eqref{eq:rhot} gives
\begin{svgraybox}
  \begin{center}
    \textbf{Box 2.11 The Lindblad equation}

    \begin{equation}
      \label{eq:Lindblad}
      \frac{d\rho}{dt} = -i[H,\rho] + \sum_{k>0} (L_k\rho L_k^{\dag} -
      \frac{1}{2}L_k^{\dag}L_k\rho - \frac{1}{2}\rho L_k^{\dag}L_k).
    \end{equation}
  \end{center}
\end{svgraybox}

The operators $L_k$ are called Lindblad operators. The first term of
Eq.~(\ref{eq:Lindblad}) is usual Hamiltonian term which generates
unitary evolutions. The other terms describe the dissipation of the
system due to interaction with the environment.

To solve Eq.~(\ref{eq:Lindblad}), it is helpful to look at the
interaction picture. Let
\begin{equation}
  \tilde{\rho}(t)=e^{iHt}\rho(t)e^{-iHt},
\end{equation}
which then gives
\begin{equation}
  \frac{d\tilde{\rho}(t)}{dt} = \sum_{k>0}( \tilde{L}_k\tilde{\rho}
  \tilde{L}_k^{\dag}-\frac{1}{2}\tilde{L}_k^{\dag}\tilde{L}_k\tilde{\rho}-\frac{1}{2}\tilde{\rho}
  \tilde{L}_k^{\dag}\tilde{L}_k ),
\end{equation}
where
\begin{equation}
  \tilde{L}_k=e^{iHt}L_ke^{-iHt}.
\end{equation}

\subsection{Master equations for a single qubit}
\label{sec:masterqubit}

We now examine some examples of the non-unitary dynamics for a single
quit. More precisely, we will discuss a qubit under amplitude
damping, phase damping, and depolarizing respectively. 


\subsubsection*{Amplitude Damping}

By studying a typical example, we will show how to derive the master
equation for a specified system. Here we consider a two-level atom
interacting with an electromagnetic environment, which is modeled as
\begin{equation}
  \label{eq:6}
  H = H_S + H_E + V,
\end{equation}
where
\begin{align}
  \label{eq:7}
  H_S & =  \frac {\omega_a} {2} \sigma_z, \\
  H_E & =  \sum_j \omega_j b_j^\dagger b_j, \\
  V   & =  \sum_j g_j (\sigma_+ b_j +\sigma_- b_j^\dagger).
\end{align}
Here $H_S$ and $H_E$ are the free Hamiltonians for the two-level atom
and the electromagnetic environment respectively, and $V$ describes
the interactions between the atom and the environment.
$\sigma_{x},\sigma_{y},\sigma_{z}$ are the Pauli matrices,
$\sigma_{\pm}=\sigma_{x}\pm i\sigma_{y}$, and $\omega_{a}$ is the
energy level splitting for the atom. $b_j$ and $b_j^\dagger$ are the
annihilation and creation operator for the $j$-th mode of the
electromagnetic field with frequency $\omega_{j}$, and $g_j$ is the
coupling strength between the atom and the $j$-th mode of the
environment.

In the interaction picture, the dynamics of the global system is
governed by
\begin{equation}
  \label{eq:8}
  \frac {d \tilde{\rho}} {dt} = -i [\tilde{V},\tilde{\rho}],
\end{equation}
where
\begin{align}
   \tilde{V}(t) & =  e^{i (H_S + H_E) t} V e^{-i (H_S +H_E) t} \nonumber\\ 
            & =  \sum_j g_j (\sigma_+ b_j e^{-i (\omega_j - \omega_a) t} +
                  \sigma_- b_j^\dagger e^{i(\omega_j
              -\omega_a)t}). \label{eq:20} 
\end{align}
First, we assume that the reservoir initially stays in a vacuum state,
i.e., the photon number is zero. Furthermore, we assume that the
condition for the Markov approximation is satisfied. Then the state of
the global system at time $t$ is approximated as
\begin{equation}
  \label{eq:10}
  \tilde{\rho}(t) = \tilde{\rho}_S(t)\otimes\rho_E
\end{equation}
with $\rho_E=\prod_j\vert 0_i\rangle\langle 0_j\vert$.

To the second order approximation, the state evolution from time $t$
to $t+\Delta t$ is
\begin{align}
   \tilde{\rho}(t+\Delta t)-\tilde{\rho}(t) 
  & =  -i
        \int_{t}^{t+\Delta t} [\tilde{V}(t^{\prime}),\tilde{\rho}(t)]
        \nonumber\\
  & \quad + (-i)^2 \int_{t}^{t+\Delta t} dt^{\prime}
      \int_{t}^{t^{\prime}} dt^{\prime\prime}
      [\tilde{V}(t^{\prime}),[\tilde{V}(t^{\prime\prime}),
    \tilde{\rho}(t)]]. \label{eq:21}
\end{align}

Inserting Eq.~(\ref{eq:20}) and Eq.~(\ref{eq:10}) into
Eq.~(\ref{eq:21}), we have
\begin{align}
  \label{eq:12}
  & \mathrel{\phantom{=}} \tilde{\rho}_S(t+\Delta
    t)-\tilde{\rho}_S(t)\nonumber\\ 
  & =  - \int_{t}^{t+\Delta t} dt^{\prime}
        \int_{t}^{t^{\prime}} dt^{\prime\prime}\nonumber\\
  & \left(
     \sum_{j} g_{j}^{2} e^{-i (\omega_{j} - \omega_{a})
    (t^{\prime}-t^{\prime\prime})} \Tr_{E} [\sigma_{+} b_{j}, 
     [\sigma_{-} b_{j}^{\dagger}, \tilde{\rho}_{S}(t) \otimes \rho_{E}]]
     +  h.c. \right)\nonumber \\
  & =  - \int_{t}^{t+\Delta t} dt^{\prime}
        \int_{t}^{t^{\prime}} dt^{\prime\prime}
   \left(\sum_{j} g(t^{\prime}-t^{\prime\prime})  [\sigma_{+}, 
     \sigma_{-}  \tilde{\rho}_{S}(t)]
     +  h.c. \right)\nonumber
\end{align}
where
\begin{equation}
  \label{eq:13}
  g(\tau) = \sum_{j} g_{j}^{2} e^{-i (\omega_{j} - \omega_{a}) \tau}.
\end{equation}
Since $g(\tau)$ is a combination of many oscillation functions, then
in many case it will decrease to zero in a characteristic time
$\tau_{c}$. We consider the case when $\Delta t \gg \tau_{c}$. Then
\begin{equation}
  \label{eq:13}
  \tilde{\rho}_{S}(t+\Delta t)-\tilde{\rho}_{S}(t)
    \simeq   -\int_{0}^{\infty} d\tau
    \int_{t}^{t+\Delta t} dt^{\prime} 
    \left(\sum_{j} g(\tau)  [\sigma_{+}, 
    \sigma_{-}  \tilde{\rho}_{S}(t)]
    +  h.c. \right)
\end{equation}
Therefore we obatin
\begin{align}
  \label{eq:14}
  \frac {d\tilde{\rho}_{S}(t)} {dt} 
 & =  -
        \int_{0}^{\infty} d\tau \left(
        g(\tau)  [\sigma_{+},
        \sigma_{-}  \tilde{\rho}_{S}(t)]
        -  g^{*}(\tau)  [\sigma_{-},
        \tilde{\rho}_{S}(t) \sigma_{+}]
        \right) \nonumber\\
  & =  \frac {1} {2} \left((\Gamma + \Gamma^{\ast})\sigma_{-} \tilde{\rho}_{S}
        \sigma_{+} - \Gamma \sigma_{+} \sigma_{-} \tilde{\rho}_{S}
        - \Gamma^{\ast} \tilde{\rho}_{S} \sigma_{+} \sigma_{-}\right),
\end{align}
where
\begin{equation}
  \label{eq:15} 
  \Gamma = 2\int_{0}^{\infty} d\tau g(\tau). 
\end{equation}
Since the imarginary part of $\Gamma$ represents the energy shift due
to the environment, we neglect its contribution here. Then
\begin{align}
  \label{eq:16}
  \Gamma & =  \int_{-\infty}^{\infty} d\tau
               \int_{0}^{\infty}d\omega \rho(\omega) g^{2}(\omega)
               e^{-i(\omega-\omega_{a})\tau} \nonumber\\
         & =  2\pi \int_{0}^{\infty} d\omega \rho(\omega)
               g^{2}(\omega) \delta(\omega-\omega_{a}) \nonumber\\
         & =  2\pi g(\omega_{a}) \rho(\omega_{a}),
\end{align}
which is the decay rate of the excited level, consistent with the
result from the Fermi golden rule.

Therefore, in the interaction picture, the master equation of
amplitude damping is given by

\begin{equation}
  \label{eq:amp}
  \frac{d\rho}{dt} = \frac{\Gamma}{2} (2\sigma_{-}\rho\sigma_{+} -
  \sigma_{+}\sigma_{-}\rho - \rho\sigma_{+}\sigma_{-}).
\end{equation}

Recall that the bloch representation of the density matrix
$\rho=\frac{1}{2}(I+\vec{r}\cdot\vec{\sigma})$. Solving this equation
for $\vec{r}(t)$ then gives

\begin{align}
  r_x(t) & = r_x(0)e^{-\frac{\Gamma}{2} t}\nonumber\\
  r_y(t) & = r_y(0)e^{-\frac{\Gamma}{2} t}\nonumber\\
  r_z(t) & = r_z(0)e^{-\Gamma t} -1 + e^{-\Gamma t}. \label{eq:9}
\end{align}

Eq.~(\ref{eq:9}) can be further written as
\begin{align}
  \rho_{00}(t) & = \rho_{00}(0) e^{-\Gamma t}, \label{eq:11} \\
  \rho_{01}(t) & = \rho_{01}(0) e^{- \frac{\Gamma}{2} t} \label{eq:22}.
\end{align}
Eq.~(\ref{eq:11}) implies that $\Gamma$ is the decay rate of the
excited state. It is worthy to note that in this case the decay rate
of the non-diagonal term $\rho_{01}$ is $\frac{\Gamma}{2}$.

Let $\gamma=1-e^{-\Gamma t}$, then one has
\begin{equation}
  \rho(t)=E_0\rho E_0^{\dag}+E_1\rho E_1^{\dag},
\end{equation}
where the Kraus operators $E_0, E_1$ are given as the following.

\begin{svgraybox}
  \begin{center}
    \textbf{Box 2.12 Kraus operators for amplitude damping}

    $E_0=\begin{pmatrix} 1 & 0 \\0 &
      \sqrt{1-\gamma}\end{pmatrix},\quad E_1=\begin{pmatrix} 0 &
      \sqrt{\gamma} \\ 0 & 0 \end{pmatrix}$.
  \end{center}
\end{svgraybox}

To give a physical explanation of the amplitude damping channel, let
us imagine that the qubit is a two-level atom, and it is initially
prepared in the excited state $|0\rangle$. Then the probability for
the atom keeping in the excited state is
\begin{equation}
  \label{eq:1}
  \langle 0|\rho|0\rangle (t)=\frac {1+r_{z}(t)} {2}=e^{-\Gamma t}.
\end{equation}
Eq.~(\ref{eq:1}) implies that the atom spontaneously decays from the
excited state to the ground state with the rate $\Gamma$. Therefore
the amplitute damping channel physically corresponds to the
spontaneous decay process in atomic physics.

\subsubsection*{Phase Damping}

In this subsection, we consider another type of interaction between
the two-level atom and the electromagnetic environment, which is
described by
\begin{equation}
  \label{eq:17}
  V = \sum_{j}  g_{j} \sigma_{z} (b_{j} + b_{j}^{\dagger}).
\end{equation}
Notice that this type of interaction does not change the system's
energy, but labels different energy levels through the environment,
which leads to the relative phase damping between the system's levels.

Similarly as the derivation for the case of amplitude damping, we get
the master equation in the Lindblad form
\begin{align}
  \frac{d\rho}{dt} & = \frac{\Gamma}{4} 
                     [2\sigma_{z} \rho \sigma_{z} - \sigma_{z}^{2} \rho
                     -\rho\sigma_{z}^{2}]  \nonumber\\
                   & = \frac{\Gamma}{2} [\sigma_{z} \rho \sigma_{z} -  \rho],
  \label{eq:phasdampn}
\end{align}
where $\Gamma$ is a coefficent, whose meaning is explained as follows.

To have a physical understanding of the phase damping noise, we
rewrite Eq.~(\ref{eq:phasdampn}) as
\begin{align}
  \label{eq:2}
  \frac {d\langle 0|\rho|0\rangle} {dt} & =  0,\\
  \frac {d\langle 0|\rho|1\rangle} {dt} & =  - \Gamma \langle 0|\rho|1\rangle.
\end{align}
Thus we have
\begin{align}
  \label{eq:3}
  \langle 0|\rho(t)|0\rangle & =  \langle 0|\rho(0)|0\rangle,\\
  \langle 0|\rho(t)|1\rangle & =  \langle 0|\rho(0)|1\rangle
                                   e^{-\Gamma t}.
\end{align}
Notice that the diagonal terms of the density matrix represent the
populations in the ground and the excited states, and the non-diagonal
terms describe the coherence between the ground state and the excited
state. Thus the phase damping channel describe a decoherencing process
without exchanging energy with the environment, and the coefficient
$\Gamma$ denotes the decay reate of coherence.

Let $\gamma=1-e^{-\Gamma t}$. Then the Kraus operators for the phase
damping channel can be written as the following.
\begin{svgraybox}
  \begin{center}
    \textbf{Box 2.13 Kraus operators for phase damping}

    $E_0=\sqrt{1-\gamma}I,\quad E_1=\begin{pmatrix} \sqrt{\gamma} & 0 \\0 &
      0\end{pmatrix},\quad E_2=\begin{pmatrix} 0 & 0 \\0 &
      \sqrt{\gamma} \end{pmatrix}$.
  \end{center}
\end{svgraybox}

\subsubsection*{Depolarizing}

In this subsection, we consider a two-level atom interacting with
three independent reservoirs, whose Hamiltonian is
\begin{align}
  \label{eq:18}
  H = H_S + \sum_{j=1}^{3} H_{E_{j}} + V_{j},
\end{align}
where
\begin{align}
  \label{eq:19}
  H_S & =  \frac {\omega_a} {2} \sigma_z, \\
  H_{E_{j}} & = \sum_{k} \omega_{jk} b^{\dagger}_{jk} b_{jk},\\
  V_{j} & = \sum_{k} g_{jk} \sigma_{j} (b_{jk}^{\dagger} + b_{jk}).
\end{align}
Approximately, the system's evolution can be understood as the sum of
the influences caused by the independent reservoirs.

Similarly we obtain  the master equation of depolarizing in the
Lindblad form

\begin{equation}
  \label{eq:depo}
  \frac{d\rho}{dt} = \frac{\Gamma}{6} \sum\limits_{j=x,y,z} (
  2\sigma_{j}\rho\sigma_{j}-\sigma_{j}\sigma_{j}\rho-\rho\sigma_{j}\sigma_{j}
  ).
\end{equation}
Intuitively, the master equation of depolarizing can be understood as
a combination of three different `phase dampings'. Here $\Gamma$ can
be understood as the decay rate from any state to its orthogonal
state.

Note that Eq.~(\ref{eq:depo}) can be simplified as
\begin{equation}
  \label{eq:4}
  \frac {d\rho} {dt} = -\Gamma (\rho - \frac {I} {2}).
\end{equation}
Eq.\eqref{eq:4} gives
\begin{equation}
  \label{eq:5}
  \rho(t) = \rho(0) e^{-\Gamma t} + (1 - e^{-\Gamma t})\frac {I} {2}.
\end{equation}
Thus the depolarizing noise is the quantum operation that depolarizes
the state into a completely mixed state. The depolarizing channel is
so simple that it is often used in theoretical investigations related
with the effect of quantum noise.

Let $\gamma=1-e^{-\Gamma t}$. Then the Kraus operators for the
depolarizing channel as the following.

\begin{svgraybox}
  \begin{center}
    \textbf{Box 2.14 Kraus operators for depolarizing}

    $E_0=\sqrt{1-\gamma}I,\quad E_1=\sqrt{\frac {\gamma}
      {3}}\sigma_{x},\quad E_2=\sqrt{\frac {\gamma}
      {3}}\sigma_{y},\quad E_3=\sqrt{\frac {\gamma}
      {3}}\sigma_{z}.$
  \end{center}
\end{svgraybox}

\section{Summary and further reading}

In this chapter, we have discussed evolution for a quantum system $S$
with Hilbert space $\mathcal{H}_{S}$, whose quantum state is described
by a density matrix $\rho_S$. In the ideal case, the evolution of the
wave function $\ket{\psi_{S}}\in\mathcal{H}_{S}$ is unitary, which is
governed by the Schr\"{o}dinger's equation. This unitary evolution
gives rise to the circuit model of quantum computation, where the
computational procedure is to `apply' single- and two-bit unitary
operations to the quantum state carrying information of the
computation. This quantum circuit viewpoint is practical as usual
Hamiltonians for an interacting systems involve only few-body
interactions, and in the most cases two-body interactions, which can
be use to carry out single- and two-qubit unitary operations.

It has been shown in quantum information theory that arbitrary
single-qubit unitary operators and a non-trivial two-qubit unitary
operator (e.g. the controlled-NOT) are enough to construct any
$N$-qubit unitary operator. Historically, this was first shown
in~\cite{DiV95}, which is extended and simplified in the follow-up
paper~\cite{BBC+95}.

Our treatment in Section 2.3 starting from two-level unitaries then
controlled-NOTs is according to Chapter 4.5 of Nielsen and Chuang's
book~\cite{NC00}. This approach is originally in~\cite{RZB+94}. It was
shown in ~\cite{DBE95} and independently in~\cite{Llo95} that almost
any two-qubit unitary operation can be used to construct any $N$-qubit
unitary.

The circuit model of quantum computing is originally due to Deutsch's
1989 work~\cite{Deu89}. The circuit diagram we used in
Fig.~\ref{fig:circuit} is not a standard one used in quantum
computation. In fact, circuit diagrams are draw with time evolution
from left to right, and standard unitaries such as single-particle
Pauli operators and controlled-NOT have their corresponding notation
used for quantum circuits in quantum computing literatures. Readers
interested in quantum circuits should refer to textbooks in quantum
computing, for instance Chapter 4 in~\cite{NC00}. It is not the goal
of this chapter to be involved too much with quantum circuit theory.
Instead, we would introduce only the very basic concepts such as
circuit size and depth, and in diagrams as Fig.~\ref{fig:circuit}, we
adopt the tradition in theoretical physics to treat time evolution
from down to up.

The idea for simulation of time evolution of many-body quantum systems
by a quantum computer dates back to Feynman's famous 1982
paper~\cite{Fey82}. The Lie product formula is due
to~\cite{Tro59}. Readers interested in quantum simulation may refer to
Chapter 4 of~\cite{NC00}, and references therein.

The theory of open quantum systems is extensively developed in the
field of quantum theory and quantum optics, where many good textbooks
are available for readers interested in this subject
(e.g.~\cite{KBD+83,GZ04}).

The Kraus operators are due to Kraus~\cite{KBD+83}. The Lindblad form
is due to Lindblad~\cite{Lin76}. Our treatment on the master equations
for amplitude damping noise is based on ~\cite{CDG98}. There are also
many literatures in quantum information science discussing these noise
and their Kraus operators. Interested reader may refer to Chapter 8
of~\cite{NC00} and references therein.

%
%
\bibliographystyle{plain}
\bibliography{Chap2}


%
%
%
\chapter{Quantum Error-Correcting Codes}
\label{cp:3} 

\abstract{Any quantum system inevitably interacts with the environment which
  causes decoherence. While the environment is generally inaccessible,
  can we protect our system against noise to maintain its quantum
  coherence? One technique developed in quantum information science,
  called the quantum error-correcting codes, does the job.  The main
  idea is to `encode' the system into a subspace of the entire
  $N$-qubit space, called the `code space', such that the errors
  caused by decoherence of the system can be `corrected'.}
  
\section{Introduction}

We have discussed in Chapter~\ref{cp:2} that the evolution of a quantum system is
in general non-unitary, which is caused by the inevitable interaction
of the system with its environment. This is some bad news for quantum
coherence, which is also the biggest obstacle for realizing large
scale quantum computer in practice. It is not the goal of this book to
discuss how to build a practical quantum computer. However, the
techniques developed in quantum information theory for fighting
against decoherence, turn out to have dramatic nice connection to
modern condensed matter physics. The topic of this chapter is to
introduce these techniques.

The central idea is to `correct' the errors induced by non-unitary
evolution. The idea of `error correction' is actually borrowed from
classical information theory that our modern life relies on every
day. That is, when we communicate with each other, through either
phones or internet, the communication channels between us are
noisy. Therefore, information transmitted inevitably encounters errors
that need to be corrected -- the simplest idea is to send the same
message multiple times. However, quantum information (carried by
quantum states) is dramatically different from class information as
they cannot be copied (no cloning theorem discussed in Chapter~\ref{cp:1}).

The breakthrough came in when it is realized that entanglement does
help with maintaining coherence. Consider a case of two qubits, where
the noise is to flip the phase of either the first qubit or the second
qubit, each with probability $\frac{1}{2}$. In other words, the Kraus
operators are $\{\frac{1}{\sqrt{2}}Z_1,\frac{1}{\sqrt{2}}Z_2\}$. Now
consider the state $\alpha\ket{00}+\beta\ket{11}$, then the evolution
of this state under the noise is always unitary, which is in fact just
$Z_1$ (or equivalently $Z_2$). This is to say, although the general
evolution of states in the total four dimensional Hilbert space
spanned by $\{\ket{00},\ket{01},\ket{10},\ket{11}\}$ is non-unitary,
the evolution of any state in the two dimensional subspace spanned by
$\ket{00},\ket{11}$ is unitary.

This simple example is indeed artificial. In general, one would like
to know for the real physical noise, whether such a subspace with
unitary evolution exists. Unfortunately, although such subspaces do
exist for some cases, for many cases they do not exist. One can
imagine another example of two qubits, where what the noise does, is
to flip the phase of either the first qubit or the second qubit, or
does nothing at all, which is a practical situation (phase flip) that
the artificial one discussed above. In other words, the Kraus
operators are $\{\frac{1}{\sqrt{3}}I,
\frac{1}{\sqrt{3}}Z_1,\frac{1}{\sqrt{3}}Z_2\}$. Now one can check that
the evolution of the states in the subspace spanned by
$\ket{00},\ket{11}$ is no longer unitary. In fact, there does not
exist nontrivial subspace (i.e. dimension $>1$) where the evolution
under the noise could be unitary.

This is not the end of the story. Surprisingly, it turns out that
measurements can help maintaining coherence. This is very
counterintuitive, as general measurements project the quantum state to
subspaces thus destroys coherence. We will explain in the next section
how this could actually work out. The idea of measurements will then
further leads to a general understanding how to `error correct' for a
known type of noise, called the `quantum error correction
criterion'. However, this elegant criterion does not directly provide
practical ways of finding subspaces that correct the errors of given
noise. There is indeed a practical method, called the stabilizer
formalism, which finds those subspaces, that we will also
introduce. Finally, we discuss the connection of stabilizer formalism
to topology, using the example of the so called `toric code'.

\section{Basic idea of error correction}
\label{sec:cp3sec2}

\subsection{Bit flip code}

Let us start to consider a simple example for a single qubit. Suppose
the noise of the systems is to flip $\ket{0}$ to $\ket{1}$ and vice
versa with probability $p$, i.e. the superoperator for this bit flip
noise is given by
\begin{equation}
\mathcal{E}_{BF}(\rho)=(1-p)\rho+pX\rho X,
\end{equation}
i.e. the Kraus operators are $\{\sqrt{1-p}I,\sqrt{p}X\}$.

Now suppose we have a single qubit pure state
$\vert\phi\rangle=\alpha\ket{0}+\beta\ket{1}$ that we hope to maintain the coherence
for unitary time evolution. However, due to the bit flip noise, we
will end up in a mixed state
\begin{equation}
\sigma=(1-p) \vert\phi\rangle \langle\phi\vert + p X \vert\phi\rangle
\langle\phi\vert X.
\end{equation}
Then the probability of failure is error due to noise is then
reasonably given as
\begin{equation}
p_{err} = 1-\langle\phi\vert\sigma\vert\phi\rangle = p
(1-\langle\phi\vert X \vert\phi\rangle^{2})=p(1-|\alpha^*\beta+\beta^*\alpha|^2),
\end{equation}
which is of order $p$ for general $\alpha,\beta$.

Now in order to maintain the coherence, we wish to correct error and
recover the original state $\ket{\phi}$.  This is too
much to hope for at the first place. As we already discussed, in
general one can not find a subspace where the evolution could be
unitary. Instead, let us try something more reasonable, that is, to
reduce the error probability by one order of magnitude. In other
words, we want to reduce the error probability from order $t$ to the
order of $t^2$. The simplest idea maybe that we copy the state for
three times. However recall that due to the no cloning theorem, this
cannot be done for unknown states. Instead of copying the state
itself, we `copy' the basis states three times.
\begin{equation}
\label{eq:qrepete}
\ket{0}\rightarrow \ket{000},\quad \ket{1}\rightarrow \ket{111}.
\end{equation}
In other words, instead of having a single qubit
$\ket{\phi}=\alpha\ket{0}+\beta\ket{1}$, we now have three qubit which is in the
state
\begin{equation}
\ket{\psi}=\alpha\ket{000}+\beta\ket{111}.
\end{equation}
Then at the receiver's end, the output state is a mixed state $\rho$
given by
\begin{eqnarray}
\rho&=&\mathcal{E}_{BF}^{\otimes 3}(\ket{\psi}\bra{\psi})\nonumber\\
&=&(1-p)^3\ket{\psi}\bra{\psi}\nonumber\\
&&+\;(1-p)^2p\left(X_1\ket{\psi}\bra{\psi}X_1+X_2\ket{\psi}\bra{\psi}X_2+X_3\ket{\psi}\bra{\psi}X_3\right)\nonumber\\
&&+\;(1-p)p^2\left(X_1X_2\ket{\psi}\bra{\psi}X_1X_2+X_2X_3\ket{\psi}\bra{\psi}X_2X_3+X_1X_3\ket{\psi}\bra{\psi}X_1X_3\right)\nonumber\\
&&+\;p^3(X_1X_2X_3\ket{\psi}\bra{\psi}X_1X_2X_3),
\end{eqnarray}
where $X_i$ is the Pauli operator acting on the $i$th qubit, for
instance, $X_1=X\otimes I\otimes I$ (and sometimes we write $XII$ for
short).

Our goal is to recover the transmitted state
$\alpha\ket{0}+\beta\ket{1}$ as much as we can. Our strategy is that whenever we
receive any of $000,001,010,100$ we would like to interpret it as $0$, and whenever
we receive any of $111,110,101,011$ we interpret it as $1$. However,
the difficulty in the quantum case is that we will need to keep the
coherence between $\ket{0}$ and $\ket{1}$, that, to recover the
superposition $\alpha\ket{0}+\beta\ket{1}$. In order to maintain the
coherence, we perform an orthogonal measurement $\mathcal{M} $ with
Kraus operators given as follows:
\begin{eqnarray}
\label{eq:proj}
\Pi_0&=&\ket{000}\bra{000}+\ket{111}\bra{111},\nonumber\\
\Pi_1&=&X_1(\ket{000}\bra{000}+\ket{111}\bra{111})X_1,\nonumber\\
\Pi_2&=&X_2(\ket{000}\bra{000}+\ket{111}\bra{111})X_2,\nonumber\\
\Pi_3&=&X_3(\ket{000}\bra{000}+\ket{111}\bra{111})X_3.
\end{eqnarray}
Then we get either
\begin{equation}
\sigma_{0} = (1-p)^3 \ket{\psi}\bra{\psi} + p^3 X_1X_2X_3\ket{\psi}\bra{\psi}X_1X_2X_3,
\end{equation}
or
\begin{equation}
\sigma_{1}=(1-p)^2pX_1\ket{\psi}\bra{\psi}X_1+(1-p)p^2X_2X_3\ket{\psi}\bra{\psi}X_2X_3,
\end{equation}
or
\begin{equation}
\sigma_{2}=(1-p)^2pX_2\ket{\psi}\bra{\psi}X_2+(1-p)p^2X_1X_3\ket{\psi}\bra{\psi}X_1X_3,
\end{equation}
or
\begin{equation}
\sigma_{3}=(1-p)^2pX_3\ket{\psi}\bra{\psi}X_3+(1-p)p^2X_1X_2\ket{\psi}\bra{\psi}X_1X_2,
\end{equation}
according to the measurement result. Note that $\sigma_{i}$s are not
normalized so that we can calculate the probability of getting each
$\sigma_{i}$ by its trace.

Now we interpret $000,001,010,100$ as $0$ and $111,110,101,011$ as
$1$, so when we get $\sigma_{0}$, we do the inverse of
Eq.~(\ref{eq:qrepete}); when we get $\sigma_{1}$, we perform $X_1$
and then the inverse of Eq.~(\ref{eq:qrepete}); when we get
$\sigma_{2}$, we perform $X_2$ and then the inverse of
Eq.~(\ref{eq:qrepete}); when we get $\sigma_{3}$, we perform $X_3$
and then the inverse of Eq.~(\ref{eq:qrepete}). Finally we get one of
the following, respectively.
\begin{eqnarray}
  \sigma_{0}^{\prime} & = & (1-p)^3 \ket{\psi}\bra{\psi} + p^3
  X_1X_2X_3\ket{\psi}\bra{\psi}X_1X_2X_3, \nonumber{}\\
  \sigma_{1}^{\prime} & = & (1-p)^{2}p \ket{\psi}\bra{\psi} + (1-p)p^{2}
  X_1X_2X_3\ket{\psi}\bra{\psi}X_1X_2X_3, \nonumber{}\\
  \sigma_{2}^{\prime} & = & (1-p)^{2}p \ket{\psi}\bra{\psi} + (1-p)p^{2}
  X_1X_2X_3\ket{\psi}\bra{\psi}X_1X_2X_3, \nonumber{}\\
  \sigma_{3}^{\prime} & = & (1-p)^{2}p \ket{\psi}\bra{\psi} +
  (1-p)p^{2}  X_1X_2X_3\ket{\psi}\bra{\psi}X_1X_2X_3.
\end{eqnarray}
That is to say, the final state we receive is
\begin{equation}
\sigma^{\prime}=\sum_{k=0}^3 \sigma_{i}^{\prime}
\end{equation}

Then the probability of failure is given by
\begin{eqnarray}
p'_{err}& = & 1- \langle\psi\vert \sigma^{\prime} \vert\psi\rangle\nonumber{}\\
 & = & p^{2}(3-2p) (1-\langle\psi\vert X_{1}X_{2}X_{3}\vert\psi\rangle^{2})\nonumber\\
&=& p^{2} (3-2p) (1-|\alpha^*\beta+\beta^*\alpha|^2),
\end{eqnarray}
which is of order $p^2$ for general $\alpha,\beta$.  For a given
$\alpha,\beta$ and $p<\frac{1}{2}$, we have $p'_{err}<p_{err}$,
meaning that we are able to reduce the error probability by adding
redundancy.

Now let us ask the question of how we would be able to reduce
the error probability from $p$ to $p^r$ for the bit flip noise.
A simple method would be to `copy' the basis states $2r+1$ times.
That is
\begin{equation}
\label{eq:rep}
\ket{0}\rightarrow\ket{0}^{\otimes (2r-1)},\quad
\ket{1}\rightarrow\ket{1}^{\otimes (2r-1)}
\end{equation}
Now large enough $r$ could get the error probability $p^r$ arbitrarily small,
so we can protect our qubits $\alpha\ket{0}+\beta\ket{1}$ almost perfectly
against bit flip noise.

\subsection{Shor's Code}

However, as discussed in Chapter~\ref{cp:2}, quantum noise are in general much
more complicated than just bit flip. For
instance, the depolarizing noise $\mathcal{E}_{DP}$ models the qubit
noise in a more general situation, where $X$, $Y$, $Z$ errors are
likely happen with equal probability. Is there any way that we can
reduce the error probability by adding redundancy also for this
kind of noise?

Before look into this question, let us look at what we could do for
the phase flip noise $\mathcal{E}_{PF}$. In this case, we show that
the code given by Eq.~\eqref{eq:rep} does not correct even a single
phase flip error.

In this case, one can
simply use a similar idea as for the bit flip noise in
Eq.~(\ref{eq:qrepete}). Recall that $HXH=Z$ and $HZH=X$, where $H$ is
the Hadamard gate, and define
\begin{eqnarray}
\ket{+}&=&H\ket{0}=\frac{1}{\sqrt{2}}(\ket{0}+\ket{1})\nonumber\\
\ket{-}&=&H\ket{1}=\frac{1}{\sqrt{2}}(\ket{0}-\ket{1}),
\end{eqnarray}
we can simply modify Eq.~(\ref{eq:qrepete}) as
\begin{equation}
\label{eq:qrepeteZ}
\ket{0}\rightarrow \ket{+++},\quad \ket{1}\rightarrow \ket{---}.
\end{equation}
In other words, instead of transmitting a qubit $\alpha\ket{0}+\beta\ket{1}$, we transmit
$\ket{\psi}=\alpha\ket{+++}+\beta\ket{---}$,
then all the other analysis goes through by replacing all the $X$s with $Z$s.

Now back to the depolarizing noise $\mathcal{E}_{DP}$,  
\begin{equation}
\mathcal{E}_{DF}(\rho)=(1-p)\rho+\frac{p}{3}(X\rho X+Y\rho Y+Z\rho Z),
\end{equation}
where the Kraus operators are $\{I,X_j,Y_j,Z_j\}$ (as discussed in Chapter~\ref{sec:masterqubit}, here we use the parameter $p$ instead of $\Gamma$).

Note that $Y\propto XZ$, so if it is
possible to tell that both $X$, $Z$ happen, then it means an $Y$ error
happens. In other words, there might be a way to combine the idea of
both Eq.~(\ref{eq:qrepete}) and Eq.~(\ref{eq:qrepeteZ}) such that one
can tell whether an $X$ and a $Z$ error happen. This indeed works as
first observed by Shor, in the following way
\begin{eqnarray}
\label{eq:shorcode}
\ket{0} &\rightarrow & \frac{1}{2\sqrt{2}}
(\ket{000}+\ket{111})^{\otimes 3} \equiv \ket{0_L}\nonumber\\
\ket{1} & \rightarrow & \frac{1}{2\sqrt{2}}
(\ket{000}-\ket{111})^{\otimes 3} \equiv \ket{1_L}.
\end{eqnarray}
Then instead of transmitting a qubit $\alpha\ket{0}+\beta\ket{1}$, we
transmit $\ket{\psi_L}=\alpha\ket{0_L}+\beta\ket{1_L}$. In other
words, in order to reduce the error probability from order $p$ to
order $p^2$, we use $9$ qubits to represent one qubit.

We leave the details of the calculation for
$\mathcal{E}_{DP}^{\otimes{9}}(\ket{\psi_L}\bra{\psi_L})$ to the
reader. We remark that the orthogonal measurement $\mathcal{M}$ we
perform will be given by projections of the form
$R_i(\ket{0_L}\bra{0_L}+\ket{1_L}\bra{1_L})R_i$, where
$R\in\{I,X,Y,Z\}$ and $i\in [1,2,\ldots 9]$. For each measurement
result, we perform $R_i$ and the reverse of
Eq.~(\ref{eq:shorcode}). At the end of the day we will successfully
reduce the error probability from order $p$ to order $p^2$.

\subsection{Other noise models}

One would ask what happens in those more practical cases such as the
phase damping noise and the amplitude damping noise as discussed in
Chapter~\ref{sec:masterqubit}. Let us first discuss the phase damping noise
\begin{equation}
\mathcal{E}_{PD}(\rho)=\sum_{j=0}^2 E_j\rho E_j^{\dag},
\end{equation}
with the Kraus operators given in Chapter~\ref{sec:masterqubit}, which we rewrite as follows (we use the parameter $p$ instead of $\gamma$)
\begin{equation}
E_0=\sqrt{1-p}I,\quad E_1=\frac{\sqrt{p}}{2}(I+Z),
\quad E_2=\frac{\sqrt{p}}{2}(I-Z).
\end{equation}
We note that the Kraus operators are linear combinations of $I$ and
$Z$, which is in some sense similar to the Kraus operators of the
phase flip noise with Kraus operators $\sqrt{1-p}I,\sqrt{p}Z$. So we
would wonder whether it is possible to reduce the error probability
for the phase damping noise using the same method as we have done for
the phase flip noise?

Let us go ahead to examine what happens when we transmit 
$\ket{\psi}=\alpha\ket{+++}+\beta\ket{---}$ instead of  $\alpha\ket{0}+\beta\ket{1}$. And compute
\begin{eqnarray}
\label{eq:PDc}
\rho&=&\mathcal{E}_{PD}^{\otimes 3}(\ket{\psi}\bra{\psi})\nonumber\\
&=&(1-\frac{3}{2}p+\frac{3}{4}p^2-\frac{1}{8}p^3)\ket{\psi}\bra{\psi}\nonumber\\
&+&(\frac{1}{2}p-\frac{1}{2}p^2+\frac{1}{8}p^3)\left(Z_1\ket{\psi}\bra{\psi}Z_1+Z_2\ket{\psi}\bra{\psi}Z_2+Z_3\ket{\psi}\bra{\psi}Z_3\right)\nonumber\\
&+&(\frac{1}{4}p^2-\frac{1}{8}p^3)\left(Z_1Z_2\ket{\psi}\bra{\psi}Z_1Z_2+Z_2Z_3\ket{\psi}\bra{\psi}Z_2Z_3+Z_1Z_3\ket{\psi}\bra{\psi}Z_1Z_3\right)\nonumber\\
&+&\frac{1}{8}p^3(Z_1Z_2Z_3\ket{\psi}\bra{\psi}Z_1Z_2Z_3).
\end{eqnarray}
It is then clear that the orthogonal measurement given by
Eq.~(\ref{eq:proj}) (replacing all $X$s by $Z$s) followed by the same
procedure of correction as for the phase flip noise works to reduce
the error probability from order $p$ to $p^2$.

We remark here that in Eq.~(\ref{eq:PDc}), the cross terms of the
forms, for instance $Z_1\ket{\psi}\bra{\psi}$ (or
$\ket{\psi}\bra{\psi}Z_1$) cancels in this special case. In general,
there would be such terms. However, this will not be a problem as when
we perform orthogonal measurements, these terms vanish. The above
example then illustrate that, in general, \textit{if we can deal with
Kraus operators $A$ and $B$, we can also deal with Kraus operators given
by any kind of superposition of $A$ and $B$}.

This then gives more meaning to the depolarizing noise with Kraus
operators $\propto I,X,Y,Z$, as they form a basis for $2\times 2$
matrices. That is to say, if we can able to reduce error probability
for the depolarizing noise (indeed we do as using the Shor's
method), then are able to reduce error probability for any qubit
quantum noise using the same method. For instance, consider the amplitude
damping noise discussed in Chapter~\ref{sec:masterqubit}, one can then rewrite the Kraus operators as (we use the parameter $p$ instead of $\gamma$)
\begin{eqnarray}
  E_0 & = & \begin{pmatrix} 1 & 0 \\0 &
    \sqrt{1-p} \end{pmatrix}=\frac{1+\sqrt{1-p}}{2}I+\frac{1-\sqrt{1-p}}{2}Z,\\
  E_1 & = & \begin{pmatrix} 0 &
    \sqrt{p} \\0 & 0 \end{pmatrix}=\frac{\sqrt{p}}{2}(X+iY),
\label{eq:ADKraus}
\end{eqnarray}
then using Shor's method to reduce the error probability from order
$p$ to $p^2$.

\section{Quantum error-correcting criteria, code distance}
\label{sec:QCC}

We have seen that how to reduce the error probability from order $p$
to $p^2$ using Shor's code. From other point of view, if only one
error happens, meaning $\mathcal{E}_{PD}\otimes I\otimes I$ (or
$\mathcal{E}_{PD}$ could be on the second or third qubit), the Shor's
code can maintain the coherence completely. Or in other words, we say
that Shor's code is capable of correcting one error.

We now wonder what is the the general case for a quantum code capable
of correcting certain types of errors. Before looking into that, let
us consider what really a quantum code is. From what we have done in
the previous section, we know that by mapping the basis vectors and
allowing all the superpositions, what we result in is a `subspace' of
the $N$-qubit Hilbert space.

\begin{svgraybox}
\begin{center}
\textbf{Box 3.1 Quantum code}

A quantum code is a subspace of the $N$-qubit Hilbert space.
\end{center}
\end{svgraybox}

For a given subspace, there are several ways
to describe the space. First of all one can choose an orthonormal
basis $\{\ket{\psi_i}\}$. Or, one can use the projection onto
the code space
\begin{equation}
\label{eq:projection}
\Pi=\sum_i\ket{\psi_i}\bra{\psi_i}.
\end{equation}

Now suppose the error of the system is characterized by the quantum noise
$\mathcal{E}=\{E_k\}$, where $E_k$s are the Kraus operators. In order
to distinguish any basis state corrupted by an error,
i.e. $E_k\ket{\psi_i}$ from any other basis state corrupted by another
error, i.e. $E_l\ket{\psi_j}$, one must have
$E_k\ket{\psi_i}\perp{E_l}\ket{\psi_j}$. 
Mathematically, this then means
\begin{equation}
\bra{\psi_i}E_k^{\dag}E_l\ket{\psi_j}=0,\quad i\neq j.
\end{equation}

Now let us see what could happen when $i=j$. In this case, in order to
maintain coherence, imagine the case that each Kraus operator has
every basis state $\ket{\psi_i}$ as its eigenvector with the same
eigenvalue, i.e. $E_k\ket{\psi_i}=c_k\ket{\psi_i}$, which is
independent of $i$. Therefore, for any state
$\ket{\psi}=\sum_i\alpha_i\ket{\psi_i}$ which is in the code space, we
have $E_k\ket{\psi}=c_k\ket{\psi}$, that is, up to an irrelevant
constant, each Kraus operator $E_k$ acts like identity on the code
space, thus maintains coherence. To summarize, we then have the
following

\begin{svgraybox}
\begin{center}
\textbf{Box 3.2 Quantum error-correcting criteria}

A quantum code with orthonormal basis $\{\ket{\psi_i}\}$ corrects the error set $\mathcal{E}=\{E_k\}$ if and only if
$$
\bra{\psi_i}E_k^{\dag}E_l\ket{\psi_j}=c_{kl}\delta_{ij}.
$$
\end{center}
\end{svgraybox}

One can equivalently formulate this criterion in terms of the
projection $\Pi$ onto the code space as given in
Eq.~(\ref{eq:projection}). Let us look at the quantity
\begin{eqnarray}
\Pi E_k^{\dag}E_l \Pi=\sum_{i,j}\ket{\psi_i}\bra{\psi_i}E_k^{\dag}E_l\ket{\psi_j}\bra{\psi_j},
\end{eqnarray}
using Eq.~(\ref{eq:projection}) one then gets 
\begin{equation}
\Pi E_k^{\dag}E_l \Pi=c_{kl}\Pi.
\end{equation}

As an example, now let us apply this criterion to check why Shor's
code is capable of correcting an arbitrary single error. Here the
correctable error set is
\begin{equation}
\label{eq:singlerror}
\mathcal{E}=\{I,X_i,Y_i,Z_i\},
\end{equation}
where $i=1,2,\ldots,9$. And now our code basis can be chosen as
$\ket{\psi_j}=\ket{j_L}$ ($j=0,1$), as given in Eq.~\eqref{eq:shorcode}.

So first we need to check
\begin{equation}
\bra{\psi_0}E_k^{\dag}E_l\ket{\psi_1}=0,
\end{equation}
for any $E_k,E_l\in\mathcal{E}$, which is pretty
straightforward. 

Then we need to check 
\begin{equation}
\bra{\psi_0}E_k^{\dag}E_l\ket{\psi_0}=\bra{\psi_1}E_k^{\dag}E_l\ket{\psi_1},
\end{equation}
for any $E_k,E_l\in\mathcal{E}$. Note that in general we no
longer have
$\bra{\psi_0}E_k^{\dag}E_l\ket{\psi_0}=\bra{\psi_1}E_k^{\dag}E_l\ket{\psi_1}=0$. Rather,
for instance, we have
$\bra{\psi_0}Z_1Z_2\ket{\psi_0}=\bra{\psi_1}Z_1Z_2\ket{\psi_1}=1$,
since $Z_1Z_2\ket{\psi_0}=\ket{\psi_0}$ and
$Z_1Z_2\ket{\psi_1}=\ket{\psi_1}$. One have similar results for
\begin{equation}
Z_2Z_3,Z_1Z_3,Z_4Z_5,Z_5Z_6,Z_4Z_6,Z_7Z_8,Z_8Z_9,Z_7Z_9.
\end{equation}
And or other choices of $E_k,E_l\in\mathcal{E},\ k\neq l$, one has
$\bra{\psi_0}E_k^{\dag}E_l\ket{\psi_0}=\bra{\psi_1}E_k^{\dag}E_l\ket{\psi_1}=0$. 

In practice, the most common noise is uncorrelated ones. That is,
those noise acting independently on each qubit. And in the most
discussed cases, the single qubit noise is chosen as the depolarizing
noise. Therefore, the quantum noise under consideration is
$\mathcal{E}_{DP}^{\otimes N}$.

In this case one often measures the `strength' of an error-correcting
code by the number of errors the code is capable of correcting. In
other words, if the code corrects $t$-errors, then it reduces the
error probability from order $p$ to $p^{t+1}$. This strength can be
measured by a parameter called code distance. In order to understand
code distance, let us first look at an $N$-qubit operator $O$ of the
form
\begin{equation}
O=O_1\otimes O_2,\ldots,\otimes O_N,
\end{equation}
where each $O_k$ acting on the $k$th qubit. We are interested in those
non-trivial $O_k$s, i.e. those $O_k$s which are different from
identity. The number of those non-trivial $O_k$s is then called the
weight of $M$, denoted by $\text{wt}(O)$. Apparently
$0\leq\text{wt}(O)\leq N$. When considering the depolarizing noise
$\mathcal{E}_{DP}^{\otimes N}$, where we want a quantum code capable
of correcting $t$-errors, it is enough to consider only Kraus operator
$O$ of weight $\leq t$ where each $O_k$ are one of the Pauli operators
$\{I,X_k,Y_k,Z_k\}$. In other words, a code is capable of correcting
$t$ errors for any $O$ with weight $\leq 2t+1$, the following holds
\begin{equation}
\bra{\psi_i}O\ket{\psi_j}=c_O\delta_{ij},
\end{equation} 
where $c_O$ is a constant that is independent of $i,j$.
Now we are ready to introduce the concept of code distance.

\begin{svgraybox}
\begin{center}
\textbf{Box 3.3 Quantum code distance}

The distance for quantum code with orthonormal basis
$\{\ket{\psi_i}\}$ is the largest possible weight
$d$ such that
$$\bra{\psi_i}O\ket{\psi_j}=c_O\delta_{ij}$$
holds for all operators $O$ with wt$(O)< d$.
\end{center}
\end{svgraybox}

In other words, the distance of a quantum code is given by the
smallest possible weight of $O$ such that
$\bra{\psi_i}O\ket{\psi_j}=c_O\delta_{ij}$ violates. If we consider
the situation where $i\neq j$, then code distance is the smallest
possible weight of $O$ such that $\ket{\psi_i},\ket{\psi_j}$ are no
longer distinguishable (i.e. orthogonal). Intuitively the code
distance measures how far one basis state in the code space is `away
from' another basis state, hence the name `distance'.

\section{The stabilizer formalism}

Quantum error correction criterion gives a `standard' of finding
quantum codes. Once the error set $\mathcal{E}$ is fixed, then the
problem of finding the corresponding quantum code reduces to solve the
equations for the unknown basis $\ket{\psi_i}$. This is in general not
a practical way. The problem is that each $N$-qubit quantum state
$\ket{\psi_i}$ is specified by $\sim 2^N$ complex parameters, which
makes the equations almost impossible to solve.

Therefore one needs to find a better way such that the states
$\ket{\psi_i}$ can be represented in a more efficient manner, i.e. by
less parameters. We know that doing this will no longer allow
$\ket{\psi_i}$s to be general $N$-qubit quantum state, but just a
restricted sets of states. However, we will see that such restricted
sets of states are with nice structure which makes things easy to
understand. This set of states that we are going to discuss will then
be so called `stabilizer' state, and the corresponding quantum code is
then called `stabilizer' code.

\subsection{Shor's code}
To establish the idea of the stabilizer formalism, let us again look
at the example of Shor's code. We have already noticed that
$Z_1Z_2\ket{0_L}=\ket{0_L}$ and $Z_1Z_2\ket{1_L}=\ket{1_L}$. This
means that for any state $\ket{\phi}=\alpha\ket{0_L}+\beta\ket{1_L}$,
one has $Z_1Z_2\ket{\phi}=\ket{\phi}$. Or one can instead write
${\Pi}Z_1Z_2\Pi=\Pi$, where
$\Pi=\ket{0_L}\bra{0_L}+\ket{1_L}\bra{1_L}$, is the projection onto
the code space.

We will say that $Z_1Z_2$ `stabilizes' Shor's code, in a sense that
the code is invariant under $Z_1Z_2$. Similarly, we know that
$Z_2Z_3,Z_4Z_5,Z_5Z_6,Z_7Z_8,Z_8Z_9$ also stabilize the
code. Furthermore, note that $X_1X_2X_3X_4X_5X_6,X_4X_5X_6X_7X_8X_9$
also stabilizes the code. To summarize, we have each row of
\begin{equation}
\label{eq:ShorStab}
\begin{array}{lllllllll}
Z & Z & I & I & I & I & I & I & I\\
I & Z & Z & I & I & I & I & I & I\\
I & I & I & Z & Z & I & I & I & I\\
I & I & I & I & Z & Z & I & I & I\\
I & I & I & I & I & I & Z & Z & I\\
I & I & I & I & I & I & I & Z & Z\\
X & X & X & X & X & X & I & I & I\\
I & I & I & X & X & X & X & X & X\\
\end{array}
\end{equation}
stabilizes the code.

Now let us observe an important factor: every two rows, viewed as
operators on nine qubits, commute with each other, and they hence have
common eigenspace. What is more, the operator of each row squares to
identity, which means it has only eigenvalues $\pm 1$. The Shor code
is nothing but the common eigenspace of these eight operators with
eigenvalue $1$ for each operators.

Let us denote each row of Eq.~\eqref{eq:ShorStab} by $g_i$, where $i=1,2,\ldots,8$. Indeed,
the product of any number of $g_i$s also stabilizes Shor's
code. Therefore, it is actually the group generated by $g_i$s
($i=1,2,\ldots,8$) that stabilizes the code. This group is indeed
abelian, as any two elements commute. And it is also straightforward
to check that the order of the group is $2^8$. This group is called
the stabilizer group (or in short, the stabilizer) of Shor's code,
denoted by $\mathcal{S}$, and the $g_i$s are the generators of
$\mathcal{S}$. That is,
\begin{equation}
\mathcal{S}=\langle g_1,g_2,g_3,g_4,g_5,g_6,g_7,g_8\rangle.
\end{equation}

We know that $\mathcal{S}$ completely specifies Shor's code as the
eigenspace of each group element with eigenvalue $1$. Indeed, the
projection on to the code space
$\ket{0_L}\bra{0_L}+\ket{1_L}\bra{1_L}$ can be written in terms of
stabilizers as
\begin{equation}
\Pi=\frac{1}{2^8}\prod_{i=1}^8(I+g_i).
\end{equation}

To see why this is the case, let us first compute
\begin{equation}
\Pi^2=\frac{1}{2^{16}}\prod_{i=1}^8(I+g_i)^2=\Pi,
\end{equation}
so $\Pi$ is a projection. The second equality holds because
$(I+g_i)^2=2(I+g_i)$ as $g_i^2=I$. Then we further check that
$\Pi(\ket{0_L}\bra{0_L}+\ket{1_L}\bra{1_L})\Pi=\ket{0_L}\bra{0_L}+\ket{1_L}\bra{1_L}$. Moreover, because the stabilizer group contains 8 independent generators, the code space is at most two dimensional. Therefore
$\Pi=\ket{0_L}\bra{0_L}+\ket{1_L}\bra{1_L}$. 

The big advantage of specifying the code by stabilizers, is that one
only needs in general order $n$ generators instead of some basis
states $\ket{\psi_i}$ each needs $\sim 2^n$ complex
parameters. Another interesting thing of this stabilizer formalism is
that one can write an Hamiltonian
\begin{equation}
H=-\sum_{i=1}^8 g_i,
\end{equation}
hence the code space is nothing but the ground state space of $H$.

Concerning the basis states $\ket{0_L}$ and $\ket{1_L}$, note that we
have $X^{\otimes 9}\ket{0_L}=\ket{0_L}$, $X^{\otimes
  9}\ket{1_L}=-\ket{1_L}$. In this sense, $X^{\otimes 9}$ acts like a
`logical $Z$' on the code basis, that let us write
$Z_L=X^{\otimes{9}}$. Or in other words, the basis state $\ket{0_L}$
is stabilized by
\begin{equation}
\mathcal{S}_0=\langle g_1,g_2,g_3,g_4,g_5,g_6,g_7,g_8,Z_L\rangle.
\end{equation}
As $g_7=X_1X_2X_3X_4X_5X_6$, one can equivalently choose
$Z_L=X_1X_2X_3$.  Similarly, one can then choose $X_L=Z_1Z_4Z_6$, in a
sense that $X_L\ket{0_L}=\ket{1_L}$, $X_L\ket{1_L}=\ket{0_L}$. And
furthermore $X_LZ_L=-Z_LX_L$, which is the usual commutation relation
for Pauli operators.

Similarly, the projection onto the the space $\ket{0_L}\bra{0_L}$ is 
\begin{equation}
\ket{0_L}\bra{0_L}=\frac{1}{2^9}\prod_{i=1}^8(I+g_i)(I+Z_L).
\end{equation}
One can also write the state $\ket{0_L}$ in terms of stabilizer
elements in $\mathcal{S}_0$ as
\begin{equation}
\ket{0_L}=\frac{1}{2^{9/2}}\sum_{g\in\mathcal{S}_0}g\ket{0}^{\otimes 9}.
\end{equation}
Indeed, it is enough to use only those elements in $\mathcal{S}_0$
which are products of Pauli $X$ operators.  That is,
\begin{equation}
\ket{0_L}=\frac{1}{2^{3/2}}(I+g_7)(I+g_8)(I+Z_L)\ket{0}^{\otimes 9}.
\end{equation}

Now let us look at the quantum error correction criterion in terms of
the stabilizer formalism. Let us look at $\Pi E_k^{\dag}E_l \Pi$ with
$E_k$s given by Eq.~\eqref{eq:singlerror}. Note that for any
$E_k^{\dag}E_l$, it either commute or anticommute with each $g_i$. Let
us first consider the case that $E_k^{\dag}E_l$ at least anticommute
with one $g_i$, and let us assume it is $g_r$. In this case,
\begin{equation}
\Pi E_k^{\dag}E_l \Pi=E_k^{\dag}E_l \Pi'\Pi
\end{equation}
where
\begin{equation}
\Pi'=\frac{1}{2^8}(I- g_r)\prod_{i=1,i\neq r}^8(I\pm g_i).
\end{equation}
Here $\pm$ means it could be either $+$ or $-$, but not both.

Note that
\begin{equation}
(I-g_r)(I+g_r)=I+g_r-g_r-I=0,
\end{equation}
therefore
\begin{equation}
\Pi E_k^{\dag}E_l \Pi=0,
\end{equation}
i.e. the quantum error correction criterion is satisfied with $c_{kl}=0$.

Now we need to consider the case that $E_k^{\dag}E_l$ commute with all
$g_i$s. Then one will have
\begin{equation}
\Pi E_k^{\dag}E_l \Pi=E_k^{\dag}E_l \Pi\Pi=E_k^{\dag}E_l \Pi.
\end{equation}
However this still not the quantum error correction condition unless
$E_k^{\dag}E_l \Pi=c_{kl}\Pi$.  It is straightforward to check that
this is indeed the case and either $c_{kl}=1$
(e.g. $E_k^{\dag}E_l=Z_1Z_2$) or $c_{kl}=0$
(e.g. $E_k^{\dag}E_l=X_1X_2$), so the quantum error correcting
condition holds. 

Let us now look at the case of two errors. We will show that Shor's
code cannot correct two errors. In this case, the error set is
\begin{equation}
\label{eq:doublerror}
\mathcal{E}=\{I,X_k,Y_k,Z_k, X_lX_m,X_lY_m,X_lZ_m,Y_lY_m,Y_lY_m,Y_lZ_m,Z_lX_m,Z_lY_m,Z_lZ_m\},
\end{equation}
where $k,l,m=1,2,\ldots,9$. 

It is still true that for any $E_k,E_l\in\mathcal{E}$, if
$E_k^{\dag}E_l$ anticommute with at least one $g_i$, then
${\Pi}E_k^{\dag}E_l\Pi=0$. However, if $E_k^{\dag}E_l$ commute with
all $g_i$s, then it is not longer true $E_k^{\dag}E_l \Pi=c_{kl}\Pi$
for certain $E_k^{\dag}E_l \Pi=c_{kl}\Pi$. For instance, choose
$E_k=Z_1$, $E_l=Z_4Z_7$, then $E_k^{\dag}E_l=Z_1Z_4Z_7=X_L$, which is
the logical $X$ of the code space. In this case,
\begin{equation}
E_k^{\dag}E_l \Pi=X_L(\ket{0_L}\bra{0_L}+\ket{1_L}\bra{1_L})=\ket{1_L}\bra{0_L}+\ket{0_L}\bra{1_L},
\end{equation}
which is no longer $c_{kl}\Pi$ for any constant $c_{kl}$. Therefore,
quantum error correction criterion is no longer satisfied, hence
Shor's code cannot correct two errors.

\subsection{The stabilizer formalism}

Let us now look at the general situation of the stabilizer
formalism. We first recall $N$-qubit Pauli operators, which are
operators of the form
\begin{equation}
O_1\otimes O_2,\ldots,\otimes O_N,
\end{equation}
where each $O_k\in\{I_{k},X_k,Y_k,Z_k\}$, is a Pauli operator acting on the $k$th qubit.

Note that all such $N$-qubit Pauli operators together form a group
that we denote by $\mathcal{P}_N$. 

\begin{svgraybox}
\begin{center}
\textbf{Box 3.4 Stabilizer code}

Let $\mathcal{S} \subset \mathcal{P}_N$ be an
abelian subgroup of the Pauli group that does not contain
$-I$, and let 
$$ Q(\mathcal{S}) = \{\ket{\psi}\ {\rm s.t.}\ P\ket{\psi} =
\ket{\psi},\ \forall P \in \mathcal{S} \} .$$
Then $Q(\mathcal{S})$ is a stabilizer code and $\mathcal{S}$ is its stabilizer.
\end{center}
\end{svgraybox}

Let 
\begin{equation}
  \mathcal{S}^\perp = \{ E \in \mathcal{P}_N,\ \text{s.t.}\ [E,S] = 0, \ \forall S \in \mathcal{S} \}.
\end{equation} 
The stabilizer code is the $+1$-eigenspace of all elements of the
stabilizer $\mathcal{S}$. The dimension of this eigenspace is $2^{M}$
where $M=N-\#$ of generators of the stabilizer $\mathcal{S}$.  The
distance for a stabilizer code is given by the following

\begin{svgraybox}
\begin{center}
\textbf{Box 3.5 Stabilizer code: dimension and distance}

Let $\mathcal{S}$ be a stabilizer with $N-M$ generators.
Then $\mathcal{S}$ encodes $M$ qubits and has distance $d$,
where $d$ is the smallest weight of a Pauli operator in
  $\mathcal{S}^\perp \setminus \mathcal{S}$. 
\end{center}
\end{svgraybox}

Let us consider an example where the code encodes more than one
qubits, which the stabilizer $\mathcal{S}$ is generated by the
following two Pauli operators.
\begin{equation}
\begin{array}{lllll}
g_1=& X & X & X & X\\
g_2=& Z & Z & Z & Z
\end{array}
\end{equation}

There are total $n=4$ qubits and $2$ generators for the stabilizer, so
this code encodes $4-2=2$ qubits. The logical $\ket{0_L}\ket{0_L}$ can
be chosen as the state stabilized by the following four Pauli
operators.
\begin{equation}
\begin{array}{lllll}
g_1=& X & X & X & X\\
g_2=& Z & Z & Z & Z\\
\bar{Z}_1=& I& Z & Z & I\\
\bar{Z}_2=& I & I & Z & Z
\end{array}
\end{equation}

Here for convenience we use the notation $\bar{Z}_i$ to denote the logical $Z$ operators
(previously denoted as $Z_L$), where $i$ refers to the $i$th encoded qubits, as
there is more than one encode qubit.

Similarly, the logical $\ket{0_L}\ket{1_L}$ can be chosen as the state
stabilized by $\{g_1,$$g_2,$$\bar{Z}_1,$
$-\bar{Z}_2\}$, and $\{g_1,g_2,-\bar{Z}_1,\bar{Z}_2\}$
stabilizes the logical $\ket{1_L}\ket{0_L}$, $\{g_1,g_2,-\bar{Z}_1,-\bar{Z}_2\}$
stabilizes the logical $\ket{1_L}\ket{1_L}$.

The distance of this code is $2$, meaning that the smallest weight
Pauli operator which commute with $g_1,g_2$ is $2$, for instance,
$\bar{Z}_1$ is such an operator with weight $2$.

\subsection{Stabilizer states and graph states}
\label{sec:cp3sec4.3}

If a stabilizer code of $N$-qubit has $N$ generators, then the
dimension of the common eigenspace of eigenvalue $1$ will be of
dimension $2^{N-N}=1$. That is, the stabilizer code contains indeed
only a unique state. Such kind of state is called stabilizer
state.

For example, the $4$-qubit version of the GHZ state 
(see Chapter~\ref{sec:ghz-state-bell} for the discussion of $3$-qubit GHZ state)
\begin{equation}
\label{eq:GHZ4}
\ket{GHZ_4}=\frac{1}{\sqrt{2}}(\ket{0000}+\ket{1111})
\end{equation}
is a stabilizer state. To see why, consider the following $4$ stabilizer
generators
\begin{equation}
\label{eq:4GHZstab}
\begin{array}{lllll}
g_1=& Z & Z & I & I\\
g_2=& I  & Z & Z & I\\
g_3=& I & I & Z & Z\\
g_4=& X & X & X & X,
\end{array}
\end{equation}
and it is straightforward to check that $g_i\ket{GHZ}_4=\ket{GHZ}_4$.

There is a special kind of stabilizer states called the graph states,
whose stabilizer generators correspond to some given graphs. We start
from a undirected graph $G$ with $n$-vertices. For the $i$th vertex,
we associate it with a stabilizer generator
\begin{equation}
\label{eq:graph}
g_i=X_i\bigotimes_{k\in\text{neighbor}\ i}Z_k,
\end{equation}
where $k\in\text{neighbor}\ i$ means all the vertices which share an
edge with $i$. It is straightforward to see that $g_i,g_j$ commute for
any $i,j$. $g_i$s then gives a stabilizer group with $N$ generators,
whose common eigenspace of eigenvalue $1$ is a single stabilizer
state, which we call a graph state.

%
\begin{figure}[h!]
\centerline{
\includegraphics{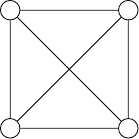}
}
%
%
\caption{A complete graph of $4$ vertices}
\label{fig:4qubit}       
\end{figure}

As an example, for the complete graph given in Fig.~\ref{fig:4qubit}
with $N=4$, the $4$ stabilizer generators are given by
\begin{equation}
\begin{array}{lllll}
\label{eq:4qubit}
{g}_1=& X & Z & Z & Z\\
{g}_2=& Z  & X & Z & Z\\
{g}_3=& Z & Z & X & Z\\
{g}_4=& Z & Z & Z & X 
\end{array}
\end{equation}
The common eigenspace of eigenvalue $1$ of these $4$ stabilizer
generators is a graph state, i.e. the graph state associated with the
complete graph of $4$ vertices. 

\section{Toric code}
\label{sec:cp3sec5}

In this section, we discuss an interesting example of stabilizer code,
namely the toric code.  We will see later in this book that this is
the simplest example of topologically ordered system. We will come
back to talk about the physics of this model in later chapters of the
book. Here we introduce the model and discuss from the viewpoint of
quantum error-correcting code. 

Consider a square lattice.
The name \textit{toric code} means 1) the square lattice is putting on
a torus; 2) it is a stabilizer quantum code.
%
\begin{figure}[h!]
\centerline{
\includegraphics[scale=1.3]{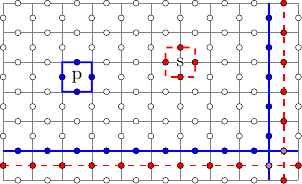}
}
%
%
\caption{Toric code. Each small circle represents a qubit, 
which are sitting on the link of the square lattice. The plaquette and star
operators are illustrated with four blue and red dots. The logical operators for the two 
encoded qubits are illustrated with blue and red lines, respectively.}
\label{fig:toric square}
\end{figure}

Fig.~\ref{fig:toric square} illustrates the layout of the toric code on
the square lattice of a torus. The solid lines gives the lattice, and
on each edge of the lattice lies a green dot which represents a
qubit. For an $r\times r$ lattice, we have $2r^2$ qubits.

There are two types of stabilizer generators.

Type I (Star type): 
\begin{equation}
Q_s=\prod_{j\in star(s)}Z_j
\end{equation}

Type II (Plaquette type):
\begin{equation}
B_p=\prod_{j\in plaquette(p)}X_j
\end{equation}

It is straightforward to check that $Q_s$ and $B_p$ commute for
any pair of $s,p$.  Although there are total $r^2+r^2=2r^2$
generators, there are indeed relations between them.
\begin{equation}
\prod_s Q_s=\prod_p B_p=I
\end{equation}

It can be shown that these are the only relations therefore the code
has dimension
\begin{equation}
2^{2r^2-(2r^2-2)}=2^2,
\end{equation}
in other words this code encodes two qubits into $2r^2$ qubits.

It seems that this code has a relative bad rate
$\frac{2}{2r^2}=\frac{1}{r^2}$ (i.e. we use $r^2$ qubits
to represent each logical qubit), which turns out to be small when $r$
goes large. However, it turns out that the error correcting property
of the code is good, as the minimum distance of the
code is $r$. This is because that the logical operators are cycles on
the torus, as shown in Fig.~\ref{fig:toric square}. 

More precisely, the corresponding logical operators are given by
\begin{eqnarray}
\bar{Z}_1=\prod_{j\in pink_v}Z_j,&\quad& \bar{X}_1=\prod_{j\in pink_h}X_j,\nonumber\\
\bar{Z}_2=\prod_{j\in green_h}Z_j,&\quad& \bar{X}_2=\prod_{j\in green_v}X_j,\nonumber\\
\end{eqnarray}
where $pink_{v}$/$green_{v}$ refer to the vertical pink/green line 
and $pink_{h}$/$green_{h}$ refer to the vertical pink/green line 
in Fig.~\ref{fig:toric square}, respectively.

Similar to the case of Shor's code, we can write a logical state in the code space, or a ground state of $H_{\text{toric}}$,
in terms of the stabilizer generators, i.e.
\begin{equation}
\label{eq:Htoric}
H_{\text{toric}}=-\sum_s Q_s -\sum_p B_p.
\end{equation}

To do so, let $\mathcal{S}_Z=\langle Q_s\rangle$ and $\mathcal{S}_X=\langle B_p,\bar{X}_1,\bar{X}_2\rangle$,
i.e. the $Z$ and $X$ part of the stabilizers, respectively. 
Then one ground state can be given by the following
\begin{equation}
\label{eq:toricg}
\ket{\psi_{\text{toric}}}=\sum_{g\in\mathcal{S}_X} g\ket{0}^{\otimes 2r^2},
\end{equation}

\begin{figure}[h!]
\centerline{
\includegraphics[scale=1.3]{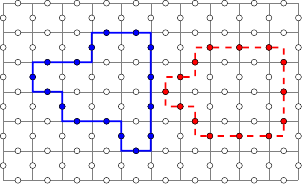}
}
%
%
\caption{Ground state of toric code on square lattice. The closed loops.}
\label{fig:toricsquareloop}
\end{figure}

There is a nice geometrical viewpoint of this ground state.  If we put
a red line on the edge representing the qubit on the edge which is in
state $\ket{1}$, then this ground state is the equal weight
superposition of all closed loops, as demonstrated in
Fig.~\ref{fig:toricsquareloop}.

We now summarize the property of toric code as follows.

\begin{svgraybox}
\begin{center}
\textbf{Box 3.6 Properties of toric code}

The toric code on a square lattice has the following properties.
\begin{enumerate}
\item  Every stabilizer generator is local (i.e. each 
generator only involves interactions of nearby qubits).
\item  The code space encodes two qubits (i.e. four-dimensional subspace).
\item  The code distance grows with $r$, as an order of
    $\sqrt{N}$ when $N$ goes arbitrarily large.
\end{enumerate}
\end{center}
\end{svgraybox}

\section{Summary and further reading}

In this chapter, we have discussed the idea for fighting again
decoherence in open quantum systems. The ultimate goal is to maintain
coherence (or unitary evolution) of the system. In practice, this
cannot in general be done perfect, but in a manner of approximation to
reduce the error caused by interaction with the environment by certain
order of magnitude. The central idea to realize this reduction is the
theory of quantum error correction.

Historically, the first quantum code was introduced by Shor in
1995~\cite{Sho95}, which is the Shor's code we discussed in
Sec.~\ref{sec:cp3sec2}. It then followed by~\cite{CS96} and~\cite{Ste96}, which introduce a framework for constructing
quantum error-correcting codes from classical linear codes, which is
now called `Calderbank-Shor-Steane (CSS) code'. The most well known
CSS code is the $7$-qubit Steane code, which is not discussed in this
chapter. Interested readers may refer to~\cite{CS96} and~\cite{Ste96} as mentioned above, or refer to some
general textbooks such as the one by Nielsen and
Chuang~\cite{nielsenchuang}.

The quantum error-correcting criterion is first proposed in 1997~\cite{KR97}, which is now called the
Knill-Laflamme condition. The stabilizer formalism is independently
proposed in ~\cite{thesis:gottesman} and in~\cite{CRSS97}. It should be
mentioned that the CSS codes are indeed a special case of the
stabilizer codes, where the stabilizers of the CSS codes contain
tensor product of only Pauli $X$ operators, or tensor product of only
Pauli $Z$ operators.

The graph state is first proposed in~\cite{SW02}. Indeed they discussed not only graph states,
but in a more general, case, graph codes. It was
further showed in that any stabilizer code is equivalent to 
a graph code in some sense (i.e. `local Clifford equivalence'), and similarly every stabilizer state is equivalent to a
graph state~\cite{Sch02}.

The toric code is first proposed in~\cite{Kit97}. More details can be found in the book by Kitaev, 
Sen, and Vyalyi~\cite{KSV:computation}. A similar model, called the Wen-plaquette
model is proposed in~\cite{Wen03}, whose physical
properties with different lattice sizes (i.e. odd by odd, odd by even
and even by even lattices) are further discussed in~\cite{KLW08}. 
There are many literatures
on topological quantum codes, we refer the reader
to the survey~\cite{Bombin13} and references therein.

A special kind of graph state, called the cluster state, was first introduced in~\cite{BR01}. Cluster state correspond to graphs of lattices (e.g. a 1D chain or a 2D square lattice). One important application of
cluster state is to be used as a resource state for one way quantum
computing, which is proposed in~\cite{RB01}. More on cluster states, graph states and their
application in one-way quantum computing can be found in a review
article~\cite{HDE+06}.

It is also realized that the cluster
state can be viewed as a valence bond solid using the tenor product
state (TPS) formalism~\cite{VC04}, which will be discussed in part IV
of this book. In 2007 it is further shown that the TPS formalism provides a powerful method to
construct resource state for one-way quantum computing~\cite{GE07},
which are better than cluster states in various circumstances, hence
triggering lots of on-going related research ever since. Readers
interested in these recent developments could refer to the review
articles~\cite{KWZ11} and ~\cite{RC12}.

The most general framework for constructing quantum error-correcting
codes known to date is the codeword stabilized (CWS) quantum code
framework~\cite{CSS+09}. The CWS framework encompasses stabilizer codes,
as well as all known examples of good codes beyond stabilizer
codes. It also has a good connection with graph codes. Interested
readers could refer to the original CWS code paper mentioned above, or
the subsequent follow-up papers~\cite{cross-2007,chen-2008,BCG+11}.  

%
%
\bibliographystyle{plain}
\bibliography{Chap3}


%
%
%

\begin{partbacktext}
\part{Local Hamiltonians, Ground States, and Many-body Entanglement}

\end{partbacktext}

%
%
%
\chapter{Local Hamiltonians and Ground States}
\label{cp:4} 

\abstract{We discuss many-body systems, where the Hamiltonian involves only few-body interactions. With the tensor product structure of the many-body Hilbert space in mind, we introduce the concept of locality. It is naturally associated with the spatial geometry of the system, 
where the most natural interaction between degrees of freedom are those `local' ones, for instance nearest neighbor interactions. 
We discuss the effect of locality on the ground-state properties. 
We then discuss ways of determining the ground-state energy of local Hamiltonians, and their hardness. Theories have been developed in quantum information science to show that even with the existence of a quantum computer, there is no efficient way of finding the ground-state energy for a local Hamiltonian in general. However, for practical cases,
special structures may lead to simpler method, such as
Hartree's mean-field theory. 
We also discuss a special kind of local Hamiltonians, called the frustration-free Hamiltonians, where the ground state is also ground states of all the local interaction terms. However, to determine whether a Hamiltonian is frustration-free is in general hard.}

\section{Introduction}

In Part I, we have introduced some basic concepts of quantum information theory that we will apply to study many-body systems. From this part on, we will focus on these systems. We will first revisit the Hilbert space of composite systems, discussing in detail the `particle basis' representation and the `occupation basis' representation of a many-body system. In many cases, the many-body Hilbert space is a tensor product of single body ones. 

A many-body system is naturally associated with a many-body Hamiltonian. We have already seen some of these Hamiltonians in Part I, such as the Ising Hamiltonian, Heisenberg Hamiltonian and the Toric Code Hamiltonian. One important property of these Hamiltonians is that they usually involve only few-body interactions. What is more, for Hamiltonians living on some lattice, the few-body interactions usually only involve degrees of freedom `near' each other on the lattice. This naturally leads to a concept of locality, that the `naturally-occurring' Hamiltonians are those `local' with respect to some spatial lattice geometry. In other words, they involve only few-body interactions of nearby degrees of freedom. Hence we call these many-body Hamiltonians as `local Hamiltonians'. Locality
has an important consequence. That is, the ground states of 
these systems exhibit special correlation/entanglement
properties compared to a `generic' (i.e. randomly chosen)
quantum state in the system Hilbert space. Exploring these properties is then a central topic of this part.

In condensed matter physics, one usually needs to consider infinitely large system (thermodynamic limit) for studying physical properties. However, in many cases, one can also read some important information from a finite system, and its `scaling' properties with the system size $N$. Important basic things naturally include the ground-state properties, for instance their correlations and entanglement properties. We will start to look into these correlation properties for ground states of local Hamiltonians for finite systems. We will also discuss consequences of system size $N$ getting large, in some places in this chapter, but mainly in the next chapter (Chapter~\ref{cp:5}).

We then move on to deal with more practical questions: given a local Hamiltonian of an $N$-body system, can we determine its ground state energy? Can we find its ground-state wave-functions, and other important properties such as degeneracy?

Anyone with some experience in quantum many-body physics knows that those questions should be very difficult in general. Although calculating ground-state properties for interacting system is so hard, we keep tackling them everyday by making good approximations and developing better algorithms. Quantum information science concurs with those hardness observations but at the same time raises a new interesting question: what if we have a quantum computer, can we compute ground-state energy for a given local Hamiltonian in an efficient way?

One seems to have some hope here because we have shown in Chapter~\ref{cp:2} that quantum computer can efficiently simulate quantum evolution of a many-body system with local Hamiltonians. Unfortunately, it is no longer the case regarding computing ground state energy for local Hamiltonians. Quantum information science develops a theory, based on some computer science ideas to show that, even if there is a quantum computer, it is very unlikely that one can efficiently calculate the ground-state energy for local Hamiltonians in general.

This is on the one hand disappointing, which seems to reveal some limitations of quantum computing. On the other hand, this is acceptable as those `general' local Hamiltonians might not be real (that it is unlikely for us to encounter them in practice). One may think that imposing a bit more structure might make things better, such as looking at a two-spatial dimensional (2D) systems with two-body nearest neighbor interactions only, or even just a one-spatial dimensional (1D) system. Unfortunately even under such restricted situations things do not get much better, which seems to reveal some intrinsic complexity of quantum many-body systems. Indeed, these system with `hard to analyze' ground state properties are closely related to glassy systems, which needs exponential long cooling time to get to their ground states.

To further understand the local Hamiltonian problem, we 
discuss another approach, based on the reduced density
matrix. This approach has been developed by the quantum
chemistry community since 1960s, with recent progress
obtained by the quantum information community. The basic
idea is that for local Hamiltonians evolving only 
few-body interactions, the ground-state energy is completely
determined by these few-body local reduced density matrices. 
Therefore, one only needs variations with local density
matrices to find the ground-state energy, instead of variations with wave-functions
on the entire Hilbert space, which saves exponentially 
number of variational parameters, in principle. 

Unfortunately,
it seems very hard to determine the conditions these
few-body local density matrices have to satisfy, in order to be
a `part' of a larger quantum systems. In other words, to 
determine whether some given local density matrices are
consistent with each other, i.e. whether they are 
the reduced density matrices of a state in a larger system, is a hard problem. And it is shown to be as hard as the local Hamiltonian problem.

Nevertheless, this gives an alternative approach for
finding the ground-state energy of local Hamiltonians.
Closely related, there is an interesting result on
the structure of these local reduced density matrices, for bosonic
systems, namely the quantum de Finetti's theorem. It
states that any local reduced density matrix for bosonic systems
in the $N\rightarrow\infty$ limit ($N$ is the number of particles in the system) is always not entangled, i.e. it is a mixture of product states. This justifies the validity
of Hatree's mean-field approximation, which always gives
the exact ground-state energy for bosonic systems, 
although the ground-state itself may be genuinely entangled.

Another interesting topic we will discuss are the frustration-free systems. We have already known that frustration-free Hamiltonians are enough to produce interesting physics such as topological order, as discussed in Chapter~\ref{sec:cp3sec5} (toric code). And for frustration-free systems, the ground state energy can be easily determined as the ground state is just the ground state of each local term of the Hamiltonian. Now the question is, can we determine whether a given system is frustration-free or not. Unfortunately, again there is no efficient way of determining this even with the existence of a quantum computer.

There is one exception though. There is a way to determine whether a Hamiltonian of spin-$1/2$ system (e.g. qubits) involving only two-body interactions is frustration-free or not. And in case it is, one can further characterize the structure of the corresponding ground space structure. It turns out that there always exists a ground state for such a system which has no entanglement at all. This means that the ground state space `lacks correlation' in a sense, so the ground-state energy as well as the ground state itself can be given by the mean-field theory. Therefore this kind of systems are relatively simple, which could not represent nontrivial strongly correlated phases in practice, whose ground states are expected to be highly entangled.

If one goes beyond spin-$1/2$ systems with two-body interactions, even frustration-free systems will have highly entangled ground states, for instance the toric code Hamiltonian. We will look at some other interesting frustration-free system and their ground state properties, such as the Affleck-Kennedy-Lieb-Tasaki (AKLT) model.

\section{Many-body Hilbert space}
\label{12quantization}

Let's start by discussing carefully the basic concept of the Hilbert space of a many-body system. The Hilbert space of a many-body system is naturally obtained by putting together Hilbert spaces of single-body systems. While this may sound straight-forward, there are two different and both commonly used ways to do it, one from the point of view of particles, one from the point of view of `modes'.

In the first approach, which we call the `particle basis' representation, one starts from a single particle Hilbert space which contains all possible states $|\psi\rangle$ of this single particle (described by the position, momentum, angular momentum, etc. of the particle).  A many-body system contains more than one, say $N$, particles, each being in a single particle state $|\psi_i\rangle$. The many-body Hilbert space is then the combination of the single particle Hilbert spaces, but usually with extra constraints. 

The constraint comes from the quantum statistics of the particles, which can be either bosons, fermions or distinguishable particles. When the particles are distinguishable, there is no constraint. The many-body Hilbert space is the tensor product of the single-body Hilbert spaces. If a single particle can be in $m$ orthogonal states, then the many-body Hilbert space is $m^N$ dimensional. Particles are distinguishable when, for example, their locations are fixed and the only degrees of freedom in the system or those internal to the particles, like spin. Therefore, in what is called a `spin system', the total Hilbert space $\mathcal{H}$ is a tensor product of the Hilbert space of individual spins $\mathbb{C}_m$
\begin{equation}
\mathcal{H}=\mathbb{C}_m^{\otimes N}.
\end{equation}
where $m$ is the dimension of a single spin Hilbert space.

When the particles are bosons, exchanging two of the particles should keep the total many-body wave function invariant. That is, if the many-body wave function contains a configuration $|\psi_1\rangle|\psi_2\rangle...|\psi_N\rangle$, it should also contain the configuration $|\psi_{S(1)}\rangle|\psi_{S(2)}\rangle...|\psi_{S(N)}\rangle$ with the same amplitude, where $S$ is an arbitrary permutation on the $N$ labels. When the particles are fermions, exchanging two of them should change the sign of the total wave function. That is, if the many-body wave function contains a configuration $|\psi_1\rangle|\psi_2\rangle...|\psi_N\rangle$, it should also contain the configuration $|\psi_{S(1)}\rangle|\psi_{S(2)}\rangle...|\psi_{S(N)}\rangle$ but with an extra sign factor $(-1)^{p(S)}$ where $p(S)$ is the parity of the permutation operation $S$. Therefore, the many-body wave functions for bosons or fermions are highly constrained and occupies a very small subspace in the $m^N$ dimensional total Hilbert space. 

A highly useful example of many-body wave function written in this form is Laughlin's wave function for quantum Hall states. Laughlin's wave function describes the motion of $N$ bosons or fermions on a two dimensional plane. Each particle can be at different spatial locations labelled by $z = x + iy$. In the simplest Laughlin state, the amplitude for the $N$ fermions to be at locations $z_1,z_2...,z_N$ is given by
\be
\Psi(z_1,z_2,...,z_N) = \prod_{N \geq i > j \geq 1} (z_i-z_j) \prod_{k=1}^N \exp(-|z_k|^2)
\ee  
which obviously gets a minus sign if two particles are exchanged. In a simple Laughlin state for $N$ bosons, the amplitude for them to be at location $z_1,z_2...,z_N$ is given by
\be
\Psi(z_1,z_2,...,z_N) = \prod_{N \geq i > j \geq 1} (z_i-z_j)^2 \prod_{k=1}^N \exp(-|z_k|^2)
\ee  
which obviously remains invariant under any exchange. 

This `particle basis' representation is extremely useful, but it also has an important flaw: one cannot write wave functions for systems where the total particle number is fluctuating, like in a superfluid or superconductor. To have a more general way to write many-body wave functions, we can switch to an `occupation basis' representation. The `occupation basis' representation starts from individual `modes' that single particles can occupy. A mode can be labeled by the position, momentum, angular momentum or other physical quantities of a single particle. A mode can be empty or occupied. If the system contains bosons, a single mode can be occupied by any number of particles; if the system contains fermions, a single mode can only be occupied by one (or zero) particle. The corresponding single mode Hilbert space is then $\infty$ dimensional or two dimensional. Usually we can assume that due to certain physical reason, it is not possible to put too many bosons in a single mode and there is an upper bound $m$. The single mode Hilbert space becomes $m$ dimensional. The many-body Hilbert space is then obtained by putting $N$ modes together and has a tensor product structure
\begin{equation}
\mathcal{H}=\mathbb{C}_m^{\otimes N}, \text{ or } \mathcal{H}=\mathbb{C}_2^{\otimes N}
\end{equation}
There is no extra constraint on the many-body Hilbert space. Any wave function is in principle allowed. The difference between bosons and fermions not encoded in the structure of the many-body Hilbert space any more; instead it is encoded in the way operators act on states in the Hilbert space.

The `particle basis' and `occupation basis' representation of many-body Hilbert space and wave function are often also called the first and second quantization of many-body quantum systems. In our following discussion, in this chapter and for the rest of the book, we will be mainly focusing on the spin system, and boson, fermion systems in the `occupation basis', so that the total Hilbert space has a tensor product structure. Occasionally, we will also use the `particle basis' representation of boson fermion systems to discuss associated interesting problems. When we do so, we will explicitly state that we are using the `particle basis' representation.

\section{Local Hamiltonians}

Consider an $N$-body system. For simplicity, we assume each degree of freedom is a qubit (i.e. a two-level spin degree of freedom), hence the single-body Hilbert space has dimension $2$, which is denoted by $\mathbb{C}_2$. Note that our discussion is readily applied to other systems with larger dimension of its single-body spaces.

The Hilbert space $\mathcal{H}$ of the $N$-body system is then the tensor product of the Hilbert space of all its subsystem, i.e.
\begin{equation}
\mathcal{H}=\mathbb{C}_2^{\otimes N}.
\end{equation}
If for each single qubit subsystem, the Hilbert space is spanned by the orthonormal basis $\{\ket{0},\ket{1}\}$, the orthonormal basis for $\mathcal{H}=\mathbb{C}_2^{\otimes N}$ can then be chosen as
\begin{equation}
\{\ket{00\ldots 0}, \ket{00\ldots 1}, \ldots \ket{11\ldots 1}\}.
\end{equation}

The Hamiltonian $H$ of the system is usually given in terms of summation of many terms, i.e.
\begin{equation}
H=\sum_j H_j,
\end{equation}
where each $H_j$ involves only few-body interactions. We say $H$ is a $k$-body Hamiltonian, if each $H_i$ involves at most $k$-body interactions, where $k$ is a constant that is independent of the system size $N$.

\subsection{Examples}

\begin{figure}[h!]
\centerline{
\includegraphics[scale=1.00]{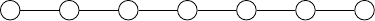}
}
%
%
\caption{A 1D lattice}
\label{fig:1D}       
\end{figure}

In general, for $k$-body interactions, there are total $N\choose k$ ways of choosing the $k$ degrees of freedom involved. In practice, it is not always true that all of them has to show up in a $k$-body Hamiltonian $H$. For instance, for a lattice spin system, the interaction usually only involves the spins that are `near each other'.  As a concrete example, consider a 1D chain as shown in Fig.~\ref{fig:1D}, the Ising model Hamiltonian in a transverse filed is given by
\begin{equation}
H^{\text{tIsing}}=-J\sum_{j} Z_jZ_{j+1}-B\sum_j X_j
\end{equation}

$H^{\text{tIsing}}$ involves only $2$-body local interactions that in this sense we call  $H^{\text{tIsing}}$ a $2$-local Hamiltonian.

\begin{figure}[htbp]
\centerline{
\includegraphics[scale=1.5]{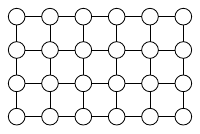}
}
%
%
\caption{A 2D square lattice}
\label{fig:2D}       
\end{figure}

For a 2D example, consider the following Hamiltonian on a 2D square lattice as shown in Fig.~\ref{fig:2D}, which has the 2D cluster state (i.e. the graph state associated with the graph as a 2D square lattice, as discussed in Chapter~\ref{sec:cp3sec4.3}) as its unique ground state, reads
\begin{equation}
H_{\text{clu}}=-\sum_{i,j} X_{i,j}Z_{i+1,j}Z_{i-1,j}Z_{i,j+1}Z_{i,j-1}
\end{equation}

$H_{\text{clu}}$ involves only $5$-body local interactions, so we call  $H_{\text{clu}}$ a $5$-local Hamiltonian.

In general, we will consider a $D$-spatial dimensional system, which in general refers to the usual Euclidean geometry in $\mathbb{R}^D$. We also discuss other manifolds such as torus, which we already encountered in Chapter~\ref{sec:cp3sec5} for toric code. We will further discuss dimensionality and locality from condensed-matter theory point of view, in Chapter~\ref{cp:5}.

\subsection{The effect of locality}

After talking about locality, which is associated with some spacial geometry, one natural question is that what is the effect of locality. Or putting in other words, what is the difference of a $k$-local Hamiltonian associated with some spacial geometry, compared with those `non-local' $k$-body Hamiltonians. For simplicity, when we talk about locality of a $D$ spatial dimensional system, we refer to the usual Euclidean geometry in $\mathbb{R}^D$.

Indeed, local Hamiltonians will be the main focus of this book, by studying the properties of their ground state space and beyond. In other words, most part of the book will deal with local Hamiltonians, which we have in mind the spatial locality for a $D$ spatial dimensional system with respect to the Euclidean geometry in $\mathbb{R}^D$. Before we look further into these systems of spatially local Hamiltonians, we would like to understand a bit what is the difference between local Hamiltonians and those non-local $k$-body Hamiltonians. 

Let us recall the toric code system on torus with a square lattice (as discussed in Chapter~\ref{sec:cp3sec5}). Now our locality refers to the $D=2$-dimensional Euclidean geometry. The system has $N=2r^2$ spins, and the ground state space is four-fold degenerate, which is a stabilizer quantum error correcting code encoding $2$ qubits. The code distance is $d$, which is of the order $\sqrt{N}$ for large $N$. It is natural to wonder whether we can do better than this. For instance, does there exist a local Hamiltonian on the square lattice such that its degenerate ground state space has larger degeneracy, but meantime maintain the a code distance as good as $\sqrt{N}$.

Intuitively, this is not possible. If the degeneracy is too large, say, exponential in $N$, then local perturbation shall be enough to destroy such a degeneracy. This can be shown for a large class of local Hamiltonians. 

For a $k$-local stabilizer Hamiltonian on a $D$-dimensional lattice, the ground state space degeneracy $R$ and the code distance $d$ satisfies a constraint
\begin{equation}
\label{eq:dbound}
\log R\leq\frac{cN}{d^{\alpha}},\ \alpha=\frac{2}{D-1},
\end{equation}
where $c$ is some constant independent of the system size $N$. This means if $\log R/N$ is a constant, i.e. the degeneracy is exponential in $N$ in a sense that $\log R/N $ is a constant, then the code distance $d$ is upper bounded by a constant $\left(\frac{cN}{\log R}\right)^{
\frac{1}{\alpha}}$. That is, the code distance cannot increase with $N$.

For $D=2$, the code distance bound Eq.(\ref{eq:dbound}) becomes $\sqrt{\frac{cN}{\log R}}$. That is,  $d$ at most can scale as the square root of the system size $N$, which the toric code does.

However, a $k$-body but non-local Hamiltonian system can perform quite differently. There are indeed $k$-body but non-local Hamiltonians constructed on the $2D$ square lattice, such that the ground state space has dimension $R$ which increases exponentially with $N$ , i.e. $\log R/N$ is come constant. Meanwhile, the code distance $d$ scales as the square root of $N$, similar to the toric code.

As to demonstrate the proof of the bound given by Eq.(\ref{eq:dbound}) as well as the construction of these highly nontrivial $k$-body but non-local Hamiltonians mentioned above is getting too much involved in the technical details of the theory quantum error-correcting code, which goes beyond the main scope of the book, so we omit those details. The main information to convey is that `locality' does have strong restriction on how such physical systems could actually behave. Throughout the book, we will only look at `local' Hamiltonians associated with Euclidean geometry on some (finite) $D$-spatial-dimensional lattice.

\section{Ground-state energy of local Hamiltonians}

Having built some general understanding for correlations in finite system, we now move into a more practical question: given a local Hamiltonian of an $N$-body system, can we determine its ground state energy? Can we find its ground state wave functions, and other important properties such as degeneracy?

Let us start from the first question to determine the ground- state energy of a $k$-local Hamiltonian. Our experience tells us to find the ground state energy of an interacting many-body system is a hard question. By `hard' here we mean to deal with the task as computationally. Imagine to exact diagonalize a Hamiltonian of $N=20$ qubits, which will
be an $\sim 10^6\times 10^6$ matrix. This is doable with today's personal computers, but even supercomputers can hardly deal with $N=30$, as the cost of computation, in terms of both memory use and computing time, grows exponentially with the number of degrees of freedom $N$.

On the other hand, we are talking about quantum computers, which can in principle exist.
And we have already demonstrated its power for simulating evolution of quantum systems. Now the question is, if there were a quantum computer, could we determine the ground state energy of local systems efficiently. By efficient here we mean an algorithm running on a quantum computer whose computing time grows only polynomially with the number of degrees of freedom $N$. If this were the case, then we can easily compute a systems of hundreds or even thousands of qubits, whose behaviour could well approximate the practical condensed matter systems in the $N\rightarrow\infty$ limit.

Unfortunately, it turns out not the case. In other words, even a quantum computer is very unlikely to compute the ground state energy of a local Hamiltonian efficiently. In order to reach this definite conclusion, an important subfield of quantum information science, namely the theory of quantum computational complexity, has been extensively developed. It is beyond the scope of this book to go into the details of such a theory, but we would like to briefly review some important practical relevant results obtained from the theory.  

\subsection{The local Hamiltonian problem}

We start to state the so called `local Hamiltonian problem' explicitly.

\begin{svgraybox}
\begin{center}
\textbf{Box 4.1 The local Hamiltonian problem}

Given a local Hamiltonian $H=\sum_j H_j$, where each $H_j$ acts 
non-trivially on at most $k$ qubits. Denote $E_0$ the ground state energy of $H$. For given $(b-a)\propto 1/\text{poly}(N)$, determine which of the following is true.

1. $E_0>b$.

2. $E_0<a$.
\end{center}
\end{svgraybox}

Some points need to be clarified. Firstly, we have not only a single Hamiltonian $H$, but in fact, a family of Hamiltonians $\{H_N\}_{N=1}^{\infty}$. Here each $H_N$ denotes the Hamiltonian for a systems of $N$ degrees of freedom. When taking a limit $N$ goes to infinity, we end up with the thermodynamic limit (an infinitely large system). 

Secondly, $b-a$ gives the precision of this problem, i.e. the error we can actually tolerate for deciding the ground state energy. We know for numerical stability reasons, it makes no sense to take $b=a$.
For technical reasons, for all the problems studied, the precision $b-a$ is set to be scale as an inverse polynomial of the system's size $N$, i.e. $b-a\propto 1/\text{poly}(N)$, where $\text{poly}$ is some polynomial function, and $\propto$ means up to some constant factor.

Under these setups, the quantum computational complexity theorem provides the following rather surprising assertion.

\begin{svgraybox}
\begin{center}
\textbf{Box 4.2 The hardness of the local Hamiltonian problem}

The local Hamiltonian problem is very unlikely to be efficiently solvable, even with the existence of a quantum computer.
\end{center}
\end{svgraybox}

As mentioned we are not digging into the details of how this result is technically shown. Rather, we would try to explain some aspects as a consequence of this result which are of more practical interest.

First of all, we would say something regarding what we meant by `very unlikely to be efficiently solvable'. We know that efficiently solvable means that there exists a polynomial size quantum circuit (i.e. circuits with $\text{poly}(N)$ gates), which answers the problem for $H_N$. So by saying `unlikely to be efficiently solvable', we mean that it is very unlikely to have such kind of quantum circuits. The underline reason is that if such a circuit exists, then it will contradict some common belief in the fundamental theory of computer science. That is, the class of problems whose solutions can efficiently verified (the so-called `{\bf NP}' class), is in fact different from the class of problem whose solutions can be efficiently found (the so-called `{\bf P}' class). This $\textbf{P}\neq\textbf{NP}$ conjecture is widely believed to be true among computer scientists, yet no rigorous proof ever found. Note that what is directly relevant to discussions in Box 4.6 is the `quantum analogy' of this $\textbf{P}\neq\textbf{NP}$ conjecture, which for technical reasons we omit the details.

We then discuss the structure of the Hamiltonian $H$. In the assertion it is only roughly said that $H$ is a local Hamiltonian, but has not yet specified its local structure (i.e. spatial dimension, nearest-neighbor etc.). 
One might think that the local structure which leads to the asserted result might be quite non-physical, in a sense that it might involve $k\geq 3$-body interactions, or interactions between degrees of freedom which are geometrically not nearest neighbors. Quite counter-intuitively, the local structure could be surprisingly simple: it can only be associated with nearest-neighbor interactions on a $2D$ square lattice, or even nearest neighbor interactions on a $1D$ chain. It worth mentioning that $1D$ result is with a designed Hamiltonian not for qubit (i.e spin $1/2$) systems, but for a spin system with spin-$23/2$ (i.e. single-spin Hilbert space dimension $12$), which does not seems quite realistic. However, this is still very surprising given the usually impression that a $1D$ systems should be relatively simple. We will discuss more about $1D$ many-body systems in later chapters.

It is interesting to note that there is another way of looking at the local Hamiltonian problem, from the viewpoint of quantum simulation. Recall that the quantum simulation problem is to find the quantum state $\rho(t)$, where the evolution is governed by the 
Schr\"{o}dinger's equation, i.e.
\begin{equation}
\label{eq:simulation}
\rho(t)=\frac{(e^{-iHt})^{\dag}\rho(0)e^{-iHt}}{\Tr\left[(e^{-iHt})^{\dag}\rho(0)e^{-iHt}\right]},
\end{equation}
where $t$ is a real number indicates the `real' time.

Now imagine that $t$ to be purely imaginary, i.e. $t=-i\beta$, where $\beta=\frac{1}{\kappa T}$. Choose 
$\rho_0=\frac{I}{\Tr I}$, then Eq.(\ref{eq:simulation}) becomes $\rho(\beta)=\frac{e^{-2\beta H}}{\Tr e^{-2\beta H}}$.
The ground state $\rho_g$ of the Hamiltonian can then be given by
\begin{equation}
\label{eq:imsimulation}
\rho_g=\lim_{\beta\rightarrow\infty}\frac{e^{-2\beta H}}{\Tr e^{-2\beta H}}.
\end{equation}

This then indicates that to find the ground state of the Hamiltonian $H$ is equivalently to `simulate' the imaginary time evolution for $\beta$ large, or alternatively, to simulate the `cooling' of the system to its zero temperature (i.e $T=0$) ground state. The local Hamiltonian problem with a given local Hamiltonian $H$ is hard then means that in the worst case, the time taken to cool the system to its ground state is `exponentially long' (here `exponential' is again, in terms of the system size $N$). In practice, there are indeed physical systems which are `hard' to cool to its ground state, for instance, the spin glasses. These systems have access to a large number of metastable states such that they are much easier ending up in some metastable states than their ground states. In other words, for those physical systems whose ground states are hard to compute with a quantum computer may be just those systems whose ground states are `not real', i.e. they never end up in their ground states in real world.

Finally we remark a bit more on the $b-a\propto 1/\text{poly}(N)$ in the local Hamiltonian problem. It is not yet known how much this condition can be relaxed, e.g. say, can we set $b-a\propto$, the number of the local interaction terms in local Hamiltonian, i.e. the number of terms in summation of $H=\sum_j H_j$, but the local Hamiltonian problem remains hard? Quite surprisingly, in the classical case, the answer is affirmative, which is given by the so-called probabilistically checkable proof (PCP) theorem. However, the quantum case remains open, while it is stated as the `quantum PCP conjecture', which means the answer is indicated to be true by some evidences. This problem has raised considerably attention in the quantum information community in recent years, as its solution will need the development of many new tools, while provide a fundamental understanding of what many-body quantum systems are the true `hard' ones.

\subsection{The quantum marginal problem}
\label{sec:marginal}

We now would like to look at the local Hamiltonian problem from another point of view, namely the variational approach. For the local Hamiltonian $H=\sum_j H_j$, the ground state energy $E_0$ can be given by
\begin{equation}
E_0=\min_{\ket{\psi}}\bra{\psi}H\ket{\psi},
\end{equation}
where the minimization is over all wave functions $\ket{\psi}$.

Now because $H$ is local, so for any $\ket{\psi}$, we can write
\begin{equation}
\bra{\psi}H\ket{\psi}=\sum_j \Tr(H_j\rho_j),
\end{equation}
where $\rho_j$ is the reduced density matrix for particles sets that $H_j$ acting non-trivially on.

Now in order to find the ground-state energy, we can do minimization over the set of $\{\rho_j\}$ instead. This at first glance seems to be much simpler than the minimization over the $n$-particle wave function $\ket{\psi}$, as the set of $\{\rho_j\}$ has much less parameters. However, there is a problem: the minimization is not over the set of all density matrices for particles sets that $H_j$ acting non-trivially on,
but over the set of all \textit{reduced} density matrices or particles sets that $H_j$ acting non-trivially on. Therefore, one has to first determine the condition such that the density matrices are indeed reduced density matrices. This is the so called quantum marginal problem.

\begin{svgraybox}
\begin{center}
\textbf{Box 4.3 The quantum marginal problem}

Given a set of local density matrices $\{\rho_j\}$, determine whether there exists an $n$-particle state $\rho$, such that $\{\rho_j\}$ are reduced density matrices of $\rho$.
\end{center}
\end{svgraybox}

To understand this marginal problem better, let us look at a simple example. Suppose we have a system with three qubits $A,B,C$. Now given a density matrices $\rho_{AB}$ of qubits $A,B$ and $\rho_{AC}$ of qubits $A,C$. We ask whether these exists a three-qubit states $\rho_{ABC}$, such that $\rho_{AB}=\Tr_C\rho_{ABC}$ and $\rho_{AC}=\Tr_B\rho_{ABC}$.

Unfortunately, even in this simple case, no analytical condition is known to tell the answer easily. Let us then try to further simplify the problem a bit. Let us assume $\rho_{AB}=\rho_{AC}$, in other words, we assume a symmetry when interchanging qubit $B$ with qubit $C$. Now we would like to explore the conditions that $\rho_{AB}$ has to satisfy to guarantee the existence of some $\rho_{ABC}$, which is also assumed to be symmtric when interchanging qubit $B$ with qubit $C$. In this sense, $\rho_{ABC}$ is also called the symmetric extension of $\rho_{AB}$.

This symmetric extension problem happen to have an elegant analytical solution. That is, a two-qubit state $\rho_{AB}$ has symmetric extension if and only if
\begin{equation}
\label{eq:symext}
\Tr(\rho_B^2)\geq\Tr(\rho_{AB}^2)-4\sqrt{\det(\rho_{AB})},
\end{equation}
where $\rho_B=\Tr_A(\rho_{AB})$.

To demonstrate that Eq.~\eqref{eq:symext} makes sense, let us first consider the case where $\rho_{ABC}$ could be a pure state. That is,  $\rho_{AB}$ is `pure symmetric extendable', and $\rho_{ABC} = \ket{\psi_{ABC}}\bra{\psi_{ABC}}$. Using Schmidit decomposition between qubits $A,B$ and qubit $C$, we can write $\ket{\psi_{ABC}}$ as
\begin{equation}
\ket{\psi_{ABC}}=\sum_{\alpha}\lambda_{\alpha}\ket{\alpha_{AB}}\ket{\alpha_C}.
\end{equation}

This means that the non-zero eigenvalues of $\rho_{AB}=\Tr_C\ket{\psi_{ABC}}\bra{\psi_{ABC}}$ is the same as those of $\rho_{B}=\rho_{C}=\Tr_{AB}\ket{\psi_{ABC}}\bra{\psi_{ABC}}$, where $\rho_{B}=\rho_{C}$ comes from the symmetry assumption between qubits $B$ and $C$. Therefore we have $\Tr(\rho_B^2)=\Tr(\rho_{AB}^2)$. And because $\rho_{AB}$ is at most rank $2$, so $\det(\rho_{AB})=0$. Therefore the equality of Eq.(\ref{eq:symext}) holds.

It is interesting to mention that the validity of Eq.~\eqref{eq:symext} is to explicitly construct a corresponding local Hamiltonian 
\begin{equation}
H=H_{AB}+H_{AC},
\end{equation}
such that 
\begin{equation}
\label{eq:symconvex}
\Tr(H_{AB}\rho_{AB})\geq 0 
\end{equation}
for any $\rho_{AB}$ satisfying Eq.~\eqref{eq:symext}.
Here $H_{AB}$ acts on the qubits $A,B$ and $H_{AC}$ acts on the qubits $A,C$. Due to symmetry between qubits $B,C$, $H_{AB}$ and $H_{AC}$ are in fact the same operator.

To further understand the meaning of Eq~\eqref{eq:symconvex},  notice that for any $\rho_{AB}$ that satisfies the equality of Eq~\eqref{eq:symconvex}, the corresponding $\rho_{ABC}$ is in fact the ground state of $H$.
It turns out that 
for any rank $4$ $\sigma_{AB}$ that satisfies the equality of Eq.~\eqref{eq:symext}, the corresponding $H_{AB}$ that has $\rho_{AB}$ as a ground state has an explicit form given by
\begin{equation}
\label{eq:H}
H_{AB}(\sigma_{AB})=\sqrt{\det\sigma_{AB}} \sigma_{AB}^{-1}- \sigma_{AB}+\sigma_B.
\end{equation}

Let us consider an example of the two-qubit state
\begin{equation}
\rho_W(p)=(1-p)\frac{{I}}{4}+p\ket{\phi}\bra{\phi},
\end{equation}
where $\ket{\phi}=\frac{1}{\sqrt{2}}(\ket{00}+\ket{11})$, and $p\in[0,1]$.

The equality of Eq.~(\ref{eq:symext}) gives
\begin{equation}
\Tr\big(\rho^2_W(p)\big)= \Tr\Big(\big(\Tr_A\rho_W(p)\big)^2\Big)+4\sqrt{\det{\rho_W(p)}},
\end{equation}
providing a unique solution of $p=\frac{2}{3}$.

Then Eq.~\eqref{eq:H} gives 
\begin{equation}
\label{eq:HABexam}
H_{AB}\left(\rho_W\left(\frac{2}{3}\right)\right)=
\begin{pmatrix}
\frac{2}{9}&0&0&-\frac{4}{9}\\
0&\frac{2}{3}&0&0\\
0&0&\frac{2}{3}&0\\
-\frac{4}{9}&0&0&\frac{2}{9}
\end{pmatrix}
\end{equation}

The ground-state space of the Hamiltonian 
$
H=H_{AB}+H_{AC}
$
is two-fold degenerate and spanned by
\begin{eqnarray}
\ket{\psi_0}&=&\frac{1}{\sqrt{6}}\left(2\ket{000}+\ket{101}+\ket{110}\right),\nonumber\\
\ket{\psi_1}&=&\frac{1}{\sqrt{6}}\left(2\ket{111}+\ket{010}+\ket{001}\right).
\end{eqnarray}
And it it straightforward to check that 
\begin{equation}
\rho_W(\frac{2}{3})=\frac{1}{2}\Tr_{C}(\ket{\psi_0}\bra{\psi_0}+\ket{\psi_1}\bra{\psi_1}).
\end{equation}
This means that the symmetric extension of $\rho_W(\frac{2}{3})$ is in fact the maximally mixed state of the ground-state space of $H\left(\rho_W(\frac{2}{3})\right)$.

We remark that the validity of Eq~\eqref{eq:symconvex} does not mean that the operator $H_{AB}$ is non-negative, which one can easily observe from the example of Eq.~\eqref{eq:HABexam}. What Eq~\eqref{eq:symconvex} says is that the operator $H=H_{AB}+H_{AC}$ is non-negative, for example the ground states $\ket{\psi_0}, \ket{\psi_1}$ have zero energy. This illustrate the concept of `frustration,' where the ground state of the $3$-qubit system with the Hamiltonian $H=H_{AB}+H_{AC}$ does not need to be also the ground state of each interaction term $H_{AB}$, $H_{AC}$. Generally local Hamiltonians are frustrated, but the special case of frustration-free Hamiltonians are also of importance that we will discuss more in Chapter~\ref{sec:ffHam}.

Although Eq.(\ref{eq:symext}) is valid for two-qubit case, it is known to be not generalizable to almost any other case, even for two-qubit marginals $\rho_{AB}, \rho_{AC}$ without symmetry between $B,C$. In fact, the real trouble is that spectra of the marginal $\rho_{AB}$ is in almost all cases, not enough to fully solve the quantum marginal problem. In other words Eq.(\ref{eq:symext}) is just a lucky situation for two qubits, and in general there is not much hope to have a simple condition which answers the question raised by the quantum marginal problem. Therefore, we now turn our hope to computers: can we design an algorithm that approximate the answer to the question, and could such an algorithm be efficient, possibly on a quantum computer.

The answer to the first question is no doubt affirmative. One can simply parametrize  $\rho_{ABC}$ in a general operator basis, and then check the condition that $\Tr_C\rho_{ABC}=\rho_{AB}$,  $\Tr_B\rho_{ABC}=\rho_{AC}$, and $\rho_{ABC}\geq 0$ to see whether such a $\rho_{ABC}$ exists. However, we know that an general $n$-qubit
density matrix has exponentially many parameters in terms of $n$, so this procedure cannot efficiently deal with large $n$ cases. Worse, even with a quantum computer, there is not much hope for an efficient algorithm, as given by the following fact shown in quantum computational complexity theory.

\begin{svgraybox}
\begin{center}
\textbf{Box 4.4 The hardness of the quantum marginal problem}

The quantum marginal problem is as hard as the local Hamiltonian problem.
\end{center}
\end{svgraybox}

This fact, however, is very natural. Because the local Hamiltonian problem and the quantum marginal problem, both used to determine the ground state energy of local Hamiltonians, are in fact to look at the same problem from different perspectives. Therefore their computational complexities should be essentially the same.

\subsection{The $N$-representability problem}

The quantum marginal problem has also been extensively studied in the field of quantum chemistry, for bosonic/fermionic systems, which correspond to states supported on the symmetric/antisymmetric subspace of the $N$-particle Hilbert space. Due to symmetry, all the two-particle reduced density matrices ($2$-RDMs) are the same, which we simply denote by $\rho_2$.
Also, the single-particle Hilbert space is no longer a qubit, but in general with a large dimension in order to have non-vanishing fermionic wave function (in the first quantization picture). The corresponding quantum marginal problem, also called the $N$-representability problem, is formulated in the following way.

\begin{svgraybox}
\begin{center}
\textbf{Box 4.5 The $N$-representability problem}

Given a two-particle bosonic/fermionic density matrices $\{\rho_2\}$, determine whether there exists an $N$-particle bosonic/fermionic state $\rho$, such that $\{\rho_2\}$ is the two-particle reduced density matrices of $\rho$.
\end{center}
\end{svgraybox}

The symmetry requirement seem to simplify the problem a bit. As an example, for the bosonic case, the conjecture given by Eq.(\ref{eq:symext}) holds, which leads to a simple condition. That is,
a $2$-matrix $\rho_2$ of a two-mode bosonic system is $3$-representable if and only if
\begin{equation}
\label{eq:iff}
\Tr\rho_1^2\geq \Tr\rho_2^2,
\end{equation}
where $\rho_1$ is the single-particle reduced density matrix of $\rho_2$.
Note that in terms of Eq.(\ref{eq:symext}), the bosonic $2$-matrix is only supported on the symmetric subspace of the two-qubit Hilbert space, thus the determinant term vanishes.

In general, the condition given in Eq.(\ref{eq:iff}) is also necessary for $3$-fermion or  $3$-boson systems with arbitrary single particle dimension. That is, if a bosonic/fermionic $2$-matrix $\rho_2$ satisfies
$\Tr\rho_1^2\geq \Tr\rho_2^2$,
where $\rho_1$ is the single-particle reduced density matrix of $\rho_2$,
then $\rho_2$ is $3$-representable.

However, Eq.(\ref{eq:iff}) is not sufficient for higher dimensional 
single particle space, 
where the equality in Eq.(\ref{eq:iff}) does not even imply that the spectra
of $\rho_2$ and $\rho_1$ are equal. 
This shows that the dimension $m$ of the single particle space is a crucial parameter
which may determine the hardness of the $N$-representability problem. When $m$ is small, the dimension of the $N$-particle Hilbert space is also relatively small. However, in the cases where $m$ is relatively large, we have

\begin{svgraybox}
\begin{center}
\textbf{Box 4.6 The hardness of the $N$-representability problem}

The $N$-representability Problem for either bosonic/fermionic system with large enough dimension of the single particle space, is as hard as the Local Hamiltonian Problem.
\end{center}
\end{svgraybox}
Here by large we mean that the dimension $m$ of the single particle space is at least twice the particle number, i.e. $m\geq 2N$, and in general $m$ grows with $N$. This condition makes perfect sense for fermions as the Pauli principle and particle-hole duality tells us that the system of $m$ single-particle states is the same as that of $m-N$ single-particle states. In fact, the validity of the results in Box 4.6 for fermions are shown by a mapping between $N$-fermion system with $2N$ single particle states and an $N$-qubit system. The bosonic case is a bit more complicated but essentially the dimension of single particle space plays the key role in the hardness conclusion. The result in Box 4.6 matches our general sense that the interacting bosonic/fermonic systems are hard to understand.

\subsection{de Finetti theorem and mean-field bosonic systems}

When considering an $N$-particle bosonic system, the single-particle space dimension $m$ could be just a constant that is independent of $N$ ($m$ can in fact even be infinite, as long as it does not grow with $N$, see the discussion below). It is natural to expect in this case that the corresponding $N$-representability should not be as hard as the case when $m$ grows with $N$. Now the question is, can we say anything about the set of $2$-RDMs? 

To examine this question, we would like to go back to the discussion of the symmetric extension problem discussed in Chapter~\ref{sec:marginal}. We want to know a little bit more regarding what kind of states could have symmetric extension. Let us start from the simplest possible case of a bipartite separable state, as given in Box 1.19 of Chapter~\ref{sec:entanglement}, i.e.
\begin{equation}
\rho_{AB}=\sum_{i}p_{i}\vert\varphi_{i_A}\rangle\langle\varphi_{i_A}\vert\otimes\vert\phi_{i_B}\rangle\langle\phi_{i_B}\vert.
\end{equation}

This separable $\rho_{AB}$ obviously has symmetric extension. In fact, then only thing one needs to do is to `copy' qubit $B$, which results in
\begin{equation}
\rho_{ABB'}=\sum_{i}p_{i}\vert\varphi_{i_A}\rangle\langle\varphi_{i_A}\vert\otimes\vert\phi_{i_B}\rangle\langle\phi_{i_{B}}\vert\otimes\vert\phi_{i_{B'}}\rangle\langle\phi_{i_{B'}}\vert.
\end{equation}
Now let us further extend $\rho_{AB}$ to a state $\rho_{ABB_1B_2}$, where $B_1,B_2$ are another two qubits, by `copying' the qubit $B$ twice, which has full symmetry between the qubits $B,B_1,B_2$. We can continue to make more $s$ copies of qubit $B$ to produce a state $\rho_{ABB_1B_2\ldots B_s}$, which has full symmetry between the qubits $B,B_1,B_2\ldots B_s$. 

We may ask the question for a given $s$, what kind of bipartite state $\rho_{AB}$ admits an $s$-copy symmetric extension. That is, there exists a state $\rho_{ABB_1B_2\ldots B_s}$, which has full symmetry between the qubits $B,B_1,B_2\ldots B_s$, such that $\rho_{AB}$ is the reduced density matrix after tracing out the qubits $B_1,B_2\ldots B_s$. Notice that if $\rho_{AB}$ admits an $s$-copy symmetric extension, then it naturally admits an $s-1$-copy symmetric extension (by tracing out $B_s$). Therefore, the set of bipartite states that admit $s$-copy symmetric extensions is a subset of those admit $s-1$ symmetric extensions. And for any $s$, this set contains the separable states as a subset.

A more interesting question is what happens if we take the limit $s\rightarrow\infty$. That is, what kind of $\rho_{AB}$ admit $s$-copy symmetric extension for any $s$. It turns out that only  separable $\rho_{AB}$ could admit all $s$-copy symmetric extension. In other words, the $s\rightarrow\infty$ limit of $s$-copy symmetric extendible states is the set of all separable states. 

This is pretty much the situation for an $N$ two-mode boson system, where all the particles are fully symmetrized.  In other words, the set of $2$-RDMs for this $N$-boson system contains only states that are very close to separable states when $N$ goes large. This observation can also be generalized to the situation of $m$ modes (i.e. $m$ single particle states) and for $k$-RDMs, whenever $m$ and $k$ are fixed. 

This is given by the following finite quantum de Finetti's theorem.
\begin{svgraybox}
\begin{center}
\textbf{Box 4.7 The finite quantum de Finetti's theorem}

The $k$-RDM $\rho_k$ of an $N$-particle bosonic state can be approximated with an 
error at most $O(m^2k/N)$ by a mixture of product states of the form $\ket{\alpha}^{\otimes k}$,
where $\ket{\alpha}$ is some single-particle bosonic state.
\end{center}
\end{svgraybox}

One immediately sees that if $m$ is a fixed constant that is independent of $N$, and if one takes the limit
$N\rightarrow\infty$, then the error $m^2k/N\rightarrow 0$ for any fixed $k$. This is to say, any $k$-RDM of a bosonic system of infinite size with finite modes can only be a mixture of product states (i.e. separable states). This is the very content of the quantum de Finetti's theorem. 

In fact, the validity of quantum de Finetti's is much more general. It applies to the situation beyond that the single-particle Hilbert space with finite dimension $m$. It is true even when the single-particle Hilbert space is `separable', which is a mathematical term meaning that the single-particle Hilbert space has countable number of basis states. That is, the single-particle Hilbert space can have infinite dimension with basis labeled by some integer $m$. This is a very general case for quantum mechanics, where observables (energy, angular momentum etc.) are with quantized eigenvalues, and the corresponding eigenvectors form a basis of the Hilbert space.

We now present the quantum de Finetti's theorem as below.
\begin{svgraybox}
\begin{center}
\textbf{Box 4.8 The quantum de Finetti's theorem}

Consider an $N$-boson system with a separable single-particle
Hilbert space $\mathcal{H}$. For any $N$-boson wave function
$\ket{\Psi_N}$ that lies in the symmetric subspace of $\mathcal{H}^{\otimes N}$, and for any constant integer $k>0$ that is independent of $N$, the $k$-RDM $\rho_k$ of $\ket{\Psi_N}$ is a mixture of product states of the form $\ket{\alpha}^{\otimes k}$, in the $N\rightarrow\infty$ limit.
\end{center}
\end{svgraybox}

The de Finetti's theorem has an immediate physical consequence -- it justifies the validity of Hartree's mean-field theory to calculate the ground-state energy of a large class of interacting bosonic systems. To be more concrete, let us consider an $n$-particle bosonic system with an interaction of
the following (generic) form
\begin{equation}
H_N=\sum_{j=1}^{N} T_j + \frac{1}{N-1}\sum_{1\leq k<l\leq N}\omega_{kl},
\end{equation}
where $T_j$ is a single-particle operator on the $j$th
boson, and $\omega_{kl}$ is a symmetric operator on the two-particle space $\mathcal{H}^{\otimes 2}$.

The Hartree's mean-field theory assumes that the (variational) ground state
is a product state with the form $\ket{\Psi_{\alpha}}=\ket{\alpha}^{\otimes N}$,
so the the ground-state
energy (per particle), denoted by $E_0^h$, is given by
\begin{equation}
\label{eq:Hartree}
\epsilon_0^h=\min_{\Psi_{\alpha}}\frac{\bra{\Psi_\alpha}H_N\ket{\Psi_\alpha}}{N}
=\frac{1}{2}\min_{\ket{\alpha}}\{\bra{\alpha}^{\otimes 2}H_2\ket{\alpha}^{\otimes 2}\}.
\end{equation}

Now for any wavefunction $\ket{\Psi}$ of the system, the
corresponding energy per particle is given by
\begin{equation}
\frac{\bra{\Psi}H_N\ket{\Psi}}{N}
=\Tr(\rho_1T)+\frac{1}{2}\Tr(\rho_2\omega)=\frac{1}{2}\Tr(\rho_2H_2).
\end{equation}

Therefore, in the thermodynamic limit, the ground-state
energy (per particle) is given by
\begin{equation}
\epsilon_0=\lim_{N\rightarrow\infty}\frac{\bra{\Psi}H_N\ket{\Psi}}{N}
=\frac{1}{2}\min_{\rho_2}\{\Tr(\rho_2H_2)\}.
\end{equation}

According to the quantum de Finetti's theorem, $\rho_2$ is a mixture of product states of the form $\ket{\alpha}^{\otimes 2}$, therefore, we only need to take minimization over all $\ket{\alpha}^{\otimes 2}$, i.e.

\begin{equation}
\epsilon_0=\frac{1}{2}\min_{\ket{\alpha}}\{\bra{\alpha}^{\otimes 2}H_2\ket{\alpha}^{\otimes 2}\}.
\end{equation}
which is exactly the same as the mean-field ground state
energy $\epsilon_0^h$ as given by Eq.~\eqref{eq:Hartree}. In other words, the mean-field ground-state energy, although comes from
a trivial wave-function, is in fact exact. We summarize this
fact below.

\begin{svgraybox}
\begin{center}
\textbf{Box 4.9 The validity of the mean-field approximation}

For a generic interacting bosonic system, the ground-state energy given by the Hartree's mean-field approximation is exact. This is a consequence of the special structure of the bosonic reduced density matrices in the thermodynamic limit (the quantum de Finetti's theorem), which does not depend on 
any specific properties of the Hamiltonian.
\end{center}
\end{svgraybox}

We remark that this result does not contradict the hardness of the $N$-presentability problem as discussed in Box 4.6. The key difference is that here the dimension of single-particle system (i.e. $m$), once chosen, is fixed, which does not grow with $N$. In other words, in the $N\rightarrow\infty$ limit, the number of particle per mode is in fact $\gg 1$, which corresponds to the so called `high density' limit in physics. 

Also, although the mean-field theory gives the exactly ground-state energy $\epsilon_0$, the ground-state wave-function $\ket{\Psi_0}$ may not be anywhere near a product state $\ket{\alpha}^{\otimes N}$. In other words, despite that the $k$-RDMs are mixture of product states, the ground-state wave-function may be genuinely entangled. 

Let us consider a concrete example. Consider a two-body
Hamiltonian
\begin{equation}
\label{eq:Hij}
H_{ij}=\ket{1_i1_j}\bra{1_i1_j}+\ket{\phi^s_{ij}}\bra{\phi^s_{ij}},
\end{equation}
where $\ket{\phi^s_{ij}}=\frac{1}{\sqrt{2}}(\ket{0_i1_j}-\ket{1_i0_j})$
is the singlet state, and $i$ ($j$) corresponds to the
$i$th ($j$th) particle. 

For the Hamiltonian
\begin{equation}
\label{H:ff}
H_0=\sum_{1\leq i<j\leq N}H_{ij}, 
\end{equation}
the ground state is two-fold degenerate, and is spanned
by $\ket{0}^{\otimes N}$
and
\begin{equation}
\ket{W_N}=\frac{1}{\sqrt{N}}
(\ket{10\ldots 00}+\ket{01\ldots 00}
\cdots+\ket{00\ldots 01}).
\end{equation}

Now further consider the Hamiltonian
\begin{equation}
H=\sum_{1\leq i<j\leq N}H_{ij}+B\sum_{j=1}^N Z_j. 
\end{equation}
For small $B<0$ (as a perturbation to $H_0$), the ground state of $H$ is then $\ket{W_N}$.

Notice that $\ket{W_N}$ is genuinely entangled. And
it is not anywhere near a product state $\ket{\alpha}^{\otimes N}$. This can be
seen from its maximal overlap with $\ket{\alpha}^{\otimes N}$,
which is given by 
\begin{equation}
\Lambda_{\max}(\ket{W_N})=\left(\frac{N-1}{N}\right)^{\frac{N-1}{2}}.
\end{equation}
This means that the geometric measure of entanglement, as discussed in Chapter~\ref{cp:1}, increases with $N$.

On the other hand, the $2$-RDM $\rho_2$ of $\ket{W_N}$ is given by
\begin{equation}
\rho_2(\ket{W_N})=
\frac{N-2}{N}\ket{00}\bra{00}
+\frac{1}{2N}\left(
\ket{01}+\ket{10}\right)\left(
\bra{01}+\bra{10}\right),
\end{equation}
which is not separable but approaches $\ket{00}\bra{00}$
when $N\rightarrow\infty$. This is consistent with the
prediction of the quantum de Finetti's theorem.

\section{Frustration-free Hamiltonians}
\label{sec:ffHam}

In this subsection, we discuss a special kind of local Hamiltonian, namely, the frustration-free Hamiltonians, which will be extensively used later in this book. We have already seen an example demonstrating the concept of `frustration' in Sec. 4.4.2. We now start from stating more formally what a frustration-fee Hamiltonian is. 

\begin{svgraybox}
\begin{center}
\textbf{Box 4.8 The frustration-free Hamiltonians}

A $k$-local Hamiltonian $H=\sum_j H_j$ is frustration-free, 
if the ground state state $\ket{\psi_0}$ of $H$ is also the ground states of each $H_j$.
\end{center}
\end{svgraybox}

\subsection{Examples of frustration-free Hamiltonians}
\label{sec:ffH}

Frustration-free Hamiltonians are widely found in practical many-body spin models. One simple example is the ferromagnetic Ising chain with an interacting Hamiltonian
\begin{equation}
H_{FIC}=-\sum_i J_iZ_iZ_{i+1},
\end{equation}
where $J_i>0$. The two-fold degenerate ground state space of $H_{FIC}$ is spanned by
\begin{equation}
\ket{0}^{\otimes N}=\ket{00\ldots 0},\quad \ket{1}^{\otimes N}=\ket{11\ldots 1},
\end{equation}
i.e. all spin up or all spin down. It is then easy to observe that both $\ket{0}^{\otimes N}$ and $\ket{1}^{\otimes N}$ are ground states of each interaction term $-J_iZ_iZ_{i+1}$.

The Hamiltonian $H_0$ given by Eq.~\eqref{H:ff} is also frustration-free, since both $\ket{0}^{\otimes N}$ and $\ket{W_N}$
are ground states of each term $H_{ij}$ as given by
Eq.~\eqref{eq:Hij}.

The toric code Hamiltonian given by the Hamiltonian $H_{toric}$ in Eq.(\ref{eq:Htoric})
is also a frustration-free one. The ground state $\ket{\psi_g}$ given in Eq.
(\ref{eq:toricg}) is the ground state of each operators $-Q_s$ and $-B_p$ for any
 $s,p$. This is straightforward to see. For $Q_s$, as $B_p$ commutes with any
 $g\in\mathcal{S}_X$, and $Q_s\ket{0}^{\otimes 2r^2}=\ket{0}^{\otimes 2r^2}$, we have
 $Q_s\ket{\psi_g}=\ket{\psi_g}$.
For $B_p$, because $\ket{\psi_g}$ sums over all $g\in\mathcal{S}_X$ where
$\mathcal{S}_X$ is a group, we have $B_p\ket{\psi_g}=\ket{\psi_g}$. This agrees with our
previous discuss in Chapter~\ref{sec:cp3sec5} that $\ket{\psi_g}$ is the stabilizer state stabilized by the
stabilizer group generated by $Q_s$ and $B_p$. In fact, any stabilizer state, with its stabilizer group generated by local Pauli operators, is the unique ground state corresponding to the local Hamiltonian given by the minus sum of all the local stabilizer generators.

Another famous frustration-free systems, namely, the Affleck-Kennedy-Lieb-Tasaki (AKLT) system, considers a spin-$1$ chain. The AKLT Hamiltonian is given by
\begin{equation}
\label{eq:HAKLT}
H_{AKLT}=\sum_j\vec{S}_j\cdot\vec{S}_{j+1}+\frac{1}{3}(\vec{S}_j\cdot\vec{S}_{j+1})^2=\sum_{j}2P_{j,j+1}^{(J=2)}-\frac{2}{3}.
\end{equation}
Here $\vec{S}_j$ is the spin operator of the $j$-th spin, and
$P_{j,j+1}^{(J=2)}$ is the projection onto the total spin $J=2$
subspace of each neighboring pair of particles.

%
\begin{figure}[h!]
\centerline{
\includegraphics[scale=2.00]{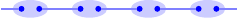}
}
%
%
\caption{A Valence-bond Solid. Each blue dot denotes a spin-$1/2$ particle. Each solid line connecting two particles are called a `bond', which represents a singlet state. Each oval contains two particles, which represents the projection of those two particles onto the spin triples subspace}
\label{fig:VBS}       
\end{figure} 

The AKLT Hamiltonian $H_{AKLT}$ is known to be frustration-free, by explicitly constructing the ground state. The idea is to use the picture of the valence-bond solid, which is illustrated in Fig.~\ref{fig:VBS}, where each bond denotes the singlet state
\begin{equation}
\ket{\text{singlet}}=\frac{1}{\sqrt{2}}(\ket{01}-\ket{10}),
\end{equation}
and each oval represents the projection onto the triplet subspace, i.e.
\begin{equation}
\label{eq:AKLTProj}
\Pi_{triplet}=\ket{+1}\bra{00}+\ket{0}\frac{1}{\sqrt{2}}(\bra{01}+\bra{10})+\ket{-1}\bra{11},
\end{equation}
where $\ket{\pm 1}$ and $\ket{0}$ are the eigenstates of spin $S_z$ operator,
corresponding to the eigenvalues $\pm 1, 0$, respectively.

This then gives us a state of a spin-$1$ chain, called the AKLT state denoted by $\ket{\psi_{AKLT}}$, when periodic boundary condition is considered. $\ket{\psi_{AKLT}}$ is a ground state of $H_{AKLT}$, because it is the ground state of each projection $P_{j,j+1}^{(J=2)}$. This is because, among the four spin-$1/2$s making up a pair of neighboring spin-$1$s, the two in the middle form a spin singlet. Therefore, the total spin of the four spin-$1/2$s, and correspondingly the total spin of the pair of neighboring spin-$1$s, can only be $0$ and $1$ but not $2$. 
Therefore, the Hamiltonian $H_{AKLT}$ is frustration-free.

$\Pi_{triplet}$ given in Eq.~\eqref{eq:AKLTProj} in fact gives a matrix product state (MPS) representation for $\ket{\psi_{AKLT}}$, which will be discussed in more detail in Chapter~\ref{chap8}. Also, $H_{AKLT}$ can be viewed as the parent Hamiltonian of the MPS state $\ket{\psi_{AKLT}}$, see Chapter~\ref{sec:parentHam}. In fact, all the MPS parent Hamiltonians discussed in Chapter~\ref{sec:parentHam} are frustration-free, with the corresponding MPS state as a ground state.

We would like to mention here that the AKLT state is an example of the symmetry-protected topologically ordered (SPT) phase, where the nontrivial order of the state is protected by the symmetry of the Hamiltonian $H_{AKLT}$. That is to say, when breaking the corresponding symmetry, there is no intrinsic long range entanglement in the AKLT state, so there exists a constant depth quantum circuit to transform $\ket{\psi_{AKLT}}$ to the product state $\ket{0}^{\otimes N}$. We will examine these symmetry and transformation in more detail in Chapter~\ref{chap10}.

\subsection{The frustration-free Hamiltonians problem}

If we know that a local Hamiltonian $H=\sum_j H_j$ is frustration-free, then its ground state energy is $E_0=\sum_j E_{0j}$, where $E_{0j}$ is the ground state energy of $H_j$, which is easy to find given that each $H_j$ acts nontrivially only on a few particles. However, for a given local Hamiltonian $H$, one first needs to determine whether it is frustration-free. In the theory of quantum computational complexity, this problem is formulated as follows.

\begin{svgraybox}
\begin{center}
\textbf{Box 4.9 The frustration-free Hamiltonian problem}

Given a local Hamiltonian $H=\sum_j H_j$, where each $H_j$ acts
non-trivially on at most $k$ qubits whose ground state energy is $0$.\\ For a given $b>c/\text{poly}(N)$, where $c>0$ is a constant, denote $E_0$ the ground state energy of $H$. Determine which of the following is true.

1. $H$ is frustration free, i.e. $E_0=0$.

2. $E_0>b$.
\end{center}
\end{svgraybox} 

When formulated in this form, it hints that the frustration-free Hamiltonian problem is quite similar  
to the local Hamiltonian problem except that $E_0$ is exactly zero, while in the latter $E_0$ is upper-bounded by some constant $a<b$. It is quite the case as given by the following result as reached in the quantum computational complexity theory. 

\begin{svgraybox}
\begin{center}
\textbf{Box 4.10 The hardness of the frustration-free Hamiltonian problem}

Given a local Hamiltonian $H=\sum_j H_j$ of an $N$-qubit system, with each $H_j$ acting nontrivially on at most $k$ particles, the problem of determining whether $H$ is frustration-free or not, is as hard as the local Hamiltonian problem for $k\geq 3$.
\end{center}
\end{svgraybox}

This observation indicates that frustration-free Hamiltonians may be already enough to characterize many kind of interesting physics. For instance, the toric code Hamiltonian gives a simple example of the so-called `topologically ordered system', which exhibit quantum phases beyond the explanation of the Landau symmetry-breaking theory. These topologically ordered systems will be the major topic in the rest chapters of this book.

\subsection{The $2$-local frustration-free Hamiltonians}

The problem can be significantly simplified when restricting to special cases. It turns out that it is easy to determine whether a $2$-local qubit-Hamiltonian $H$ is frustration-free or not. By easy we mean that for an $N$-qubit systems, there is an algorithm with running time polynomial in $N$, which determines whether $H$ is frustration-free or not. However, we know that for the local Hamiltonian problem, even the $k=2$ case is hard. In other words, although to determine whether a $2$-local qubit-Hamiltonian $H$ is frustration-free or not is easy; on the other hand, if we know $H$ is not frustration-free, then determining the ground state energy of $H$ to some precision is still hard.

To see how to determine whether a $2$-local qubit-Hamiltonian $H$ is frustration-free or not, we give a procedure which finds a special kind of ground state for $H$. To do so, we start from a simple fact that if a local Hamiltonian $H=\sum_j H_j$ is frustration-free, then for $L=\bigotimes_{i=1}^N L_i$, where each $L_i$ is a invertible operator acting on a single qubit $i$, the Hamiltonian
\begin{equation}
H'=L^{\dag}HL=\sum_j L^{\dag}H_jL
\end{equation}
is also frustration-free, because $L^{-1}\ket{\psi}$ is a ground state of $L^{\dag}H_jL$ if and only if
$\ket{\psi}$ is a ground state of $H_j$. Note that $H'$ does not have the same spectra as those of $H$, just that the frustration-free property of $H$ remains after the transformation $L$.

To understand more about the effect of $L$, let us start from the two-particle case, i.e. $N=2$. Due to Schmidt decomposition, any $2$-qubit state can be written as, in some basis
\begin{equation}
\ket{\psi_{AB}}=\sum_{\alpha=0}^{1}\sqrt{\lambda_{\alpha}}\ket{\alpha_A}\ket{\alpha_B}.
\end{equation}
There are then two nontrivial cases: 1. if one of $\lambda_{\alpha}$ is zero, then up to the transformation $L$, $\ket{\psi_{AB}}$ is essentially a product state $\ket{00}$; 2. none of $\lambda_{\alpha}$ is zero, then up to the transformation $L$ $\ket{\psi_{AB}}$ is essentially a singlet state $\frac{1}{\sqrt{2}}(\ket{01}-\ket{10})$.

Now let us move to the case of $n=3$. After some mathematics, which we omit here, one can show that up to $L$, there are essentially four possibilities to write a general three-qubit state $\ket{\psi_{ABC}}$:

\begin{enumerate}
\item a product state, i.e. $\ket{000}$.
\item a tensor product of $\ket{0}$ and a singlet state $\frac{1}{\sqrt{2}}(\ket{01}-\ket{10})$.
\item a GHZ state $\ket{GHZ}=\frac{1}{\sqrt{2}}(\ket{000}+\ket{111})$.
\item a W state $\ket{W_3}=\frac{1}{\sqrt{3}}(\ket{001}+\ket{010}+\ket{100}$.
\end{enumerate}

For case $3$, suppose $\ket{GHZ}$ is a ground state of some $2$-local frustration-free qubit Hamiltonian $H_{GHZ}$, then $\ket{GHZ'}=\frac{1}{\sqrt{2}}(\ket{000}-\ket{111})$ is also the ground state of the same $H_{GHZ}$, because the $2$-RDMs of $\ket{GHZ}$ and $\ket{GHZ'}$ are exactly the same. In other words, $\ket{000}$ must be the ground state of $H_{GHZ}$.

For case $4$, suppose $\ket{W}$ is a ground state of some $2$-local frustration-free qubit Hamiltonian $H_W$, then $\ket{000}$ must also be the ground state of $H_{W}$. In fact, any of the $2$-RDMs of $\ket{W}$ is supported on the two-dimensional subspace spanned by $\ket{1}{\sqrt{2}}(\ket{01}+\ket{10})$ and $\ket{00}$, which contains the subspace that any of the $2$-RDMs of $\ket{000}$ is supported on (which is nothing but $\ket{00}$). Or in another viewpoint, a $2$-local frustration-free Hamiltonian only `sees' the information of the range of the $2$-RDMs of its ground states, i.e. independent on any details of the $2$-RDMs beyond just its range.

As a result, in all the four possible cases of $N=3$, there always exists a ground state for any $2$-local frustration-free qubit Hamiltonian, which is either a product state, or a tensor product of a single qubit state $\ket{0}$ and a singlet state $\frac{1}{\sqrt{2}}(\ket{01}-\ket{10})$, up to certain transformation $L$. This result generalizes to the case $N>3$ by some induction argument, whose technical details are omitted here. We summarize this result as follows.

\begin{svgraybox}
\begin{center}
\textbf{Box 4.11 A ground state for $2$-local frustration-free qubit Hamiltonian}

for any $2$-local frustration-free qubit Hamiltonian $H$, there always exits a ground state which is either a product state, or a tensor product of some single qubit states and some singlet states, up to certain transformation $L$.
\end{center}
\end{svgraybox}

\begin{figure}[htbp]
\centerline{
\includegraphics[scale=1.25]{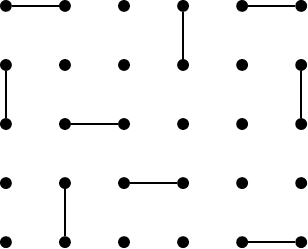}
}
%
%
\caption{A possible pattern of a ground state for a $2$-local frustration-free qubit Hamiltonian}
\label{fig:pattern}       
\end{figure}

A possible pattern of a ground state for a $2$-local frustration-free qubit Hamiltonian is illustrated in Fig.~\ref{fig:pattern}. Here each dot black denotes a single qubit state, and each solid line linking two black dots denote any two-qubit entangled state. This is not translational invariant though. In practice, if translational invariance is taken into account, we will end up with tensor product of only single qubit states, or only two-qubit entangled states, but not both. Physically, that result is pretty much saying that some kind of mean-field method always works perfect, if a $2$-local qubit Hamiltonian is frustration-free. The existence of such a single solutions provides an algorithm to efficiently determine whether a $2$-local qubit Hamiltonian is frustration-free or not.

In fact, it is not only that one can determine whether a $2$-local qubit-Hamiltonian $H$ is frustration-free or not within a reasonable amount of computational cost, but also the structure of the entire ground space can be characterized. That is, the ground state space of any $2$-local frustration-free qubit Hamiltonian can be spanned by some (maybe nonorthogonal) basis $\{\ket{\psi_i}\}$ where each basis state is is either a product state, or a tensor product of some single qubit states and some singlet states, up to certain transformation $L$. Furthermore, these basis state share similar tensor product structures, for example, if $\ket{\psi_1}$ is a tensor product of a singlet state of the first and second qubit, and single-qubit states of all the remaining qubits, up to some transformation $L_1$, then if $\ket{\psi_2}$ is a tensor product of a singlet state of the first and second qubit, and single-qubit states of all the remaining qubits, up to some transformation $L_2$, etc.

Finally, it needs to be mentioned that the results on $2$-local qubit-Hamiltonians are not extendable for $2$-local non-qubit  Hamiltonians. In other words, one no longer hopes to always find a ground state whose structure is pretty much a product state (i.e. a tensor product of single-particle states and some two-particle states). On example is that the AKLT state, which is in fact the unique ground state of $H_{AKLT}$, possesses a quite complicated entanglement structure. Note that the AKLT states does not hint anything about the hardness of the $2$-local frustration-free Hamiltonian problem in the case of systems with single-particle
dimension $m=3$, where its ground state is easy to construct.

What is known to date in quantum computational complexity theory, for how hard a the frustration-free Hamiltonian problem is, for $2$-local Hamiltonians, is the following result. 
\begin{svgraybox}
\begin{center}
\textbf{Box 4.12 The hardness of the frustration-free Hamiltonian problem for $2$-local Hamiltonians}

Given a $2$-local Hamiltonian $H=\sum_j H_j$ of an $N$-particle system with single-particle
dimension $m$, the problem of determining whether $H$ is frustration-free or not, is as hard as the Local Hamiltonian Problem for $m\geq 5$. For $m\geq 3$, the problem is hard to solve on a classical computer.
\end{center}
\end{svgraybox}
However, the hardness with quantum computers for the cases of $m=3,4$ remains unknown. That is, there might exist some other $2$-local Hamiltonian for spin-$1$ systems, such that whether it is frustration-free or not is hard to determine even with a quantum computer.

\section{Summary and further reading}

In this chapter, we introduced the concept of local Hamiltonians, and discussed related problems studied in the quantum information theory. We started with `re-emphasizing' the tensor product structure which is already discussed in Chapter~\ref{cp:1}. Traditionally many-body physics has this tensor product structure in mind, however is not emphasized. Quantum information science, however, systematically studies this structure and naturally extends this structure as a tool to study correlation and entanglement in many-body systems.

Equipped with the tensor product structure of the Hilbert space, it is then natural to discuss locality of a Hamiltonian. Usually a local Hamiltonian is associated with some spatial geometry.
It turns out that this spatial locality does play a crucial role in the study of properties for local Hamiltonians. For Hamiltonians involving only $k$-body interactions, the ground-state property is quite differently for those with interactions `localized' with respect to some $D$-dimensional lattice. In viewing the ground state space as a quantum error-correcting code, a $k$-local stabilizer Hamiltonian on a $D$-dimensional lattice, the ground state space degeneracy $R$ and the code distance $d$ satisfies a constraint as given in Eq.(\ref{eq:dbound}), that is, if $\log R\propto N$, then the distance $d$ can only be some constant independent of $N$.  This result is obtained in ~\cite{BT08}, which is further extended to more general cases beyond stabilizer quantum code in~\cite{Bra11}. In contrast, a general Hamiltonian
involving only $k$-body interactions could have $d\propto\sqrt{N}$, even if $\log R\propto N$. This result is obtained in~\cite{TZ07}, and further studied in more detail with different systems in~\cite{KP09}.

Given a $k$-local Hamiltonian $H$, the local Hamiltonian problem concerns determining the ground-state energy to certain precision. It is one of the most extended studied problems in quantum computational complexity theory, with a general assertion that the local Hamiltonian problem is hard even with the existence of a quantum computer. This observation is first proposed in~\cite{KSV02}, which shows that the local Hamiltonian problem is hard for some $5$-local Hamiltonian.
Following the original work~\cite{KSV02}, considerable progress has been made. It is shown that the $3$-local Hamiltonian problem is hard~\cite{KR03}, followed by~\cite{KKR04} 
showing that the $2$-local Hamiltonian problem is hard. 

Further taking into account of spatial geometry, it is shown that the local Hamiltonian problem is hard for a Hamiltonian involving only nearest-neighbour interaction on a $2D$ square lattice~\cite{OT08}. A surprising fact was discovered in~\cite{AGIK09} that the local Hamiltonian problem remains hard even for a Hamiltonian associated with a $1D$ chian, involving only nearest-neighbour interactions. Review articles on quantum computational complexity includes~\cite{AN02} and~\cite{Osb12}.

The $N$-representability problem has been studied in quantum chemistry for several decades, see e.g.~\cite{Col63}. For the history and of the quantum marginal problem, we refer to~\cite{Kly06}. It is shown that the quantum marginal problem is hard, even with the existence of a quantum computer ~\cite{Liu06}. The hardness of the $N$-representability problem is shown in~\cite{LCV07}. The hardness of the $N$-representability problem is also looked at in~\cite{WMN10}.

The notion of symmetric extendibility for a bipartite quantum state was introduced in~\cite{doherty2002distinguishing} as a test for entanglement, as a state without symmetric extension is evidently entangled. The condition of symmetric extension for two-qubit state as given in Eq.~\eqref{eq:symext} is conjectured in~\cite{myhr2009spectrum} and proved in~\cite{chen2014symmetric}.

The original de Finetti's theorem is a theorem in probability theory, which is named in honour of Italian statistician Bruno de Finetti. The theorem states that an infinite exchangeable sequence of Bernoullli
random variable is a mixture of independent and identically distributed Bermoulli random variables. The finite version of de Finetti's theorem is due to~\cite{diaconis1980finite}. The quantum version of de Finetti's theorem dates back to 1960s paper~\cite{stormer1969symmetric,hudson1976locally}. The development of quantum information theorem raises great attention 
of the quantum de Finetti's theorem and its finite version, due to its applications in many aspects. For a recent review we direct the readers to~\cite{harrow2013church} and references
therein. For the validity of Hartree's mean-field theory
for bosonic systems, we refer to~\cite{lewin2014derivation} and references
therein.

The ALKT Hamiltonian is originally discussed by Affleck, Kennedy, Lieb and Tasaki in 1987~\cite{AKLT87}. The computational complexity of frustration-free Hamiltonians was first studies in~\cite{Bra06}, which showed that for qubit Hamiltonians, the frustration-free Hamiltonian problem is easy for the $2$-local case, and is hard for the $4$-local case. 

It is further shown in~\cite{CCD+10} that for any $2$-local frustration-free qubit Hamiltonian $H$, there always exits a ground state which is either a product state, or a tensor product of some single qubit states and some singlet states, up to certain transformation $L$. The ground-state space structure for any $2$-local frustration-free qubit Hamiltonian is characterized in~\cite{JWZ10}. Recently it is shown that the frustration-free Hamiltonian problem is hard for the $3$-local case~\cite{GN13}.

For the $2$-local case of the frustration-free Hamiltonian problem, it is shown that the $m=5$ care is hard with the existence of a quantum computer~\cite{EG08}. It is known in 1979 already that the $m=3$ case is hard with a classical computer~\cite{GJ79}. However, whether this case (and $m=4$) is also hard with a quantum computer remains unknown. It is worth mentioning that the $2$-local Hamiltonian constructed in~\cite{AGIK09} is in fact frustration-free.

For review articles on frustration-free Hamiltonians, we refer to~\cite{Dan08}. Frustration-free Hamiltonians also play important role in the area of the so-called `measurement-based quantum computing'. Related review particles include~\cite{RW12} and~\cite{KWZ12}.

%
%
\bibliographystyle{plain}
\bibliography{Chap4}

%
%
%
\chapter{Gapped Quantum Systems and Entanglement Area Law}
\label{cp:5} 

\abstract{We discuss quantum systems of large system size $N$, in particular the $N\rightarrow\infty$ limit (thermodynamic limit). We call them the quantum many-body systems. We ask about the physical properties of these quantum many-body systems in the thermodynamic limit. Our main focus in this chapter will be many-body systems with an energy gap and the entanglement properties of their ground states. We introduce the entanglement area law and its constant correction -- the topological entanglement entropy. We discuss the information-theoretic meaning of topological entanglement entropy, and generalize it to construct a `universal entanglement detector', whose values on the gapped ground states provide non-trivial information of the system.}

\section{Introduction}

The solid or liquid materials we encounter everyday are macroscopic systems comprised of a large number of quantum particles, like bosons, fermions or spins. The number of particles in the systems are so large that amazing macroscopic quantum phenomena emerge, like superfluidity, superconductivity and topological order, which are not possible with a small number of quantum particles. These are the systems of interest in the study of quantum condensed matter physics.

From our experience in the last chapter, it seems impossible to theoretically study quantum systems of this size, as solving quantum systems of a few tens of particles are already extremely hard. However, the quantum materials we are interested in comprise a special set of all quantum system where the notion of dimensionality, locality, and thermodynamic limit play an important role and allow generic features of such quantum systems to be attainable. We will start to look at these concepts in section~\ref{sec:qMany-body} and discuss how they determine important properties like the correlation, gap and entanglement in the system.

We will then focus on studying gapped systems, in particular the 
entanglement properties of their ground states. We will demonstrate with examples that different gapped systems may exhibit different features in their ground-state entanglement pattern, such that they belong to different `order'. Although the readers may be familiar with the term `symmetry breaking order', the terms of `topological order' and even `symmetry-protected topological (SPT) order' may not sound familiar. We will formally define the concept of `quantum phase' in Chap~7 and discuss a general theory based on local transformations. Here in this chapter, we will just use those 
terms together with the corresponding examples, hoping to give the readers some feelings about their meaning through looking at concrete examples.

In section~\ref{sec:area}, we discuss a general structure of ground-state entanglement for gapped systems, namely the entanglement area law, which states that the entropy of the reduced density matrix of some connected area of the system is proportional to the boundary length of the area. If the system is `topologically ordered', then there will be also a subleading constant term of the entropy apart from the area law, which is the so called `topological entanglement entropy', denoted by $S_\text{topo}$. The existence of such an topological entanglement entropy for some gapped ground states then indicates that the system is topologically ordered. In other words, a non-zero $S_\text{topo}$ detects topological order.

In section~\ref{sec:irr-corr}, we develop an information-theoretic viewpoint for the topological entanglement entropy $S_\text{topo}$. We show that $S_\text{topo}$ essentially captures the `irreducible tripartite correlation' $C_{tri}(\rho_{ABC})$ (as discussed in Chapter 1.4.2) for gapped systems for the areas $A,B,C$. The relevant general quantity in quantum information theory is the conditional mutual information $I(A{:}C|B)$, which is the quantum mutual information of the parts $A$ and $C$, conditioned on the existence of the part $B$. When choosing large enough areas $A,B,C$ of the system, and $A,C$ are far from each other, a non-zero $I(A{:}C|B)$ hence indicates a non-trivial kind of many-body entanglement. This then generalizes the topological entanglement entropy, which can be also used to detect other orders (e.g. symmetry breaking orders, SPT orders) of the system (even without knowing the symmetry of the system). 

In section~\ref{sec:QECC}, we discuss the property of the degenerate ground-state space of a gapped system from the viewpoint of quantum error-correcting codes (QECC), which has been discussed in Chapter~\ref{cp:3}. We know that toric code is a QECC with a macroscopic distance. We show that, for systems with symmetry constraints and `symmetry breaking orders' and even 'SPT orders', if we only consider the errors that respect the symmetry of the system, then the corresponding degenerate ground-state spaces are also QECCs with macroscopic distances.

We briefly discuss gapless systems in Sec.~\ref{sec:ent_gapless}, where the area law is violated. We show that the conditional mutual information $I(A{:}C|B)$ depends on the shapes of the areas $A,B,C$, which can also provide information for critical systems (e.g. central charge), if one calculates  $I(A{:}C|B)$ for different area shapes. In this sense, $I(A{:}C|B)$ is a `universal entanglement detector' for both gapped and gapless systems, which contains non-trivial information of the orders of the systems. 

\section{Quantum many-body systems}
\label{sec:qMany-body}

In this section, we discuss the concepts of dimensionality, locality, and thermodynamic limit and how they determine important properties like the correlation, gap and entanglement in the system.

\subsection{Dimensionality and locality}

Condensed matter systems usually live in a space of fixed dimension. For example, a sodium crystal is composed of ions forming a three dimensional lattice, a graphene sheet is made of carbon atoms in a two-dimensional (2D) lattice and nanowires are effectively one-dimensional (1D). The electrons in these systems are confined to move within their dimensions. 

Systems in the same dimension can have different geometry or topology. For example, in one dimension, the system can be in an open chain with two end points or a closed ring with no boundary; in two dimension, the system can be in a disc with a one dimensional boundary or in a sphere or torus with no boundary but different topology; similarly in three dimension system can be on a cube with a two dimensional boundary or we can imagine hyperthetically putting the system in a closed three dimensional manifold by closing the boundary. Closed three dimensional manifold can also have different topology.

\begin{figure}[htbp]
\begin{center}
\centerline{
\includegraphics[width=1.2in]{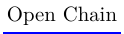}~~~~~~~~
\includegraphics[width=1.2in]{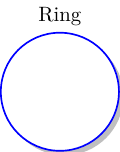}~~~~~~~~
\includegraphics[width=1.2in]{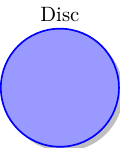}
}
\centerline{
\includegraphics[width=1.2in]{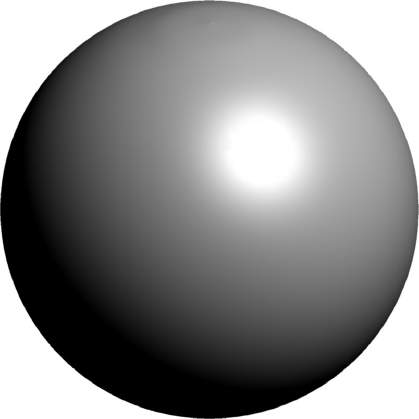}~~~~~~~~
\includegraphics[width=1.2in]{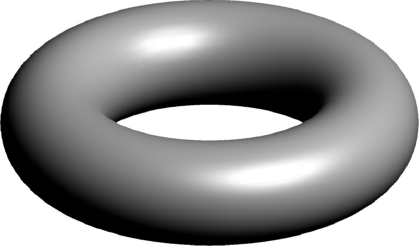}~~~~~~~~
\includegraphics[width=1.2in]{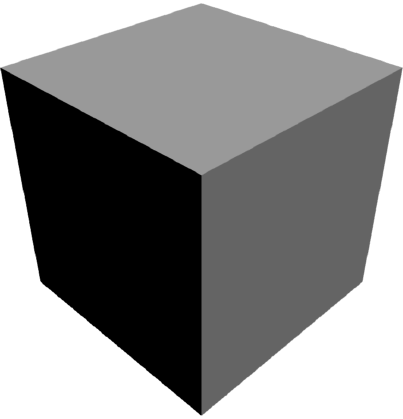}
}
\caption{Manifolds with geometry or topologies in one, two and three dimensions.} 
\label{topologies}
\end{center}
\end{figure}

The dimensionality not only confines the motion of the quantum particles within the system, it also puts restrictions on how particles interact with each other. In condensed matter systems, particles can only interact with one another if they are within certain distance. The strength of interaction decays to zero if the particles are sufficiently far apart. Therefore, the notion of locality is more strict than in the quantum systems we considered in the last chapter. Besides the request that only a few (for example 2 or 3) quantum particles can be involved in an interaction, the particles also have to be close enough. Exactly how close the particles have to be to interact depends on the physical details of the system. In general, we consider systems with a finite interaction range. That is the interaction strength decay to zero beyond a fixed length scale while the total system size $N$ can be taken to be infinity. Therefore, in the following discussion, by local interaction, we refer to interactions involving not only a finite number of particles but also within a finite range. The total Hamiltonian $H$ of the system is a sum of such local interaction terms
\be
H=\sum_j H_j.
\ee

\subsection{Thermodynamic limit and universality}

For quantum systems of the size of $10^{23}$ (or even much smaller), it is impossible to know all the details of the system. However, in most cases, we are not interested in most details. What we care most about are some global generic features like whether the material is conducting, whether it is a superfluid, etc. Such properties are observed on the macroscopic scale of the system and does not depend on a lot of details at the microscopic scale. Therefore, in the study of quantum condensed matter systems we are generally interested only in what happens when the system size $N$ goes to infinity, the so-called thermodynamic limit, and how the system respond to external probes on a macroscopic length and time scale. In particular, we will be investigating the physical properties such as their gap $\Delta$, correlation $\xi$, and entanglement $S$, which determine the electronic, magnetic, or optical properties of the system. In the thermodynamic limit, generic features appear for these quantities in a quantum many-body systems. Such generic features are said to be `universal' for the quantum many-body systems, which do not depend on much of the details of the system.

\subsection{Gap}
\label{sec:gap}

In the limit of the system size $N\rightarrow\infty$, one important property of the Hamiltonian $H$ is the gap $\Delta$.
Denote $H_N$ the system Hamiltonian with system size $N$.
The system is called gapped if one of the following case is true.
\begin{itemize}
\item[(1)]~  As $N \to \infty$, the ground state degeneracy $m_N$ of $H_N$ is upper bounded by a finite integer $m$, and the gap $\Delta_N$ between the ground states and the first excited states of $H_N$ is lower bounded by a finite positive number $\Delta$. 
\item[(2)]~  As $N \to \infty$, there are a finite number $m$ of lowest energy states which have energy separations $\epsilon$ among themselves, which is exponentially small in $N$, and the energy separation of these lowest energy states to all the other states is lower bounded by a finite number $\Delta$ for arbitrary $N$. 
\end{itemize}

\begin{figure}[htbp]
\begin{center}
\includegraphics[width=2in]{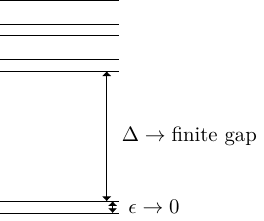}
\caption{
The energy spectrum of a gapped system: 
$\epsilon$ vanishes as system size increases, while
$\Delta$ approaches a finite non-zero value (which is called the energy gap).
} 
\label{gap}
\end{center}
\end{figure}

The energy levels in these two cases are illustrated in Fig.\ref{gap} (notice that case (1) is in fact case (2) with $\epsilon=0$). In both cases (1) and (2), $m$ is said to be the ground state degeneracy of the system in thermodynamic limit, even though in case (2), the ground states are not exactly degenerate for any finite system size $N$. A more formal definition of gap will be discussed in Chapter~\ref{chap7}.

\begin{figure}[htbp]
\begin{center}
\includegraphics[width=4.0in]{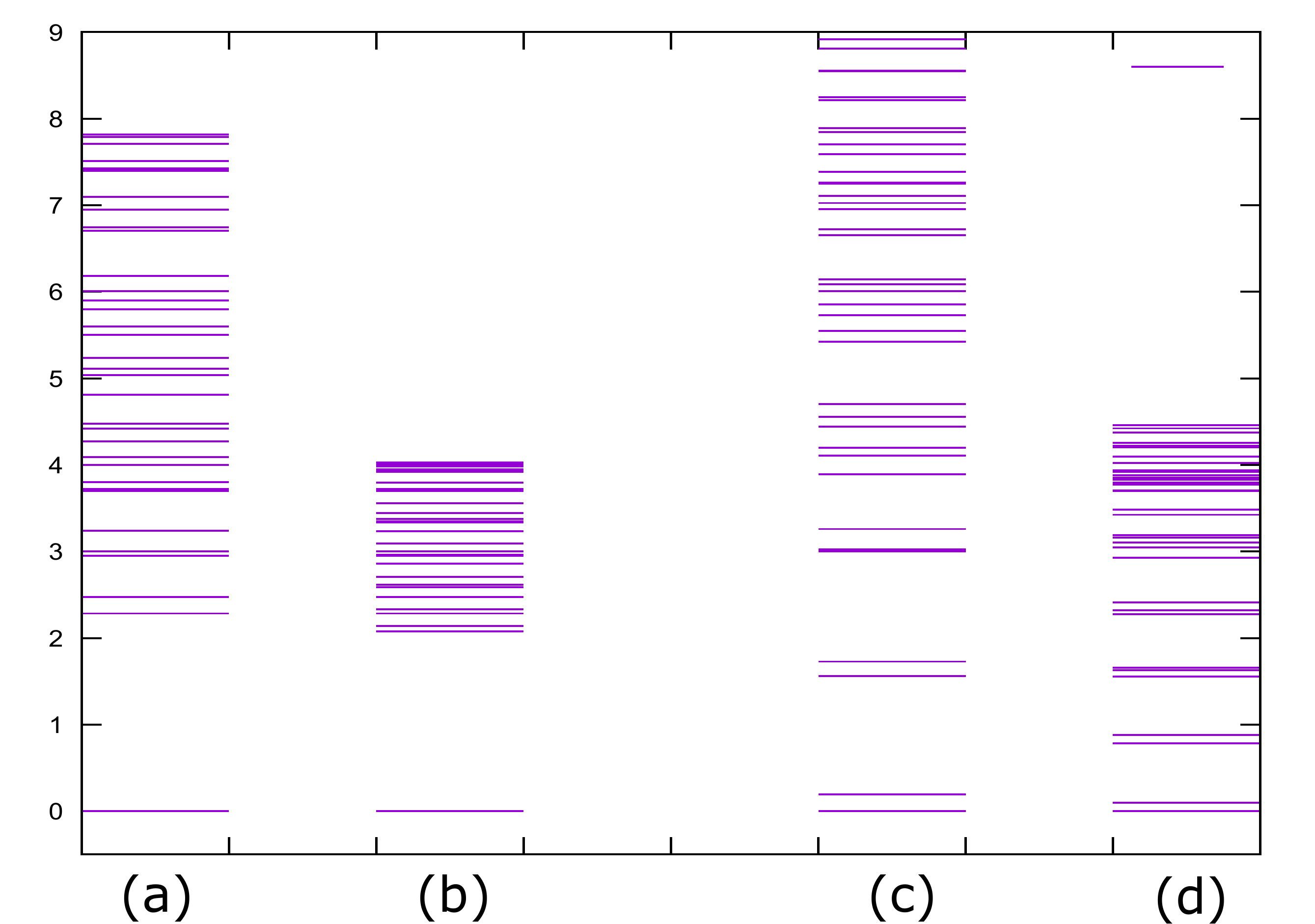}
\caption{Change in energy level spacings with system size in gapped (a and b) and gapless (c and d) systems. 
(a) The first 100 energy levels of
transverse Ising model with $J=1, B=0.5$ on a ring of 8 sites.
(b) The first 100 energy levels with $J=1, B=0.5$ on a ring of 16 sites.
(c) The first 100 energy levels with $J=1, B=1$ on a ring of 8 sites.
(d) The first 100 energy levels with $J=1, B=1$ on a ring of 16 sites.
In (a) and (b) the ground states are nearly two-fold degenerate (within the thickness of the line).
} 
\label{tIsing}
\end{center}
\end{figure}

The transverse Ising model on a 1D chain, with Ising coupling between nearest neighbor pairs, is described by the Hamiltonian 
\be
H^{\text{tIsing}}= -J\sum_{j}Z_jZ_{j+1} - B\sum_jX_j
\ee
is gapped as long as $J \neq B$. When $|J|>|B|$, the ground state degeneracy is $m=2$ (see Fig. \ref{tIsing}(a,b); when $|J|<|B|$, the ground state is nondegenerate with $m=1$. In particular, when $B=0$, the ground space is spanned by
\be
\ket{0}^{\otimes N}=\ket{00...0}, \ \ket{1}^{\otimes N}=\ket{11...1}
\ee
and when $J=0$, the ground state is
\be
\ket{+}^{\otimes N} =\left(\frac{1}{\sqrt{2}}\ket{0}+\frac{1}{\sqrt{2}}\ket{1}\right)^{\otimes N}
\ee

Sometimes we say a state is a gapped quantum state without explicitly identifying the interactions in the Hamiltonian of the system. In such cases we are implying that a Hamiltonian with local interactions can be constructed which has the state as a gapped ground state.

If we cannot find a finite set of exponentially close lowest energy states
which are finitely separated from all excited states, then the system is called
gapless. The most generic energy spectrum of a gapless system has a continuum
of energy levels above the ground states with energy spacing between them being
polynomially small in system size $N$. Simple examples of gapless system
include the Ising model at critical point, i.e. when $|J|=|B|$ (see Fig.
\ref{tIsing}(c,d)). Also, the spin-$1/2$ Heisenberg model 
\be
H_{\text{Heisenberg}}= -J\sum_j {\vec{S}_j\cdot \vec{S}_{j+1}}
\ee 
on a one dimensional
chain with nearest neighbor coupling is gapless,
where $\vec{S}_j=(X_j,Y_j,Z_j)$ is the spin operator acting
on the $j$th spin.

\subsection{Correlation}

Correlation in a quantum many body system is usually measured between local operators. Suppose that $O_1$ and $O_2$ are operators acting on finite regions $R_1$ and $R_2$ separated by a distance $r$, then the (connected) correlation function between $O_1$ and $O_2$ is defined as
\be
C^{O_1,O_2}(r)=\<O_1O_2\>-\<O_1\>\<O_2\>
\ee 
where $\<\cdot\>$ denotes taking average in the ground state (at zero temperature) or the thermal state (at finite temperature) of the system. 

The behavior of the correlation function as $r$ goes to infinity is an important indicator of the physical properties of the system. 

If a gapped system at zero temperature has a unique ground state, then all correlation functions of the ground state decay exponentially with $r$. 
\be
C(r)\sim e^{-r/\xi}
\ee
where $\xi$ is called the correlation length of the system. Therefore, a gapped quantum system with non-degenerate ground state has a finite correlation length at zero temperature. For example, in the Ising model with $|J|<|B|$, the unique ground state has a finite correlation length. In particular at the point of $J=0$, the ground state is a total product state of all spins pointing in the $x$ direction and has correlation length $\xi=0$. States with finite correlation length $\xi$ (i.e. a constant $\xi>0$ that is independent of the system size $N$) for all local operators are called short range correlated states.

On the other hand, if the system is gapless, for example a system with a fermi surface, the correlation function in the ground state decays polynomially with $r$
\be
C(r)\sim \frac{1}{r^{\alpha}}
\ee
with $\alpha>0$. As inverse polynomial functions decay slower than any inverse exponential functions, such systems are said to have infinite correlation lengths $\xi$. For example the correlation functions all decay polynomially in the ground state of the Heisenberg model. Correlation functions for different operators decay with different coefficient $\alpha$.

In the case where the system is gapped but has ground state degeneracy, the behavior of correlation functions depends on the origin of the ground state degeneracy. Let us consider the examples in two different cases: the Ising model and the toric code model. In the Ising model with $|J|>|B|$, the ground state is two fold degenerate. There is a set of basis states, for example the $\ket{0}^{\otimes N}$ and the $\ket{1}^{\otimes N}$ state at $B=0$, which individually have finite correlation length for all operators $O$. However, if we measure correlation length in the full ground space of operators $Z$, we find that
\be
C^{Z,Z}(r)=\<Z_iZ_{i+r}\>_2-\<Z_i\>_2\<Z_i\>_2 \stackrel{r \to \infty}{\rightarrow} \text{constant}
\ee
where $\<\cdot\>_2$ denotes taking average in the two dimensional ground space. In particular, when $B=0$, $C^{Z,Z}(r)\stackrel{r \to \infty}{\rightarrow} 1$. 

Such constant correlation functions are closely related to the face that the Hamiltonian of the system is invariant under the symmetry of spin flipping $\ket{0} \leftrightarrow \ket{1}$ while the two short range correlated ground states are not. In fact, the two short range correlated ground states are mapped into each other under this symmetry transformation.  This phenomena is called symmetry breaking and is going to be explained further in detail in Chapter~\ref{chap7}. Here we just want to mention that constant correlation functions for operators (operator $Z$ in this example) which break the symmetry of the system (spin flip in this example) is the most important indicator of symmetry breaking.

However, in the case of the toric code model as discussed in Chapter~\ref{sec:cp3sec5}, the situation is very different. The ground space of the toric code Hamiltonian
\be
H_{\text{toric}}= -\sum_s \prod_{j\in \text{star}(s)} Z_j - \sum_p \prod_{j\in \text{plaquette}(p)} X_j.
\ee
as given in Eq.(\ref{eq:Htoric}) is four-fold degenerate. The correlation length of any state in this four dimensional space is $0$. This is very different from the Ising model and is closely related to the fact that the ground state degeneracy in the toric code model has a topological original. The notion of topological order is going to be discussed in more detail in Chapter~\ref{chap7}.

\subsection{Entanglement}

As we have seen from Chapter 1, the entanglement property of systems with $3$ or $4$ quantum degrees of freedom has already become extremely complicated. For condensed matter systems with $~10^{23}$ degrees of freedom, it is impossible and in most cases not necessary to understand the entanglement structure of the many-body system exactly. The philosophy in studying many-body entanglement in condensed matter systems is to again focus on the scaling behavior of certain entanglement quantities of the system in the thermodynamic limit.

Entanglement quantities which has been extensively used in such studies include: (1) entanglement entropy (mutual information or other entanglement measures) with respect to a bipartition of the system in the limit of the size of both regions going to infinity (2) entanglement entropy (geometric entanglement, negativity or other entanglement measures) of two local regions in the system as the distance of the two regions going to infinity. 

In the following discussion, we are going to focus on entanglement quantities of the first type, which has been shown to be able to reveal much of the universal properties of the system. In particular, in a large class of physically interesting systems, the bipartite entanglement is found to be proportional to the area of the boundary between the two parts of the system, satisfying the so-called area law. On top of that, we discuss the sub-leading correction term to this area law behavior -- the 
topological entanglement entropy $S_\text{topo}$. Unlike the two types of entanglement entropy mentioned above, $S_\text{topo}$ is essentially the quantum conditional mutual information $I(A{:}C|B)$ for three large regions of the system. We will further explore the meaning of  $I(A{:}C|B)$, based on which we design entanglement detectors to detect different orders in the system. 

Just like correlation functions, the behavior of many-body entanglement quantities in the thermodynamic limit are closely related to the physical properties of the system. For example, they behave differently in systems with or without a gap. Moreover, many-body entanglement quantities measures the `quantum correlation' in the system that is not detected by the classical correlation function and play an important role in the study of topological orders, which will be the main focus of later chapters. 

Let us summarize some basic properties of many-body entanglement in different types of systems. The fact that degrees of freedom can interact only locally with each other in condensed matter systems puts a strong constraint on the amount and form of entanglement that can be present in  many-body systems.

\section{Entanglement area law in gapped systems}
\label{sec:area}

\subsection{Entanglement area law}
\label{subsec:area_law}

\begin{figure}[htbp]
\begin{center}
\includegraphics[width=2.00in]{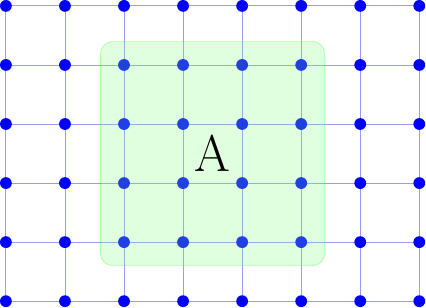}
\caption{2D square lattice with subregion A} 
\label{2D_sq}
\end{center}
\end{figure}

Consider, for example, a system on a two dimensional square lattice as shown in Fig. \ref{2D_sq} with local interactions. 
If the system is gapped (i.e. $\Delta>0$), then the many-body entanglement in the ground state satisfies a surprising property called the `entanglement area law'. More specifically, suppose we take a subregion A (as shown in Fig. \ref{2D_sq}) of size $L^2$ from the whole system and calculate the entanglement entropy $S=Tr(\rho_A\ln \rho_A)$ for this region. This calculation is done in the limit of total system size going to infinity. The number of degrees of freedom in this region is proportional to $L^2$, so the maximum entanglement entropy we can get (and actually will get for a generic many-body entangled state) scales as $L^2$. However, the calculation for a gapped ground state always gives an entanglement entropy $S$ which is proportional to the length of the boundary of the region, which grows as $~L$. 

\begin{svgraybox}
\begin{center}
\textbf{Box 5.1 Entanglement area law for gapped systems}

For a gapped system in 2D, we have
\be
S_A \sim \alpha L
\label{area_law}
\ee
\end{center}
\end{svgraybox}
Therefore, a gapped ground state in a locally interacting system always contains much less entanglement than a generic quantum many-body entangled state. 

The term `area law' is better suited to describe three dimensional system where the entanglement entropy of a subregion in a gapped ground state scales as the surface area of the region rather than the volume of the region. The basic idea applies to systems in any dimension though, which says that the entanglement entropy of a subregion scales as the size of the boundary rather than the size the bulk of the system. In particular, in one dimension, the boundary of a subregion -- a segment of the chain -- contains only two points. Therefore, entanglement entropy of a segment is bounded by a constant $S_A \leq \text{constant}$ in a one dimensional gapped system. In two dimension, the entanglement entropy scales as the linear size of the subregion $S_A \sim \alpha L$ while in three dimension, the entanglement entropy scales as the linear size squared $S_A \sim \alpha L^2$.

\begin{figure}[htbp]
\begin{center}
\includegraphics[width=2.00in]{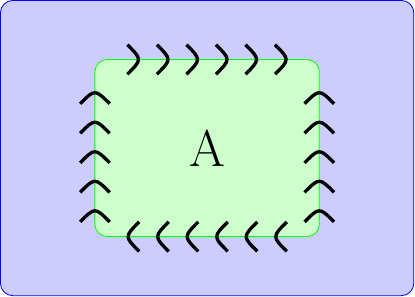}
\caption{In a local gapped quantum system, entanglement between a subregion A and the rest of the system is due to local entanglement along the boundary and hence scales as the size of the boundary.} 
\label{2D_sq_vb}
\end{center}
\end{figure}

The existence of such an `area law' in many-body entanglement depends crucially on the locality of the interactions and the existence of a gap in the system. An intuitive way to understand why the `area law' holds for gapped quantum systems is to realize that quantum correlation is generated by interactions. If the local degrees of freedom in different regions of the system do not interact at all, the ground state would be a total product state and hence no entanglement exists. If the local degrees of freedom interact locally in the system, then between the subregion $A$ and the rest of the system, only degrees of freedom close to the boundary can be interacting with each other. Moreover, in a gapped system, correlations exist in the system only under a finite length scale. Intuitively, this means that degrees of freedom in the system can only `feel' those within a finite region around it. Therefore, the entanglement between a subregion $A$ and the rest of the system is only due to the entanglement between degrees of freedom along the boundary. Pictorially, we can imagine the subregion $A$ and the rest of the system being `sewed' together by entangled pairs along the boundary as shown in Fig.\ref{2D_sq_vb}. The number of degrees of freedom along the boundary scales as the size of the boundary, hence the entanglement satisfies the `area law'. Of course, if the locality condition is removed, the area law no longer holds. If the system is not gapped, the `area law' will also be violated, but only mildly, as we discuss in the next section.

While the scaling of entanglement entropy with boundary size is a universal feature for gapped systems, the coefficient of the area law scaling $\alpha$ is not universal and depends strongly on the details of the interactions in the system. In one dimension, the constant bound on the entanglement entropy of a segment is also not universal. For example, in the one dimensional Ising model, when $J=0$, the ground state is a total product state and $S_A=0$ for any subregion. When $|J|\ll|s|$, the ground state is still unique and gapped. However, a segment in the chain would in general be entangled with with rest of the system and $S_A$ attains a larger value as $|J|$ increases.

The existence of such an `area law' also makes it possible to have an efficient description of many-body entangled states in gapped quantum systems, as discussed in Chapter 9 in terms of tensor product states. 

\subsection{Topological entanglement entropy}
\label{sec:topo_ent}

Other than revealing the gapped/gapless nature of the system, entanglement entropy can provide more detailed information about the order in the quantum state if we look at it more carefully. In particular for a gapped quantum system, if the system has nontrivial topological order, then the entanglement entropy of a region contains a sub-leading constant term apart from the leading area law term.

\begin{svgraybox}
\begin{center}
\textbf{Box 5.2 Entanglement area law for topologically ordered systems}

For a topologically ordered system in 2D
\be
S_A \sim \alpha L-\gamma
\label{Topo_S}
\ee
with $\gamma>0$.
\end{center}
\end{svgraybox}
Such a term indicates the existence of certain long-range entanglement structure that originates from the topological nature of the system (see Chapter~\ref{chap7} for a detailed discussion on long/short-range entanglement). $\gamma$ is called the topological entanglement entropy of the system.

\begin{figure}[htbp]
\begin{center}
\centerline{
\includegraphics[width=1.6in]{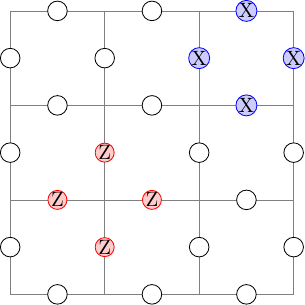}~~~~~~~~~~
\includegraphics[width=2in]{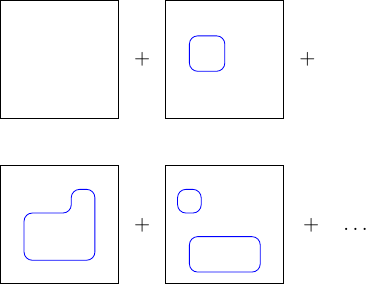}
}
\centerline{
	(a) ~~~~~~~~~~~~~~~~~~~~~~~~~~~~~~~~~~~~~~~~~~~~~~~~~~~~~~~~~~~~~~~~~~~
	(b) 
}
\caption{Hamiltonian (a) and ground state wave function (b) of the toric code model.} 
\label{TC_SN}
\end{center}
\end{figure}

While we have not defined what a topological order is, we are going to illustrate the topological entanglement entropy with the simple example of the toric code model. The value of $\gamma$ can be easily derived from a simple understanding of the ground state wave function of the system. The Hamiltonian of the toric code reads,
\be
H_{\text{toric}}= -\sum_s \prod_{j\in \text{star}(s)} Z_j - \sum_p \prod_{j\in \text{plaquette}(p)} X_j
\ee
where $Z_j$ and $X_j$ are Pauli operators acting on the qubits living on the links of, for example, a square lattice as shown in Fig.\ref{TC_SN}. If we interpret the $\ket{0}$ state as a link with no string and the $\ket{1}$ as a link with a string, then the first term in the Hamiltonian $\prod_{j\in \text{star}(s)} Z_j$ requires that there is always an even number of strings going through a vertex. In othe words, the strings always form closed loops. The second term $\prod_{j\in \text{plaquette}(p)} X_j$ creates, annhilates, or moves closed loops around each plaquette. Therefore, the ground state wave function is an equal weight superposition of all closed loop configurations $\mathcal{C}$,
\be
\ket{\psi_{\text{toric}}} = \sum_{\mathcal{C}} \ket{\mathcal{C}}
\ee
$\mathcal{C}$ includes the vacuum configuration, small loop configurations, large loop configurations and multiple loop configurations, as shown in Fig.\ref{TC_SN} (b). 

\begin{figure}[htbp]
\begin{center}
\includegraphics[width=2.00in]{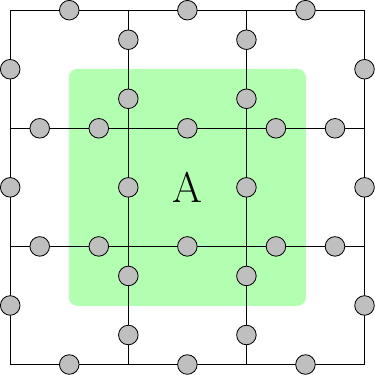}
\caption{When taking out a subregion $A$ from the lattice to calculate the entanglement entropy, we split the sites on the boundary links into two. The total wave function $\ket{\psi_{\text{toric}}}$ generalizes to the new lattice in the natural way.} 
\label{TC_cut}
\end{center}
\end{figure}

From such a string-net picture of the ground state wave function, we can easily calculate the entanglement entropy of a subregion in the system. To make the boundary more symmetric, we split the sites on the boundary links into two sites (see Fig.\ref{TC_cut}). The wave function $\ket{\psi_{\text{toric}}}$ generalizes to the new lattice in the natural way (by identifying the $\ket{0}$ and $\ket{1}$ state on sites on the same link). The new wave function (still denoted by $\ket{\psi_{\text{toric}}}$) has the same entanglement entropy. 

When the total system is divided into regions in $A$ and out of $A$ (denoted by $\bar{A}$), we can effectively view the system as a bipartite system with two parts $A$ and $\bar{A}$. According to the Schmidt decomposition as discussed in Chapter~\ref{sec:Sch}, with respect to the bipartition $A$ and $\bar{A}$,
we can decompose the ground state wave function as $\ket{\psi_{\text{toric}}}=\sum_{q}\ket{\psi^{\text{in}}_q}\ket{\psi^{\text{out}}_q}$, where $\ket{\psi^{\text{in}}_q}$ are wave functions of spins inside $A$ and $\ket{\psi^{\text{out}}_q}$ are wave functions of spins outside $A$ (i.e. in $\bar{A}$). They are connected by spins on the boundary $q_1,...q_L$. A
simple decomposition can be obtained using the string picture. For any $q_1, ..., q_L$, with $q_m= 0,1$, and $\sum_m q_m$ even, we can define a wave function $\psi^{\text{in}}_{q_1,...,q_L}$ on the spins inside of $A$: Let $X$ denote a particular spin configuration inside $A$, $\psi^{\text{in}}_{q_1,...,q_L}(X) = 1$ if (a) the strings in $X$ form closed loops and (b) $X$ satisfies the boundary condition that there is a string on $i_m$ if $q_m = 1$, and no string if
$q_m = 0$. Similarly, we can define a set of wave functions $\psi^{\text{out}}_{r_1,...,r_L}$on the spins outside of $A$.

If we glue $\psi^{\text{in}}$ and $\psi^{\text{out}}$ together - setting $q_m = r_m$ for
all $m$ - the result is $\psi$. Formally, this means that
\be
\ket{\psi_{\text{toric}}}=\sum_{q_1+...+q_L \text{even}}\ket{\psi^{\text{in}}_{q_1,...q_L}}\ket{\psi^{\text{out}}_{q_1,...q_L}}
\ee

It is not hard to see that the functions $\{\ket{\psi^{\text{in}}_{q_1,...q_L}}: \sum_m q_m \text{even}\}$, and $\{\ket{\psi^{\text{out}}_{r_1,...r_L}}: \sum_m r_m \text{even}\}$ are orthonormal up to an irrelevant normalization factor. Therefore, the density matrix for the region A is an equal weight mixture of all the $\{\ket{\psi^{\text{in}}_{q_1,...q_L}}: \sum_m q_m \text{even}\}$.
There are $2^{L-1}$
such states. The entropy is therefore
\be
S_{A,\text{toric}}=(L-1)\log 2
\ee
That is, the topological entanglement entropy for toric code model $\gamma=\log 2$.

The value of $\gamma$ is closely related to the kind of topological order in the system and is the same for quantum systems having the same topological order, independent of all other details of the system. Therefore, it provides a universal quantum number to characterize the topological order in a system, a concept which we are going to explain in much more detail in later chapters.

In a generic quantum system with non-zero correlation length, the calculation of topological entanglement entropy may not be as straight forward as in the case for the toric code model. This is because the topological entanglement entropy $\gamma$ is only a subleading term in the entanglement entropy of a subregion $S_A$. It can be hard to separate this term from the leading `area law' term and various other non-universal contributions to $S_A$ from finite size effects in actual calculations. To properly extract this universal value, the following two schemes can be used for calculation.

\begin{figure}[htbp]
\begin{center}
\centerline{
\includegraphics[width=2.00in]{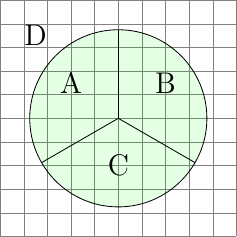}~~~~~~
\includegraphics[width=2.00in]{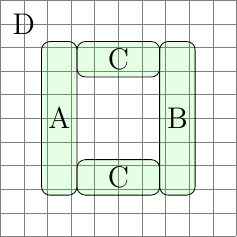}
}
\centerline{
	(a) ~~~~~~~~~~~~~~~~~~~~~~~~~~~~~~~~~~~~~~~~~~~~~~~~~~~~~~~~~~~~~~~~~~
	(b) 
}
\caption{Schemes for calculating topological entanglement entropy from a wave function.} 
\label{Stop}
\end{center}
\end{figure}

In the first scheme, the subregion $\bar{D}$ is divided into three parts $A$, $B$ and $C$, as shown in Fig.~\ref{Stop}(a). The topological entanglement entropy $\gamma$ can be calculated in the limit of both total system size and the size of $A$, $B$, $C$ going to infinity as
\be
\label{eq:topent1}
\gamma=S_{AB}+S_{BC}+S_{AC}-S_A-S_B-S_C-S_{ABC}
\ee

In the second scheme, a ring shape region $\bar{D}$ is taken which is divided into three parts $A$, $B$, and $C$, as shown in Fig.~\ref{Stop}(b). The topological entanglement entropy $\gamma$ can be calculated in the limit of both total system size and the size of $A$, $B$, $C$ going to infinity as
\be
\label{eq:topent2}
2\gamma= S_{AB}+S_{BC}-S_{B}-S_{ABC}
\ee

It can be checked that with this linear combination of entanglement entropy of different regions, the contributions from the `area law' part and other non-universal part which depends on the details of the shape of the regions are all cancelled out. Only the universal $\gamma$ value is retained in the thermodynamic limit. The factor of $2$ in Eq.~\eqref{eq:topent2} compared to 
Eq.~\eqref{eq:topent1} is due to the fact that the region $\bar{D}$ in Fig.\ref{Stop} (b) in fact has two boundaries (one inner boundary and one outer boundary).

In the following, we will mainly focus on 
the meaning and generalizations of the quantity given in the r.h.s. of Eq.~\eqref{eq:topent2}, and denote it by $S_\text{topo}$. And without confusion, we will just call $S_\text{topo}$ the `topological entanglement entropy'.
\begin{svgraybox}
\begin{center}
\textbf{Box 5.3 Topological entanglement entropy}

The topological entanglement entropy $S_\text{topo}$
is given by
\be
S_\text{topo}=S_{AB}+S_{BC}-S_{B}-S_{ABC},
\ee
where $A,B,C$ are parts of the ring shape region $\bar{D}$, as shown in Fig.~\ref{Stop}(b).
\end{center}
\end{svgraybox}

\section{Generalizations of topological entanglement entropy}
\label{sec:irr-corr}

We have learned that the topological entanglement entropy $\gamma$ is a universal quantity for many-body ground states of gapped systems. If the system is topologically ordered, then it ground state has a nonzero $\gamma$. The other direction is also true, that is, if a ground state of a gapped system has a nonzero $\gamma$, the system is topologically ordered. Here we further examine the meaning of $\gamma$, which essentially
characterizes `irreducible many-body correlation' (a concept introduced in Chapter~\ref{sec:irr}) in the system. We start to look at $\gamma$ from an information-theoretic viewpoint, which will lead to generalizations of the concept of $\gamma$ to also study gapped systems without topological order. To do so, instead of only considering a quantum system with topological order, we will consider general many-body quantum systems, with local Hamiltonians. 

\subsection{Quantum conditional mutual information}
\label{sec:conmutual}

Notice that Eq.~\eqref{eq:topent1} looks familiar - it is in fact the `trial' version of tripartite entanglement as given in Eq.(1.77). This suggests an information-theoretic meaning of Eq.~\eqref{eq:topent1}, which captures the `true' tripartite correlation between the parts $ABC$ that is not contained in bipartite systems $AB$, $BC$ and $AC$. 

As discussed in Chapter 1.4.2, the trouble of using 
the r.h.s. of Eq.~\eqref{eq:topent1} as a measure of the `true' tripartite correlation is that it could be negative. For instance, if the 
wave function is the $N$-qubit GHZ state $\frac{1}{\sqrt{2}}(\ket{0}^{\otimes N}+\ket{1}^{\otimes N})$ (a direct generalization of the $3$-qubit GHZ state as discussed in Chapter 1.4.2), then r.h.s. of Eq.~\eqref{eq:topent1} is $-1$. However, for topologically ordered systems, $\gamma$ is always positive.
Therefore, the r.h.s. of Eq.~\eqref{eq:topent1} is a good measure of the `true' tripartite correlation in topologically ordered systems.

For Eq.~\eqref{eq:topent2}, if the mutual information between the parts $A$ and $C$ vanishes, i.e.
\be
I(A{:}C)=S_A+S_C-S_{AC}=0,
\ee
then the r.h.s. of Eq.~\eqref{eq:topent2} is identical to the r.h.s. Eq.~\eqref{eq:topent1}. Notice that however, the area $ABC$ has different geometry in Fig.\ref{Stop} (a) and Fig.\ref{Stop} (b). 

For the geometry of
Fig.\ref{Stop} (b), the areas $A$ and $C$ are geometrically `far from' each other. Therefore, there will be not much correlation between them when the total system size and the size of $A$, $B$, $C$ go to infinity. Consequently,
similarly to Eq.~\eqref{eq:topent1}, Eq.~\eqref{eq:topent2} also gives a good measure of the `true' tripartite correlation of the parts $A,B,C$ for topologically ordered system. 

Different from Eq.~\eqref{eq:topent1}, the r.h.s. of Eq.~\eqref{eq:topent2} is always non-negative, for any tripartite quantum state $\rho_{ABC}$. It in fact measures the correlation of the parts $A,C$ conditioned on the existence of the part $B$. This quantity is in fact the conditional mutual information of the parts $A,C$ and denoted by $I(A{:}C|B)$. 

\begin{svgraybox}
\begin{center}
\textbf{Box 5.4 Quantum conditional mutual information}

For any tripartite state $\rho_{ABC}$, the quantum
conditional mutual information $I(A{:}C|B)$ (i.e. the
quantum mutual information between the parts $A,C$, conditioned on the existence of the part $B$), is given by  
\be
\label{eq:QCM}
I(A{:}C|B)=S_{AB}+S_{BC}-S_{B}-S_{ABC}.
\ee
\end{center}
\end{svgraybox}

It is known that $I(A{:}C|B)$ is  always non-negative, given by the strong subaddtivity in quantum information theory.

\begin{svgraybox}
\begin{center}
\textbf{Box 5.5 Strong subadditivity}

The inequality
$
\label{eq:SSA}
I(A{:}C|B)\geq 0
$
is valid for any tripartite state $\rho_{ABC}$. 
\end{center}
\end{svgraybox}

Recall that as discussed in Chapter 1, in the most general case, the `true' tripartite correlation of $\rho_{ABC}$ is measured by 
\be
C_{tri}(\rho_{ABC})=S(\rho^*_{ABC})-S(\rho_{ABC}),
\ee
where $\rho^*_{ABC}$ is the maximum entropy state among all the tripartite states $\sigma_{ABC}$ that satisfy the reduced density matrix constraint $\sigma_{AB}=\rho_{AB}$, $\sigma_{BC}=\rho_{BC}$,  $\sigma_{AC}=\rho_{AC}$. 

If we consider the case when the parts $A$ and $C$ are geometrically `far from' each other, hence there is not much correlation between them, then $\rho_{AB}$ and $\rho_{AC}$ may be enough to determine $\rho^*_{ABC}$ without the information of $\rho_{AC}$, and in fact it is generically the case. In this case, we can redefine
\be
\rho^*_{ABC}=\text{argmax}(S(\sigma_{ABC})|\sigma_{AB}=\rho_{AB},\sigma_{BC}=\rho_{BC}).
\ee

Now we apply the strong subadditivity inequality to 
\begin{equation}
S({\rho_{AB}})+S({\rho_{AC}})-S({\rho_B})-S({\rho^{*}_{ABC}})\geq 0
\end{equation}

This reduces to
\begin{equation}
S({\rho^{*}_{ABC}})-S(\rho_{ABC})\leq S({\rho_{AB}})+S({\rho_{AC}})-S({\rho_B})-S(\rho_{ABC}).
\end{equation}

\begin{svgraybox}
\begin{center}
\textbf{Box 5.6 Inequality for quantum
conditional mutual information}

The following inequality holds
\begin{equation}
\label{eq:CtriI}
C_{tri}(\rho_{ABC})\leq I(A{:}C|B),
\end{equation}
where the equality holds when ${\rho^{*}_{ABC}}$ is a quantum Markov state, i.e.
\begin{equation}
\label{eq:Markov}
{\rho^{*}_{ABC}}=(I_{A}\otimes \mathcal{N}_{B\rightarrow BC})\rho_{AB}.
\end{equation}
\label{irrcor}
\end{center}
\end{svgraybox}

Here $\mathcal{N}_{B\rightarrow BC}$ is a quantum operation
acting on the part $B$ only. Eq.~\eqref{eq:Markov} means that
the state of part $BC$ comes from a quantum operation acting
on part $B$ which does not depend on part $A$. In this sense
the parts $A,C$ are only correlated conditionally on the existence of part $B$. More explicitly, $\rho_{ABC}$ has 
the form
\begin{equation}
\rho_{ABC}=(I_A\otimes\rho_{BC}^{1/2})
[(I_A\otimes\rho_{B}^{-1/2})\rho_{AB}(I_A\otimes\rho_{B}^{-1/2})\otimes I_C]
(I_A\otimes\rho_{BC}^{1/2}).
\end{equation}

When we consider a many-body system where each of the parts
$ABC$ has the size going to infinity, the correlation between
two far-apart parts 
$AC$ should be independent of some local factors such
as the shape of the parts $ABC$. In this sense, the correlation between $AC$ (conditioned on the existence
of $B$) is a universal quantity
that does not depend much on the 
details of the system. It is believed that this is indeed
the case for gapped systems. Or in other words, $I(A{:}C|B)$ captures the true tripartite correlation $C_{tri}(\rho_{ABC})$
for gapped systems. 

We believe that the following is true for any gapped system.

\begin{svgraybox}
\begin{center}
\textbf{Box 5.7 $C_{tri}(\rho_{ABC})$ vs. $I(A{:}C|B)$ for gapped systems 
}
$$C_{tri}(\rho_{ABC})=I(A{:}C|B)$$ for gapped quantum systems in thermodynamic limit, where the parts $A,B,C$ are large, and $A,C$ are far from each other.

\end{center}
\end{svgraybox}

We then propose to use $I(A{:}C|B)$ to detect non-trivial 
many-body entanglement in quantum systems. Here `non-trivial' intuitively means some `long-range' correlation that is 
contained in the parts $A,C$ conditioned on the existence
of $B$. We will later clarify the precise meaning of
`non-trivial' in Chapter 7 using the language of local transformations. 

\begin{svgraybox}
\begin{center}
\textbf{Box 5.8 $I(A{:}C|B)$ as a detector for non-trivial
many-body entanglement
}

A nonzero $I(A{:}C|B)$ for certain large areas $A,B,C$ (larger than the correlation length of the system) with $A,C$ far from each other, is a good detector for non-trivial many-body entanglement. 
\end{center}
\end{svgraybox}

For a topologically ordered system, if one chooses $A,B,C$ as parts of the ring shape region $\bar{D}$, as shown in Fig.~\ref{Stop}(b), then we know that $I(A{:}C|B)$ is nothing but the topological entanglement entropy $S_\text{topo}$, which detects non-trivial topological order when it is nonzero. In this sense, we say that $I(A{:}C|B)$ is a generalization of $S_\text{topo}$, that can be also used to study 
quantum systems without topological order. A related theory based on local transformations will be
discussed in Chapter~\ref{chap7}.

To see that $I(A{:}C|B)$ detects phases of different kinds in gapped systems, we will discuss some simple examples. We will start from a system with topological order, then move on to systems without topological order.

\subsection{Toric code in a magnetic field}
\label{sec:toricmag}

We start from a system with topological order.
We consider the toric code system in an external magnetic
field along the direction $\vec{h}$, with the Hamiltonian
\begin{equation}
H_{\text{toric}}(\vec{h})=-H_{\text{toric}}-\sum_i (h_x X_i
+h_y Y_i+h_z Z_i),
\end{equation}
where $H_{\text{toric}}$ is the toric code Hamiltonian as 
given in Eq.~\eqref{eq:Htoric} and $\vec{h}=(h_x,h_y,h_z)$.

In this concrete example, to illustrate the relationship between $C_{tri}(\rho_{ABC})$
and $I(A{:}C|B)$ (in this case $S_{\text{topo}}$), we will calculate them for a small system on an $L_1\times L_2$ square lattice. Calculations of $S_{\text{topo}}$ with the tensor network method will be discussed in Chapter~\ref{sta_TN}.

\begin{figure}[h!]
\centerline{
\includegraphics[width=2.00in]{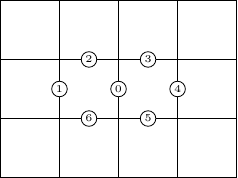}
}
\caption{A $3\times 4$ square lattice. Qubits sit on each link.} 
\label{fig:3by4lattice}
\end{figure}

An example of a $3\times 4$ square lattice is given in Fig.~\ref{fig:3by4lattice}. Qubits sit on each link, and with periodic boundary condition there are a total of $24$ qubits. 

\begin{figure}[h!]
\centerline{
\includegraphics[width=2.50in]{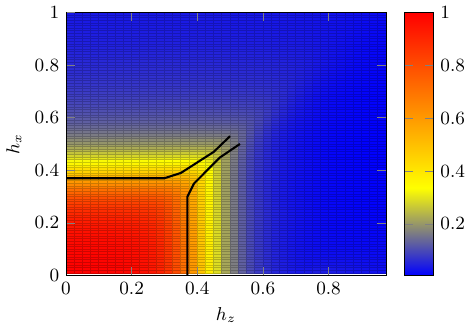}
}
\caption{For the $3\times 4$ square lattice with periodic boundary condition, the calculation of $C_{tri}(\rho_{ABC})$ for different values of $h_x,h_z$.} 
\label{fig:hXZ}
\end{figure}

We take the qubit $1$ as our part $A$, qubit $4$ as our part $C$,
qubits $2,3,5,6$ as our part $B$, and the rest of qubits as part $D$. For $h_y=0$, we calculate $C_{tri}(\rho_{ABC})$ for different values of $h_x,h_z$. The results are shown in Fig.~\ref{fig:hXZ}.

From Fig.~\ref{fig:hXZ}, we clearly see that there are two different phases: the red region corresponding to the `topological phase' (for small values of $h_x,h_z$), and the blue region  corresponding to the `trivial phase' (i.e. $C_{tri}(\rho_{ABC})=0$, for large values of $h_x,h_y$). In other words, a crucial feature of topological order is the non-vanishing irreducible tripatite correlation contained in the state $\rho_{ABC}$.

\begin{figure}[h!]
\centerline{
\includegraphics[width=2.50in]{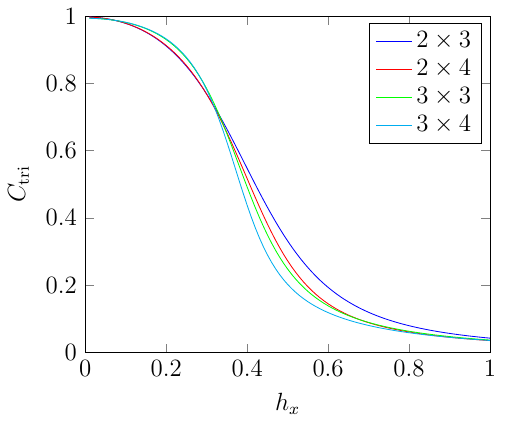}
}
\caption{$C_{tri}(\rho_{ABC})$ for different values of $h_x$, and with different lattice size. The horizontal 
axis is the magnetic field $h_x$. The vertical axis is $C_{tri}(\rho_{ABC})$ for the ground state of $H^{\text{toric}}(\vec{h})$.} 
\label{fig:hXirr}
\end{figure}

\begin{figure}[h!]
\centerline{
\includegraphics[width=2.50in]{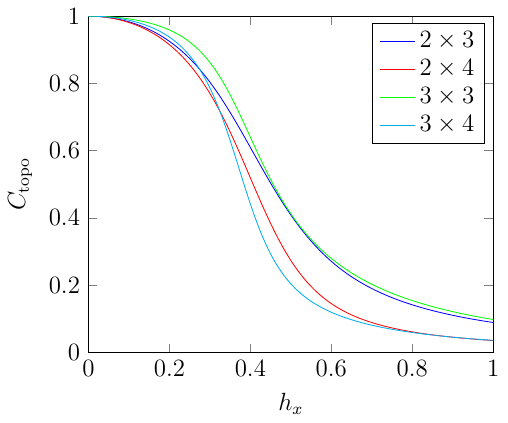}
}
\caption{$S_\text{topo}$ for different values of $h_x$. The horizontal 
axis is the magnetic field $h_x$. The vertical axis is $S_\text{topo}$ for the ground state of $H^{\text{toric}}(\vec{h})$.} 
\label{fig:hXELW}
\end{figure}

To compare with $S_\text{topo}$, we calculate both  $C_{tri}(\rho_{ABC})$ and $S_\text{topo}$ for different lattice size.
For $h_y=h_z=0$, the results are shown in Fig.~\ref{fig:hXirr}
for $C_{tri}(\rho_{ABC})$ and in Fig.~\ref{fig:hXELW}
for $S_\text{topo}$, for different values of $h_x$.

From Fig.~\ref{fig:hXirr}
$C_{tri}(\rho_{ABC})$ and Fig.~\ref{fig:hXELW}, it is clear that
for any value of $h_x$, $S_\text{topo}$ is an upper bound of $C_{tri}(\rho_{ABC})$ (i.e. $C_{tri}(\rho_{ABC})\leq S_\text{topo}$). For system this small (with at most $24$ qubits and the part $ABC$ contain only $6$ qubits), the result of $C_{tri}(\rho_{ABC})$ (compared to $S_\text{topo}$) does seem to give a better prediction of the behavior of the system. In Fig.~\ref{fig:hXirr}, all the four line intersect at a point that corresponding to approximately $h_x=0.34$ as the phase transition point, while the system is known to have a second order phase transition at approximately $h_x=0.328$.

However, once the system gets large, with both the total number of particles and the number of particles in $ABC$ going to infinity, we expect $C_{tri}(\rho_{ABC})=S_\text{topo}=2$ for 
$0\leq h_x<0.328$ (as discussed in Section~\ref{sec:topo_ent} for $h_x=0$) and $C_{tri}(\rho_{ABC})=S_\text{topo}=0$ for 
$h_x>0.328$. The reason we get $C_{tri}(\rho_{ABC})$ or $S_\text{topo}$ equal to $1$, not $2$, for the small system discussed above, is that the system is too small that can only capture the irreducible correlation contributed by the $X$ loop, but not the $Z$ loop.

\begin{figure}[h!]
\centerline{
\includegraphics[width=2.50in]{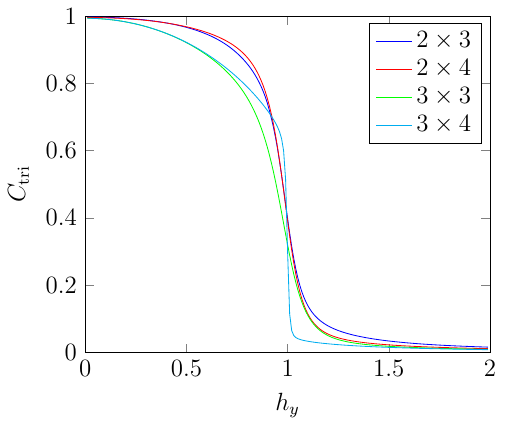}
}\caption{$C_{tri}(\rho_{ABC})$ for different values of $h_y$. The horizontal 
axis is the magnetic field $h_y$. The vertical axis is $C_{tri}(\rho_{ABC})$ for the ground state of $H^{\text{toric}}(\vec{h})$.} 
\label{fig:hYirr}
\end{figure}

\begin{figure}[h!]
\centerline{
\includegraphics[width=2.50in]{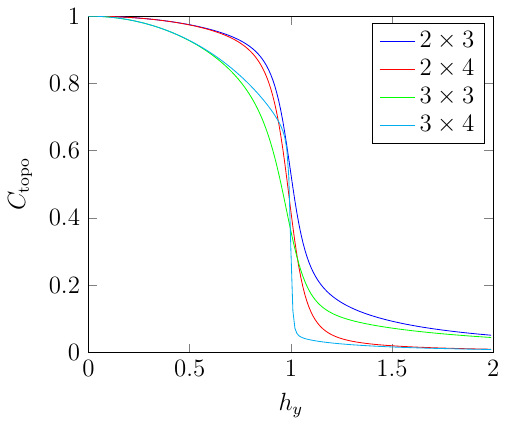}
}
\caption{$S_\text{topo}$ for different values of $h_y$. The horizontal 
axis is the magnetic field $h_y$. The vertical axis is $S_\text{topo}$ for the ground state of $H^{\text{toric}}(\vec{h})$.} 
\label{fig:hYELW}
\end{figure}

For $h_x=h_z=0$, the results are shown in Fig.~\ref{fig:hYirr}
for $C_{tri}(\rho_{ABC})$ and in Fig.~\ref{fig:hYELW}
for $S_\text{topo}$, for different values of $h_y$. Again,
these results demonstrate that
$C_{tri}(\rho_{ABC})\leq S_\text{topo}$. And this system
is known to have a first order phase transition at approximately $h_y=1$. This transition can be clearly seen from 
both $C_{tri}(\rho_{ABC})$ and $S_\text{topo}$, even for such
a small system.

For both the case of $h_y=h_z=0$ and $h_x=h_z=0$, we see 
that when the system size increases, the behaviors of
$C_{tri}(\rho_{ABC})$ and $S_\text{topo}$ become
more similar. In large systems, $C_{tri}(\rho_{ABC})$ is
very hard to calculate while $S_\text{topo}$ is easier to
get. Therefore, we proposed in Section~\ref{sec:conmutual} to 
use $I(A{:}C|B)$ (instead of $C_{tri}(\rho_{ABC})$) to detect non-trivial many-body entanglement in quantum systems.

\subsection{The transverse-field Ising model}
\label{sec:t-Ising}

We will then considers systems without topological order.
Our first such example is the 1D
transverse-field Ising model, with the Hamiltonian
\be
H^{\text{tIsing}}(B)=-\sum_{i}Z_iZ_{i+1}-B\sum_i X_i
\ee
for $B>0$ (we choose $J=1$).
This system has no topological order for any value of
$B$.

For $B=0$, the ground-state space is two-fold
degenerate (even for a finite system with $n$ particles) and is spanned by
$\{\ket{0}^{\otimes N}, \ket{1}^{\otimes N}\}$.
For $B<1$, the ground-state space is still two-fold
degenerate but only in the thermodynamic limit.
It is well-known that the system encounters a quantum phase transition at $B=1$. 

Notice that the Hamiltonian $H^{\text{tIsing}}$ has a 
$\mathbb{Z}_2$ symmetry that is given by
$\bar{X}=\prod_i X_i$. That is,
$[H^{\text{tIsing}}(B),\bar{X}]=0$. Therefore, the ground-state
space must also have the same $\mathbb{Z}_2$ symmetry.
That is, for the projection onto the ground-state space $P_g(B)$,
$[P_g(B),\bar{X}]=0$. 

For $B<1$, the system is said to be in a `symmetry breaking' phase
with symmetry breaking order, in a sense that although $P_g(B)$ is symmetric, the short range correlated states within $P_g(B)$ are not. 

\begin{figure}[htbp]
\begin{center}
\centerline{
\includegraphics[width=2.00in]{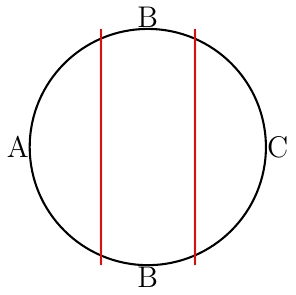}~~~~~~
\includegraphics[width=1.80in]{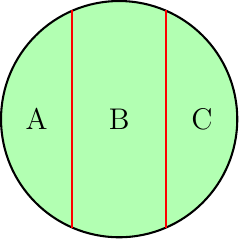}
}
\centerline{
	(a) ~~~~~~~~~~~~~~~~~~~~~~~~~~~~~~~~~~~~~~~~~~~~~~~~~~~~~~~~~~
	(b)
}
\caption{(a) Cutting of a 1D Chain (b) Cutting of a 2D sphere}
\label{fig:Cut1} 
\end{center}
\end{figure}

When the system size is finite and generically when $B\neq 0$, the two ground states splits in energy. The symmetry principle of the system dictates that the lowest energy state must be symmetric. The special feature of the `symmetry breaking' order is then encoded in a rich many-body entanglement structure of the lowest energy state. 
To probe such kind of entanglement, we can use $I(A{:}C|B)$ as a probe, with a proper choice of the parts $A,B,C$. For a 
1D system with a periodic boundary condition (i.e. a ring), we can choose the parts $A,B,C$ as illustrated in Fig.~\ref{fig:Cut1}(a). Similarly, for a 2D system on a sphere, we can 
choose the parts $A,B,C$ as illustrated in Fig.~\ref{fig:Cut1}(b). 

The key point here is to cut the entire system into only three parts, with parts $A,C$ far
from each other. With respect to this cutting, we name
the corresponding $I(A{:}C|B)$ the `tri-topological entanglement entropy' and denote it $S^\text{t}_\text{topo}$, which is given by 
\begin{equation}
\label{eq:tritopo}
S^\text{t}_\text{topo}=S_{AB}+S_{BC}-S_{B}-S_{ABC}.
\end{equation}

\begin{figure}[h!]
\centerline{
\includegraphics[width=2.50in]{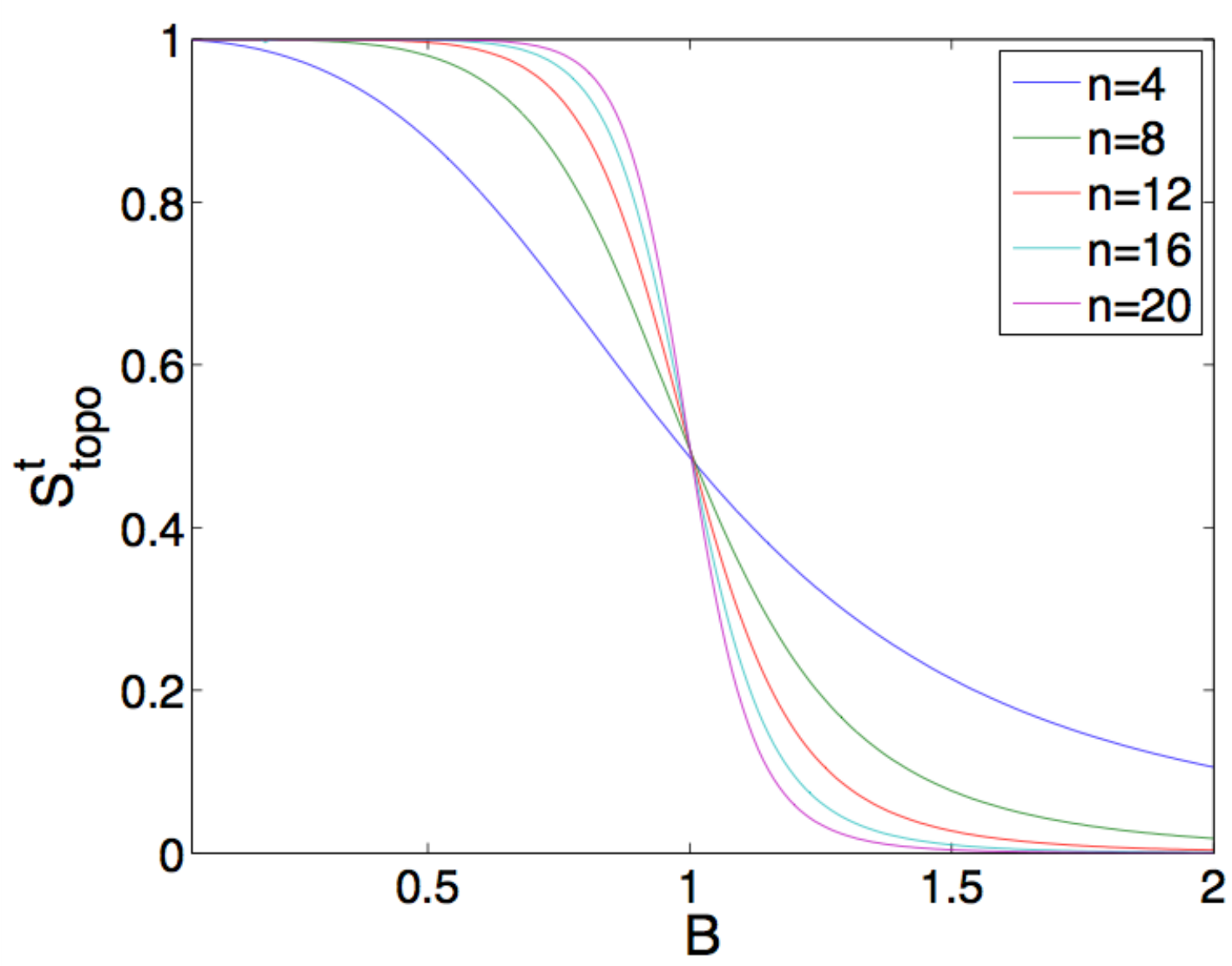}
}
\caption{$S^\text{t}_\text{topo}$ for the transverse-field Ising model. The horizontal 
axis is the magnetic field $B$. The vertical axis is $S^\text{t}_\text{topo}$ for the ground state of $H^{\text{tIsing}}(B)$.} 
\label{Ising}
\end{figure}

We choose the area $A,C$ and each connected component of the area $B$ to have $1,2,3,4,5$ qubits, respectively. So we compute $S^\text{t}_\text{topo}$ for a total of $n=4,8,12,16,20$ qubits, for the corresponding ground state of the Hamiltonian $H^{\text{tIsing}}(B)$. The results are shown in Fig.~\ref{Ising}.
The five curves intersect at the
well-known phase transition point $B=1$. In the limit of $N\rightarrow\infty$, we will expect $S^\text{t}_\text{topo}=1$ for $0\leq B<1$, and $S^\text{t}_\text{topo}=0$ for $B>1$.

This example shows us that $S^\text{t}_\text{topo}$ nicely signals
the two different quantum phases and the phase transitions
for the transverse Ising model. The system has no topological order, but for the symmetry breaking phase, the exact symmetric ground state of a finite system exhibits the feature of non-trivial many-body entanglement with $S^\text{t}_\text{topo}=1$. This feature is similar to the one contained in the symmetry ground state for $B=0$, which is nothing but a GHZ state for the $N$-qubit system:
\be
\ket{GHZ}=\frac{1}{\sqrt{2}}(\ket{0}^{\otimes N}+\ket{1}^{\otimes N}). 
\label{eq:GHZ}
\ee
And it is straightforward to see that $S^\text{t}_\text{topo}=1$ for $\ket{GHZ}$. This then provides an example that $I(A{:}C|B)$ is used to detect symmetry breaking order.

\begin{figure}[h!]
\centerline{
  \includegraphics[width=2.50in]{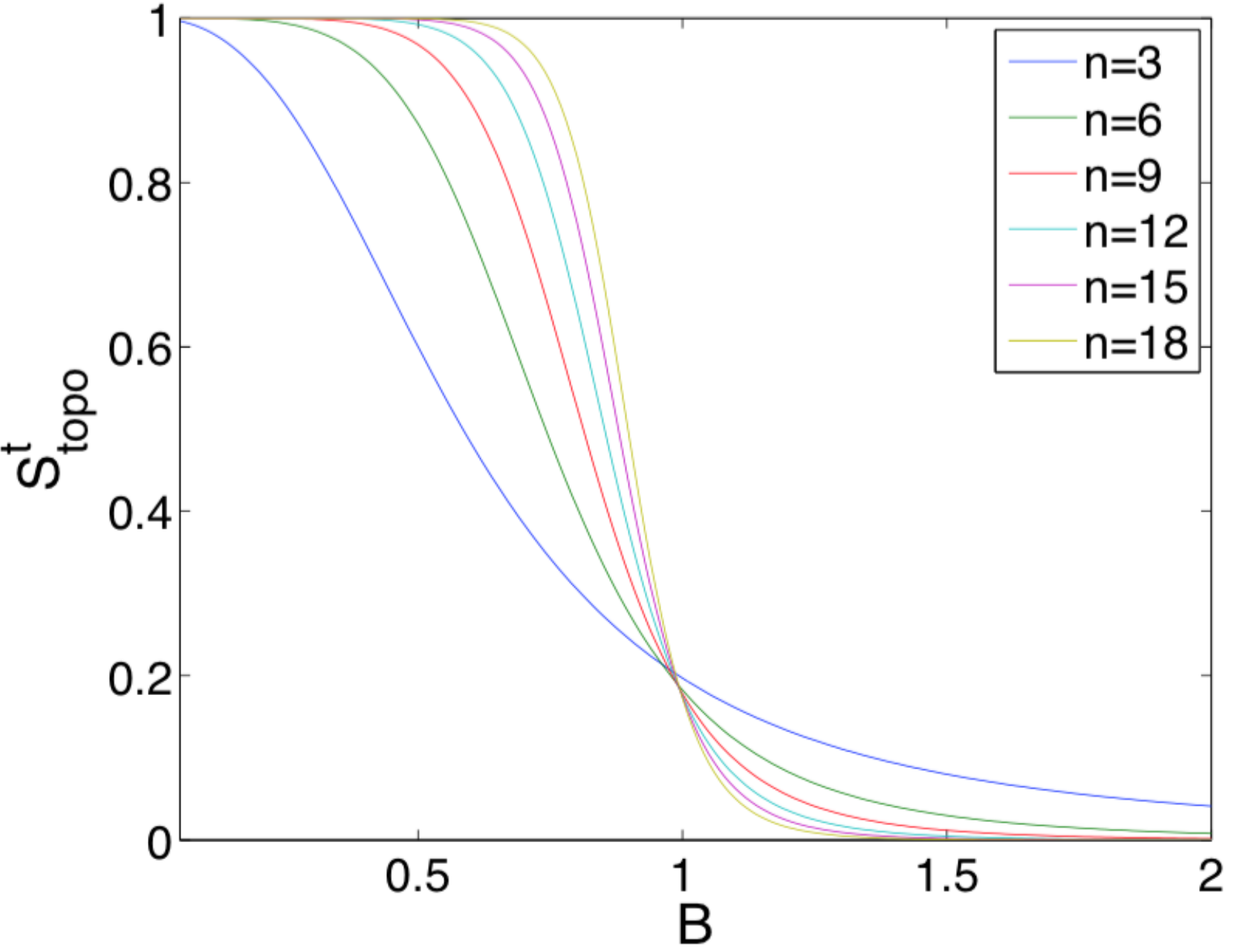}
  }
  \caption{$S^\text{t}_\text{topo}$ of the transverse-field Ising model with open boundary
    condition and the $A,B,C$ regions as chosen in
    Fig.~\ref{fig:cutting}(a). The horizontal 
axis is the magnetic field $B$. The vertical axis is $S^\text{t}_\text{topo}$ for the ground state of $H^{\text{tIsing}}(B)$.}
  \label{fig:IsingOpen}
\end{figure}

It is important to note that the choice of the regions $A,B,C$ should
respect the locality of the system. If we consider one-dimensional
systems with open boundary condition, we can choose the $A,B,C$ regions
as shown in Fig.~\ref{fig:cutting}(a). For the transverse-field Ising model with
open boundary condition, this choice will give a similar diagram of
$S^\text{t}_\text{topo}$ as in Fig.~\ref{Ising}, which is given in
Fig.~\ref{fig:IsingOpen}. 

\begin{figure}[h!]
\centerline{
  \includegraphics[width=2.00in]{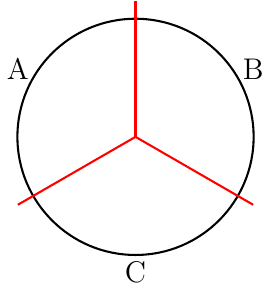}
  }
  \caption{$A,B,C$ cutting on a 1D ring.}
  \label{fig:CuttinRring}
\end{figure}

\begin{figure}[h!]
\centerline{
  \includegraphics[width=2.50in]{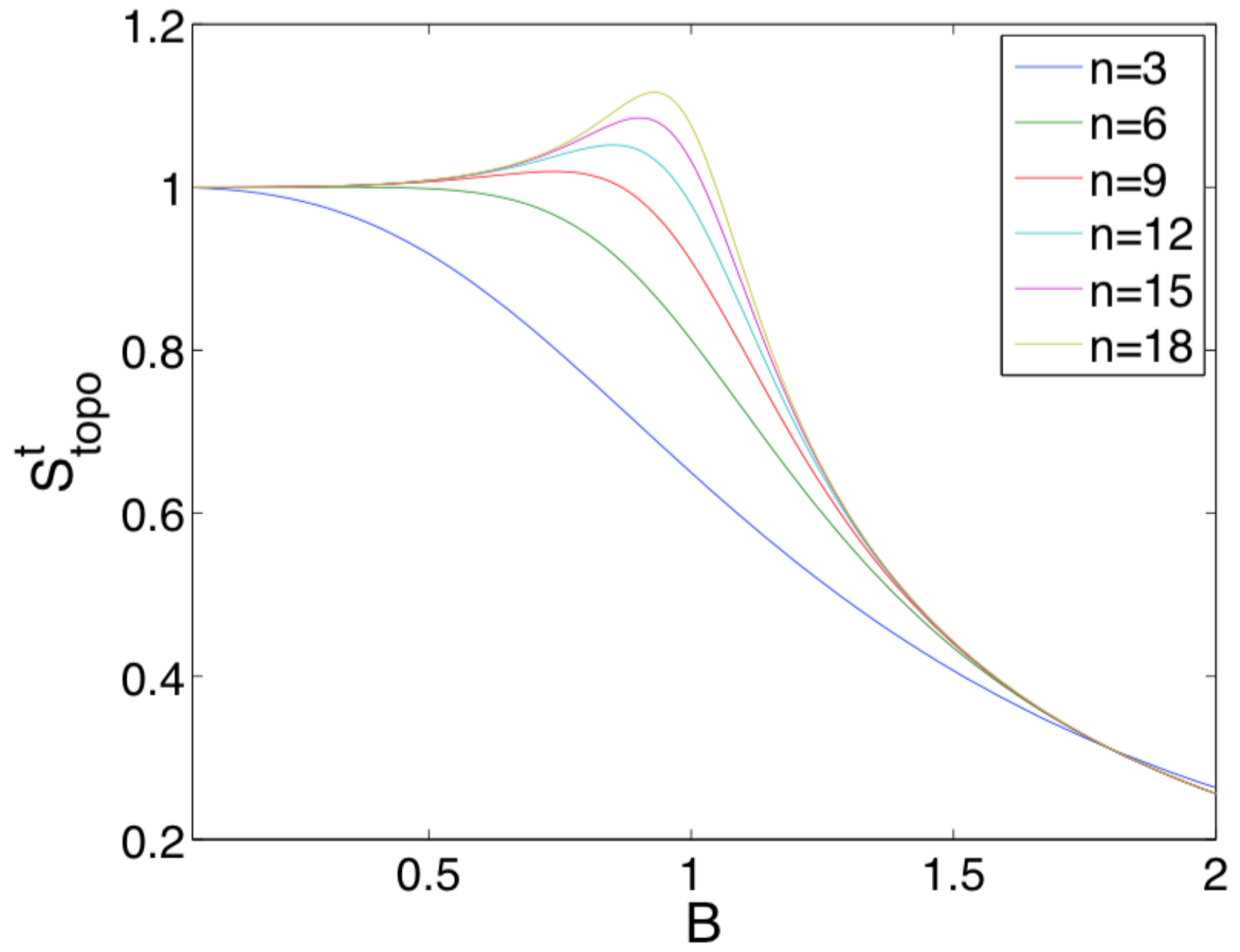}
  }
  \caption{$S^\text{t}_\text{topo}$ of the transverse-field Ising model with periodical
    boundary condition and the $A,B,C$ regions as chosen in
    Fig.~\ref{fig:CuttinRring}. The horizontal 
axis is the magnetic field $B$. The vertical axis is $S^\text{t}_\text{topo}$ for the ground state of $H^{\text{tIsing}}(B)$.}
  \label{fig:IsingRing}
\end{figure}

However, if the partition in Fig.~\ref{fig:cutting}(a) is used for the
Ising model with periodical boundary condition, as given in
Fig~\ref{fig:CuttinRring}, the behaviour of $S^\text{t}_\text{topo}$ will be
very different. In fact, in this case $S^\text{t}_\text{topo}$ reflects
nothing but the 1D area law of entanglement, which will diverge at the
critical point $B=1$ in the thermodynamic limit. For a finite
system as illustrated in Fig.~\ref{fig:IsingRing}, $S^\text{t}_\text{topo}$
does not clearly signal the two different quantum phases and the phase
transition.

Notice that, we do not need to know the symmetry of the system or the associated symmetry breaking order parameter to calculate $S^\text{t}_\text{topo}$. Still, a non-zero $S^\text{t}_\text{topo}$ indicates the existence of symmetry breaking order in the system. In fact, $S^\text{t}_\text{topo}$ probes the symmetry breaking property hidden in the exact ground state of a finite system, which does not break any symmetry, in a form of many-body entanglement that the symmetric ground state exhibits. 

\subsection{The transverse-field cluster model}
\label{sec:t-CS}

Our next example is another 1D system. We consider a 1D graph, which corresponds to a graph state as discussed in
Chapter 3. The generators of the stabilizer group is given
by $\{Z_{j-1}X_jZ_{j+1}\}$. The corresponding stabilizer state
is called the `1D' cluster state, which is the unique ground state of the Hamiltonian 

\begin{equation}
  H_{{clu}}=-\sum_j Z_{j-1}X_jZ_{j+1}.
\end{equation}

For a 1D ring without boundary, the ground state of $H_{clu}$ is the unique graph state stabilized by $\{Z_{j-1}X_jZ_{j+1}\}$. For a chain with boundary, where the summation index $j$ runs
from $2$ to $N-1$, the ground state is then $4$-fold degenerate. There
is a slight difference between even and odd $N$, but the details do not matter to our discussion. For convenience, we will just assume $N$ is even.

It is straightforward to see that
the two commuting logical operators of this code can be all chosen as the form of tensor products of
$X_j$s, which are given by
\begin{equation}
  \bar{X}_1=\prod_k X_{2k-1},\quad \bar{X}_2=\prod_k X_{2k},
\end{equation}
with $k$ runs from $1$ to $N/2$.

Another way to view $\bar{X}_1$ and $\bar{X}_2$ is that they generate
the group $\mathbb{D}_2=\mathbb{Z}_2\times\mathbb{Z}_2$ that preserves
the `topological order' of the system. Any local
perturbation respecting the symmetry cannot lift the ground state degeneracy (in the thermodynamic
limit). In this sense, the system is said to have `symmetry-protected topological (SPT) order' (we will have more detailed discussions of SPT orders in Chapter~\ref{chap10}, and we will also see this cluster state model again in Chapter~\ref{sec:1DSPT}).

One way to view this symmetry protection is to add a magnetic field
along the $X$ direction to the system, which does not break the $\mathbb{D}_2$ symmetry. That is, $X_j$ commutes with 
$\bar{X}_1$ and $\bar{X}_2$. The corresponding
Hamiltonian then reads
\begin{equation}
  \label{eq:Hamclu}
  H_{clu}(B)=-\sum_{j} Z_{j-1}X_jZ_{j+1}-B\sum_j X_j.
\end{equation}
It is known that there is a phase transition at $B=1$ (for periodic
boundary condition). 

It is interesting to compare the system $H_{clu}(B)$ with a
symmetry breaking ordered Hamiltonian
\begin{equation}
  H_{syb}(B)=-\sum_{j} Z_{j-1}Z_{j+1}-B\sum_j X_j,
\end{equation}
with the same symmetry $\mathbb{D}_2$ given by $\bar{X}_1$,
$\bar{X}_2$. 

As $B$ goes from $0$ to $\infty$, both $H_{clu}(B)$ and $H_{syb}(B)$ go through phase transitions, indicating nontrivial order in both the cluster state Hamiltonian $H_{clu}(0)$ and the Ising Hamiltonian $H_{syb}(0)$. However, $H_{clu}(0)$ and $H_{syb}(0)$ have different orders, one with symmetry protected topological order and one with symmetry breaking order. While we have not explained the exact meaning of these orders, let's see their difference with entanglement measures first.  

Denote the symmetric ground state of $H_{syb}(B)$ by
$\ket{\psi_{syb}(B)}$. Then $\ket{\psi_{syb}(0)}$ is a stabilizer
state stabilized by $Z_{j-1}Z_{j+1}$ ($j=2,\ldots,N-2$) and
$\bar{X}_1$, $\bar{X}_2$.
Similarly, we denote the symmetric ground state of $H_{clu}(B)$ by
$\ket{\psi_{clu}(B)}$. Then $\ket{\psi_{clu}(0)}$ is a stabilizer
state stabilized by $Z_{j-1}X_jZ_{j+1}$ ($j=2,\ldots,N-2$) and
$\bar{X}_1$, $\bar{X}_2$.

However, the two systems $H_{clu}(B)$ 
to $H_{syb}(B)$ are very different. One may see 
this from the fact that the ground-state space 
of $H_{syb}(B)$ remains to be four-fold degenerate even if closing the boundary. However, the ground state of 
$H_{clu}(B)$ is non-degenerate with a periodic boundary 
condition. That is, as already mentioned, $\ket{\psi_{clu}(0)}$ is in fact stabilized by
$Z_{j-1}X_jZ_{j+1}$ with a periodic boundary
condition. 

\begin{figure}[htbp]
\begin{center}
\centerline{
\includegraphics[width=3.50in]{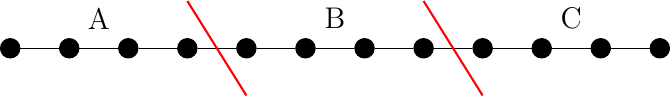}
}
\centerline{
	(a)
}
\centerline{
\includegraphics[width=3.50in]{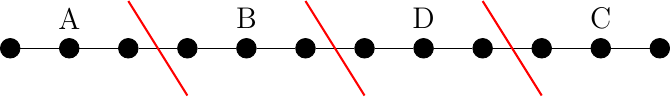}
}
\centerline{
	(b)
}
\caption{(a) Cutting a 1D chain into $A,B,C$ parts; (b) Cutting a 1D
    chain into $A,B,C,D$ parts.} 
\label{fig:cutting}
\end{center}
\end{figure}

In order to detect the non-trivial quantum order
in the system of $H_{clu}(B)$, which should be different
from the symmetry breaking order of $H_{syb}(B)$,
we will again use $I(A{:}C|B)$ with some properly chosen cuttings.
There are two kinds of cuttings introduced in Fig.~\ref{fig:cutting}.
Fig.~\ref{fig:cutting}$(a)$ cuts the system into three parts, and we
denote the corresponding topological entanglement entropy by
$S^\text{t}_\text{topo}$. Fig.~\ref{fig:cutting}$(b)$ cuts the system into four
parts, and we denote the corresponding quantum 
mutual information $I(A{:}C|B)$ by $S^\text{q}_\text{topo}$. 

\begin{figure}[h!]
\centerline{
  \includegraphics[width=2.50in]{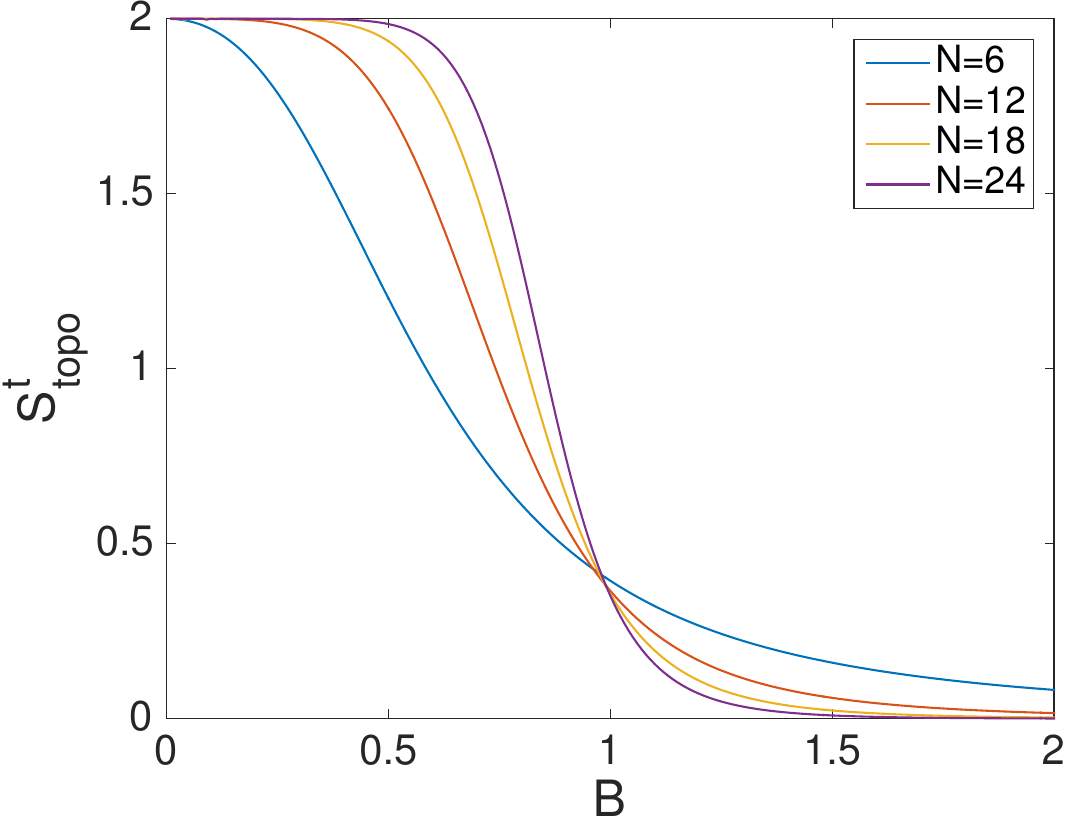}
  }
  \caption{$S^\text{t}_\text{topo}$ for the ground state of $H_{clu}$. The horizontal 
axis is the magnetic field $B$. The vertical axis is $S^\text{t}_\text{topo}$ for the ground state of $H_{clu}(B)$.}
  \label{fig:cluster}
\end{figure}

We first examine $S^\text{t}_\text{topo}$. For the ideal state of $B=0$,
$S^\text{t}_\text{topo}=2$ for both $\ket{\psi_{clu}(0)}$ and
$\ket{\psi_{syb}(0)}$. When $B$ increases, for $\ket{\psi_{clu}(B)}$,
$S^\text{t}_\text{topo}$ signals a phase transition. To demonstrate 
this, we perform an exact diagonalization of the Hamiltonian $H_{clu}(B)$,
and calculate $S^{t}_{topo}$ for the corresponding ground state. 
We do the calculation with $6,12,18,24$ qubits, where each part of $A,B,C$
contains $2,4,6,8$ qubits respectively. The results are shown in
Fig.~\ref{fig:cluster}. In the limit of $N\rightarrow\infty$, we will expect $S^\text{t}_\text{topo}=2$ for $0\leq B<1$, and $S^\text{t}_\text{topo}=0$ for $B>1$.

\begin{figure}[h!]
\centerline{
  \includegraphics[width=2.50in]{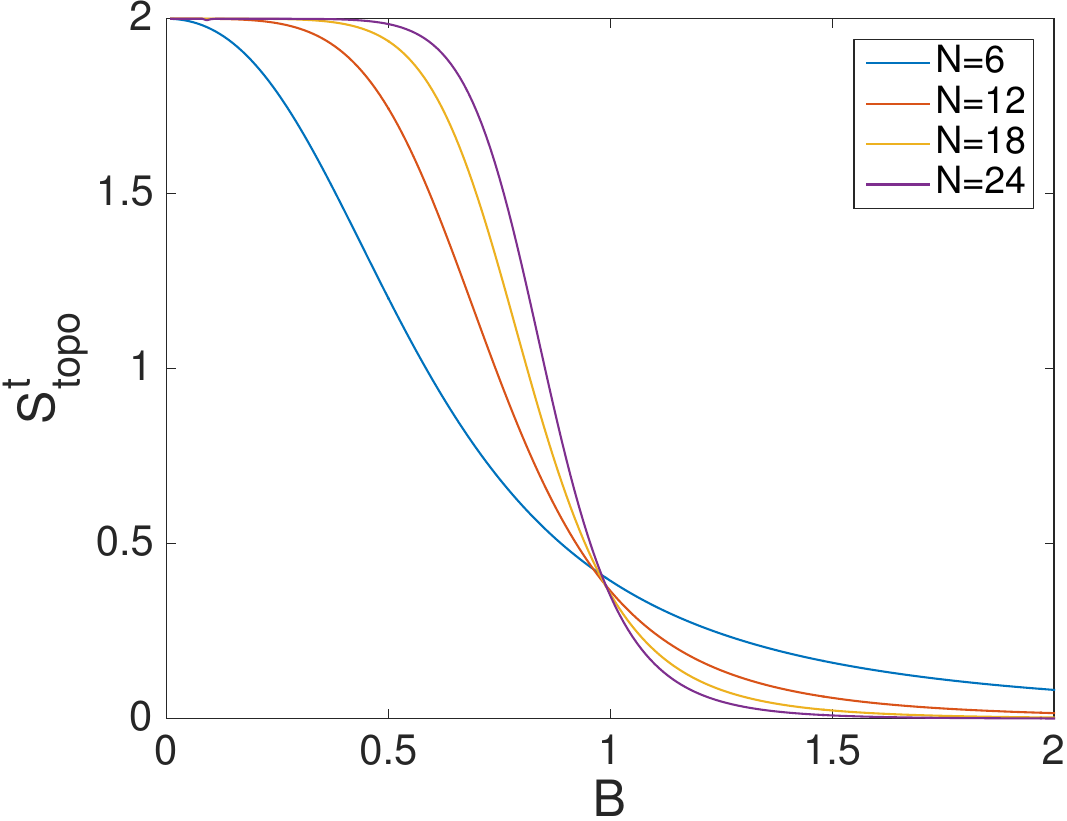}
  }
  \caption{$S^\text{t}_\text{topo}$ for the ground state of $H_{syb}$. The horizontal 
axis is the magnetic field $B$. The vertical axis is $S^\text{t}_\text{topo}$ for the ground state of $H_{syb}(B)$.}
  \label{fig:ising}
\end{figure}

However, the symmetry breaking order hidden in the exact symmetric
ground state $\ket{\psi_{syb}(B)}$ can also be detected by
$S^\text{t}_\text{topo}$. In fact, for the same calculation with $6,12,18,24$
qubits, one gets a very similar figure, as shown in
Fig.~\ref{fig:ising}. Again, in the limit of $N\rightarrow\infty$, we will expect $S^\text{t}_\text{topo}=2$ for $0\leq B<1$, and $S^\text{t}_\text{topo}=0$ for $B>1$.

\begin{figure}[h!]
\centerline{
  \includegraphics[width=2.50in]{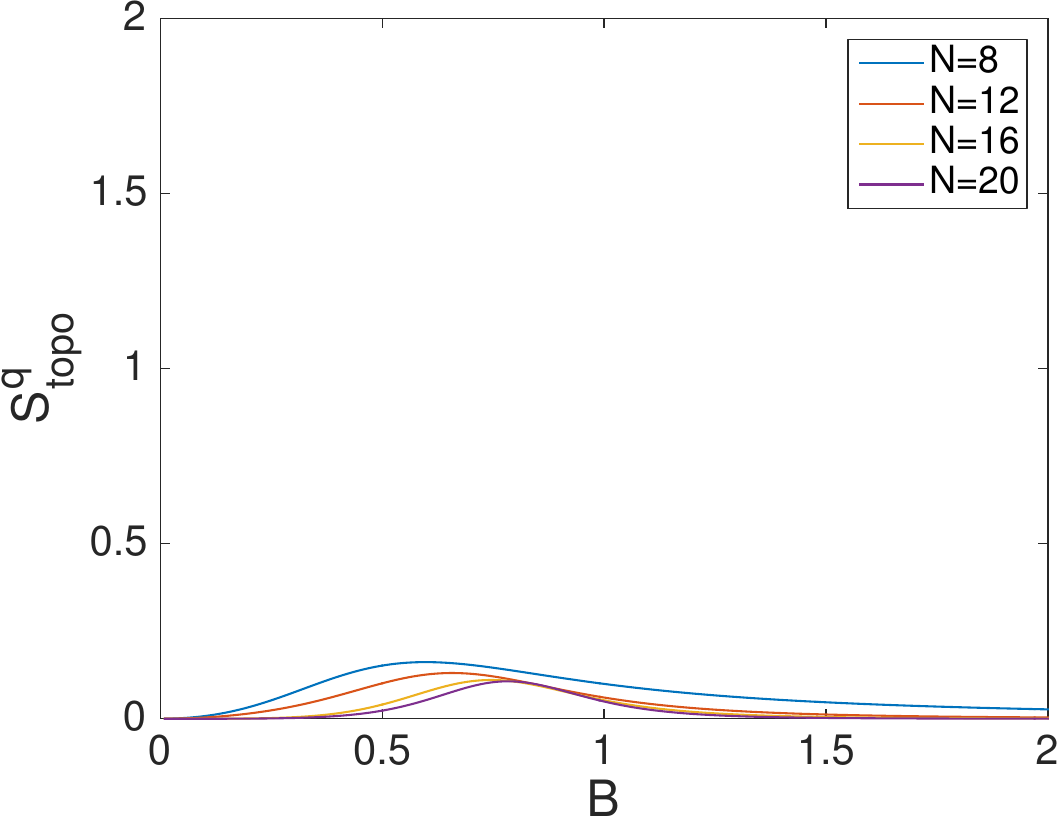}
  }
  \caption{$S^\text{q}_\text{topo}$ for the ground state of $H_{syb}$. The horizontal 
axis is the magnetic field $B$. The vertical axis is $S^\text{q}_\text{topo}$ for the ground state of $H_{syb}(B)$.}
  \label{fig:isingD}  
\end{figure}

To distinguish SPT orders from a symmetry breaking one, we can instead
use $S^\text{q}_\text{topo}$. Since the topological entanglement entropy is only
carried in the entire wave function of the exact symmetric ground
state for symmetry breaking orders, computing
$S^\text{q}_\text{topo}$ on its reduced density matrix of parts $ABC$ returns
nearly zero value (due to finite size effect) that do not signal any topological phase, as shown
in Fig.~\ref{fig:isingD}. Here we do the calculation with $12,16,20,24$ qubits, 
where each part of $A,B,C,D$
contains $3,4,5,6$ qubits respectively.

\begin{figure}[h!]
\centerline{
  \includegraphics[width=2.50in]{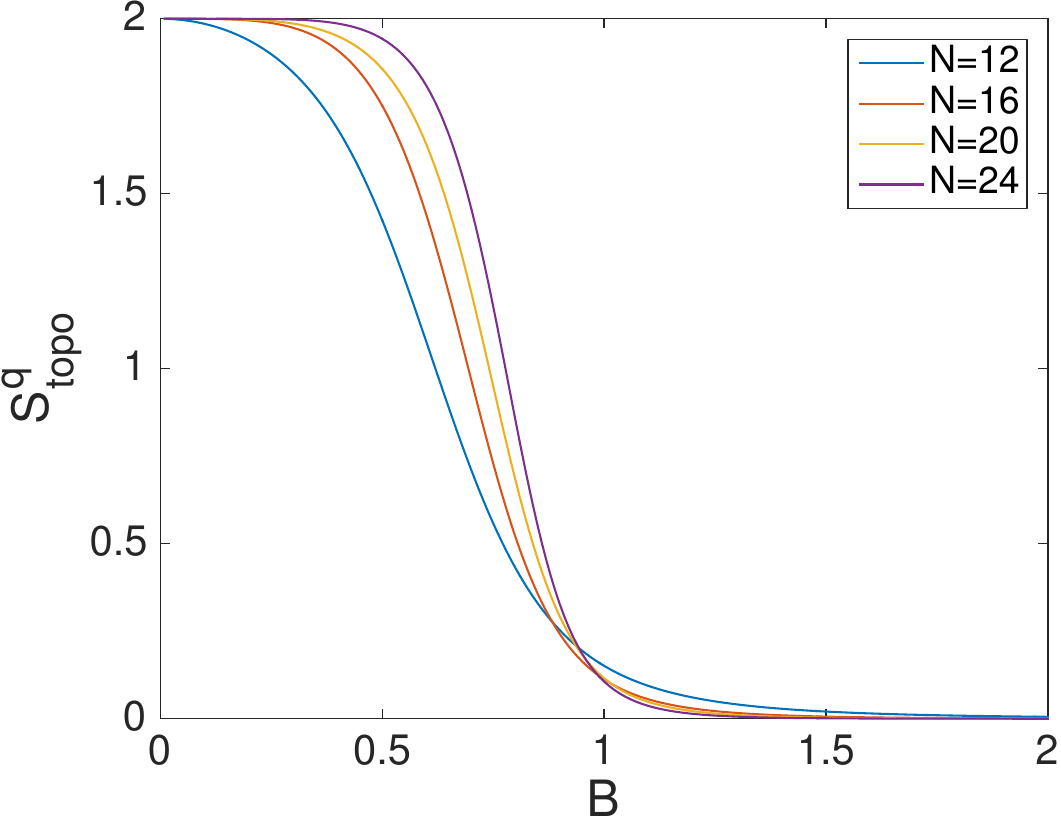}
  }
  \caption{$S^\text{q}_\text{topo}$ for the ground state of $H_{clu}$. The horizontal 
axis is the magnetic field $B$. The vertical axis is $S^\text{q}_\text{topo}$ for the ground state of $H_{clu}(B)$.}
  \label{fig:clusterD} 
\end{figure}

However, $S^\text{q}_\text{topo}=2$ for $\ket{\psi_{clu}(0)}$, because the
`topology' of the SPT states is essentially carried on the boundary,
tracing out part of the bulk has no effect on detecting the
topological order. For $\ket{\psi_{clu}(B)}$, $S^\text{q}_\text{topo}$ signals
the topological phase transition, as shown in Fig.~\ref{fig:clusterD}. Again, in the limit of $N\rightarrow\infty$, we will expect $S^\text{q}_\text{topo}=2$ for $0\leq B<1$, and $S^\text{q}_\text{topo}=0$ for $B>1$.

Notice that, similar to the symmetry breaking case, we do not need to know the symmetry of the system to calculate $S^\text{t}_\text{topo}$ and $S^\text{q}_\text{topo}$. Still, non-zero $S^\text{t}_\text{topo}$ and $S^\text{q}_\text{topo}$ indicate the existence of SPT order in the system. In this sense, $S^\text{t}_\text{topo}$ and $S^\text{q}_\text{topo}$ probe the SPT property hidden in the exact ground state of a finite system, which does not break any symmetry, in a form of many-body entanglement that the symmetric ground state exhibits. 

\subsection{Systems with mixed orders}

There could also be systems containing mixed orders of symmetry breaking, SPT and topological orders. 
Our third example will be such a system with mixed orders. 
We consider a stabilizer group generated by
$Z_{j-1}X_{j}X_{j+1}Z_{j+2}$ with $j$ running from $2$ to $N-2$. 
On a 1D chain with boundary,
i.e. for $j=2,3,\ldots, N-2$, the Hamiltonian
$-\sum_jZ_{j-1}X_{j}X_{j+1}Z_{j+2}$ has $8$-fold ground-state
degeneracy.

The ground-state as an error-correcting code has logical operators $\bar{X}_1=\prod_k
X_{3k-2}$, $\bar{X}_2=\prod_k X_{3k-2}$, $\bar{X}_3=\prod_k X_{3k}$.
Therefore, if one adds a magnetic field along the $X$ direction, i.e.
\begin{equation}
  H_{ZXXZ}(B)=-\sum_j Z_{j-1}X_{j}X_{j+1}Z_{j+2}-B\sum_j X_j,
\end{equation}
the orders of the system (either SPT or symmetry breaking) will be
protected when $B$ is small.

\begin{figure}[h!]
\centerline{
  \includegraphics[width=2.50in]{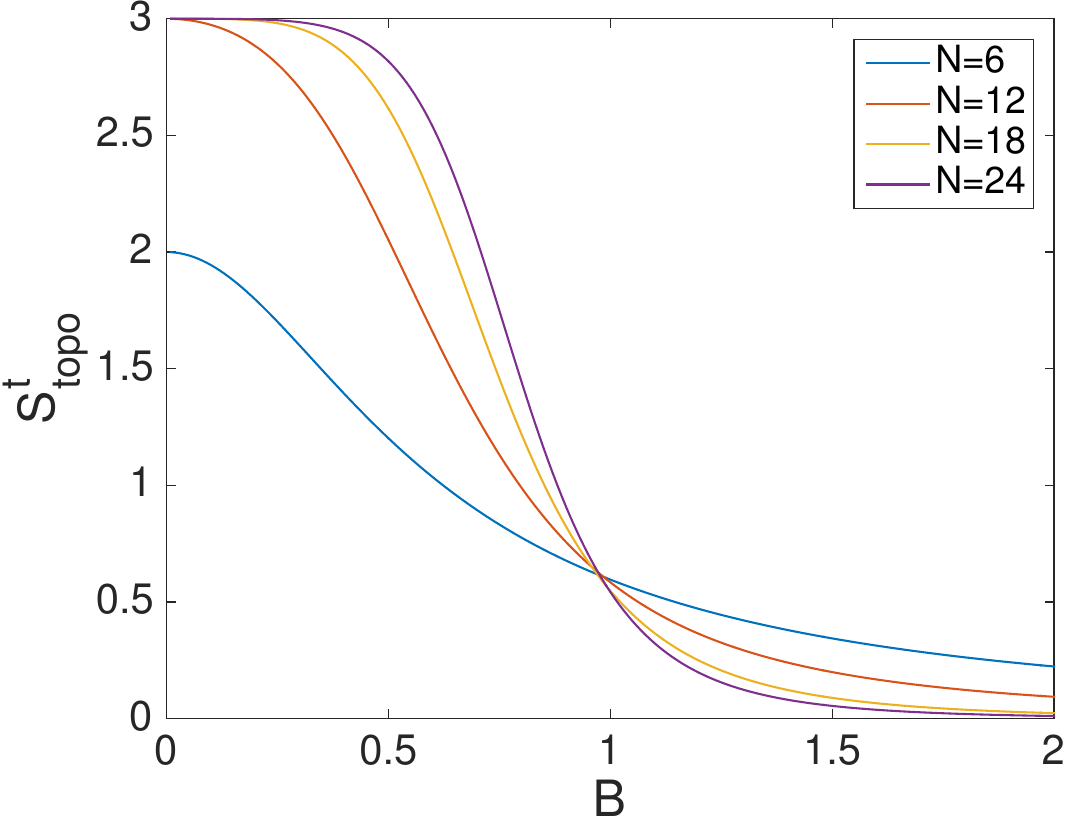}
  }
  \caption{$S^\text{t}_\text{topo}$ the ground state of $H_{ZXXZ}(B)$. The horizontal 
axis is the magnetic field $B$. The vertical axis is $S^\text{t}_\text{topo}$ for the ground state of $H_{ZXXZ}(B)$. For
    $N=6$, the maximum value of $S^\text{t}_\text{topo}$ is $2$ due to that the
    system size is too small.}
  \label{fig:ZXXZ1}   
\end{figure}

\begin{figure}[h!]
\centerline{
  \includegraphics[width=2.50in]{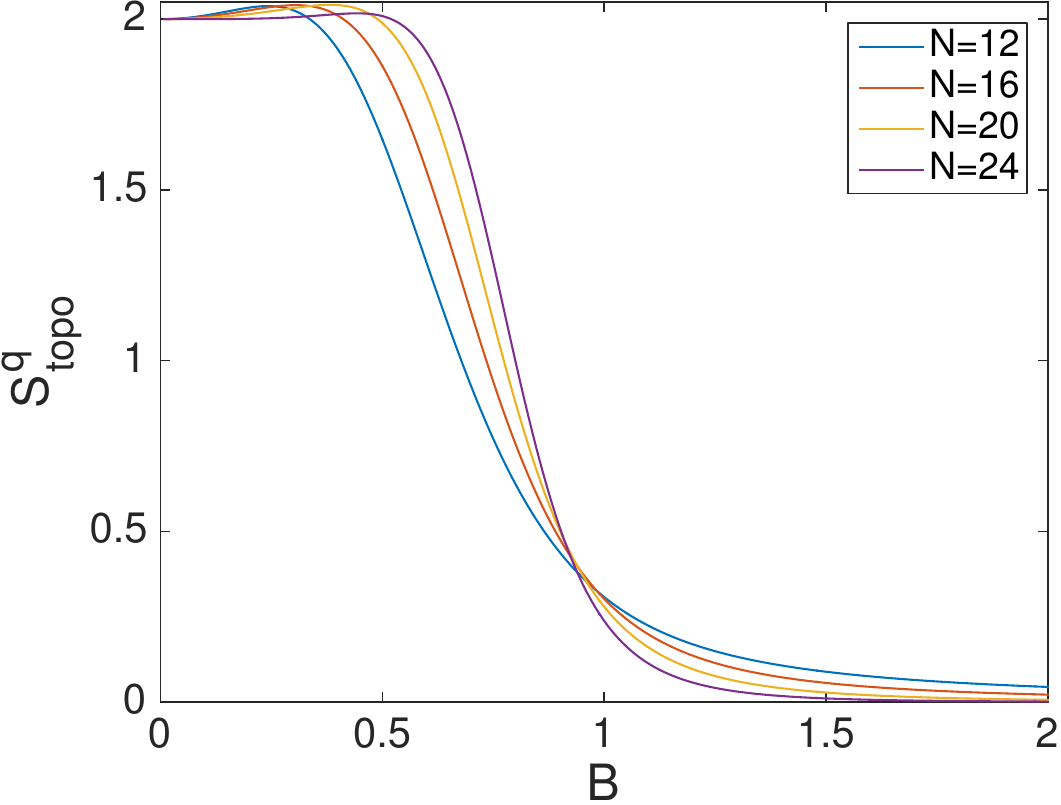}
  }
  \caption{$S^\text{q}_\text{topo}$ for the ground state of $H_{ZXXZ}(B)$. The horizontal 
axis is the magnetic field $B$. The vertical axis is $S^\text{q}_\text{topo}$ for the ground state of $H_{ZXXZ}(B)$.}
  \label{fig:ZXXZ2}    
\end{figure}

It turns out that the system combines a $\mathbb{Z}_2$
symmetry breaking order and a $\mathbb{D}_2$ SPT-order. This can be
seen from the fact that for $B=0$, the symmetric ground state has
$S^\text{t}_\text{topo}=3$ and $S^\text{q}_\text{topo}=2$. $S^\text{t}_\text{topo}$ probes both the
symmetry breaking order and the SPT order, as illustrated in
Fig.~\ref{fig:ZXXZ1}. $S^\text{q}_\text{topo}$ probes only the SPT order, as illustrated in Fig.~\ref{fig:ZXXZ2}. In the limit of $N\rightarrow\infty$, we will expect $S^\text{t}_\text{topo}=3$ and $S^\text{q}_\text{topo}=2$ for $0\leq B<1$, and $S^\text{t}_\text{topo}=S^\text{q}_\text{topo}=0$ for $B>1$.

Again, we do not need to know the symmetry of the system to calculate $S^\text{t}_\text{topo}$ and $S^\text{q}_\text{topo}$. $S^\text{t}_\text{topo}$ and $S^\text{q}_\text{topo}$ probe the symmetry breaking and/or SPT property hidden in the exact ground state of a finite system, which does not break any symmetry, in a form of many-body entanglement that the symmetric ground state exhibits.  

\subsection{$I(A{:}C|B)$ as a detector of non-trivial many-body entanglement}

From our previous discussions, we observe that to use $I(A{:}C|B)$ to
detect quantum phase and phase transitions, it is crucial to choose
the areas $A,C$ that are {\it{far from each other}}. Here `far' is
determined by the locality of the system. For instance, on an 1D
chain, the areas $A,C$ in Fig.~\ref{fig:Cut1}(a) and Fig.~\ref{fig:cutting} are far from each
other, but in Fig.~\ref{fig:CuttinRring} are not.

One may also generalize the idea of different types of topological
entanglement entropy to higher spatial dimensions. For instance, in
2D, a straightforward way is to replace the chain by a cylinder with
boundary, then use the similar cuttings as in Fig.~\ref{fig:cutting}.

\begin{figure}[htbp]
\begin{center}
\centerline{
\includegraphics[width=2.00in]{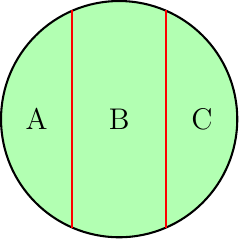}~~~~~~
\includegraphics[width=2.00in]{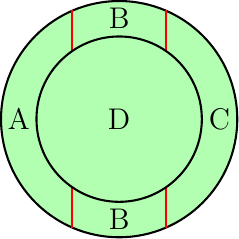}
}
\centerline{
	(a) ~~~~~~~~~~~~~~~~~~~~~~~~~~~~~~~~~~~~~~~~~~~~~~~~~~~~~~~~~~~~~~~~~~
	(b) 
}
\caption{Cuttings of a 2D disk: (a) into three parts $A,B,C$  (b) into four parts $A,B,C,D$} 
\label{fig:cutting2D}
\end{center}
\end{figure}

One may also consider a disk with boundary. For any gapped ground
state (one may need to avoid the situation of a gapless boundary by
adding symmetric local terms to the Hamiltonian), still using
$I(A{:}C|B)$, one can consider two
kinds of cuttings, as given in Fig.~\ref{fig:cutting2D}.
Similar to the 1D case, the cutting of Fig.~\ref{fig:cutting2D}(a)
probes both the symmetry breaking orders and the SPT orders, and the
cutting of Fig.~\ref{fig:cutting2D}(a) probes only SPT orders.

To summarize, we have shown that a nonzero $I(A{:}C|B)$ for certain large areas $A,B,C$ (larger than the correlation length of the system) with $A,C$ far from each other, is a good detector for non-trivial many-body entanglement. With different choices of the cuttings for $A,B,C$, $I(A{:}C|B)$
can detect different quantum orders and signal different kinds
of quantum phase transitions. 

We have discussed three kinds of different cuttings that leads to different (generalized) topological entanglement
entropy based on $I(A{:}C|B)$, i.e. $S_\text{topo}$, $S^\text{t}_\text{topo}$
and $S^\text{q}_\text{topo}$. For a product state (trivial order), all three of $S_\text{topo}$, $S^\text{t}_\text{topo}$
and $S^\text{q}_\text{topo}$ are zero. A nonzero of any one of the three indicates some non-trivial order in the system. And for probing symmetry breaking and/or SPT orders, we do not need to know the symmetry of the system to calculate these quantities (for the exact ground state that does not break any symmetry, for any finite system). We summarize their use to detect different
kinds of orders in the table below. 


\begin{svgraybox}
\begin{center}
\textbf{Box 5.9 Detecting quantum orders by $I(A{:}C|B)$}

    \begin{tabular}{  p{3cm}  | p{3cm} | p{3cm} }
    \hline
     Order of the quantum system & Nonzero $I(A{:}C|B)$ & Zero $I(A{:}C|B)$\\ \hline 
    {Trivial} Order &  & $S_\text{topo}$, $S^\text{t}_\text{topo}$, $S^\text{q}_\text{topo}$\\ \hline
    {Topological} Order & $S_\text{topo}$, $S^\text{q}_\text{topo}$ & $S^\text{t}_\text{topo}$\\ \hline
    {Symmetry}-Breaking Order & $S^\text{t}_\text{topo}$ & $S_\text{topo}$, $S^\text{q}_\text{topo}$\\ \hline
    {Symmetry}-Protected {{Topological}} Order& $S^\text{t}_\text{topo}$, $S^\text{q}_\text{topo}$ & $S_\text{topo}$\\
    \hline
    \end{tabular}
\end{center}
\end{svgraybox}

\section{Gapped ground states as quantum-error-correcting codes}
\label{sec:QECC}

In Chap.3, we have discussed the properties of the toric code. We know that the distance of the toric code grow as $\sqrt{N}$,
where $N$ is the number of qubits in the system. 

In this section, we also discuss the properties of the ground- state space of other systems, from the viewpoint of quantum error-correcting code.

Let us first consider the transverse Ising model $H^{\text{tIsing}}(B)$. For $B=0$, the ground-state space
is two-fold degenerate and is spanned by
$\{\ket{0}^{\otimes N},\{\ket{1}^{\otimes N}\}$. Denote
this space by $V_{\text{tIsing}}$.

Notice that the quantum error-correcting code $V_{\text{tIsing}}$ spanned by $\{\ket{0}^{\otimes N},\{\ket{1}^{\otimes N}\}$ has only distance $1$, since one can choose another orthonormal basis
\be
\ket{GHZ_{\pm}}=\frac{1}{\sqrt{2}}(\ket{0}^{\otimes N}\pm\ket{1}^{\otimes N}),
\ee
and we have
\be
\bra{GHZ_{+}}Z_i\ket{GHZ_{-}}=1,
\ee
for any qubit $i$.

However, if we only consider the code's ability to correct bit flip errors (i.e. $X_i$), the code actually has a `large distance'. That, for any orthonamal basis $\ket{\psi_0},\ket{\psi_1}$ of $V_{\text{tIsing}}$, if 
\be
\bra{\psi_0}O_X\ket{\psi_1}\neq 0
\ee
holds for any operator $O_X$ that is a tensor product of $X_i$s, one must
have $O_X=X^{\otimes N}$ (e.g. $X^{\otimes N}\ket{0}^{\otimes N}=\ket{1}^{\otimes N}$).

Or, one can view $V_{\text{tIsing}}$ as a stabilizer code, whose stabilizer group is generated by $Z_iZ_{i+1}$ for $i=1,2,\ldots,N-1$. And the logical operator which is a tensor product of $X_i$s is given by $X^{\otimes N}$. This means that this stabilizer code has `X-distance' $N$.

In this sense, $V_{\text{tIsing}}$ for correcting the $X$-only errors, is an analogy of the classical repetition code of $N$ bits with codewords $\{00\ldots 0,11\ldots 1\}$ for correcting bit flip error that sends $0\leftrightarrow 1$. And the distance of the classical repetition code of $N$ bits is $N$, which is the minimal number of bit flips needed to transform $00\ldots 0$ to $11\ldots 1$. In this sense, we say that $V_{\text{tIsing}}$ has a large `classical' distance, which is `macroscopic' that grows with the system size $N$.

In fact, the error-correcting properties of $V_{\text{tIsing}}$ goes much beyond of just the `classical code with large distance', given its quantum nature. Recall that the system has a $\mathbb{Z}_2$ symmetry that is given
by $X^{\otimes N}$. In fact, with respect to any local operator $L$ that does not break this $\mathbb{Z}_2$ symmetry, i.e. $[L,X^{\otimes N}]=0$, the code $V_{\text{tIsing}}$ has a `macroscopic' distance. That is, we will need to apply $L$ to number of local sites that grows the system size, to transform any orthonamal basis state $\ket{\psi_0}$ to the other state $\ket{\psi_1}$.

For $0<B<1$, the ground-state space of $H^{\text{tIsing}}(B)$ is also two-fold degenerate, and with an error-correcting property that is very similar to the case of $B=0$. We summarize the property as below.

\begin{svgraybox}
\begin{center}
\textbf{Box 5.10 The ground-state space of symmetry breaking orders}

For errors that do not break the symmetry, the degenerate ground-state space of a symmetry breaking ordered system is a quantum error-correcting code with a macroscopic distance. 

\end{center}
\end{svgraybox}

We now consider the transverse field cluster mode $H_{clu}(B)$. For $B=0$ and with open boundary condition, the ground-state space is $4$-fold degenerate. Denote this space by $V_{clu}$. As a stabilizer code, the stabilizer group of $V_{clu}$ is generated by $Z_{i-1}X_iZ_{i+1}$ for $i=2,\ldots, N-1$. The code only has distance $1$, as $Z_1$ ($Z_N$) is a logical operator that commutes with all the $Z_{i-1}X_iZ_{i+1}$. 

However, if we only consider the code's ability to correct bit flip errors (i.e. $X_i$), the code actually has a `large distance'. In fact, the logical operators that are tensor products of $X_i$s are given by $\bar{X}_1$ and $\bar{X}_2$, therefore it has `$X$-distance' $N/2$, which is
a macroscopic distance that is half of the system size.
In this sense, we say that $V_{clu}$ is quantum error-correcting code with `classical' distance $N/2$.

Similar to the symmetry breaking case, the error-correcting property $V_{clu}$ goes beyond just a quantum code with large `classical' distance. Since the system has a $\mathbb{Z}_2\times\mathbb{Z}_2$ symmetry that is given
by $\bar{X}_1,\bar{X}_2$, for any local operator $L$ that does not break this $\mathbb{D}_2$ symmetry, i.e. $[L,X_1]=0$ and $[L,X_2]=0$, the code $V_{\text{tIsing}}$ has a `macroscopic' distance. We summarize this property as below.

\begin{svgraybox}
\begin{center}
\textbf{Box 5.11 The ground-state space of SPT orders}

For errors that do not break the symmetry, the degenerate ground-state space of a SPT ordered system is a quantum error-correcting code with a macroscopic distance. 

\end{center}
\end{svgraybox}

We now summarize the error-correcting property of different gapped systems as below. Here by `classical code', we mean that the quantum code has an orthonormal basis that can be chosen as product states (e.g. $V_{\text{tIsing}}$). And by `classical' distance, we actually mean that the distance is with respect to certain symmetry, which is an analogy as the distance for classical codes.

\begin{svgraybox}
\begin{center}
\textbf{Box 5.11 Gapped ground states as quantum-error-correcting codes}

    \begin{tabular}{  p{3cm}  | p{3cm}  | p{4cm} }
    \hline
     & Degenerate Ground-State Space & Code Distance \\ \hline 
    {Topological} Order & { Quantum} Code & {Macroscopic} { Quantum} Distance  \\ \hline
    {Symmetry}-Breaking Order & {`Classical'} Code & {Macroscopic} {`Classical'} Distance \\ \hline
    {Symmetry}-Protected {{Topological}} Order& {Quantum} Code & {Macroscopic} {`Classical'} Distance \\
    \hline
    \end{tabular}
\end{center}
\end{svgraybox}

\section{Entanglement in gapless systems}
\label{sec:ent_gapless}

In a critical or gapless system, the area law can be violated, but usually only mildly by a term that scales as the logarithm of the size of the subregion. For example, at one dimensional critical points described by conformal field theory, the entanglement entropy of a segment of length $L$ in the chain scales as
\be
S_A \sim \frac{c+\bar{c}}{6} \log{L}
\ee
where $c$ and $\bar{c}$ are the central charges of the conformal field theory. As before, the entanglement entropy is calculated as the total system size goes to infinity. Compared to 1D gapped systems where $S_A$ is bounded by a constant, $S_A$ for 1D conformal critical points is unbounded, but only grows very slowly with the segment size.

Also for gapless free fermion system in $D$ spatial dimensions with a Fermi surface, the entanglement entropy of a region of linear size $L$ scales as
\be
S_A \sim \alpha L^{D-1}\log{L}
\ee
Apart from the `area law' scaling part $\alpha L^{D-1}$, the entanglement entropy also contains a logarithmic part $\log(L)$. The violation of area law comes from the existence of low energy excitations which carries correlations as it propagates. The correlation length is infinite in the system which generates more entanglement across the boundary than in the gapped case. However, the locality constraint still exists and and keeps the entanglement content in the state far from maximum ($\sim L^D$).

The change in entanglement content from an `area law' to beyond as one moves from gapped phases to phase transitions has become a useful tool in detecting phase transitions. In fact, not only entanglement entropy, but also many other different entanglement measures have been found to exhibit diverging behavior as a phase transition point is approached and therefore can be used as a probe for detecting phase transitions. While numerical and experimental challenges remain to calculate or measure entanglement in a system, one special advantage of such a probe is that it works for both symmetry breaking and topological systems. Some conventional probes of phase transitions, like order parameter, apply only to symmetry breaking phases and fail for topological phases. Entanglement measures, however, are generic probes independent of the nature of the phase transition. 

In fact, we can also use the quantum conditional mutual information $I(A{:}C|B)$ to detect non-trivial entanglement structures in gapless systems. Once we choose the large enough areas $A,B,C$ with $A,C$ far from each other, a nonzero  $I(A{:}C|B)$ also provides information for gapless systems. Unlike the area law, $I(A{:}C|B)$ does not diverge for critical systems. And due to the dependence with $L$, the area ratio of the $A,B,C$ parts will give different values of $I(A{:}C|B)$. 

As an example, at the transition point $B=1$ for the transverse-field Ising model $H_{\text{tIsing}}$, where the system is gapless, the five curves
in Fig.~\ref{Ising} intersect at $S^\text{t}_\text{topo}\sim 0.5$. However, this value of  $S^\text{t}_\text{topo}$ is not a constant, which depends on the shape of the areas $A,B,C$. Define the ratio 
\begin{equation}
r=\frac{\#\ \text{in each of the area A,C}}
{\#\ \text{in each connected component of $B$}}
\end{equation}
where $\#$ means the number of qubits.

\begin{figure}[h!]
\centerline{
\includegraphics[width=2.50in]{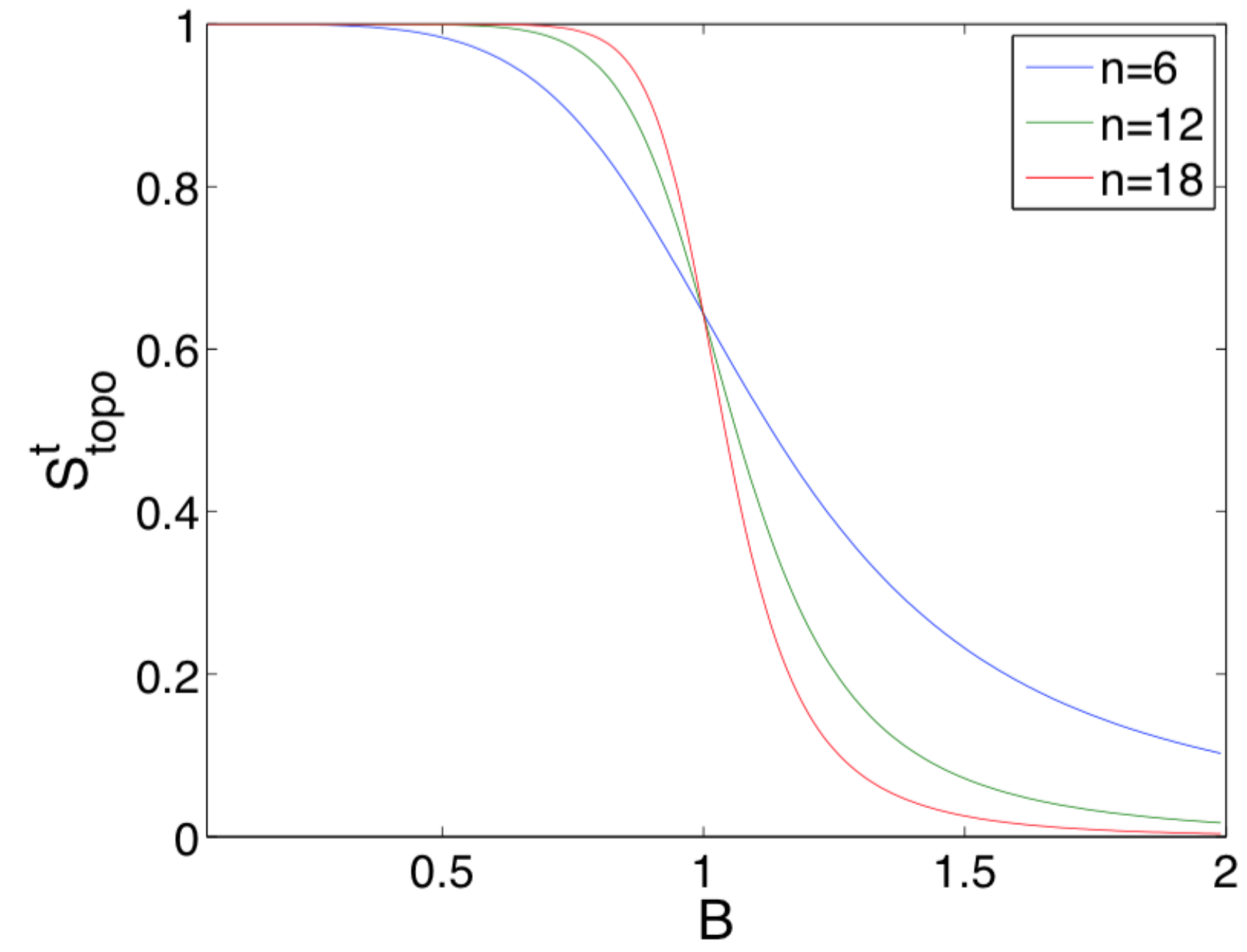} 
}
\caption{$S^\text{t}_\text{topo}$ for the transverse-field Ising model. The ratio $r=2:1$, and the system sizes are $N=6,12,18$ qubits. The horizontal 
axis is the magnetic field $B$. The vertical axis is $S^\text{t}_\text{topo}$ for the ground state of $H^{\text{tIsing}}(B)$.} 
\label{Ising1} 
\end{figure}

If we choose the ratio $r=2:1$, and to have $1,2,3$ qubits for each connected component of the area $B$, then we can compute $S^\text{t}_\text{topo}$ for total $N=6,12,18$ qubits, as shown in Fig.~\ref{Ising1}. 

\begin{figure}[h!]
\centerline{
\includegraphics[width=2.50in]{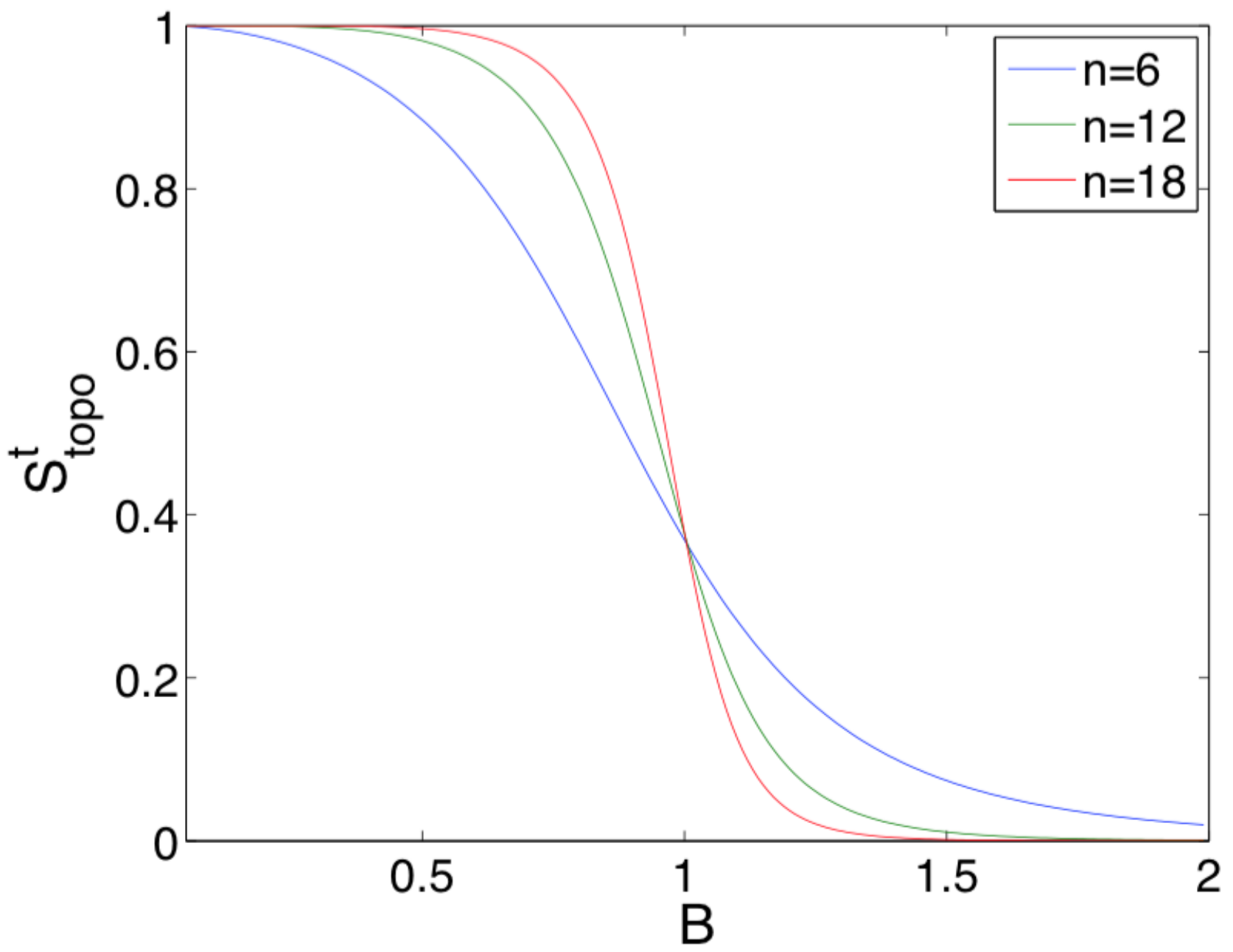} 
}
\caption{$S^\text{t}_\text{topo}$ for the transverse-field Ising model. The ratio $r=1:2$, and the system sizes are $N=6,12,18$ qubits. The horizontal 
axis is the magnetic field $B$. The vertical axis is $S^\text{t}_\text{topo}$ for the ground state of $H^{\text{tIsing}}(B)$. } 
\label{Ising2} 
\end{figure}

And if we choose $r=1:2$, and to have $2,4,6$ qubits for each connected component of the area $B$, then we can compute $S^\text{t}_\text{topo}$ for total $N=6,12,18$ qubits, as shown in Fig.~\ref{Ising2}. 

This ratio dependence is typical in critical systems. And the results for $I(A{:}C|B)$ with various ratios of the areas of $A,B,C$ are consistent with the conformal field theory (CFT) calculation. In other words, by varying the ratio of the areas of $A,B,C$, $I(A{:}C|B)$ can provide information for gapless/critical systems, for instance the value of the central charge.

\section{Summary and further reading}

In this chapter, we consider quantum systems in the limit of system size $N\rightarrow\infty$, which effectively describe macroscopic condensed matter systems containing $\sim 10^{23}$ degrees of freedom. New ideas and notions need to be introduced to study such many-body quantum systems. In particular, we introduce the concept of dimensionality, locality, thermodynamic limit, universality, gap, correlation, and many-body entanglement to characterize properties of quantum many-body systems. 

The behavior of quantum many-body entanglement is of special importance which we discuss in detail in different cases. For gapped quantum many-body systems, their many-body entanglement is found to satisfy an `area law', which imposes a strong constraint on the amount and form of many-body entanglement contained in the system. Gapless systems can violate this `area law', but usually only mildly with a logarithmic correction term. Moreover, in gapped quantum systems a constant subleading term exists in the entanglement entropy which is closely related to the topological order in the system.

We study the meaning of topological entanglement entropy from an information-theoretic viewpoint. This allows us to build a link between the topological entanglement entropy, the quantum conditional mutual information $I(A{:}C|B)$, and the irreducible three-party correlation discussed in Chapter 1. The proof of the strong subadditivity inequality $I(A{:}C|B)\geq 0$ is given in~\cite{lieb2002proof}. The structure of states that satisfy the equality is discussed in~\cite{hayden2004structure}, which are quantum Markov states with the form given in Eq.~\eqref{eq:Markov}.

We show that for large enough areas $A,B,C$ and $A,C$ far from each other,  non-zero $I(A{:}C|B)$ indicates non-trivial orders of the system. Calculating $I(A{:}C|B)$ for different choices of the areas $A,B,C$ could then detect different orders for gapped system (e.g. symmetry breaking, SPT, topological orders). And for probing symmetry breaking and/or SPT orders, we do not need to know the symmetry of the system to calculate $I(A{:}C|B)$ (for the exact ground state that does not break any symmetry, for any finite system). For gapless systems, the value of $I(A{:}C|B)$ depends on the shapes of $A,B,C$, which also contains information of the critical system (e.g. central charge). In this sense, $I(A{:}C|B)$ is a `universal entanglement detector' for both gapped and gapless system, which contains non-trivial information of the orders for the systems. 

In one dimensional systems, the existence of an area law in gapped quantum systems has been established as a rigorous mathematical theorem first by Hastings in \cite{H0724}. The constant bound on the entanglement entropy of a segment of the chain scales exponentially with the correlation length in Hastings proof, which has been subsequently tightened to a polynomial scaling in the work by Arad, Landau and Vazirani\cite{ALV1245}. In two or higher dimensions, a full proof of the `area law' does not exist yet but it has been supported by a large amount of numerical evidence. For a more detailed review of the subject, see \cite{ECP1077}.

In gapless systems, the entanglement `area law' is violated. Such a violation is particularly well understood at one dimensional critical points described by conformal field theory (CFT). In \cite{HLW9443}, it was proposed that the scaling of entanglement entropy in one dimensional critical systems is logrithmic and the scaling coefficient is related to the central charge of the CFT. In \cite{VLR0302}, numerical calculation for some one dimensional critical points was carried out which clearly demonstrated such a relation. For a more systematic discussion about entanglement entropy in CFT, see \cite{CC0905}. The scaling of entanglement entropy in higher dimensional gapless/critical systems is less well understood. \cite{ECP1077} also reviews what we currently know about such systems.

Various entanglement measures have become popular tools in studying quantum phase transitions. For summary of how to use entanglement measures to detect quantum phase transitions, see \cite{AFO0876}.

The idea of topological entanglement entropy was proposed in \cite{KP0604,LW0605} and two different schemes for calculation were provided. It has been used in numerical calculations to successfully identify nontrivial topological orders in physical systems. For example, see \cite{IHM1172,JWB1202}.

The information-theoretic aspects of topological entanglement entropy and its relationship to irreducible many-body correlation are discussed in~\cite{liu2014irreducible,MaxEnt}. The generalizations of topological entanglement entropy to study symmetry breaking orders are discussed in~\cite{MaxEnt,LIT} and to study SPT orders and mixed orders are discussed in~\cite{zeng2014topological}. The error-correcting properties of the SPT ground-states are discussed in~\cite{DB09,zeng2014topological}. The generalized topological entanglement entropy of critical systems are discussed in~\cite{LIT}, where the results for the transverse-field Ising model is shown to be consistent with the CFT calculation given in~\cite{PhysRevLett.102.170602,de2015entanglement}.

Recently, it has been realized that more detailed information about topological order can be extracted from the entanglement structure of the system than just a single number of entanglement entropy. It has been proposed in \cite{LH0804,PTB1039} that, the entanglement spectrum, i.e. the eigenvalue spectrum of the reduced density matrix, has meaning of its own. In fact for a gapped topologically ordered system, the `low energy' sector of the entanglement spectrum should reflect the nature of the low energy excitations on the edge of the system. Moreover, it was realized that entanglement entropy of systems on nontrivial manifolds, like cylinder or torus, can provide more information about the quasiparticle content of the topological system than that calculated on a plane \cite{ZGT1251}.

%
%
\bibliographystyle{plain}
\bibliography{Chap5}

%
%
%

\begin{partbacktext}
\part{Topological Order and Long-Range Entanglement}

\end{partbacktext}

%
%
%
\chapter{Introduction to Topological Order}
\label{chap6} 

\abstract{
In primary school, we were told that there are four states of matter: solid,
liquid, gas, and plasma.  In college, we learned that there are much more than
four states of matter. For example, there are ferromagnetic states as revealed
by the phenomenon of magnetization and superfluid states as defined by the
phenomenon of zero-viscosity.  The various phases in our colorful world are
extremely rich. So it is amazing that they can be understood systematically by
the symmetry breaking theory of Landau.  However, in last 20 -- 30 years, we
discovered that there are even more interesting phases that are beyond Landau
symmetry breaking theory.  In this chapter, we discuss new `topological'
phenomena, such as topological degeneracy, that reveal the existence of those
new phases -- topologically ordered phases.  Just like zero-viscosity defines
the superfluid order, the new `topological' phenomena define the topological
order at macroscopic level.
}

\section{Introduction}

\subsection{Phases of matter and Landau's symmetry breaking theory}

Although all matter is formed by only three kinds of particles: electrons,
protons and neutrons, matter can have many different properties and appear in
many different forms, such as solid, liquid, conductor, insulator, superfluid,
magnet, etc. According to  the principle of emergence in condensed matter
physics, the rich properties of materials originate from the rich ways in which
the particles are organized in the materials.  Those different organizations of
the particles are formally called `orders'.

For example, particles have a random distribution in a liquid (see Fig.
\ref{cryliq1}a), so a liquid remains the same as we displace it by an arbitrary
distance.  We say that a liquid has a `continuous translation symmetry'.
After a phase transition, a liquid can turn into a crystal.  In a crystal,
particles organize into a regular array (a lattice) (see Fig. \ref{cryliq1}b).
A lattice remains unchanged only when we displace it by a particular set of
distances
(integer times of the lattice constant), so a crystal has only `discrete
translation symmetry'.  The phase transition between a liquid and a crystal is
a transition that reduces the continuous translation symmetry of the liquid to
the discrete symmetry of the crystal. Such a change in symmetry is called
`spontaneous symmetry breaking'.  We note that the equation of motions that
govern the dynamics of the particles respects the continuous translation
symmetry for both cases of liquid and crystal.  However, in the case of
crystal, the stronger interaction makes the particles to prefer being separated
by a fixed distance and a fixed angle.  This makes particles to break the
continuous translation symmetry down to discrete translation symmetry
`spontaneously' in order to choose a low energy configuration (see Fig.
\ref{symmbrk}).  Therefore, the essence of the difference between liquids and
crystals is that the organizations of particles have different
\emph{symmetries} in the two phases.

Liquid and crystal are just two examples. In fact, particles can organize in
many ways which lead to many different orders and many different types of
materials.  Landau's symmetry breaking theory~\cite{L3726,GL5064,LanL58}
provides a general and a systematic understanding of these different orders.
It points out that different orders actually correspond to different symmetries
in the organizations of the constituent particles.  As a material changes from
one order to another order (\ie, as the material undergoes a phase transition),
what happens is that the symmetry of the organization of the particles changes.

\begin{svgraybox}
\begin{center}
\textbf{Box 6.1 Landau's symmetry breaking theory (classical)}

If a classical system with a symmetry is in a symmetry breaking phase, then its
degenerate minimal-free-energy states (or minimal-energy states at zero
temperature) break the symmetry.  Two systems with the same symmetry belong to
different phases if their minimal-free-energy states have different symmetries.
\end{center}
\end{svgraybox}

Landau's symmetry breaking theory is a very successful theory. For a long time,
physicists believed that Landau's symmetry breaking theory describes all
possible phases in materials, and all possible (continuous) phase transitions.

\begin{figure}[t]
\centerline{
\includegraphics[scale=0.5]{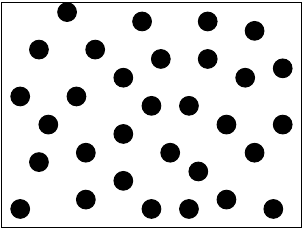}
\hfil
\includegraphics[scale=0.5]{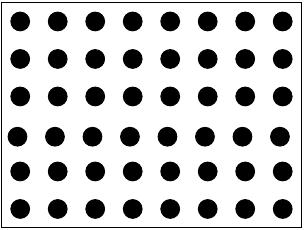}
}
\centerline{
\hfil
(a)
\hfil\hfil
(b)
\hfil
}
\caption{
(a) Particles in liquids do not have fixed relative positions.
They fluctuate freely and have a random but uniform distribution.
(b) Particles in solids form a fixed regular lattice.
}
\label{cryliq1}
\end{figure}

\begin{figure}[tb]
\centerline{
\includegraphics[scale=0.6]{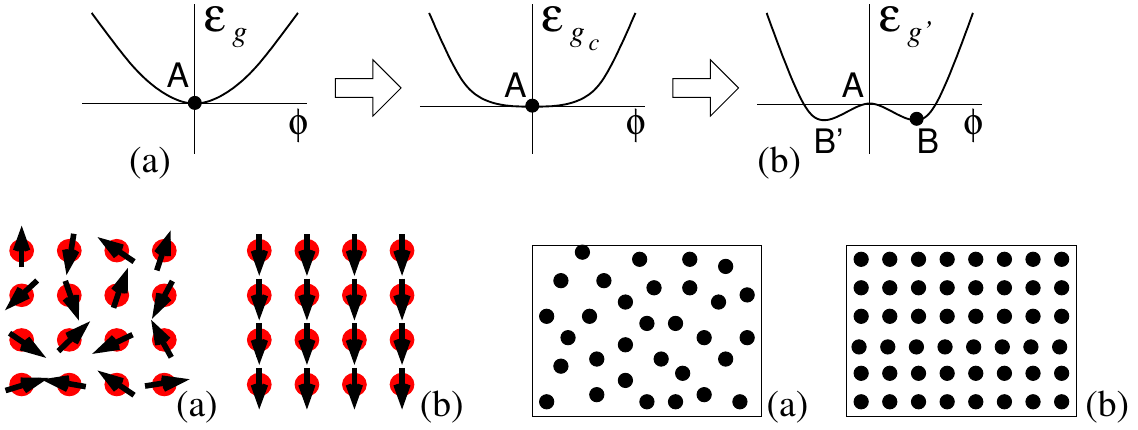}
}
\caption{ (a) Disordered states that do not break the symmetry.  (b) Ordered
states that spontaneously break the symmetry.  The energy function
$\veps_g(\phi)$ has a symmetry $\phi \to -\phi$:
$\veps_g(\phi)=\veps_g(-\phi)$.  However, as we change the parameter $g$, the
minimal energy state (the ground state) may respect the symmetry (a), or may
not respect the symmetry (b). This is the essence of spontaneous symmetry
breaking.  }
\label{symmbrk}
\end{figure}

\subsection{Quantum phases of matter and transverse-field Ising model}

Quantum phases of matter are phases of matter at zero temperature.  So quantum
phases correspond to the ground states of the quantum Hamiltonians that govern
the systems.  In this book, we mainly discuss those quantum phases of
matter.  Crystal, conductor, insulator, superfluid, and magnets can exist at
zero temperature and are examples of quantum phases of matter.

Again, physicists used to believe that Landau symmetry breaking theory also
describes all possible quantum phases of matter, and all possible (continuous)
quantum phase transitions.  (Quantum phase transitions, by definition, are zero
temperature  phase transitions.) For example, the superfluid is described by a
$U(1)$ symmetry breaking.

The simplest example to demonstrate the Landau symmetry breaking theory for
quantum phases is the transverse-field Ising model on a 1-dimensional chain.
The total Hilbert space of the transverse-field Ising model is formed
by 1/2 spins (qubits) on each site. The Hamiltonian is given by
\begin{align}
 H^{\text{tIsing}}= - \sum_i ( Z_i Z_{i+1} + B X_i),
\end{align}
where $X_i,Y_i,Z_i$ are the Pauli matrices acting on the $i^\text{th}$ spin.
The Hamiltonian has a spin-flip symmetry, $\up \leftrightarrow \down$,
generated by $\otimes_i X_i$: $[H, \otimes_i X_i]=0$.

One way to obtain the ground state of the  transverse-field Ising model is to
use the variational approach.  To design the variational trial wave function,
we note that when $B=0$ the ground states are two-fold degenerate and are given
by $\otimes_i |\up\>_i$ and $\otimes_i |\down\>_i$.  When $B>>1$ the ground
states is given by $\otimes_i (|\up\>_i+|\down\>_i)/\sqrt 2$.
Thus we choose our trial wave function as
\begin{align}
 |\Psi_\phi\> = \otimes_i \Big[\cos(\phi/2) |\up\>_i+\sin(\phi/2)|\down\>_i\Big],
\end{align}
where $\phi$ is the variational parameter.
The average energy per site is given by
\begin{align}
\veps(\phi)= \frac{\<\Psi_\phi|H|\Psi_\phi\>}{N_\text{site}}
= -[\cos^2(\phi/2) -\sin^2(\phi/2)]^2 - 2 B  \cos(\phi/2) \sin(\phi/2) .
\end{align}
We note that the spin-flip transformation $\otimes_i X_i$ changes $|\Psi_\phi\>
\to |\Psi_{\pi-\phi}\>=\otimes_i X_i |\Psi_\phi\>$. So $\veps(\phi)$ satisfies
$\veps(\phi)=\veps(\pi-\phi)$ due to the spin-flip symmetry.

\begin{figure}[tb]
\centerline{
\includegraphics[scale=1.2]{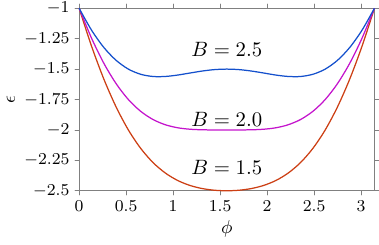}
}
\caption{
The variational energies $\veps(\phi)$ for $B=1.5,2,2.5$.
}
\label{ising}
\end{figure}

In Fig. \ref{ising}, we plot the variational energy $\veps(\phi)$ for
$B=1.5,2,2.5$. We see that there is a symmetry breaking transition at $B=2$.
For $B>2$, the energy is minimized at $\phi=\pi/2$ and the trial ground state
does not break the spin-flip symmetry $\phi\to \pi-\phi$.  For $B<2$, the
energy is minimized at two places $\phi=\pi/2\pm \Del \phi$, which give rise to
two degenerate ground states $|\Psi_{\pi -\Del \phi}\>$ and $|\Psi_{\pi +\Del
\phi}\>$.  Each of the ground state breaks the spin-flip symmetry.

\subsection{Physical ways to understand symmetry breaking in quantum theory}

The above understanding of symmetry breaking in quantum system is not
satisfactory.  It is based on a calculational trick -- the variational
approach, rather than physical measurements in real or numerical experiments.
So what is the physical ways to understand symmetry breaking in quantum theory?

Here, we will concentrate on numerical experiments.  One of a  numerical
experiment is the energy spectrum of transverse Ising model for $J=1,\ B=0.5$
(see Fig. \ref{tIsing}(a,b)).  The ground states have a near two-fold
degeneracy, with exponentially small energy splitting in large system size
limit.  The appearance of such a  near two-fold degeneracy is a very remarkable
phenomenon.  

The  transverse-field Ising model has a $\mathbb{Z}_2$ spin-flip
symmetry $|\uparrow\> \leftrightarrow |\downarrow\>$.  But such a
$\mathbb{Z}_2$ symmetry has only one dimensional representations and cannot
give rise to two-fold degeneracy.  So, the near two-fold degeneracy is not
the exact degeneracy protected by symmetry, because they do not belong to a
single irreducible representation of the $\mathbb{Z}_2$ symmetry group.  One
may wonder,  the near two-fold degeneracy has nothing to do with the
$\mathbb{Z}_2$ spin-flip symmetry. However, this is not true.  If we explicitly
break the the $\mathbb{Z}_2$ symmetry by adding a term $B_z \sum_i Z_i$ to the
Hamiltonian of transverse Ising model, then the  near two-fold degeneracy will
be destroyed.  Therefore, the  near two-fold degeneracy is protected by the
symmetry despite they do not belong to a single irreducible representation of
the symmetry group ({\it i.e.} they are not the exact degeneracy protected by
symmetry).  We note that the above emergence of ground state degeneracy happens
on spaces with any shape, such as a ring $S^1$ or a segment $I$.  This way, we
find that

\begin{svgraybox}
\begin{center}
\textbf{Box 6.2 Symmetry breaking in quantum theory I}

A quantum system with a finite symmetry group is in a symmetry breaking phase
at zero-temperature, iff it has robust emergent nearly degenerate ground states
that belong to atleast two different irreducible representations of the
symmetry group, on any shapes of space. Here, the term ``robust'' means that
emergent ground state degeneracy is robust against any perturbations that
preserve the symmetry.

\end{center}
\end{svgraybox}

We also note that, in quantum theory, $|\Psi_+\>=(|\Psi_{\pi -\Del
\phi}\>+|\Psi_{\pi +\Del \phi}\>)/2$ is also a ground state which does not
break the spin-flip symmetry.  In fact, for finite systems, $|\Psi_+\>$
represents the true ground state of the system. Such a true ground state
$|\Psi_+\>$ does not break any symmetry.  Thus, the symmetry breaking state of
a system is not characterized by the symmetry breaking breaking properties of
its true ground state. On the other hand, we note that  $|\Psi_+\>$ has a
GHZ-type of quantum entanglement.  Therefore the symmetry breaking state of a
system is characterized by the  GHZ-type of quantum entanglement in this true
ground state:

\begin{svgraybox}
\begin{center}
\textbf{Box 6.3 Symmetry breaking in quantum theory II}

If a quantum system with a finite symmetry group is in a symmetry breaking
phase at zero-temperature, then its true ground state has a GHZ-type of quantum
entanglement.

\end{center}
\end{svgraybox}

\subsection{Compare a finite-temperature phase with a zero-temperature phase}

It is interesting to compare a finite-temperature phase, \emph{liquid},
with a zero-temperature phase, \emph{superfluid}.  A liquid is described by a
random \emph{probability} distributions of particles (such as atoms), while a
superfluid is described by a quantum wave function which is the
\emph{superposition} of a set of random particle configurations:
\begin{align}
 |\Phi_\text{superfluid}\>=
\sum_\text{random configurations}
\left  |
\bmm \includegraphics[scale=0.25]{Chapters/Chap6/liquid} \emm
\right  \>
\end{align}
The superposition of many different particle positions are called quantum
fluctuations in particle positions.

Since Landau's symmetry breaking theory suggests that all quantum phases are
described by symmetry breaking, thus we can use group theory to classify all
those symmetry breaking phases: All symmetry breaking quantum phases are
classified by a pair of mathematical objects $(G_H,G_\Phi)$, where $G_H$ is the
symmetry group of the Hamiltonian and $G_\Phi$ is the symmetry group of the
ground state.  For example,  the symmetry breaking phase of the
transverse-field Ising model is labeled by $(\mathbb{Z}_2,\{1\})$, where $\mathbb{Z}_2$ is the
symmetry group of the Hamiltonian, and $\{1\}$ is the trivial group that
describe the symmetry of the ground state. 

\section{Topological order}

\subsection{The discovery of topological order}

However, in late 1980s, it became clear that Landau symmetry breaking theory
did not describe all possible phases.  In an attempt to explain high
temperature superconductivity, the chiral spin state was
introduced~\cite{KL8795,WWZ8913}. At first, physicists still wanted to use
Landau symmetry breaking theory to describe the chiral spin state. They
identified the chiral spin state as a state that breaks the time reversal and
parity symmetries, but not the spin rotation symmetry~\cite{WWZ8913}.  This
should be the end of story according to Landau symmetry breaking description of
orders.

But, it was quickly realized that there are many different chiral spin states
that have exactly the same symmetry~\cite{Wtop}. So symmetry alone was not
enough to characterize and distinguish different chiral spin states.  This means
that the chiral spin states must contain a new kind of order that is beyond the
usual symmetry description.  The proposed new kind of order was named
`topological order'~\cite{Wrig}. (The name `topological order' was motivated
by the low energy effective theory of the chiral spin states which is a
Chern-Simons theory~\cite{WWZ8913} -- a topological quantum field theory
(TQFT)~\cite{W8951}).  New quantum numbers (or new topological probes), such as
ground state degeneracy~\cite{Wtop,WNtop}  and the non-Abelian geometric phase
of degenerate ground states~\cite{Wrig,KW9327}, were introduced to
characterize/define the different topological orders in chiral spin states.

\begin{figure}[tb]
\centerline{
\hfil
$\bmm
\includegraphics[height=1.4in]{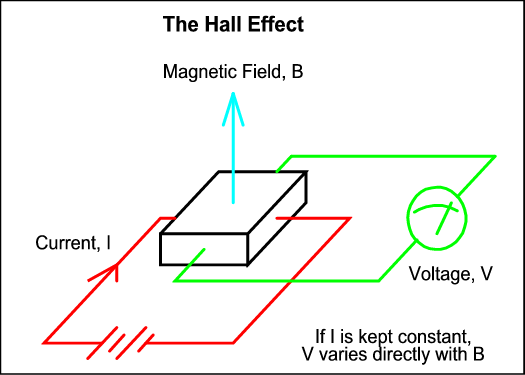}
\emm $
\hfil
$\bmm
\includegraphics[scale=0.4]{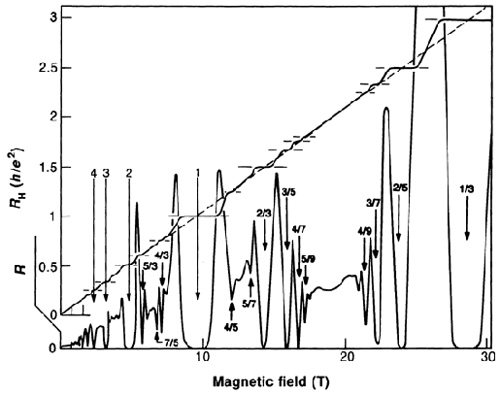}
\emm $
}
\caption{
2D electrons in strong magnetic field may form FQH states.  Each FQH state has
a quantized Hall coefficient $R_H$.
}
\label{Hall}
\end{figure}

But experiments soon indicated that chiral spin states do not describe
high-temperature superconductors, and the theory of topological order became a
theory with no experimental realization. However, the similarity~\cite{KL8795}
between chiral spin states and fractional quantum Hall (FQH)
states~\cite{TSG8259,L8395} allows one to use the theory of topological order to
describe different FQH states.

FQH states are gapped ground states of 2D electrons under strong magnetic
field.  FQH states have a property that a current density will induce an
electric field in the transverse direction: $E_y=R_H j_x$ (see Fig.
\ref{Hall}).  It is an amazing discovery that the Hall coefficient $R_H$ of a
FQH state is precisely quantized as a rational number $\frac{p}{q}$ if we
measure the Hall coefficient $R_H$ in unit of $\frac{h}{e^2}$: $
R_H=\frac{p}{q}\frac{h}{e^2} $ (see Fig. \ref{Hall})~\cite{TSG8259}.  Different
quantized $R_H$ correspond to different FQH states. Just like the chiral spin
states, different FQH states all have the same symmetry and cannot be
distinguished by symmetry breaking.  So there is no way to use different
symmetry breaking to describe different FQH states, and FQH states must contain
\emph{new} orders.  One finds that the new orders in quantum Hall states can
indeed be described by topological orders~\cite{WNtop}. So the theory of topological order
does have experimental realizations.

We would like to point out that before the topological-order
understanding of FQH states, people have tried to use the notions of
off-diagonal long-range order and order parameter from Ginzburg-Landau theory
to describe FQH states~\cite{GM8752,R8986,ZHK8982,EI9137}. Such an effort
leads to a Ginzburg-Landau Chern-Simons effective theory for FQH
states~\cite{ZHK8982,EI9137}. At same time, it was also realized that the  order
parameter in the Ginzburg-Landau Chern-Simons is not gauge invariant and is not
physical.  This is consistent with the topological-order understanding of FQH
states which suggests that FQH has no off-diagonal long-range order and cannot
be described by local order parameters.  So we can use effective theories
without order parameters to describe FQH states, and such  effective theories
are pure Chern-Simons effective
theories~\cite{WNtop,BW9045,FZ9117,FK9169,WZ9290,FS9333}. The deeper
understanding gained from pure Chern-Simons effective theories leads to a
K-matrix classification~\cite{WZ9290,BM0535} of all Abelian topologically
ordered states (which include all Abelian FQH states).

FQH states were discovered in 1982~\cite{TSG8259} before the introduction of the
concept of topological order.  But FQH states are not the first experimentally
discovered topologically ordered states.  The real-life superconductors, having
a $\mathbb{Z}_2$ topological order~\cite{W9141,Wsrvb,HOS0497}, were first experimentally
discovered topologically ordered states.\footnote{Note that real-life
superconductivity can be described by the Ginzburg-Landau theory with a
\emph{dynamical} $U(1)$ gauge field.  The condensation of charge $2e$ electron
pair break the $U(1)$ gauge theory into a $\mathbb{Z}_2$ gauge theory at low energies. A
$\mathbb{Z}_2$ gauge theory is an effective theory of $\mathbb{Z}_2$ topological order.
Thus a real-life superconductor has a  $\mathbb{Z}_2$ topological order.
In many textbook, superconductivity is described by the Ginzburg-Landau theory
without the dynamical $U(1)$ gauge field, which fails to describe the real-life
superconductors with dynamical electromagnetic interaction.
Such a textbook superconductivity is described by a $U(1)$ symmetry breaking.}
(Ironically, the
Ginzburg-Landau symmetry breaking theory was developed to describe
superconductors, despite the real-life superconductors are not symmetry
breaking states, but topologically ordered states.) 

\section{A macroscopic definition of topological order}

\begin{figure}[tb]
\centerline{
\includegraphics[scale=0.6]{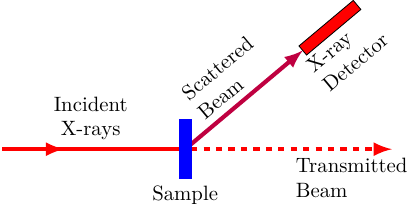}
\hfil
\includegraphics[scale=0.3]{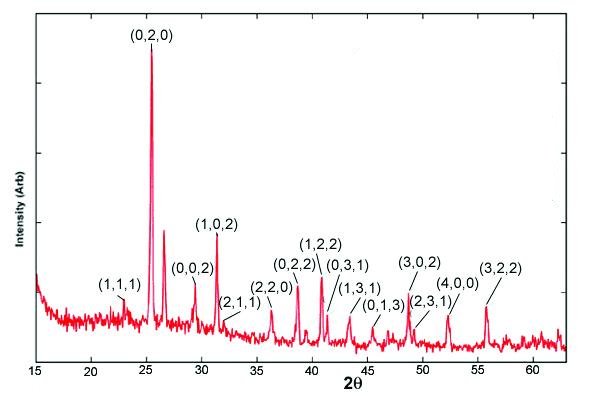}
}
\caption{
A X-ray diffraction pattern defines/probes the crystal order.
}
\label{xray}
\end{figure}

In the above, we have described topological order as a new order which is not a
symmetry breaking order.  But what \emph{is} topological order?  Here, we would like
to point out that \emph{to define a physical concept (such as symmetry breaking
order or topological order) is to design experiments or numerical calculations
that allow us to probe and characterize the  concept}.  For example, the concept
of superfluid order, is defined by zero viscosity and the quantization of
vorticity, and the concept of crystal order is defined by X-ray diffraction
experiment (see Fig.  \ref{xray}).

\begin{table}[tb]
 \centering
 \begin{tabular}{ |c|c|}
 \hline
\textbf{Order} & \textbf{Experimental probes} \\
 \hline
Crystal order & X-ray diffraction\\
 \hline
Ferromagnetic order & Magnetization\\
 \hline
Anti-ferromagnetic order & Neutron scattering\\
 \hline
Superfluid order & Zero-viscosity \& vorticity quantization\\
 \hline
 \hline
Topological order & Topological degeneracy, \\
(Global dancing pattern) & non-Abelian geometric phase \\
 \hline
 \end{tabular}
\caption{
Symmetry breaking orders can be probed/defined through linear responses.
But topological order cannot be
probed/defined through  linear responses.
We need topological probes to define topological orders.
}
\label{tab1}
\end{table}

\begin{figure}[tb]
\centerline{
\includegraphics[scale=1.0]{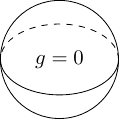}~~~~~~~~~~~~~
\includegraphics[scale=1.0]{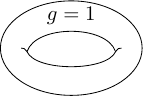}~~~~~~~~~~~~~
\includegraphics[scale=1.0]{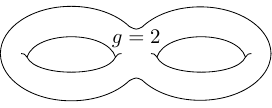}
}
\centerline{
	Deg.$=1$ ~~~~~~~~~~~~~~~~~~~~~~~~~~~
	Deg.$=D_1$ ~~~~~~~~~~~~~~~~~~~~~~~~~~~~~~~~~~~~~~~~~~~~~~~~~~~~
	Deg.$=D_2$ 
}
\caption{The topological ground state degeneracies of topologically ordered
states depend on the topology of the space, such as the genus $g$ of two
dimensional closed surfaces.}
\label{g0g1g2}
\end{figure}

The experiments that we use to define/characterize superfluid order and crystal
order are linear responses, such as viscosity and X-ray diffraction.
Linear responses are easily accessible in
experiments and the symmetry breaking order that they define are easy to
understand (see Table \ref{tab1}).  However, topological order is such a new and elusive order that it
cannot be probed/defined by any linear responses.  To probe/define topological
order we need to use very unusual `topological' probes.
In 1989, it was conjectured that topological order
can be completely defined/characterized by using only two topological
properties (at least in 2+1 dimensions)~\cite{Wrig}:\\
(1) \emph{Topological ground state degeneracies} on closed spaces
of various topologies.  (see Fig. \ref{g0g1g2})~\cite{Wtop}.\\
(2) \emph{Non-Abelian geometric phases}\cite{WZ8411} of those degenerate ground states
from deforming the spaces (see Fig. \ref{modtrns})~\cite{Wrig,KW9327}.\\
It was through such topological probes that we introduce the concept
of topological order.  Just like zero viscosity and the quantization of
vorticity define the concept of superfluid order, the topological degeneracy
and the non-Abelian geometric phases of the degenerate ground states define the
concept of topological order.

\begin{svgraybox}
\begin{center}
\textbf{Box 6.4 Topological order}

Topological order can be probed/defined by topological degeneracy and
non-Abelian geometric phases of the ground states.
\end{center}
\end{svgraybox}

\subsection{What is `topological ground state degeneracy'}

Topological ground state degeneracy, or simply, topological degeneracy is a
phenomenon of quantum many-body systems, that the ground state of a gapped
many-body system become degenerate in the large system size limit.
The topological degeneracy has the following characters:\\
\begin{enumerate}
\item For a finite system, the  topological degeneracy is not exact. The low
energy ground states have a small energy splitting.
\item The topological degeneracy becomes exact when the system size becomes
infinite.
\item The above property is robust against any local perturbations.  In other
words, the  topological degeneracy \emph{cannot be lifted by any local
perturbations} as long as the system size is
large~\cite{Wtop,WNtop,WZ9817,HWcnt}.
\item The topological degeneracy for a given system usually is different for
different topologies of space~\cite{HR8529}.  For example, for a $\mathbb{Z}_2$
topologically ordered state in two dimensions~\cite{RS9173,W9164}, the
topological degeneracy is $D_g=4^g$ on genus $g$ Riemann surface (see Fig.
\ref{g0g1g2}).
\end{enumerate}

People usually attribute the ground state degeneracy to symmetry.  But
topological degeneracy, being robust against any local perturbations that can
break all the symmetries, is not due to symmetry. So the very existence of
topological degeneracy is a surprising and amazing phenomenon.  Such an amazing
phenomenon defines the notion of topological order.  As a comparison, we know
that the existence of zero-viscosity is also an amazing phenomenon, and such an
amazing phenomenon defines the notion of superfluid order.  So topological
degeneracy,
playing the role of zero-viscosity in superfluid order, implies the existence
of a new kind of quantum phase -- topologically ordered phases.

\subsection{What is `non-Abelian geometric phase of topologically degenerate
states'}

\begin{figure}[tb]
\centerline{
\includegraphics[scale=0.5]{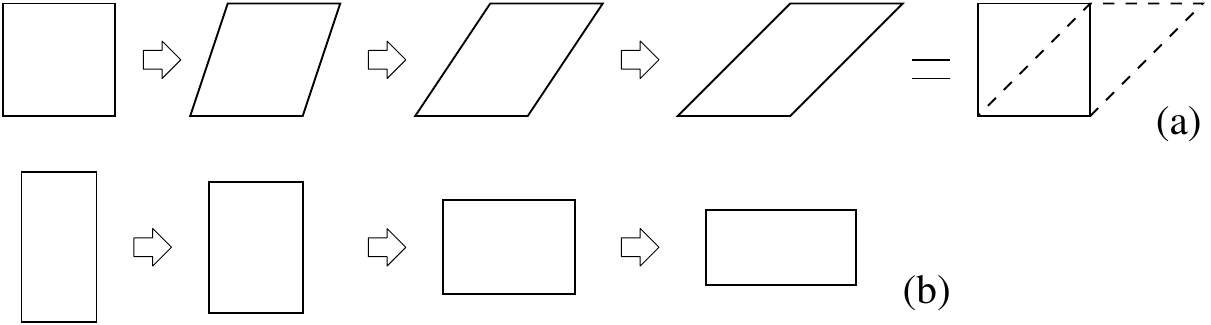}
}
\caption{
(a) The shear deformation of a torus generate a (projective) non-Abelian
geometric phase $T$, which is a generator of
a projective representation modular transformation.
The last shear-deformed torus is the same as the original
torus after a coordinate transformation:
$x\to x+y$, $y\to y$.
(b) The squeezing deformation of a torus generate a (projective) non-Abelian
geometric phase $S$, which is the other generator of
a projective representation modular transformation.
The last squeeze-deformed torus is the same as the original
torus after a coordinate transformation:
$x\to y$, $y\to -x$.
}
\label{modtrns}
\end{figure}

However, the ground state degeneracy is not enough to completely
characterize/define topological order.  Two different topological orders may
have exactly the same topological degeneracy on space of any topology.  We
would like to find, as many as possible, quantum numbers associated with the
degenerate ground states, so that by measuring these quantum numbers we can
completely characterize/define topological order.  The non-Abelian geometric
phases  of topologically degenerate states are such quantum
numbers~\cite{Wrig,KW9327}.

The non-Abelian geometric phase is a unitary matrix $U$ that can be calculated
from an one parameter family of gapped Hamiltonians $H_g$, $g\in [0,1]$,
provided that $H_0=H_1$~\cite{WZ8411}. $U$ is a one by one matrix if there is
only one ground state below the gap. $U$ is $n$ dimensional if the ground state
degeneracy is $n$ for all $g\in [0,1]$.

To use non-Abelian geometric phases to characterize/define topological order,
let us put the many-body state on a torus~\cite{Wrig,KW9327,ZGT1251,ZV1224}, and
perform a `shear' deformation of the torus to obtain a one parameter family
of gapped Hamiltonians that form a loop (\ie $H_0=H_1$) (see Fig.
\ref{modtrns}a).  The  non-Abelian geometric phase obtained this way is denoted
as $T$.  Similarly, a `squeezing' deformation of the torus gives rise to
another non-Abelian geometric phase $S$.  Both $S$ and $T$ are $D_1$
dimensional unitary matrices where $D_1$ is the topological degeneracy on
torus.  For different deformation paths that realize the loops in Fig.
\ref{modtrns}, $S$ and $T$ may be different.  However, because the ground state
degeneracy is robust, the difference is \emph{only} in the total phase factors.
Since the two deformations in Fig.  \ref{modtrns} generate the modular
transformations, thus $S$ and $T$ generate a projective representation of the
modular transformations.  $S$ and $T$ contain information about the topological
properties of the topologically ordered states, such as fractional
statistics~\cite{Wrig,KW9327,TZQ1251,ZMP1233,CV1308}. It was conjectured that

\begin{svgraybox}
\begin{center}
\textbf{Box 6.5  A complete characterization of topological order}

$S$ and $T$ (plus the path dependent  total phase factor)
provides a complete characterization and definition of topological orders in
2+1 dimensions~\cite{Wrig,KW9327}. 

\end{center}
\end{svgraybox}

\section{A microscopic picture of topological orders}

\subsection{The essence of fractional quantum Hall states}

C. N. Yang once asked: the microscopic theory of fermionic superfluid and
superconductor, BCS theory, capture the essence of the  superfluid and
superconductor, but what is this essence?  This question led him to develop the
theory of off-diagonal long range order~\cite{Y6294} which reveal the essence of
superfluid and superconductor. In fact long range order is the essence of any
symmetry breaking order.

Similarly, we may ask: Laughlin's theory~\cite{L8395} for FQH effect capture the
essence of the FQH effect, but what is this essence?  Our answer is that the
topological order
(defined by the topological ground state degeneracy and the non-Abelian
geometric phases of those degenerate ground states) is the essence of FQH
effect.

One may disagree with the above statement by pointing out that the  essence of
FQH effect should be the quantized Hall conductance.  However, such an opinion
is incorrect, since even after we break the particle number conservation (which
breaks the quantized Hall conductance), a FQH state is still a non-trivial
state with  topological degeneracy and non-Abelian geometric phases.
The non-trivialness of FQH state does not rely on any symmetry (except the
conservation of energy).  In fact, the topological degeneracy and the
non-Abelian geometric phases discussed above are the essence of FQH states
which can be defined even without any symmetry.  They provide a
characterization and definition of topological order that does not rely on any
symmetry.  We would like to point out that the topological entanglement entropy
is another way to characterize the topological order  without any symmetry (see
Chapter \ref{sec:topo_ent})~\cite{KP0604,LWtopent}.

\subsection{Intuitive pictures of topological order}

Topological order is a very new concept that describes quantum entanglement in
many-body systems.  Such a concept is very remote from our daily experiences
and it is hard to have an intuition about it.  So before we define topological
order in general terms (which can be abstract), let us first introduce and
explain the concept through some intuitive pictures.

\begin{figure}[tb]
.\hfil
\includegraphics[height=1.1in]{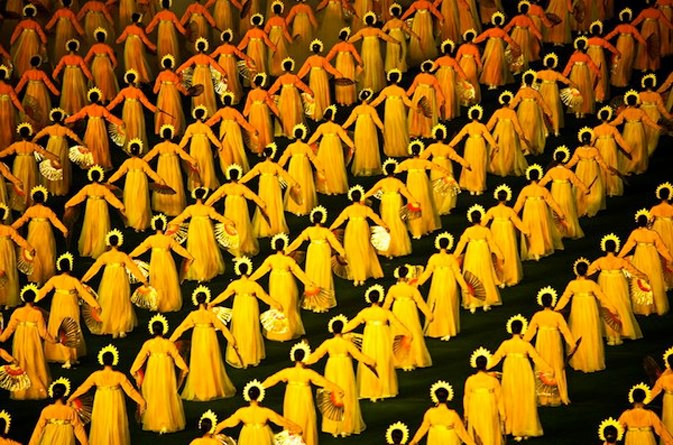}
\hfil
\includegraphics[height=1.1in]{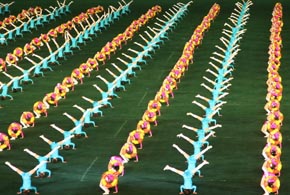}\\
.\hfil
Ferromagnet
\hfil
~~~~~ Anti-ferromagnet\\[2mm]
.\hfil
\includegraphics[height=1.1in]{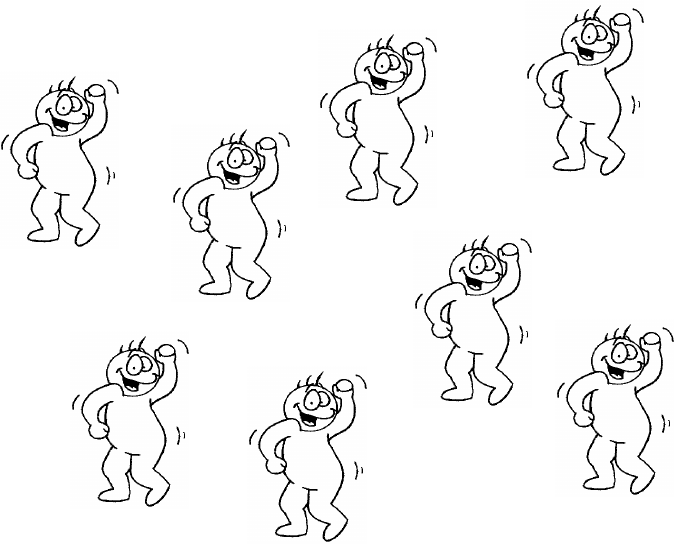}
\hfil
\hfil
\includegraphics[height=1.1in]{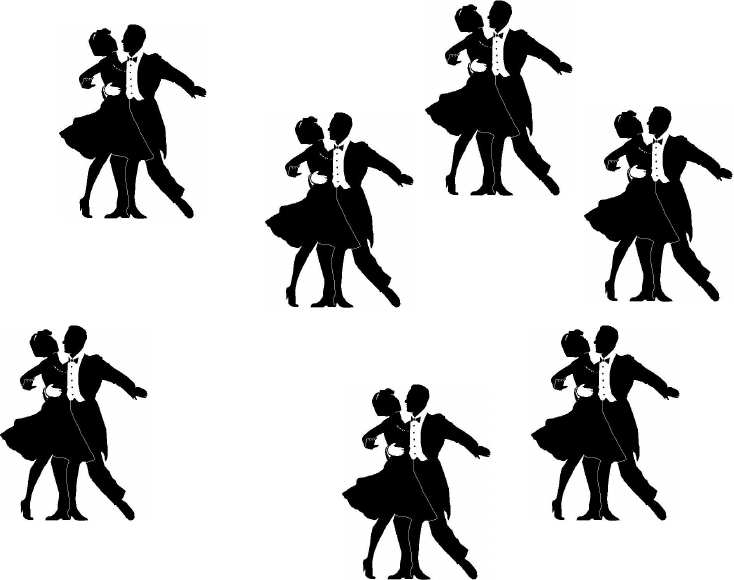}\\
.\hfil
Superfluid of bosons
\hfil
~~~~ Superfluid of fermions
\caption{
The dancing patterns for
the symmetry breaking orders.
}
\label{symmpic}
\end{figure}

We can use dancing to gain an intuitive picture of topological order. But
before we do that, let us use dancing picture to describe the old symmetry
breaking orders (see Fig.  \ref{symmpic}).  In the symmetry breaking orders,
every particle/spin (or every pair of particles/spins) dance by itself, and
they all dance in the same way.  (The `same way' of dancing represents a
long-range order.)  For example, in a ferromagnet, every electron has a fixed
position and the same spin direction.  We can describe an anti-ferromagnet by
saying every pair of electrons has a fixed position and the two electrons in a
pair have opposite spin directions.  In a boson superfluid, each boson is
moving around by itself and doing the same dance, while in a fermion
superfluid, fermions dance around in pairs and each pair is doing the same
dance.

\begin{figure}[tb]
.\hfil
\includegraphics[height=1.2in]{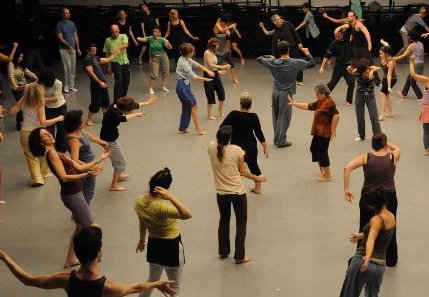}
\hfil
\hfil
\includegraphics[height=1.2in]{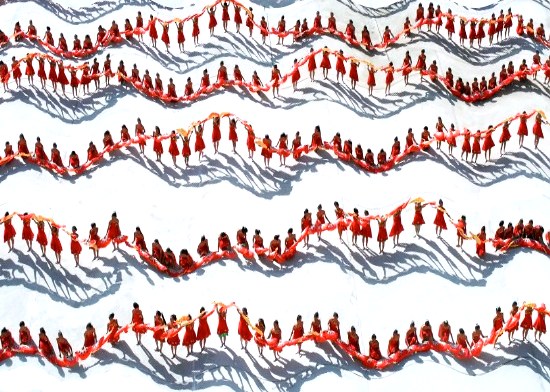}\\
.\hfil
~~~~~~~~~~~~~ FQH state
\hfil
\hfil
~~~~~
~~~~~
String liquid (spin liquid)
\caption{
The dancing patterns for the topological orders.
}
\label{toppic}
\end{figure}

We can also understand topological orders through such dancing pictures.
Unlike fermion superfluid where fermions dance in pairs, a topological order is
described by a global dance, where every particle (or spin) is dancing with
every other particle (or spin) in a very organized way:  (a) all
spins/particles dance following a set of \emph{local} dancing `rules' trying
to lower the energy of a \emph{local} Hamiltonian.  (b) If all the
spins/particles follow the local dancing `rules', then they will form a
global dancing pattern, which correspond to the topological order.
(c) Such a global pattern of collective dancing is a pattern of quantum
fluctuation which corresponds to a pattern of \emph{long-range entanglement}.
(A more rigorous definition of long-range entanglement will be given in Chapter~\ref{chap7}.)

For example in FQH liquid,
the electrons dance following the following local
dancing rules:\\
(a) electron always dances anti-clockwise which implies that
the electron wave function only depend on the electron coordinates $(x,y)$
via $z=x+\ii y$.\\
(b) each electron always takes exact three steps to dance around any
other electron,
which implies that the phase of the
wave function changes by $6\pi$ as we move an electron around
any other electron.\\
The above two local dancing rules fix a global dance pattern which correspond
to the Laughlin wave function $\Phi_\text{FQH} = \prod_{i<j}
(z_i-z_j)^3$~\cite{L8395}.  Such an collective dancing gives rise to the
topological order (or long-range entanglement) in the FQH state.

\begin{figure}[tb]
\centerline{
\includegraphics[height=1.2in]{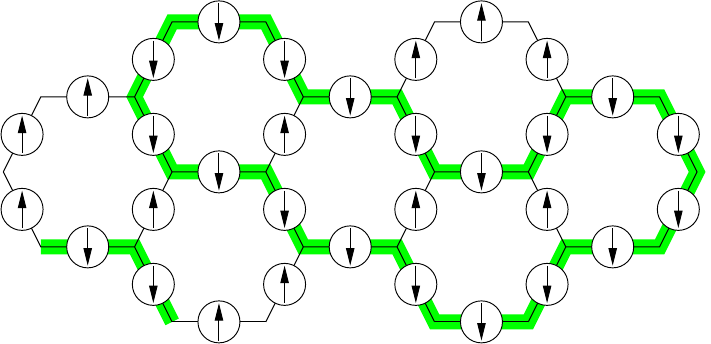}
}
\caption{
The strings in a spin-1/2 model.
In the background of up-spins, the down-spins form closed strings.
}
\label{strspin}
\end{figure}

\begin{figure}[tb]
\centerline{
\includegraphics[scale=0.15]{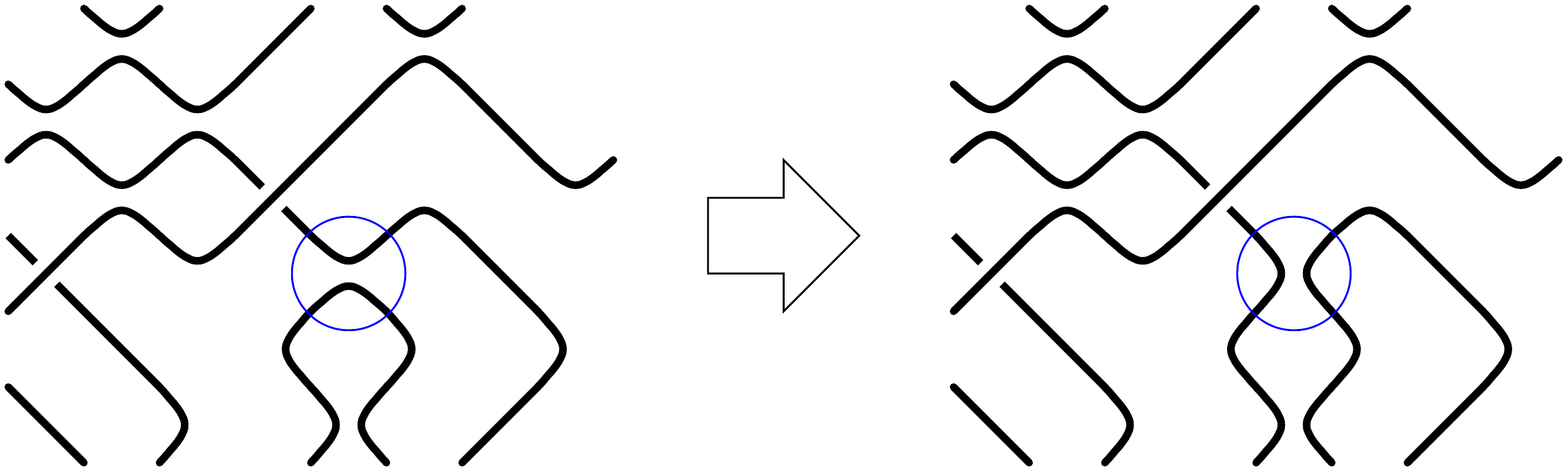}
}
\caption{
In string liquid, strings can move freely,
including reconnecting the strings.
}
\label{strnetSa}
\end{figure}

In additional to FQH states, some spin liquids also contain topological
orders~\cite{WWZ8913,RS9173,Wsrvb,MLB9964,MS0181}. (Spin liquids refer to ground
states of quantum spin systems that do not \emph{spontaneously} break the spin
rotation and the translation symmetries in the spin Hamiltonians.) In those
spin liquids, the spins `dance' following the follow local dancing rules:\\
(a) Down spins form
closed strings with no open ends, in the background of up-spins (see Fig.  \ref{strspin}).\\
(b) Strings can otherwise move freely, including reconnect freely (see Fig.
\ref{strnetSa}). \\
The global dance formed by the spins following the above dancing rules gives us
a quantum spin liquid which is a superposition
of all closed-string configurations~\cite{K032}:
$|\Phi_\text{string}\>=\sum_\text{closed string pattern} \left |\bmm
\includegraphics[height=0.3in]{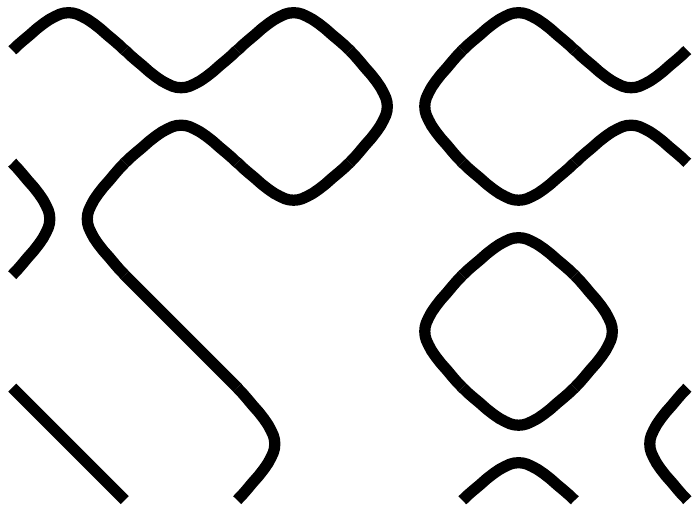}\emm\right \> $.  Such a state is called
a string or string-net condensed state~\cite{LWstrnet}. The collective dancing
gives rise to a non-trivial topological order and a pattern of long range
entanglement in the spin liquid state.

\begin{svgraybox}
\begin{center}
\textbf{Box 6.6 Microscopic picture of topological order}

Topological orders correspond to global correlated dances which are produced by
various local dancing rules.  The global correlated dances produce patterns of
long-range entanglement, which is the microscopic origin of topological order.
\end{center}
\end{svgraybox}

\section{What is the significance of topological order?}

The above descriptions of topological order is intuitive and not concrete. It
is not clear if the topological order (the global dancing pattern or the
long-range entanglement) has any experimental significance.  In order for the
topological order to be a useful concept, it must have new experimental
properties that are different from any symmetry breaking states. Those new
experimental properties should indicate the non-trivialness of the topological
order.  In fact, the concept of topological order should be defined by the
collection of those new experimental properties.

Indeed, topological order does have new  characteristic properties.  Those
properties of topological orders reflect the significance of topological order:

\enu{
\item The finite-energy defects of topological order (\ie the quasiparticles)
can carry fractional statistics~\cite{H8483,ASW8422} (including non-Abelian
statistics\cite{Wnab,MR9162}) and fractional charges~\cite{JR7698,L8395} (if
there is a symmetry).  Such a property allows us to use topologically ordered
states as a medium for topological quantum memory~\cite{DKL0252} and topological
quantum computations~\cite{K032}. Fractional statistics and fractional charges
also provide us ways to experimentally detect topological orders.

\item Some topological orders have gapless boundary
excitations~\cite{H8285,Wedge,M9020}. Such gapless boundary excitations are
topologically protected, which cannot be \\gapped/localized by any impurities on
the boundary.  Those topologically protected gapless modes lead to perfect
conducting boundary channels even with magnetic impurities~\cite{KDP8094}.  This
property may lead to device applications.

\item Topologically ordered states and their gapless generalization, quantum
ordered states~\cite{W0275}, can produce emergent gauge theory.  Those states
can gives rise to new kind of waves (\ie the gapless collective excitations
above the ground
states)\cite{Wlight,SM0204,Walight,Wqoem,MS0312,HFB0404,LWuni,LWqed,CMS1235}
that satisfy the Maxwell equations or the Yang-Mills equations~\cite{YM5491}.
The new kind of waves can be probed/studied in practical experiments, such as
neutron scattering experiments~\cite{MS0312}.  (For details, see Chapter~\ref{chap11}.)
}

\begin{svgraybox}
\begin{center}
\textbf{Box 6.7 The significance of topological order}

Topological order can produce quasiparticles with fraction quantum numbers and
fractional statistics, robust gapless boundary states, and emergent gauge
excitations.
\end{center}
\end{svgraybox}

In the following, we will study some examples of topological orders and reveal
their amazing topological properties.

\section{Quantum liquids of unoriented strings}

Our first example of topological order is a quantum liquid of qubits, where
qubits organize into unoriented strings.
Quantum liquids of unoriented strings are simplest  topologically ordered
states.
The strings in quantum liquids of unoriented strings can be realized in a
spin-1/2 model.  We can view up-spins as background and lines of down-spins as
the strings (see Fig. \ref{strspin}). Clearly, such string is unoriented.  The
simplest  topologically ordered state in such spin-1/2 system is given by the
equal-weight superposition of all closed strings~\cite{K032}:
$|\Phi_{\mathbb{Z}_2}\>=\sum_\text{all closed strings} \left |\bmm
\includegraphics[height=0.3in]{Chapters/Chap6/strnetS}\emm\right \> $. Such a wave function
represents a global dancing pattern that correspond to a non-trivial
topological order.

As we have mentioned before, the  global dancing pattern is determined by local
dancing rules.  What are those local rules that give rise to the global
dancing pattern $|\Phi_{\mathbb{Z}_2}\>=\sum_\text{all closed strings} \left |\bmm
\includegraphics[height=0.3in]{Chapters/Chap6/strnetS}\emm\right \> $?
The first rule is that, in the
ground state, the down-spins are always connected with no open ends.
To describe the second rule,
we need to introduce
the amplitudes of close strings in the
ground state:
$\Phi\bpm
\includegraphics[height=0.3in]{Chapters/Chap6/strnetS}\epm$.
The ground state is given by
\begin{align}
\sum_\text{all closed strings}
\Phi\bpm
\includegraphics[height=0.3in]{Chapters/Chap6/strnetS}\epm
 \left |\bmm
\includegraphics[height=0.3in]{Chapters/Chap6/strnetS}\emm\right \>.
\end{align}
Then the second rule relates the amplitudes of close strings in the ground state
as we change the strings locally:
\begin{align}
\label{Z2rl}
 \Phi
\bpm \includegraphics[height=0.2in]{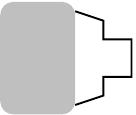} \epm  =&
\Phi
\bpm \includegraphics[height=0.2in]{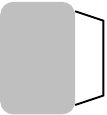} \epm ,
&
 \Phi
\bpm \includegraphics[height=0.2in]{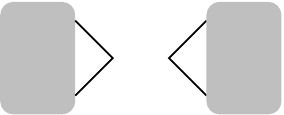} \epm  =&
\Phi
\bpm \includegraphics[height=0.2in]{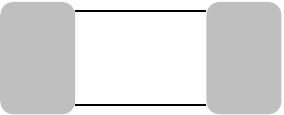} \epm,
\end{align}
In other words, if we locally deform/reconnect the strings as in Fig.
\ref{strnetSa}, the amplitude (or the ground state wave function) does not
change.

The first rule tells us that
the amplitude of a string configuration only depend on the topology
of the string configuration.
Starting from a single loop, using the local deformation and the
local reconnection in Fig. \ref{strnetSa}, we can generate all closed string
configurations with any number of loops.  So all those  closed string
configurations have the same amplitude.  Therefore, the local dancing rule
fixes the wave function to be the equal-weight superposition of all closed
strings: $|\Phi_{\mathbb{Z}_2}\>=\sum_\text{all closed strings} \left |\bmm
\includegraphics[height=0.3in]{Chapters/Chap6/strnetS}\emm\right \> $.
In other words,  the local dancing rule
fixes the global dancing pattern.

If we choose another local dancing rule, then we will get a different global
dancing pattern that corresponds to a different topological order.  One of the
new choices is obtained by just modifying the sign in \eqn{Z2rl}:
\begin{align}
\label{Semrl}
 \Phi
\bpm \includegraphics[height=0.2in]{Chapters/Chap6/Xi1} \epm  =&
\Phi
\bpm \includegraphics[height=0.2in]{Chapters/Chap6/Xi} \epm ,
&
 \Phi
\bpm \includegraphics[height=0.2in]{Chapters/Chap6/XijklX} \epm  =&
- \Phi
\bpm \includegraphics[height=0.2in]{Chapters/Chap6/XijX} \epm  .
\end{align}
We note that each local reconnection operation changes the number of loops by
1.  Thus the new local dancing rules gives rise to a wave function which has a
form $|\Phi_\text{Sem}\>=\sum_\text{all closed strings} (-)^{N_\text{loops}}
\left |\bmm \includegraphics[height=0.3in]{Chapters/Chap6/strnetS}\emm\right \> $, where
$N_\text{loops}$ is the number of loops.  The wave function
$|\Phi_\text{Sem}\>$ corresponds to a different global dance and a different
topological order.

\begin{figure}[tb]
\centerline{
\includegraphics[height=1.2in]{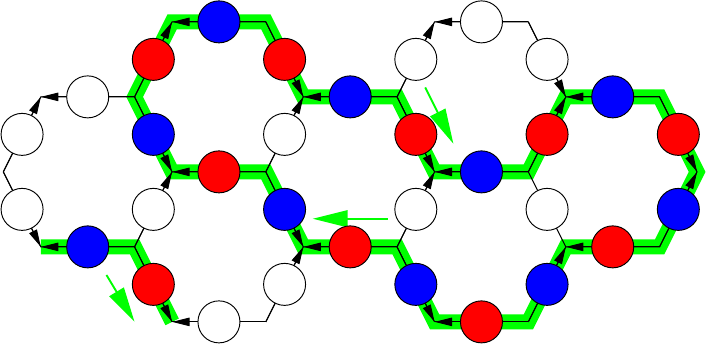}
}
\caption{
The oriented strings in a spin-1 model.  In the background of $S_z=0$ spins
(the white dots), the $S_z=1$ spins (the red dots) and the $S_z=-1$ spins (the
blue dots) form closed strings.
}
\label{strspinO}
\end{figure}

In the above, we constructed two quantum liquids of unoriented strings in
a spin-1/2 model.  Using a similar construction, we can also obtain a quantum
liquid of oriented strings which gives rise to waves satisfying Maxwell
equation as discussed before (see Chapter 11).  To obtain  quantum liquid of
oriented strings, we need to start with a spin-1 model, where spins live on
the links of honeycomb lattice (see Fig. \ref{strspinO}).  Since the honeycomb
lattice is bipartite, each link has an orientation from the A-sublattice to the
B-sublattice (see Fig. \ref{strspinO}).  The oriented strings is formed by
alternating  $S_z=\pm 1$ spins on the background of $S_z=0$ spins.  The string
orientation is given be the orientation of the links under the $S_z=1$ spins
(see Fig. \ref{strspinO}).  The superposition of the oriented strings gives
rise to quantum liquid of oriented strings.

\section{The emergence of fractional quantum numbers and
Fermi/fractional statistics}

Why the two  wave functions of unoriented strings, $|\Phi_{\mathbb{Z}_2}\>$ and
$|\Phi_\text{Sem}\>$, have non-trivial topological orders?  This is because
the two  wave functions give rise to non-trivial topological properties.  The
two  wave functions correspond to different topological orders since they give
rise to different topological properties.  In this section, we will discuss two
topological properties: emergence of fractional statistics and topological
degeneracy on compact spaces.

\subsection{Emergence of fractional angular momenta}

The two topological states in two dimensions contain only closed strings, which
represent the ground states.  If the wave functions contain open strings (\ie
have non-zero amplitudes for open string states), then the ends of the open
strings will correspond to point-like topological excitations above the ground
states.  Although an open string is an extended object, its middle part merge
with the strings already in the ground states and is unobservable.  Only its
two ends carry energies and correspond to two point-like particles.

We note that such a point-like particle from an end of string cannot be created
alone.  Thus an end of string correspond to a topological point defect, which
may carry fractional quantum numbers.  This is because an open string as a
whole always carry non-fractionalized quantum numbers.  But an open string
corresponds to \emph{two} topological point defects from the two ends.  So we
cannot say that each end of string carries non-fractionalized quantum numbers.
Some times, they do carry fractionalized quantum numbers.

Let us first consider the defects in the $|\Phi_{\mathbb{Z}_2}\>$ state.  To understand
the fractionalization, let us first consider the spin of such a defect to see
if the spin is fractionalized or not~\cite{FFN0683,Wang10}. An end of string
can be represented by
\begin{align}
\big |\bmm \includegraphics[scale=0.33]{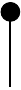}\emm \big \>_\text{def}
=
\big |\bmm \includegraphics[scale=0.33]{Chapters/Chap6/def1}\emm \big \>+
\big |\bmm \includegraphics[scale=0.33]{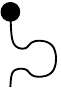}\emm \big \>+
\big |\bmm \includegraphics[scale=0.33]{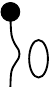}\emm \big \>+ ...
.
\end{align}
which is an equal-weight superposition of all
string states obtained from the
deformations and the reconnections of $\bmm \includegraphics[scale=0.33]{Chapters/Chap6/def1}\emm$.

Under a $360^\circ$ rotation, the end of string is changed to $\big |\bmm
\includegraphics[scale=0.33]{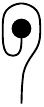}\emm \big \>_\text{def} $, which is an equal
weight superposition of all string states obtained from the
deformations and the reconnections of $\bmm \includegraphics[scale=0.33]{Chapters/Chap6/def3}\emm$.
Since $\big |\bmm \includegraphics[scale=0.33]{Chapters/Chap6/def1}\emm \big \>_\text{def} $
and $\big |\bmm \includegraphics[scale=0.33]{Chapters/Chap6/def3}\emm \big \>_\text{def} $ are
alway different, $\big |\bmm \includegraphics[scale=0.33]{Chapters/Chap6/def1}\emm \big
\>_\text{def} $ is not an eigenstate of $360^\circ$ rotation and does not carry
a definite spin.

To construct the  eigenstates of $360^\circ$ rotation, let us make a
$360^\circ$ rotation to $\big |\bmm \includegraphics[scale=0.33]{Chapters/Chap6/def3}\emm \big
\>_\text{def}$.  To do that, we first use the string reconnection move in Fig.
\ref{strnetSa}, to show that $\big |\bmm \includegraphics[scale=0.33]{Chapters/Chap6/def3}\emm
\big \>_\text{def} = \big |\bmm \includegraphics[scale=0.33]{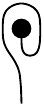}\emm \big \>_\text{def} $.  A
$360^\circ$ rotation on $\big |\bmm \includegraphics[scale=0.33]{Chapters/Chap6/def2}\emm \big
\>_\text{def} $ gives us $\big |\bmm \includegraphics[scale=0.33]{Chapters/Chap6/def1}\emm \big \>_\text{def} $.

We see that the $360^\circ$ rotation
exchanges $\big |\bmm \includegraphics[scale=0.33]{Chapters/Chap6/def1}\emm \big \>_\text{def} $
and $\big |\bmm \includegraphics[scale=0.33]{Chapters/Chap6/def3}\emm \big \>_\text{def} $.
Thus the  eigenstates of
$360^\circ$ rotation are given by
$\big |\bmm \includegraphics[scale=0.33]{Chapters/Chap6/def1}\emm \big \>_\text{def} + \big |\bmm
\includegraphics[scale=0.33]{Chapters/Chap6/def3}\emm \big \>_\text{def} $ with
eigenvalue 1, and by $\big |\bmm
\includegraphics[scale=0.33]{Chapters/Chap6/def1}\emm \big \>_\text{def} - \big |\bmm
\includegraphics[scale=0.33]{Chapters/Chap6/def3}\emm \big \>_\text{def} $ with eigenvalue $-1$.
So the particle $\big |\bmm \includegraphics[scale=0.33]{Chapters/Chap6/def1}\emm \big \>_\text{def} +
\big |\bmm \includegraphics[scale=0.33]{Chapters/Chap6/def3}\emm \big \>_\text{def} $ has a spin 0 (mod
1), and the particle $\big |\bmm \includegraphics[scale=0.33]{Chapters/Chap6/def1}\emm \big \>_\text{def}
- \big |\bmm \includegraphics[scale=0.33]{Chapters/Chap6/def3}\emm \big \>_\text{def} $ has a spin 1/2
(mod 1).

\subsection{Emergence of Fermi and fractional statistics}

\begin{figure}[tb]
\centerline{
\includegraphics[scale=0.35]{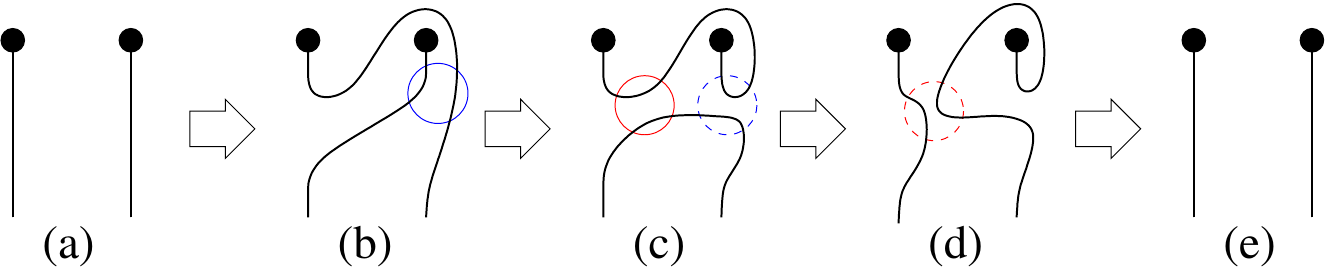}
}
\caption{
Deformation of strings and two reconnection moves, plus an exchange of two ends
of strings and a $360^\circ$ rotation of one of the end of string, change the
configuration (a) back to itself.  Note that from (a) to (b) we exchange the
two ends of strings, and from (d) to (e) we rotate of one of the end of string
by $360^\circ$.
The combination of those moves do not generate any phase.
}
\label{exch}
\end{figure}

If one believes in the spin-statistics theorem, one may guess that the particle
$\big |\bmm \includegraphics[scale=0.33]{Chapters/Chap6/def1}\emm \big \>_\text{def} + \big |\bmm
\includegraphics[scale=0.33]{Chapters/Chap6/def3}\emm \big \>_\text{def} $ is a boson and the particle
$\big |\bmm \includegraphics[scale=0.33]{Chapters/Chap6/def1}\emm \big \>_\text{def} - \big |\bmm
\includegraphics[scale=0.33]{Chapters/Chap6/def3}\emm \big \>_\text{def} $ is a fermion.  This guess is
indeed correct.  Form Fig. \ref{exch}, we see that we can use deformation of
strings and two reconnection moves to generate an exchange of two ends of
strings and a $360^\circ$ rotation of one of the end of string.  Such
operations allow us to show that Fig. \ref{exch}a and  Fig. \ref{exch}e have
the same amplitude, which means that an exchange of two ends of strings
followed by a $360^\circ$ rotation of one of the end of string do not generate
any phase.  This is nothing but the spin-statistics theorem.

The emergence of Fermi statistics in the $|\Phi_{\mathbb{Z}_2}\>$
state of a purely bosonic spin-1/2 model
indicates that the state is a topologically ordered state.  We also see
that the $|\Phi_{\mathbb{Z}_2}\>$ state has a bosonic quasi-particle $\big |\bmm
\includegraphics[scale=0.33]{Chapters/Chap6/def1}\emm \big \>_\text{def} + \big |\bmm
\includegraphics[scale=0.33]{Chapters/Chap6/def3}\emm \big \>_\text{def} $, and a fermionic quasi-particle
$\big |\bmm \includegraphics[scale=0.33]{Chapters/Chap6/def1}\emm \big \>_\text{def} - \big |\bmm
\includegraphics[scale=0.33]{Chapters/Chap6/def3}\emm \big \>_\text{def} $.  The bound state of the above
two particles is a boson (not a fermion) due to their  mutual semion
statistics.  Such quasi-particle content agrees exactly with the $\mathbb{Z}_2$ gauge
theory which also has three type of non-trivial quasiparticles excitations, two
bosons and one fermion.  In fact, the low energy effective theory of the
topologically ordered state $|\Phi_{\mathbb{Z}_2}\>$ is the $\mathbb{Z}_2$ gauge theory and we
will call  $|\Phi_{\mathbb{Z}_2}\>$ a $\mathbb{Z}_2$ topologically ordered state.

Next, let us consider the defects in the $|\Phi_\text{Sem}\>$ state.
Now
\begin{align}
\big |\bmm \includegraphics[scale=0.33]{Chapters/Chap6/def1}\emm \big \>_\text{def}
=
\big |\bmm \includegraphics[scale=0.33]{Chapters/Chap6/def1}\emm \big \>+
\big |\bmm \includegraphics[scale=0.33]{Chapters/Chap6/def1a}\emm \big \>-
\big |\bmm \includegraphics[scale=0.33]{Chapters/Chap6/def1b}\emm \big \>+ ...
.
\end{align}
and a similar expression for $\big |\bmm \includegraphics[scale=0.33]{Chapters/Chap6/def3}\emm
\big \>_\text{def}$, due to a
change of the local dancing rule
for reconnecting the strings (see \eqn{Semrl}).
Using the string reconnection move in Fig.
\ref{strnetSa}, we find that $\big |\bmm \includegraphics[scale=0.33]{Chapters/Chap6/def3}\emm
\big \>_\text{def} = - \big |\bmm \includegraphics[scale=0.33]{Chapters/Chap6/def2}\emm \big
\>_\text{def} $.  So a $360^\circ$ rotation, changes $(\big |\bmm
\includegraphics[scale=0.33]{Chapters/Chap6/def1}\emm \big \>_\text{def}, \big |\bmm
\includegraphics[scale=0.33]{Chapters/Chap6/def3}\emm \big \>_\text{def} )$ to $( \big |\bmm
\includegraphics[scale=0.33]{Chapters/Chap6/def3}\emm \big \>_\text{def}, -\big |\bmm
\includegraphics[scale=0.33]{Chapters/Chap6/def1}\emm \big \>_\text{def} )$.  We find that
$\big |\bmm \includegraphics[scale=0.33]{Chapters/Chap6/def1}\emm \big \>_\text{def} + \ii
\big |\bmm \includegraphics[scale=0.33]{Chapters/Chap6/def3}\emm \big \>_\text{def} $ is the
eigenstate of the $360^\circ$ rotation with eigenvalue $-\ii$, and $\big
|\bmm \includegraphics[scale=0.33]{Chapters/Chap6/def1}\emm \big \>_\text{def} - \ii \big
|\bmm \includegraphics[scale=0.33]{Chapters/Chap6/def3}\emm \big \>_\text{def} $ is the other
eigenstate of the $360^\circ$ rotation with eigenvalue $\ii$.  So the
particle $\big |\bmm \includegraphics[scale=0.33]{Chapters/Chap6/def1}\emm \big \>_\text{def}
+ \ii \big |\bmm \includegraphics[scale=0.33]{Chapters/Chap6/def3}\emm \big \>_\text{def} $
has a spin $-1/4$, and the particle $\big |\bmm
\includegraphics[scale=0.33]{Chapters/Chap6/def1}\emm \big \>_\text{def} - \ii \big |\bmm
\includegraphics[scale=0.33]{Chapters/Chap6/def3}\emm \big \>_\text{def} $ has a spin $1/4$.
The spin-statistics theorem is still valid for $|\Phi_\text{Sem}\>_\text{def}$
state, as one can see form Fig. \ref{exch}.  So, the particle $\big |\bmm
\includegraphics[scale=0.33]{Chapters/Chap6/def1}\emm \big \>_\text{def} + \ii\big |\bmm
\includegraphics[scale=0.33]{Chapters/Chap6/def3}\emm \big \>_\text{def} $ and particle $\big
|\bmm \includegraphics[scale=0.33]{Chapters/Chap6/def1}\emm \big \>_\text{def} - \ii\big
|\bmm \includegraphics[scale=0.33]{Chapters/Chap6/def3}\emm \big \>_\text{def} $ have
fractional statistics with statistical angles of semion: $\pm \pi/2$.  Thus the
$|\Phi_\text{Sem}\>$ state contains a non-trivial topological order.  We will
call such a topological order a double-semion topological order.

It is amazing to see that the long-range quantum entanglement in string liquid
can gives rise to fractional spin and fractional statistics, even from a purely
bosonic model.  Fractional spin and Fermi statistics are two of most mysterious
phenomena in natural.  Now, we can understand them as merely a phenomenon of
long-range quantum entanglement.  They are no longer mysterious.

\begin{svgraybox}
\begin{center}
\textbf{Box 6.8 Fractional quantum numbers and fractional statistics}

Fractional quantum numbers and fractional statistics can be determined from the
global dancing pattern (\ie the pattern of long-range entanglement) in the
ground state.
\end{center}
\end{svgraybox}

\section{Topological degeneracy of unoriented string liquid}
\label{sec:deg}

The $\mathbb{Z}_2$ and the  double-semion topological states (as well as many other
topological states) have another important topological property: topological
degeneracy~\cite{Wtop,Wrig}.  Topological degeneracy is the ground state
degeneracy of a gapped many-body system that is robust against any local
perturbations as long as the system size is large.
We like to make a few remarks.
\enu{
\item Topological degeneracy can be used as protected qubits which allows us to
perform topological quantum computation~\cite{K032}.
\item It is believed that the appearance of topological degeneracy implies the
topological order (or long-range entanglement) in the ground
state~\cite{Wtop,Wrig}.
\item Many-body states with topological degeneracy are described by topological
quantum field theory at low energies~\cite{W8951}.
}

The simplest topological  degeneracy appears when we put topologically ordered
states on compact spaces with no boundary.  We can use the global dancing
pattern to understand the  topological  degeneracy.  We know that the local
dancing rules determine the global dancing pattern.  On a sphere, the  local
dancing rules determine a unique global dancing pattern.  So the ground state
is non-degenerate.  However on other  compact spaces, there can be several
global dancing patterns that all satisfy the  local dancing rules. In this
case, the ground state is degenerate.

\begin{figure}[tb]
\centerline{
\includegraphics[height=1.4in]{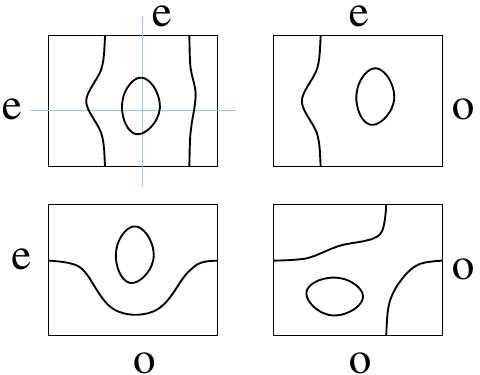}
}
\caption{
 On a torus, the closed string configurations
can be divided into four sectors, depending on even or
odd number of strings crossing the x- or y-axises.
}
\label{z2eo}
\end{figure}

For the $\mathbb{Z}_2$ topological state on torus, the local dancing rule relate the
amplitudes of the string configurations that differ by a string reconnection
operation in Fig. \ref{strnetSa}.  On a torus, the closed string configurations
can be divided into four sectors (see Fig. \ref{z2eo}), depending on even or
odd number of strings crossing the x- or y-axises.  The string reconnection
move only connect the string configurations within each sector.  So the
superposition of the string configurations in each sector represents a
different global dancing pattern.  Most importantly, we cannot distinguish the
four global dancing patterns locally by examine a local region of the system,
since they all follow the same local dancing rule.  As a result, the four
global dancing patterns degenerate ground states.  Therefore, the local dancing
rule for the  $\mathbb{Z}_2$ topological order gives rise to four fold degenerate ground
state on torus~\cite{Wsrvb}. Similarly,  the double-semion topological order
also gives rise to four fold degenerate ground state on torus.

\begin{svgraybox}
\begin{center}
\textbf{Box 6.9 Topological degeneracy}

The topological degeneracy is determined from the global dancing pattern (\ie
the pattern of long-range entanglement) in the ground state.
\end{center}
\end{svgraybox}

\section{Topological excitations and string operators}

In the last a few sections, we have used simple intuitive pictures to explain
several important properties of topologically ordered states.  We stress that
those topological properties are results of long-range entanglement in the
ground state. In this section, we will use more rigorous approach
to obtain those  topological properties

\subsection{Toric code model and string condensation}

\begin{figure}[tb]
\centerline{
\includegraphics[height=1.5in]{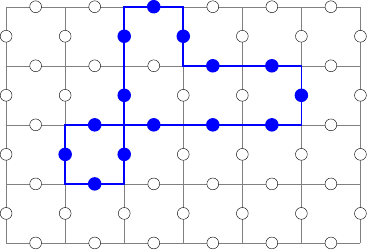}
}
\caption{
The toric code model with spin-1/2 spins on the links.  A light dot represents a
up-spin and a dark dot a down-spin.
A closed string state is shown.
}
\label{z2Ham}
\end{figure}

The $\mathbb{Z}_2$ topological order from the condensation of unoriented strings can be
realized by the toric code model (see Chapter~\ref{sec:cp3sec5})~\cite{K032},
which is formed by spin-1/2 spins on links of square lattice.  The Hamiltonian
is given by
\begin{align}
\label{HZ2}
H_{\mathbb{Z}_2}&=-U\sum_{\v s}Q_{\v s}
-g \sum_{\v p}B_{\v p}
,\ \ \ \ \
B_{\v p} \equiv
\prod_{j\in \text{plaquette}(p)} X_{\v j},\ \ \
Q_{\v s} \equiv
\prod_{j\in \text{star}(s)} Z_{\v j} .
\end{align}
Here $\v p$ labels the plaquettes and $\prod_{j\in \text{plaquette}(p)} X_{\v j}$ is
the product of the four Pauli operators $X_{\v j}$ on the four edges of the
plaquette $\v p$.  $\v s$ labels the vertices and $\prod_{j\in \text{star}(s)}
Z_{\v j}$ is the product of the four Pauli operators $Z_{\v j}$ on the four
legs of the vertex $\v s$.

If we view an up-spin as a state with no string and string as line of
down-spins, we find that the $U$-terms enforce the first dancing rule to make
spins to form closed strings in the ground state.  Since  $\prod_{j\in \text{star}(s)} Z_{\v j}=1$ for all closed string states (including the no string
state), all closed string states have the same low energy.  Every end of open
string will cost an energy $+2U$.

\begin{figure}[tb]
\centerline{
 \includegraphics[height=1.2in]{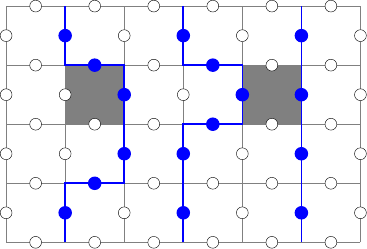}~~~~~~~~~~~~
  \includegraphics[height=1.2in]{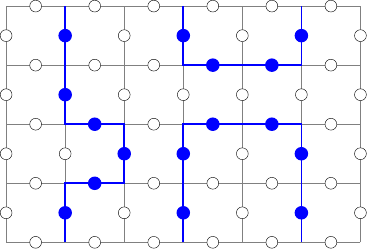}
}
\caption{ Applying the $\prod_{j\in \text{plaquette}(p)} X_{\v j}$ operator to the
shaded squares will change the shape of the string, or reconnect the strings.
}
\label{z2sec}
\end{figure}

If we only have the $U$-terms, the ground states will be highly degenerate
which include all the closed string states. The $g$-term enforce the second
dancing rule where only a particular `dance' (\ie superposition) of closed
string corresponds to the ground state.  We note that the operator
$\prod_{j\in \text{plaquette}(p)} X_{\v j}$ creates/annhilates a small loop of
string around a square.  So the operator will change the shape of the strings
or reconnect the strings (see Fig.  \ref{z2sec}). Due to the minus sign in the
$g$-term, The change of the shapes and the reconnection of the strings will not
change the amplitude.  Thus, the different shapes and different connections of
closed strings in the ground state will have the same amplitude.
Therefore, the ground state of $H_{\mathbb{Z}_2}$ satisfies the two dancing rules and
describe the $\mathbb{Z}_2$ topological order.

In fact since $[Q_{\v s}, B_{\v p}]=0$, $H_{\mathbb{Z}_2}$ is exactly soluble.  The
exact eigenstates of $H_{\mathbb{Z}_2}$ are the common eigenstates of $Q_{\v s}$ and
$B_{\v p}$ where the eigenvalues of $Q_{\v s}$ is $q_{\v s}=\pm 1$ and the
eigenvalues of $B_{\v p}$ is $b_{\v p}=\pm 1$.  The energy of an eigenstate is
given by $-U\sum_{\v s} q_{\v s} -g\sum_{\v p} b_{\v p}$.  The ground state is
given by $|\Psi_\text{grnd}\>=|q_{\v s} =b_{\v p}=1\>$. We can show that the
state $|q_{\v s} =b_{\v p}=1\>$ is an equal weight superposition of all closed
string states: $|q_{\v s} =b_{\v p}=1\> =\sum|\text{all closed-strings}\>$.

Using the arguments in Section \ref{sec:deg}, we see that the ground states of
$H_{\mathbb{Z}_2}$ have a four-fold degeneracy on torus.  The  four-fold degeneracy can
also be understood through the following argument.  We note that there are
operator identities $\prod_{\v s} Q_{\v s}=1$ and $\prod_{\v p} F_{\v p}=1$ if
the square lattice form a torus.  Therefore, the number of independent quantum
numbers $b_{\v p}=\pm 1,\ q_{\v s}=\pm 1$ on torus is $2^N_\text{site}
2^N_\text{site}/4$ where $N_\text{site}$ is the number of sites.  The number of
states on torus $2^N_\text{site} 2^N_\text{site}$.  So the number of
independent labels $b_{\v p}=\pm 1,\ q_{\v s}=\pm 1$ is 1/4 of the number of
states. Each label correspond to 4 states.  Since, the energy is a function of
$b_{\v p}, q_{\v s}$. The degeneracy of each energy eigenvalue (including the
ground states) is 4.  Such a four-fold degeneracy is a topological degeneracy,
which implies that the  ground states of $H_{\mathbb{Z}_2}$ have a nontrivial
topological order.

Next, we will discuss the quasiparticle excitations in the $\mathbb{Z}_2$ topologically
ordered state described by $H_{\mathbb{Z}_2}$.  In particular, we will discuss their
nontrivial statistics.  But before we do that, we would like to have a general
discussion of topological excitations.  Only topological excitations can have
nontrivial statistics and fractional quantum numbers.

\subsection{Local and topological excitations}

\begin{figure}[tb]
  \centering
  \includegraphics[scale=0.5]{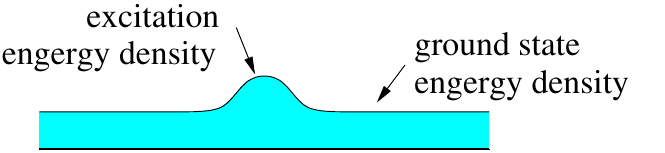}
  \caption{The energy density distribution of a particle-like excitation.}
  \label{exceng}
\end{figure}

Topological orders (or patterns of long range entanglement) can be
characterized by the appearance of the `topological excitations'.  In this
section, we will discuss/define the notion of topological excitations.

First we define the notion of `particle-like' excitations.  Consider a system
with translation symmetry.  The ground state has a uniform energy density.  If
we have a state with an excitation, we can measure the energy distribution of
the state over the space.  If for a local area, the energy density is higher
than ground state, while for the rest area the energy density is the same as
ground state, one may say that there is a `particle-like' excitation, or a
quasiparticle, in this area (see Fig.  \ref{exceng}).  Quasiparticles defined
like this can be further divided into two classes.  The first class can be created
or annihilated by local operators, such as a spin flip.  So the first class of
the particle-like excitations is called local quasiparticle excitations.  The
second class cannot be  created or annihilated by any finite number of local
operators (in the infinite system size limit).  In other words, the higher
local energy density cannot be created or removed by \emph{any} local operators
in that area.  The second class of the particle-like excitations is called
topological quasiparticle excitations.

From the notions of local quasiparticles and topological quasiparticles, we can
also introduce a notion of topological quasiparticle types, or simply,
quasiparticle types.  We say that local quasiparticles belong to the trivial
type, while topological quasiparticles belong to nontrivial types.  Also two
topological quasiparticles are of the same type if and only if they differ by
local quasiparticles.  In other words, we can turn one topological
quasiparticle into the other one of the same type by applying some local
operators.

\begin{svgraybox}
\begin{center}
\textbf{Box 6.10 Topological excitation}

A topological excitation is a particle-like excitation with localized energy,
that cannot be created/annihlate by any local operators near the excitation.
\end{center}
\end{svgraybox}

The $\mathbb{Z}_2$ topologically ordered state described by $H_{\mathbb{Z}_2}$ have nontrivial
topological excitations.  In fact, it has three types of nontrivial topological
excitations. In the following, we will discuss those topological excitations.

\subsection{Three types of quasiparticles}

The first type of  topological excitations, denoted as $e$, corresponds to ends
of strings which we have discussed before.  In the $\mathbb{Z}_2$ model $H_{\mathbb{Z}_2}$, the
ground state is described by $|\Psi_\text{grnd}\>=|q_{\v s} =b_{\v p}=1\>$. If we change one $q_{\v
s}$ from 1 to $-1$, we will create a  topological excitation of the first type.
We see that, to create a  topological excitation of the first type, we break
the first dancing rule -- the closed string condition.

In contrast, to create a  topological excitation of the second type, denoted as
$m$, we keep the  first dancing rule, but break the second dancing rule -- the
equal amplitude condition.  If there is a  topological excitation of the second
type at $\v x$, it wave function given by
\begin{align}
 \Phi(X_\text{open})=0,\ \ \ \
 \Phi(X_\text{closed})=(-)^{W_{\v x}(X_\text{closed})},
\end{align}
where $X_\text{open}$ represents string configurations with open ends, and
$X_\text{closed}$ represents closed string configurations.  Here $W_{\v
x}(X_\text{closed})$ is the number of times that the closed strings wind around
$\v x$.  In the $\mathbb{Z}_2$ model $H_{\mathbb{Z}_2}$, if we change one $b_{\v p}$ from 1 to
$-1$, we will create a  topological excitation of the second type.

The third type of  topological excitations, denoted as $\eps$, corresponds to
the bond states of one $e$ and one $m$.  The above three nontrivial topological
excitations plus the trivial one are the four types of  topological excitations
in the $\mathbb{Z}_2$ topologically ordered state.

\subsection{Three types of string operators}

\begin{figure}[tb]
  \centering
  \includegraphics[scale=1.0]{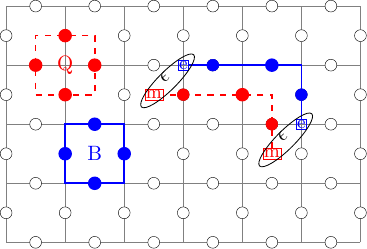}
  \caption{The black lines are the type-I strings and the shaded lines are the
type-II strings.  Here, the open type-I string and the open type-II string are
related by a displacement $(1/2,1/2)$.  A type-I string operator is a product
of $X_i$'s on a type-I string.  A type-II string operator is a product of
$Z_i$'s on a type-II string.
}
  \label{z2opstr}
\end{figure}

As we have stressed that, although the excitations $e$, $m$, and $\eps$ have
local energy distributions, they cannot be created by local operators.
However, we can create a pair of $e$ via a non-local string operator.
Similarly, we can also create a pair of $m$ or $\eps$ via other non-local
string operators.

First let us introduce a notion of type-I string.  A  type-I string is a string
formed by the links of the square lattice which connects the vertices of the
square lattice (see Fig. \ref{z2opstr}).  A type-I string operator $
W_\text{type-I}$ is a product of $X_i$'s on a type-I string  (see Fig.
\ref{z2opstr}):
\begin{align}
 W_\text{type-I}=\prod_{i\in \text{type-I string}} X_i .
\end{align}
A  type-I string operator creates an open string, and creates two $e$'s at its
two ends (see Fig. \ref{z2opstr}).

A  type-II string is a string formed by the lines that connects the squares of
the square lattice (see Fig. \ref{z2opstr}).  A type-II string operator $
W_\text{type-II}$ is a product of $Z_i$'s on a type-II string  (see Fig.
\ref{z2opstr}):
\begin{align}
 W_\text{type-II}=\prod_{i\in \text{type-II string}} Z_i .
\end{align}
A  type-II string operator creates two $m$'s at its two ends (see Fig.
\ref{z2opstr}).  This is because the type-II string operator anti-commutes with
the two $B_{\v p}$ operators at its two ends and commute with other $B_{\v p}$
operators.  So a type-II string operator flips the sign of $b_{\v p}$ at its
two ends and hence creates to  $m$ excitations.
It is interesting to note that $B_{\v p}$ is
a small loop of  type-I string operator and $Q_{\v I}$ is a small loop of
type-II string operator (see Fig.  \ref{z2opstr}).

A type-III string operator $ W_\text{type-III}$ is a product of a type-I string
operator $W_\text{type-I}$ and a type-II string operator $ W_\text{type-II}$:
\begin{align}
 W_\text{type-III}= \prod_{i\in \text{type-I string}} X_i
\prod_{i\in \text{type-II string}} Z_i
,
\end{align}
where the type-II string is obtained by displacing the type-I string by
$(1/2,1/2)$ (see Fig.  \ref{z2opstr}).  A  type-III string operator creates two
$\eps$'s at its two ends (see Fig.  \ref{z2opstr}).

\begin{svgraybox}
\begin{center}
\textbf{Box 6.11 String operator and topological excitation}

A pair of topological excitation can be created by an open string operator at
the two ends of the open string.
\end{center}
\end{svgraybox}

Although the string operator
is non-local, it creates a two point-like energy distribution (\ie two
quasiparticles) at its two ends. In other words, the closed string operators
without ends commute with the Hamiltonian and leave the ground state
unchanged:
\begin{align}
&
 [ W^\text{closed}_\text{type-I},H_{\mathbb{Z}_2}]=
 [ W^\text{closed}_\text{type-II},H_{\mathbb{Z}_2}]=
 [ W^\text{closed}_\text{type-III},H_{\mathbb{Z}_2}]=0,
\nonumber\\
&
W^\text{closed}_\text{type-I} |\Psi_\text{grnd}\>=
W^\text{closed}_\text{type-II} |\Psi_\text{grnd}\>=
W^\text{closed}_\text{type-III} |\Psi_\text{grnd}\>=
|\Psi_\text{grnd}\> .
\end{align}

We also note that, on a torus, a closed string operator (such as
$W^\text{closed}_\text{type-I}$) can map a degenerate ground state of $H_{\mathbb{Z}_2}$
to another ground state if the closed string operator winds all the way around
the torus. This is because $W^\text{closed}_\text{type-I}$ can change the
number of closed strings going around the torus by an odd number.  Since the
closed string operator that winds all the way around the torus contains $L$
local operators where $L$ is the linear size of the system, therefore, we can
use a product of $L$ local operators to mix the different degenerate ground
states.  But we cannot use a product of any finite numbers of local operators
to mix the different degenerate ground states in the $L\to \infty$ limit.  So
the `code distance' (see Section \ref{sec:QCC}) for the  degenerate ground
states is large (of order $L$).  This large `code distance' is why the ground
state degeneracy is robust against any local perturbations, since a local
perturbation always contains a finite number of local operators which cannot mix
the  degenerate ground states.

\begin{figure}[tb]
  \centering
  \includegraphics[scale=1.0]{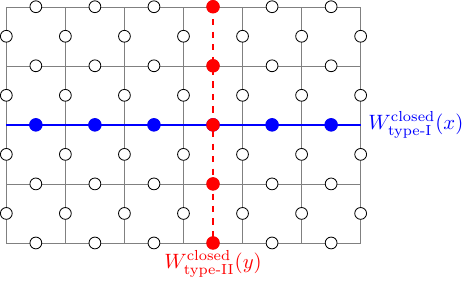}
  \caption{The type-I and type-II closed string operators.
}
  \label{strtorus}
\end{figure}

Let $W^\text{closed}_\text{type-I;x}$ be the type-I closed string operator
that winds around the torus once in $x$-direction.  Let
$W^\text{closed}_\text{type-I;y}$ be the type-I closed string operator that
winds around the torus once in $y$-direction.  Similarly, we can also define
$W^\text{closed}_\text{type-II;x}$ and $W^\text{closed}_\text{type-II;y}$.
We find that
\begin{align}
\label{clsdalg}
 W^\text{closed}_\text{type-I;x} W^\text{closed}_\text{type-II;y}
&= - W^\text{closed}_\text{type-II;y} W^\text{closed}_\text{type-I;x},
\nonumber\\
 W^\text{closed}_\text{type-I;y} W^\text{closed}_\text{type-II;x}
&= - W^\text{closed}_\text{type-II;x} W^\text{closed}_\text{type-I;y}.
\end{align}
So the closed string operators form two independent algebra $\hat A\hat B=-\hat
B\hat A$.  Since the algebra $\hat A\hat B=-\hat B\hat A$ has only one
two-dimensional irreducible representation, the algebra of the closed string
operators \eqn{clsdalg} has only  one four-dimensional irreducible
representation.  Since all closed string operators commute with $H_{\mathbb{Z}_2}$, all
the eigenvalues of $H_{\mathbb{Z}_2}$ are four-fold degenerate.

\subsection{Statistics of ends of strings}

\begin{figure}[tb]
  \centering
  \includegraphics[scale=0.8]{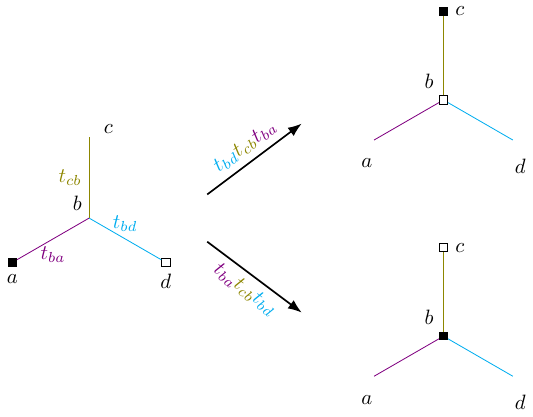}
  \caption{There are two ways to move a two-particle state (with the two
particles at site-$a$ and site-$d$) to another two-particle state (with the two
particles at site-$b$ and site-$c$). The ways differ by an exchange of the two
particles.
}
  \label{hopalg}
\end{figure}

\begin{figure}[tb]
  \centerline{
  \includegraphics[scale=1.0]{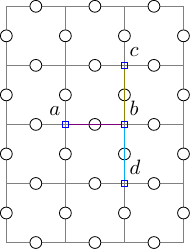}~~~~~~~~~~~~~~~~~~
  \includegraphics[scale=1.0]{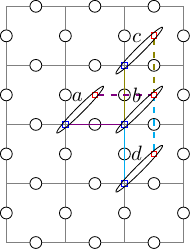}
  }
  \centerline{
	(a) ~~~~~~~~~~~~~~~~~~~~~~~~~~~~~~~~~~~~~~~~~~~~~~~~~~~~~~~~~~
	(b)
}
  \caption{
(a) Type-$e$ particles hoping among the four sites $a$, $b$, $c$, $d$.
(b) Type-$\eps$ particles hoping among the four sites $a$, $b$, $c$, $d$.
}
  \label{z2hopalg}
\end{figure}

We have seen that a pair of topological excitations $e$ can be
created by a type-I open string operator $W_\text{type-I}$.  Since
$W_\text{type-I}$ is a product of bosonic spin operators, It can only create
bosonic excitations. So a pair of $e$ must be bosonic.  But what is the
statistics of a single $e$.  Since the bound state of two $e$'s is a boson, the
statistics of a single $e$ can be bosonic, fermionic, or semionic.  To go
further, we need a new way to calculate the statistics of a single $e$.

To obtain a new way  to calculate the statistics of a particle-like excitation,
we note that the statistics of a particle is determined by hopping operators of
the particle~\cite{LW0316}. Let $|i \cdots \>$ be a state with the particle at
site $i$, where $\cdots$ describes the location of other particles.  The
hopping operator $\hat t_{ji}$ moves the particle at site-$i$ to site-$j$:
\begin{align}
 |j \cdots\> = \hat t_{ji} |i \cdots\>.
\end{align}
From Fig. \ref{hopalg}, we see that, starting from a two-particle state with
the two particle at site-$a$ and site-$d$, there are two ways to move the two
particles to site-$b$ and site-$c$.  The two ways of hopping differ by an
exchange of the two particles.  Therefore, the statistics of the particle can
be determined by the algebra of the hopping operators. If the hopping operator
satisfies
\begin{align}
\hat t_{bd}
\hat t_{cb}
\hat t_{ba}
=
\e^{\ii \th}
\hat t_{ba}
\hat t_{cb}
\hat t_{bd},
\end{align}
then the statistics of the particle is given by
$\e^{\ii \th}$.

So, to calculate the statistics of $e$, we need to know the algebra of the
hopping operator for $e$.  This can be easily done since the hopping operator
$\hat t_{ji}$ for $e$ is nothing but the type-I open string operator
$W_\text{type-I}$ that connect the site-$i$ and site-$j$.  Therefore, the
algebra of the open string operator determine the statistics of the string
ends.

For type-$e$ particles, hoping among the four sites
$a$, $b$, $c$, $d$ in Fig. \ref{z2hopalg}a,
their hopping operators are given by
$\hat t_{ba} = \si^x_1 $,
$
\hat t_{cb}
= \si^x_2
$,
$
\hat t_{bd}
= \si^x_3
$.
We note that the sites for the type-$e$ particles are vertices of the lattice.
We find
$
\hat t_{bd}
\hat t_{cb}
\hat t_{ba}
=
\hat t_{ba}
\hat t_{cb}
\hat t_{bd}
$. Thus the type-$e$ particles (\ie the ends of type-I string) are a boson.`

Similarly, we can calculate the statistics of type-$\eps$
particles.
For type-$\eps$ particles, hoping among the four sites
$a$, $b$, $c$, $d$ in Fig. \ref{z2hopalg}b,
their hopping operators are given by
$
\hat t_{ba}
=  \si^x_1\si^z_2
$,
$
\hat t_{cb}
= \si^x_2 \si^z_4
$,
$
\hat t_{bd}
= \si^x_3 \si^z_5
$.
We note that the sites for the type-$e$ particles are
represented by the ellipses in Fig. \ref{z2hopalg}b.
We find
$
\hat t_{bd}
\hat t_{cb}
\hat t_{ba}
=
-
\hat t_{ba}
\hat t_{cb}
\hat t_{bd} .
$
The type-$\eps$ particles (\ie the ends of type-III strings) are fermions.

Using the same method, we can show that the type-$m$ particles are bosons.  We
note that the sites for the type-$m$ particles are at the center of squares.

\begin{svgraybox}
\begin{center}
\textbf{Box 6.12 Hopping algebra and statistics of the ends of string}

An open string operator can be viewed as a hopping operator for its ends.  The
statistics of the ends of string is determined by the hopping algebra of the
open string operators.
\end{center}
\end{svgraybox}

\section{Summary and further reading}

In this chapter, we introduced a macroscopic definition of topological order in
terms of the topological degeneracy and the non-Abelian geometric phases of the
ground states.  We also discussed some microscopic pictures of topological
orders, the global dance, that lead to simple microscopic many-body wave
functions, realizing the topologically ordered states.  Through those wave
functions, we calculate some physical properties, such as fraction quantum
number and fractional statistics, of topological order.

The simple local dancing rules \eqn{Z2rl} and \eqn{Semrl} can be generalized,
which allow strings to have different types and allow three strings to join at
a point. The generalized  local dancing rules can be quantitatively described
by a complex tensor $F^{ijm,\al}_{kln,\bt}$.  Not all the tensors
$F^{ijm,\al}_{kln,\bt}$ can lead to a global dancing pattern. Only the tensors
that satisfy certain conditions can lead to valid global dancing patterns (\ie
well defined many-body wave functions). By find all those valid tensors, we can
obtain a systematic theory for a class of topological order in 2+1 dimensions
with gapped boundary.  We can even calculate the topological properties of the
topological order, such as the ground state degeneracy and the fractional
statistics, from the valid tensors $F^{ijm,\al}_{kln,\bt}$.  For more details,
see Refs.~\cite{LW0510,H0904,CGW1038,H1171,WW1132,KK1251,HW1232,HWW1314}.

We can also use sequences of integers $\{S_a\}$, $a=2,3,\cdots$ (the pattern of
zeros), to quantitatively describe local dancing rules in FQH wave functions.
Again, not all the  sequences $\{S_a\}$ give rise to valid global dances.  Only
the  sequences that satisfy certain conditions can lead to valid global dancing
patterns (\ie well defined many-body wave functions). By find all those valid
sequences, we obtain a quite systematic theory for a class of FQH states.  We
can even calculate the topological properties of the FQH states, such as the
ground state degeneracy and the fractional charges, from the valid
sequences.  For more details, see Refs.
~\cite{SL0604,BKW0608,SL0701,WW0808,WW0809,ABK0816,S0802,SY0802,BH0802,BW0932,S1002,BW1001a,BW1001}.

%
%
\bibliographystyle{plain}
\bibliography{all}

%
%
%
\chapter{Local Transformations and Long-Range Entanglement}
\label{chap7} 

\abstract{To understand the origin of the topological phenomena discussed in the previous chapters, we need a microscopic theory for topological order. It was realized that the key microscopic feature of topologically ordered systems is the existence of long-range many-body entanglement in the ground-state wave function. Useful tools from quantum information theory to characterize many-body entanglement are local transformations, including local unitary (LU) transformation and stochastic local (SL) transformation. In this chapter, we apply these tools to the study of gapped quantum phases and phase transitions and establish the connection between topological order and long/short range entanglement. This allows us to obtain a general theory to study topological order and symmetry breaking order within the same framework. This leads to a basic understanding of the structure of the full quantum phase diagram.}


\section{Introduction}

After the experimental discovery of superconducting order via zero-resistance
and Meissner effect, it took 40  years to obtain the microscopic
understanding of superconducting order through the condensation of fermion
pairs. However, we are luckier for topological orders. After the
theoretical  discovery of topological order via the topological degeneracy and
the non-Abelian geometric phases of the degenerate ground states, it
took us only 20 years to obtain the microscopic understanding of topological
order: topological order is due to long-range entanglement and different topological
orders come from different patterns of long-range entanglement. In this
section, we will explain such a microscopic understanding.
 
This chapter is structured as follows. In section \ref{Qphase_TO}, we review the general idea of a quantum phase. We start from an intuitive picture of systems with very different physical properties being in different phases and arrive at a definition of phase based on phase transitions. In section \ref{Qphase_LU}, we introduce the concept of
local unitary (LU) transformation. Based on the idea of defining quantum phase in terms of phase transitions, we show that quantum states are in the same phase if and only if they are connected through LU transformations. We present two equivalent forms of LU transformation: the LU time evolution and the LU quantum circuit, which are useful in different circumstances. 
In section~\ref{sec:liquid}, we develop a general framework to study topological order, in the thermodynamic limit. We introduce the concept of `gapped quantum liquid', and show that topological
orders are in fact stable gapped quantum liquids. Classifying topological order hence corresponds to classifying stable gapped quantum liquids.
In section~\ref{sec:symb}, we show that symmetry breaking orders for on-site symmetry are also gapped quantum liquids, but with unstable ground-state degeneracy. The universality classes of generalized local unitary (gLU) transformations contain both topologically ordered states and symmetry breaking states.
In section~\ref{sec:SL}, we introduce the concept of stochastic local (SL) transformations. We show that the universality classes
of topological orders and symmetry breaking orders can be distinguished by SL transformations: small SL transformations can convert the symmetry breaking classes to the trivial class of product states with finite probability of success, while the topological-order classes are stable against any small SL transformations, demonstrating a phenomenon of emergence of unitarity.  This
allows us to give a definition of long-range entanglement based on SL transformations, under which only topologically ordered states are long-range entangled.
In section \ref{TO_MBET}, we discuss the situations where the system has certain symmetries and we obtain a generic structure of the possible phase diagram when symmetries are taken into account.

\section{Quantum phases and phase transitions}
\label{Qphase_TO}

Generally speaking, a phase is a collection of condensed matter systems with qualitatively the same but possibly quantitatively different properties, like density, magnetization, conductance, etc. As a system evolves within a phase, for example by changing temperature or external magnetic field, its property changes smoothly. However, when we reach a critical temperature or magnetic field, something dramatic could happen in the system and its property changes qualitatively as the system transits into a different phase. This is the point of phase transition. Different phases are hence separated by singular phase transition points where some physical observables of the system diverges.

\begin{figure}[htbp]
\begin{center}
\includegraphics[width=2.50in]{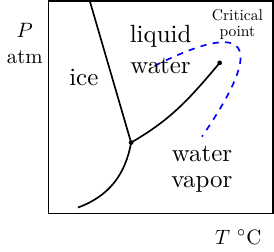}
\caption{Phase diagram of water.
}

\label{Water_PD}
\end{center}
\end{figure}

Therefore, two systems are in the same phase if and only if they can evolve
into each other smoothly without inducing singularity in any local physical
observable.  In this sense, liquid water and water vapor belong to the same
phase because the two can change into each other smoothly by following, for
example, the dashed line past the critical point in the phase diagram, as shown
in \ref{Water_PD}.  Note that in order to show two systems are in different
phases, we need to explore all possible paths of evolution and show there is no
smooth way to connect the two phases.

A similar definition holds for quantum systems as well. One special aspect of quantum many-body systems is that even at zero temperature, there can be different phases and phase transitions can happen without adding heat to the system. In our following discussions, we will focus mostly on quantum systems at zero temperature with a finite energy gap between the ground state and all the excited states. 

\begin{figure}[htbp]
\begin{center}
\includegraphics[width=2.50in]{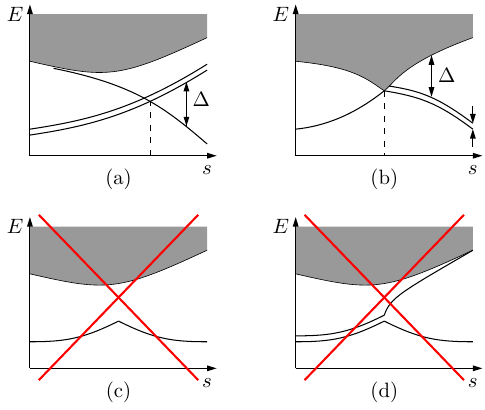}
\caption{
Energy spectrum of a gapped system as a function of
a parameter $s$ in the Hamiltonian.
(a,b) For gapped system, a quantum phase
transition can happen only when energy gap closes.
(a) describes a first order  quantum phase
transition (caused by level crossing).
(b) describes a continuous quantum phase
transition which has a continuum of gapless
excitations at the transition point.
(c) and (d) cannot happen for generic states.
A gapped system may have ground state degeneracy, where
the energy splitting $\eps$ between the ground states
vanishes when system size $L\to \infty$: $\lim_{L\to \infty}
\eps=0$. The energy gap $\Del$ between ground and excited states on the other hand
remains finite as $L\to \infty$.
}

\label{trans}
\end{center}
\end{figure}

For gapped quantum systems, quantum phase transition at zero temperature is
closely related to gap closing in the system. Consider a local Hamiltonian
$H(0)$, with ground state $\ket{\psi(0)}$ and a finite gap $\Delta(0)$ above
the ground state. Expectation value of any physical observable $O$ is given by
$\<O\>(0)=\bra{\psi(0)}O\ket{\psi(0)}$. Suppose that we smoothly change certain
parameter $g$  in the Hamiltonian so that the system follows a path $H(g)$. The
ground state $\ket{\psi(g)}$ and the expectation value of the physical
observable $\<O\>(g)=\bra{\psi(g)}O\ket{\psi(g)}$ will change accordingly.  It
is generally believed that, as long as the gap of the system $\Delta(g)$
remains finite, $\<O\>(g)$ will change smoothly.  Roughly speaking, when
$\Delta(g)>0$, we can use perturbation theory to calculate the change in
$\<O\>(g)$ as we change $g$ by a small amount, which will give rise to a smooth
dependence. Only when the gap $\Delta(g)$ closes can there be singularity in
any physical quantity. The possible and impossible situations are depicted in
Fig. \ref{trans}. Therefore, for gapped quantum systems at zero temperature,
two systems $H(0)$ and $H(1)$ are within the same phase if and only if there
exists a smooth path $H(g), 0\leq g \leq 1$ connecting the two and has a finite
gap for all $g$.

A question which is of general interest in condensed matter physics and which we will try to address is: what quantum phases could possibly exist at zero temperature in local gapped quantum systems? That is, for the class of local gapped quantum systems, how many sets can we group them into such that systems within a set can be smoothly connected and systems in different sets can not? Here we are considering quantum systems with arbitrary local degrees of freedom: bosons, fermions, spins... (spin and bosonic degrees of freedom have no intrinsic difference from each other, as in both cases operators on degrees of freedom at different spatial locations commute with each other. In our following discussion, we may use the words interchangably.) We also allow arbitrary form of local interaction between them, as long as the interaction involves a finite number of parties and affects a finite region in the lattice. 

We want to emphasize that quantum phase is a property of a class of
Hamiltonians, not of a single Hamiltonian. We call such a class of Hamiltonian
an H-class. Usually we are considering an H-class of fermionic or bosonic degrees of freedom, of a certain dimension and with possible symmetry
constraints. For example we can consider two dimensional fermionic systems with charge conservation symmetry or three dimensional bosonic systems with no special symmetry.  For a specific H-class, we can ask whether the Hamiltonians in it are separated into different
groups by phase transition and hence form different phases. Two Hamiltonians in
an H-class are in the same/different phase if they can/cannot be connected
\emph{within the H-class} without going through phase transition. We see that
without identifying the class of Hamiltonians under consideration, it is not
meaningful to ask which phase a Hamiltonian belongs to. Two Hamiltonians can
belong to the same/different phases if we embed them in different H-classes. 
We will see examples of this kind below.

For an H-class with certain symmetry constraints, one mechanism leading to
distinct phases is symmetry breaking. Starting from
Hamiltonians with the same symmetry, the ground states of them can have
different symmetries, hence resulting in different phases. This symmetry
breaking mechanism for phases and phase transitions
is well understood with Landau' symmetry breaking theory. 

However, it has been realized that quantum systems at zero temperature can be
in different phases even without breaking any symmetry. Such phases are often
said to be `topological'. Fractional quantum Hall is one of the first and most
important systems found to have topological order. It was realized that,
different fractional quantum Hall systems at different filling fractions all
have the same symmetry in the ground state, yet there must be a phase
transition if the system is to go from one to another. In one dimension, the
spin-1 chain $H=\sum \v S_i\cdot \v S_{i+1}$ is another example of gapped
topological phase, which does not break any symmetry of the system and is
separated from a trivial phase. More recently, the exciting discovery of
topological insulators and superconductors offers another class of topological
phases with interesting topological features.

So we would like to have a theory beyond Landau's symmetry breaking theory for a more complete understanding of the quantum phase diagram at zero temperature.


\section{Quantum phases and local unitary transformations}
\label{Qphase_LU}

Quantum phase and phase transitions are usually discussed in terms of the Hamiltonian of the system. For example for gapped quantum systems at zero temperature, two systems are in the same phase if and only if their Hamiltonians can be connected smoothly without closing gap. On the other hand, gapped quantum phases at zero temperature can be equally well studied in terms of their ground states. In this section, we describe how to determine the phase relation between two systems from their ground states.

In the following we may say that a quantum state $\ket{\psi}$ is gapped. Note that when we say so, we are always assuming that there exists a gapped Hamiltonian which has the state as its ground state. There can be multiple Hamiltonians satisfying this requirement, but their difference is not important, as their zero temperature property is completely determined by $\ket{\psi}$.

\subsection{Quantum phases and local unitary evolutions in ground states}

Suppose that we have two gapped quantum systems with Hamiltonians $H(0)$ and $H(1)$ and ground states $\ket{\Phi(0)}$ and $\ket{\Phi(1)}$ respectively. We want to determine from the ground states when the two systems are in the same phase. In order to have a well defined problem, we need to specify the H-class containing both $H(0)$ and $H(1)$. In this section, we will be considering H-classes with either bosonic or fermionic degrees of freedom, of a specific dimension and with no particular symmetry constraint. The symmetry constrained case is considered later. Note that systems in the same H-class can have different local Hilbert spaces, e.g. spin $1/2$ or spin $3/2$ on each site. In general, we are allowed to change the local Hilbert space by adding or removing local bosonic (fermionic) degrees of freedom in a bosonic (fermionic) system in the process of evolution.
 
From the Hamiltonians, we know that they are in the same phase iff there exists a gapped smooth path $H(g), 0\leq g \leq 1$ connecting them in the H-class. Such a smooth connection in Hamiltonians induces an adiabatic evolution connecting the ground states. That is, if we change the Hamiltonian $H(g)$ very slowly (compared to the inverse gap of the system), then the ground state follows an adiabatic evolution which begins with $\ket{\Phi(0)}$ and ends with $\ket{\Phi(1)}$. Therefore, we see that: if two gapped quantum states are in the same phase $|\Phi(0)\> \sim |\Phi(1)\>$ then they can be connected by an adiabatic evolution that does not close the energy gap.

Given two states, $|\Phi(0)\>$ and $|\Phi(1)\>$, determining the existence of such a gapped adiabatic connection can be hard. We would like to have a more operationally practical equivalence relation between states in the same phase. Here we would like to show that 

\begin{svgraybox}
\begin{center}
\textbf{Box 7.1 The same quantum phase}

Two gapped states $|\Phi(0)\>$
and $|\Phi(1)\>$ are in the same phase, if and only if they are
related by a local unitary (LU) evolution. 
\end{center} 
\end{svgraybox}
We define a local unitary(LU) evolution as an unitary operation generated by time evolution of a local Hamiltonian for a finite time. That is,
\begin{align}
\label{LUdef}
 |\Phi(1)\> \sim |\Phi(0)\> \text{\ iff\ }
 |\Phi(1)\> =  \cT[e^{-i\int_0^1 dg\, \t H(g)}] |\Phi(0)\>
\end{align}
where $\cT$ is the path-ordering operator and $\t
H(g)=\sum_{\v i} O_{\v i}(g)$ is a sum of local Hermitian
operators. Note that $\t H(g)$ is in general different from
the adiabatic path $H(g)$ that connects the two states.

First, we have shown in the above that if 
two states $|\Phi(0)\>$ and $|\Phi(1)\>$
are in the same phase, then we can find an 
adiabatic path $H(g)$ between the states. It has been shown that, the existence of a
gap prevents the system to be excited to higher energy
levels and leads to a local unitary evolution, the
Quasi-adiabatic Continuation,
that maps from one state to the other. That is,
\begin{equation}
\label{lcltrn}
 |\Phi(1)\> = U |\Phi(0)\>, \ \ \ \ \ \
U=\cT[e^{-i\int_0^1 dg\, \t H(g)}]
\end{equation}
The exact form of $\t H(g)$ can be found from $H(g)$. For details see ``summary and further reading'' section at the end of this chapter.

On the other hand, the reverse is also true: \emph{if two
gapped states $|\Phi(0)\>$ and $|\Phi(1)\>$ are related by a
local unitary evolution, then they are in the same
phase. } Since $|\Phi(0)\>$ and $|\Phi(1)\>$ are related by
a local unitary evolution, we have $ |\Phi(1)\>  =
\cT[e^{-i\int_0^1 dg\, \t H(g)}] |\Phi(0)\> $.  Let us
introduce the partial evolution operator
\begin{align}
|\Phi(s)\>  = U(s)
|\Phi(0)\> ,\ \ \ \ \ U(s)=
\cT[e^{-i\int_0^s dg\, \t H(g)}] .
\end{align}
Assume that $|\Phi(0)\>$ is a ground state of $H(0)$, then
$|\Phi(s)\>$ is a ground state of $H(s)=U(s)HU^\dag(s)$. If $H(s)$ remains local and gapped for all $s \in [0,1]$, then we have found an adiabatic connection between $|\Phi(0)\>$ and $|\Phi(1)\>$.

To see this, first let us show that $H(s)$ is  a local Hamiltonian.
Since $H$ is a local Hamiltonian, it has a form
$H=\sum_{\v i} O_{\v i}$ where $O_{\v i}$ only acts on a
cluster whose size is $\xi$.  $\xi$ is called the range of
interaction of $H$.  We see that $H(s)$ has a form
$H(s)=\sum_{\v i} O_{\v i}(s)$, where
$O_{\v i}(s) = U(s) O_{\v i} U^\dag(s)$.
To show that $O_{\v i}(s)$ only acts on a cluster of a
finite size, we note that for a local system
described by $\t H(g)$, the propagation velocities of its
excitations have a maximum value $v_{max}$.  Since $O_{\v
i}(s)$ can be viewed as the time evolution of $O_{\v i}$ by
$\t H(t)$ from $t=0$ to $t=s$, we find that $O_{\v i}(s)$
only acts on a cluster of size $\xi+\t \xi+s
v_{max}$, where $\t\xi$ is the range of
interaction of $\t H$.  Thus $H(s)$ is indeed a local
Hamiltonian. Secondly, if $H$ has a finite energy gap, then it is easy to see that $H(s)$ also have a
finite energy gap for any $s$ because $H(s)$ is obtained from $H$ by a unitary transformation. 

Therefore, $H(0)$ and $H(1)$ are connected by a smooth local gapped path $H(s)$. As $s$ goes for $0$ to $1$,
the ground state of the local Hamiltonians, $H(s)$, goes
from $|\Phi(0)\>$ to $|\Phi(1)\>$.  Thus the two states
$|\Phi(0)\>$ and $|\Phi(1)\>$ belong to the same phase.
This completes our argument that states related by a local
unitary evolution belong to the same phase.

The finiteness of the evolution time is very important in the above discussion. Here `finite' means the evolution time does not grow with system size, and in the thermodynamic limit, phases remain separate under such evolutions. On the other hand, if the system size under consideration is finite, there is a critical time limit above which phase separation could be destroyed. The time limit depends on the propagation speed of interactions in the Hamiltonian. 

Thus through the above discussion, we show that: Two gapped ground states, $|\Phi(0)\>$ and $|\Phi(1)\>$, belong to the same
phase if and only if they are related by a local unitary evolution \Eqn{LUdef}.

The relation \Eqn{LUdef} defines an equivalence relation between
$|\Phi(0)\>$ and $|\Phi(1)\>$. The equivalence classes
of such an equivalence relation represent different quantum phases.
So the above result implies that the equivalence classes
of the LU evolutions are the universality classes of
quantum phases for gapped states.

\subsection{Local unitary evolutions and local unitary quantum circuits}

\begin{figure}[htbp]
\begin{center}
\includegraphics[scale=0.8]{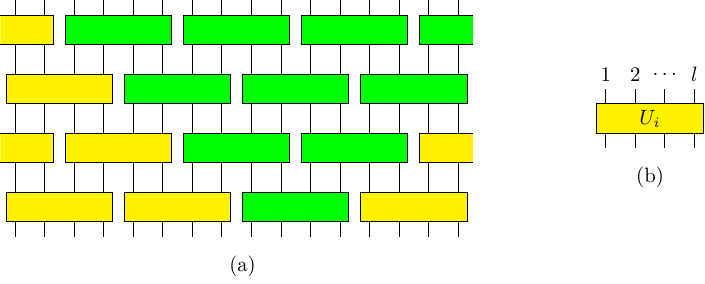}
\end{center}
\caption{
(a) A graphic representation of a
quantum circuit, which is formed by
(b) unitary operations on
patches of finite size $l$. The green shading
represents a causal structure.
}
\label{qc}
\end{figure}

The LU evolutions introduced here is closely related to
\emph{quantum circuits with finite depth}.  To define
quantum circuits, let us introduce  piece-wise local unitary
operators.  A piece-wise local unitary operator has a form $
U_{pwl}= \prod_{i} U_i$ where $\{ U_i \}$ is a set of
unitary operators that act on non overlapping regions. The
size of each region is less than some finite number $l$. The
unitary operator $U_{pwl}$ defined in this way is called a
piece-wise local unitary operator with range $l$.  A quantum
circuit with depth $M$ is given by the product of $M$
piece-wise local unitary operators.

\begin{svgraybox}
\begin{center}
\textbf{Box 7.2 Local unitary (LU) transformation}

An LU transformation, as shown 
in Fig.~\ref{qc}, is given by a finite number of layers (i.e.\
the number of layers is 
a constant that is independent of the system size) of piecewise
local unitary transformations
\begin{equation*}
 U^M_{circ}= U_{pwl}^{(1)} U_{pwl}^{(2)} \cdots U_{pwl}^{(M)}
\end{equation*}
where each layer has a form
\begin{equation*}
U_{pwl}= \prod_{i} U^i.
\end{equation*}.

Here $\{ U^i \}$ is a set of unitary operators that act on non-overlapping
regions. The size of each region is less than a finite number $l$.
\end{center} 
\end{svgraybox} 

In quantum information theory, it is known that finite time
unitary evolution with local Hamiltonian (LU evolution
defined before) can be simulated with constant depth quantum
circuit and vice-verse. The simulation of LU evolution by a LU quantum circuit proceeds as follows.

Consider the LU evolution generated by a local Hamiltonian $\cT[e^{-i\int_0^1 dg\, \t H(g)}]$. First group local terms in $\t H(g)$ into $m$ sets $\t H^i(g)$, $i=1,...,m$, such that local terms in each set $\t h^i_k(g)$ commute with each other. 
\be
\t H(g)=\t H^1(g) + \t H^2(g) + ... \t H^m(g)= \sum_k \t h^1_k(g) + \sum_k \t h^2_k(g) + ... + \sum_k \t h^m_k(g)
\ee
Because all terms in $\t H(g)$ are local, such a grouping can always be achieved with a finite number of groups. For example, if $\t H(g)$ is composed of nearest neighbor two-body interaction terms $h_{i,i+1}$ on a one dimensional chain, $h_{2i,2i+1}$ commute with each other and $h_{2i-1,2i}$ commute with each other. Therefore, $m=2$ is enough.

Although $\t H^i(g)$ in general does not commute with $\t H^{i'}(g)$, we can simulate the unitary evolution generated by $\t H(g)$ with Trotter decomposition. In particular, divide the evolution time into $N$ small intervals $\delta t$. Evolve with each $\t H^i(0)$ separately for time $\delta t$. Then evolve with each $\t H^i(\delta t)$ separately for time $\delta t$... Repeat the process for $N$ times. That is, we simulate the LU evolution generated by $\t H(g)$ as
\be
\cT[e^{-i\int_0^1 dg\, \t H(g)}] \approx \left(\prod^m_{i=1}e^{i \t H^i(0)\delta t}\right)\left(\prod^m_{i=1}e^{i \t H^i(\delta t)\delta t}\right)...\left(\prod^m_{i=1}e^{i \t H^i(1)\delta t}\right)
\ee
As shown in Chapter~\ref{cp:2}, the approximation becomes more and more accurate with larger and larger $N$.

In this way, we have decomposed the LU evolution into $Nm$ layers of unitary transformations. While $N$ is a large number, it remains finite for infinite system size. Therefore, the number of layers $Nm$ is also finite. Each layer can be further decomposed into local pieces. This step is exact as local terms in each $\t H^i(g)$ commute with each other.
\be
e^{i\t H^i(g) \delta t} = \prod_k e^{i\t h^i_k(g)\delta t}
\ee
Therefore, the LU evolution can be simulated with a piece-wise local quantum unitary circuit, as shown in Fig. \ref{qc}. Further more, the quantum circuit has only a constant number of layers, i.e. a constant depth. 

The equivalence relation defined using LU evolution
\eqn{LUdef} can therefore be equivalently stated in terms of constant
depth quantum circuits:
\begin{equation}
\label{PhiUcPhi}
|\Phi(1)\> \sim |\Phi(0)\> \text{ iff }
 |\Phi(1)\> = U^M_{circ} |\Phi(0)\>
\end{equation}
where $M$ is a constant independent of system size. Because
of their equivalence, we will use the term `Local Unitary (LU)
Transformation' to refer to both local unitary evolution and
constant depth quantum circuit in general. Similar to LU evolution, we are allowed to add or remove local degrees of freedom in an LU quantum circuit, although this step is not explicitly shown in Fig. \ref{qc}.

The idea of using LU transformation to study gapped phases can be easily generalized to
study topological orders and quantum phases with symmetries
(see section \ref{TO_MBET}). One difference between the LU evolution and the LU quantum circuit is that the quantum circuit breaks translation symmetry explicitly while the LU evolution does not. Therefore, the LU transformation defined through LU evolution
\Eqn{LUdef} is more general and can be used to study systems with translation symmetry. The
LU quantum circuit has a more clear and simple causal
structure. Although it cannot be used to study systems with translation symmetry, it can be applied to study topological orders and quantum phases with other (e.g. internal) symmetries.

\subsection{Local unitary quantum circuits and wave function renormalization}

As an application of the notion of LU quantum circuits,
we would like to describe a wave function renormalization
group flow. The idea of wave
function renormalization group flow is to use LU operators to remove
entanglement at small length scales, simplify the wave function and reach a fixed point form of wave function at a large enough length scale. As LU transformations map between states within the same phase, the wave function renormalization group flow is expected to flow every gapped quantum states to the fixed point wave function in the phase it belongs to.

\begin{figure}[htbp]
\begin{center}
\includegraphics[scale=0.8]{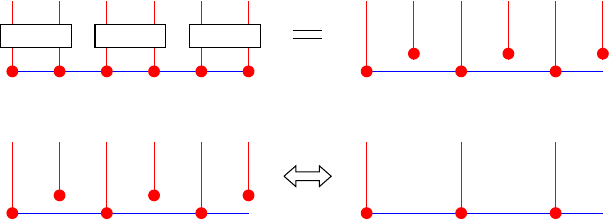}
\end{center}
\caption{
A piece-wise local unitary transformation can transform
some degrees of freedom in a state $|\Phi\>$ into a direct
product.  Removing/adding the degrees of freedom in the form
of direct product defines an additional equivalence relation
between quantum states.
}
\label{entre}
\end{figure}

To implement such a renormalization flow on wave functions, first we can use a LU transformation $U$ to transform some
degrees of freedom in a state into direct product (see Fig.
\ref{entre}). We can then remove those degrees of freedom in the
form of direct product. Such a procedure does not change
the phase the state belongs to. The reverse process of adding
degrees of freedom in the form of direct product states also does not
change the phase.  We call the local
transformation in Fig. \ref{entre} which involves changing the degrees
of freedom a generalized local unitary (gLU) transformation.
It is clear that a generalized local unitary transformation
inside a region $A$ does not change the reduced density
matrix $\rho_A$ for the region $A$.  This is the reason why
we say that (generalized) local unitary transformations
cannot change entanglement structure at large length scale and the quantum phase of the system.

\begin{figure}[htbp]
\begin{center}
\includegraphics[scale=0.8]{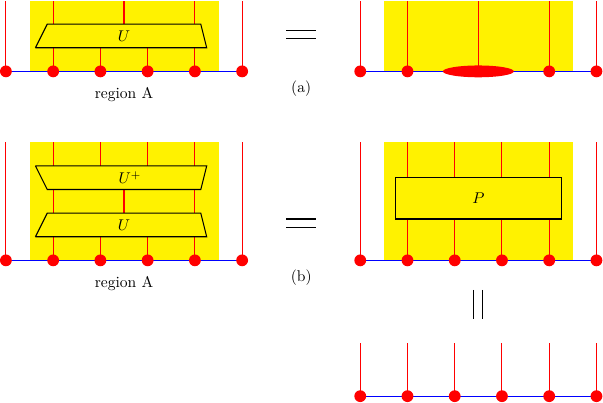}
\end{center}
\caption{
(a) A gLU transformation $U$ acts in
region A of a state $|\Phi\>$, which reduces the
degree freedom in region A to those contained only in the
support space of $|\Phi\>$ in region A.
(b) $U^\dag U=P$ is a projector that does not change the
state $|\Phi\>$.
}
\label{gLUT}
\end{figure}

Let us define the gLU transformation $U$ more carefully
and in a more general setting.
Consider a state $|\Phi\>$.  Let $\rho_A$ be the reduced
density matrix of $|\Phi\>$ in region $A$. Let $|\psi_{i}\>$,
$i=1,...,D_A$ be a basis of the total Hilbert space $V_A$ in region A, where
$D_A$ is the dimension of $V_A$. $\rho_A$ may act
in a subspace of $V_A$,
which is called the support space $V^{sp}_A$ of region $A$.
The dimension $D^{sp}_A$ of $V^{sp}_A$ is called the support
dimension of region $A$. Now the Hilbert space $V_A$ in
region A can be written as $V_A=V_A^{sp}\oplus \bar
V_A^{sp}$. Let $|\t \psi_{i}\>$, $i=1,...,D^{sp}_A$ be a
basis of this support space $V^{sp}_A$, $|\t \psi_{i}\>$,
$i=D^{sp}_A+1,...,D_A$ be a basis of $\bar V^{sp}_A$. We can introduce a LU
transformation $U^{full}$ on the full $D_A$ dimensional Hilbert space which rotates the basis
$|\psi_{i}\>$ to $|\t \psi_{i}\>$.  We note that in the new
basis, the wave function only has non-zero amplitudes on the
first $D^{sp}_A$ basis vectors.  Thus, in the new basis
$|\t \psi_{i}\>$, we can reduce the range of the label $i$
from $[1,D_A]$ to $[1,D^{sp}_A]$ without losing any
information. This motivates us to introduce the gLU
transformation $U$ as composed of two parts: 1. a rotation from the basis of the full Hilbert space $|\psi_{i}\>$,
$i=1,...,D_A$ to the basis of the support space
$|\t\psi_{i}\>$, $i=1,...,D^{sp}_A$ with a rectangular matrix $U'$ is given by
$U'_{ij}=\<\t\psi_{i}|\psi_{j}\>$. 2. a unitary transformation restricted to the support space alone.  
We also regard the
inverse of $U$, $U^\dag$, as a gLU transformation.  A LU
transformation is viewed as a special case of gLU
transformation where the degrees of freedom are not changed.
Clearly $U^\dag U=P$ and $U U^\dag=P'$ are two projectors.
The action of $P$ does not change the state $|\Phi\>$ (see
Fig. \ref{gLUT}(b)). 

We note that
despite the reduction of
degrees of freedom, a gLU transformation defines an
equivalent relation.  Two states related by a gLU
transformation belong to the same phase.  The
renormalization flow induced by the gLU transformations
always flows within the same phase. Therefore, in general, we are allowed to use such gLU transformations in the wave function renormalization scheme as long as they are unitary on the support space of a local region in the wave function.

After applying several rounds of the wave function renormalization procedure, nonuniversal local entanglement structures at larger and larger length scales are removed and the wave function is expected to flow to a simplified fixed point form which remains invariant under the renormalization transformation. Note that under the renormalization flow, the degrees of freedom in the system can change and so does the lattice structure of the system. Therefore, the fixed point wave function is not a single wave function, but rather a set of wave functions having the same form on lattice structures of different length scales.

Let us consider some simple examples of model wave functions which are fixed points under a wave function renormalization group flow.

\begin{figure}[htbp]
\begin{center}
\includegraphics[scale=0.8]{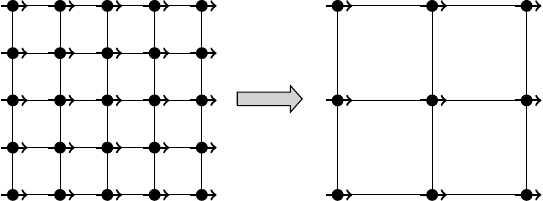}
\end{center}
\caption{Wave function renormalization group transformation on a product state. The form of the product state remains invariant under this transformation. 
}
\label{ps_rg}
\end{figure}

The simplest example is a total product state, for example the Ising paramagnet where all the spins point to the $+x$ direction.
\be
|\Phi^+\>=\otimes_{\v i} (|\up\>_{\v i} + |\down\>_{\v i})
\ee
As all the spins are already disentangled from each other, to renormalize the state to a doubled length scale, we simply remove the redundant degrees of freedom, as shown in Fig.\ref{ps_rg}. After the renormalization step, the wave function is still a total product state of spins in the $+x$ direction. Therefore, the product state is a fixed point under the wave function renormalization group flow.

A nontrivial example of fixed point wave function is given by the toric code model. Remember that for a toric code model defined on a squre lattice with spins on the links, the ground state wave function is an equal weight superposition of all closed loop configurations, where spin $0$ corresponds to no string and spin $1$ corresponds to having string on a link. The toric code wave function is the fixed point of the following renormalization flow.

\begin{figure}[htbp]
\begin{center}
\includegraphics[scale=0.8]{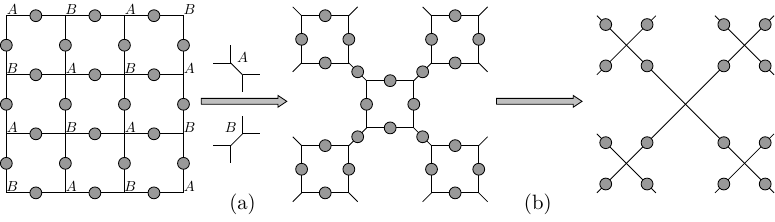}
\end{center}
\caption{Wave function renormalization group transformation on the toric code wave function. The form of the wave function remains invariant under this transformation. 
}
\label{TC_rg}
\end{figure}

First we divide the lattice into A and B sublattices and add an extra spin at each vertex in the state $\ket{0}$. Then apply a unitary transformation $U_1$ to the spins around each vertex. For vertices in sublattice A, apply a $\sigma_x$ operator to the added spin if the up and left links carry an odd number of strings and do nothing otherwise. For vertices in sublattice B, apply a $\sigma_x$ operator to the added spin if the up and right links carry an odd number of strings and do nothing otherwise. From Fig.\ref{TC_rg} we can see that such an operation splits the degree four vertex into two degree three ones and the added spin is on the link between the two vertices. The unitary transformation is applied such that the number of strings going through each vertex is still even and the ground state wave function is still an equal weight superposition of all closed loop configurations, now on the modified lattice as shown in the middle of Fig.\ref{TC_rg}. Now apply a unitary transformation $U_2$ on the eight spins around each square (four on the diagonal links $abcd$ and four on the square boundaries $ijkl$). From the previous discussion we know that we only need to describe the action of this transformation on the support space of the eight spins. In the support space, the four diagonal spins $abcd$ always carry an even number of strings. For each fixed configuration of $abcd$, $ijkl$ is in a superposition of two configurations, both satisfying the constraint at the four vertices and differing by a loop around the square. For example if $abcd$ are all $0$, then $ijkl$ is in a superposition of $0000$ and $1111$. Now for each fixed configuration of $abcd$, apply a transformation to $ijkl$ and map the state to $0000$. Because $abcd$ remain invariant during this process, the resulting states are still orthogonal to each other, even though the state of $ijkl$ become the same. From this we can see that $U_2^{\dg}U_2$ is identity on the support space of the eight spins and $U_2$ a gLU as defined before. After this step, the $ijkl$ spins are totally decoupled from everything else and can be removed. In this way, we have shrunk the square bubbles to a point. The resulting state (with the diagonal spins) live on a renormalized lattice and is still an equal weight superposition of all closed loop configurations as in the whole renormalization process we did not break the closed loop constraint and did not change the amplitude of any loop configuration. Therefore, the toric code wave function is a fixed point under this renormalization scheme .


\section{Gapped Hamiltonians and topological order}
\label{sec:liquid}

In this section, we will discuss the relationship between gapped Hamiltonians and
topological order.  We first point out that the
topologically ordered systems are not arbitrary gapped systems, but belong to a
special kind of gapped quantum systems, called \emph{gapped quantum liquids}. We
will discuss the concept of gapped quantum liquids. 

We remark that the notion of gapped quantum liquids can also be applied to solve the problem
of taking the thermodynamic limit for systems without translation
symmetry.  In general, in the presence of strong randomness, the thermodynamic
limit is not well defined (without impurity average). We show that for gapped
quantum liquids, the thermodynamic limit is well defined even without impurity
average. Consequently, the notions of quantum phases and quantum phase
transitions are well defined for gapped quantum liquids.

\subsection{Gapped quantum systems and gapped quantum phases}

Topologically ordered systems are gapped quantum systems.
We have discussed the idea of gapped quantum systems in Chapter~\ref{cp:5}. 
Here we would like to clarify the concepts of gapped quantum systems 
in a more formal manner.

Since a gapped system may
have gapless excitations on the boundary (such as quantum Hall systems), so to
discuss gapped Hamiltonians, we put the Hamiltonian on a space with no
boundary.  Also, system with certain sizes may contain non-trivial excitations
(such as a spin liquid state of spin-1/2 spins on a lattice with an odd number
of sites), so we need to specify that the system has a certain sequence of sizes when we take the thermodynamic limit. These observations lead to the following notion.

\begin{svgraybox}
\begin{center}
\textbf{Box 7.3 Gapped quantum system}

Consider a local Hamiltonian of a qubit system on a graph with no boundary, with
finite spatial dimension $D$.  If there is a sequence of sizes of the system
$N_k$, $N_k\to \infty$, as $k\to \infty$, such that the size-$N_k$ system has
the following `gap property' (as given in Box 7.4), then the system, defined by the Hamiltonian sequence $\{H_{N_k}\}$, is said to be gapped.  Here
$N_k$ can be viewed as the number of qubits in the system.
\end{center}
\end{svgraybox}

The notion of `gap property' is given below.

\begin{svgraybox}
\begin{center}
\textbf{Box 7.4 Gap property}

There is a fixed $\Delta$ (i.e.\ independent of $N_k$) such that 
 (1) the size-$N_k$ Hamiltonian has no eigenvalue in an energy window of size $\Delta$; 
 (2) the number of eigenstates below the energy window does not depend on 
     $N_k$;  
 (3) the energy splitting of those eigenstates below the energy window
approaches zero as $N_k \to \infty$. 
\end{center} 
\end{svgraybox}  
Note that the notion of `gapped quantum system' is not for a
single Hamiltonian. It is a property of a sequence of Hamiltonians,
$\{H_{N_k}\}$, in the large size limit $N_k \to \infty$ (i.e. an `H-class' as discussed previously).  In the rest of this chapter, the
term `a gapped quantum system' refers to a sequence of Hamiltonians
$\{H_{N_k}\}$, which satisfy the gap property.

Now we introduce the notion of ground-state degeneracy and ground-state space.
\begin{svgraybox}
\begin{center}
\textbf{Box 7.5 Ground-state degeneracy and ground-state space}

The number of eigenstates below the energy window is the ground-state
degeneracy of the gapped system $\{H_{N_k}\}$.  
The states below the
energy window span the \emph{ground-state space}, which is denoted as
$\mathcal{V}_{N_k}$.
\end{center} 
\end{svgraybox} 

\begin{figure}[h!]
\vskip-\baselineskip
\centerline{\footnotesize\unitlength 0.8\unitlength%
\begin{tabular}{cccc}
\begin{picture}(50,80)(0,-40)
\put(25,33){\makebox(0,0){$H_{N_1}$}}
\put(35,0){\makebox(0,0){$LU$}}
\put(25,0){\vector(0,1){24}}
\put(25,0){\vector(0,-1){24}}
\put(25,-33){\makebox(0,0){$H'_{N_1}$}}
\end{picture}
&
\begin{picture}(50,80)(0,-40)
\put(25,33){\makebox(0,0){$H_{N_2}$}}
\put(35,0){\makebox(0,0){$LU$}}
\put(25,0){\vector(0,1){24}}
\put(25,0){\vector(0,-1){24}}
\put(25,-33){\makebox(0,0){$H'_{N_2}$}}
\end{picture}
&
\begin{picture}(50,80)(0,-40)
\put(25,33){\makebox(0,0){$H_{N_3}$}}
\put(35,0){\makebox(0,0){$LU$}}
\put(25,0){\vector(0,1){24}}
\put(25,0){\vector(0,-1){24}}
\put(25,-33){\makebox(0,0){$H'_{N_3}$}}
\end{picture}
&
\begin{picture}(50,80)(0,-40)
\put(25,33){\makebox(0,0){$H_{N_4}$}}
\put(35,0){\makebox(0,0){$LU$}}
\put(25,0){\vector(0,1){24}}
\put(25,0){\vector(0,-1){24}}
\put(25,-33){\makebox(0,0){$H'_{N_4}$}}
\end{picture}
\end{tabular}}
\caption{
The two rows of Hamiltonians describe two gapped quantum systems.  The two rows
connected by LU transformations represent the equivalence relation between the
two gapped quantum systems, whose equivalence classes are gapped quantum
phases. 
}
\label{gapLUH}
\end{figure}
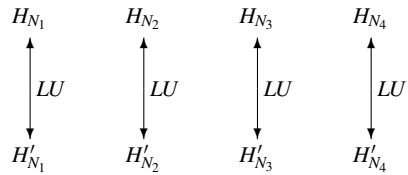

Now we discuss the concept of gapped quantum phase. Recall that as discussed in Sec.~\ref{Qphase_LU},
two gapped systems connected by an LU transformation can deform into each other smoothly without closing the energy gap, and thus belong to the same phase. We summarize this observation in a more formal manner as below.
\begin{svgraybox}
\begin{center}
\textbf{Box 7.6 Gapped quantum phase}

Two gapped quantum systems $\{H_{N_k}\}$ and $\{H'_{N_k}\}$ are equivalent if
the \emph{ground-state spaces} of $H_{N_k}$ and $H'_{N_k}$ are connected by
LU transformations for all $N_k$. The equivalence classes of the above
equivalence relation are the gapped quantum phases (see Fig. \ref{gapLUH}).
\end{center} 
\end{svgraybox} 

It is highly desired to identify topological orders as gapped quantum phases,
since both concepts do not involve symmetry.  In the following, we will show
that gapped quantum phases, sometimes, are not well behaved in the
thermodynamic limit.  As a result, it is not proper to associate topological
orders with all gapped quantum phases.  To fix this problem, we will introduce the
concept of gapped quantum liquid phase.

\subsection{Gapped quantum liquid system and gapped quantum liquid phase}

We start by examining the question of why gapped quantum systems may not be well-behaved in the thermodynamic
limit. This is because the Hamiltonians with different sizes may not be related in a way based on our notion of gapped quantum systems
(i.e. the way as shown in Fig. \ref{gapLUH}).  As a
result, we are allowed to choose totally different $H_{N_k}$ and $H_{N_{k+1}}$
as long as the Hamiltonians have the same ground-state degeneracy.  For example,
one can be topologically ordered and the other can be symmetry breaking.
  
To overcome this problem, we choose a subclass of  gapped quantum systems which are
well-behaved in the thermodynamic limit.  Those gapped quantum systems are
`shapeless' and can `dissolve' any product states on
additional sites to increase its size.
Such gapped quantum systems are called \textit{gapped quantum liquid systems}. 

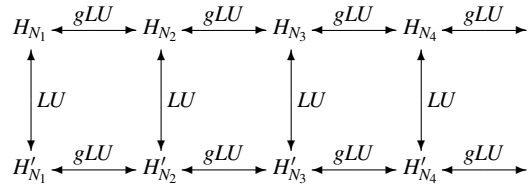
\begin{figure}[hbtp]
\begin{center}
\vskip-\baselineskip
\centerline{\footnotesize\unitlength 0.8\unitlength%
\begin{tabular}{ccccccccc}
\begin{picture}(50,80)(20,-30)
\put(25,33){\makebox(0,0){$H_{N_1}$}}
\put(35,0){\makebox(0,0){$LU$}}
\put(25,0){\vector(0,1){24}}
\put(25,0){\vector(0,-1){24}}
\put(25,-33){\makebox(0,0){$H'_{N_1}$}}
\end{picture}
&
\begin{picture}(4,80)(44,-30)
\put(25,40){\makebox(0,0){$gLU$}}
\put(25,33){\vector(1,0){20}}
\put(25,33){\vector(-1,0){20}}
\put(25,-26){\makebox(0,0){$gLU$}}
\put(25,-33){\vector(1,0){20}}
\put(25,-33){\vector(-1,0){20}}
\end{picture}
&
\begin{picture}(50,80)(20,-30)
\put(25,33){\makebox(0,0){$H_{N_2}$}}
\put(35,0){\makebox(0,0){$LU$}}
\put(25,0){\vector(0,1){24}}
\put(25,0){\vector(0,-1){24}}
\put(25,-33){\makebox(0,0){$H'_{N_2}$}}
\end{picture}
&
\begin{picture}(4,80)(44,-30)
\put(25,40){\makebox(0,0){$gLU$}}
\put(25,33){\vector(1,0){20}}
\put(25,33){\vector(-1,0){20}}
\put(25,-26){\makebox(0,0){$gLU$}}
\put(25,-33){\vector(1,0){20}}
\put(25,-33){\vector(-1,0){20}}
\end{picture}
&
\begin{picture}(50,80)(20,-30)
\put(25,33){\makebox(0,0){$H_{N_3}$}}
\put(35,0){\makebox(0,0){$LU$}}
\put(25,0){\vector(0,1){24}}
\put(25,0){\vector(0,-1){24}}
\put(25,-33){\makebox(0,0){$H'_{N_3}$}}
\end{picture}
&
\begin{picture}(4,80)(44,-30)
\put(25,40){\makebox(0,0){$gLU$}}
\put(25,33){\vector(1,0){20}}
\put(25,33){\vector(-1,0){20}}
\put(25,-26){\makebox(0,0){$gLU$}}
\put(25,-33){\vector(1,0){20}}
\put(25,-33){\vector(-1,0){20}}
\end{picture}
&
\begin{picture}(50,80)(20,-30)
\put(25,33){\makebox(0,0){$H_{N_4}$}}
\put(35,0){\makebox(0,0){$LU$}}
\put(25,0){\vector(0,1){24}}
\put(25,0){\vector(0,-1){24}}
\put(25,-33){\makebox(0,0){$H'_{N_4}$}}
\end{picture}
&
\begin{picture}(4,80)(44,-30)
\put(25,40){\makebox(0,0){$gLU$}}
\put(25,33){\vector(1,0){20}}
\put(25,33){\vector(-1,0){20}}
\put(25,-26){\makebox(0,0){$gLU$}}
\put(25,-33){\vector(1,0){20}}
\put(25,-33){\vector(-1,0){20}}
\end{picture}
\end{tabular}}
\caption{
The two rows define two gapped quantum liquid systems via gLU transformations.
The two rows connected by LU transformations represent the equivalence relation
between two gapped quantum liquid systems, whose equivalence classes are
gapped quantum liquid phases.
}
\label{liquidLUH}
\end{center}
\end{figure}

\begin{svgraybox}
\begin{center}
\textbf{Box 7.7 Gapped quantum liquid system}

A gapped quantum liquid system is
a gapped quantum system, described by the sequence $\{H_{N_k}\}$, 
with two additional properties: 
(1) $0<c_1<(N_{k+1}-N_{k})/N_k < c_2 $ where $c_1$ and $c_2$ are constants that do not depend on the system size; 
(2) the ground-state spaces of $H_{N_k}$ and $H_{N_{k+1}}$ are connected by a generalized local unitary
(gLU) transformation (see Fig. \ref{liquidLUH}).  
\end{center} 
\end{svgraybox}

We need to explain the concept of gLU transformation. For the system $H_{N_k}$, we first need to add  
$N_{k+1}-N_k$ qubits. We would like to do this addition `locally'.
That is, the distribution of the added
qubits may not be uniform in space but maintains a finite density
(number of qubits per unit volume). We call this `local addition'
(LA) transformation. We then discuss
how to write Hamiltonians after adding particles to the system,
as given below.

\begin{figure}[htbp] 
\begin{center} 
\includegraphics[scale=1.2]{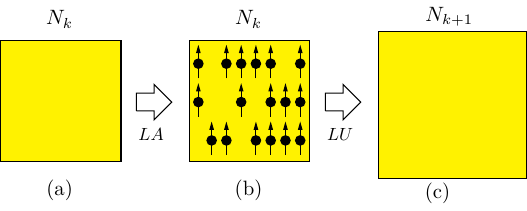} 
\end{center}
\caption{Two systems (a) and (c), with size $N_k$ and $N_{k+1}$, are
described by $H_{N_k}$ and $H_{N_{k+1}}$ respectively.  (a) $\to$ (b) is an LA
transformation where we add $N_{k+1}-N_k$ qubits to the system $H_{N_k}$ to
obtain the Hamiltonian $H_{N_k} + \sum_i Z_i$ for the combined system (b).
Under the LA transformation, the ground states of $H_{N_k}$ is tensored with a
product state to obtain the ground states of $H_{N_k}+ \sum_i Z_i$.  In (b)
$\to$ (c), we transform the ground-state space of $H_{N_k} + \sum_i Z_i$ to
the ground-state space of $H_{N_{k+1}}$ via an LU transformation.
} 
\label{NkNk} 
\end{figure}

\begin{svgraybox}
\begin{center}
\textbf{Box 7.8 Local addition (LA) transformation}

For adding $N_{k+1}-N_k$ qubits to the system $H_{N_k}$ locally,
we consider the Hamiltonian $H_{N_k} + \sum_{i=1}^{N_{k+1}-N_k} Z_i$ 
for the combined system (see Fig. \ref{NkNk}b), where $Z_i$ 
is the Pauli $Z$ operator acting on the
$i^\text{th}$ qubit.  This defines an LA transformation from
$H_{N_k}$ to $H_{N_k} + \sum_{i=1}^{N_{k+1}-N_k} Z_i$. 
\end{center} 
\end{svgraybox}

Now we are ready to discuss the notion of gLU transformation.

\begin{svgraybox}
\begin{center}
\textbf{Box 7.9 gLU transformation}

If for any LA transformation from $H_{N_k}$ to $H_{N_k} +
\sum_{i=1}^{N_{k+1}-N_k} Z_i$, the ground-state space of $H_{N_k} +
\sum_{i=1}^{N_{k+1}-N_k} Z_i$ can be transformed into  the ground-state
space of $H_{N_{k+1}}$ via an LU transformation, then we say $H_{N_k}$ and
$H_{N_{k+1}}$ are connected by a gLU transformation.  
\end{center} 
\end{svgraybox}

Fig.~\ref{NkNk} illustrates how we transform $H_{N_k}$ to $H_{N_{k+1}}$ via a
gLU transformation. 

\begin{figure}[htbp] 
\begin{center} 
\includegraphics[scale=1.2]{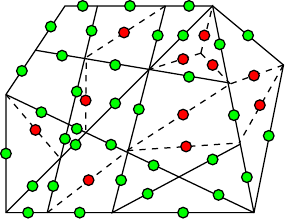} \end{center}
\caption{
Toric code as a gapped quantum liquid: toric code of $N_k$ qubits on an arbitrary 
2D lattice, where the green dots represent qubits sitting on the link of the lattice (given by 
solid lines). By
adding $N_{k+1}-N_{k}$ qubits (red dots), the gLU transformation 
$H_{N_k}\to H_{N_{k+1}}$ `dissolves' the red qubits in the new lattice (with both
the solid lines and dashed lines).
} 
\label{Toric1} 
\end{figure}

According to our notion, the sequence of following Hamiltonians
\begin{equation}
H^\text{trivial-liquid}_{N_k}=-\sum_{i=1}^{N_k}Z_i,
\end{equation}
gives rise to a gapped quantum liquid system. 
The topologically-ordered toric code Hamiltonian $H^\text{toric}_{N_k}$ is also
a gapped quantum liquid, as illustrated in Fig.~\ref{Toric1}. This reveals one
important feature of a gapped quantum liquid -- the corresponding
lattice in general does not have a `shape'
(i.e.\ the system can be defined on an arbitrary lattice with a meaningful thermodynamic limit).

To have an example of a gapped quantum system that is not 
a gapped quantum liquid, 
consider another sequence of Hamiltonians
\begin{equation}
\label{Hnonliquid}
H^\text{non-liquid}_{N_k}=-\sum_{i=1}^{N_k-1}Z_i.
\end{equation}
It describes a gapped quantum system with two-fold degenerate ground states
(coming from the $N_k^\text{th}$ qubit which carries no energy).  However, such
a gapped quantum system is not a gapped quantum liquid system. Because
the labelling of the $N_{k+1}$ qubit is essentially arbitrary,
for some LA transformations, the map from 
$H_{N_k} + \sum_{i=1}^{N_{k+1}-N_k} Z_i$ to $H_{N_{k+1}}$
cannot be local.

Through the above example, we see that a gapped quantum system may not
have a well defined thermodynamic limit (because the low energy property --
the degenerate ground states, is given by an isolated qubit which is not a
thermodynamic property). Similarly, gapped quantum phase (as given in
Box 7.6) is not a good concept, since it is not always a thermodynamic
property.  In contrast, gapped quantum liquid system and
gapped quantum liquid phase (given below in Box 7.10) are
good concepts, because they are always related to thermodynamic properties.

\begin{svgraybox}
\begin{center}
\textbf{Box 7.10 Gapped quantum liquid phase}

Two gapped quantum liquid systems $\{H_{N_k}\}$ and $\{H'_{N_k}\}$ are
equivalent if the ground-state spaces of
$H_{N_k}$ and $H'_{N_k}$ are connected by LU transformations
for all $N_k$. The equivalence classes of this relation
are the gapped quantum liquid phases (see Fig. \ref{liquidLUH}).
\end{center} 
\end{svgraybox}

\subsection{Topological order}

Using the notion of gapped quantum liquid phase, we can discuss
the concept of
topological order in a more formal way.  First, we introduce the
concept of `stable gapped quantum system'.
\begin{svgraybox}
\begin{center}
\textbf{Box 7.11 Stable gapped quantum system}

If the ground-state degeneracy of a gapped quantum system
is stable against any local perturbation (in the large $N_k$ limit),
then the gapped quantum system is stable.
\end{center} 
\end{svgraybox}

An intimately related fact to this concept
is that the ground-state space of a stable gapped quantum system 
(in the large $N_k$ limit) is a quantum
error-correcting code with macroscopic distance. 
This is to say, for
any orthonormal basis $\{\ket{\Phi_i}\}$ of the ground-state space, for any local operator
$M$, we have 
\begin{equation}
\label{eq:qcode}
\bra{\Phi_i}M\ket{\Phi_j}=C_M\delta_{ij},
\end{equation}
where $C_M$ is a constant which only depends on $M$ (see the discussions in Chapter~\ref{cp:3} and \ref{cp:5}).

Note that a gapped quantum liquid system may not be a stable gapped quantum
system. A symmetry breaking system is an example, which is a  gapped quantum
liquid system but not a stable gapped quantum system (the ground-state
degeneracy can be lifted by symmetry breaking perturbations).  Also a stable
gapped quantum system may not be a gapped quantum liquid system.  A non-Abelian
quantum Hall states with traps that trap
non-Abelian quasiparticles is an example.  Since the ground state with traps
contain non-Abelian quasiparticles, the resulting degeneracy is robust against
any local perturbations.  So the system is a stable gapped quantum system.  However,
for such a system, $H_{N_k}$ and $H_{N_{k+1}}$ are not connected via gLU transformations,
hence it is not a gapped quantum liquid system.

Now we can introduce the notion of topological order (or different phases of 
topologically ordered states):
\begin{svgraybox}
\begin{center}
\textbf{Box 7.12 Topological order} 

Topological orders are stable gapped quantum liquid phases.
\end{center} 
\end{svgraybox}

We remark that we in fact associate different topological orders as different equivalence classes.
One of these equivalence classes represents the trivial (topological) order.
Here we put trivial and non-trivial topological orders together to have a simple definition. 
This is similar to symmetry transformations, which usually include both trivial and non-trivial transformations, so that we can say symmetry transformations form a group. Similarly, if we include the trivial one, then we can say that topological orders form a monoid under the stacking operation.

The first order phase-transition point
is also an unstable gapped quantum liquid system,
which is with accidental degenerate
ground states.  

\begin{svgraybox}
\begin{center}
\textbf{Box 7.13 First-order phase transition for gapped quantum liquid systems} 

A deformation of a gapped quantum liquid system experiences a first order phase
transition if the Hamiltonian remains gapped along the deformation path and if
the ground-state degeneracy at a point on the deformation path is different
from its neighbours.  That point is the transition point of the first order
phase transition.  
\end{center} 
\end{svgraybox}

From the above discussions, we see that topological orders are the universality
classes of stable gapped quantum liquid systems that are separated by gapless
quantum systems or unstable gapped quantum systems.  Moving from one
universality class to another universality class by passing through a gapless
system corresponds to a continuous phase transition.  Moving from one
universality class to another universality class by passing through an unstable
gapped system corresponds to a first order phase transition.

\begin{figure}[htbp] 
\begin{center} 
\includegraphics[scale=1.0]{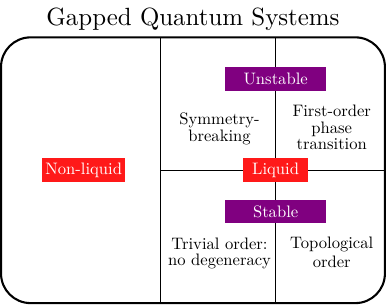} \end{center}
\caption{
Summary of gapped quantum systems: gapped quantum systems
include gapped quantum liquid systems, and systems that are not liquid (nonliquid).
For gapped quantum liquids, there are stable systems (including the trivial systems
given by e.g. the Hamiltonian $H^\text{non-liquid}_{N_k}$ and the topologically ordered
systems) and unstable systems (including symmetry breaking systems and first-order
phase transitions).
} 
\label{Liquid} 
\end{figure}

We summarize the different kinds of gapped quantum systems in Fig.~\ref{Liquid}.

\section{Universality classes of many-body wave functions}

We would like to emphasize that the topological order is a notion of
universality classes of local Hamiltonians (or more precisely, gapped quantum
systems). In the following, we will introduce the universality classes of
many-body wave functions.  We can also use the universality classes of
many-body wave functions to understand topological orders.

\subsection{Gapped quantum liquid}

\begin{svgraybox}
\begin{center}
\textbf{Box 7.14 Gapped quantum state}

A gapped quantum system is given by a sequence of Hamiltonians $\{H_{N_k}\}$.
Let $\mathcal{V}_{N_k}$ be the  ground-state space of $H_{N_k}$.
The sequence of ground-state spaces $\{\mathcal{V}_{N_k}\}$
is referred to as a gapped quantum state. 
\end{center} 
\end{svgraybox}

Note that a gapped quantum state is not described by a single wave function,
but by a sequence of ground-state spaces $\{\mathcal{V}_{N_k}\}$.
Similarly,
\begin{svgraybox}
\begin{center}
\textbf{Box 7.15 Gapped quantum liquid}

The sequence of ground-state spaces $\{\mathcal{V}_{N_k}\}$ of a gapped quantum
liquid system given by $\{H_{N_k}\}$ is referred to as a gapped quantum liquid. 
\end{center} 
\end{svgraybox}

\begin{figure}[hbtp]
\begin{center}
\vskip-\baselineskip
\centerline{\footnotesize\unitlength 0.8\unitlength%
\begin{tabular}{ccccccccc}
\begin{picture}(50,80)(20,-30)
\put(25,33){\makebox(0,0){$\{\mathcal{V}_{N_1}\}$}}
\put(35,0){\makebox(0,0){$LU$}}
\put(25,0){\vector(0,1){24}}
\put(25,0){\vector(0,-1){24}}
\put(25,-33){\makebox(0,0){$\{\mathcal{V}'_{N_1}\}$}}
\end{picture}
&
\begin{picture}(4,80)(44,-30)
\put(25,40){\makebox(0,0){$gLU$}}
\put(25,33){\vector(1,0){20}}
\put(25,33){\vector(-1,0){20}}
\put(25,-26){\makebox(0,0){$gLU$}}
\put(25,-33){\vector(1,0){20}}
\put(25,-33){\vector(-1,0){20}}
\end{picture}
&
\begin{picture}(50,80)(20,-30)
\put(25,33){\makebox(0,0){$\{\mathcal{V}_{N_2}\}$}}
\put(35,0){\makebox(0,0){$LU$}}
\put(25,0){\vector(0,1){24}}
\put(25,0){\vector(0,-1){24}}
\put(25,-33){\makebox(0,0){$\{\mathcal{V}'_{N_2}\}$}}
\end{picture}
&
\begin{picture}(4,80)(44,-30)
\put(25,40){\makebox(0,0){$gLU$}}
\put(25,33){\vector(1,0){20}}
\put(25,33){\vector(-1,0){20}}
\put(25,-26){\makebox(0,0){$gLU$}}
\put(25,-33){\vector(1,0){20}}
\put(25,-33){\vector(-1,0){20}}
\end{picture}
&
\begin{picture}(50,80)(20,-30)
\put(25,33){\makebox(0,0){$\{\mathcal{V}_{N_3}\}$}}
\put(35,0){\makebox(0,0){$LU$}}
\put(25,0){\vector(0,1){24}}
\put(25,0){\vector(0,-1){24}}
\put(25,-33){\makebox(0,0){$\{\mathcal{V}'_{N_3}\}$}}
\end{picture}
&
\begin{picture}(4,80)(44,-30)
\put(25,40){\makebox(0,0){$gLU$}}
\put(25,33){\vector(1,0){20}}
\put(25,33){\vector(-1,0){20}}
\put(25,-26){\makebox(0,0){$gLU$}}
\put(25,-33){\vector(1,0){20}}
\put(25,-33){\vector(-1,0){20}}
\end{picture}
&
\begin{picture}(50,80)(20,-30)
\put(25,33){\makebox(0,0){$\{\mathcal{V}_{N_4}\}$}}
\put(35,0){\makebox(0,0){$LU$}}
\put(25,0){\vector(0,1){24}}
\put(25,0){\vector(0,-1){24}}
\put(25,-33){\makebox(0,0){$\{\mathcal{V}'_{N_4}\}$}}
\end{picture}
&
\begin{picture}(4,80)(44,-30)
\put(25,40){\makebox(0,0){$gLU$}}
\put(25,33){\vector(1,0){20}}
\put(25,33){\vector(-1,0){20}}
\put(25,-26){\makebox(0,0){$gLU$}}
\put(25,-33){\vector(1,0){20}}
\put(25,-33){\vector(-1,0){20}}
\end{picture}
\end{tabular}}
\caption{
The two rows define two gapped quantum liquids via gLU transformations.
The two rows connected by LU transformations represent the equivalence relation
between two gapped quantum liquids, whose equivalence classes are
gapped quantum liquid phases. The ground-state spaces $\mathcal{V}_{N_k}$ and $\mathcal{V}'_{N_k}$ of two  equivalent quantum liquids are connected by the LU transformations.
}
\label{liquidLUVk}
\end{center}
\end{figure}
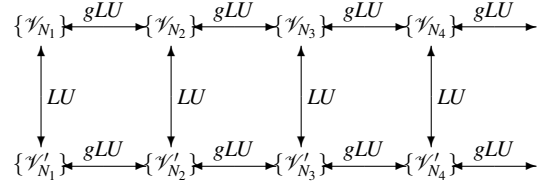

Now we are ready to introduce the concept of gapped quantum liquid phase in terms
of ground-state subspaces, which is indeed the same as the notion of gapped quantum liquid phase
given in Box 7.10 in terms of Hamiltonians.
\begin{svgraybox}
\begin{center}
\textbf{Box 7.16 Gapped quantum liquid phase and topologically ordered phase}

Two gapped
quantum liquids, given by two sequences of ground-state spaces
$\{\mathcal{V}_{N_k}\}$ and $\{\mathcal{V}'_{N_k}\}$ (on graphs with no boundary), are
equivalent if they can be connected via LU transformations.
The equivalence classes of
gapped quantum liquids are gapped quantum liquid phases (See Fig.~\ref{liquidLUVk}).
\end{center} 
\end{svgraybox}

To study the universality classes of many-body wave functions, a natural idea
is from the LU transformations as discussed in Sections~\ref{Qphase_LU} and~\ref{sec:liquid}.  
We will analyze
the classes of wave functions under LU transformations, or more generally, gLU
transformations.

As discussed above, the gLU transformations define an equivalence relation among
many-body ground-state spaces.  The equivalence classes defined by such an
equivalence relation will be called the gLU classes.  The gLU classes of gapped
quantum liquids correspond to gapped quantum liquid phases.

We now ask the following question.
\begin{svgraybox}
\begin{center}
\textbf{Box 7.17 gLU classes}

Since the notion of the gLU classes does not require symmetry, do the gLU classes of gapped quantum liquid have a one-to-one correspondence
with topological orders (as given in Box 7.12)?
\end{center} 
\end{svgraybox}
We will show that the answer is no, i.e.\ there are \emph{unstable} gapped quantum
liquids. Only the gLU classes for \emph{stable} gapped quantum liquids have a
one-to-one correspondence with topological orders.

\subsection{Symmetry breaking order}
\label{sec:symb}

Example of  unstable gapped quantum liquids are given by symmetry breaking
states. These unstable gapped quantum liquids are in a different gLU class
from the trivial phase, and thus are non-trivial gapped quantum liquid phases.

Let us consider an example of the unstable gapped quantum liquids, the 1D
transverse Ising model with the Hamiltonian (with periodic boundary condition)
\begin{equation}
H^\text{tIsing}_{N_k}(B)=-\sum_{i=1}^{N_k}Z_iZ_{i+1}-B\sum_{i=1}^{N_k}X_i,
\end{equation}
where $Z_i$ and $X_i$ are the Pauli $Z/X$ operators
acting on the $i$th qubit.
The Hamiltonian $H^\text{tIsing}_{N_k}(B)$ has a $\mathbb{Z}_2$ symmetry, which is given by 
$\prod_{i=1}^{N_k}X_i$.
The gapped ground states are non-degenerate for $B>1$. For $0\leq B<1$, the
gapped ground states are two-fold degenerate.  The degeneracy is unstable
against perturbation that breaks the $\mathbb{Z}_2$ symmetry.

The phase for $B>1$ is a trivial gapped liquid phase.  The phase for $0<B<1$ is
a non-trivial gapped liquid phase.  This is due to a very simple reason: the two
phases have different group state degeneracy, and the ground-state degeneracy
is an gLU invariant.  Gapped quantum liquids with different ground-state degeneracy always belong to different gapped liquid phases.

Now, let us make a more non-trivial comparison. Here we view
$H^\text{tIsing}_{N_k}(B)$ (with $0<B<1$) as a gapped quantum system (rather
than a gapped quantum liquid system).  We compare it with another gapped
quantum system $H^\text{non-liquid}_{N_k}$ (see \eqn{Hnonliquid}) discussed
before.  Both gapped systems have two-fold degenerate ground states.  Do the
two systems belong to the same gapped quantum phase (as given in Box 7.6)?

Consider $H^\text{tIsing}_{N_k}(B)$ for any $0<B<1$ and any size $N_k<\infty$. The (symmetric) exact ground
state $\ket{\Psi^+_{N_k}(B)}$ is an adiabatic continuation of the GHZ state
\begin{equation}
\ket{GHZ^+_{N_k}}=\frac{1}{\sqrt{2}}(\ket{0}^{\otimes N_k}+\ket{1}^{\otimes N_k}),
\end{equation}
i.e.\  $\ket{\Psi^+_{N_k}(B)}$ is in the same gLU class of $\ket{GHZ^+_{N_k}}$. 
There is another state $\ket{\Psi_-(B)}$ below the energy window $\Delta$ which is an adiabatic continuation of the state
\begin{equation}
\ket{GHZ^-_{N_k}}=\frac{1}{\sqrt{2}}(\ket{0}^{\otimes N_k}-\ket{1}^{\otimes N_k}).
\end{equation}
The energy splitting of $\ket{\Psi^+_{N_k}(B)}$ and $\ket{\Psi^-_{N_k}(B)}$ approaches zero as $N_k \to \infty$.

However, we know that the GHZ state $\ket{GHZ^+_{N_k}}$ (hence $\ket{\Psi^+_{N_k}(B)}$) and
the product state $\ket{0}^{\otimes N_k}$ belong to two different gLU classes.
Both states are regarded to have the same trivial topological order.  So gLU
transformations assign GHZ states, or symmetry breaking many-body wave functions, 
to non-trivial classes.  Therefore by studying the gLU classes of
gapped quantum liquids, we can study both the topologically ordered states and
the symmetry breaking states.

To be more precise, the ground-state space of $H^\text{tIsing}_{N_k}(B)$
($0<B<1$) contain non-trivial GHZ states.  On the other hand, the ground-state
space of $H^\text{non-liquid}_{N_k}$ contain only product states. There is
no GHZ states.  That make the two systems  $H^\text{tIsing}_{N_k}(B)$ and
$H^\text{non-liquid}_{N_k}$ to belong to two different gapped quantum phases,
even though the two systems have the same ground-state degeneracy.

We now discuss the concept of `gapped symmetry breaking quantum system'.
\begin{svgraybox}
\begin{center}
\textbf{Box 7.18 Gapped symmetry breaking quantum system}

A gapped symmetry breaking system is a gapped quantum liquid system 
with certain symmetry and degenerate ground states, where
the symmetric ground states have the GHZ-form of entanglement.
\end{center} 
\end{svgraybox}

We recall that as discussed in Chapter~\ref{sec:QECC}, the ground-state space of a gapped symmetry breaking quantum
system is a `classical' error-correcting code with macroscopic distance,
correcting errors that do not break the symmetry. This is to say, for any
orthonormal basis $\{\ket{\Phi_i}\}$ of the ground-state space, for any local
operator $M_s$ that does not break symmetry, we have 
\begin{equation}
\bra{\Phi_i}M_s\ket{\Phi_j}=C_{M_s}\delta_{ij},
\end{equation}
where $M_s$ is a constant that only depends on $M_s$.

Here by `classical' we mean the following. 
For the ground-state space,
there exists a basis $\{\ket{\Phi_i}\}$ 
that is connected by symmetry. 
In this basis, the ground-state space is a classical error-correcting code of macroscopic distance,
in the sense that for any local operator $M$, we have 
\begin{equation}
\label{eq:ccode}
\bra{\Phi_i}M\ket{\Phi_j}=0,\ i\neq j.
\end{equation}
Notice that Eq~\eqref{eq:ccode} does not contain the coherence condition for
$i=j$, which is the requirement to make the space a `quantum' code.

The transverse Ising mode is an example of such a special case with
$\mathbb{Z}_2$ symmetry. The basis that is connected by the $\mathbb{Z}_2$
symmetry are $\ket{\Psi^{\pm}_{N_k}(B)}$.  And it is obvious that
$\bra{\Psi^+_{N_k}(B)}M\ket{\Psi^-_{N_k}(B)}=0,\ i\neq j$, for $N_k\rightarrow\infty$.

\begin{figure}[htbp] 
\begin{center} 
\includegraphics[scale=1.0]{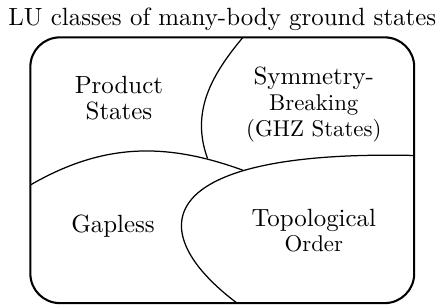} \end{center}
\caption{
LU classes for ground states (many-body wave functions) of local Hamiltonians.
} 
\label{LU1} 
\end{figure}

We have now shown that gapped liquid phases also contain
symmetry breaking phases. We summarize 
the LU classes for ground states of local Hamiltonians
in Fig.~\ref{LU1}.

\subsection{Stochastic local transformations and long-range entanglement}
\label{sec:SL}

We have seen that the non-trivial equivalence classes of many-body wave
functions under the gLU transformations contain both topologically ordered
phases and symmetry breaking phases (described by the symmetric many-body wave
functions with GHZ-form of entanglement).  In this section, we will introduce
the generalized stochastic local (gSL) transformations, which are local
invertible transformations that are not necessarily unitary.  The term
`stochastic' means that these transformations can be realized by generalized
local measurements with finite probability of success. 

We show that the many-body wave functions for symmetry breaking phases (i.e.\ the
states of GHZ-form of entanglement) are convertible to the product states under
the gSL transformations with a finite probability, while the
topological ordered states are not. 
This allows discuss the notion of
long-range entanglement under which only topologically ordered states are
long-range entangled.  We further show that the topological orders are stable
against small stochastic local transformations, while the symmetry breaking
orders are not.



The idea for using gSL transformations is simple.  The topologically
\emph{stable} degenerate ground states for a topologically ordered system is
not only stable under real-time evolutions (which are described by gLU
transformations), they are also stable and are the fixed points under
imaginary-time evolutions.  The imaginary-time evolutions of the ground states
are given by the gSL transformations (or local non-unitary transformations),
therefore the topological orders are robust under (small) gSL transformations.

On the other hand, the states of GHZ-form of entanglement are not robust under
small gSL transformations, and can be converted into product states with a
finite probability.  Thus, there is no emergence of unitarity for symmetry-
breaking states.

To discuss gSL transformations, we recall from Chapter~\ref{sec:OpenQS} that the most general form of
quantum operations are
completely-positive trace-preserving maps. A quantum
operation $\mathcal{E}$ acting on any density matrix $\rho$ has the form
\begin{equation}
\mathcal{E}(\rho)=\sum_{k=1}^r A_k\rho A_k^{\dag},
\end{equation} 
with 
\begin{equation}
\sum_{k=1}^r A_k^{\dag}A_k=I,
\end{equation} 
where $I$ is the identity operator.

The operators $A_k$ are called Kraus operators
of $\rho$ and satisfies
\begin{equation}
A_k^{\dag}A_k\leq I.
\end{equation}
This means that the operation $A_k\rho A_k^{\dag}$
can be realized with probability $\Tr(A_k\rho A_k^{\dag})$
for a normalized state $\Tr\rho=1$. In the following 
we will drop the label $k$ for the measurement outcome. 

We will now definite gSL transformations along a similar line 
as the definition of gLU transformations. Let us first define
a layer of SL transformation that has a form
\begin{equation*}
 W_{pwl}= \prod_{i} W^i
\end{equation*}
where $\{ W^i \}$ is a set of invertible operators that act on non-overlapping
regions, and each $W^i$ satisfies
\begin{equation}
\label{eq:trace}
W^{i\dag}W^i\leq I.
\end{equation}
The size of each region is less than a finite number $l$. The
invertible operator $W_{pwl}$ defined in this way is called a layer of
piecewise local stochastic transformation with a range $l$.  

A stochastic local (SL) transformation is then given by a finite layers of piecewise
local invertible transformation:
\begin{equation*}
 W^M_{circ}= W_{pwl}^{(1)} W_{pwl}^{(2)} \cdots W_{pwl}^{(M)}
\end{equation*}
We note that such a transformation does not change the degree of freedom of
the state.

Similarly to the gLU transformations, we can also have a transformation that can change the degree of freedom of
the state, by a tensor product of the state with another product state $
 \ket{\Psi} \to 
\Big(\otimes_i\ket{\psi_i}\Big) 
\otimes \ket{\Psi}
$,
where $\ket{\psi_i}$ is the wave function for the $i^\text{th}$ qubit.  A finite
combination of the above two types of transformations is 
then a generalized stochastic local (gSL) transformation. Here 
we use the notion $\ket{\Psi}$ to represent a sequence of states $\{\ket{\Psi}_{N_k}\}$.

We remark that, although it is similar to the gLU transformations, gSL transformations are more subtle to deal with. 
First of all, notice that gSL transformations do not preserve the norm of quantum states (i.e.\ not trace-preserving, as given by Eq.~\eqref{eq:trace}).
Furthermore, as we are dealing with thermodynamic limit ($N_k\rightarrow\infty$), we are applying gSL transformations on a system of infinite dimensional Hilbert space. In this case, even if each $W^i$ is invertible, $W_{pwl}= \prod_{i} W^i$ may be non-invertible due to 
the thermodynamic limit. We will discuss these issues in more detail
in the next subsection.

It is known in fact that the SL convertibility in 
infinite dimensional systems is subtle, and to avoid
technical difficulties dealing with the infinite dimensional Hilbert space,
we would instead use $\epsilon$-convertibility instead to talk about the exact convertibility of states under gSL. For simplicity
we will omit the notation `$\epsilon$' and still name it `gSL convertibility'.

\begin{svgraybox}
\begin{center}
\textbf{Box 7.19 Convertibility by gSL transformation}

We say that $\ket{\Psi}$ is convertible to $\ket{\Phi}$ by a gSL transformation, if for any $\epsilon>0$, there exists an integer $N$, a probability $0<p<1$, and gSL transformations $W_{N_k}$, such that for any $N_k>N$, $W_{N_k}$ satisfy the condition 
\begin{equation}
\left\|\frac{W_{N_k}(\ket{\Psi_{N_k}}\bra{\Psi_{N_k}})W^{\dag}_{N_k}}{\Tr\left(W_{N_k}(\ket{\Psi_{N_k}}\bra{\Psi_{N_k}})W^{\dag}_{N_k}\right)}
-\frac{\ket{\Phi}\bra{\Phi_{N_k}}}{\Tr(\ket{\Phi_{N_k}}\bra{\Phi_{N_k}})}\right\|_{\text{tr}}<\epsilon,
\end{equation}
where $\|\cdot\|_{\text{tr}}$ is the trace norm and
\begin{equation}
\label{eq:prob}
\frac{\Tr(W_{N_k}\ket{\Psi_{N_k}}\bra{\Psi_{N_k}}W^{\dag}_{N_k})}{\Tr(\ket{\Psi_{N_k}}\bra{\Psi_{N_k}})}>p.
\end{equation}
\end{center} 
\end{svgraybox}

The idea underlying the definition in Box 7.19 is that $\ket{\Psi}$ can be transformed to any neighbourhood of $\ket{\Phi}$, though not $\ket{\Phi}$ itself, and these neighbourhood states become indistinguishable from $\ket{\Phi}$ in the thermodynamic limit.

Using the idea of gSL transformations, we can have a  definition for short-range
and long-range entanglement.
\begin{svgraybox}
\begin{center}
\textbf{Box 7.20 Short/long-range entanglement}

A state is short-range entangled (SRE) if it is convertible to a product
state by a gSL transformation. Otherwise the state is long-range entangled (LRE).
\end{center} 
\end{svgraybox}

Under this definition, the states which can be transformed to
product states by gLU transformations are SRE. However, the
SRE states under gSL transformations will also include some
of the states that cannot be transformed to 
product states by gLU transformations. 

As an example, the state 
\begin{equation}
\ket{GHZ^+_{N_k}(a)}=a\ket{0}^{\otimes N_k}+b\ket{1}^{\otimes N_k}
\end{equation} 
with $|a|^2+|b|^2=1$
cannot be transformed 
to product states under gLU transformations.
However if one allows gSL transformations,
then all the $\ket{GHZ^+_{N_k}(a)}$ are convertible to $\ket{GHZ^+_{N_k}(1)}$, i.e.\ the
product state $\ket{0}^{\otimes N_k}$. To see
this, one only needs to apply the gSL transformation
\begin{equation}
\label{eq:SLeg}
W_{N_k}=\prod_{i=1}^{N_k} O_i,
\end{equation}
where $O_i$ is the invertible operator
\begin{equation}
\begin{pmatrix}
1 & 0 \\ 0 & \gamma
\end{pmatrix}
\end{equation}
acting on the $i$ the qubit, and $0<\gamma<1$.
And we have 
\begin{equation}
\begin{pmatrix}
1 & 0 \\ 0 & \gamma
\end{pmatrix}^{\dag}\begin{pmatrix}
1 & 0 \\ 0 & \gamma
\end{pmatrix}\leq \begin{pmatrix}
1 & 0 \\ 0 & 1
\end{pmatrix}=I.
\end{equation}

That is
\begin{eqnarray}
\label{eq:GHZc}
W_{N_k}\ket{GHZ^+_{N_k}(a)}=
a\ket{0}^{\otimes N_k}+b\gamma^{N_k}\ket{1}^{\otimes N_k}=\ket{\Gamma_{N_k}(a)}.
\end{eqnarray}
Obviously, the right hand side of Eq.~\eqref{eq:GHZc}
can be arbitrarily close to the product state $\ket{0}^{\otimes N_k}$ 
as long as $N_k$ is large enough. Furthermore, $\Tr(\ket{\Gamma_{N_k}(a)}\bra{\Gamma_{N_k}(a)})>|a|^2$
for any $N_k$.
Therefore, according to Box 7.19, $\ket{GHZ^+_{N_K}(a)}$ 
is convertible to the product state $\ket{0}^{\otimes N_k}$ by the gSL transformation $W_{N_k}$.

If $\ket{\Psi}$ is convertible to $\ket{\Phi}$ by a gSL transformation, we write
\begin{equation}
\ket{\Psi}\xrightarrow{\text{gSL}}\ket{\Phi}.
\end{equation}
Notice that $\ket{\Psi}\xrightarrow{\text{gSL}}\ket{\Phi}$ does not mean
$\ket{\Phi}\xrightarrow{\text{gSL}}\ket{\Psi}$. 
For example, while we have
\begin{equation}
\ket{GHZ^+_{N_k}(a)}\xrightarrow{\text{gSL}}\ket{0}^{\otimes N_k},
\end{equation}
$\ket{0}^{\otimes N_k}$ is not gSL convertible to 
$\ket{GHZ^+_{N_k}(a)}$.

That is, 
the gSL convertibility is not an equivalence relation. 
It instead defines a partial order (in terms of set theory)
on all the quantum states. That is,
if $\ket{\Psi}\xrightarrow{\text{gSL}}\ket{\Phi}$
and $\ket{\Phi}\xrightarrow{\text{gSL}}\ket{\Omega}$,
then $\ket{\Psi}\xrightarrow{\text{gSL}}\ket{\Omega}$.
And there exists $\ket{\Psi}$ and $\ket{\Phi}$
that is not comparable under gSL, i.e.\
neither $\ket{\Psi}$ is gSL convertible to
$\ket{\Phi}$, nor is $\ket{\Phi}$ gSL convertible to
$\ket{\Psi}$. Based on this partial order we can 
further define equivalent classes.

\begin{svgraybox}
\begin{center}
\textbf{Box 7.21 gSL equivalent states}

We say that two states $\ket{\Psi}$ and $\ket{\Phi}$
are equivalent under gSL transformations if they
are convertible to each other by gSL transformations.
That is, $\ket{\Psi}\xrightarrow{\text{gSL}}\ket{\Phi}$
and $\ket{\Phi}\xrightarrow{\text{gSL}}\ket{\Psi}$.
\end{center} 
\end{svgraybox}

Under this notion, all the states
$\ket{GHZ^+_{N_k}(a)}$ are in the same gSL class
unless $a=0,1$. The product states
with $a=0,1$ are not in the same gSL class,
but any $\ket{GHZ^+_{N_k}(a)}$ is convertible 
to the product states by gSL transformations.
The converse is not true, that a product state 
is not convertible to $\ket{GHZ^+_{N_k}(a)}$ 
with $a \neq 0,1$ by gSL transformations. 

That is to say, the states with GHZ-form of entanglement
are indeed `more entangled' than product states, but they are 
`close enough' to produce states under gSL transformations.
Furthermore, the topological entanglement entropy
$S^\text{t}_\text{topo}$ for these types of states are unstable under 
small gSL transformations.
In this sense, we can still treat the GHZ-form of entanglement
as product states, i.e.\ states with no long-range entanglement.


We can now define topologically ordered states based on gSL 
transformations (notice that Box 7.12 defines topological order
through properties of the Hamiltonian).

\begin{svgraybox}
\begin{center}
\textbf{Box 7.22 Topologically ordered states}

Topologically ordered states are LRE gapped quantum liquids.  
In other words, a ground state $\ket{\Psi}$ of a gapped Hamiltonian has a
nontrivial topological order if it is not convertible to a product state by
any gSL transformation.
\end{center} 
\end{svgraybox}

Not all LRE states can be transformed into each other
via gSL transformations.  Thus LRE states can belong to different phases: i.e.\
the LRE states that are not connected by gSL transformations belong to different
phases.  When we restrict ourselves to LRE gapped quantum liquids, those different
phases are nothing but the topologically ordered
phases.

\begin{svgraybox}
\begin{center}
\textbf{Box 7.23 Topologically ordered phases} 

Topologically ordered phases
are equivalence classes of LRE gapped quantum liquids under the gSL
transformations.
\end{center} 
\end{svgraybox}

We now consider the property of the topological entanglement entropy $S_\text{topo}$ (discussed in Chapter~\ref{sec:irr-corr}) under the local transformations. It is known that $S_\text{topo}$ an LU invariant. And it is also believed that in general the quantum conditional mutual information $I(A{:}C|B)$ (hence the generalized topological entanglement entropy $S_\text{topo}^\text{t}$ or $S_\text{topo}^\text{q}$ as discussed in Chapter 5) is also an LU invariant. We summarize this observation as below. 

\begin{svgraybox}
\begin{center}
\textbf{Box 7.24 (Generalized) topological entanglement entropy under gLU transformations} 

For large enough areas $A,B,C$ and $A,C$ far from each other,
the quantum conditional mutual information $I(A{:}C|B)$ is invariant under gLU transformations. Consequently, the topological entanglement entropy $S_\text{topo}$ and the generalized topological entanglement entropy $S_\text{topo}^\text{t}$ are $S_\text{topo}^\text{q}$ are all gLU invariants. 
\end{center} 
\end{svgraybox}

However, in general $I(A{:}C|B)$ is not an gSL invariant. However, for small gSL transformations, the topological entanglement entropy $S_\text{topo}$ stands out, which will remain unchanged. We believe the following observation is true, which provides a support
to the above picture and notion of topologically ordered phases.

\begin{svgraybox}
\begin{center}
\textbf{Box 7.25 $S_\text{topo}$ under small gSL transformations} 

The topological entanglement entropy $S_\text{topo}$
for topological order is stable under small gSL transformations.
Furthermore, $S_\text{topo}$ is an invariant for any gSL equivalence class
of topological orders.
\end{center} 
\end{svgraybox}

Similarly, for symmetry breaking orders, we have
\begin{svgraybox}
\begin{center}
\textbf{Box 7.26 $S^\text{t}_\text{topo}$ under small gSL transformations} 

The generalized entanglement entropy $S^\text{t}_\text{topo}$
for symmetry breaking orders is stable 
under small gSL transformations that do not break symmetry, but unstable
under small gSL transformations that break the symmetry.
Furthermore, $S^\text{t}_\text{topo}$ is not an invariant for any gSL equivalence class
of symmetry breaking orders.
\end{center} 
\end{svgraybox}

As an example, in the transverse Ising model, the gSL transformation which
transforms $\ket{GHZ^+_{N_k}(a)}$ of different $a$ breaks the $\mathbb{Z}_2$ symmetry.
However, $\ket{GHZ^+_{N_k}(a)}$ of different $a$ are in the same gSL equivalent
class, yet with different topological entanglement entropy. 

The second sentence of Box 7.26 is more subtle,
as the topological entanglement entropy $S_\text{topo}$ for topological order
is not an invariant of gSL transformations (as a finite probability $p$ as given 
in Eq.~\eqref{eq:prob} may not exist). This is because that unlike gLU transformations, gSL 
transformations can be taken arbitrarily close to a non-invertible transformation. For instance, take the gSL transformation $W_{N_k}$ as given in Eq.~\eqref{eq:SLeg}. If we allow $\gamma$
to be arbitrarily close to zero, then for any wave function, applying 
$W_{N_k}$ is `as if' we are just projecting everything to $\ket{0}^{N_k}$,
which should not protect any topological order.

On the other hand, the option to choose $\gamma$ arbitrarily small does not mean any quantum state is gSL convertible to a product state. The key point here is the existence of a finite probably $p$ that is independent of system size $N_k$, as given in Box 7.19. For states with GHZ-form of entanglement, we know that we can always find such a finite probability $p$. 

However, for topological ordered states, there does not exist such a finite probability $p$. In fact,
we have $p\rightarrow 0$ when $N_k\rightarrow\infty$, and furthermore the speed of $p$ approaching $0$ may be exponentially fast in terms of the growth of $N_k$. Therefore $S^\text{tri}_\text{topo}$ shall remain invariant within any gSL equivalent class.

The above idea is further supported by the results known for the geometric measure of entanglement for topological ordered states.
More precisely,  let us divide the system to $m$ non-overlapping local parts, as 
illustrated in Fig.~\ref{qc} for one layer. Label each part by $i$ and write the Hilbert
space of the system by $\mathcal{H}=\bigotimes_{i=1}^M\mathcal{H}_i$. Now
for any normalized wave function $\ket{\Psi}\in\mathcal{H}$, the goal
is to determine how far $\ket{\Psi}$ is from a normalized product state 
\begin{equation}
\ket{\Phi}=\otimes_{i=1}^M\ket{\phi_i}
\end{equation}
with $\ket{\phi_i}\in\mathcal{H}_i$.

Recall that, as discussed in Chapter~\ref{cp:1}, the  geometric measure of entanglement $E_G(\ket{\Psi})$ is then revealed by the maximal overlap
\begin{equation}
\Lambda_{\max}(\ket{\Psi})=\max_{\ket{\Phi}}|\langle\Phi\ket{\Psi}|,
\end{equation}
and is given by
\begin{equation}
E_G(\ket{\Psi})=-\log\Lambda^2_{\max}(\ket{\Psi}).
\end{equation}
Notice that for $\Lambda_{\max}(\ket{\Psi})$, the maximum is also taken for all the partition
of the system into local parts. 

For a topologically ordered state $\ket{\Psi}$, 
$E_G(\ket{\Psi})$ is proportional to the number of qubits in the system. This means that
the probability to project $\ket{\Psi}$ to any product state is exponentially small in terms
of the system size $N_k$. Therefore one shall not expect $\ket{\Psi}$ to be convertible to
any product state with a finite probability $p$.

In contract, the geometric entanglement for states with GHZ-form of entanglement is
a constant independent of the system size $N_k$. As an example,
for the state
$
\ket{GHZ^+_{N_k}},
$
the maximal overlap
$
\Lambda_{\max}(\ket{GHZ_{+}})=\frac{1}{2},
$
with the maximum at either $\ket{0}^{\otimes N_k}$ or $\ket{1}^{\otimes N_k}$, hence the geometric measure of entanglement is
$
E_G(\ket{GHZ^+_{N_k}})=1.
$
And it remains to be the case for the entire symmetric-breaking phase, which indicates that these GHZ-form states are convertible to product states with some finite probability $p$.  

\section{Symmetry-protected topological order}
\label{TO_MBET}

In the above discussions, we have defined phases without any
symmetry consideration.  The $\t H(g)$ or $U_{pwl}$ in the
LU transformation does not need to have any symmetry and can
be the sum / product of any local operators.  In this case, two
Hamiltonians with an adiabatic connection are in the same
phase even if they may have different symmetries.  Also, all
states with short-range entanglement belong to the same
phase (under the LU transformations that do not have any
symmetry).

On the other hand, we can consider only Hamiltonians $H$
with certain symmetries and define phases as the equivalent
classes of symmetric local unitary transformations:
\begin{equation}
\begin{array}{llll}
   |\Psi\>\sim &\cT\Big( e^{-i \int_0^1 d g\; \t H(g)} \Big)
|\Psi\> \
 \text{ or }\ |\Psi\> \sim U^M_{circ} |\Psi\>
\end{array}
\end{equation}
where $\t H(g)$ or $U^M_{circ}$ has the same symmetries as
$H$. \footnote{We note that the symmetric local unitary transformation
in the form $\cT\Big( e^{-i \int_0^1 d g\; \t H(g)}
\Big)$ always connect to the identity transformation
continuously.  This may not be the case for the
transformation in the form $U^M_{circ}$.  To rule out that
possibility, we define symmetric local unitary
transformations as those that connect to the identity
transformation continuously.}

The equivalence classes of the symmetric LU transformations
have very different structures compared to those of LU
transformations without symmetry.  Each equivalence class of
the symmetric LU transformations is smaller and there are
more kinds of classes, in general.

In particular, states with short range entanglement can belong to
different equivalence classes of the symmetric LU
transformations even if they do not spontaneously break any symmetry of
the system. (In this case, the ground states have the same symmetry.)  We say those states
have Symmetry Protected Topological orders. Haldane
phase in spin-1 chains and the spin-0 chains are
examples of states with the same symmetry which belong to
two different equivalence classes of symmetric LU
transformations (with spin rotation symmetry).
Band and topological
insulators are
other examples of states that have the same symmetry and at
the same time belong to two different equivalence classes of
symmetric LU transformations (with time reversal symmetry). Systems with symmetry protected topological order cannot have ground state degeneracy, fractional charge and statistics, nor nonzero topological entanglement entropy. They can, however, have gapless edge excitations which are protected by symmetry.

We are now ready to summarize what we have learned and obtain a general structure of the quantum phase diagram of gapped systems at zero temperature.

\begin{figure}[htbp]
\begin{center}
\includegraphics[scale=0.9]{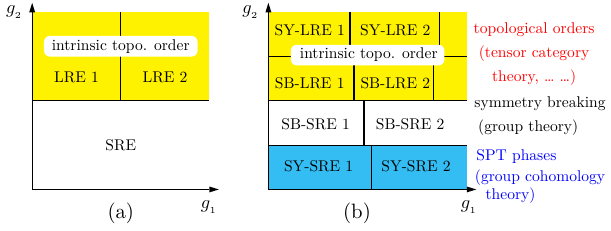}
\end{center}
\caption{
(a) The possible phases for a Hamiltonian $H(g_1,g_2)$ without
any symmetry.
(b) The possible phases for a Hamiltonian $H_\text{symm}(g_1,g_2)$ with
some symmetries.
The shaded regions in (a) and (b) represent the phases with
short range entanglement (i.e. those ground states
can be transformed into a direct product state via a generic LU
transformations that do not have any symmetry.)
}
\label{topsymm}
\end{figure}

Fig. \ref{topsymm} compares the structure of phases for
systems without any symmetry and systems with some symmetry
in more detail.  

For a system without any symmetry, all the
short-range-entangled (SRE) states (i.e. those ground states
can be transformed into a direct product state via a generic
LU transformations that do not have any symmetry) are in the
same phase (SRE in Fig. \ref{topsymm}(a)).  On the other
hand, long range entanglement (LRE) can have many different
patterns that give rise to different `intrinsic' topological phases (LRE
1 and LRE 2 in Fig.  \ref{topsymm}(a)).  The different `intrinsic'
topological orders usually give rise to quasi particles with
different fractional statistics.

For a system with some symmetries,
the phase structure can be much more complicated.
The short-range-entangled states no longer belong to
the same phase, since the equivalence relation is described
by more special symmetric LU transformations:\\
(A) States with short range entanglement belong to
different equivalence classes of the symmetric LU
transformations if they break symmetry in different ways.
They correspond to the symmetry breaking (SB)
short-range-entangled phases SB-SRE 1 and SB-SRE 2 in
Fig.  \ref{topsymm}(b).  They are Landau's symmetry
breaking states.\\
(B) States with short range entanglement can belong to
different symmetry protected topological phases if they do not break any symmetry of
the system. They correspond to
the symmetric (SY) short-range-entangled phases SY-SRE 1 and
SY-SRE 2 in Fig.  \ref{topsymm}(b).  

Also, for a system with some symmetries,
the long-range-entangled states are
divided into more classes (more phases):\\
(C) Symmetry breaking and long range entanglement can appear
together in a state, such as SB-LRE 1, SB-LRE 2, etc. in
Fig.  \ref{topsymm}(b). The topological superconducting
states are examples of such phases.\\
(D) Long-range-entangled states that do not break any
symmetry can also belong to different phases such as the
symmetric long-range-entangled phases SY-LRE 1, SY-LRE 2,
etc. in Fig.  \ref{topsymm}(b). They are called the Symmetry Enriched Topological Phases. The many different $\mathbb{Z}_2$
symmetric spin liquids with spin rotation, translation,
and time-reversal symmetries are examples of those
phases.
Some time-reversal symmetric topological orders, called
topological Mott-insulators or fractionalized
topological insulators, also belong to this case.

Having obtained the general structure of the phase diagram, our next goal is to find out all the entries in the diagram, or in other words, to classify all possible
phases in strongly correlated systems, especially the topological ones. In the next two chapter, we will study topological phases in one and two dimensions, with the help of tensor network representations. 

\section{A new chapter in physics}

Our world is rich and complex. When we discover the inner working of our world
and try to describe it, we ofter find that we need to invent new mathematical
language describe our understanding and insight.  For example, when Newton
discovered his law of mechanics, the proper mathematical language was not
invented yet.  Newton (and Leibniz) had to develop calculus in order to
formulate the law of mechanics.  For a long time, we tried to use theory of
mechanics and calculus to understand everything in our world.

As another example, when Einstein discovered the general equivalence principle
to describe gravity, he needed a  mathematical language to describe his theory.
In this case, the needed mathematics, Riemannian geometry, had been developed,
which leaded to the theory of general relativity.  Following the idea of
general relativity, we developed the gauge theory. Both general relativity and
gauge theory can be described by the mathematics of fiber bundles.  Those
advances led to a beautiful geometric understanding of our world based on
quantum field theory, and we tried to  understand everything in our world in
term of quantum field theory.

It appears that we are at another turning point.  In a study of quantum
matter, we find that long-range entanglement can give rise to many new quantum
phases.  So long-range entanglement is a natural phenomenon that can happen in
our world. This greatly expand our understanding of possible quantum phases,
and bring the research of quantum matter to a whole new level.  To gain a
systematic understanding of new quantum phases and long-range entanglement, we
like to know what mathematical language should we use to describe long-range
entanglement?  The answer is not totally clear. But early studies suggest that
tensor category and group cohomology should be a part of the mathematical frame
work that describes long-range entanglement.  The further progresses in this
direction will lead to a comprehensive understanding of long-range
entanglement and topological  quantum matter.

However, what is really exciting in the study of quantum matter is that it
might lead to a whole new point of view of our world. This is because
long-range entanglement can give rise to both gauge interactions and Fermi
statistics.  In contrast, the geometric point of view can only lead to gauge
interactions. So maybe we should not use geometric pictures, based on fields
and fiber bundles, to understand our world.  Maybe we should use entanglement
pictures to understand our world.  This way, we can get both gauge interactions
and fermions from a single origin -- qubits.  We may live in a truly quantum
world.  So, quantum entanglement represents a new chapter in physics.

\section{Summary and further reading}

In this chapter, we start to establish a microscopic theory for topological order. We use tools from quantum information theory to characterize many-body entanglement. We begin from the fundamental notion of gapped quantum phase and phase transition and explore its implication on the structure of the ground-state wave function. We find that an equivalence relation can be established between ground states of gapped quantum systems in the same phase in terms of a local unitary (LU) transformation which takes the form of either a finite time unitary evolution with a local Hamiltonian or a finite depth quantum circuit. Such a LU transformation gives rise to a renormalization group flow on gapped quantum states which can be used to simplify the wave functions and flow the states to fixed points.

We develop a general framework to study topological order, in thermodynamic limit. We introduce the concept of `gapped quantum liquid', and show that topological
orders are stable gapped quantum liquids. Classifying topological order hence corresponds to classifying stable gapped quantum liquids. We show that symmetry breaking orders for on-site symmetry are also gapped quantum liquids, but with unstable ground-stable degeneracy. The universality classes of generalized local unitary (gLU) transformations contains both topologically ordered states and symmetry breaking states.

We introduce the concept of stochastic local (SL) transformations, and show that the universality classes
of topological orders and symmetry breaking orders can be distinguished by SL: small SL transformations can convert the symmetry breaking classes to the trivial class of product states with finite probability of success, while the topological-order classes are stable against any small SL transformations, demonstrating a phenomenon of emergence of unitarity. Based on the small SL transformations, we 
give a definition of long-range entanglement (LRE), under which only topologically ordered states are long-range entangled. This then implies that the key to topological order is the existence of LRE in the ground-state wave function which cannot be changed under LU and small SL transformation.
This allows us to obtain a general theory to study topological order and symmetry breaking order within a same framework. 
Based on such an understanding, a general structure of the quantum phase diagram is obtained which contains much more possibilities than that given by the conventional symmetry- breaking theory on phase and phase transitions. 

The idea of local unitary transformation and wave funcion renormalization has been used in various studies of quantum states. In~\cite{LW0510},
the wave function renormalization for string-net states is discussed, which can reduce the string-net wave
functions to very simple forms~\cite{LW0605}.
In~\cite{VCL0501}, the local unitary transformations
described by quantum circuits was used to define a
renormalization group transformations for states and
establish an equivalence relation in which states are
equivalent if they are connected by a local unitary
transformation.  Such an approach was used to classify 1D
matrix product states.
In~\cite{Vidal0705}, the local unitary
transformations with disentanglers was used to perform a renormalization
group transformations for states, which give rise to the
multi-scale entanglement renormalization ansatz (MERA) in
one and higher dimensions.  The disentanglers and the
isometries in MERA can be used to study quantum phases and
quantum phase transitions in one and higher dimensions. For a class of exactly solvable Hamiltonians which come from the stabilizer codes in quantum computation, topological order has also been classified using local unitary circuits~\cite{Yoshida1115}.

In establishing the equivalence relation in terms of local unitary transformations, the quasi-adiabatic continuation plays an important role and was proved in \cite{HW0541}. It proves that any local observable changes smoothly when one gapped Hamiltonian is changed into another without closing gap and provides an explicit local unitary transformation between their ground states. An important idea used in the proof is the existence of an upper bound on the interaction propagation velocity in a gapped quantum system, which was derived in \cite{LR7251}. An improved version of the local unitary transformation can be found in \cite{BHM1012}. In \cite{BHV0601} it was further shown that in order to connecte different topological phases, the quantum circuit needs to have a depth which scales at least linearly with system size. On the other hand, the simulability of local unitary evolution by a local unitary quantum circuit was shown in \cite{Lloyd9673}. \cite{CGW1038} gives a general discussion of the local unitary equivalence condition, the relation between topological order and long/short range entanglement, and wave function renormalization.

The concept of gapped quantum liquid is introduced in~\cite{zeng2015gapped}. The discussions in Chap.7.4-7.5 are mainly based on (with some parts taken from)~\cite{zeng2015gapped}. The cubic code of the Haah model provides an example of
gapped quantum system that is not a gapped quantum liquid
system~\cite{haah2011local}. Denote
$H^\text{Haah}_{N_k}$ the Hamiltonian of the cubit code of size $N_k$. There exists a sequence of the linear sizes of
the cube: $L_k\rightarrow\infty$, where the ground-state degeneracy is two,
provided that $L_k=2^k-1$ (or $L_k=2^{2k+1}-1$) for any integer $k$, and
correspondingly $N_k=L_k^3$.  However, 
$H^\text{Haah}_{N_{k+1}}$ cannot be connected by a gLU transformation~\cite{haah2014bifurcation,swingle2014renormalization}. 
Topological quantum liquid is also discussed in therms of the $s$ source framework~\cite{swingle2014renormalization}, 
which is shown to obey area law and have $s\leq 1$. The cubic code can be described by the generalized $s$-source framework~\cite{haah2014bifurcation,swingle2014renormalization}.  Also a stable gapped quantum system may not be a gapped quantum liquid system. A non-Abelian
quantum Hall states~\cite{MR9162,W9102} with traps as discussed in~\cite{LW1384} that trap non-Abelian quasiparticles is an example.

The definition of topological order given in Box 7.11 also include the trivial order. Under this definition we can say that topological orders form a monoid under the stacking operation~\cite{KW1458}. And when we restrict ourselves to LRE gapped quantum liquids, the different
phases are nothing but the topologically ordered
phases~\cite{Wtop,WNtop,Wrig,KW9327,furukawa2006systematic,furukawa2007reduced,nussinov2009symmetry,
schuch2010peps}. The relationship between topological order and quantum error-correcting codes is discussed in~\cite{bravyi2010topological}. 

The term
`stochastic' means that these transformations can be realized by generalized
local measurements with finite probability of success, which is introduced in~\cite{bennett2000exact}. The `convertibility by gSL transformation' given in Box 7.19 borrows the idea of ~\cite{owari2008varepsilon} to use $\epsilon$-convertibility instead to talk about the exact convertibility of states under gSL.

As another example, we can see how to convert a ground state of any 1D gapped quantum liquid to a product state by gSL transformations. Hence there is no long-range entangled states (i.e.\ no topological order) in 1D systems (for details, see Chapter~\ref{sec:no_sym}). We may use the isometric form of the matrix product state representation  ~\cite{SPC5139} 
\begin{equation}
\sum_{\alpha}\ket{\alpha,\ldots,\alpha}\otimes\ket{\omega_{D_{\alpha}}}^{\otimes N_k}
\end{equation} 
(for more details, see Chapter~\ref{chap8}).
This state is the convertible to a product state by gSL transformations via two steps: the first step is an gLU transformation to convert the $\ket{\omega_{D_{\alpha}}}^{\otimes N_k}$ part to a product state and end up with a GHZ state. The the next step is to apply the gSL transformation $W_{N_k}$ as given in Eq.~\eqref{eq:GHZc}, which transforms the GHZ state to a product state with a finite probability.

The geometrical entanglement for topological ordered states is discussed in~\cite{orus2014geometric}, which shows that for a topologically ordered state $\ket{\Psi}$, 
$E_G(\ket{\Psi})$ is proportional to the number of qubits in the system. And the geometrical entanglement case for the symmetric-breaking phase is discussed in~\cite{wei2005global}, which indicates that these GHZ-form states are convertible to product states with some finite probability $p$.  

Examples of `intrinsic' topologically ordered systems include quantum Hall
systems\cite{WN9077}, chiral spin liquids,\cite{KL8795,WWZ8913} $\mathbb{Z}_2$ spin
liquids,\cite{RS9173,W9164,MS0181} quantum double model\cite{K0302} and
string-net model\cite{LW0510}. 
Examples of symmetry protected topological phases include the Haldane
phase\cite{H8364} of spin-1 chain\cite{GW0931,PBT1225} and
topological
insulators\cite{KM0501,BZ0602,KM0502,MB0706,FKM0703,QHZ0824}.
The topological superconducting
states are examples of topologically ordered phases with symmetry breaking.\cite{RG0067,KLW0902}
Examples of symmetry enriched topological phases include various $\mathbb{Z}_2$ spin liquids with spin rotation, translation,
and time-reversal symmetries\cite{W0213,KLW0834,KW0906} and topological Mott-insulators or fractionalized
topological insulators with time reversal symmetry\cite{RQH0801,ZRV0931,PB1076,YK1011,MQK1009,SBM1076}.

%
%
\bibliographystyle{plain}
\bibliography{Chap7}

%
%
%

\begin{partbacktext}
\part{Gapped Topological Phases and Tensor Networks}

\end{partbacktext}

\chapter{Matrix Product State and 1D Gapped Phases}
\label{chap8} 

\abstract{Based on the general notions introduced in the previous chapters, including local unitary transformations and short / long range entanglement, we study gapped phases in one spatial dimension in this chapter. Our goal is to understand what short / long range entangled phases exist in 1D and for this purpose, a useful tool is the matrix product state representation. The matrix product state representation provides an efficient description of the ground state wave function of 1D gapped systems. We introduce this formalism in this chapter and discuss its various properties. By mapping matrix product states to their fixed point form through renormalization group transformations, we show that there is actually no long range entangled phase, hence no intrinsic topological order, in one dimensional spin systems.
}


\section{Introduction}
\label{sec:intro_MPS}

Having established the general structure of the quantum phase diagram and the criteria for classifying gapped quantum phases, we would like to apply it to condensed matter systems of interest. In this chapter, we consider one dimensional gapped boson / spin systems, and try to find all possible short / long range entangled phases. 
Completely classifying strongly correlated boson / spin systems seems to be a
hard task as in general strongly interacting quantum
many-body systems are very hard to solve. Instead of starting from the Hamiltonian, we focus on the ground state wave function of the system which encodes all the important low energy property of the phase. It was realized that the many body entanglement pattern in 1D gapped ground states has very nice structural properties, allowing a complete understanding of the quantum phases they correspond to.

In particular, it has been shown that 1D gapped ground states can be well represented using the Matrix Product State representation. On the one hand, matrix product states
capture the essential features of 1D gapped ground states,
like an entanglement area law and
a finite correlation length, and provide an
efficient description of the wave function. On the
other hand, generic matrix product states satisfying a
condition called `injectivity' are all gapped ground states
of local 1D Hamiltonians. Therefore, studying
this class of matrix product states will enable us to give a full
classification of 1D gapped systems.

Now the question of what gapped phases exist in 1D boson / spin systems can be restated as what equivalence classes of matrix product states exist under local unitary
transformations. To answer this question, we first introduce the definition and basic properties of matrix product states in section \ref{sec:MPS}, including their entanglement property, gauge degree of freedom, parent Hamiltonian, etc. Next, in section \ref{sec:RG_MPS}, we describe a way to perform wave function renormalization group transformation on matrix product states and obtain a simple fixed point form. Using such a procedure, we are able to show in section \ref{sec:no_sym} that there are no long range entangled phases in 1D boson / spin systems and all short range entangled states belong to the same phase. In other words, there is no intrinsic topological order in 1D. Note that there is no fundamental difference between spin and boson systems in our discussion, as they are both composed of local degrees of freedom which commute with each other.

In this chapter we focus only on boson / spin systems without special symmetry constraint. In systems with symmetry, the phase diagram is more interesting as there are various symmetry protected topological phases, which we will discuss in Chapter \ref{chap10}. Also, the discussion about 1D gapped phases in fermion systems is deferred to Chapter \ref{chap10}, as 1D fermion systems can be mapped to 1D boson / spin systems with an extra $\mathbb{Z}_2$ symmetry through Jordan Wigner transformation.


\section{Matrix product states}
\label{sec:MPS}

\subsection{Definition and examples}

Matrix product states describe many-body entangled states of spins living on a one dimensional chain.
\begin{svgraybox}
\begin{center}
\textbf{Box 8.1 Matrix Product State}

A matrix product state (MPS) of a chain of $N$ spins is described as
\begin{equation}
\label{MPS}
|\psi\rangle = \sum_{i_1,i_2,...,i_N}
\Tr (A^{[1]}_{i_1}A^{[2]}_{i_2}...A^{[N]}_{i_N})|i_1i_2...i_N\rangle
\end{equation}
\end{center}
\end{svgraybox}
$i_k=1...d$, $A^{[k]}_{i_k}$'s are $D\times D$ matrices on site $k$ with $D$ being the dimension of the MPS. $d$ is the dimension of the physical Hilbert space at each site and is called the physical dimension. $D$ is the dimension of the matrices used in the matrix product representation which does not correspond to physical Hilbert spaces. $D$ is called the inner dimension of the MPS. We suppress the normalization of the wave functions here. The representation is efficient as with fixed $D$ for a state of $N$ spins, the number of parameters involved is at most $ND^2$ as compared to $d^N$ in the generic case. If the set of matrices does not depend on site label $k$, then the state represented is translation invariant.

Taking the trace of all matrices $A^{[k]}$ corresponds to periodic boundary condition on the one dimensional chain. If the chain has open boundary condition, it may be more convenient to use a slightly different form of MPS.
\begin{equation}
\label{MPS_ob}
|\psi\rangle = \sum_{i_1,i_2,...,i_N}
\bra{l}A^{[1]}_{i_1}A^{[2]}_{i_2}...A^{[N]}_{i_N}\ket{r}|i_1i_2...i_N\rangle
\end{equation}
where $\ket{l}$ and $\ket{r}$ are two $D$ dimensional vectors giving the left and right boundary conditions in the state.

If $D=1$, i.e. if $A$ are numbers, then $\ket{\psi}$ is a product state. For example, if $d=2$ and $A^{[i]}_{1}=\frac{1}{\sqrt{2}}$ and $A^{[i]}_{2}=\frac{1}{\sqrt{2}}$, then $|\psi\rangle$ describes a product state of two level spins of the form
\be
|\psi\rangle = \left[\frac{1}{\sqrt{2}}\left(\ket{1}+\ket{2}\right)\right] \left[\frac{1}{\sqrt{2}}\left(\ket{1}+\ket{2}\right)\right]...\left[\frac{1}{\sqrt{2}}\left(\ket{1}+\ket{2}\right)\right]
\ee

However, if $D \geq 2$, then $\ket{\psi}$ would in general be an entangled state of many spins. As the simplest example, consider matrices
\be
A_0= \begin{pmatrix} 1 & 0 \\ 0 & 0 \end{pmatrix}, A_1= \begin{pmatrix} 0 & 0 \\ 0 & 1 \end{pmatrix}
\ee
which are independent of site. Then the matrix product state they produce is the many-body entangled GHZ state introduced in Eq.~\eqref{eq:GHZ} (unnormalized),
\be
\ket{GHZ}=\ket{0}^{\otimes N} + \ket{1}^{\otimes N}
\ee

One of the most important examples of matrix product states is the AKLT state describing an anti-ferromagnetic ground state of spin $1$ chains, introduced in section \ref{sec:ffH}. For spin $1$ chains, $d=3$ and the three basis states can be chosen as the eigenstate of spin in the $z$ direction with $S_z=-1$, $0$ and $1$. The matrices defining the (unnormalized) AKLT state $\ket{\psi}_{AKLT}$ (as discussed in Chapter~\ref{sec:ffHam}) are site independent and are given by
\be
A_{-1}=-(X-iY)/\sqrt{2}, \ A_{0}=-Z, \ A_{1}=(X+iY)/\sqrt{2}
\label{AKLT_Sz}
\ee
We can also choose a different set of basis states for spin $1$ as
\be
\ket{x}=\frac{1}{\sqrt{2}}\left(-\ket{-1}+\ket{1}\right), \ \ket{y}=\frac{-i}{\sqrt{2}}\left(\ket{-1}+\ket{1}\right), \ \ket{z}=-\ket{0}
\ee
where $\ket{x}$, $\ket{y}$ and $\ket{z}$ are eigenvalue $0$ eigenstates of the spin in the $x$, $y$ and $z$ directions respectively. In this basis, the matrices for the AKLT state take the nice form
\be
A_x=X, \ A_y=Y, \ A_z=Z
\label{AKLT_xyz}
\ee

\begin{figure}[htbp]
\centering
         \subfigure[][]{%
\includegraphics{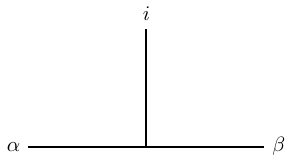}}%
         \hspace{12pt}%
         \subfigure[][]{%
           \includegraphics{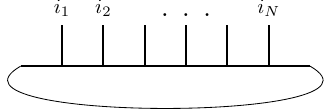}
 }%

\caption{Graphical representation of (a) the set of matrices (b) a matrix product state.} 
\label{fig:MPS}
\end{figure}

Graphically, the set of matrices are represented as shown in Fig.\ref{fig:MPS} (a) where the vertical bond denotes the physical index and the two horizontal bonds denote the inner indices. An MPS is then represented as in Fig.\ref{fig:MPS} (b), with all the inner indices contracted between neighboring sites.

\subsection{Double tensor}
\label{sec:DT}

An important mathematical construction in the MPS description is the double tensor.

\begin{svgraybox}
\begin{center}
\textbf{Box 8.2 The double tensor}

The double tensor of a matrix product state is defined as
\be
\mathbb{E}_{\alpha\gamma,\beta\chi}=\sum_{i}
A_{i,\alpha\beta} \times (A_{i,\gamma\chi})^*
\label{DT}
\ee
\end{center}
\end{svgraybox}

\begin{figure}[htbp]
\centerline{
~~~~~~
\includegraphics[width=0.80in]{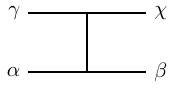}~~~~~~
\includegraphics[width=1.80in]{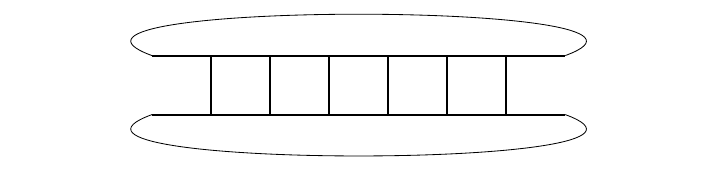}
\includegraphics[width=1.80in]{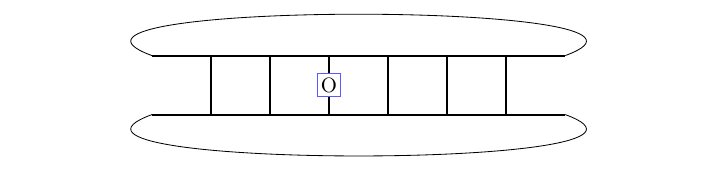}
}
\centerline{
	(a) ~~~~~~~~~~~~~~~~~~~~~~~~~~~~~~~~~~~~~~~~~~~~~~
	(b) ~~~~~~~~~~~~~~~~~~~~~~~~~~~~~~~~~~~~~~~~~~~~~~
	(c)
}
\caption{Graphical representation of (a) the double tensor (b) the norm (c) expectation value of local observable (un-normalized by norm) in a matrix product state.} 
\label{dT_MPS}
\end{figure}

If we combine $\al$ with $\ga$ and $\bt$ with $\chi$ and treat $\mathbb{E}$ as a matrix, then we can write
\be
\mathbb{E}=\sum_i A_i \otimes A_i^*
\ee
Graphically, it is represented as in Fig.\ref{dT_MPS} (a) where the phyiscal indices of the lower and upper set of matrices are contacted.

On the other hand, if we combine $\al$ with $\bt$ and $\ga$ with $\chi$, $\mathbb{E}$ can be thought of as a different matrix, rotated $90$ degrees. In this perspective, $\mathbb{E}$ is a Hermitian matrix with non-negative eigenvalues. If $\al$, $\bt$, $\ga$, $\chi$ are of dimension $D$, then the eigenvalue decomposition of $\mathbb{E}$ with respect to indices $\al\bt$ and $\ga\chi$ yields at most $D^2$ positive eigenvalues. If we think of $\al$ and $\bt$ as the input indices and $\ga$ and $\chi$ as the output indices, $\mathbb{E}$ actually describes the non-unitary evolution of an open quantum systems with Kraus operators $A_i$, as introduced in section \ref{sec:OpenQS}.  

If two set of matrices $A_i$ and $B_i$ are related by a a unitary transformation $U$ on the physical index $i$:
\begin{align}
B_{i,\alpha\beta}= \sum_j U_{ij} A_{j,\alpha\beta} .
\end{align}
Then they give rise to the same double tensor, which can be seen from
\be
\mathbb{E}_B=\sum_{j} B_{j,\alpha\beta} \times (B_{j,\gamma\chi})^* = \sum_{ii'j} U_{ij}U^*_{i'j} A_{i,\alpha\beta} \times (A_{i',\gamma\chi})^*=\sum_{i} A_{i,\alpha\beta} \times (A_{i,\gamma\chi})^*=\mathbb{E}_A
\ee

The reverse is also true and is the most important property of the double tensor: A double tensor $\mathbb{E}$ uniquely determines the matrices $A_i$ up to some unitary transformation on the physical index $i$. That is, if
\begin{align}
\mathbb{E}_{\alpha\gamma,\beta\chi}=\sum_{i} A_{i,\alpha\beta} \times (A_{i,\gamma\chi})^* = \sum_{i} B_{i,\alpha\beta} \times (B_{i,\gamma\chi})^*
\end{align}
then $A_{i,\alpha\beta}$ and $B_{i,\alpha\beta}$ are related by a unitary transformation $U$ on the physical index $i$:
\begin{align}
B_{i,\alpha\beta}= \sum_j U_{ij} A_{j,\alpha\beta} .
\end{align}
Therefore, states described by $A_i$ and $B_i$ can be related by unitary transformations on each physical index which only change basis for each spin without affecting the entanglement structure in the sate. 
\begin{svgraybox}
\begin{center}
\textbf{Box 8.3 Unitary equivalence of MPS with the same double tensor}

If two matrix product states have the same double tensor, then the two states can be mapped to each other by unitary transformations on each physical index.

\end{center}
\end{svgraybox}

This property is useful for applying renormalization transformation on the state, as discussed in section \ref{sec:RG_MPS}.

A straight-forward way to obtain one possible form of $A_i$ from $\mathbb{E}$ is to think of $\mathbb{E}$ as a matrix with left index $\al\bt$ and right index $\gamma\chi$. As discussed before, in this perspective, $\mathbb{E}$ is a Hermitian matrix with non-negative eigenvalues. Find the eigenvalues $\eta_i$ of $\mathbb{E}$ and the corresponding eigenvectors $v_i$. That is
\be
\mathbb{E}_{\al\bt,\gamma\chi}=\sum_i \eta_i v_{i,\al\bt}\times v^*_{i,\gamma\chi} 
\ee
Then define $A_{i,\al\bt}=\sqrt{\eta_i}v_{i,\al\bt}$, which are exactly the matrices we are looking for with physical index $i$ and inner indices $\al\bt$ and satisfies
\be
\mathbb{E}_{\alpha\gamma,\beta\chi}=\sum_{i}
A_{i,\alpha\beta} \times (A_{i,\gamma\chi})^*
\ee
All other possible forms of matrices giving rise to the same $\mathbb{E}$ are related to this particular form of $A_i$ by a unitary transformation on the physical index $i$.

\subsection{Calculation of norm and physical observables}

The double tensor is useful for the calculation of the norm and physical observables of the matrix product state. 

The norm of an MPS is given by
\be
\braket{\psi}{\psi}=\Tr(\mathbb{E} \times \mathbb{E}...\times \mathbb{E})=\Tr(\mathbb{E}^N)
\ee
as shown graphically in Fig.\ref{dT_MPS} (b).

The expectation value of measuring any local observable $O$ (on the $k$th spin for example) on the state is equal to
\be
\<O\>=\frac{\Tr(\mathbb{E}^{k-1}\mathbb{E}[O]\mathbb{E}^{N-k})}{\braket{\psi}{\psi}}, \ \text{where} \  \mathbb{E}[O]=\sum_{i,j} O_{i,j} A_i \otimes A_j^*
\ee
The numerator is graphically shown as in Fig.\ref{dT_MPS} (c). Note that as matrix multiplication takes time $\sim D^3$, the calculation of any physical observable is efficient (polynomial in inner dimension $D$ and linear in system size $N$) for MPS.

\subsection{Correlation length}

From this we can see that the double tensor is directly related to an important quantity for many-body systems, the correlation length. In fact, an MPS has a finite correlation length if the largest eigenvalue of $\mathbb{E}$ is nondegenerate. This is shown as follows:

WLOG, we can set the largest eigenvalue of $\mathbb{E}$ to be $1$ and hence the norm goes to a finite value (dimension of the eigenspace) as $N$ goes to infinity. The correlation function between two operators $O_1$ and $O_2$ becomes
\begin{equation}
\langle O_1 O_2 \rangle -\langle O_1 \rangle \langle O_2 \rangle=
\frac{\Tr(\mathbb{E}^{N-L-2}\mathbb{E}[O_1]\mathbb{E}^{L}\mathbb{E}[O_2])}{\Tr(\mathbb{E}^N)}-
\frac{\Tr(\mathbb{E}^{N-1}\mathbb{E}[O_1])\Tr(\mathbb{E}^{N-1}\mathbb{E}[O_2])}{\Tr^2(\mathbb{E}^N)}
\end{equation}
Denote the projection onto the eigenspace of eigenvalue $\lambda$ as $P_{\la}$. At large system size $N$, the correlation function becomes
\be
\frac{\Tr(P_1\mathbb{E}[O_1](\sum_{\lambda}\lambda P_{\lambda})^{L} \mathbb{E}[O_2])}{\Tr (P_1)} -\frac{\Tr(P_1\mathbb{E}[O_1])\Tr(P_1\mathbb{E}[O_2])}{\Tr^2 (P_1)}
\ee
When $L$ is large, we keep only the first order term in $(\sum_{\lambda}\lambda P_{\lambda})^{L}$ and the correlation function goes to 
\be
\frac{\Tr(P_1\mathbb{E}[O_1]P_1\mathbb{E}[O_2])}{\Tr (P_1)}-\frac{\Tr(P_1\mathbb{E}[O_1])\Tr(P_1\mathbb{E}[O_2])}{\Tr^2 (P_1)}
\ee
If $P_1=\ket{v_1}\bra{v_1}$ is one dimensional, the two terms both become 
\be
\langle v_1|\mathbb{E}[O_1]| v_1\rangle\langle v_1|\mathbb{E}[O_2]|v_1\rangle
\ee
and cancel each other for any $O_1,O_2$ and the second order term in $(\sum_{\lambda}\lambda P_{\lambda})^{L}$ dominates which decays as $\lambda^L$. For $\lambda<1$, the correlator goes to zero exponentially and the matrix product
state as finite correlation length. On the other hand, if $P_1$ is more than one dimensional, the first order term has a finite contribution independent of $L$: 
\be
\frac{\sum_{i,j} \langle v_i|\mathbb{E}[O_1]| _j\rangle\langle v_j|\mathbb{E}[O_2]|v_i\rangle}{\Tr(P_1)} - \frac{\langle v_i|\mathbb{E}[O_1]|v_i \rangle \langle
v_j|\mathbb{E}[O_2]|v_j\rangle}{\Tr^2(P_1)}
\ee
where $v_i,v_j$ are eigenbasis for $P_1$. Therefore, degeneracy of the largest eigenvalue of the double tensor implies non-decaying correlation. To describe quantum states with finite correlation length, the double tensor must have a largest eigenvalue which is non-degenerate and the correlation length $\xi$ is given by
\be
\xi = -1/\ln{\la_2}
\ee
where $\la_2$ is the second largest eigenvalue and $\la_2 <1$. Here $\xi$ is measured in units of lattice spacing.

\begin{svgraybox}
\begin{center}
\textbf{Box 8.4 MPS with finite correlation length}

A matrix product state has finite correlation length if and only if the largest eigenvalue of its double tensor is non-degenerate.
\end{center}
\end{svgraybox}

\subsection{Entanglement area law}

Double tensor is important in studying not only the correlation length but also the \mbet structure of an MPS. In fact, $\mathbb{E}$ uniquely determines the state up to a local change of basis on each site and hence contains all the entanglement information of the state. First, we will show with the help of double tensor that \mbet in a matrix product state satisfies an exact area law. Actually, if we take a continuous segment out of the chain, the reduced density matrix has rank at most $D^2$. 

Suppose that we cut the chain into the left half with site $1$ to $k$ and the right half with site $k+1$ to $N$. If we think of the doulbe tensors on each site as matrices with left index $\al$, $\gamma$ and right index $\bt$, $\chi$, then the double tensor of the left half of the chain is the product of all double tensors from sites $1$ to $k$. Similarly, the double tensor of the right half of the chain is the product of all double tensors from sites $k+1$ to $N$.
\be
\mathbb{E}_l=\prod^k_{i=1} \mathbb{E}_i, \ \mathbb{E}_r=\prod^N_{i=k+1} \mathbb{E}_i
\ee
The entanglement between the left and right half of the chain is faithfully captured by $\mathbb{E}_l$ and $\mathbb{E}_r$. Now we can decompose double tensors $\mathbb{E}_l$ and $\mathbb{E}_r$ back into matrices and find an upper bound on entanglement. In order to do this, we rotate the double tensors 90 degrees and think of them as matrices with left index $\al$, $\bt$ and right index $\gamma$, $\chi$. $\mathbb{E}_l$ and $\mathbb{E}_r$ are both $D^2\times D^2$ Hermitian matrices in this perspective with non-negative eigenvalues. When we perform the eigenvalue decomposition and obtain the matrices $A^l_i$ and $A^r_i$ as described previously, we find that there are at most $D^2$ nonzero eigenvalues hence $D^2$ nonzero $A^l_i$ and $A^r_i$. This is saying that under separate unitary transformations on the left and right half of the chain, the number of phyiscal degrees of freedom can be reduced to $\leq D^2$ on both sides.

\begin{svgraybox}
\begin{center}
\textbf{Box 8.5 Entanglement area law in matrix product states}

The entanglement entropy between the left and right half of the chain (in fact between any segment and the rest of the chain) is upper bounded by $2\ln{D}$.
\be
S \leq 2\ln(D)
\ee
\end{center}
\end{svgraybox}

With $D$ being constant, the MPS satisfies an exact entanglement `area law' in one dimension.

On the other hand, it is not true that every one-dimensional state satisfying an exact area law can be written exactly as a matrix product state with finite inner dimension $D$.  For example, consider a 1D chain of boson modes where each mode can host any integer number $i$ of bosons. Consider the state that is composed of nearest-neighbor dimers between boson modes $2k-1$ and $2k$ of the form
\be
\ket{\psi}=\sum_{i} \alpha_i \ket{ii}
\ee

This state satisfies area law as long as $\sum_i \alpha_i\ln{\alpha_i}$ is finite. But this does not necessarily mean that there is a finite number of $i$'s. As long as $\alpha_i$ decays fast enough with $i$, the entanglement of a segment will be bounded. However, if $i$ is unbounded, the reduced density matrix of a segment will have an infinite rank and therefore not possible to represent with a finite dimensional MPS.

The situation is not too bad though. It has been shown that for any 1D state satisfying an area law, the necessary inner dimension to approximately describe the state scales only polynomially with system size. Therefore the matrix product state representation is still efficient. Moreover it has been proven that all gapped ground states of 1D local Hamiltonians satisfy an area law, therefore the matrix product representation for such states is always efficient. The power of matrix product states is limited to one spatial dimension though. To represent a gapped two-dimensional quantum state satisfying an area law using matrix product formalism would require in general matrices of exponential size. Therefore, we need more general constructions -- the tensor product states -- to deal with higher dimensional systems.

\subsection{Gauge degree of freedom}

The matrix product state representation is not unique. 

\begin{svgraybox}
\begin{center}
\textbf{Box 8.6 Gauge degree of freedom of MPS representation}

An MPS represented by a set of matrices $\{A_i\}$ is equally well represented by $\{B_i=MA_iM^{-1}\}$, for any invertible matrix $M$. 

\end{center}
\end{svgraybox}

This is true because
\be
\Tr(B_{i_1}B_{i_2}...B_{i_N})= \Tr(MA_{i_1}M^{-1}MA_{i_2}M^{-1}...MA_{i_N}M^{-1})=\Tr(A_{i_1}A_{i_2}...A_{i_N})
\ee
This property can be generalized to site-dependent $M$ and $A$'s as well.

This gauge degree of freedom will play an important role in our understanding of symmetry protected topological orders in one dimension.

\subsection{Projected entangled pair picture}

Matrix product states have another name -- the Projected Entangled Pair State (PEPS). It comes from the following construction. (The construction applies to higher dimensional tensor product states as well.)

\begin{figure}[htbp]
\begin{center}
\includegraphics[width=4.00in]{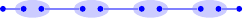}
\caption{Projected Entangled Pair State(PEPS)} 
\label{PEPS}
\end{center}
\end{figure}

Consider a chain of maximally entangled pairs as shown in Fig. \ref{PEPS}. Suppose that they connect into a ring,
Each pair of connected dots represents a maximally entangled pair of spins in state
\be
\ket{\psi}=\frac{1}{\sqrt{D}}\sum_{\al=1}^D \ket{\al\al}
\ee
where $D$ is the dimension of each spin. Each shaded big circle represents a projection $P$ (a mapping) from two spins of dimension $D$ to a physical degree of freedom of dimension $d$ (a physical spin)
\be
P=\sum_{i,\al,\bt} A_{i,\al,\bt} \ket{i}\bra{\al\bt}
\ee
where the summation is over $i=1...d$, and $\al,\bt=1..D$. Direct calculation shows that after the projection, we obtain a many-body entangled state of physical spins which can be written as
\be
\ket{\psi}=\sum_{i_1,i_2...i_N} \Tr(A_{i_1}A_{i_2}...A_{i_N}) \ket{i_1i_2...i_N}
\ee
which is exactly the matrix product states given in \Eqn{MPS}. Here $A_i$ is treated as a matrix with row index $\al$ and column index $\bt$. In this projected entangled pair construction of matrix product states, the spins in maximally entangled pairs are said to be virtual and the spins obtained after projection are physical. 

The PEPS and MPS (or more generally tensor product states (TPS)) formalisms are totally equivalent. But sometimes, one picture is more convenient and intuitive than the other. 


\subsection{Canonical form}
\label{sec:canon_form_MPS}

A canonical form exists for the matrices in an MPS representation, which provides much insight into the structure of the many-body state. We are not going to prove but only state the result in this section. We focus on the case where the matrices are not site dependent and hence the state is translational invariant.

The matrices $A_i$'s in an MPS representation can be put into a `canonical' form which is block diagonal
\begin{equation}
A_i = \begin{bmatrix} A^{(0)}_i & & \\ & A^{(1)}_i  & \\ & & \ddots \end{bmatrix}
\label{canon_form}
\end{equation}
where the double tensor for each block $\mathbb{E}^{(k)} =\sum_i A_i^{(k)} \otimes (A_i^{(k)})^*$ has a positive non-degenerate largest eigenvalue $\lambda_k >0$. Note that each $\mathbb{E}^{(k)}$ can have eigenvalues with the same magnitude as $\la_k$, in the form $\la_i e^{i2\pi/p}$, $p \in \mathbb{Z}_{q_k}$.

There are several implications that can be directly read from this `canonical form'. First the matrix product state $\ket{\psi}$ represented by $A_i$ can be written as a superposition of $\ket{\psi^{(k)}}$'s, represented by matrices $A^{(k)}_i$. 
\be
\ket{\psi}=\sum_k \ket{\psi^{(k)}}
\ee
If $\mathbb{E}^{(k)}$ has only one eigenvalue with magnitude $\la_k$, then $\ket{\psi^{(k)}}$ is short range correlated (with finite correlation length). If $\mathbb{E}^{(k)}$ has other eigenvalues with the same magnitude as $\la_k$, then $\ket{\psi^{(k)}}$ can be further decomposed into states $\ket{\psi^{(k)}_{p}}$ with block translation symmetry of block size $q_k$ and finite correlation length. $\ket{\psi^{(k)}_{p}}$ is related to $\ket{\psi^{(k)}_{1}}$ by translation of $p$ sites.
\be
\ket{\psi}=\sum_k \sum^{q_k}_{p=1} \ket{\psi^{(k)}_{p}}
\ee
Therefore, the canonical form directly yields a decomposition of the MPS into a finite (and minimum) number of short range correlated states.

\subsection{Injectivity}

If the canonical form of an MPS contains only one block and the double tensor has only one eigenvalue with largest magnitutde, then the MPS is said to be `injective'. Otherwise, the MPS is said to be `noninjective'. Injective MPS hence has only one component in the canonical decomposition and is short range correlated. 

For example, the matrices for the GHZ state $\ket{00...0} + \ket{11...1}$ contain only one block and the double tensor is
\be
\mathbb{E}_{GHZ}=\begin{pmatrix} 1 & 0 \\ 0 & 0 \end{pmatrix} \otimes \begin{pmatrix} 1 & 0 \\ 0 & 0 \end{pmatrix} + \begin{pmatrix} 0 & 0 \\ 0 & 1 \end{pmatrix} \otimes \begin{pmatrix} 0 & 0 \\ 0 & 1 \end{pmatrix} = \begin{pmatrix} 1 & 0 & 0 & 0 \\ 0 & 0 & 0 & 0 \\ 0 & 0 & 0 & 0 \\ 0 & 0 & 0 & 1 \end{pmatrix}
\ee
which has two-fold degenerate largest eigenvalue $1$. Therefore, the matrix product representation of the GHZ state is `noninjective'. 

On the other hand, the matrices for the AKLT state contain only one block as well but the double tensor is
\be
\mathbb{E}_{GHZ}=X \otimes X^* + Y \otimes Y^* + Z \otimes Z^* = \begin{pmatrix} 1 & 0 & 0 & 2 \\ 0 & -1 & 0 & 0 \\ 0 & 0 & -1 & 0 \\ 2 & 0 & 0 & 1 \end{pmatrix}
\ee
which has a single eigenvalue with the largest magnitude $3$. The eigenvalue with the second largest magnitude is $-1$. Therefore, the matrix product representation of the AKLT state is `injective'.

Here the terminology is related to the injectiveness of the following map
\be
\Gamma_L: W \mapsto \sum^{d}_{i_1,...i_L=1} \Tr(WA_{i_1}...A_{i_L})|i_1...i_L\> 
\ee
$\Gamma_L$ being injective means that for different $W$, $\sum^{d}_{i_1,...i_L=1} \Tr(WA_{i_1}...A_{i_L})|i_1...i_L\>$ are always different. An MPS is injective if there exists a finite $L_0$ such that $\Gamma_{L}$ is an injective map for $L\leq L_0$. Therefore, the previous definition of injectivity in terms of canonical form is equivalent to the following.

\begin{svgraybox}
\begin{center}
\textbf{Box 8.7 Injective matrix product states}

A matrix product state described by matrices $A_i$ is injective if there exists a finite $L_0$ such that the set of matrices
\be
\t{A}_{I_{L_0}}=A_{i_1}...A_{i_{L_0}}
\ee
spans the the whole space of $D\times D$ matrices. 
\end{center}
\end{svgraybox}

If this condition is satisfied for $L_0$, then obviously it is satisfied for all $L>L_0$. 

In the case of GHZ state, no matter how big a segment we take, the set of matrices for a segment of length $L$ always contain only two matrices
\be
A_{0...0}= \begin{pmatrix} 1 & 0 \\ 0 & 0 \end{pmatrix}, A_{1...1}= \begin{pmatrix} 0 & 0 \\ 0 & 1 \end{pmatrix}
\ee
These two matrices cannot span the whole space of $2\times 2$ matrices, therefore the MPS representation of the GHZ state is noninjective.

For AKLT state, there are three matrices on a single site
\be
A_x=X, \ A_y=Y, \ A_z=Z
\ee
which do not span the $4$ dimensional space of $2 \times 2$ matrices. However, on two sites there are nine matrices
\be
A_{xx}=A_{yy}=A_{zz}=I, A_{xy}=-A_{yx}=iZ, A_{yz}=-A_{zy}=iX, A_{zx}=-A_{xz}=iY
\ee
which do span the whole space of $2 \times 2$ matrices. Obviously, on segments of length larger than $2$, the set of matrices also span the whole space of $2 \times 2$ matrices. Therefore, the MPS representation of the AKLT state is injective, which is consistent with the conclusion obtained by examining the canonical form of the MPS.

Actually, the `injectivity' property is generically true for random matrix product states. The number of matrices on a segment of length $L$ is $d^L$, which grows exponentially with $L$. Generically, for a fixed inner dimension $D$, it is always possible to span the whole space of $D \times D$ matrices with $d^L$ matrices for a large enough $L$, unless the matrices are designed to have special structures like in the GHZ state. If we pick an MPS randomly, then it always satisfies the injective condition.

Therefore, `injectivity' plays an important role in our study of matrix product state. First, a random matrix product state is always injective; secondly, any MPS has a canonical decomposition into a finite number of injective components; moreover, injective MPS enjoys very nice properties like finite correlation length. In the next section, we discuss another nice property of injective MPS: the existence of a local gapped Hamiltonian which has the MPS has its unique ground state.

\subsection{Parent Hamiltonian}
\label{sec:parentHam}

We set out to study MPS because it describes gapped ground states of 1D local Hamiltonians. However, up to now, it is unclear what the Hamiltonian is for a given MPS and what kind of gapped ground state is the MPS. We address this question in this section.

From the decomposition obtained from the `canonical form', a `parent Hamiltonian' can be constructed which has the MPS as a gapped ground state, thus making contact with usual condensed matter studies.

In particular, if the MPS is injective, that is if there is only one component in the decomposition of $\ket{\psi}$, then the parent Hamiltonian has $\ket{\psi}$ as a unique gapped ground state. As a single block MPS has finite correlation length, we find

\newpage
\begin{svgraybox}
\begin{center}
\textbf{Box 8.8 Parent Hamiltonian for MPS with finite correlation length}

A parent Hamiltonian can be constructed for a finite dimensional matrix product state with finite correlation length, such that the matrix product state is the unique gapped ground state of the parent Hamiltonian.

\end{center}
\end{svgraybox}

The procedure for constructing the parent Hamiltonian is as follows: 
\begin{enumerate}
\item{take a large enough but finite segment of length $l$ of the chain}
\item{calculate the reduced density matrix $\rho_l$ of this segment}
\item{write the projection operator $P_l$ onto the support space of $\rho_l$}
\end{enumerate}
The parent Hamiltonian is the MPS is then the sum of all such local projectors
\be
H=\sum_i (1-P^i_l)
\ee
where $P^i_l$ is the projector applied to the segment centered around site $i$. 

Obviously, each term in the Hamiltonian has energy $0$ on the matrix product state. Therefore, the MPS is a frustration free ground state of the parent Hamiltonian (as discussed in Chapter~\ref{sec:ffHam}). Is the MPS a unique and gapped ground state of the parent Hamiltonian? The answer is yes and this is guaranteed by the injectivity of the MPS. It can be shown that if the MPS is injective, then as long as $l$ is large enough, the projectors $P^i_l$ impose strong enough constraints such that $H$ has a unique ground state and a finite energy gap. 

The parent Hamiltonian for the AKLT state can be obtained in this way. Written in terms of spin variables, the Hamiltonian reads
\be
H=\sum_i \vec{S}_i\vec{S}_{i+1} + \frac{1}{3}(\vec{S}_i\vec{S}_{i+1})^2
\ee
which takes the same form as that given in Eq. \ref{eq:HAKLT} in Chapter~\ref{sec:ffH}. Note that because the ground state is unique and gapped, if the Hamiltonian has certain symmetry, then the ground state also has it. 

On the other hand, if $\ket{\psi}$ can be decomposed into a set of short range correlated $\ket{\psi^{(k)}}$'s, then it cannot be the unique gapped ground state of a local Hamiltonian. A parent Hamiltonian can be constructed which is still gapped but has a degenerate ground space spanned by all $\ket{\psi^{(k)}}$'s. To construct such a parent Hamiltonian, first notice that on large enough segments, the support space of the reduced density matrices $\rho^{(k)}_l$ and $\rho^{(k')}_{l}$ are orthogonal to each other. The parent Hamiltonian can then be written as
\be
H=\sum_i \sum_k (1-\left(P^i_l\right)^{(k)})
\ee
where $\left(P^i_l\right)^{(k)}$ are projectors onto the support space of $\rho^{(k)}_l$ centered around site $i$. Obviously, $\ket{\psi}$ is a frustration-free ground state of $H$ but not the unique one. Actually, all superpositions of $\ket{\psi^{(k)}}$'s are ground states of $H$. It can still be proved that $H$ has a finite energy gap above this ground subspace. Therefore, noninjective MPS is one of the degenerate ground states of a local gapped Hamiltonian. One important consequence of ground state degeneracy is that if the Hamiltonian has a certain symmetry, each of the $\ket{\psi^{(k)}}$'s does not have to. They can be related to each other by the symmetry transformation.


\section{Renormalization group transformation on MPS}
\label{sec:RG_MPS}

The wave function renormalization group transformation, as discussed in the last chapter, aims to remove short range entanglement structures from a quantum state and extract the universal properties of the phase from the fixed point of the renormalization flow. The key to a successful renormalization procedure is in choosing the right local unitary operators which removes short range entanglement in an optimal way. The matrix product representation provides us with an efficient way to find such local unitaries and implement the renormalization procedures.

\begin{figure}[htbp]
\begin{center}
\includegraphics[width=4.5in]{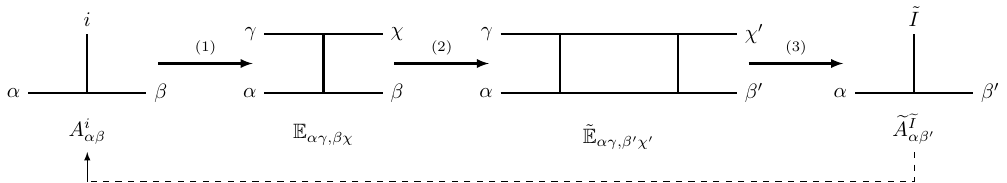}
\end{center}
\caption{Quantum state renormalization group transformation on matrix product state.}
\label{QSRG_MPS}
\end{figure}

Consider matrix product state
\begin{equation}
\label{MPS}
|\psi\rangle = \sum_{i_1,i_2,...,i_N}
\Tr (A^{i_1}A^{i_2}...A^{i_N})|i_1i_2...i_N\rangle
\end{equation}
where $i_k=1...d$ with $d$ being the physical dimension of a spin at each site, $A^{i_k}$'s are $\chi\times \chi$
matrices related to the physical state $\ket{i_k}$ with $\chi$ being the inner dimension of the MPS.

To implement the wave function renormalization group transformation on the matrix product state, first construct the double tensor
\be
\mathbb{E}_{\alpha\gamma,\beta\chi}=\sum_{i} A^i_{\alpha\beta} \times (A^i_{\gamma\chi})^*
\label{1dqsrg-i}
\ee
as shown in Fig.\ref{QSRG_MPS} step (1).
Treat $\mathbb{E}$ as a $\chi^2\times \chi^2$ matrix with row index $\alpha\gamma$ and
column index $\beta\chi$. Combine the double tensor of the two sites together into
\be
\tilde{\mathbb{E}}_{\alpha\gamma,\beta'\chi'}=\sum_{\beta\chi}\mathbb{E}_{\alpha\gamma,\beta\chi} \mathbb{E}_{\beta\chi,\beta'\chi'}
\ee
as shown in Fig.\ref{QSRG_MPS} step (2).
Then think of
$\tilde{\mathbb{E}}_{\alpha\gamma,\beta'\chi'}$ as a matrix
with row index $\alpha\beta'$ and column index
$\gamma\chi'$. It is easy to see that with such a
recombination, $\tilde{\mathbb{E}}$ is a positive matrix and
can be diagonalized
\be
 \tilde{\mathbb{E}}_{\alpha\gamma,\beta'\chi'}
=\sum_{\tilde i} \lambda_{\tilde i} V_{\tilde i,\al\bt'} V^*_{\tilde i,\ga\chi'} ,
\ee
where we have kept only the non-zero eigenvalues $\lambda_{\tilde i}(>0)$ and the corresponding eigenvectors $V_{\tilde i,\al\bt'}$.
$\tilde A$ is then given by
\be
 \tilde A^{\tilde i}_{\alpha\beta'}=\sqrt{\lambda_{\tilde i}} V_{\tilde i,\al\bt'}
 \label{1dqsrg-f}
\ee
which are the matrices representing the renormalized state, which form the basis for the next round of renormalization transformation.

\begin{figure}[htbp]
\begin{center}
\includegraphics[width=3.5in]{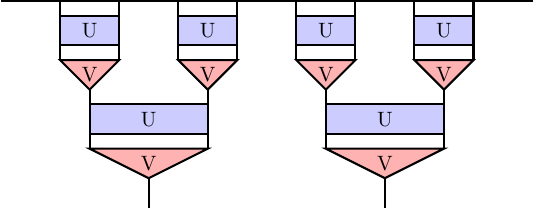}
\end{center}
\caption{Quantum state renormalization group transformation on 1D quantum states.}
\label{QSRG_1D}
\end{figure}

These steps apply one round of renormalization procedure on the matrix product state by applying local unitaries to each pair of neighboring sites, removing local entanglement between them and combining the remaining degrees of freedom of the two sites into one. To see how this is achieved, notice that an important property of $\mathbb{E}$ is that it uniquely determines the matrices, and hence the state, up to a local change of basis on each site, as discussed in section \ref{sec:DT}. That is, if 
\be
\mathbb{E}_{\alpha\gamma,\beta\chi}=\sum_{i}
A^i_{\alpha\beta} \times (A^i_{\gamma\chi})^*=\sum_{i} B^j_{\alpha\beta} \times (B^j_{\gamma\chi})^*
\ee
then $A^i_{\alpha\beta}$ and $B^j_{\alpha\beta}$ are related by a unitary transformation $U$:
\be
B^j_{\alpha\beta}= \sum_i U_{ji} A^i_{\alpha\beta}
\ee 
Therefore, in decomposing $\tilde{\mathbb{E}}$ into $\tilde{A}$, we have implemented a unitary on every two sites and the wave function has been transformed as 
\begin{align}
|\psi\rangle \rightarrow  |\tilde{\psi}\rangle= U_{1,2}\otimes
U_{3,4} \otimes ... \otimes  U_{2i-1,2i} \otimes ... |\psi\rangle.
\end{align}
$\tilde{\mathbb{E}}$ contains all the information about the
entanglement of the two sites with the rest of the system
but not any detail of entanglement structure among the
two sites. By setting the range of $\tilde i$ to be over only the nonzero $\lambda$'s, we have reduced the physical dimension of the two sites to only those necessary for describing the entanglement between them and the rest of the system. Local entanglement among the two sites has been optimally removed. Repeating this procedure several times correspond to a (generalized) local unitary transformation on the quantum state as shown in Fig.\ref{QSRG_1D} and the matrices flow from $A^{(0)}$ to $A^{(1)}$,..., until the fixed point form of $A^{(\infty)}$ from which the universal properties of the state can be determined.

\section{No intrinsic topological order in 1D bosonic systems}
\label{sec:no_sym}

Let's apply this wave function renormalization group transformation to gapped, short range correlated matrix product states and determine what quantum phases exist in bosonic systems. First when no symmetry is required for the class of system, we want to know what
kind of long range entanglement exists and thereby classify intrinsic
topological orders in 1D gapped boson systems. We will show
that:
\begin{svgraybox}
\begin{center}
\textbf{Box 8.9 Classification of gapped 1D bosonic systems without symmetry}

All gapped 1D bosonic systems belong to the same phase
if no symmetry is required.
\end{center}
\end{svgraybox}

In other words, there is no intrinsic topological order in 1D bosonic systems and all gapped quantum states are short range entangled. 

To obtain such a result, we use the fact that gapped 1D bosonic
states are described by short-range correlated matrix product
states. Then one can show that all short-range correlated
matrix product states can be mapped to product states with
LU transformations. Therefore there is no intrinsic
topological order in 1D.

Consider a generic system without any symmetry whose gapped ground state is described as an MPS with matrices $A^{i}$. In general, the system may not have translation symmetry and $A^{i}$ can vary from site to site. For simplicity of notation, we will not write the site label for the matrices explicitly. As we are interested in matrix product states with a finite correlation length and as the gapped ground state of a local Hamiltonian, we only need to consider the so-called `injective' matrix product states. That is, the canonical form of the MPS contains only one block and the double tensor $\mathbb{E}_{\alpha\gamma,\beta\chi}$ has only one eigenvalue with the largest magnitutde (set to be $1$) when treated as a matrix with row index $\alpha\gamma$ and colomn index $\beta\chi$. The corresponding left eigenvector is $\Lambda^l_{\alpha\gamma}$ and the right eigenvector is $\Lambda^r_{\beta\chi}$
\be
\mathbb{E}_{\alpha\gamma,\beta\chi}=\Lambda^l_{\alpha\gamma}(\Lambda^r_{\beta\chi})^* + ...
\ee
where all the other terms in this decomposition has norm smaller than $1$. Let's label this starting point of renormalization group transformation as $\mathbb{E}^{(0)}$.

\begin{figure}[htbp]
\begin{center}
\includegraphics[width=4.5in]{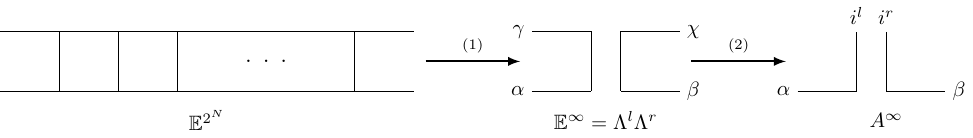}
\end{center}
\caption{Renormalization fixed point for injective matrix product states.}
\label{QSRG_fp}
\end{figure}

Apply the renormliazation group transformation to this matrix product state and we can see that the fixed point has a very simple form. Note that in each step of the renormalization group transformation, the double tensor changes as
\be
\mathbb{E}^{(n+1)}=\mathbb{E}^{(n)}\mathbb{E}^{(n)}
\ee
Therefore,
\be
\mathbb{E}^{(N)}=\left(\mathbb{E}^{(0)}\right)^{2^N}
\ee
In this procedure, the terms in $\mathbb{E}$ with $<1$ eigenvalues all decay exponentially. After repeating the renormalization process a finite number
of times, $\mathbb{E}^{(N)}$ will be arbitrarily
close to a fixed point form $\mathbb{E}^{(\infty)}$
with only one nonzero eigenvalue $1$ and
\be
\mathbb{E}^{(\infty)}_{\alpha\gamma,\beta\chi}=\Lambda^l_{\alpha\gamma} (\Lambda^r_{\beta\chi})^*
\ee
This process is shown in Fig.\ref{QSRG_fp} step (1). 

Now we can decompose $\mathbb{E}^{(\infty)}$ into matrices to find the fixed point wave function as shown in Fig.\ref{QSRG_fp} step (2). Because $\mathbb{E}^{\infty}$ is positive when treated as a matrix with row index $\alpha\beta$ and colomn index $\gamma\chi$, $\Lambda^l$ ($\Lambda^r$) is also positive when treated as a matrix with row index $\alpha$ ($\beta$) and column index $\gamma$ ($\chi$). $\Lambda_l$ and $\Lambda_r$ can be decomposed as
\be
\Lambda^l_{\alpha\gamma}=\sum_{i} \lambda_{i} v^{i}_{\alpha} (v^{i}_{\gamma})^*, \Lambda^r_{\beta\chi}=\sum_{j} \eta_{j} u^{j}_{\beta} (u^{j}_{\chi})^*
\ee
where $i,j=1,...,D$, $\lambda_{i},\eta_{j}>0$ and $\{v^i\}$, $\{u^j\}$ are two sets of orthonormal vectors. 

It then follows that fixed point matrices $A^{(\infty)}$ of the following form can give rise to the fixed point double tensor $\mathbb{E}^{(\infty)}$
\be
\left(A^{(\infty)}\right)^{i^l,i^r}_{\alpha\beta} = \sqrt{\lambda_{i^l}\eta_{i^r}} v^{i^l}_{\alpha} u^{i^r}_{\beta}
\label{Avv}
\ee

\begin{figure}[htbp]
\begin{center}
\includegraphics[width=4.5in]{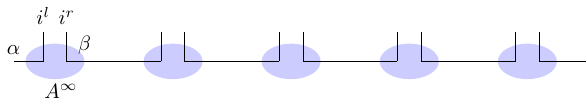}
\end{center}
\caption{Valence bond structure of fixed point wave function for short range correlated 1D quantum states.}
\label{wf_fp}
\end{figure}

%
From this structure of $A^{(\infty)}$, we can see that at fixed point the physical degrees of freedom on each site splits into two parts labeled by $i^l$ and $i^r$. Moreover, $i^l$ is only entangled with degrees of freedom to the left of the site and $i^r$ is only entangled with those to the right of the site. This can be seen more clearly when we put the fixed point matrices $A^{(\infty)}$ together and find the ground state wave function, as shown in Fig.\ref{wf_fp}. The total wave function takes a valence bond structure and is the tensor product of entangled pairs between neighboring sites
\be
\ket{\psi^{(\infty)}} = \prod_k \ket{EP_{k_r,k+1_l}} = \prod_k \left(\sum_{i^r,i^l} \sqrt{\lambda_{i^l}\eta_{i^r}} \left(\sum_{\alpha} v^{i^l}_{\alpha}u^{i^r}_{\alpha}\right) \ket{i^ri^l}\right)_{k,k+1}
\ee
The form of the entangled pair $\ket{EP_{k_r,k+1_l}}$ between site $k$ and $k+1$ looks rather complicated. But this is of no importance as we are free to apply a local unitary transformation and change it to any other state between the right spin on site $k$ and the left spin on site $k+1$.

\begin{figure}[htbp]
\begin{center}
\includegraphics[width=4.0in]{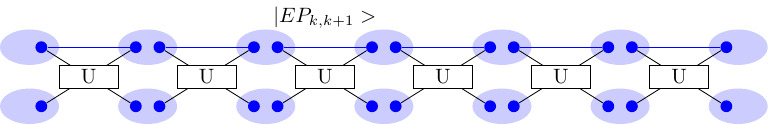}
\end{center}
\caption{Disentangling fixed point wave function (upper layer,
product of entangled pairs) into direct product state (lower
layer) with LU transformations.}
\label{fig:FP}
\end{figure}

In particular, we can
disentangle these pairs by applying one layer of local
unitary transformations between every neighboring sites and
map the state to a product state (Fig. \ref{fig:FP}).

Through these steps we have shown that all SRC
matrix product states can be mapped to product
states with LU transformations and have only short range entanglement. Therefore, there is
no topological order in 1D gapped bosonic system.

\section{Summary and further reading}

In this chapter, we introduce the matrix product state representation of 1D states and use it to classify 1D gapped phases in boson / spin systems without symmetry constraint. First, we discuss in detail the definition and basic properties of matrix product states, including their entanglement property, gauge degree of freedom and parent Hamiltonian. In particular, matrix product states satisfying the `injectivity' condition have a finite correlation length and is the unique gapped ground state of a local parent Hamiltonian. This set of matrix product states form the basis for our classification of 1D gapped phases. We describe a renormalization group transformation based on local unitary circuits and apply it to flow any `injective' matrix product state to a simple fixed point form. By analyzing the structure of all possible fixed point states, we show that all gapped phases in 1D are short range entangled and there is no intrinsic topological order in 1D boson / spin systems. We leave the discussion of fermion phases and phases with symmetry constraint to Chapter \ref{chap10}.

The one-dimensional AKLT state is the earliest example of matrix product states studied\cite{AKL8799}. Generalizations of this model were discussed in terms of `Finitely Correlated States'\cite{FNW9243,FNW9411} where it was shown that matrix product states with a finite correlation length are all gapped ground states of local Hamiltonians. A more detailed study of matrix product states, including its canonical form, is given in \cite{PVW0701}. It was later realized that, the powerful numerical method of Density Matrix Renormalization Group (DMRG)\cite{White9263} can be interpreted as a variational calculation with matrix product state ansatz\cite{OR9537,DMN9857}. Recent efforts have put the efficiency of the DMRG algorithm on more rigorous footing. It has been shown that all gapped ground states of one dimensional local Hamiltonians satisfy an entanglement area law\cite{H0724,AKL13arxiv} and that the necessary inner dimension to approximately describe these state scales only polynomially with system size\cite{SWV0804}. Therefore, the matrix product representation for such states is always efficient. Moreover, it has been rigorously proven that a polynomial time algorithm exist to find the matrix product state representation of 1D gapped states, although the algorithm differs from DMRG\cite{LVV13arxiv}.

The fact that the double tensor of a matrix product state (or a tensor product state in general) uniquely determines the state up to a local change of basis was proved in \cite{NC2000} in the form of the unitary degree of freedom in the operator sum representation of quantum channels $\mathcal{E}$ which is defined in terms the matrices as $\mathcal{E}(X)=\sum_i A_i X A^{\dg}_i$.

The renormalization group transformation on matrix product states described in this chapter was proposed in \cite{VCL0501}, where a partial classification of 1D matrix product states were obtained. It was shown in \cite{CGW1128, SPC1139} using matrix product states that no intrinsic topological order exist in 1D boson / spin systems.

%
%
\bibliographystyle{plain}
\bibliography{Chap8}

\chapter{Tensor Product States and 2D Gapped Phases}
\label{chap9}

\abstract{Tensor product state is a natural generalization of matrix product state to two and higher dimensions. It is similar to matrix product states in many ways, like satisfying the entanglement area law and having a projected entangled pair interpretation. However, it is also different from matrix product states in that it represents not only states with short-range entanglement but topologically ordered states with long-range entanglement as well. In this chapter, we first introduce the basic properties of tensor product states and then proceed to discuss how it represents symmetry breaking phases and topological phases with explicit examples. In particular, we are going to focus on the general structural properties of the local tensors which are responsible for the corresponding symmetry breaking or topological order. Other forms of tensor network state, including the tree tensor network state and MERA (multiscale entanglement renormalization ansatz), are also discussed.
}

\section{Introduction}

The matrix product state representation had a great success in 1D both analytically and numerically. Can we achieve the same kind of success in two and higher dimension? Tensor product states (TPS) provide a natural generalization of matrix product states to higher dimensions by placing higher rank tensors, instead of matrices, on each lattice site. The representation hence obtained is similar to matrix product states in many ways. For example, TPS satisfies the entanglement area law. As ground states in general dimensions are found to obey the area law, TPS is expected to provide a good representation of them. 

On the other hand, it is much harder to achieve analytical rigorousness and numerical efficiency with TPS. It has not been proven that the TPS representation of gapped ground states is always efficient. Also it is not easy to identify which subset of TPS correspond to gapped ground states and which subset to gapless ones. On the numerical side, variational simulation using TPS requires the contraction of a two dimensional tensor network, which is in general computationally hard. Computation accuracy needs to be sacrificed in order to achieve efficiency.

Despite all this, tensor product state is also more interesting than matrix product state because it can describe long-range entangled states, apart from short ranged ones. Simple tensor product representation exists for a large class of topologically ordered states. A better understanding of how topological order, as a global feature, can emerge out of local tensors provides a deeper understanding of the special entanglement structure in such phases.

In this chapter, we start by introducing the definition and basic properties of tensor product states in section \ref{sec:TPS}. In particular, we compare and contrast it to what we already know about matrix product states. We then move on to discuss how to represent different 2D phases using tensor product states. In section \ref{sec:TN_SB} symmetry breaking phases are discussed with the example of Ising model, and in section \ref{sec:TN_TO} topological phases is discussed with the example of toric code. We focus in particular on the structural properties of the local tensors which is responsible for the corresponding symmetry breaking or topological order. Matrix product states and tensor product states are special examples of tensor network representations. In section \ref{sec:TN}, we briefly introduce other forms of tensor network states, including the tree tensor network state and MERA (the multiscale entanglement renormalization ansatz).

\section{Tensor product states}
\label{sec:TPS}

The idea of introducing extra inner indices to efficiently represent many-body entangled states can be generalized to higher dimensions. In describing one dimensional many-body entangled states, a set of matrices were used whose left and right indices encode the entanglement to the left and right part of the chain. To describe two and higher dimensional many-body entangled states, we need tensors with three or more inner indices to represent entanglement in a higher dimensional space. Such higher dimensional generalizations of matrix product states are in general called the 'Tensor Product States'. Tensor product states share many properties with their matrix product counterparts, like an entanglement area law and a projected entangled pair picture. However, some properties of matrix product states do not generalize to tensor product states and in general we know much less about tensor product states than matrix product states. In this section, we summarize what we know and what we do not know about tensor product states.

\subsection{Definition and examples}
\label{sec:def_TPS}

\begin{svgraybox}
\begin{center}
\textbf{Box 9.1 Tensor product state}

A tensor product state in a many-body spin system is represented as
\begin{equation}
|\psi\rangle=\sum_{i_1,i_2,...i_m...}\text{tTr}(T^{i_1}T^{i_2}...T^{i_m}...)|i_1
i_2 ... i_m...\rangle \label{TPS}
\end{equation}
\end{center}
\end{svgraybox}

Here $i_k=1...d$, with $d$ being the physical dimension of each spin in the system. $T_i$'s are tensors living on each site of a lattice with three or more inner indices. They are usually connected according to the underlying lattice structure of the system and \text{tTr} represents tensor contraction. Here by tensor we mean in general a set of numbers labeled by several indices. A vector is a tensor with one index and a matrix is a tensor with two indices. WLOG, the word `tensor' is usually used when there are three or more indices. Two tensors can be contracted if we match their corresponding indices, multiply their values and sum over the matched indices. For vectors or matrices, such an operation corresponds to vector or matrix multiplication.

\begin{figure}[htbp] \centering
\includegraphics[width=3.0in]{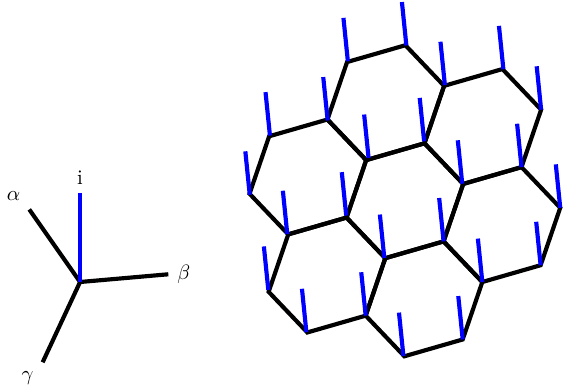}
\caption{Left: tensor $T$ representing a
2D quantum state on hexagonal lattice. $i$ is the physical index,
$\alpha,\beta,\gamma$ are inner indices. Right: a tensor product
state where each vertex is associated with a tensor. The inner
indices of the neighboring tensors connect according to the
underlying hexagonal lattice.} \label{fig:TPS}
\end{figure}

For example, consider a two-dimensional spin model on a hexagonal lattice with
one spin (or one qudit) living at each vertex. The state can be represented by
assigning to every vertex a set of tensors
$T^{i}_{\alpha\beta\gamma}$, where $i$ labels the local physical
dimension and takes value from $1$ to $d$. $\alpha,\beta,\gamma$ are
inner indices along the three directions in the hexagonal lattice
respectively. The dimension of the inner indices is $D$. Fig. \ref{fig:TPS} gives a side view of a local tensor and a tensor product state with inner indices in the
horizontal plane and the physical indices pointing in the vertical
direction.

Note that a 'Tensor Product State' is different from a 'Product State'. By 'Product State', we mean that the wave function is a product of wave functions on each individual spin
\be
|\psi_{\text{product state}}\rangle=|\psi_1\rangle \otimes ... \otimes |\psi_N\rangle
\ee
However, a 'Tensor Product State' is in general an entangled state. By 'Tensor Product State', we mean that the wave function is written as in Eq. \ref{TPS} in terms of local tensors. When a product state is written in terms of the tensor product formalism, we only need trivial local tensors with all inner indices being one dimensional. 

Nontrivial tensor product states are many-body entangled. For example, the GHZ state
\be
|\psi_{\text{GHZ}}\rangle = \frac{1}{\sqrt{2}} \left(|00...0\rangle + |11...1\rangle \right)
\ee
can be represented with tensors (ignore normalization of the wave function)
\be
T^0_{0...0}=1,\ T^1_{1...1}=1,\ \text{all other terms are $0$}
\ee

More interestingly, some highly nontrivial topological states can also be represented in a very simple way using tensors. Toric code is an example. Consider a toric code model defined on a two dimensional square lattice with one spin $1/2$ per each link. The Hamiltonian of the toric cdoe is a sum of vertex and plaquette term (as introduced in chapter \ref{cp:3} and chapter \ref{cp:5})
\be
H_{\text{toric code}}= -\sum_s \prod_{j\in \text{star}(s)} Z_j - \sum_p \prod_{j\in \text{plaquette}(p)} X_j
\ee
and the ground state wave function is an equal weight superposition of all closed loop configurations 
\be
\ket{\psi}_{\text{toric code}} = \sum_{C} \ket{C}
\label{TC_wf}
\ee
This wave function can be represented as a tensor product state with two sets of tensors, one at each vertex and one on each link. The one at the vertex $T$ has four two-dimensional inner indices but no physical index
\be
\begin{array}{ll}
T_{ijkl}=1,  & \text{\ if\ } i+j+k+l=0 \text{\ mod \ }2; \\ \nonumber
T_{ijkl}=0,  & \text{\ if\ } i+j+k+l=1 \text{\ mod \ }2; 
\end{array}
\ee
The one on each link $t$ has two two-dimensional inner indices and one two-dimensional physical index
\be
t^{0}_{00}=t^{1}_{11}=1, \text{\ all other terms are $0$}
\ee
The tensors connect according to the underlying square lattice as shown in Fig. \ref{fig:TPS_TC}.

\begin{figure}[htbp] \centering
\includegraphics[width=2.0in]{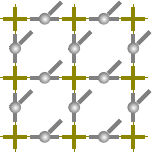}
\caption{Tensor product representation of the toric code state.} \label{fig:TPS_TC}
\end{figure}

It is easy to see why these tensors give rise to the wave function in Eq. \ref{TC_wf} by interpreting the $0$ inner index as no string and the $1$ inner index as with a string. $t$ then connects the physical spin state with the string and $T$ enforces the constraint that all strings form closed loops. All closed loop configurations enter the wave function with an equal amplitude.

\subsection{Properties}

\subsubsection{Properties similar to matrix product state}

Tensor product states are similar to matrix product states in terms of the formulation of double tensor, entanglement area law, gauge degree of freedom, and PEPS description although it is not clear when it describes a short range correlated gapped state and when it does not.

For tensor product states, we can similarly define a double tensor which can be used for the calculation of the norm and local observables on the state.
\be
\mathbb{T}_{\alpha...\gamma,\t{\alpha}...\t{\gamma}}=\sum_i T^{i}_{\alpha...\gamma}\left(T^{i}_{\t{\alpha}...\t{\gamma}}\right)^*
\ee
The norm of the tensor product state can be calculated by contracting all the double tensors according to the tensor network structure of the tensor product state
\be
\langle \psi|\psi\rangle = \text{tTr}(\mathbb{T}[1]\mathbb{T}[2]...\mathbb{T}[m]...)
\ee
where $\mathbb{T}[m]$ is the double tensor at the $m$th lattice site.

The expectation value of a local observable $O$ at site $n$ is given by
\be
\langle O\rangle = \frac{\text{tTr}(\mathbb{T}[1]\mathbb{T}[2]...\mathbb{T}_O[n]...\mathbb{T}[m]...)}{\text{tTr}(\mathbb{T}[1]\mathbb{T}[2]...\mathbb{T}[n]...\mathbb{T}[m]...)}
\ee
where
\be
\left(\mathbb{T}_O\right)_{\alpha...\gamma,\t{\alpha}...\t{\gamma}}= \sum_{i,j} O_{ij} T^{i}_{\alpha...\gamma}\left(T^{j}_{\t{\alpha}...\t{\gamma}}\right)^*
\ee

Tensor product states also enjoy the nice property of having an entanglement area law. In fact, for a tensor product state with inner dimension $D$, the rank of the reduced density matrix of a subregion is bounded by $D^n$, where $n$ is the number of indices connecting the subregion with the rest of the system. As $n$ scales linearly with the boundary $L$ of the subregion, the entanglement entropy of a subregion also scales linearly with with $L$.

\begin{svgraybox}
\begin{center}
\textbf{Box 9.2 Entanglement area law for tensor product states}

In a finite dimensional tensor product state, the entanglement entropy of a subregion scales linearly with the length $L$ of the boundary of the subregion
\be
S \sim \alpha L
\ee
\end{center}
\end{svgraybox}

Therefore, tensor product states could provide a nice description of gapped quantum systems in two and higher dimensions.

Similar to the matrix product state representation, the tensor product state representation also has a gauge degree of freedom. In particular,
\be
T'_{\al'\bt'\ga'}=\sum_{\al,\bt,\ga}M_{\al'\al}N_{\bt'\bt}O_{\ga'\ga}T_{\al\bt\ga}
\ee
represents the same state as $T$ if the invertible matrices $M$,$N$,$O$ cancel out for each pair of connected indices.

The projected entangled pair state (PEPS) representation of tensor product states can be constructed analogously as for matrix product states. Starting from a two or higher dimensional lattice with maximally entangled virtual pairs $|\psi\rangle=\frac{1}{\sqrt{D}}\sum^D_{\al=1}|\al\al\rangle$ between nearest neighbor sites, apply a mapping $P$ from virtual spins at each site to the physical Hilbert space 
\be
P=\sum_{i,\alpha,...,\gamma} T^i_{\alpha,...,\gamma} |i\rangle\langle \al...\gamma|
\ee
The wave function obtained in this procedure is the tensor product state represented by tensors $T^i_{\al,...,\gamma}$. Equivalently, we are start from other maximally entangled state which are all local unitary equivalent to each other. 

An interesting many-body entangled state that can be understood in the PEPS representation is the AKLT state (in two or higher dimensions). For a lattice with degree $n$ vertices, put singlet pairs $|\psi\rangle=\frac{1}{\sqrt{2}}(|01\rangle-|10\rangle)$ onto each link. Then project the $n$ spin $1/2$'s at each vertex to the spin $n/2$ space. The wave function obtained in this way is called the spin $n/2$ AKLT state. From this construction we can see that it is naturally invariant under global spin rotation symmetry.

\subsubsection{Properties different from matrix product states}

On the other hand, tensor product states are also different from matrix product states in many ways. 

First of all, there is no known efficient way to extract the correlation length of the state from the tensors. Therefore it is not easy to identify tensor product states which are gapped ground states of local Hamiltonians. An analogous notion of injectivity can be defined for tensor product states. An injective tensor product state satisfies that for a large enough region with $L$ sites inside, the following map is injective
\be
\Gamma_L: W \mapsto \sum^{d}_{i_1,...i_L=1} \text{tTr}(WT^{i_1}...T^{i_L})|i_1...i_L\> 
\ee
where $W$ is a tensor which contracts to all the open inner indices around the boundary of the region. 
Injectivity is still a generic property for tensor product states on two and three dimensional (or any finite dimensional) lattice, however, they are no longer directly related to finite correlation length. In fact, it is known that there are injective tensor product states whose correlation functions only decay polynomially. Therefore, it is not clear which subset of tensor product states describe short range correlated, gapped quantum states. 

Moreover, one major difficulty with using tensor product states for numerical simulation is that the contraction of tensor networks in two and higher dimensions is in general not efficient. Usually an approximate renormalization algorithm is used, but the error is not always well bounded.

\subsubsection{Approximate calculation of local observables}

In order to calculate expectation value of local observables in a tensor product state, we need to contract two-dimensional tensor networks. Unlike in the one-dimensional case, the contraction of a two-dimensional tensor network is not efficient in general. Approximate methods have been developed to efficiently evaluate the tensor contraction for physical tensor networks of interest and can be implemented as follows.

\begin{figure}[htbp] 
\centerline{
\includegraphics[width=4.5in]{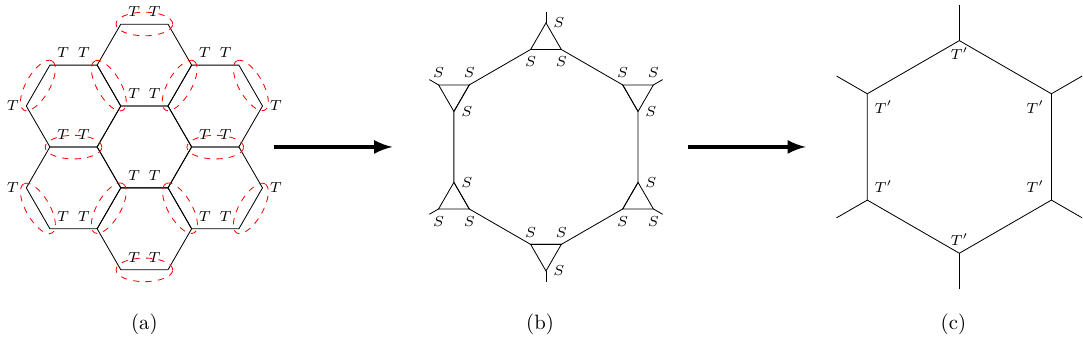}
}
\caption{Tensor renormalization on hexagonal lattice.} \label{RG_hex_a}
\end{figure}

\begin{figure}[htbp] 
\centerline{
\includegraphics[width=2.0in]{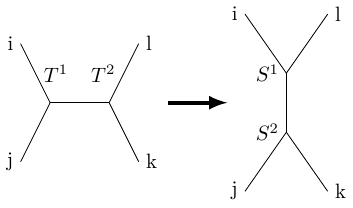}~~~~~~
\includegraphics[width=2.5in]{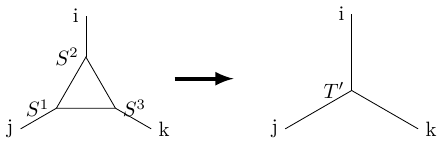}
}
\centerline{
	(a) ~~~~~~~~~~~~~~~~~~~~~~~~~~~~~~~~~~~~~~~~~~~~~~~~~~~~~~~~~~
	(b) 
}
\caption{Individual steps of tensor renormalization on hexagonal lattice.} \label{RG_hex_b}
\end{figure}

Consider, for example, a tensor network on a honeycomb lattice with one three-indexed tensor $T_{ijk}$ per each site. First, combine each pair of tensors in dashed circles in Fig. \ref{RG_hex_a}(a). Apply a singular value decomposition in the perpendicular direction and obtain tensors $S^1$ and $S^2$ which satisfy
\be
\sum_{m} T^1_{ijm}T^2_{mkl} = \sum_{n} S^1_{iln}S^2_{njk}
\ee
In order to keep the computation efficient, we need to keep an upper bound $D_{cut}$ on the dimension of the indices. In this decomposition step, we may need to cut off on the dimension of $n$ if the number of nonzero singular values exceeds $D_{cut}$. The most natural way to do this cut-off is to throw away dimensions with singular values of smallest weights. This also guarantees the best approximation of the original tensor network by the transformed one.

After such a step, the hexagonal lattice changes into a structure depicted in Fig.\ref{RG_hex_a}(b). Then combine the three $S$ tensors around a triangle into a $T'$ tensor
\be
\sum_{lmn} S^1_{iln}S^2_{ljm}S^3_{mnk} = T'_{ijk}
\ee
No approximation is necessary in this step.
After this combination, the lattice is transformed back into a regular hexagonal structure as shown in Fig.\ref{RG_hex_a}(c), but with lattice constant $\sqrt{3}$ times that of the original lattice in (a) and only one third the number of tensors. With these steps, we have finished one round of tensor renormalization group transformation and the resultant $T'$ tensor can be used as the starting point for the next round of RG. After $N$ rounds of RG transformaiton, we can reduce a tensor network with, for example, $3^{N+1}$ tensors to one with only $3$ tensors, which is then trivial to evaluate. Therefore, this RG scheme provides us with an efficient, although approximate way to contract big tensor networks.
 
A similar RG procedure can also be devised for a square lattice tensor network.

\begin{figure}[htbp] \centering
\includegraphics{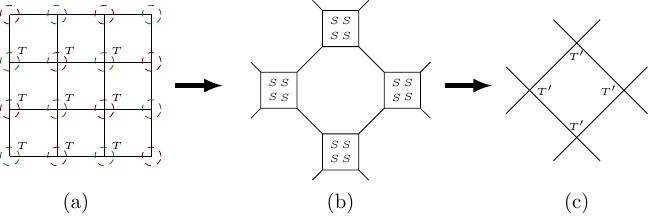}
\caption{Tensor renormalization on square lattice.} \label{RG_sqr_a}
\end{figure}

\begin{figure}[htbp]
\centerline{
\includegraphics[width=2.0in]{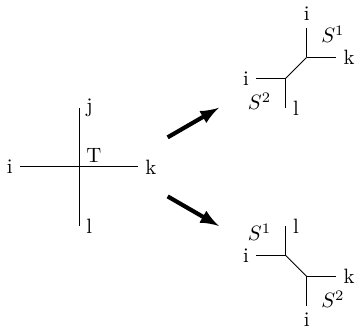}~~~~~~
\includegraphics[width=2.2in]{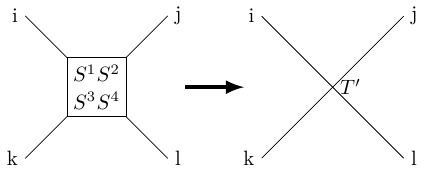}
}
\centerline{
	(a) ~~~~~~~~~~~~~~~~~~~~~~~~~~~~~~~~~~~~~~~~~~~~~~~~~~~~~~~~~~
	(b) 
}
\caption{Individual steps of tensor renormalization on square lattice.} \label{RG_sqr_b}
\end{figure}

As shown in Fig.\ref{RG_sqr_b}(a), first take a local tensor $T_{ijkl}$ and do a singular value decomposition into two tensors $S^1$ and $S^2$. The two ways of decomposition are applied to the two sublattice (red and green) in the square lattice as shown in Fig.\ref{RG_sqr_a}(a).
\be
T_{ijkl}=\sum_m S^1_{jkm}S^2_{mil} \  \text{or} \ \ T_{ijkl}=\sum_m S^1_{ijm}S^2_{mkl}
\ee
After the decomposition, the lattice structure is transformed into Fig.\ref{RG_sqr_a}(b). Similarly, we may need to cut-off the dimension of $m$ at $D_{cut}$ to keep the computation efficient.

Then combine four $S$ tensors around a square into a $T'$ tensor, as shown in Fig.\ref{RG_sqr_b}(b)
\be
\sum_{mnpq} S^1_{iqm}S^2_{jmn}S^3_{knp}S^4_{lpq} = T'_{ijkl}
\ee
No cut-off is necessary in this step. After the combination, the lattice is transformed back into a square lattice structure (Fig.\ref{RG_sqr_a}(c)), with $\sqrt{2}$ times the lattice constant and half the number of tensors as the original lattice (Fig.\ref{RG_sqr_a}(a)). The resultant tensor $T'$ serves as the input of the next round of RG process. Similar to the hexagonal case, this provides us with a way to efficiently approximately contract a 2D tensor network on a square lattice.


\section{Tensor network for symmetry breaking phases}
\label{sec:TN_SB}

The transverse field Ising model is a prototypical example of a gapped symmetry breaking phase and we are going to use it for the study of the tensor network representation of symmetry breaking phases. We are going to see how symmetry breaking is encoded in the structure of the local tensors representing the ground states. As a simple and interesting example of short range entangled states, it will be compared to later when we discuss the long range entangled cases.

\subsection{Ising model}

On a 2D lattice of two-level spins, the Hamiltonian of the transverse field Ising model takes the form
\be
H^{\text{tIsing}}=-J\sum_{<ij>} Z_iZ_j - B\sum_i X_i
\ee
where $<ij>$ are nearest neighbor pairs of spins on the lattice. The system has a $\mathbb{Z}_2$ symmetry of
\be
U=\prod_i X_i
\ee
which flips the spins between $\ket{\uparrow}$ and $\ket{\downarrow}$.

When $J=1$ and $B=0$, the system is in a symmetry breaking phase with two degenerate ground states $\ket{\uparrow\uparrow...\uparrow}$ and $\ket{\downarrow\downarrow...\downarrow}$. Each state breaks the $\mathbb{Z}_2$ symmetry and maps into each other under the symmetry transformation. Their superposition, $\ket{\uparrow\uparrow...\uparrow}+\ket{\downarrow\downarrow...\downarrow}$, is however symmetric under $\mathbb{Z}_2$ and has a tensor network representation as
\be
T^{\uparrow}_{000}=1, \ T^{\downarrow}_{111}=1, \ \ \ \text{all other terms being $0$}
\ee
on a hexagonal lattice and 
\be
T^{\uparrow}_{0000}=1, \ T^{\downarrow}_{1111}=1, \ \ \ \text{all other terms being $0$}
\ee
on a square lattice. Here in $T^i_{\al\bt...}$, $i$ is the physical index and $\al\bt...$ are the inner indices.

When $J=0$ and $B=1$, the system is in a simple symmetric phase with a unique ground state $\ket{\rightarrow\rightarrow...\rightarrow}$ where $\ket{\rightarrow}=\left(\ket{\uparrow}+\ket{\downarrow}\right)/\sqrt{2}$. The tensor product representation for this state is simply $T^{\uparrow}=1$, $T^{\downarrow}=1$.

\subsection{Structural properties}

The tensor for the symmetric ground state in the symmetry breaking phase has some interesting structural properties. 

First of all, we can see that it is not injective. The notion of injectivity was defined in chapter \ref{chap6}. The tensor on any local region is supported on only two dimensions of the inner indices $\ket{00...0}$ and $\ket{11...1}$. All other dimensions are $0$. In fact, the tensor can be decomposed into two blocks $T_a$ and $T_b$
\be
T^{\uparrow}_{a,00...0}=1, \  \text{all other terms being $0$}
\ee
and
\be
T^{\downarrow}_{b,11...1}=1, \  \text{all other terms being $0$}
\ee
$T_a$ and $T_b$ are each supported on orthogonal inner dimensions and there are no cross terms in the tensor.

Moreover, the tensor has a $\mathbb{Z}_2$ symmetry which acts as $X$ on the physical index and $X$ on the inner indices. The two blocks map into each other under the symmetry and the whole tensor is invariant. When the tensor network is contracted togehter, the $X$ transformation on the inner indices cancel in pairs and the full state is invariant under $\prod_i X_i$ on all the physical indices, as expected.

\subsection{Symmetry breaking and the block structure of tensors}

While the meaning of symmetry breaking is straight-forward in classical system, this concept is more subtle in
the quantum setting. The tensor network representation provides better insight into the notion of quantum symmetry breaking.

A classical system is in a symmetry breaking phase if each possible ground state has lower
symmetry than the total system. For example, the classical Ising model with Hamiltonian
\be
H_c=\sum_{<ij>} n^in^j
\ee
has a spin flip symmetry between spin up $\uparrow$ and spin down $\downarrow$. Here $n=1$ for spin up $\uparrow$ and $n=-1$ for spin down $\downarrow$. However neither of its ground
states $\uparrow\uparrow...\uparrow$ and $\downarrow\downarrow...\downarrow$ has this symmetry. Therefore, the meaning of symmetry breaking in classical systems is obvious.

However, in the quantum Ising model discussed above, at $J=1$ and $B=0$, the ground space contains not only the two states of $\ket{\uparrow\uparrow...\uparrow}$ and $\ket{\downarrow\downarrow...\downarrow}$, but also any superposition of them. While each of $\ket{\uparrow\uparrow...\uparrow}$ and $\ket{\downarrow\downarrow...\downarrow}$ breaks the $\mathbb{Z}_2$ symmetry, their superposition $\ket{\uparrow\uparrow...\uparrow}+\ket{\downarrow\downarrow...\downarrow}$ is invariant under this symmetry. This is the GHZ state as we discussed before. In fact, if we move away from the exactly solvable point by adding symmetry preserving perturbations (such as transverse field $B\sum_i X_i$) and solve for the ground state with finite system size, we will always get a state symmetric under this spin flip symmetry. Only in the thermodynamic limit does the ground space become two dimensional. How do we tell then whether the ground states of the system spontaneously break the symmetry?

With the tensor network representation (including the matrix product state representation), the symmetry breaking pattern can be easily seen from the tensors (matrices). Suppose that we solved the ground state of a system with certain symmetry at finite size and found a unique minimum energy state which has the same symmetry. To see whether the system is in the symmetry breaking phase, we can write this minimum energy state in the tensor product state representation. The tensors in the representation can be put into a block form
\be
T^i_{\al\bt...}= T^i_{a,\al_a\bt_a...} \oplus T^i_{b,\al_b\bt_b...} \oplus ...
\ee
where $\al_a$, $\al_b$, etc. span orthogonal sub-dimensions of $\al$, $\bt_a$, and $\bt_b$, etc. span orthogonal sub-dimensions of $\bt$. When the tensor network is contracted, only tensors of the same block contract with each other. Tensors of different blocks are supported on orthogonal dimensions and their contraction is $0$. This decomposition of $T^i$ is such that $T_a^i$, $T^i_b$ etc. each represents a short range correlated state. Then if in the thermodynamic limit, the block form of the tensor contains only one block, this minimum energy state is short range
correlated and the system is in a symmetric phase. However,
if the block form splits into more than one block with
equal amplitude, then we say the symmetry of the system
is spontaneously broken in the ground states. 

We can see that the tensors for the  $J=1,B=0$ ground states contain two blocks
while the tensors for the $J=0,B=1$ ground state contain only one block. Therefore, we say the symmetry is spontaneously broken in the former and not broken in the latter.

The symmetry breaking interpretation of the block form can be understood as follows. Each block $T_k$ represents a short range correlated state $\ket{\psi_k}$. Note that here by correlation we always mean connected correlation $<O_1O_2>-<O_1><O_2>$. Therefore, the symmetry breaking states like $\ket{\uparrow\uparrow...\uparrow}$ and $\ket{\downarrow\downarrow...\downarrow}$ both have short range correlation. Two different short range correlated state $\ket{\psi_k}$ and $\ket{\psi_{k'}}$ have zero overlap $\langle \psi_k | \psi_{k'} \rangle = 0$ and any local observable has zero matrix element between them $\langle \psi_k |O | \psi_{k'} \rangle = 0$. The ground state represented by $T^i$ is an equal weight superposition of them $\ket{\psi}=\sum_k \ket{\psi_k}$. Actually the totally mixed state $\rho=\sum_k \ket{\psi_k}\bra{\psi_k}$ has the same energy as $\ket{\psi}$ as $\langle \psi_k |H| \psi_{k'} \rangle = 0$ for $k \neq k'$. Therefore, the ground space is spanned by all $\ket{\psi_{k}}$'s. Consider the operation which permutes the $\ket{\psi_k}$'s. This operation keeps ground space invariant and can be a symmetry of the system. However,
each short range correlated ground state is changed under this operation. Therefore, we say that the ground
states spontaneously break the symmetry of the system.

\begin{svgraybox}
\begin{center}
\textbf{Box 9.3 Symmetry breaking and block structure of tensors}

Tensors representing the Ising symmetry breaking ground state have multiple blocks with each block representing different short range correlated spin configurations.

\end{center}
\end{svgraybox}


\section{Tensor network for topological phases}
\label{sec:TN_TO}

While the quantum symmetry breaking phases have their classical counterparts and can be understood at least qualitatively using a classical picture, the topological phases are intrinsically quantum and demand an inherently quantum approach of study. Tensor networks can be used to represent a large class of topological states and provide a promising tool for both analytical and numerical study of topological phases. The fact that states with long range entanglement can be represented with local tensors is very surprising and in this section we are going to discuss, with the example of the toric code model, how the topological property of the state is manifested in the tensors.

\subsection{Toric code model}

Recall the toric code tensor network introduced in section \ref{sec:def_TPS}. The Hamiltonian reads
\be
H_{\text{toric code}}= -\sum_s \prod_{j\in \text{star}(s)} Z_j - \sum_p \prod_{j\in \text{plaquette}(p)} X_j
\label{H_Z2}
\ee
and the ground state wave function can be represented as a tensor product state with $T$ tensors at the vertex and $t$ tensors on the links
\be
\begin{array}{ll}
T_{ijkl}=1,  & \text{\ if\ } i+j+k+l=0 \text{\ mod \ }2; \\ \nonumber
T_{ijkl}=0,  & \text{\ if\ } i+j+k+l=1 \text{\ mod \ }2; 
\end{array}
\label{T_tc}
\ee
The one on each link $t$ has two two-dimensional inner indices and one two-dimensional physical index
\be
t^{0}_{00}=t^{1}_{11}=1, \text{\ all other terms are $0$}
\label{tt_tc}
\ee
The tensors connect according to the underlying square lattice as shown in Fig. \ref{fig:TPS_TC}.

\begin{figure}[htbp] \centering
\includegraphics[width=2.0in]{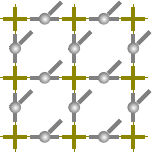}
\caption{Tensor product representation of the toric code state.} \label{fig:TPS_TC}
\end{figure}

\subsection{Structural properties}
\label{Z2sym_TC}

\begin{figure}[htp] 
\centerline{
\includegraphics[width=1.6in]{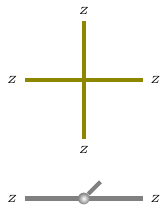}~~~~~~
\includegraphics[width=2.0in]{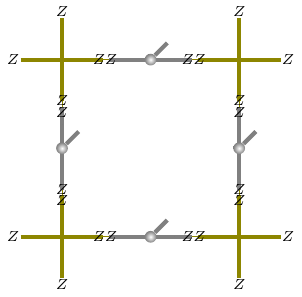} 
}
\centerline{
	(a) ~~~~~~~~~~~~~~~~~~~~~~~~~~~~~~~~~~~~~~~~~~~~~~~~~~~~~~~~~~
	(b) 
}
\caption{Symmetry of
the toric code tensor. (a) Each local tensor remains invariant under the
action of $Z \otimes Z \otimes Z$ on its inner indices. (b) A region of tensor network is invariant under $\prod_i Z_i$ on its open inner indices.} \label{fig:symm}
\end{figure}

The most important property of the above tensor is that it has certain inner
symmetry, that is, the tensor is invariant under some
non-trivial operations on the inner indices, as shown in
Fig. \ref{fig:symm} (a).

$Z$ does nothing to the tensor when the index is $0$ and
changes the sign of the tensor when the index is $1$. In the
$T$ tensor of the ideal toric code (Eq.\ref{T_tc}), only even configurations
of the inner indices are non-zero. Hence applying $Z$ at the
same time to all four inner indices doesn't change the
tensor. That is, $Z \otimes Z \otimes Z \otimes Z$ is a symmetry of
the tensor. Similarly, in the $t$ tensor (Eq.\ref{tt_tc}), the two inner indices are either both $0$ or both $1$. Therefore, applying $Z$ at the same time to both inner indices does not change the tensor. That is, $Z\times Z$ is a symmetry of the tensor. Note that this symmetry operation does not act on the physical indices at all and is purely an inner property of the tensor.

As $Z$ squares to identity,
we will say that the tensor has a $\mathbb{Z}_2$ symmetry. 
Note that we can insert a set of unitary
operators $U,U^{\dagger}$ between any connected links in a
tensor network without affecting the result of tensor
contraction and hence the quantity represented by the tensor
network. Therefore, the $\mathbb{Z}_2$ symmetry could take any 
form which is local unitary equivalent to $Z \otimes Z \otimes Z \otimes Z$.

This symmetry property is true not only for each local tensor but for any region in the tensor network as well. As shown in Fig.\ref{fig:symm} (b), when the local tensors are put together, the symmetry transformation on the contracted inner indices cancel in pairs and the symmetry transformation on the outer un-contracted inner indices are left behind. Therefore, this piece of tensor network has also a $\mathbb{Z}_2$ symmetry given by $\prod_i Z_i$ over all its outer inner indices. 

This $\mathbb{Z}_2$ symmetry is closely related to the closed loop constraint of the state. 
Due to this symmetry, the tensor network cannot be `injective', because only even configurations can be nonzero on each any piece of tensor network. But the tensor does span the full space which is even under this $\mathbb{Z}_2$ transformation. 

\begin{svgraybox}
\begin{center}
\textbf{Box 9.4 $\mathbb{Z}_2$ injectivity of toric code tensor}

The tensor representing the toric code ground state has an inner $\mathbb{Z}_2$ symmetry and it spans the full space which is $\mathbb{Z}_2$ invariant. The tensor is hence said to be $\mathbb{Z}_2$ injective.

\end{center}
\end{svgraybox}

\subsection{Topological property from local tensors}

Encoded in this $\mathbb{Z}_2$ symmetry of the tensor are some interesting topological properties of the toric code wave function. 

\begin{figure}[htp]
\centerline{
\includegraphics[width=2.0in]{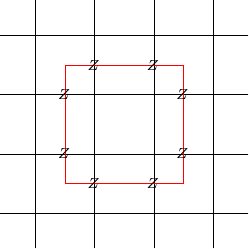}~~~~~~
\includegraphics[width=2.0in]{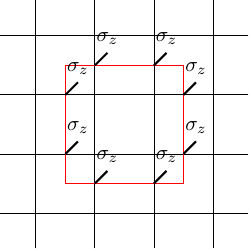} 
}
\centerline{
	(a) ~~~~~~~~~~~~~~~~~~~~~~~~~~~~~~~~~~~~~~~~~~~~~~~~~~~~~~~~~~
	(b) 
}
\caption{The tensor network is invariant under (a) acting $Z$ on all inner indices surrounding a region (b) acting $Z$ on all the spin-$1/2$'s in a loop. Physical indices are not shown unless they are acted upon.} \label{loop}
\end{figure}

Consider a tensor network as shown in Fig.\ref{loop}. For clarity, we are not drawing the physical indices unless they are acted upon. Insert $Z$ operators on the inner indices around a region, as shown in Fig.\ref{loop} (a). As discussed above, as long as the $Z$ operators act on all the outer inner indices of a region, the tensors are invariant and the state represented remains the same. Due to the one-to-one correspondence between the inner configurations and the physical configurations, such an action on the inner indices translates into a physical action on the spin-$1/2$'s in the wave function, as shown in Fig.\ref{loop} (b). Therefore, the toric code wave function is invariant under a loop operator of $Z$'s around a region. This is the so-called `Wilson-loop' operator and is a hall-mark for topological phases.

\begin{figure}[htp] \centering
\includegraphics[width=4.0in]{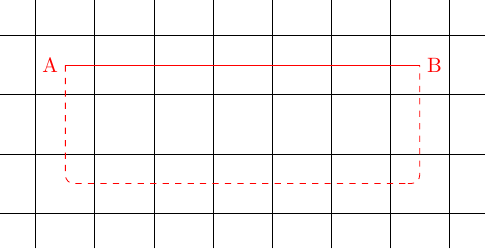} \caption{Acting $Z$ operators along a defect line changes the state only near the two end points, which does not depend on the exact path of the defect line.} \label{string}
\end{figure}

If we insert $Z$ operators not around a full loop but only along a defect line (solid line in Fig.\ref{string}), then the tensor network does change, but only near the two end points ($A$ and $B$). This is easy to understand by noting that if we complete the circle by inserting $Z$'s along another defect line connecting the same end points (dashed line in Fig.\ref{string}), the tensor network goes back to itself. As the two lines can be very far away from each other except at the end points, their effects cannot cancel anywhere else. Therefore, in the middle part of the defect line, the state represented remains the same. Such a string operator hence creates two local excitations in the system near the end points, which are actually the `charge' particle of the corresponding $\mathbb{Z}_2$ gauge theory. We are free to move the defect line around and the excitations remain the same as long as we keep the end points fixed. From the string operator it is easy to see that when the `charge' particles braid with each other, the resulting statistics is trivial.

Now imagine putting the tensor network onto a torus. The ground space of the toric code model is four fold degenerate and the state represented with tensors in Eq.\ref{T_tc} and \ref{tt_tc} is one of them. Insert the $Z$ operators long a nontrivial loop of the torus. As the nontrivial loop does not enclose any region, the tensor network does change. However, as discussed above, we can move the location of the loop around without effecting the resulting state. This is because two nontrivial loops in the same direction on the torus always enclose a region and keeps the original tensor network invariant. Therefore, the effect of a single nontrivial loop is the same no matter where the loop is. 

Because of this, the resulting state must have the same energy or any local observable as the original state because in calculating them we can always move the inserted loop to be very far away from the location of the operator. That is, the loop operator maps between degenerate ground states of the toric code model which cannot be distinguished from each other with any local operator. Translated to the physical spin-$1/2$, this corresponds to a nontrivial loop of $Z$ operators which is one of the logical operators that rotates the degenerate ground space on the torus.

\subsection{Stability under symmetry constraint}
\label{sta_TN}

The inner $\mathbb{Z}_2$ symmetry is essential for keeping the stability of the topological order represented by the tensor network state. A slight violation of the symmetry at each local tensor can immediately destroy the topological order of the state, as shown in the example below.

Let us break the $\mathbb{Z}_2$ symmetry by assigning a small and equal weight
$\epsilon$ to all odd configurations in the $T$ tensor, which now reads,
\be
\begin{array}{ll}
T_{ijkl}=1,  & \text{\ if\ } i+j+k+l=0 \text{\ mod \ }2; \\ \nonumber
T_{ijkl}=\epsilon,  & \text{\ if\ } i+j+k+l=1 \text{\ mod \ }2; 
\end{array}
\label{TPS_Z2_e}
\ee
We keep the $t$ tensor invariant.
\be
t^{0}_{00}=t^{1}_{11}=1, \text{\ all other terms are $0$}
\ee
Note that even though each $t$ tensor is still $\mathbb{Z}_2$ symmetric, a piece of tensor network involving both $T$ and $t$ tensors will in general break the symmetry.

When $\epsilon=0$, this is reduced to the tensors in the ideal
toric code TPS. When $\epsilon$ is non-zero, odd configurations are allowed at each vertex, which correspond to end of strings.
The wave function with nonzero $\epsilon$ then contains all possible string configurations,
closed loop or open string. The weight of each string
configuration is exponentially small in the number of end of
strings contained. 
 \begin{equation}
|\psi^{\epsilon}_{\text{toric code}}\rangle = \sum_{C'}
\epsilon^{N(C')} \ket{C'} \label{Phi_z2_e}
\end{equation}
where the summation is over all possible string
configurations $C'$ (both closed and open) and $N(C')$ is the number of end of
strings in a particular configuration $C'$.

To see how topological order of the state changes as
$\epsilon$ varies from $0$, we can calculate the
topological entanglement entropy $\gamma$ of the state as defined in Eq. \ref{eq:topent1} in section \ref{sec:topo_ent}. As is shown below for
any finite value of $\epsilon$, $\gamma$ goes to zero when system size goes to
infinity.  Hence topological order is
unstable under this kind of variation. Let us first go through the process of the calculation and then discuss the implication of the result.

\begin{figure}[htp]
\centerline{
\includegraphics[width=2.0in]{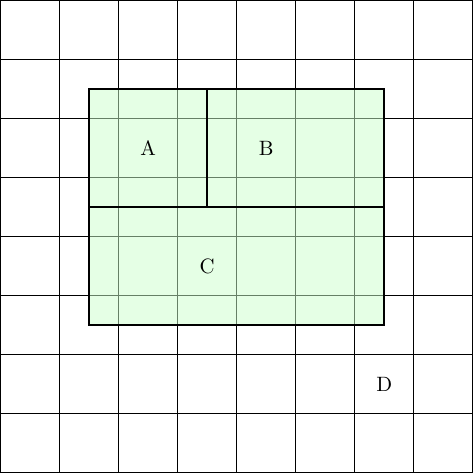}~~~~~~~~~~~~
\includegraphics[width=0.9in]{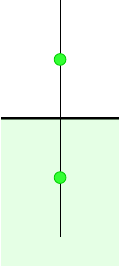} 
}
\centerline{
	(a) ~~~~~~~~~~~~~~~~~~~~~~~~~~~~~~~~~~~~~~~~~~~~~~~~~~~~~~~~~~
	(b) 
}
\caption{(a) Dividing the system into four regions for the calculation of topological entanglement entropy; (b) for simplicity of calculation, we double the number of spins per link and cut between them when dividing regions.} \label{TC_e_Stp}
\end{figure}

To calculate topological entanglement entropy, we first divide the lattice into four regions as shown in Fig.\ref{TC_e_Stp} (a) and use the formula
\be
\gamma = S_{AB}+S_{BC}+S_{CA}-S_A-S_B-S_C-S_{ABC}
\ee
Let us calculate the entanglement entropy to each region. For simplicity of calculation, when dividing the lattice, we double the number of spins per each link and cut between them, as shown in Fig.\ref{TC_e_Stp} (b). We require that the two spins per each link are either both in the $\ket{0}$ state or both in the $\ket{1}$ state, therefore, they represent a continuous string configuration on each link. Correspondingly, in the tensor network representation, we double the $t$ tensor per each link.

Without the closed loop constraint, a region with m boundary links has $2^m$ different boundary configurations. Rewriting the wave function according different boundary configuration $b_i$ as
\be
|\psi^{\epsilon}_{\text{toric code}}\rangle = \sum_{b} \beta_{b} \ket{\phi^{out}_{b}}\ket{\phi^{in}_{b}}
\label{psi_z2_e} 
\ee
Because different $b$'s are orthogonal to each other, we have obtained the Schmidt-decomposed form of the wave function and all we need to know to calculate entanglement entropy are the $\beta_b$'s and the norm.

To calculate the norm, form the double tensor $\mathbb{T}$ and $\mathbb{S}$ as
\be
\mathbb{T}_{ijkl,i'j'k'l'} = T_{ijkl} \times T^*_{i'j'k'l'}, \ \mathbb{S}_{ij,i'j'} = \sum_n t^{n}_{ij} (t^{n}_{i'j'})^*
\ee
Combine each $\mathbb{T}$ with the four $\mathbb{S}$ around it, we obtain the double tensor $\mathbb{T}'$
\be
\begin{array}{ll}
\mathbb{T}'_{ijkl,i'=i\ j'=j\ k'=k\ l'=l} = 1, & \text{\ if\ } i+j+k+l=0 \text{\ mod \ }2; \\ \nonumber
\mathbb{T}'_{ijkl,i'=i\ j'=j\ k'=k\ l'=l} = \epsilon^2, & \text{\ if\ } i+j+k+l=1 \text{\ mod \ }2
\end{array}
\ee
Contracting the $\mathbb{T}'$ tensors on each site gives us the norm of the wave function. It happens that such a contraction can be done easily with a change of basis for the inner indices. For each pair of inner indices $ii'$, $jj'$, $kk'$, $ll'$, apply transformation
\be
\ket{00}+\ket{11} \to \ket{\tilde{0}}, \  \ket{00}-\ket{11} \to \ket{\tilde{1}}
\label{trans_DT}
\ee
$\mathbb{T}'$ is transformed into
\be
\mathbb{T}'_{\tilde{0}\tilde{0}\tilde{0}\tilde{0}}=\frac{1+\epsilon^2}{2}, \mathbb{T}'_{\tilde{1}\tilde{1}\tilde{1}\tilde{1}}=\frac{1-\epsilon^2}{2},\ \text{all other terms are zero}
\ee
Obviously, this tensor network can be contracted easily and gives the norm of the wave function
\be
norm = \langle \psi^{\epsilon}_{\text{toric code}}|\psi^{\epsilon}_{\text{toric code}}\rangle = 2^N(1+\epsilon^2)^N +  2^N(1-\epsilon^2)^N 
\ee
where $N$ is the total systems size.

In a similar way, we can calculate $|\beta_b|^2$. To do so, we fix the boundary configuration and replace the double tensor $\mathbb{T}'$ on the boundary with
\be
\begin{array}{ll}
\mathbb{T}'_{ijkl,i'=i\ j'=j\ k'=k\ l'=l} = 1, & \text{\ if\ } i+j+k+l=0 \text{\ mod \ }2, \ i=0 \text{\ or\ }1; \\ \nonumber
\mathbb{T}'_{ijkl,i'=i\ j'=j\ k'=k\ l'=l} = \epsilon^2, & \text{\ if\ } i+j+k+l=1 \text{\ mod \ }2, \ i=0 \text{\ or\ }1
\end{array}
\ee
where $i$ corresponds to the link divided by the boundary. Then apply the same transformation as given in Eq.\ref{trans_DT} and contract the tensor network, we find
\be
|\beta_b|^2 = \frac{2^N}{2^{m}} \left((1+\epsilon^2)^N +  (1-\epsilon^2)^N + (1+\epsilon^2)^{N_i}(1-\epsilon^2)^{N_o}+ (1-\epsilon^2)^{N_i}(1+\epsilon^2)^{N_o}\right)
\ee
where $N_i$ is the number of vertices inside a region and $N_o$ is the number of vertices outside the region. Taking the limit of large system size $N_i \to \infty$, $N \to \infty$
\be
|\beta_b|^2 / norm = \frac{1}{2^{m}}
\ee
The entanglement entropy of a region is
\be
S =  m
\ee
which satisfy the area law. Therefore, topological entanglement entropy is $0$.

At first sight this
may be a surprising result, as we are only changing the
tensors locally and we are not expected to change the global
entanglement pattern of the state. However, when we write
out the wave function explicitly we will see that we have
actually induced global changes to the state.  The
wave function in Eq.\ref{psi_z2_e} can be expanded in powers
of $\epsilon$ as 
\begin{equation}
|\psi^{\epsilon}_{\text{toric code}}\rangle=
|\psi_{\text{toric code}}\rangle + \epsilon^2 \sum_{v_i,v_j}
|\psi^{v_i,v_j}_{\text{toric code}}\rangle + ... \label{Phi_vivj}
\end{equation}
where the $v$'s are any vertices in the lattice.
$|\psi^{v_i,v_j}_{\text{toric code}}\rangle$ is an excited
eigenstate of the toric code Hamiltonian (Eq.
\ref{H_Z2}) which minimizes energy of all local terms except
the vertex terms at $v_i$, $v_j$ and is hence an equal
weight superposition of all configurations with end of
strings at $v_i$ and $v_j$. Note that end of strings always
appear in pairs.  $v_i$, $v_j$ can be separated by any
distance and the number of local operations needed to take
$|\psi_{\text{toric code}}\rangle$ to
$|\psi^{v_i,v_j}_{\text{toric code}}\rangle$ scales with this
distance.

On the other hand, with arbitrary local perturbation to the
dynamics, the Hamiltonian reads \begin{equation} H' =
H_{TC} + \eta \sum_{u} h_u   \label{H'}
\end{equation}
where $h_u$'s are any local operator and $\eta$ is small. The
perturbed ground state wave function will contain terms like
$|\psi^{v_i,v_j}_{\text{toric code}}\rangle$ but only with weight
$~\eta^{distance(v_i,v_j)}$. When $v_i$, $v_j$ are separated by a
global distance, the weight will be exponentially small. Hence a
constant, finite weight $\epsilon^2$ for all
$|\psi^{v_i,v_j}_{\text{toric code}}\rangle$ as required in Eq.
\ref{Phi_vivj} is not possible. Therefore, while we are only
modifying the tensors locally, we introduce global `defects' to the
state, which cannot be the result of any local perturbation to the
Hamiltonian. We can, of course, design a Hamiltonian $H_{\epsilon}$ which has $|\Phi^{\epsilon}_{TC}\rangle$ as its exact ground state. However, $H_{\epsilon}$ will not be able to smoothly connect to $H_{TC}$ as $\epsilon \to 0$. 

Therefore, the $\mathbb{Z}_2$ inner symmetry is essential in maintaining the topological order represented by the tensor network. As soon as such a symmetry is broken, the topological order is lost. On the other hand, if only variations preserving the symmetry is allowed to be added to the tensor, the topological order is always stable. That is, the tensor network state has topological entanglement entropy $\gamma=1$ as long as the variation is small enough. This is because all such variation can be generated with local \textit{physical} operations and topological order is always stable against such \textit{physical} actions.

\begin{svgraybox}
\begin{center}
\textbf{Box 9.5 Stability of topological order in TPS}

In the tensor product representation of the toric code wave function, the $\mathbb{Z}_2$ symmetry is essential for the stability of topological order. Any variation in the tensor that breaks the $\mathbb{Z}_2$ symmetry can destroy the topological order immediately.

\end{center}
\end{svgraybox}


\section{Other forms of tensor network representation}
\label{sec:TN}

Beside matrix product states and tensor product states, other forms of tensor network representations have also been deviced and applied to study many-body systems with different forms of many-body entanglement. We discuss two examples in this section: the Multiscale Entanglement Renormalization Ansatz (MERA) and the Tree Tensor Network State.

\subsection{Multiscale entanglement renormalization ansatz}

The Multiscale Entanglement Renormalization Ansatz (MERA) provides a tensor network approach to study gapless systems in one spatial dimension. As discussed previously, matrix product states all satisfy an entanglement area law, hence incapable of describing gapless systems which contain a logrithmic violation of the area law. MERA utilizes a multi-layer structure to properly represent the entanglement in a gapless system and therefore has become a useful tool in the analytical and numerical study of such systems.

\begin{figure}[htbp] \centering
\includegraphics[width=3.5in]{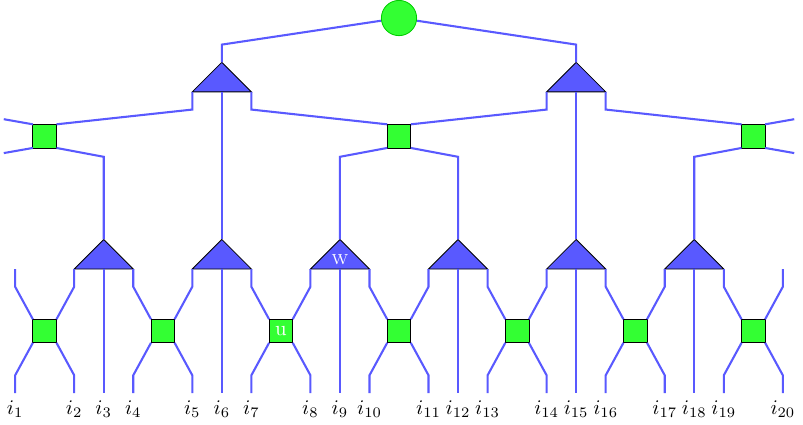}
\caption{The Multiscale Entanglement Renormalization Ansatz (MERA).} \label{fig:MERA}
\end{figure}

The tensors in a typical MERA are organized as shown in Fig.\ref{fig:MERA}. The physical indices of the tensor network $i_1,...,i_N$ are at the lowest layer and all the other indices are inner indices. The $u$ tensors are called disentanglers while the $w$ tensors are called the isometries. They are organized in a way to mimic the renormalization group (RG) transformation process of the gapless state. As we go from the $k$th layer to the $k+1$th layer, the number of lattice sites are reduced by a fixed fraction ($2/3$ as shown in Fig.\ref{fig:MERA}). The gapless state is an RG fixed point and the tensor network contains an infinite number of layers, which scales logrithmically with the system size. The amount of entanglement contained in such a tensor network may go beyond an area law and can adequately describe a gapless state.

\begin{figure}[htbp] \centering
\includegraphics[width=3.0in]{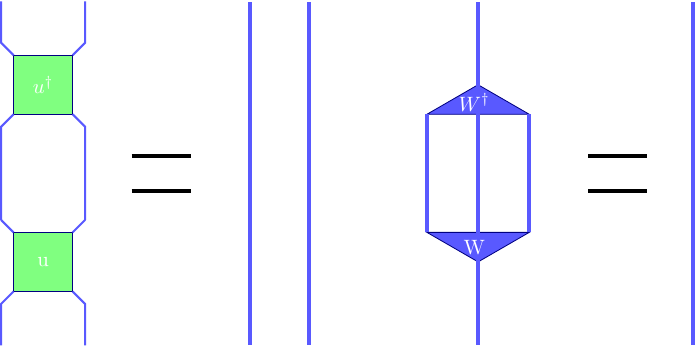}
\caption{Property of disentanglers and isometries in MERA.} \label{fig:uw_MERA}
\end{figure}

The disentangler $u$ is chosen to satisfy
\be
\sum_{\gamma\delta} u_{\alpha\beta,\gamma\lambda} \times \left(u_{\alpha'\beta',\gamma\lambda}\right)^* = \delta_{\alpha\alpha'}\delta_{\beta\beta'}
\ee 
and the isometry $w$ satisfies
\be
\sum_{\beta\gamma\lambda} w_{\alpha,\beta\gamma\lambda} \times \left(w_{\alpha',\beta\gamma\lambda}\right)^* = \delta_{\alpha\alpha'}
\ee
as illustrated in Fig.\ref{fig:uw_MERA}. 

\begin{figure}[htbp] \centering
\includegraphics[width=3.5in]{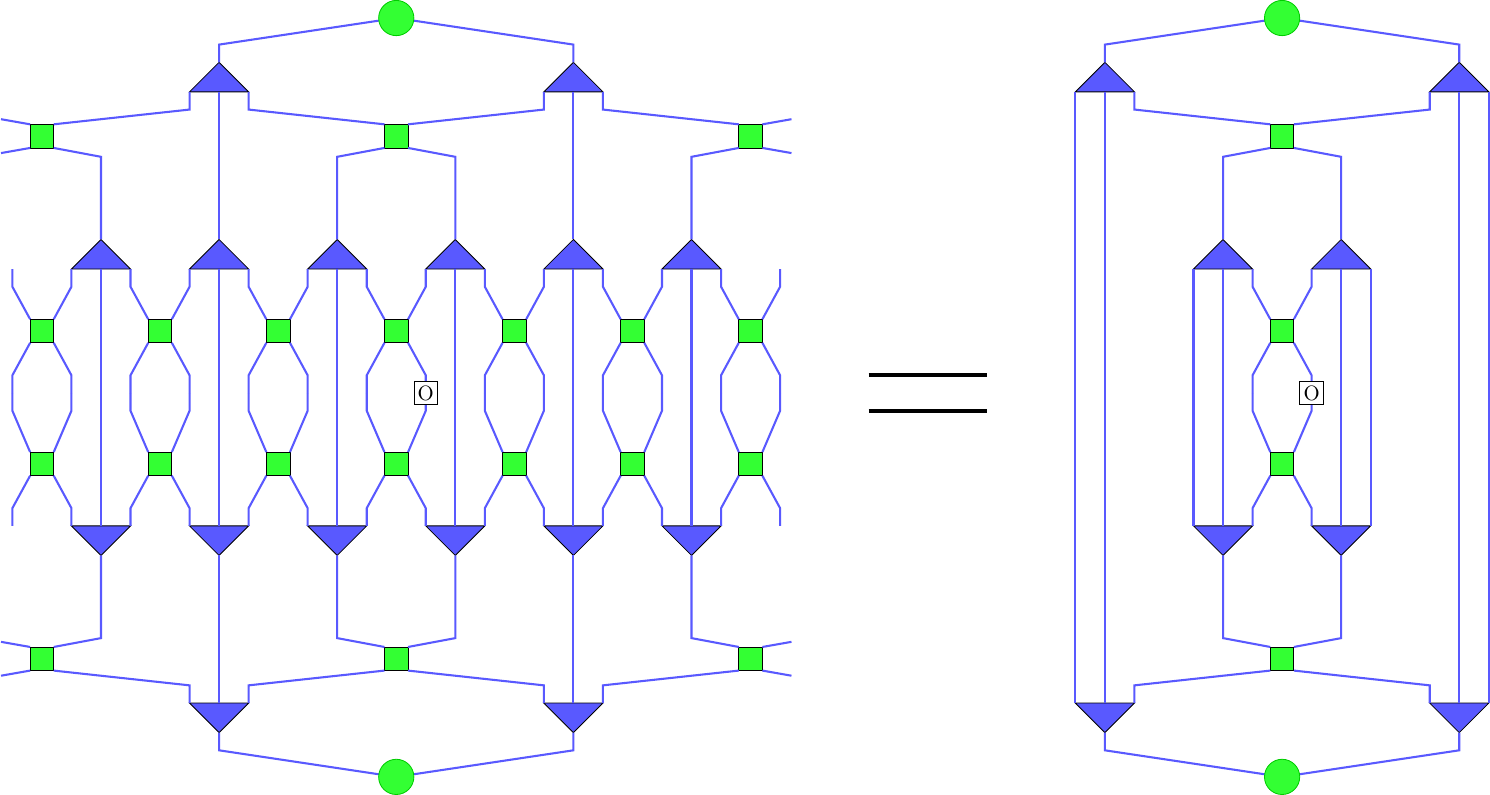}
\caption{Property of disentanglers and isometries in MERA.} \label{fig:O_MERA}
\end{figure}

Using this property, the tensor network for calculating the norm and local observables of a MERA can be reduced and efficiently contracted as illustrated in Fig.\ref{fig:O_MERA}.

\subsection{Tree tensor network state}

\begin{figure}[htbp] \centering
\includegraphics[width=2.0in]{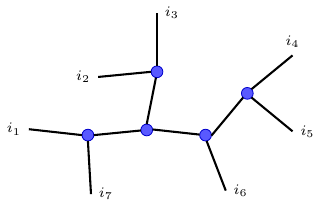}
\caption{The Tree Tensor Network.} \label{fig:TTN}
\end{figure}

Another important tensor network state has a tree structure as shown in Fig.\ref{fig:TTN} and is called the Tree Tensor Network. The tree tensor network is made up of local tensors with three indices and they are connected in such a way that there are no loops in the tensor network. 

The open indices $i_1,...,i_N$ on the outer edge of the graph are the physical indices of the state it represents. A simple counting shows that in order to represent an $N$ spin state, we need $N-2$ tensors. When a link in this tensor network is broken, the graph breaks into two parts. Therefore, the Schmidt rank of such a bipartition is bounded by the dimension $\chi$ of the indices. 

A nice property of the tree tensor network state is that several manipulations of the network can be implemented in an efficient way. For example, the calculation of reduced density matrix and the simulation of time evolution on the state. The required computation time for these tasks scales at most linearly in the number $N$ of spins and as a small polynomial in the dimension $\chi$ of the inner indices.


\section{Summary and further reading}

In this chapter, we introduce the tensor product state representation and study how it can be used to represent different phases in two dimension. First, we discuss the definition and basic properties of tensor product states. While tensor product states share many nice properties with matrix product states, including an entanglement area law, the correspondence between `injectivity' and finite correlation length breaks down, making it much less clear which tensor product states are gapped and which ones not. Numerically, it is also much harder to contract two dimensional tensors. An approximate algorithm is introduced to do the computation efficiently. The tensor product representation of the Ising model and the toric code model is discussed in detail as examples of symmetry breaking phases and topologically ordered phases. In particular, the block structure of the local tensor is found to be closely related to the symmetry breaking order while some internal symmetry of the local tensor is found to be essential for the existence of topological order.

The extension of the DMRG (matrix product) formalism to higher dimensions was used for the calculation of classical partition functions in e.g. \cite{NO9866,NHO0109}. On the quantum side, the representation of two dimensional AKLT state in a tensor product form was discussed in \cite{AKL8877} and variational parameters were introduced into the tensors in \cite{NKZ9703,SM98arxiv,HOA9907} to simulate more general spin systems. A renormalization algorithm based on tensor product state (also called the `projected entangled pair state') was proposed in \cite{VC04arXiv}, with various properties of the tensor product representation discussed in \cite{VWP0601}. The fact that injective tensor product states are unique but not necessarily gapped ground states is pointed out in \cite{PVC0850}. The approximate method for the contraction of 2D tensor networks was proposed in \cite{LN0701}.

A review of the matrix product states and tensor product states formalism, including various numerical algorithms based on them, is given in \cite{VMC0843}.

The tensor network representation of the toric code wave function was discussed first in \cite{Verstraete2006}. Later it was found that a much larger class of topological wave functions -- the string-net states -- can all be represented with tensor networks.\cite{GLS0918,BAV0919}. The gauge symmetry in tensors for topologically ordered states was emphasized in \cite{SW10arXiv,SCP1053,Buerschaper1447,SWB14arxiv}, where various topological properties was derived simply from the local tensors. 

The stability of the topological order under variation of the toric code tensor was studied in \cite{CZG1019}, where the necessary symmetry condition is demonstrated. Our discussion in section \ref{sta_TN} follows closely this paper and details of the computation can be found therein. 

A wave function renormalization algorithm for tensor product states was discussed in \cite{CGW1038}, where it is shown that the algorithm can flow a tensor to its fixed point form from which the symmetry breaking or topological order contained in the state can be identified.

The idea of Multiscale Entanglement Renormalization Ansatz was proposed in \cite{Vidal0705}. The Tree Tensor Network was first studied in \cite{SDV0620}.


%
%
\bibliographystyle{plain}
\bibliography{Chap9}

\chapter{Symmetry Protected Topological Phases}
\label{chap10} 

\abstract{Short range entangled states can all be connected to each other through local unitary transformations and hence belong to the same phase. However, if certain symmetry is required, they break into different phases. First of all, the symmetry can be spontaneously broken in the ground state leading to symmetry breaking phases. Even when the ground state remains symmetric, there can be different Symmetry Protected Topological (SPT) phases, whose nontrivial nature is reflected in their symmetry protected degenerate or gapless edge states. In this chapter, we discuss these phases in detail. Using the matrix product state formalism, we completely classify SPT phases in 1D boson / spin systems. By mapping 1D fermion systems to spins through Jordan Wigner transformation, we obtain a classification for fermionic SPT phases as well. In 2D, the tensor product representation falls short of firmly establishing a complete classification. But we present a exactly solvable construction of SPT phase with $\mathbb{Z}_2$ symmetry, which can be generalized to any internal symmetry and in any dimension.}


\section{Introduction}
Symmetry protected topological (SPT) phases are gapped quantum phases with topological properties protected by symmetry. The ground states of SPT phases contain only short-range entanglement and can
be smoothly deformed into a totally trivial product state if the symmetry requirement is not enforced in the system. However, with symmetry, the nontrivial SPT order is manifested in the existence of gapless edge states on the boundary of the system which cannot be removed as long as symmetry is not broken. What symmetry protected topological phases exist and what nontrivial properties do they have? This is the question that we are going to address in this chapter.

First we focus on one dimensional SPT phases in section\ref{sec:1DSPT}. We start by introducing some simple examples of nontrivial SPT orders in 1D. To have a more complete understanding of 1D bosonic SPT phases, the matrix product state representation provides us with a powerful tool. In fact, we can obtain a complete classification of boson / spin SPT phases systems by studying the form of symmetric fixed point states using the matrix product formalism. A one to one correspondence is found between bosonic SPT phases and the projective representations of group $G$. Note that there is no fundamental difference between spin and boson systems in our discussion, as they are both composed of local degrees of freedom which commute with each other.

By mapping 1D fermion system to 1D spin systems through Jordan Wigner transformation, we obtain a classification of 1D fermionic SPT phases as well. An important difference of fermion systems compared to bosonic ones is that fermion system has an intrinsic $\mathbb{Z}_2$ symmetry related to fermion parity conservation. Such a symmetry cannot be broken, not even spontaneously. The bosonic $\mathbb{Z}_2$ symmetry breaking phase when mapped back to fermion chains through inverse Jordan Wigner transformation results in a topological phase with nontrivial edge state, as we explain in section \ref{sec:fSPT}.

With a good understanding of SPT phases in 1D, we move on to construct SPT phases in 2D interacting boson / spin systems in section \ref{sec:2DSPT}. We generalize the short range entanglement structure in 1D SPT phases to 2D and design the symmetry action per each site such that the system always has gapless excitations on the edge unless symmetry is explicitly or spontaneously broken. Such a construction actually generalize to any dimension and any internal symmetry, as we discuss in section \ref{sec:nDSPT}, providing a systematic understanding of SPT phases in interacting boson and spin systems.


\section{Symmetry protected topological order in 1D bosonic systems}
\label{sec:1DSPT}

\subsection{Examples}

Let's start by introducing some simple models with nontrivial SPT order in 1D.

\begin{figure}[htbp]
\begin{center}
\includegraphics[width=3.5in]{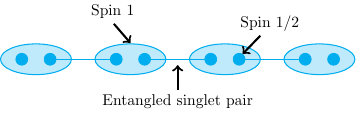}
\end{center}
\caption{Projected entangled pair structure of the AKLT wave function.}
\label{AKLT_PEPS}
\end{figure}

The AKLT model on a spin 1 chain discussed in Chap. \ref{chap8} is a prototypical example. The Hamiltonian of the AKLT model is
\be
H_{AKLT}=\sum_i \vec{S}_i\cdot \vec{S}_{i+1} + \frac{1}{3} \left(\vec{S}_i\cdot \vec{S}_{i+1}\right)^2
\ee
where $\vec{S}$ is the spin $1$ spin operator. This Hamiltonian is obviously invariant under the $SO(3)$ spin rotation symmetry generated by $S^x$, $S^y$ and $S^z$. The ground state wave function of this Hamiltonian can be explicitly constructed using a simple projected entangled pair picture. As shown in Fig. \ref{AKLT_PEPS}.

Each lattice site (big oval) contains two spin $1/2$s (small circle), which form singlet pairs (connected bonds) $|\uparrow\downarrow\rangle - |\downarrow\uparrow\rangle$ with another spin $1/2$ on a neighboring site. By projecting the two spin $1/2$s on each lattice site to a spin $1$, we obtain the ground state wave function $H_{AKLT}$. 

On a ring with periodic boundary condition, the ground state preserves spin rotation symmetry and is unique and gapped. On a chain with boundary, on the other hand, there are isolated spin $1/2$s at each end of the chain which are not coupled with anything and give rise to a two fold degenerate edge state. The full ground state on an open chain is hence four fold degenerate. The degenerate edge state is stable as long as spin rotation symmetry is preserved. In particular, spin $1/2$s transform under spin rotation in a very special way with a $2\pi$ rotation around any axis giving rise to a $-1$ phase factor. Because of this, the edge state cannot be smoothly connected to a trivial spin $0$, which gets a phase factor of $1$ under $2\pi$ rotation, without closing the bulk gap. With a gapped symmetric bulk and degenerate edge states protected by spin rotation symmetry, the AKLT model is hence in a nontrivial SPT phase.

\begin{svgraybox}
\begin{center}
\textbf{Box 10.1 SPT order of the AKLT model}

The 1D AKLT model has nontrivial symmetry protected topological order protected by $SO(3)$ spin rotation symmetry, as indicated by its degenerate spin $1/2$ edge state.

\end{center}
\end{svgraybox}

The 1D cluster state on a spin $1/2$ chain discussed in Chap. \ref{cp:5} provides another example of nontrivial SPT order. The Hamiltonian of the 1D cluster state is
\begin{equation}
  H_{{clu}}=-\sum_j Z_{j-1}X_jZ_{j+1}.
\end{equation}
Here, $Z$ and $X$ are Pauli operators for the spin $1/2$s. As explained in Chap. \ref{cp:5}, for a 1D ring without boundary, the ground state of $H_{clu}$ is the unique graph state stabilized by $\{Z_{j-1}X_jZ_{j+1}\}$. For a chain with boundary, where the summation index $j$ runs from $2$ to $N-1$, the ground state is then $4$-fold degenerate. 

This $4$-fold degeneracy is a result of two edge states, each being $2$-fold degenerate protected by a $\mathbb{Z}_2\times \mathbb{Z}_2$ symmetry. The $\mathbb{Z}_2\times \mathbb{Z}_2$ symmetry is generated by
\begin{equation}
  \bar{X}_1=\prod_k X_{2k-1},\quad \bar{X}_2=\prod_k X_{2k},
\end{equation}
Any local perturbation to the system cannot lift the degeneracy as long as this $\mathbb{Z}_2\times \mathbb{Z}_2$ symmetry is preserved. To see this, first we notice that the effective Pauli $\tilde{X}$ and $\tilde{Z}$ operators on the $2$-fold degenerate edge states (at the left end of the chain for example) can be chosen as $\tilde{X}=X_1Z_2$ and $\tilde{Z}=Z_1$, which commute with all the bulk Hamiltonian terms and anti-commute with each other. Next, we find that the effective action of $\bar{X}_1$ and $\bar{X}_2$ on the edge state is the same as $\tilde{X}$ and $\tilde{Z}$ because
\be
\bar{X}_1 \prod_{k=2}^{\infty} Z_{2k-2}X_{2k-1}Z_{2k} = X_1Z_2 = \tilde{X} , \ \ \bar{X}_2 \prod_{k=1}^{\infty} Z_{2k-1}X_{2k}Z_{2k+1} = Z_1 = \tilde{Z}
\ee
From this we can see that the $\mathbb{Z}_2\times \mathbb{Z}_2$ symmetry acts on the edge state in a very special way: the two $\mathbb{Z}_2$'s anti-commute with each other! Because of this, the edge state must be at least two fold degenerate and the degeneracy cannot be removed without breaking the symmetry or going through a bulk phase transition. This demonstrates the nontrivial-ness of the SPT order in the cluster state.

\begin{svgraybox}
\begin{center}
\textbf{Box 10.2 SPT order of the cluster state model}

The 1D cluster state model has nontrivial symmetry protected topological order protected by a $\mathbb{Z}_2\times \mathbb{Z}_2$ symmetry, as indicated by its two fold degenerate edge state.

\end{center}
\end{svgraybox}

Through the AKLT model and the cluster state model, we see some common features of SPT order in 1D: the bulk wave function is gapped and symmetric while the edge state must be degenerate because it transforms in a nontrivial way under the symmetry. This picture can be generalized to all kinds of symmetries and we want to understand what 1D SPT phases exist in general with any given symmetry. The matrix product formalism again provides a powerful tool in addressing this question. In the following, we are going to follow a procedure similar to Chap. \ref{chap8} and completely classify SPT phases in 1D interacting boson / spin systems using a renormalization group transformation on matrix product states.

When the class of systems under consideration has certain
symmetry, the equivalence classes of states are defined in
terms of LU transformations that do not break the symmetry.
Therefore, when applying the renormalization procedure, we
should carefully keep track of the symmetry and make sure
that the resulting state has the same symmetry at each step.
Due to such a constrain on local unitary equivalence, we
will see that gapped ground states which do not break the
symmetry of the system divide into different universality
classes corresponding to different symmetry protected
topological orders. We will first study in detail the case of
on-site unitary symmetries.  Then we will also discuss systems with time reversal (anti-unitary) symmetry.
Finally, we shall study translational invariant (TI) systems, with the
possibility of having on-site symmetry or parity symmetry.

\subsection{On-site unitary symmetry}
\label{usymm}

A large class of systems are invariant under on-site
symmetry transformations. For example, the Ising model is symmetric under
the $\mathbb{Z}_2$ spin flip transformation and the Heisenberg model is
symmetric under $SO(3)$ spin rotation transformations. In this section, we will
consider the general case where the system is symmetric
under $u(g) \otimes...\otimes u(g)$ with
$u(g)$ being a unitary representation of a symmetry
group $G$ on each site and satisfy
\begin{equation}
u(g_1)u(g_2)=u(g_1g_2)
\end{equation}

We will focus on the case where the on-site symmetry is the only symmetry required for the class of system. In particular, we do not require translational symmetry for the systems. We will classify possible phases for
different $G$ when the ground state is invariant (up to a
total phase) under such
on-site symmetry operations and is gapped (i.e. short-range
correlated). Specifically, the ground state $|\phi\rangle$ satisfies
\begin{align}
\label{u0act}
u(g) \otimes...\otimes
u(g) |\psi\rangle= \alpha^L(g)
|\psi\rangle
\end{align}
where $|\alpha(g)|=1$ is a one-dimensional representation of $G$ and $L$ is the system size.

Now we will try to classify these symmetric ground states using symmetric LU transformations and we find that:

\begin{svgraybox}
\begin{center}
\textbf{Box 10.3 Classification of bosonic symmetry protected topological phases}

For 1D bosonic systems with ONLY an
on-site symmetry of group $G$,
the gapped phases that do not
break the symmetry are labeled by the projective representations of the group $G$.
\end{center}
\end{svgraybox}
This result applies when the 1D representations $\alpha(G)$ form a finite group, when $G=U(1)$, further classification according to different $\alpha(U(1))$ exist.

\subsubsection{Symmetric RG transformation and fixed point}
\label{sec:symRG}

We will again use the fact that all gapped states can be represented
as short range correlated matrix product states and use the renormalization
flow discussed in section \ref{sec:RG_MPS} to simplify the matrix
product states. We find that
\begin{enumerate}
\item{With symmetric LU transformations, all gapped bosonic states with on-site symmetry can be mapped to the fixed point form shown in Fig.\ref{wf_fp}.}
\item{In the fixed point wave function, each of the two degrees of freedom on a site carries a projective representation of the symmetry.}
\end{enumerate}

In order to compare different equivalent classes under
\emph{symmetric} LU transformations, it is
important to keep track of the symmetry while doing
renormalization.

First, in the renormalization procedure we group two sites
together into a new site. The on-site symmetry
transformation becomes $u(g) \otimes u(g)$,
which is again a linear representation of
$G$. The next step in RG procesure applies a unitary
transformation $w_1$ to the support space of new site. This is actually itself composed of two steps. First we
project onto the support space of the new site, which is the
combination of two sites in the original chain. This is an
allowed operation compatible with symmetry $G$ as the
reduced density matrix $\rho_{2}$ is invariant under
$u(g) \otimes u(g)$, so the support space form a linear
representation for $G$. The projection of
$u(g) \otimes u(g)$ onto the support space
$P_2(u(g) \otimes u(g))P_2$ hence remains a linear
representation of $G$. In the next step, we do some unitary
transformation $w_1$ within this support space which
relabels different states in the space. The symmetry
property of the state should not change under this
relabeling. In order to keep track of the symmetry of the
state, the symmetry operation needs to be redefined as
\be
u^{(1)}(g)=w_1P_2(u(g) \otimes u(g))P_2(w_1)^{\dagger}
\ee
After this redefinition, the symmetry operations
$u^{(1)}(g)$ on each new site form a new
linear representation of $G$. By redefining $u^{(N)}(g)$ at
each step of the RG transformation, we keep track of the
symmetry of the system. Finally at the fixed point (i.e. at a large RG step $N$), we
obtain a state described by
$\left(A^{(\infty)}\right)_{i^l,i^r}$ which is
again given by the fixed point form \eqn{Avv}. The symmetry transformation on each site is given by $u^{(\infty)}(g)$.

One may want to proceed to disentangle each pair in the fixed point state and map the state to a total product state. However, it is not always possible to do so without breaking symmetry. Consider the case where the entangled pairs in the fixed point state are spin $1/2$ singlets. The total state is invariant under $SO(3)$ symmetry, but there does not seem to be a way to disentangle the singlet without breaking symmetry. Actually, all product states of two spin $1/2$'s necessarily break spin rotation symmetry! As we show in the following, this is a very general observation and is related inherently to the nontrivial SPT order in the state. 

But first, let's look more closely at the fixed point matrices we obtained. In fact, the form of the fixed point is already simple enough that we can extract useful information about the universal properties of the phase from it. The fact that the fixed point state is invariant under $u^{(\infty)}(g)$ requires special transformation property of $\left(A^{(\infty)}\right)_{i^l,i^r}$  under the symmetry. We are going to derive this transformation property in the following and see that how projective representations of the symmetry group emerge in the MPS representation.

\begin{figure}[htbp]
\begin{center}
\includegraphics[width=3.0in]{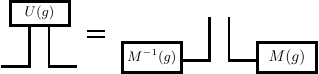}
\end{center}
\caption{Symmetry transformation of fixed point matrices under on-site symmetry $G$.}
\label{A_sym}
\end{figure}

Because $\left(A^{(\infty)}\right)_{i^l,i^r}$ is injective, the transformed matrices must be equivalent to the original ones by a gauge transformation on the inner indices. That is (we are omitting the fixed point label $\infty$ in the following)
\be
 \sum_{j^lj^r} u_{i^li^r,j^lj^r}(g) A_{j^lj^r} = \al(g)M^{-1}(g)A_{i^li^r} M(g)
 \label{Eqn:LU_A}
\ee
with invertible matrices $M(g)$ and $\al(g)$ is a 1D representation of $G$.
Since $\frac{u(g)}{\alpha(g)}$
is also a linear unitary representation of
$G$, we can
absorb $\alpha(g)$ into $u(g)$ and rewrite
\eqn{Eqn:LU_A} as
\begin{equation}
\label{Eqn:LU_A_C}
\sum_{j^lj^r} u_{i^li^r,j^lj^r}(g) A_{j^lj^r}=
M^{-1}(g)A_{i^li^r}M(g)
\end{equation}
We note that matrix elements $A_{i^li^r,\al\bt}$ is
non-zero only when $\alpha=i^l$, $\beta=i^r$ and the
full set of $\{A_{i^li^r}\}$ form a complete basis in the
space of $D\times D$ dimensional matrices. Such a symmetry transformation property of the fixed point matrices can be represented graphically as in Fig.\ref{A_sym}.

$M(g)$ does not necessarily form a linear representation of $G$.
But the fixed point form of the matrices requires that $M(g)$ be a so-called `projective' representation, as on the one hand
\begin{align}
 & \sum_{j^lj^r} u_{i^li^r,j^lj^r}(g_1g_2) A_{j^lj^r} \\ \nonumber
= & \sum_{j^lj^rk^lk^r}
u_{i^li^r,k^lk^r}(g_1)u_{k^lk^r,j^lj^r}(g_2) A_{j^lj^r} \\ \nonumber
=& \sum_{k^lk^r}
u_{i^li^r,k^lk^r}(g_1) M^{-1}(g_2)A_{k^lk^r}M(g_2) \\ \nonumber
=&
M^{-1}(g_2)M^{-1}(g_1)A_{i^li^r}M(g_1)M(g_2) \\ \nonumber
\end{align}
and on the other hand
\begin{align}
\sum_{j^lj^r} u_{i^li^r,j^lj^r}(g_1g_2) A_{j^lj^r}
=M^{-1}(g_1g_2)A_{i^li^r}M(g_1g_2)
\end{align}
Therefore
\begin{align}
M^{-1}(g_2)M^{-1}(g_1)A_{i^li^r}M(g_1)M(g_2)
=M^{-1}(g_1g_2)A_{i^li^r}M(g_1g_2)
\end{align}
for all $i^li^r$. However, the set of matrices
$\{A_{i^li^r}\}$ form a complete basis in the space of
$D\times D$ dimensional matrices. Therefore, $M(g_1)M(g_2)$ can differ from $M(g_1g_2)$ by at most a phase factor
\begin{align}
\label{MNproj}
 M(gh)&=\om(g,h) M(g) M(h),
\end{align}
with $|\om(g_1,g_2)|=1$. Therefore, $M(g)$ are a set of matrices labeled by group elements and satisfy the group multiplication rule up to a phase factor. That is, $M(g)$ form a projective representation of the symmetry group $G$.

The transformation law of the fixed point matrices is related to the transformation law of the degrees of freedom in the fixed point wave function.

\begin{figure}[htbp]
\begin{center}
\includegraphics[width=3.5in]{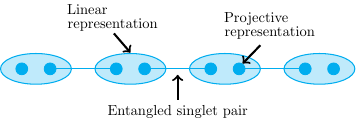}
\end{center}
\caption{Fixed point wave function with on-site symmetry.}
\label{fp_sym}
\end{figure}

Let us rewrite \eqn{Eqn:LU_A_C} as
\begin{equation}
\label{Eqn:LU_A_C1}
M(g)(\sum_{j^lj^r} u_{i^li^r,j^lj^r}(g) A_{j^lj^r}) M^{-1}(g)=
A_{i^li^r}
\end{equation}
We note that
\begin{equation}
\label{Eqn:LU_A_C2}
M(g)(\sum_{j^lj^r}
(Nl^{-1})_{j^l,i^l}
Nr_{i^r,j^r}
A_{j^lj^r}) M^{-1}(g)=
A_{i^li^r}
\end{equation}
where the matrices $Nl$ and $Nr$ are given by
\begin{align}
Nl_{\al\bt} =   M_{\al\bt}
\frac
{\sqrt{\la_\al}}
{\sqrt{\la_\bt}}
,\ \ \ \
Nr_{\al\bt} =   M_{\al\bt}
\frac
{\sqrt{\eta_\bt}}
{\sqrt{\eta_\al}}
 .
\end{align}
Since the set of matrices $\{A_{i^li^r}\}$ form a complete
basis in the space of $D\times D$ dimensional matrices,
we find
\begin{align}
\label{uMN}
u_{i^li^r,j^lj^r}(g)=
Nl^{-1}_{j^l,i^l}(g)
Nr_{i^r,j^r}(g) .
\end{align}
That is, the symmetry transformation acts on the two degrees of freedom on each site separately and in a projective way similar to $M(g)$.
\be
\begin{array}{lll}
Nl^{-1}(g_1)Nl^{-1}(g_2)&  = & \om^{-1}_{g_1,g_2} Nl^{-1}(g_1g_2) \\ \nonumber
Nr(g_1)Nr(g_2) & = & \om_{g_1,g_2} Nr(g_1g_2)
\end{array}
\ee
Therefore, in the fixed point wave function as shown in Fig.\ref{fp_sym}, the two degrees of freedom on a site each carry a projective representation of the symmetry and form a singlet state with another degree of freedom on a neighboring site. Note that the symmetry representation on each full site is still linear and the total wave function is invariant under the symmetry.

\subsubsection{Example}

Let's look at some examples of gapped states with on-site symmetry and find their corresponding fixed point structure.

The symmetric phase of the Ising model provides a trivial example. At the exactly solvable limit, the Hamiltonian is $H=-\sum_i X_i$ and the ground state wave function is a total product state of spins pointing in the $+x$ direction
\be
\ket{\psi}=\ket{+} \otimes \ket{+} \otimes ... \otimes \ket{+}
\ee
which is invariant under the $\mathbb{Z}_2$ symmetry of $X \otimes X \otimes...\otimes X$.
This wave function is already in the fixed point form where each site contains two spins in state $\ket{+}$ and each spin forms a linear representation of the $\mathbb{Z}_2$ symmetry -- a trivial projective representation with $\om(g,h)=1$. Also there is no entanglement between neighboring sites.

Let's see how the nontrivial SPT order is manifested in the AKLT model
\be
H_{AKLT}=\sum_i \vec{S}_i\cdot \vec{S}_{i+1} + \frac{1}{3} \left(\vec{S}_i\cdot \vec{S}_{i+1}\right)^2
\ee
which is invariant under the $SO(3)$ spin rotation symmetry generated by $S^x$, $S^y$ and $S^z$. Its unique gapped ground state is given by the matrix product representation
\be
A_x = X, \ A_y = Y, A_z = Z
\ee
where the basis states $\ket{x}$, $\ket{y}$ and $\ket{z}$ written in the $S_z$ basis are
\be
\ket{x}=\frac{1}{\sqrt{2}}\left(\ket{1}-\ket{-1}\right), \ \ket{y}=\frac{-i}{\sqrt{2}}\left(\ket{1}+\ket{-1}\right), \ \ket{z} = -\ket{0}
\ee
The AKLT state is $SO(3)$ symmetric and, as we will see, in a nontrivial way with nontrivial projective representations in its fixed point. To see this, first we construct its double tensor
\be
\mathbb{E}=\frac{1}{3} [X \otimes X^* + Y \otimes Y^* + Z \otimes Z^*] = \frac{1}{3}\begin{pmatrix} 1 & 0 & 0 & 2 \\ 0 & -1 & 0 & 0 \\ 0 & 0 & -1 & 0 \\ 2 & 0 & 0 & 1 \end{pmatrix}
\ee
Normalization factor is added to ensure that the largest eigenvalue is $1$.
The fixed point double tensor is then
\be
\mathbb{E}^{(\infty)} = \lim_{N \to \infty} \mathbb{E}^N = \frac{1}{2} \begin{pmatrix} 1 & 0 & 0 & 1 \\ 0 &0 & 0 & 0 \\ 0 & 0 & 0 & 0 \\ 1 & 0 & 0 & 1 \end{pmatrix} 
\ee
which can be decomposed as
\be
A^{(\infty)}_s = I_2, \  A^{(\infty)}_x = X, \ A^{(\infty)}_y = Y, \ A^{(\infty)}_z = Z
\ee
Here among the four fixed point basis states, $\ket{s}$ is a spin $0$ state and $\ket{x}$, $\ket{y}$, $\ket{z}$ form a spin $1$. $I_2$ is the $2\times 2$ identity matrix. We can then compute the symmetry transformation matrices $M(g)$ on the inner indices by rotating the physical spins. For example, under the rotation around $z$ axis for an angle $\theta$, the matrices change into
\be
\t A^{(\infty)}_s = I_2, \  \t A^{(\infty)}_x = \cos{\theta}X -\sin{\theta}Y, \ \t A^{(\infty)}_y = \cos{\theta}Y + \sin{\theta}X, \ \t A^{(\infty)}_z = Z
\ee
from which we can see that $M(R_z(\theta)) = e^{i\theta Z/2}$. Similar calculation for all the other rotations show that
\be
M(R_{\vec{n}} (\theta)) = e^{i\frac{\theta}{2}(n_xX+n_yY+n_zZ)}
\ee 
That is, the symmetry transformation on the inner indices is generated by an effective spin $1/2$. Correspondingly, in the fixed point wave function, the two degrees of freedom on each site are spin $1/2$'s and they form spin singlets between neighboring sites. This is similar to the structure shown in Fig.\ref{AKLT_PEPS}, except that at fixed point we do not need to project the two spin $1/2$ per each site to a spin $1$ any more.

The most important property of $M(g)$ is that, it forms a projective rather than linear representation of the $SO(3)$ group which can be seen from $2\pi$ rotations
\be
M(R_{\vec{n}}(2\pi)) = e^{i\frac{2\pi}{2}(n_xX+n_yY+n_zZ)} = -I
\ee
Rotation by $2\pi$ is equivalent to the identity operation while the matrix representation $M(R_{\vec{n}}(2\pi))$ is only equivalent to the identity matrix up to a minus sign. Therefore, nontrivial sign factors $\om(g,h)$ occur in composing $M(g)$ and $M(h)$. For example
\be
M(R_{\vec{n}}(\pi))M(R_{\vec{n}}(\pi)) = - M(R_{\vec{n}}(0))
\ee
And $M(R_{\vec{n}}(\theta))$ generated by spin $1/2$ spin operators form a projective representation of the $SO(3)$ rotation symmetry.

In the ground state, such a projective representation is most clearly seen when we cut the system open and put it on an open chain. On an open chain, there are isolated spin $1/2$'s at either end of the chain which do not form singlets with other spin $1/2$'s. They give rise to a total of four fold ground state degeneracy on an open chain as long as spin rotation symmetry is preserved. That is, the projective representation leads to degenerate edge states on an open chain protected by the symmetry. Of course, if the symmetry is broken, by for exmaple adding a magnetic field, the degeneracy will be removed.

This is a generic feature of one dimensional bosonic systems with on-site symmetry protected topological orders, as we discuss for an arbitrary group $G$ in the next section.


\subsubsection{Projective representation and edge state}
\label{prorep}

Let's first define projective representation for a general group $G$ more carefully.
Operators $u(g)$ form a projective representation of symmetry group $G$ if
\begin{align}
 u(g_1)u(g_2)=\om(g_1,g_2)u(g_1g_2),\ \ \ \ \
g_1,g_2\in G.
\end{align}
Here $\om(g_1,g_2) \in U(1)$, the factor system of the projective representation, satisfies
\begin{align}
 \om(g_2,g_3)\om(g_1,g_2g_3)&=
 \om(g_1,g_2)\om(g_1g_2,g_3),
\end{align}
for all $g_1,g_2,g_3\in G$, which comes from the associativity condition of the representation $[u(g_1)u(g_2)]u(g_3)=u(g_1)[u(g_2)u(g_3)]$
If $\om(g_1,g_2)=1$, this reduces to the usual linear representation of $G$.

On the other hand, not all projective representations with $\om(g_1,g_2)\neq 1$ are nontrivial. Notice that a different choice of pre-factor for the representation matrices
$u'(g)= \bt(g) u(g)$ will lead to a different factor system
$\om'(g_1,g_2)$:
\begin{align}
\label{omom}
 \om'(g_1,g_2) =
\frac{\bt(g_1g_2)}{\bt(g_1)\bt(g_2)}
 \om(g_1,g_2).
\end{align}
Therefore, if a factor system satisfies $\om(g_1,g_2)=\frac{\bt(g_1)\bt(g_2)}{\bt(g_1g_2)}$, then by redefining the pre-factor of the matrices, we can reduce the factor system to $1$ and hence the projective representation to a linear one. Only $\om(g_1,g_2)$'s which cannot be reduced to $1$ in this way are nontrivial. Moreover, if two factor systems $\om'(g_1,g_2)$ and $\om(g_1,g_2)$ can be related as in \eqn{omom}, then their corresponding representation matirces $u'(g)$ and $u(g)$ differ only by a pre-factor and belong to the same class of projective representation.

Suppose that we have one projective representation $u_1(g)$
with factor system $\om_1(g_1,g_2)$ of class $\om_1$ and
another $u_2(g)$ with factor system $\om_2(g_1,g_2)$ of
class $\om_2$, obviously $u_1(g)\otimes u_2(g)$ is a
projective presentation with factor system\\
$\om_1(g_1,g_2)\om_2(g_1,g_2)$. The corresponding class
$\om$ can be written as a sum $\om_1+\om_2$. Under such an
addition rule, the equivalence classes of factor systems
form an Abelian group, which is called the second cohomology
group of $G$ and denoted as $H^2(G,U(1))$. The identity
element $\om_0$ of the group is the class that contains the linear representation of the group.

Here are some simple examples:
\begin{enumerate}
\item{cyclic groups $\mathbb{Z}_n$ do not have non-trivial projective representation. Hence for $G=\mathbb{Z}_n$, $H^2(G,U(1))$ contains only the identity element.}
\item{a simple group with non-trivial projective representation is the Abelian dihedral group $D_2=\mathbb{Z}_2\times \mathbb{Z}_2$. For the four elements of the group $(0/1,0/1)$, consider representation with Pauli matrices $g(0,0)=\begin{bmatrix} 1 & 0 \\ 0 & 1 \end{bmatrix}$, $g(0,1) = \begin{bmatrix} 0 & 1 \\ 1 & 0 \end{bmatrix}$, $g(1,0) = \begin{bmatrix} 1 & 0 \\ 0 & -1 \end{bmatrix}$, $g(1,1) = \begin{bmatrix} 0 & -i \\ i & 0 \end{bmatrix}$. It can be check that this gives a non-trivial projective representation of $D_2$.}
\item{when $G=SO(3)$, $H^2(G,U(1))=\mathbb{Z}_2$. The two elements
correspond to integer and half-integer representations of
$SO(3)$ respectively.}
\item{when $G=U(1)$, $H^2(G,U(1))$ is trivial:
$H^2(U(1),U(1))=Z_1$.  We note that $\{ \e^{\imth m \th} \}$
form a representation of $U(1)=\{ \e^{\imth \th} \}$ when
$m$ is an integer.  But $\{ \e^{\imth m \th} \}$ will form
a projective representation of $U(1)$ when $m$ is not an
integer.  But under the equivalence relation \Eqn{omom}, $\{
\e^{\imth m \th} \}$ correspond to the trivial projective
representation, if we choose $\bt(g)=\e^{-\imth m \th}$. Note
that $\bt(g)$ can be a discontinuous function over the group
manifold.}
\end{enumerate}

An important property of nontrivial projective representations is that the representation space must be at least two dimensional. That is, there are no one-dimensional nontrivial projective representations. This has direct physical consequence for nontrivial SPT states. When the nontrivial SPT states are put on an open chain, there are isolated spins carrying projective representations at each end of the chain. These spins are always of dimension larger than $1$, hence giving rise to a ground state degeneracy at each end of the chain. This degeneracy is stable as long as the symmetry of the system is not broken. This property holds not only at the fixed point, but at any point in the symmetry protected topological phase. Therefore, we see that the most distinctive property of 1D SPT phases with on-site symmetry is the existence of nontrivial edge state around a gapped and nondegenerate bulk, which is protected by the symmetry of the system.


\subsubsection{Equivalence between symmetric fixed point states}
\label{equiv}

In section \ref{sec:symRG} we have shown that all gapped bosonic states with on-site symmetry $G$ can be mapped to a fixed point form with symmetric LU transformations. If we can further determine the equivalence relation between different symmetric fixed points under symmetric LU transformation, we would be able to obtain a complete classification of SPT phases. This is what we are going to do in this section.

From the discussion in section \ref{sec:symRG}, we know that the fixed point state symmetric under on-site symmetry of group $G$ takes the form
\begin{equation}
|\psi\rangle^{(\infty)} = |EP_{1,2}\rangle |EP_{2,3}\rangle ... |EP_{k,k+1}\rangle ...
\end{equation}
where $|EP_{k,k+1}\rangle$ is an entangled pair between the right spin on site $k$ and the left spin on site $k+1$(see Fig. \ref{wf_fp}). Each entangled pair is invariant under a linear symmetry transformation of the form $u^{[k],r}(g)\otimes u^{[k+1],l}(g)$
\begin{equation}
u^{[k],r}(g)\otimes u^{[k+1],l}(g) |EP_{k,k+1}\rangle = |EP_{k,k+1}\rangle
\end{equation}
But $u^{[k],r}(g)$ or $u^{[k+1],l}(g)$ alone might not form
a linear representation of $G$. They could in general be a
projective representation of $G$. If $u^{[k],r}(g)$ is a
projective representation corresponding to class $\om$ in
$H^2(G,U(1))$, then $u^{[k+1],l}$ must correspond to class
$-\om$. $\om$ does not vary from site to site and labels a particular symmetric fixed point state.

Now we want to show that symmetric fixed point states with the
same $\om$ can be connected through symmetric LU
transformations and hence belong to the same phase while
those with different $\om$ cannot and belong to different phases.

First, suppose that two symmetric fixed point states
$|\phi_1\rangle$ and $|\phi_2\rangle$ are related with the
same $\om$, i.e.
\begin{align}
u^{[k],r}_1(g)\otimes u^{[k+1],l}_1(g) |EP_{k,k+1}\rangle_1 = |EP_{k,k+1}\rangle_1 \\ \nonumber
u^{[k],r}_2(g)\otimes u^{[k+1],l}_2(g) |EP_{k,k+1}\rangle_2 = |EP_{k,k+1}\rangle_2
\end{align}
where $|EP_{k,k+1}\rangle_{1(2)}$ is an entangled pair of spins on Hilbert space $\mathcal{H}^{[k],r}_{1(2)}
\otimes \mathcal{H}^{[k+1],l}_{1(2)}$. $u^{[k],r}_{1(2)}(g)$
is a projective representation of $G$ corresponding to $\om$
on $\mathcal{H}^{[k],r}_{1(2)}$ and $u^{[k+1],l}_{1(2)}(g)$
a projective representation corresponding to $-\om$ on $\mathcal{H}^{[k+1],l}_{1(2)}$. As $u^{[k],r}_{1}(g)$ and $u^{[k],r}_{2}(g)$ ($u^{[k+1],l}_{1}(g)$ and $u^{[k+1],l}_{2}(g)$) belong to the same $\om$, we can choose their pre-factor such that they have the same factor system. In the following discussion, we will assume WLOG that this is true.

We can think of $|EP_{k,k+1}\rangle_1$ and $|EP_{k,k+1}\rangle_2$ as living together in a joint Hilbert space $(\mathcal{H}^{[k],r}_{1} \oplus \mathcal{H}^{[k],r}_{2})\otimes(\mathcal{H}^{[k+1],l}_{1} \oplus \mathcal{H}^{[k+1],l}_{2})$.  The symmetry representation on this joint Hilbert space can be defined as
\begin{align}
 u^{[k],r}(g) \otimes u^{[k+1],l}(g) 
= (u^{[k],r}_1(g) \oplus u^{[k],r}_2(g))\otimes(u^{[k+1],l}_1(g) \oplus u^{[k+1],l}_2(g))
\end{align}
As $u^{[k],r}_1(g)$ and $u^{[k],r}_2(g)$ (also
$u^{[k+1],l}_1(g)$ and $u^{[k+1],l}_2(g)$) both correspond
to $\om$ ($-\om$), their direct sum
$u^{[k],r}(g)$($u^{[k+1],l}(g)$) is also a projective
representation corresponding to $\om$($-\om$). Therefore, we have a linear representation of $G$ on each site $k$, $u^{[k],l}(g) \otimes u^{[k],r}(g)$ and both $|EP_{k,k+1}\rangle_1$ and $|EP_{k,k+1}\rangle_2$ are symmetric under $u^{[k],r}(g) \otimes u^{[k+1],l}(g)$.

Now we can perform a LU transformation on the joint Hilbert space and rotate continuously between $|EP_{k,k+1}\rangle_1$ and $|EP_{k,k+1}\rangle_2$. That is,
\begin{equation}
U(\theta)= \cos(\frac{\theta}{2})I-i\sin(\frac{\theta}{2})(|a\rangle\langle b| +|b\rangle\langle a| )
\end{equation}
where $|a\rangle = |EP_{k,k+1}\rangle_1$, $|b\rangle =
|EP_{k,k+1}\rangle_2$ and $\theta$ goes from $0$ to $\pi$.
By doing this locally to each pair, we can map
$|\phi_1\rangle$ to $|\phi_2\rangle$ (and vice verse) with
LU transformations without breaking the on-site symmetry of
group $G$. Therefore, $|\phi_1\rangle$ and $|\phi_2\rangle$
belong to the same phase if they are related with the same
$\om$.

On the other hand, if $|\phi_1\rangle$ and $|\phi_2\rangle$
are related to different $\om_1$ and $\om_2$ respectively, they cannot be connected by any LU transformation that does not break the symmetry. In fact, no matter what symmetric LU transformation we apply on the state, as long as the system remains gapped and short range correlated, we can always perform an RG transformation to the resulting state and find the projective representation on the edge state. Because the classes of projective representations are discrete, they cannot jump from one to another under symmetric LU transformation. Therefore, $|\phi_1\rangle$ and $|\phi_2\rangle$ cannot be related through symmetric LU transformation and hence belong to different SPT phases.

In this way, we are able to classify one dimensional gapped bosonic phases with on-site symmetry and find that: 
\begin{svgraybox}
\begin{center}
\textbf{Box 10.4 1D bosonic SPT with on-site symmetry}

Symmetry protected topological phases in one-dimensional bosonic systems with on-site symmetry of group $G$ have a one to one correspondence with the classes of projective representations of $G$, labeled by group elements of $H^2(G,U(1))$.
\end{center}
\end{svgraybox}
This result applies when the 1D representations $\alpha(G)$ form a finite group, when $G=U(1)$, further classification according to different $\alpha(U(1))$ exist.


\subsection{Time reversal symmetry}
\label{TR}
Time reversal, unlike other symmetries, is represented by anti-unitary operator $T$, which is equivalent to the complex conjugate operator $K$ followed by a unitary operator $u$. The classification of gapped 1D time reversal invariant phases follows closely the cases discusses before. In this section, we will highlight the differences and give our conclusion.

First, a state $|\psi\rangle$ is called time reversal invariant if 
\be
u \otimes u ... \otimes u K |\psi\rangle = \bt |\psi\rangle
\ee 
where $|\bt|=1$. But for anti-unitary $T$, the global phase $\bt$ is arbitrary and in particular we can redefine $|\psi'\rangle=\sqrt{\bt}|\psi\rangle$, such that $u \otimes u ... \otimes u K |\psi'\rangle = |\psi'\rangle$. Therefore, in the following discussion, we will assume WLOG that $\bt=1$.

Time reversal symmetry action on each site can belong to two different types with $T^2=uu^*=I$ or $-I$ respectively. For example, on a spin $1/2$, time reversal acts as $T=iYK$ and hence $T^2=-I$, while on a spin $1$, $T=ie^{i\pi S_y}K$ and $T^2=I$.
However, as long as we are considering systems without translational symmetry,
$T^2=I$ or $-I$ does not make any difference as we can always
take block size $2$ so that on the renormalized site, $T^2$
is always equal to $I$. WLOG, we will consider only the case with $T^2=I$ on each site.

Using argument similar to the case
of on-site unitary symmetry, we can keep track and redefine
symmetry operations as we do renormalization. Finally, at
the fixed point we have a state described by matrices
$A^{(\infty)}_{i^li^r}$ which is invariant under
time reversal operation
$T^{(\infty)}=u^{(\infty)}K$, that is,
\begin{align}
\label{Eqn:TR_A}
& \sum_{j^lj^r} u_{i^li^r,j^lj^r} A^*_{j^lj^r}=
M^{-1}A_{i^li^r}M
\end{align}
where the fixed-point label $\infty$ has been omitted. 

Solving this equation we find,\\
(a)$MM^*=e^{i\theta}I$. As $M$ is invertible, $e^{i\theta}=m=\pm 1$.\\
(b)$u = u^l \otimes u^r$. where $u^l$ and $u^r$ acts on $i^l$ and $i^r$ respectively, $u^l(u^l)^*=m I$ and $u^r(u^r)^*=m I$, $m=\pm 1$.
In the fixed point wave function, each entangled pair is time reversal invariant
\begin{equation}
(u^{[k],r}\otimes u^{[k+1],l})K |EP_{k,k+1}\rangle = |EP_{k,k+1}\rangle
\end{equation}

The definition of projective representation can be generalized to the anti-unitary case, from which we can see that the two cases with $m=\pm 1$ correspond to two different projective representations and hence two SPT phases with time reversal symmetry. Suppose that we have a group of symmetry actions $\tau(g)$, $g \in G$, some of which can be anti-unitary. We give a label to each symmetry operator $s(g)$, where $s(g)=1$ if $\tau(g)$ is unitary and $s(g)=-1$ if $\tau(g)$ is anti-unitary. $s(g)$ satisfies $s(g_1)s(g_2)=s(g_1g_2)$. Factor system of this representation $\om(g_1,g_2)$ is again given by
\be
\tau(g_1)\tau(g_2)=\om(g_1,g_2)\tau(g_1g_2)
\ee
but satisfies a modified consistency condition due to associativity
\be
\om(g_1,g_2)\om(g_1g_2,g_3)=\om^{s(g_1)}(g_2,g_3)\om(g_1,g_2g_3)
\ee
By changing the pre-factor of $\tau(g)$ by $\bt(g)$, we find that two factor systems are equivalent up to
\be
\om(g_1,g_2) \sim \om'(g_1,g_2)=\om(g_1,g_2)\frac{\bt(g_1g_2)}{\bt(g_1)\bt^{s(g_1)}(g_2)}
\ee
According to this definition, $MM^*= -1$ corresponds to a nontrivial projective representation of time reversal while $MM^*=1$ corresponds to a trivial one. Similar to the unitary case, nontrivial projective representations of time reversal cannot be one-dimensional, givigin rise to a nontrivial edge degeneracy in the fixed point state with the protection of time reversal symmetry.
Moreover, we can show that the time reversal invariant fixed point states can be mapped into each other if and only if they are related to the same $m$ value. Therefore, our classification result for time reversal symmetry is:

\begin{svgraybox}
\begin{center}
\textbf{Box 10.5 1D bosonic SPT with time reversal symmetry}

For 1D gapped boson / spin systems with ONLY time reversal symmetry,
there are two phases that do not break the symmetry.
\end{center}
\end{svgraybox}

\subsection{Translation invariance}

In this section, we would like to discuss translational invariant (TI) systems whose ground states are gapped and also translational invariant. The renormalization procedure discussed in section \ref{sec:RG_MPS} breaks translation symmetry and hence can not be used to study topological phases with translation symmetry. In this section, we will use the time evolution formulation of LU transformation and find a smooth path of gapped TI Hamiltonian whose adiabatic evolution connects two
states within the same TI phase.

\subsubsection{Translation invariance only}
\label{sec:TI}

First, as an example, we consider the case of TI only and show that there is only one gapped TI phase. Each translational invariant MPS is described(up to local change of basis) by a double tensor $\mathbb{E}$
\begin{equation}
\mathbb{E}_{\alpha\gamma,\beta\chi} = \sum_i A_{i,\alpha\beta} \otimes A^*_{i,\gamma\chi}
\end{equation}
Note that here the matrices and the double tensor are site independent.
The MPS is short-range correlated if $\mathbb{E}$ has a non-degenerate largest
eigenvalue 1. $\mathbb{E}$ can be written as
\begin{equation}
\mathbb{E}_{\alpha\gamma,\beta\chi} = \mathbb{E}^0_{\alpha\gamma,\beta\chi} + \mathbb{E}'_{\alpha\gamma,\beta\chi} = \Lambda^l_{\alpha\gamma}\Lambda^r_{\beta\chi}+\mathbb{E}'_{\alpha\gamma,\beta\chi}
\label{E=E0+E'}
\end{equation}
where $\Lambda^l$ ($\Lambda^r$) is the left (right) eigenvector of eigenvalue $1$ and $\mathbb{E}'$ is of eigenvalue less than $1$. As we have discussed previously, with a suitable choice of basis, 
\be
\begin{array}{l}
\Lambda^l_{\alpha\gamma}=\lambda_{\alpha}\delta_{\alpha\gamma}, \ \lambda_{\alpha}>0 \\ \nonumber
\Lambda^r_{\beta\chi}=\eta_{\beta}\delta_{\beta\chi}, \ \eta_{\beta}>0
\end{array}
\ee
Obviously, $\mathbb{E}^0$ is a valid double tensor and represents a state in the fixed point form.

Next we show that we can smoothly change $\mathbb{E}$ to $\mathbb{E}^0$ by turning down the $\mathbb{E}'$ term to $0$ from $t=0$ to $t=T$ as
\begin{equation}
\mathbb{E}(t)=\mathbb{E}^0+(1-\frac{t}{T})\mathbb{E}'
\label{Et}
\end{equation}
We will demonstrate that this process corresponds to an LU time evolution preserving translation symmetry.

Every $\mathbb{E}(t)$ represents a TI SRC MPS state. To see this, note that if we recombine the indices $\alpha\beta$ as row index and $\gamma\chi$ as column index and denote the new matrix as $\hat{\mathbb{E}}$, then both $\hat{\mathbb{E}}$ and $\hat{\mathbb{E}}^0$ are positive semidefinite matrices. But then every $\hat{\mathbb{E}}(t)$ is also positive semidefinite, as for any vector $|v\rangle$
\begin{equation}
\begin{array}{lll}
\langle v|\hat{\mathbb{E}}(t)|v\rangle & = & \langle v|\hat{\mathbb{E}}^0|v\rangle +(1-\frac{t}{T})\langle v|\hat{\mathbb{E}'}|v\rangle \\ \nonumber
 & = & (1-\frac{t}{T})\langle v|\hat{\mathbb{E}}|v\rangle + \frac{t}{T} \langle v|\hat{\mathbb{E}}^0|v\rangle > 0
\end{array}
\end{equation}
$\mathbb{E}(t)$ is hence a valid double tensor and the state represented can be determined by decomposing $\mathbb{E}(t)$ back into matrices $A_i(t)$. Such a decomposition is not unique. $A_i(t)$ at different time is determined only up to a local unitary on the physical index $i$.  But WLOG, we can choose the local unitary to be continuous in time, so that $A_i(t)$ vary continuously with time and reach the fixed point form at $t=T$(up to local change of basis). The state represented $|\psi(t)\rangle$ hence also changes smoothly with $t$ and is a pure state with a finite correlation length as all eigenvalues of $\mathbb{E}(t)$ expect for $1$ are diminishing with $t$. Therefore, $\mathbb{E}(t)$ represents a smooth path in TI SRC MPS that connects any state to a fixed point state(up to local change of basis).

How do we know that no phase transition happens along the
path? This is because for every state $|\psi(t)\rangle$, we
can find a parent Hamiltonian which changes smoothly with
$t$ and has the state as a unique gapped ground
state. Following the discussion in chapter \ref{chap6}, we choose a sufficiently large but
finite $l$ and set the parent Hamiltonian to be
\be
H(t)=-\sum_k h(t)_{k,k+l}
\ee
where $h(t)_{k,k+l}$ is the
projection onto the support space of the reduced density
matrix on site $k$ to $k+l$ at time $t$. Note that this
Hamiltonian is translation invariant. For large enough $l$,
$h(t)_{k,k+l}$ will always be $D\times D$ dimensional. As
the state changes continuously, its reduced density matrices
of site $k$ to $k+l$ changes smoothly. Because the dimension
of the space does not change, $h(t)_{k,k+l}$ also changes
smoothly with time. Moreover, it can be shown that $H(t)$ is
always gapped as the second largest eigenvalue of
$\mathbb{E}(t)$ never approaches $1$.
Therefore, by evolving the Hamiltonian adiabatically from
$t=0$ to $t=T$, we obtain a local unitary
transformation connecting any state to the
fixed point form, and in particular without breaking the
translation symmetry.

Because any TI fixed point state can be disentangled into
product state in a TI way, we find that 
\begin{svgraybox}
\begin{center}
\textbf{Box 10.6 1D bosonic SPT with translation symmetry}

All translation invariant 1D
gapped ground states are in the same phase, if no other
symmetries are required
\end{center}
\end{svgraybox}

\subsubsection{Translation invariance and on-site symmetry}

If the system is TI and has on-site symmetry, we need to maintain the on-site symmetry while doing the smooth deformation. We will not present the detailed derivation here but only summarize what we have learned.

First we can show that:
\begin{svgraybox}
\begin{center}
\textbf{Box 10.7 Nonexistence of short range correlated ground states}

For a 1D spin system with translation and an
on-site projective symmetry $u(g)$, the symmetric ground state
cannot be short-range correlated, if the projective
symmetry $u(g)$ corresponds to a non-trivial element in
$H^2(G,U(1))$.
\end{center}
\end{svgraybox}

The reason is as follows. If a 1D state with translation
symmetry is short-range correlated, it can be
represented by a TI MPS. Suppose that we perform the RG transformation described in section \ref{sec:RG_MPS} and flow the state to a fixed point form with on-site symmetry $\t u(g)$. With a proper choice of block size $n$ in each RG step, we can make $u(g)$ and $\t u(g)$
to be the same type of projective representation described by
$\om_{sym}\in H^2(G,U(1))$. The fixed point matrices then transform as
\begin{equation}
\sum_{j^lj^r} \t u_{i^li^r,j^lj^r}(g) A_{j^lj^r}=
M^{-1}(g)A_{i^li^r}M(g)
\end{equation}
Because $M^{-1}(g)$ and $M(g)$ form projective representations of class $\om$ and $-\om$,
$\om_{sym}$ has to be $0$, that is, the trivial element in  $H^2(G,U(1))$. So, if
$\om_{sym}\neq 0$, the 1D TI state cannot be short-range
correlated. In other words,
1D spin systems with translation and an
on-site projective symmetry
are always gapless or have degenerate ground states that
break the symmetries.

If the ground state of the 1D spin system does not break
the on-site symmetry and the translation symmetry, then
ground state is not short-range correlated and is gapless.  If
the ground state of the 1D spin system breaks the on-site
symmetry or the translation symmetry, then the ground state
is degenerate.

As an application of the above result, we find that:
\begin{svgraybox}
\begin{center}
\textbf{Box 10.8 Nonexistence of short range correlated half integer spin chain}

1D half-integer-spin systems with translation and the $SO(3)$
spin rotation symmetry
are always gapless or have degenerate ground states.
\end{center}
\end{svgraybox}

Note that this condition is not necessary in systems without translation symmetry. Indeed, without TI, $A^{[k]}_{j^lj^r}$ may depend on site label $k$ and when transformed under symmetry as
\be
\sum_{j^lj^r} \t u^{[k]}_{i^li^r,j^lj^r}(g) A^{[k]}_{j^lj^r}=
(M^{[k]}(g))^{-1}A_{i^li^r}M^{[k+1]}(g)
\ee
$M^{[k]}(g)$ and $M^{[k+1]}(g)$ may form projective representations of different types. Therefore, it is possible to have SRC non-translation invariant state with projective symmetry action per site. In particular, 1D half-integer spin chains with $SO(3)$ spin rotation symmetry can be gapped and SRC if the spins are dimerized (form singlets between site $2i-1$ and $2i$) and break translation symmetry explicitly.

On the other hand, to have a gapped TI 1D state with an on-site symmetry, the
symmetry must act linearly (i.e. not projectively).  In this
case, new labels are needed for SPT states besides the projective representation class. In particular, for 1D bosonic systems of $L$ sites with translation and
an on-site linear symmetry of group $G$, a gapped
state that does not break the two symmetries
must transform as
\begin{align}
\label{u0act_TI}
u(g) \otimes...\otimes
u(g) |\psi_L\rangle= [\alpha(g)]^L
|\psi_L\rangle
\end{align}
for all values of $L$ that is large enough.
Here $u(g)$ is the linear representation of $G$ acting
on the physical states in each site and $\al(g)$ is a
one-dimensional linear representation of $G$. Due to translation symmetry, the symmetric LU transformations cannot change 1D representation
$\al(g)$. So the different SPT phases are also distinguished
by the 1D representations $\al$ of $G$. 

Similar to the derivation in the previous section, we find that SRC MPS with translation and on-site linear symmetry can be mapped to a fixed point form which transform under the on-site symmetry as
\begin{align}
\label{Eqn:LU_ATI}
 \sum_{j^lj^r} \t u_{i^li^r,j^lj^r}(g) A_{j^lj^r}
& =
\al(g)
M^{-1}(g)A_{i^li^r} M(g)
\end{align}
where $\t u_{i^li^r,j^lj^r}(g)$ is the on-site linear symmetry at fixed point, $\al(g)$ is the 1D representation of $G$ and $M(g)$ is a projective representation of $G$. Both $\al(g)$ and the class of $M(g)$ have to be the same for two SPT states to be connectable through symmetric LU transformations.

\begin{svgraybox}
\begin{center}
\textbf{Box 10.9 1D bosonic SPT with translation and on-site symmetry}

For 1D boson / spin systems with only translation and
an on-site linear symmetry $G$, all the phases of gapped
states that do not break the two symmetries are classified
by a pair $(\om,\al)$ where $\om\in H^2(G,U(1))$ label
different types of projective representations of $G$ and
$\al$ label different 1D representations of $G$.
\end{center}
\end{svgraybox}

Here are a few concrete examples:

If we choose the symmetry group to be $G=\mathbb{Z}_n$, we find: 
For 1D spin systems with only translation and
on-site $\Z_n$ symmetry, there are $n$ phases for gapped
states that do not break the two symmetries.

This is because $\Z_n$ has no projective representations
and has $n$ different 1D representations.
As an example, consider the following model
\begin{align}
 H=\sum_i [ - h \si^z_i - \si^x_{i-1} \si^y_i \si^z_{i+1} ] ,
\end{align}
where $\si^{x,y,z}$ are the Pauli matrices.  The model has a
$\mathbb{Z}_2$ symmetry generated by $\si^z$.  The two
different $\mathbb{Z}_2$ symmetric phases correspond the $h\to
\infty$ phase and the $h\to -\infty$ phase of the model.

If we choose the symmetry group to be $G=SO(3)$, we find: 
For 1D integer-spin systems with only translation and
$SO(3)$ spin rotation symmetry, there are two phases for gapped
states that do not break the two symmetries.

This is because $SO(3)$ has only one 1D representation and
$H^2(SO(3),U(1))=\mathbb{Z}_2$. A nontrivial example in this classification is given by the AKLT state in spin-1 chains and a trivial example is given by the direct product state with spin-0 on each site. 

On the other hand, if $\al(g)$ does not form a 1D representation of $G$, then the state cannot be both symmetric and short range correlated. 

Let us apply the above result to a boson system with $p/q$
bosons per site.  Here the bosons number is conserved and
there is an $U(1)$ symmetry.  Certainly, the system is well
defined only when the number of sites $L$ has a form $L=J q$
(assuming $p$ and $q$ have no common factors).  For such an
$L$, we find that $\al_L(g)=\al_0(g)^{J}=\al_0(g)^{L/q}$,
where $\al_0(g)$ is the generating 1D representation of the
$U(1)$ symmetry group.  So \eqn{u0act_TI} is \emph{not}
satisfied for some large $L$.  Therefore 

\begin{svgraybox}
\begin{center}
\textbf{Box 10.10 Nonexistence of gapped boson state at fractional filling}

A 1D state of
conserved bosons with fractional bosons per site must be
gapless, if the state does not break the $U(1)$ and the
translation symmetry.

\end{center}
\end{svgraybox}

In higher dimensions, the situation
is very different.  A 2D state of conserved bosons with
fractional bosons per site can be gapped, and, at same time,
does not break the $U(1)$ and the translation symmetry.
2D fractional quantum Hall states of bosons
on lattice provide examples for such kind of states.

Results discussed in this section apply not only to unitary on-site symmetry, but to anti-unitary time reversal symmetry as well.

\subsubsection{Translation invariance and parity symmetry}

In this section, we will consider the case of parity
symmetry for translational invariant system. The parity
operation $P$ for a spin (boson) chain is in general composed of two
parts: $P_1$, exchange of sites $n$ and $-n$; $P_2$, on-site
unitary operation $u$ where $u^2=I$.
\footnote{The $\mathbb{Z}_2$ operation $u$ is necessary in the definition of parity if we want to consider for example, fixed point state with $|EP\rangle=|00\rangle+|11\rangle$ be to parity symmetric. The state is not invariant after exchange of sites, and only maps back to itself if in addition the two spins on each site are also exchanged with $u$.}

Following previous discussions, it is possible to show that the matrices describing the SRC state with translation and parity symmetry can be deformed to a fixed point form, which satisfies:
\begin{equation}
\sum_{j^lj^r} u_{i^li^r,j^lj^r} A_{j^lj^r}^T =\pm M^{-1}A_{i^li^r}M
\label{PA_A}
\end{equation}
for some invertible matrix $M$ and $u^2=I$, where we have used that the
1D representation of parity is either $(1,1)$ or $(1,-1)$. We label the two 1D representations with $\alpha(P)=\pm 1$.
Here $M$ satisfies $M^{-1}M^T=e^{i\theta}$. But $M=(M^T)^T=e^{2i\theta}M$, therefore, $e^{i\theta}=\pm 1$ and correspondingly $M$ is either symmetric $M=M^T$ or antisymmetric $M=-M^T$. We will label this sign factor as $\beta(P)=\pm 1$.

Solving this equation gives that $u=\alpha(P) v(u^l\otimes u^r)$,
where $v$ is the exchange operation of the two spins
$i^l$ and $i^r$ and $u^l$,$u^r$ act on $i^l$,$i^r$
respectively. $(u^l)^T=\beta(P)u^l$ and $(u^r)^T=\beta(P)u^r$.
It can then be shown that each entangled pair
$|EP_{k,k+1}\rangle$ must be symmetric under parity
operations and satisfies $u^r_k\otimes
u^l_{k+1}|EP_{k+1,k}\rangle = \alpha(P) |EP_{k,k+1}\rangle$. There are hence four different symmetric phases corresponding to $\alpha(P)=\pm 1$ and $\beta(P)=\pm 1$.
We can show similarly
as before that fixed points within each class can be mapped
from one to the other with TI LU
transformation preserving the parity symmetry. On the other hand, fixed points in different classes can not be connected without breaking the symmetries. Therefore,
there are four
parity symmetric TI phases.

\begin{svgraybox}
\begin{center}
\textbf{Box 10.11 1D bosonic SPT with translation and parity symmetry}

For 1D boson / spin systems with only translation and
parity symmetry, there are four phases for gapped
states that do not break the two symmetries.

\end{center}
\end{svgraybox}

As an example, consider the following model
\begin{align}
 H=\sum_i [ -B S^z_i + \v S_i\cdot \v S_{i+1} ] ,
\end{align}
where $\v S_i$ are the spin-1 operators.  The model has a
parity symmetry. The $B=0 $ phase and the $B\to +\infty$
phase of the model correspond to two of the four phases discussed above.
The $B=0$ state is in the
same phase as the AKLT state. In the fixed-point state for
such a phase, $|EP_{k,k+1}\rangle = |\up\down\>-
|\down\up\>$. The parity transformation exchange the first
and the second spin, and induces a minus sign: $P:
|EP_{k,k+1}\rangle \to -|EP_{k,k+1}\rangle $.  The $B\to
+\infty$ state is the $S^z=1$ state.  Its entangled pairs are
$|EP_{k,k+1}\rangle = |\up\up\>$ which do not change sign
under the parity transformation.  Thus the stability of the
Haldane/AKLT state is also protected by the parity
symmetry.

\begin{figure}[htbp]
\begin{center}
\includegraphics[width=3.0in]{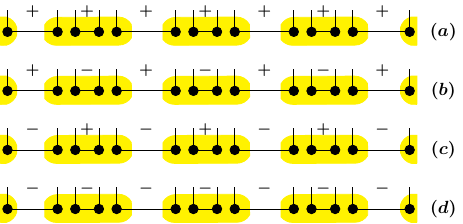}
\end{center}
\caption{
Representative states of the four parity symmetric phases, each corresponding to (a) $\alpha(P)=1$, $\beta(P)=1$ (b) $\alpha(P)=-1$, $\beta(P)=1$ (c) $\alpha(P)=-1$, $\beta(P)=-1$ (d) $\alpha(P)=1$, $\beta(P)=-1$. $+$ stands for a parity even entangled pair (e.g. $|00\rangle+|11\rangle$), $-$ stands for a parity odd entangled pair (e.g. $|01\rangle-|10\rangle$). Each site contains four virtual spins.}
\label{fig:4Pstates}
\end{figure}

To understand why there are four parity symmetric phases instead of two (parity even/parity odd), we give four representative states in Fig. \ref{fig:4Pstates}, one for each phase. Connected pair of black dots denotes an entangled pair. $+$ stands for a parity even pair, for example $|00\rangle+|11\rangle$, and $-$ stands for a parity odd pair, for example $|01\rangle-|10\rangle$. Each rectangle corresponds to one site, with four spin degrees of freedom on each site. The four states are all translational invariant. If the parity operation is defined to be exchange of sites together with exchange of spins $1$ and $4$, $2$ and $3$ on each site, then states (a) and (d) are parity even while (b) and (c) are parity odd. But (a) and (d) (or (b) and (c)) are different parity even (odd) states and cannot be mapped to each other through local unitary transformations without breaking parity symmetry. Written in the matrix product representation, the matrices of the four states will transform with $\alpha(P)=\pm 1$ and $\beta(P)=\pm 1$ respectively. Therefore, the parity even/odd phase breaks into two smaller phases and there are in all four phases for parity symmetric systems.

\subsection{Summary of results for bosonic systems}

Here we summarize the classification of topological phases in 1D bosonic systems with different symmetries in Table \ref{Table:result}.
 
 \begin{table*}[h]
 \centering
 \footnotesize
 \begin{tabular}{ |c|c|c| }
 \hline
 \multirow{2}{*}{Symmetry}             & No. or Label of & \multirow{2}{*}{Example System}\\
 																			& Different Phases &																\\ \hline\hline
 None 		& 		1 		&  \\ \hline
 On-site  &  \multirow{2}{*}{$\om \in H^2(G,U(1))$}  & On-site $\mathbb{Z}_n$ or $SU(2)$: 1 phase \\
 Symmetry of Group G (*)   &               & On-site $SO(3)$/$D_2$ on integer spin: 2 phases \\ \hline
 Time Reversal(TR)    &  2                              &          \\ \hline
 Translational Invariance(TI)    &  1                              &          \\ \hline
 TI+On-site Linear   & $\om \in H^2(G,U(1))$   &  TI+On-site $\Z_n$: n phases \\
 Symmetry of Group G     &  and $\alpha(G)$ &  TI+On-site $SO(3)$ on integer spin: 2 phases \\ \hline
 TI+ On-site  Projective  &   \multirow{2}{*}{0}  & TI+On-site $SO(3)$ or $D_2$ on\\
 Symmetry of Group G &             & half-integer spin: no gapped phase \\ \hline
 TI+Parity             &   4                  &                              \\ \hline
 \multirow{2}{*}{TI+TR} & 2 if $T^2=I$  & TI+TR on integer spin: 2 phases     \\
                        & 0 if $T^2=-I$ & on half-integer spin: no gapped phase \\ \hline
 \end{tabular}
 \caption{Summary of classification result for 1D gapped spin system with symmetric ground states. TI stands for translational invariance. TR stands for time reversal symmetry. $H^2(G,U(1))$ is the second cohomology group of group $G$ over complex number $U(1)$. $\alpha(G)$ is a 1D representation of $G$. 
 (*): this result applies when $\alpha(G)$ form a finite group, when $G=U(1)$, further classification according to different $\alpha(U(1))$ exist.
 }
 \label{Table:result}
 \end{table*}


\section{Topological phases in 1D fermion systems}
\label{sec:fSPT}

Although our previous discussions have been focused on boson/spin systems, it actually also applies to fermion systems. Because in 1D fermion systems and spin systems can be mapped to each other through Jordan Wigner transformation, we can classify fermionic phases by classifying the corresponding spin phases, as we discuss in this section. We are not going to study the fermionic topological phases in detail and related references are given in Summary and Further Reading.

Specifically, for a class of fermion systems with certain symmetry we are going to do the following

1. identify the corresponding class of spin systems by mapping the symmetry to spin 

2. classify possible spin phases with this symmetry, including symmetry breaking and symmetry protected topological phases

3. map the spin phases back to fermions and identify the fermionic order 

In the following we are going to apply this strategy to 1D fermion systems in four cases: no symmetry(other than fermion parity), time reversal symmetry for spinless fermions, time reversal symmetry for spin half integer fermions, and $U(1)$ symmetry for fermion number conservation. One special property of fermionic systems is that it always has a fermionic parity symmetry. That is, the Hamiltonian is a sum of terms composed of even number of fermionic creation and annihilation operators. Therefore, the corresponding spin systems we classify always have an on-site $\mathbb{Z}_2$ symmetry. Note that this approach can only be applied to systems defined on an open chain. For system with translation symmetry and periodic boundary condition, Jordan Wigner transformation could lead to non-local interactions in the spin system. 

\subsection{Jordan Wigner transformation}

First, let us briefly summarize the procedure of Jordan Wigner transformation for mapping 1D fermion systems to 1D spin systems. 

Consider the simplest case where each site $k$ in the fermion system contains one fermion mode with creation and annhilation operator $a^{\dg}_k$ and $a_k$. The local Hilbert space is two dimensional and can be mapped to a spin $1/2$ degree of freedom
\be
\ket{\Omega_k} \to \ket{0_k}, \ a^{\dg}_i\ket{\Omega_k} \to \ket{1_k}
\ee
where $\ket{\Omega_k}$ is the unoccupied fermionic state on site $k$, $\ket{0_k}$ and $\ket{1_k}$ are spin states with $\pm \frac{1}{2}$ spin in $z$ direction.

The mapping between operators, however, has to be non-local to preserve the anti-commutation relation between fermionic operators on different sites. In particular,
\be
a_k \to \frac{1}{2}\prod_{j<k} Z_j (X_k+iY_k), \ a^{\dg}_k \to \frac{1}{2}\prod_{j<k} Z_j (X_k-iY_k)
\ee
It can be checked explicitly that the operator algebra is preserved under this mapping. The fermion occupation number on site $k$ is mapped to
\be
a^{\dg}_ka_k \to \frac{1}{2}(I_k-Z_k)
\ee
and the total fermion parity operator is mapped to
\be
P_f = \prod_k (1-2a^{\dg}_ka_k)  \to \prod_k Z_k
\ee
Local Hamiltonian terms in the middle of the 1D fermionic chain are mapped to local Hamiltonian terms in the spin system. For example, fermion hopping terms are mapped as
\be
a_k^{\dagger}a_{k+1} + a^{\dagger}_{k+1}a_k \to \frac{1}{2} X_kX_{k+1} + Y_kY_{k+1}
\ee
Terms across the boundary however may become nonlocal. For example
\be
a_N^{\dagger}a_1+a^{\dagger}_1a_N = -\frac{1}{2} \prod_k Z_k (X_NX_1+Y_NY_1)
\ee
Therefore, with periodic boundary condition, local fermion models do not exactly map into local spin models. However, if the total fermion parity ($\prod_k Z_k$) is fixed, the boundary term becomes local, as can be seen from the previous example. Therefore, if we are considering gapped fermion systems with nondegenerate ground state, which has a fixed fermion parity, the Jordan-Wigner transformation is effectively local.  

If the local Hilbert space on site $i$ is larger than two dimensional, we can always embed it into a larger (finite dimensional) Hilbert space of the form $H_f\otimes H_b$, where $H_f$ is the two dimensional Hilbert space corresponding to a fermion mode and $H_b$ is a bosonic Hilbert space. Note that the classification we are considering is stable with respect to addition of local degrees of freedom. Therefore, embedding into a larger local Hilbert space is allowed. After this embedding, Jordan Wigner transformation proceeds as described above on the $H_f$ sector only.

\subsection{Fermion parity symmetry only}

For a 1D fermion system with only fermion parity symmetry, how many gapped phases exist? 

To answer this question, first we do a Jordan-Wigner transformation and map the fermion system to a spin chain. The fermion parity operator $P_f = \prod (1-2a^{\dagger}_ia_i)$ is mapped to an on-site $\mathbb{Z}_2$ operation. On the other hand, any 1D spin system with an on-site $\mathbb{Z}_2$ symmetry can always be mapped back to a fermion system with fermion parity symmetry(expansion of local Hilbert space maybe necessary). As the spin Hamiltonian commute with the $\mathbb{Z}_2$ symmetry, it can be mapped back to a proper physical fermion Hamiltonian. Therefore, the problem of classifying fermion chains with fermion parity is equivalent to the problem of classifying spin chains with $\mathbb{Z}_2$ symmetry.

There are two possibilities in spin chains with $\mathbb{Z}_2$ symmetry: (1) the ground state is symmetric under $\mathbb{Z}_2$. As $\mathbb{Z}_2$ does not have non-trivial projective representation, there is one symmetric phase. (If translational symmetry is required, systems with even number of fermions per site are in a different phase from those with odd number of fermions per site. This difference is somewhat trivial and we will ignore it.) (2) the ground state breaks the $\mathbb{Z}_2$ symmetry. The ground state will be two-fold degenerate. Each short-range correlated ground state has no particular symmetry and they are mapped to each other by the $\mathbb{Z}_2$ operation. There is one such symmetry breaking phases. These are the two different phases in spin chains with $\mathbb{Z}_2$ symmetry.

This tells us that there are two different phases in fermion chains with only fermion parity symmetry. But what are they? First of all, fermion states cannot break the fermion parity symmetry. All fermion states must have a well-defined parity. Does the spin symmetry breaking phase correspond to a real fermion phase?

The answer is yes and actually the spin symmetry breaking phase corresponds to a $\mathbb{Z}_2$ symmetric fermion phase. Suppose that the spin system has two short-range correlated ground states $|\psi_0\>$ and $|\psi_1\>$. All connected correlations between spin operators decay exponentially on these two states. Mapped to fermion systems, $|\psi^f_0\>$ and $|\psi^f_1\>$ are not legitimate states because they don't have fixed fermion parity but $|\t{\psi}^f_0\>=|\psi^f_0\> + |\psi^f_1\>$ and $|\t{\psi}^f_1\>=|\psi^f_0\> - |\psi^f_1\>$ are. They have even/odd fermion parity respectively. In spin system, $|\t{\psi}_0\>$ and $|\t{\psi}_1\>$ are not short range correlated states but mapped to fermion system they are. To see this, note that any correlator between bosonic operators on the $|\t{\psi}^f_0\>$ and $|\t{\psi}^f_1\>$are the same as that on $|\psi_0\>$ and $|\psi_1\>$ and hence decay exponentially. Any correlator between fermionic operators on the $|\t{\psi}^f_0\>$ and $|\t{\psi}^f_1\>$ gets mapped to a string operator on the spin state, for example $a^{\dagger}_ia_j$ is mapped to $(X-iY))_iZ_{i+1}...Z_{j-1}(X-iY)_j$, which also decays with separation between $i$ and $j$. Therefore, the symmetry breaking phase in spin chain corresponds to a fermionic phase with symmetric short range correlated ground states. 

\begin{figure}[ht]
\begin{center}
\includegraphics[scale=1.5]{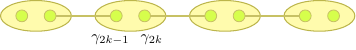}
\end{center}
\caption{Majorana chain model.
}
\label{MC}
\end{figure}

This phase can be realized in Kitaev's Majorana chain model. Consider a 1D fermionic chain with one fermion mode per site. Denote the creation and annihilation operator of the fermion mode on site $k$ as $a^{\dagger}_k$ and $a_k$. To understand the special property of the Majorana chain model, it is helpful to represent each fermion mode as two Majorana fermion modes
\be
\gamma_{2k-1} = a^{\dagger}_k + a_k, \ \ \gamma_{2k} = i(a^{\dagger}_k-a_k)
\ee
such that the $\gamma$'s are all Hermitian and $\gamma^2=I$. Now suppose that each Majorana mode couples to another Majorana mode on a neighboring site as shown in Fig.\ref{MC}. The Hamiltonian of this system reads
\be
H = \sum_k i\gamma_{2k}\gamma_{2k+1}
\ee

If we recombine the Majorana modes into fermion modes as
\be
b^{\dagger}_k = \frac{1}{2}(\gamma_{2k} -i\gamma_{2k+1}), b_k = \frac{1}{2}(\gamma_{2k}+i\gamma_{2k+1})
\ee
then we can map the Hamiltonian into the form
\be
H = 2b^{\dagger}_kb_k - 1
\ee
The Hamiltonian decouples into individual terms for each $b_k$ mode and it is easy to see that the ground state is the vacuum state for all such modes.

Now we are ready to see the most interesting feature of this model: with periodic boundary condition, all the Majorana modes are coupled in pairs and the system has a unique ground state (the vacuum state for all $b_k$ modes); with open boundary condition however, the two modes on the boundary are not coupled to anything, as shown in Fig.\ref{MC} and leaves a two fold degeneracy in the ground state. The degenerate ground states are $|\t{\psi}^f_0\>=|\psi^f_0\> + |\psi^f_1\>$ and $|\t{\psi}^f_1\>=|\psi^f_0\> - |\psi^f_1\>$ discussed above.

Note that while Fig.\ref{MC} has a similar structure to Fig.\ref{fp_sym} for bosonic SPT states, they have one important difference: each dot in Fig.\ref{fp_sym} represents a well defined Hilbert space but each dot in Fig.\ref{MC} does not. The dots in Fig.\ref{MC} represent Majorana modes and only by combining pairs of them do we have a well define Hilbert space of dimension two.

To summarize, the symmetry breaking phase of the spin chain corresponds to a topological phase in the fermion chain with Majorana edge modes. On the other hand, the symmetric phase in the spin chain corresponds to a topologically trivia phase for the fermions. A representative Hamiltonian in this phase can be written as
\be
H = \sum_k 2a^{\dagger}_ka_k -1 = \sum_k i\gamma_{2k-1}\gamma_{2k}
\ee
The ground state of this Hamiltonian is the vacuum state for all $a_k$ modes which is unique and gapped with both closed and open boundary conditions. As the two fermion phases have different edge states, they cannot be connected under any physical fermionic perturbation without closing gap and going through phase transition.


\begin{svgraybox}
\begin{center}
\textbf{Box 10.12 1D fermionic gapped phases with fermion parity symmetry}

For 1D fermion system with only fermion parity symmetry, there are two gapped phases, one with Majorana edge mode and one without.

\end{center}
\end{svgraybox}
 
\subsection{Fermion parity and $T^2=1$ time reversal}

Now consider the more complicated situation where aside from fermion parity, there is also a time reversal symmetry. Time reversal acts as an anti-unitary $T=UK$ on each site. In this section we consider the case where $T^2=1$(spinless fermion). 

So now the total symmetry for the fermion system is the $\mathbb{Z}_2$ fermion parity symmetry $P_f$ and $T^2=1$ time reversal symmetry. $T$ commutes with $P_f$. The on-site symmetry group is a $\mathbb{Z}_2\times \mathbb{Z}_2$ group and has four elements $G=\{I,T,P_f,TP_f\}$. Mapped to spin system, the symmetry group structure is kept. 

The possible gapped phases for a spin system with on-site symmetry $G=\{I,T,P_f,TP_f\}$ include both symmetry breaking and symmetric phases. If we use $G'$ to label the unbroken symmetry subgroup on the SRC ground state, then the possibilities are: 

(1) $G'=G$. Following discussion in previous sections we find that it has four different projective representations. Examples of the four representations are a.$\{I, K, Z, KZ\}$, b. $\{I,iYK, Z, iYKZ\}$, c. $\{I,iYKZ\otimes I, I\otimes Z, iYKZ\otimes Z\}$ d. $\{I, K, Y, KY\}$. There are hence four different symmetric phases. (If translational symmetry is required, the number is multiplied by $2$ due to $\alpha(\mathbb{Z}_2)$) 

(2) $G'=\{I,P_f\}$ with no non-trivial projective representation, the time reversal symmetry is broken. There is one such phase.  (If translational symmetry is required, there are two phases) 

(3) $G'=\{I,T\}$, with two different projective representations(time reversal squares to $\pm I$ on boundary spin). The $\mathbb{Z}_2$ fermion parity is broken. There are two phases in this case.  

(4) $G'=\{I,TP_f\}$, with two different projective representations. The fermion parity symmetry is again broken. Two different phases. 

(5) $G'=I$, no projective representation, all symmetries are broken.

Mapped back to fermion systems, fermion parity symmetry is never broken. Instead, the $P_f$ symmetry breaking spin phases are mapped to fermion phases with Majorana boundary mode on the edge as discussed in the previous section. Therefore the above spin phases correspond in the fermion system to: 

(1) Four different symmetric phases 

(2) One time reversal symmetry breaking phase. 

(3) Two symmetric phases with Majorana boundary mode 

(4) Another two symmetric phases with Majorana boundary mode. 

(5) One time reversal symmetry breaking phase. 

Among all these cases, (1)(3)(4) contains the eight symmetric phases for time reversal invariant fermion chain with $T^2=1$.

\begin{svgraybox}
\begin{center}
\textbf{Box 10.13 1D fermionic gapped phases with $T^2=1$ time reversal}

For 1D fermion system with $T^2=1$ time reversal symmetry and fermion parity symmetry, there are eight different gapped phases.

\end{center}
\end{svgraybox}

\subsection{Fermion parity and $T^2 \neq 1$ time reversal}

When $T^2 \neq I$, the situation is different. This happens when we take the fermion spin into consideration and for a single particle, time reversal is defined as $e^{i\pi S_y}K$. With half integer spin, $\left(e^{i\pi S_y}K\right)^2=-I$. Note that for every particle the square of time reversal is $-I$, however when we write the system in second quantization as creation and annihilation operator on each site, the time reversal operation defined on each site satisfies  $T^2=P_f$. Therefore, the symmetry group on each site is a $Z_4$ group $G=\{I,T,P_f,TP_f\}$. To classify possible phases, we first map everything to spin.

The corresponding spin system has on-site symmetry $G=\{I,T,P_f,TP_f\}$. $T^2=P_f$, $P_f^2=I$. The possible phases are: 

(1) $G'=G$, with two possible projective representations, one with $T^4=I$, the other with $T^4=-I$. Example for the latter includes $T=(1/\sqrt{2})(X+Y)K$. Therefore, there are two possible symmetric phases. (If translational symmetry is required, there are four phases.) 

(2) $G'=\{I,P_f\}$, the time reversal symmetry is broken. One phase. (If translational symmetry is required, there are two phases.) 

(3) $G'=I$, all symmetries are broken. One phase.

Therefore, the fermion system has the following phases: 

(1) Two symmetric phases 

(2) One time reversal symmetry breaking phase

(3) One time reversal symmetry breaking phase with Majorana boundary mode. 

Among all these cases, (1) contains the time reversal symmetry protected topological phase. 

\begin{svgraybox}
\begin{center}
\textbf{Box 10.14 1D fermionic gapped phases with $T^2=P_f$ time reversal}

For 1D fermion system with $T^2=P_f$ time reversal symmetry and fermion parity symmetry ($P_f$), there are two different gapped phases.

\end{center}
\end{svgraybox}

\subsection{Fermion number conservation}

Consider the case of a gapped fermion system with fixed fermion number. This
corresponds to an on-site $U(1)$ symmetry, $e^{i\theta N}$. Mapped to spins,
the spin chain will have an on-site $U(1)$ symmetry. This symmetry cannot be
broken and $U(1)$ does not have a non-trivial projective representation. One
thing special about $U(1)$ symmetry though, is that it has an infinite family
of 1D representations. The fermion number per
site is a good quantum number and labels different phases. Therefore, mapped back to fermions,
there is an infinite number of phases with different average number of fermions per site.

\begin{svgraybox}
\begin{center}
\textbf{Box 10.15 1D fermionic gapped phases with $U(1)$ symmetry}

For 1D fermion system with $U(1)$ charge conservation symmetry, there is an infinite number of phases with different average number of fermions per site.

\end{center}
\end{svgraybox}


\section{2D symmetry protected topological order}
\label{sec:2DSPT}

Having understood symmetry protected topological order in 1D, we can ask are there similar phases in two and higher dimensions? That is, we want to know if there exist gapped phases in two and higher dimension with short range entanglement in the bulk and symmetry protected nontrivial edge state on the boundary. From the study of free fermion system, we know that there are indeed such phases such as topological insulators and superconductors. These phases are gapped in the bulk and have gapless edge states as along as certain symmetries are preserved. However, it is not clear from the study of such free fermion system what SPT phases exist in general interacting system. In particular, it is not even clear whether SPT order can exist in bosonic system where without interaction no nontrivial topological order can emerge. In this section, we describe two strongly interaction bosonic models with nontrivial SPT order. The first one -- the 2D AKLT model -- is a straight forward generalization of the 1D AKLT model to 2D. Similar to the 1D version, the 2D AKLT state has spin rotation symmetry. However, its gapless edge state is protected only when translation symmetry is also preserved. The second model -- the CZX model -- demonstrates that translation symmetry is not always necessary for nontrivial SPT order to exist in 2D bosonic systems. That is, the gapless edge state of the CZX model is robust even in the presence of disorder, as long as certain internal symmetry is preserved. This is similar to what happens in topological insulators and superconductors in free fermion systems. 

\subsection{2D AKLT model}
\label{sec:2DAKLT}

\subsubsection{Bulk definition and boundary state}

\begin{figure}[ht]
\begin{center}
\includegraphics[scale=1.2]{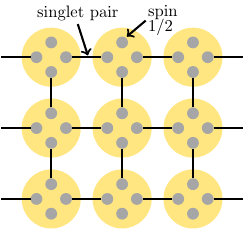}
\end{center}
\caption{The 2D AKLT model which is short range entangled and symmetric under spin rotation symmetry. Each site contains four spin $1/2$s. Two spin $1/2$s connected by a bond form a singlet under spin rotation symmetry. On a lattice with boundary, the boundary degrees of freedom are a chain of spin $1/2$s.
}
\label{2D_AKLT}
\end{figure}

This simple picture of 1D SPT phases, in particular the 1D AKLT state, can be generalized to two dimension to give the 2D AKLT model. Consider the 2D state in Fig. \ref{2D_AKLT}.

Every site contains four spin $1/2$s. Each spin $1/2$ forms a projective representation of the spin rotation symmetry $SO(3)$, but the four spins on each site together form a linear representation of $SO(3)$. Two spins on neighboring sites which are connected by a bond form a singlet pair $|01\rangle - |10\rangle$. Similar to the 1D case, the total state is invariant under spin rotation. The state is short range entangled and can be the gapped ground state of a simple Hamiltonian 
\be
H=\sum_{<ab>} X_aX_b + Y_aY_b + Z_aZ_b
\ee
where $<ab>$ labels the pairs of spins connected by a bond. 

If the system is defined on a disk with boundary, there will be free spin $1/2$s at each site on the boundary, as shown in Fig. \ref{2D_AKLT}. These boundary spins can couple to each other, for example through nearest neighbor Heisenberg interaction,
\be
H=\sum_i X_iX_{i+1} + Y_iY_{i+1} + Z_iZ_{i+1}
\ee
With such a coupling the boundary is in a gapless state. Of course, other types of coupling terms can also exist. The question is then, is the gapless edge state protected? Correspondingly, does the 2D AKLT state possess nontrivial SPT order.

If only spin rotation symmetry is considered, then the answer is no. Indeed, if we introduce modulation to the coupling strength of the Heisenberg interaction on the boundary
\be
H=\sum_{i} J_1 (X_{2i}X_{2i+1}+Y_{2i}Y_{2i+1}+Z_{2i}Z_{2i+1}) + J_2 (X_{2i+1}X_{2i+2}+Y_{2i+1}Y_{2i+2}+Z_{2i+1}Z_{2i+2})
\ee
the spin $1/2$s on the boundary would become `dimerized' and gapped, without breaking spin rotation symmetry. This is easiest to understand in the limit of $J_1>0$, $J_2=0$. Then every pair of $2i$th spin and $2i+1$th spin couple into a singlet pair. The $2i+1$th spin and the $2i+2$th spin are decoupled from each other. The total state is hence gapped and preserves spin rotation symmetry. Therefore, the 2D AKLT state does not have nontrivial SPT order protected by spin rotation symmetry alone.

However, the story changes once translation symmetry is added to the picture. If translation symmetry is also preserved, the boundary as a spin $1/2$ chain with translation symmetry is always gapless. Therefore, the 2D AKLT state has translation symmetry and spin rotation symmetry preserved gapless edge state and hence nontrivial SPT order under these symmetries.

\begin{svgraybox}
\begin{center}
\textbf{Box 10.16 SPT order of 2D AKLT state}

2D AKLT state has nontrivial symmetry protected topological order protected by translation and spin rotation symmetry.

\end{center}
\end{svgraybox}

Such a construction can be generalized to all kinds of internal symmetries. Consider an on-site symmetry of group $G$. On each site, instead of spin $1/2$s, we would have four degrees of freedom which carry projective representations of $G$. A pair of degree of freedom connected by a bond have projective representation $\omega$ and $-\omega$ respectively. Together they form an entangled state which is a linear representation of $G$. Therefore, the bulk of the system is gapped and symmetric. On the boundary, each site contains one projective representation. If translation symmetry is preserved, each boundary degree of freedom is well defined and the projective representation they carry do label different SPT phases. If each site forms a nontrivial projective representation of $G$, then translation symmetry requires that the boundary be gapless, indicating nontrivial SPT order in the bulk. On the other hand, in the absence of translation symmetry, the boundary degrees of freedom can be combined. As projective representations form an additive group (the second cohomology group $\mathcal{H}^2(G,U(1))$ of $G$), combining boundary spins would change the projective representations from one class to another and in particular, to the trivial class. Therefore, without translation symmetry, all 2D states with a bond form as shown in Fig. \ref{2D_AKLT} belong to the same phase. 

\subsubsection{Tensor network representation}

Such a bond state has a simple tensor network representation. More interestingly, the way the tensor transforms under symmetry contains important information about the SPT order of the state. This is what we are going to discuss in this section.

\begin{figure}[ht]
\begin{center}
\includegraphics[scale=1.2]{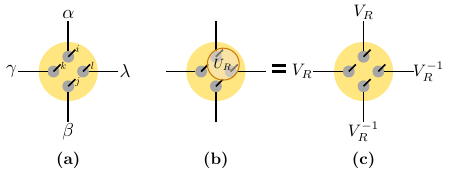}
\end{center}
\caption{The tensor network representation of the 2D AKLT state.
}
\label{T_AKLT}
\end{figure}

Consider the 2D AKLT example. The tensor on each site is composed of four parts $(t^i_{\alpha})_u$, $(t^j_{\beta})_d$, $(t^k_{\gamma})_l$, $(t^l_{\lambda})_r$, as shown in Fig. \ref{T_AKLT} (a). Each part contains one two-dimensional physical index $i$, $j$, $k$ or $l$, and one two-dimensional inner index $\alpha$, $\beta$, $\gamma$, or $\lambda$. The nonzero terms are
\be
\begin{array}{llll}
(t^0_{0})_u = 1, & (t^1_{1})_u = 1; & (t^0_{1})_d = -1, & (t^1_{0})_d = 1 \\ \nonumber
(t^0_{0})_l = 1, & (t^1_{1})_l = 1; & (t^0_{1})_r = -1, & (t^1_{0})_r = 1 \\ \nonumber
\label{eq:T_AKLT}
\end{array}
\ee
All other terms in the tensor are zero. With such a tensor, it is straight forward to check that two spin $1/2$'s connected by a bond are in the singlet state $|01\rangle-|10\rangle$.

Now let's see how the tensor transform under spin rotation symmetry. Apply a spin rotation transformation to the physical spins
\be
U_R = \prod_i e^{i\frac{\theta}{2}(n_xX_i+n_yY_i+n_zZ_i)}
\ee
as shown in Fig.\ref{T_AKLT} (b). If such a transformation is applied to all spins in the system, the wave function remains invariant. On the other hand, if we consider the action of the transformation on each individual tensor, the tensor may not remain invariant. They can change by some gauge transformation under the symmetry operation, similar to the 1D case we discussed previously. In particular, the tensor given in Eq. \ref{eq:T_AKLT} change by unitary transformations $V_R$, $V_R$, $V^{-1}_R$, $V^{-1}_R$ on the up, left, down, right inner indices respectively
\be
V_R=e^{-i\frac{\theta}{2}(n_xX^*+n_yY^*+n_zZ^*)}
\ee
$V_R$ corresponds to the rotation of a spin $1/2$, which forms a projective representation of the $SO(3)$ symmetry group.

Of course, this tensor only represents a very special point in the SPT phase with zero correlation length. In general, wave functions in the same SPT phase can have a finite correlation length and the tensors representing them maybe more complicated and have larger bond dimension. However, for any tensor in the same SPT phase as the 2D AKLT model, we expect them to transform under symmetry in a similar way. In particular, if we apply symmetry operator $U_R$ to the physical indices, the tensors would transform by a gauge transformation $\tilde{V}_R$ (or $\tilde{V}^{-1}_R$) on each inner index and $\tilde{V}_R$ (or $\tilde{V}^{-1}_R$) forms a projective representation of the $SO(3)$ symmetry group. That is, $\tilde{V}_R$ (or $\tilde{V}^{-1}_R$) represents spin rotation on half integer spins and satisfies
\be
\tilde{V}_R(2\pi)=-1
\ee
In general, if we take not just one tensor, but a piece of tensor network on a connected region and apply symmetry transformation, this would induce gauge transformation $\tilde{V}_R$ (or $\tilde{V}^{-1}_R$) on each of the inner index on the boundary of this region. Such a gauge transformation on the tensors corresponds to the symmetry transformation on the boundary degrees of freedom if we physically open a boundary to the system.

\subsection{2D CZX model}
\label{CZX}

On the other hand, SPT phases are known to exist in two and higher dimensions without the protection of translation symmetry, for example in topological insulators. The simple bond picture above therefore cannot account for their SPT order. In order to have nontrivial SPT order, we need to generalize the 2D AKLT state in two ways: (1) the local entanglement structure is not bonds between two spins, but rather plaquettes among four spins on sites around a square. This alone is not enough to construct new SPT order. We also need (2) symmetry transformation on each site does not factorize into separate operations on each of the four spins. That is, the total linear symmetry operation on each site is not a tensor product of four projective representations as otherwise the state can be reduced to a bond state.

Following this line of thought, we construct the CZX model in this section. The CZX model has an on-site $\mathbb{Z}_2$ symmetry that does not factorize into projective representations and the symmetry protected topological order of the state is robust against disorder. The boundary effective degrees of freedom in CZX model has an effective $\mathbb{Z}_2$ symmetry which cannot be written in an on-site form. Moreover, the boundary cannot be in a gapped symmetric state under the effective symmetry. In other words, the boundary must either break the $\mathbb{Z}_2$ symmetry or have gapless excitations. This is different from the 2D AKLT state discussed above(Fig.\ref{2D_AKLT}). In the 2D AKLT state, the boundary degrees of freedom are the boundary spins with projective representations. The effective symmetry is still on-site. Several boundary spins can form a singlet if their projective representations add up to a linear representation. Therefore, in the 2D AKLT state, the boundary can be in a gapped symmetric state under on-site symmetry simply by breaking translation symmetry. However, in the CZX model, this is not possible.

\subsubsection{Bulk definition}

In this section, we construct the CZX model explicitly which turns out to have nontrivial SPT order protected only by on-site $\mathbb{Z}_2$ symmetry.

\begin{figure}[ht]
\centerline{
\includegraphics[scale=1.2]{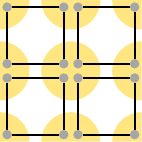}~~~~~~
\includegraphics[scale=1.2]{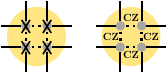}~~~~~~
\includegraphics[scale=1.2]{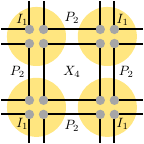}
}
\centerline{
	(a) ~~~~~~~~~~~~~~~~~~~~~~~~~~~~~~~~~~~~~~~~~~~~~~
	(b) ~~~~~~~~~~~~~~~~~~~~~~~~~~~~~~~~~~~~~~~~~~~~~~
	(c)
}
\caption{CZX model (a) each site (circle) contains four spins (dots) and the spins in the same plaquette (square) are entangled. (b) on-site $\mathbb{Z}_2$ symmetry is generated by $U_{CZX}=X_1 X_2 X_3 X_4CZ_{12}CZ_{23}CZ_{34}CZ_{41}$ (c) a local term in the Hamiltonian, which is a tensor product of one $X_4$ term and four $P_2$ terms as defined in the main text.
}
\label{CZX_model}
\end{figure}

Consider a square lattice with four two-level spins per site, as shown in Fig. \ref{CZX_model}(a) where sites are represented by circles and spins are represented by dots. We denote the two levels as $|0\>$ and $|1\>$. The system has an on-site $\mathbb{Z}_2$ symmetry as given in Fig. \ref{CZX_model}(b). It is generated by
\be
U_{CZX}=U_X U_{CZ}
\ee 
where
\be
U_X=X_1\otimes X_2\otimes X_3\otimes X_4
\ee
$X_i$ is Pauli $X$ operator on the $i$th spin and 
\be
U_{CZ}=CZ_{12}CZ_{23}CZ_{34}CZ_{41}
\ee
where $CZ$ is the controlled-$Z$ operator on two spins defined as
\be
CZ=|00\>\<00|+|01\>\<01|+|10\>\<10|-|11\>\<11|
\ee
As defined, $CZ$ does nothing if at least one of the spins is in state $|0\>$ and it adds a minus sign if both spins are in state $|1\>$. Different $CZ$ operators overlap with each other. But because they commute, $U_{CZ}$ is well defined. Note that $U_{CZ}$ cannot be decomposed into separate operations on the four spins and the same is true for $U_{CZX}$. $U_{X}$ and $U_{CZ}$ both square to $I$ and they commute with each other. Therefore, $U_{CZX}$ generates a $\mathbb{Z}_2$ group.

The Hamiltonian of the system is defined as a sum of local terms around each plaquette. Plaquettes are represented by squares in Fig. \ref{CZX_model}. $H=\sum H_{p_i}$, where the term around the $i$th plaquette $H_{p_i}$ acts not only on the four spins in the plaquette but also on the eight spins in the four neighboring half plaquettes as shown in Fig. \ref{CZX_model}(c)
\be
H_{p_i}=-X_4\otimes P_2^{u}\otimes P_2^{d}\otimes P_2^{l}\otimes P_2^{r}
\ee
where $X_4$ acts on the four spins in the middle plaquette as
\be
X_4=|0000\>\<1111|+|1111\>\<0000|
\ee
and $P_2$ acts on the two spins in every neighboring half plaquette as
\be
P_2=|00\>\<00|+|11\>\<11|
\ee
$P_2^{u}$, $P_2^{d}$, $P_2^{l}$, $P_2^{r}$ acts on the up, down, left and right neighboring half plaquettes respectively. For the remaining four spins at the corner, $H_{p_i}$ acts as identity on them. The $P_2$ factors ensure that each term in the Hamiltonian satisfies the on-site $\mathbb{Z}_2$ symmetry defined before. 

All the local terms in the Hamiltonian commute with each other, therefore it is easy to solve for the ground state. If the system is defined on a closed surface, it has a unique ground state which is gapped. In the ground state, every four spins around a plaquette are entangled in the state 
\be
|\psi_{p_i}\>=|0000\>+|1111\>
\ee
and the total wavefunction is a product of all plaquette wavefunction. If we allow any local unitary transformation, it is easy to see that the ground state can be disentangled into a product state, just by disentangling each plaquette separately into individual spin states. Therefore, the ground state is short range entangled. However, no matter what local unitary transformations we apply to disentangle the plaquettes, they necessarily violate the on-site symmetry and in fact, the plaquettes cannot be disentangled if the $\mathbb{Z}_2$ symmetry is preserved, due to the nontrivial SPT order of this model which we will show in the next sections.

It can be checked that this ground state is indeed invariant under the on-site $\mathbb{Z}_2$ symmetry. Obviously this state is invariant under $U_X$ applied to every site. It is also invariant under $U_{CZ}$ applied to every site. To see this note that between every two neighboring plaquettes, $CZ$ is applied twice, at the two ends of the link along which they meet. Because the spins within each plaquette are perfectly correlated (they are all $|0\>$ or all $|1\>$), the effect of the two $CZ$'s cancel each other, leaving the total state invariant. 

Therefore, we have introduced a 2D model with on-site $\mathbb{Z}_2$ symmetry whose ground state does not break the symmetry and is short-range entangled. In particular, this on-site symmetry is inseparable as discussed in the introduction and therefore cannot be characterized by projective representation as in the 2D AKLT state. We can add small perturbation to the system which satisfies the symmetry and the system is going to remain gapped and the ground state short range entangled and symmetric. It seems that the system is quite trivial and boring. However, we are going to show that surprising things happen if the system has a boundary and because of these special features the system cannot be smoothly connected to a trivial phase even if translation symmetry is not required.

\subsubsection{Boundary description}

\begin{figure}[ht]
  \centering
         \subfigure[][]{%
\includegraphics[scale=1.0]{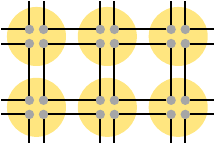}}%
         \hspace{12pt}%
         \subfigure[][]{%
           \includegraphics[scale=0.5]{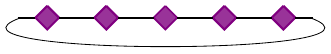}
 }%
                  \hspace{12pt}%
         \subfigure[][]{%
\includegraphics[scale=0.5]{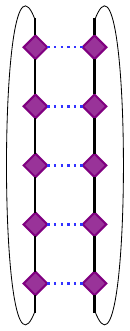}}%

\caption{(a)CZX model on a disk with boundary (b) boundary effective degrees of freedom form a 1D chain which cannot have a SRE symmetric state (c) two boundaries together can have a SRE symmetric state which is a product of entangled pairs between effective spins connected by a dashed line.
}
\label{CZX_boundary}
\end{figure}

The non-trivialness of this model shows up on the boundary. Suppose that we take a simply connected disk from the lattice, as shown in Fig.\ref{CZX_boundary}(a). 

The reduced density matrix of spins in this region is invariant under on-site symmetry in this region. The reduced density matrix is a tensor product of individual terms on each full plaquette, half plaquette and corner of plaquette respectively. On a full plaquette
\be
\rho_4=(|0000\>+|1111\>)(\<0000|+\<1111|)
\ee
On a half plaquette
\be
\rho_2=|00\>\<00|+|11\>\<11|
\ee
On a corner of a plaquette
\be
\rho_1=|0\>\<0|+|1\>\<1|
\ee
The state of spins on the plaquettes totally inside this region is completely fixed. But on the boundary there are free degrees of freedom. However, unlike in the 2D AKLT state, only part of the total Hilbert space of the spins on the boundary is free. In particular, two spins in a half plaquette on the boundary are constrained to the two-dimensional subspace $|00\>\<00|+|11\>\<11|$ and form an effective spin degree of freedom if we map $|00\>$ to $|\t{0}\>$ and $|11\>$ to $|\t{1}\>$. 

In Fig. \ref{CZX_boundary}(b), we show the effective degrees of freedom on the boundary as diamonds on a line. Projecting the total symmetry operation on the disk to the space supporting reduced density matrix, we find that the effective symmetry operation on the boundary effective spins is 
\be
\t{U}_{CZX}=\prod_{i=1}^N \t{X}_i\prod_{i=1}^N \t{CZ}_{i,i+1}
\label{tU}
\ee
with Pauli $\t{X}$ on each effect spin and $\t{CZ}$ operation between neighboring effective spins. The boundary is periodic and $\t{CZ}_{N,N+1}$ acts on effective spin $N$ and $1$. This operator generates a $\mathbb{Z}_2$ symmetry group.

This is a very special symmetry on a 1D system. First it is not an on-site symmetry. In fact, no matter how we locally group sites and take projections, the symmetry operations are not going to break down into an on-site form. Moreover, no matter what interactions we add to the boundary, as long as it preserves the symmetry, the boundary cannot have a gapped symmetric ground state. 

We can start by considering some simple cases. The simplest interaction term preserving this symmetry is $Z_iZ_{i+1}$. This is an Ising interaction term and we know that the ground state of 
\be
H=\sum_{i} Z_iZ_{i+1}
\ee
breaks the $\mathbb{Z}_2$ symmetry. In the transverse Ising model, the system goes to a symmetric phase if magnetic field in the $x$ direction is increased. However, $X_i$ breaks the $\mathbb{Z}_2$ symmetry $\t{U}_{CZX}$ on the boundary and therefore cannot be added to the Hamiltonian. Because $X_i$ is mapped into $Z_{i-1}X_iZ_{i+1}$ under the $\mathbb{Z}_2$ transformation, a possible symmetric Hamiltonian reads
\be
H=\sum_i X_i + Z_{i-1}X_iZ_{i+1}
\ee
Direct calculation shows that this Hamiltonian has a gapless spectrum. This is actually the transverse field cluster model discussed in Eq. \ref{eq:Hamclu} in section \ref{sec:t-CS}. $B=1$ in order to satisfy the $\Z_2$ symmetry and the Hamiltonian is known to be gapless.

In fact, we are going to prove that, as long as the $\mathbb{Z}_2$ symmetry is preserved, the boundary cannot have SRE symmetric ground state (actually a more generalized version of it) in the next section. This is one special property that differs the CZX model from the 2D AKLT in Fig.\ref{2D_AKLT}. In the 2D AKLT state, the symmetry operations on the boundary are just projective representations on each site. Without translational invariance, there can always be a SRE symmetric state with this symmetry. 

The special property on the boundary only shows up when there is an isolated single boundary. If we put two such boundaries together and allow interactions between them, everything is back to normal. As shown in Fig.\ref{CZX_boundary}(c), if we have two boundaries together, there is indeed a SRE symmetric state on the two boundaries. The state is a product of entangled pairs of effective spins connected by a dashed line. The entangled pair can be chosen as $|\t{0}\t{0}\>+|\t{1}\t{1}\>$. In contrast to the single boundary case, we can locally project the two effective spins connected by a dashed line to the subspace $|\t{0}\t{0}\>\<\t{0}\t{0}|+|\t{1}\t{1}\>\<\t{1}\t{1}|$ and on this subspace, the symmetry acts in an on-site fashion.

This result should be expected because if we have two pieces of sheet with boundary and glue them back into a surface without boundary, we should have the original SRE 2D state back. Indeed if we map the effective spins back to the original degrees of freedom $|\t{0}\>\to|00\>$ and $|\t{1}\> \to |11\>$, we see that the SRE state between two boundaries is just the a chain of plaquettes $|0000\>+|1111\>$ in the original state.

This model serves as an example of non-trivial SPT order in 2D SRE states that only needs to be protected by on-site symmetry. In order to prove the special property on the boundary of CZX model and have a more complete understanding of possible SPT orders in 2D SRE states with on-site symmetry, we are going to introduce a mathematical tool called Matrix Product Unitary Operator. We will show that 2D SPT phases are related to elements in $\mathcal{H}^3(G,U(1))$ which emerge in the transformation structure of the matrix product unitary operators. The discussion in the next section is general, but we will work out the CZX example explicitly for illustration.

\subsubsection{Boundary property}

\subsubsection{Matrix product unitary operators and its relation to 3 cocycle}
\label{MPUO_H3}

In this section, we discuss the matrix product unitary operator (MPUO) formalism and show how the effective symmetry operation on the boundary of CZX model can be expressed as MPUO. Moreover, we are going to relate MPUO of a symmetry group to the 3-cocycle of the group and in particular, we are going to show that the CZX model corresponds to a nontrivial 3-cocycle of the $\mathbb{Z}_2$ group.

A matrix product operator acting on a 1D system is given by,
\be
O=\sum_{\{i_k\},\{i_k'\}}Tr(T^{i_1,i'_1}T^{i_2,i'_2}...T^{i_N,i'_N})|i'_1i'_2...i'_N\>\<i_1i_2...i_N|
\ee
where for fixed $i$ and $i'$, $T^{i,i'}$ is a matrix with index $\alpha$ and $\beta$. Here we want to use this formalism to study symmetry transformations, therefore we restrict $O$ to be a unitary operator $U$. Using matrix product representation, $U$ does not have to be an on-site symmetry. $U$ is represented by a rank-four tensor $T^{i,i'}_{\alpha,\beta}$ on each site, where $i$ and $i'$ are input and output physical indices and $\alpha$, $\beta$ are inner indices. The matrix product unitary operators also have a canonical form, similar to matrix product states discussed in Chap. \ref{chap8}.

In particular, the symmetry operator $U_{CZX}$ (we omit the $\sim$ label for effective spins in following discussions) on the boundary of the CZX model can be represented by tensors
\be
\begin{array}{l}
T^{0,1}(CZX) = |0\>\<+|, \\
T^{1,0}(CZX) = |1\>\<-|, \\
\text{other terms are zero}
\end{array}
\ee
where $|+\>=|0\>+|1\>$ and $|-\>=|0\>-|1\>$.  It is easy to check that this tensor indeed gives $U_{CZX}=CZ_{12}...CZ_{N1}X_1...X_N$.

The other element in the $\mathbb{Z}_2$ group--the identity operation--can also be represented as MPUO with tensors
\be
\begin{array}{l}
T^{0,0}(I)=|0\>\<0|, \\
T^{1,1}(I)=|0\>\<0|, \\
\text{other terms are zero}
\end{array}
\ee
These two tensors are both in the canonical form, following a similar definition as in Chap.\ref{chap8}.

\begin{figure}[ht]
\begin{center}
\includegraphics[scale=1.2]{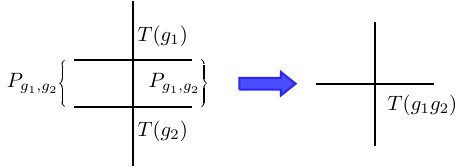}
\end{center}
\caption{Reduce combination of $T(g_2)$ and $T(g_1)$ into $T(g_1g_2)$. 
}
\label{P12}
\end{figure}

If two MPUO $T(g_2)$ and $T(g_1)$ are applied subsequently, their combined action should be equivalent to $T(g_1g_2)$. However, the tensor $T(g_1,g_2)$ obtained by contracting the output physical index of $T(g_2)$ with the input physical index of $T(g_1)$, see Fig. \ref{P12}, is usually more redundant than $T(g_1g_2)$ and might not be in the canonical form. It can only be reduced to $T(g_1g_2)$ if certain projection $P_{g_1,g_2}$ is applied to the inner indices (see Fig. \ref{P12}).

$P_{g_1,g_2}$ is only defined up to an arbitrary phase factor $e^{i\theta(g_1,g_2)}$. If the projection operator on the right side $P_{g_1,g_2}$ is changed by the phase factor $e^{i\theta(g_1,g_2)}$, the projection operator $P^{\dg}_{g_1,g_2}$ on the left side is changed by phase factor $e^{-i\theta(g_1,g_2)}$. Therefore the total action of $P_{g_1,g_2}$ and $P^{\dg}_{g_1,g_2}$ on $T(g_1,g_2)$ does not change and the reduction procedure illustrated in Fig.\ref{P12} still works. In fact, $P_{g_1,g_2}$ is unique up to a phase factor (on the unique block in the canonical form of $T(g_1,g_2)$).

Let us illustrate how the reduction is done for the symmetry group $(I,U_{CZX})$. For example, if we apply $U_{CZX}U_{CZX}$ the totally action should be equivalent to $I$. However the tensor $T(CZX,CZX)$ is given by
\be
\begin{array}{l}
T^{0,0}(CZX,CZX)=|01\>\<+-|, \\
T^{1,1}(CZX,CZX)=|10\>\<-+|, \\
\text{other terms are zero}
\end{array}
\ee
This tensor is reduced to $T(I)$ if projection
\be
P_{CZX,CZX} = (|01\>-|10\>)\<0|
\ee
and its Hermitian conjugate are applied to the right and left of $T(CZX,CZX)$ respectively.\footnote{The mapping actually reduces $T(CZX,CZX)$ to $-T(I)$. But this is not a problem as we can redefine $\t{T}(CZX)=iT(CZX)$ and the extra minus sign would disappear.} Adding an arbitrary phase factor $e^{i\theta(CZX,CZX)}$ to $P_{CZX,CZX}$ does not affect the reduction at all. By writing $P_{CZX,CZX}$ in the above form, we have made a particular choice of phase. 

Below we list the (right) projection operators for all possible combinations of $g_1$ and $g_2$ of this $\mathbb{Z}_2$ group.
\be
\begin{array}{lll}
P_{I,I} & = & |00\>\<0| \\
P_{CZX,I} & = & |00\>\<0|+|10\>\<1| \\
P_{I,CZX} & = & |00\>\<0|+|10\>\<1| \\
P_{CZX,CZX} & = & (|01\>-|10\>)\<0| \\
\end{array}
\ee
Note that in giving $P_{g_1,g_2}$ we have picked a particular choice of phase factor $e^{i\theta(g_1,g_2)}$. In general, any phase factor is allowed.

\begin{figure}[ht]
\begin{center}
\includegraphics[scale=1.0]{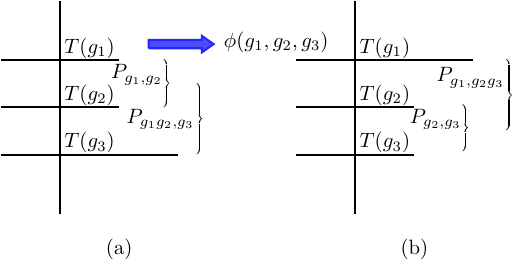}
\end{center}
\caption{Different ways to reduce combination of $T(g_3)$, $T(g_2)$ and $T(g_1)$ into $T(g_1g_2g_3)$. Only the right projection operators are shown. Their combined actions differ by a phase factor $\phi(g_1,g_2,g_3)$.
}
\label{P123}
\end{figure}

Nontrivial phase factors appear when we consider the combination of three MPUO's. See Fig. \ref{P123}. 

There are two different ways to reduce the tensors. We can either first reduce the combination of $T(g_1)$, $T(g_2)$ and then combine $T(g_3)$ or first reduce the combination of $T(g_2)$,$T(g_3)$ and then combine $T(g_1)$. The two different ways should be equivalent. More specifically, they should be the same up to phase on the unique block of $T{g_1,g_2,g_3}$. Denote the projection onto the unique block of $T(g_1,g_2,g_3)$ as $Q_{g_1,g_2,g_3}$. We find that
\be
\begin{array}{l}
Q_{g_1,g_2,g_3}(I_3\otimes P_{g_1,g_2})P_{g_1g_2,g_3}= \\
\phi(g_1,g_2,g_3) Q_{g_1,g_2,g_3}(P_{g_2,g_3}\otimes I_1)P_{g_1,g_2g_3}
\end{array}
\ee
From this we see that the reduction procedure is associative up to a phase factor $\phi(g_1,g_2,g_3)$ which satisfies
\be
\frac{ \phi(g_2,g_3,g_4) \phi(g_1,g_2g_3,g_4)\phi(g_1,g_2,g_3) }
{\phi(g_1g_2,g_3,g_4)\phi(g_1,g_2,g_3g_4)}
=1
\ee
From the definition of cocycles given in section \ref{Gcoh}, we see that $\phi(g_1,g_2,g_3)$ forms a 3-cocycle of group $G$. 

Let's calculate $\phi(g_1,g_2,g_3)$ explicitly for the group generated by $U_{CZX}$.
\be
\begin{array}{ll}
\phi(I,I,I)=1 & \phi(I,I,CZX)=1 \\
\phi(I,CZX,I)=1 & \phi(CZX,I,I)=1 \\
\phi(I,CZX,CZX)=1 & \phi(CZX,CZX,I)=1\\
\phi(CZX,I,CZX)=1 & \phi(CZX,CZX,CZX)=-1
\end{array}
\label{phi_CZX}
\ee
We can check that $\phi$ is indeed a 3-cocycle. The last term shows a nontrivial $-1$. This minus one cannot be removed by redefining the phase of $P_{g_1,g_2}$ in any way. Therefore $\phi$ corresponds to a nontrivial 3-cocycle for the $\mathbb{Z}_2$ group.

What does this nontrivial mathematical structure imply about the physics of the CZX model? In the next section we are going to answer this question by proving that MPUO related to a nontrivial 3-cocycle cannot have a short range entangled symmetric state. That is, the boundary of the CZX model cannot have a gapped symmetric ground state. It either breaks the symmetry or is gapless.

\subsubsection{Nontrivial 3-cocycle of MPUO and nonexistence of SRE symmetric state}
\label{cocycleSRE}

In this section we will show that a symmetry defined by a MPUO on a 1D chain can have a SRE symmetric state only if the MPUO corresponds to a trivial 3-cocycle. Therefore, the boundary of the CZX model must be gapless or have symmetry breaking. For this proof, we will be using the matrix product state representation of SRE states.

Suppose that the symmetry on a 1D chain is represented by tensors $T^{i,i'}_{\alpha,\beta}(g)$. Assume that it has a SRE symmetric state represented by matrices $A^i_{\lambda,\eta}$ which is also single-blocked and in the canonical form. 

\begin{figure}[ht]
\begin{center}
\includegraphics[scale=1.0]{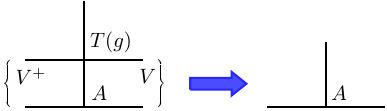}
\end{center}
\caption{Reduction of the combination of $T(g)$ and $A$ into $A$. Here $T^{i,i'}(g)$ is a MPUO, $A^i$ is a matrix product state symmetric under $T^{i,i'}(g)$.
}
\label{PgA}
\end{figure}

Because the state represented by $A^i$ is symmetric under $T^{i,i'}$, the set of matrices obtained by acting $T^{i,i'}$ on $A_i$ can be related to $A_i$ through a gauge transformation.
\be
A^i= V^{\dg} (\sum_{i'} T^{i,i'}(g)A^{i'}) V
\label{TA}
\ee
where $V^{\dg}V=I$ and $V$ is unique on the single block of $\sum_{i'} T^{i,i'}(g)A^{i'}$ up to phase. This is saying that we can reduce the MPS obtained from $\sum_{i'} T^{i,i'}(g)A^{i'}$ back to the original form $A^i$ by applying $V^{\dg}$ and $V$ to the left and right of the matrices respectively. See Fig. \ref{PgA}.

For a fixed representation of the SRE state $A^{i}$ and fixed representation of the MPUO symmetry $T(g)$, $V$ is fixed up to phase. We can pick a particular choice of phase for $V$.

\begin{figure}[ht]
\begin{center}
\includegraphics[scale=1.0]{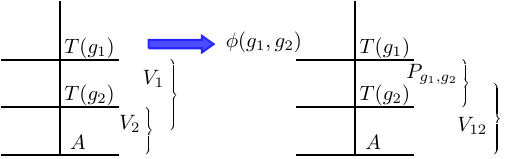}
\end{center}
\caption{Two ways of reducing the combination of $T(g_2)$, $T(g_1)$ and $A$ into $A$. Only the right projection operators are shown. Their combined actions differ by a phase factor $\varphi(g_1,g_2)$.
}
\label{Pg12A}
\end{figure}

Now we consider the combined operation of $T(g_1)$ and $T(g_2)$ on $A$. See Fig.\ref{Pg12A}.

We can either first combine $T(g_2)$ and $A$ and then combine $T(g_1)$ and $A$ or first combine $T(g_1)$ and $T(g_2)$ and then combine $T(g_1g_2)$ and $A$. The right projection operator for these two methods differ by a phase factor $\varphi(g_1,g_2)$. This phase factor can be arbitrarily changed by changing the phase of $P_{g_1,g_2}$. For following discussions, we fix the phase of $P_{g_1,g_2}$ and hence $\varphi(g_1,g_2)$.

\begin{figure}[ht]
\begin{center}
\includegraphics[scale=1.0]{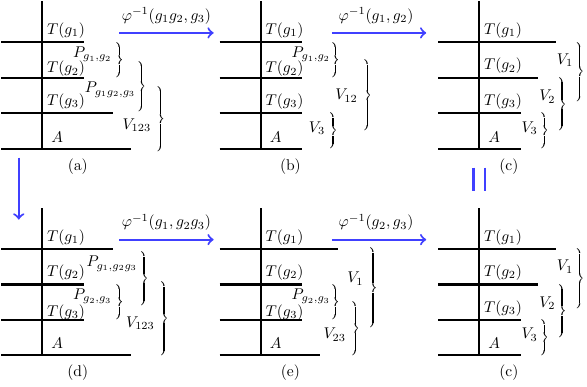}
\end{center}
\caption{Different ways of reducing the combination of $T(g_3)$, $T(g_2)$, $T(g_1)$ and $A$ into $A$. Only the right projection operators are shown. Their combined actions differ by a phase factor written on the arrow. 
}
\label{Pg123A}
\end{figure}

This is all the freedom we can have. If we are to combine three or more $T$'s with $A$, different reduction methods differ by a phase factor but the phase factor are all determined by $\varphi(g_1,g_2)$. Consider the situation in Fig. \ref{Pg123A}, where we are to combine $T(g_3)$, $T(g_2)$ and $T(g_1)$ with $A$. 

To change the reduction procedure in Fig.\ref{Pg123A}(a) to that in Fig.\ref{Pg123A}(c), we can either go through step (b) or steps (d) and (e). If we go through step (b), the phase difference in the right projection operators is
\be
\varphi^{-1}(g_1g_2,g_3)\varphi^{-1}(g_1,g_2)
\ee
On the other hand, if we go through steps (d) and (e), the phase difference in the right projection operators is
\be
\phi(g_1,g_2,g_3)\varphi^{-1}(g_1,g_2g_3)\varphi^{-1}(g_2,g_3)
\ee

But these two procedures should be equivalent as the initial and final configurations are the same whose phases have been fixed previously. Therefore, we find that
\be
\phi(g_1,g_2,g_3)=\frac{\varphi(g_1,g_2g_3)\varphi(g_2,g_3)}{\varphi(g_1g_2,g_3)\varphi(g_1,g_2)}
\ee
For the $\phi(g_1,g_2,g_3)$ given in Eq.\ref{phi_CZX}, we can check explicitly that such a equation cannot be satisfied for any $\varphi$. Therefore we found a contradiction. This shows that the boundary of the CZX model must be either gapless or breaks symmetry. Therefore, 

\begin{svgraybox}
\begin{center}
\textbf{Box 10.17 SPT order of 2D CZX model}

2D CZX model has nontrivial symmetry protected topological order protected by a on-site unitary $\mathbb{Z}_2$ symmetry.

\end{center}
\end{svgraybox}

as we promised in section \ref{CZX}.

\subsubsection{Tensor network representation}

Finally we want to discuss the tensor network representation of the CZX wave function. As we will see, the tensors representing the state have a very simple form and its transformation under the $\mathbb{Z}_2$ symmetry encodes important information about the nontrivial SPT order.

\begin{figure}[ht]
\begin{center}
\includegraphics[scale=1.0]{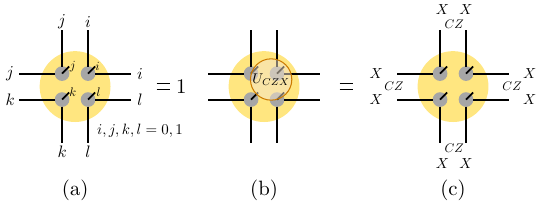}
\end{center}
\caption{The tensor network representation of the 2D CZX state.
}
\label{T_CZX}
\end{figure}

The tensor shown in Fig. \ref{T_CZX} represents the ground state of the CZX model, which is the tensor product of entangled plaquettes in the state $|0000\rangle + |1111\rangle$. The tensor is composed of four parts, as shown in Fig. \ref{T_CZX} (a), each containing one two-dimensional physical index (slanted) and two two-dimensional inner indices (horizontal and vertical). Each part of the tensor is nonzero and equal to $1$ if and only if the physical index and two inner indices take on the same value ($0$ or $1$). Otherwise, the tensor is zero.

It is interesting to see how the tensor transform under the $\mathbb{Z}_2$ symmetry. As shown in Fig. \ref{T_CZX} (b) and (c), applying the $\mathbb{Z}_2$ symmetry to the physical index on each site induces gauge transformation on the inner indices. The gauge transformation is
\be
V=(X\otimes X) CZ
\ee
It looks similar to the transformation of the AKLT tensor. However, one important difference for the gauge transformation in the CZX tensor is that $V$ does not form a representation of the $\mathbb{Z}_2$ symmetry group, not even projectively. In fact,
\be
V^2 = - Z \otimes Z
\ee
On the whole tensor, the action of $V^2$ is trivial, due to the identification of pairs of inner indices connected to the same physical index. Therefore the tensor remains invariant if we apply the $\mathbb{Z}_2$ symmetry twice, as expected. However, the action of $V^2$ is not identify on inner indices in each direction alone. 

More interestingly, we notice a close relation between $V$ and the boundary symmetry action $\t{U}_{CZX}$ given in Eq. \ref{tU}. In general, for any state in the same SPT phase, we expect the same relation to apply. That is, the gauge transformation of the tensor network in a region under the $\mathbb{Z}_2$ symmetry operation corresponds to the effective symmetry action on the boundary degrees of freedom when the system has an edge.

\section{General construction of SPT phases}
\label{sec:nDSPT}

The above discussion regarding CZX model can be generalized to arbitrary symmetry groups and to arbitrary dimensions. In order to do this, let us define group cohomology in more generality.

\subsection{Group cohomology}
\label{Gcoh}

For a group $G$, Let $\om_n(g_1,...,g_n)$ be a function from $n$ group
elements to a $U(1)$ phase factor. Group elements $g\in G$ can act on $\om$. In particular, if $g$ is a unitary operation, the action is trivial
\be
g\cdot \om = \om
\ee
If $g$ is an anti-unitary operation, like time reversal, the action is nontrivial
\be
g\cdot \om = \om^*
\ee

Let $\mathcal{C}^n(G,M)=\{\om_n \}$ be the space of all such functions.  
Note that $\mathcal{C}^n(G,M)$ is an Abelian group under the function multiplication 
\be
\om''_n(g_1,...,g_n)= \om_n(g_1,...,g_n) \om'_n(g_1,...,g_n)
\ee
We define a map $d_n$ from $\mathcal{C}^n[G,U(1)]$ to $\mathcal{C}^{n+1}[G,U(1)]$:
\be
\begin{array}{ll}
(d_n \om_n) (g_1,...,g_{n+1})= & g_1\cdot \om_n (g_2,...,g_{n+1})\om_n^{(-1)^{n+1}} (g_1,...,g_{n}) \\ 
 & \times \prod_{i=1}^n\om_n^{(-1)^i} (g_1,...,g_{i-1},g_ig_{i+1},g_{i+2},...g_{n+1})
\end{array}
\label{dn}
\ee
Let
\begin{align}
 \mathcal{B}^n(G,M)=\{ \om_n| \om_n=d_{n-1} \om_{n-1}|  \om_{n-1} \in \mathcal{C}^{n-1}(G,M) \}
\end{align}
and
\begin{align}
 \mathcal{Z}^n(G,M)=\{ \om_{n}|d_n \om_n=1,  \om_{n} \in \mathcal{C}^{n}(G,M) \}
\end{align}
$\mathcal{B}^n(G,M)$ and $\mathcal{Z}^n(G,M)$ are also Abelian groups
which satisfy $\mathcal{B}^n(G,M) \subset \mathcal{Z}^n(G,M)$ where
$\mathcal{B}^1(G,M)\equiv \{ 1\}$.
The $n$-cocycle of $G$ is defined as
\begin{align}
 \mathcal{H}^n(G,M)= \mathcal{Z}^n(G,M) /\mathcal{B}^n(G,M) 
\end{align}

Let us discuss some examples. 
When $n=1$, Eq.\ref{dn} reads
\begin{align}
 (d_1 \om_1)(g_1,g_2)= \om_1^{s_1}(g_2)\om_1(g_1)/\om_1(g_1g_2)
\end{align}
where $s_1=1$ if $g_1$ is unitary and $s_1=-1$ if $g_1$ is anti-unitary.
We see that
\begin{align}
 \mathcal{Z}^1(G,U(1))=\{  \om_1| \om_1^{s_1}(g_2)\om_1(g_1)=\om_1(g_1g_2) \} .
\end{align}
In other words, $\mathcal{Z}^1(G,U(1))$ is the set formed by all the 1D representations
of $G$.  Since $\mathcal{B}^1(G,U(1))\equiv \{ 1\}$ is trival.
$\mathcal{H}^1(G,U(1))=\mathcal{Z}^1(G,U(1))$ is also the set of all the 1D representations of
$G$.

When $n=2$, Eq.\ref{dn} reads
\be
(d_2 \om_2)(g_1,g_2,g_3)= \om_2^{s_1}(g_2,g_3) \om_2(g_1,g_2g_3)/\om_2(g_1g_2,g_3)\om_2(g_1,g_2)
\ee
we see that
\be
\mathcal{Z}^2(G,U(1))=\{  \om_2| \om_2^{s_1}(g_2,g_3) \om_2(g_1,g_2g_3)=\om_2(g_1g_2,g_3)\om_2(g_1,g_2)\} 
\ee
and
\be
\mathcal{B}^2(G,U(1))=\{ \om_2|\om_2(g_1,g_2)=\om_1^{s_1}(g_2)\om_1(g_1)/\om_1(g_1g_2)\}
\ee
The 2-cocycle
$\mathcal{H}^2(G,U(1))=\mathcal{Z}^2(G,U(1))/\mathcal{B}^2(G,U(1))$ classify the
projective representations discussed in section \ref{prorep}.

When $n=2$, Eq.\ref{dn} reads
\be
(d_3 \om_3)(g_1,g_2,g_3,g_4)= \frac{ \om_3{s_1}(g_2,g_3,g_4) \om_3(g_1,g_2g_3,g_4)\om_3(g_1,g_2,g_3) }
{\om_3(g_1g_2,g_3,g_4)\om_3(g_1,g_2,g_3g_4)}
\ee
we see that
\be
\mathcal{Z}^3(G,U(1))=\{  \om_3|\frac{ \om_3^{s_1}(g_2,g_3,g_4) \om_3(g_1,g_2g_3,g_4)\om_3(g_1,g_2,g_3) }
{\om_3(g_1g_2,g_3,g_4)\om_3(g_1,g_2,g_3g_4)}=1\}
\ee
and
\be
\mathcal{B}^3(G,U(1))=\{ \om_3| \om_3(g_1,g_2,g_3)=\frac{
\om_2^{s_1}(g_2,g_3) \om_2(g_1,g_2g_3)}{\om_2(g_1g_2,g_3)\om_2(g_1,g_2)}\}
\label{3coboundary}
\ee
which give us the 3-cocycle
$\mathcal{H}^3(G,U(1))=\mathcal{Z}^3(G,U(1))/\mathcal{B}^3(G,U(1))$.

\subsection{SPT model from group cohomology}

Now we can discuss the general construction of SPT phases in $n$ dimension with $\mathcal{H}^{n+1}$ group cocycle.

When $n=0$, we have a quantum mechanical system with symmetry $G$. We cannot talk about quantum phases in $0$ dimension, but the system can have a unique symmetric ground state $|\psi\>$. It transforms under symmetry operator $U_g$ as
\be
U_g|\psi\> = \om_1(g)|\psi\>
\ee
where $\om_1(g)$ is a 1D representation of $G$. Therefore, a symmetric $0$ dimensional quantum state is labeled by $\om_1(g) \in \mathcal{H}^1(G,U(1))$. When moving between states labeled by different $\om_1$, there must a level crossing -- the $0$ dimensional analogue of phase transition.

When $n=1$, as we discussed in section \ref{sec:1DSPT}, different SPT phases are labeled by projective representations with inequivalent factor system $\om_2(g_1,g_2) \in \mathcal{H}^2(G,U(1))$. Given a $\om_2$, we can construct a state in the corresponding SPT phase as follows.

\begin{figure}[ht]
\begin{center}
\includegraphics{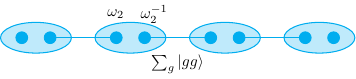}
\end{center}
\caption{Model 1D state with SPT order corresponding to 2-cocycle $\om_2 \in \mathcal{H}^2(G,U(1))$.
}
\label{om2SPT}
\end{figure}

Every lattice site (big oval) contains two spins (small circle), each with basis state $|g\>$, $g\in G$. Symmetry operation on the left and right spin is given by
\be
U^l_h|g\> = \om_2(g^{-1}h^{-1},h)|hg\>, \ U^r_h|g\> = \om^{-1}_2(g^{-1}h^{-1},h)|hg\>
\ee
Each pair of connected spins are in the maximally entangled state
\be
|\psi\> = \sum_{g} |gg\>
\ee
It is straight-forward to check that $U^l_g$ and $U^r_g$ form projective representations with factor systems $\om$ and $\om^{-1}$ respectively. Therefore, each lattice site contains a linear representation of the symmetry and the total wave function as a product of $|\psi\> = \sum_{g}|gg\>$ is invariant under the global symmetry. When the system has a boundary, the edge state carries projective representations of the symmetry with factor systems $\om$ and $\om^{-1}$, as expected for an SPT phase.

Similarly, when $n=2$, the construction of the CZX model can be generalized to arbitrary symmetry groups. Given a 3-cocycle $\om_3 in \mathcal{H}^3(G,U(1))$, we can construct a state in the corresponding SPT phase as follows.

\begin{figure}[ht]
\begin{center}
\includegraphics[scale=1.2]{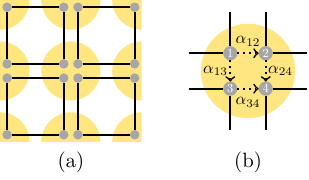}
\end{center}
\caption{Model 2D state with SPT order corresponding to 3-cocycle $\om_3 \in \mathcal{H}^3(G,U(1))$.
}
\label{om3SPT}
\end{figure}

Every lattice site (big oval) contains four spins (small circle), each with basis state $|g\>$, $g\in G$. Symmetry operation on the four spins of each site is given by
\be
U_h |g_1,g_2,g_3,g_4\>= \frac{\om_{3}(g_1^{-1}g_2,g_2^{-1}h^{-1},h)\om_{3}(g_2^{-1}g_4,g_4^{-1}h^{-1},h)}{\om_{3}(g_1^{-1}g_3,g_3^{-1}h^{-1},h)\om_{3}(g_3^{-1}g_4,g_4^{-1}h^{-1},h)} |hg_1,hg_2,hg_3,hg_4\>
\ee
Note that this symmetry operation on four spins does not decompose into a tensor product of operators on the four spins individually.

Each four spins connected in a square are in the maximally entangled state
\be
|\psi\> = \sum_{g} |gggg\>
\ee
It is straight-forward to check using the property of $\om_3$ that 1. each lattice site contains a linear representation of the symmetry; 2. the total wave function as a product of $|\psi\> = \sum_{g}|gggg\>$ is invariant under the global symmetry; 3. when the system has a boundary, the symmetry transformation on the boundary is described by a MPUO whose local transformation as given in Fig.\ref{P123} is related to $\om_3{g_1,g_2,g_3}$. Reference for the proof of these facts can be found in summary and further reading. 

The proof outline in section\ref{cocycleSRE} regarding the nonexistence of SRE edge states with nontrivial 3-cocycle applies to the general case. Therefore, the above construction gives a trivial / nontrivial SPT phase if we started from a trivial / nontrivial 3-cocycle.

Note that the CZX model is not written in this 'canonical' form, but gives rise to the same SPT phase and the same edge physics as the MPUO on the boundary transform with the same $\om_3 \in \mathcal{H}^3(\mathbb{Z}_2,U(1))$.

\begin{svgraybox}
\begin{center}
\textbf{Box 10.18 General construction of SPT models from group cocycle}

Using different group cocycles from $\mathcal{H}^{n+1}$,  $n$-dimensional boson / spin models with symmetry protected topological orders of internal (on-site unitary or time reversal) symmetry can be constructed. 

\end{center}
\end{svgraybox}


\section{Summary and further reading}

In this chapter, we study symmetry protected topological phases in strongly interacting boson / spin systems. In one dimension, a complete classification can be obtained. In particular, for SPT order with internal symmetry, it was found that the edge of a nontrivial SPT phase is always degenerate, carrying a nontrivial projective representation of the symmetry. As 1D fermion system can be mapped to 1D boson system through Jordan Wigner transformation, we get the classification for 1D fermion system as a bonus. Generalizing our understanding of 1D SPT to higher dimension, a systematic construction of SPT phases is presented where symmetry action is related to group cocycles. In particular, we prove that, in the 2D SPT phases we constructed, their edge state is always gapless unless the symmetry is broken, establishing the nontrivial SPT order in the model.

The first and most well understood 1D SPT phase is the spin $1$ chain with anti-ferromagnetic Heisenberg interaction. It was first proposed by Haldane that, unlike spin $1/2$ Heisenberg chains, the spin $1$ chain is gapped\cite{H8364}. Moreover, it was found to have
degenerate edge spin-1/2 states\cite{HKA9081,GGL9114,Ng9455}
and non-trivial string order parameter\cite{NR8909,KT9204}, indicating its nontrivial order. These properties of the `Haldane phase' were established rigorously by the exactly solvable AKLT point in the phase\cite{AKL8799}, whose ground state has a simple projected entangled pair structure.

Following the example of the AKLT state, the idea of symmetry protected topological order was generalized to other symmetries\cite{GW0931,PBT1225}. It was realized that the SPT order is closely related to the entanglement structure of the system and projective representations of the edge state\cite{PTB1039,GW0931}, based on which order parameters for detecting SPT order were found in \cite{PT1241,HPC1202}. The classification discussed in this chapter follows the work in \cite{CGW1107} and was also derived in \cite{SPC1139}.

A full classification of 1D bosonic phases with symmetry, including both the symmetry breaking and the symmetric phases, can be found in \cite{CGW1128, SPC1139}. It was observed that 1D gapped spin phases with on-site symmetry of group $G$ are basically labeled by (1) the unbroken symmetry subgroup $G'$, (2) the projective representations of $G'$.
Also the gaplessness of 1D translational invariant spin $1/2$ chains was proved in \cite{LSM6107} and was generalized to higher dimensions in \cite{Hastings0431}.

For 1D fermion systems, the existence of the so-called `Majorana chain' with Majorana edge modes at the end of the chain was proposed by Kitaev \cite{K0131}. The classification of 1D fermionic topological phases (with various symmetries) was obtained in \cite{FK1009,FK1103,TPB1102} and is consistent with the result obtained in this chapter using Jordan Wigner transformation.

The matrix product operator, used in section \ref{sec:2DSPT}, was introduced in \cite{MCP1012}.

The 2D AKLT model was first introduced in \cite{AKL8877}. In section \ref{sec:2DAKLT} we used a slightly different version of this model. In \cite{AKL8877}, all the spins on a single lattice site are projected onto their symmetric subspace. For example, on a square lattice model, the four spin $1/2$'s are projected onto the total spin 2 subspace. In the version in section \ref{sec:2DAKLT}, no projection is done.The model with and without this projection are supposed to be in the same phase and have the same SPT order.

The CZX model was introduced in \cite{CLW1141} and the general construction of SPT phases using group cocycle was discussed in \cite{CGL1204,CGL1314}. 

We focused mostly on interacting boson / spin systems in our discuss. On the other hand, SPT phases have been extensively studied in fermion systems. In particular, topological insulators in 2D and 3D free fermion systems have not only been theoretically predicted\cite{KM0502,BHZ0657,FKM0703,MB0706,Roy0922}, but also experimentally realized\cite{Konig2007,Hsieh2008,Hsieh2009,Chen2009}. Moreover, SPT phases in free fermion systems have been completely classified\cite{SRF0825,Kitaev2009}. However, a complete understanding of SPT phases in interacting fermion systems is much harder. For recent progress see for example \cite{Gu2014a,KTT14arXiv}.


%
%
\bibliographystyle{plain}
\bibliography{Chap10}

%
%
%

\begin{partbacktext}
\part{Outlook}

\end{partbacktext}

%
%
%
\chapter{A Unification of Information and Matter}
\label{chap11} 

\abstract{
In this book, our discussions on many-body quantum systems have been
concentrated on gapped topological states.  After the introduction of the
concept of long-range quantum entanglement and the discovery of related
mathematical theories (such as tensor category theory), a systematic
understanding of all gapped states in any dimensions is emerging, which include
topological orders, and SPT orders.  However, our understanding of highly
entangled gapless states is very limited.  We do not even know where to start,
to gain a systematic understanding of highly entangled gapless states.  This
will be the next big challenge in condensed matter physics.  In this chapter,
we will study some examples of  highly entangled gapless states.  We will show
that long-range entangled qubits can provide a unified origin of light and
electrons (or more generally, gauge interactions and Fermi statistics): light
waves (gauge fields) are fluctuations of long-range entanglement, and electrons
(fermions) are defects of long-range entanglement.  Since gauge bosons and
fermions represent almost all the elementary particles, the above results
suggest that the space formed by long-range entangled qubits may be an origin
of all matter.  In other word, (quantum) information unifies matter.  This
happen to be the central theme of this book: a theory of quantum information
and quantum matter.  }

\section{Four revolutions in physics}

We have a strong desire to understand everything from a single or very few
origins. Driven by such a desire, physics theories were developed through the
cycle of discoveries, unification, more discoveries, bigger unification.  Here,
we would like review the development of physics and its four
revolutions\footnote{Here we do not discuss the revolution for thermodynamical
and statistical physics.}. We will see that the history of physics can be
summarized into three stages: 1) all matter is formed by particles; 2) the
discovery of wave-like matter; 3) particle-like matter = wave-like matter.  It
appears that we are now entering into the fourth stage: matter and space =
information (qubits), where qubits emerge as the origin of everything.

\subsection{Mechanical revolution}

\begin{figure}[tb]
\centerline{
\includegraphics[height=2in]{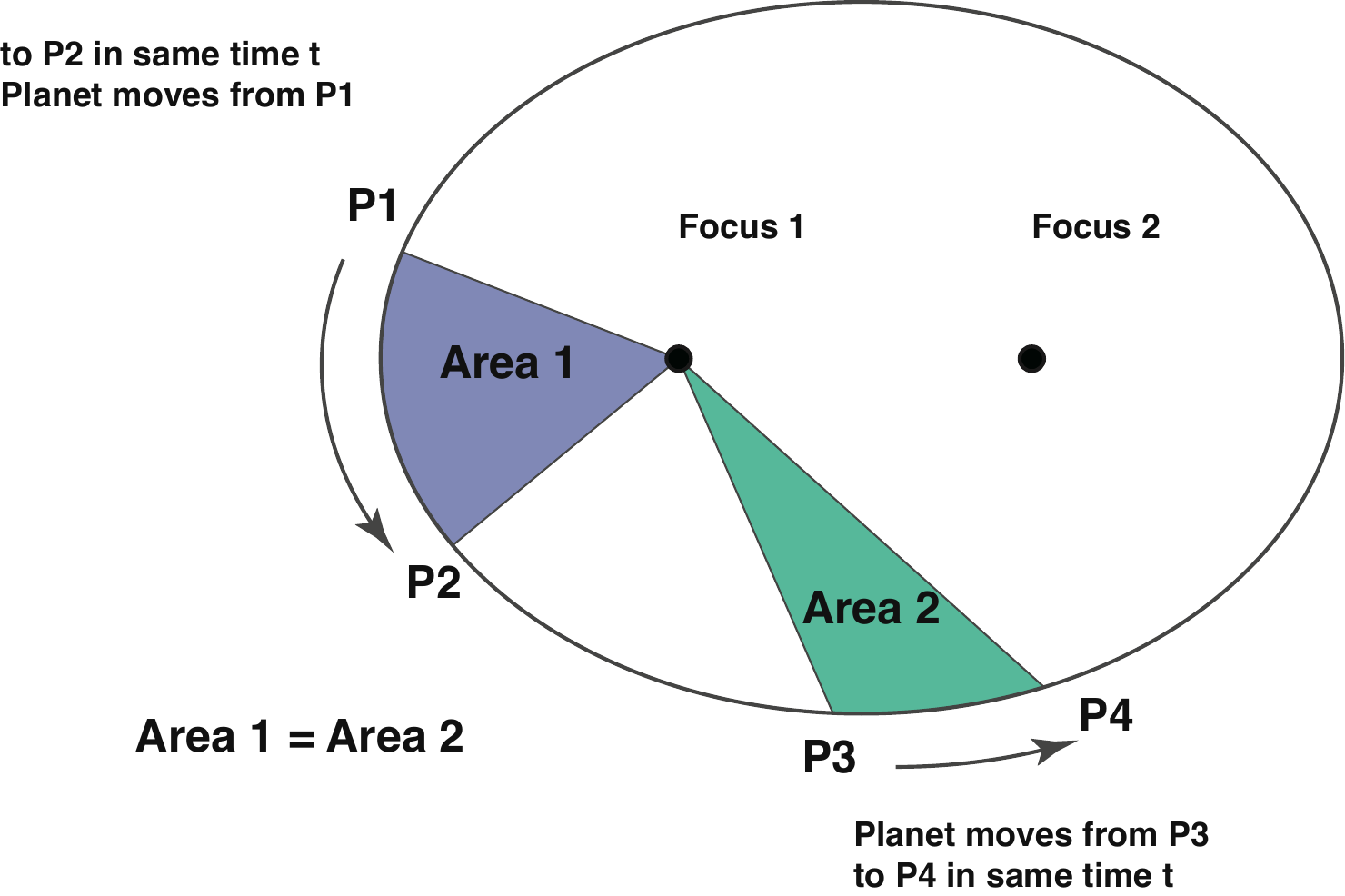} }
\caption{
Kepler's Laws of Planetary Motion: 1) The orbit of a planet is an ellipse with
the Sun at one of the two foci.  2) A line segment joining a planet and the Sun
sweeps out equal areas during equal intervals of time. 3) The square of the
orbital period of a planet is proportional to the cube of the semi-major axis
of its orbit.
}
\label{Kepler2Diagram}
\end{figure}

\begin{figure}[tb]
\centerline{
\includegraphics[height=1.0in]{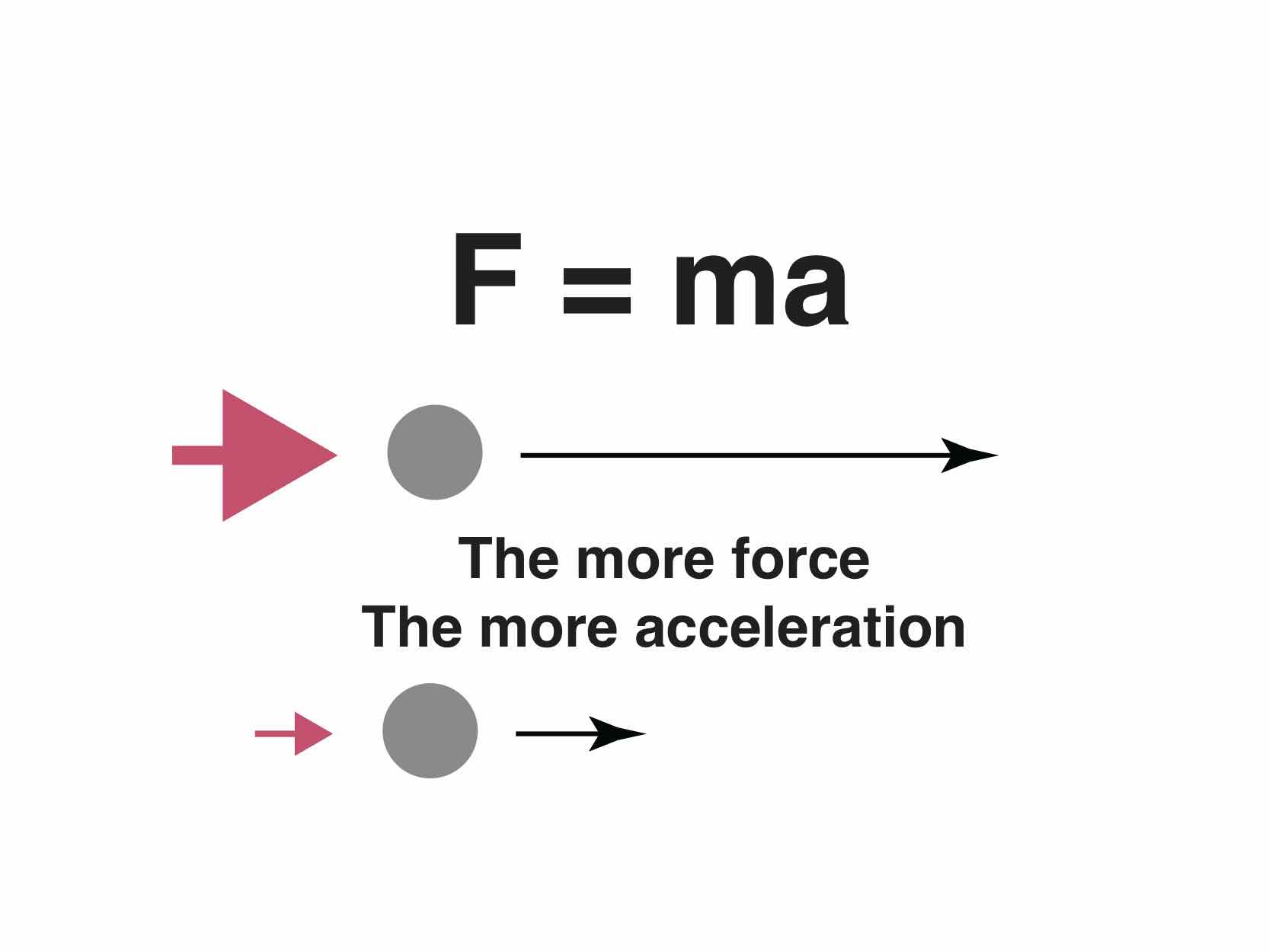}  ~~~~~~~~~
\includegraphics[height=1.0in]{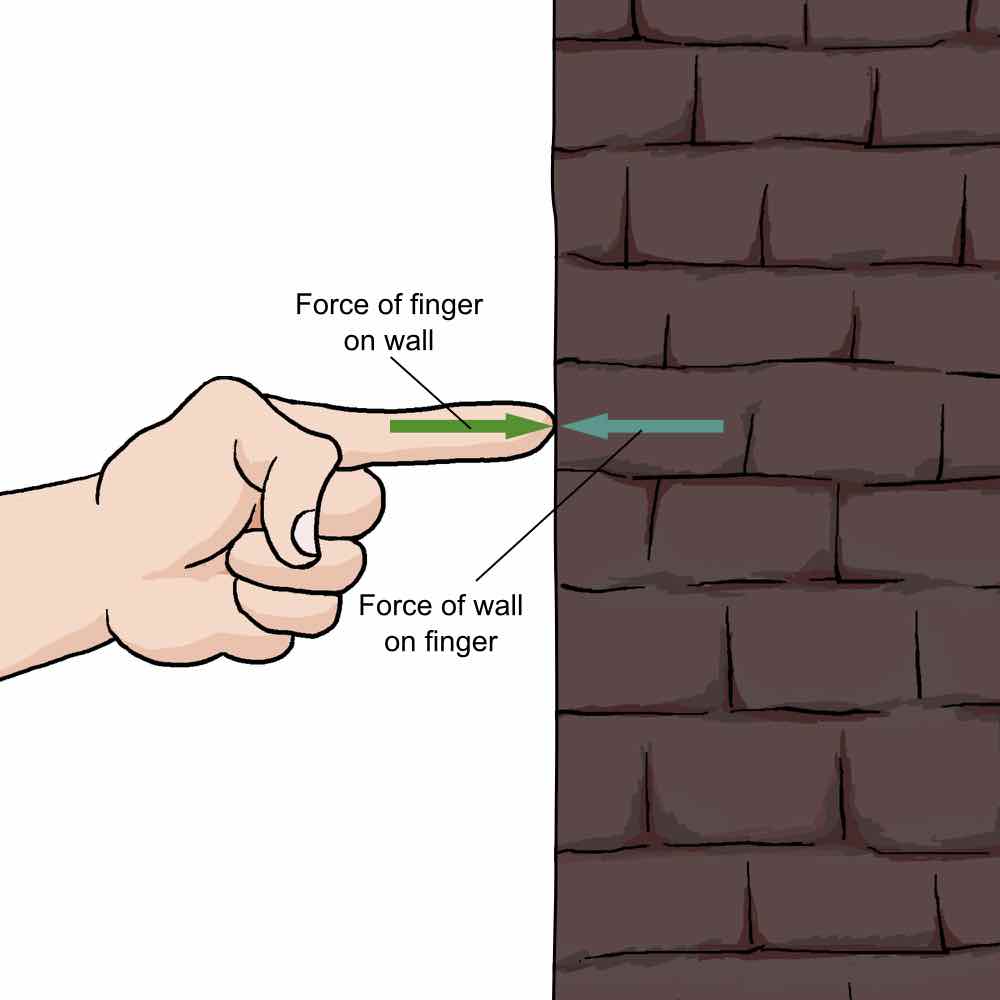}  ~~~~~~~~~
\includegraphics[height=1.0in]{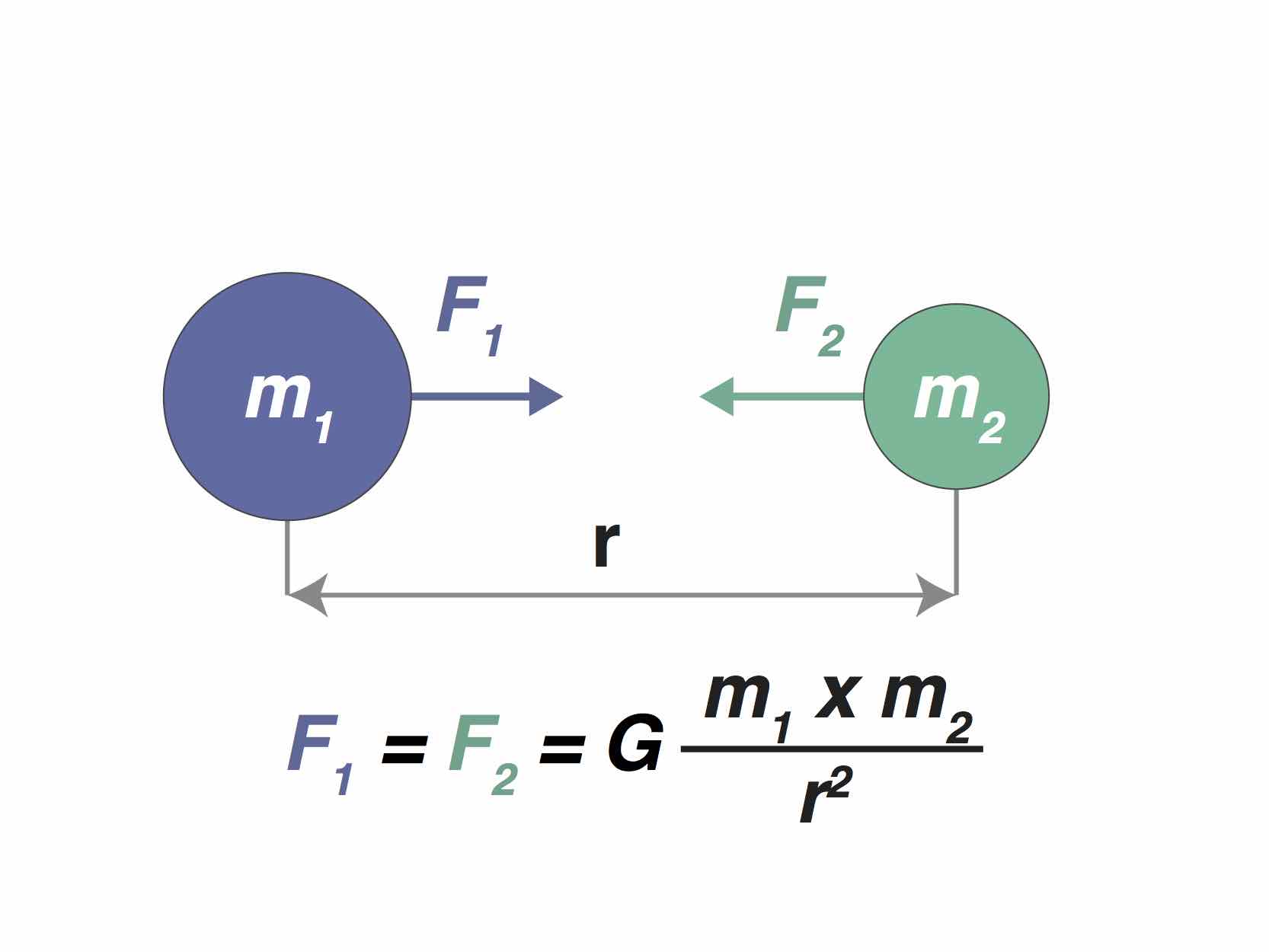}
}
\centerline{
	(a) ~~~~~~~~~~~~~~~~~~~~~~~~~~~~~~~~~~~~~~~~~~~~~~
	(b) ~~~~~~~~~~~~~~~~~~~~~~~~~~~~~~~~~~~~~~~~~~~~~~
	(c)
}
\caption{
	Newton laws: (a) the more force the more acceleration, no force
	no  acceleration. (b) action force = reaction force.
	(c) Newton's universal gravitation:
$F=G\frac{m_1m_2}{r^2}$, where $G=6.674\times 10^{-11}
\frac{\text{m}^3}{\text{kg\ s}^2}$.
}
\label{newtonlaw}
\end{figure}

\begin{figure}[tb]
\centerline{ \includegraphics[height=2.3in]{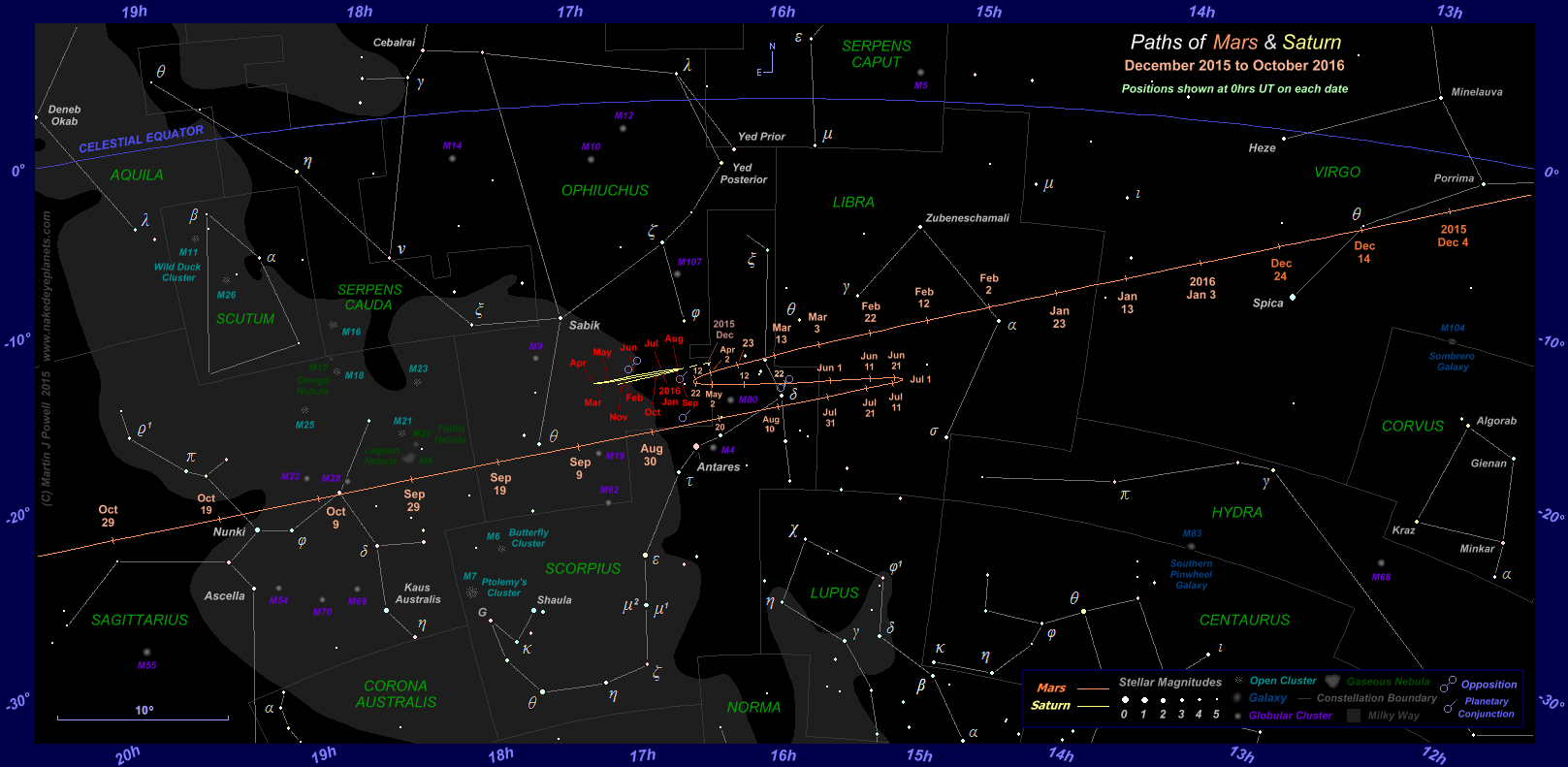}}

\centerline{\includegraphics[height=2.0in]{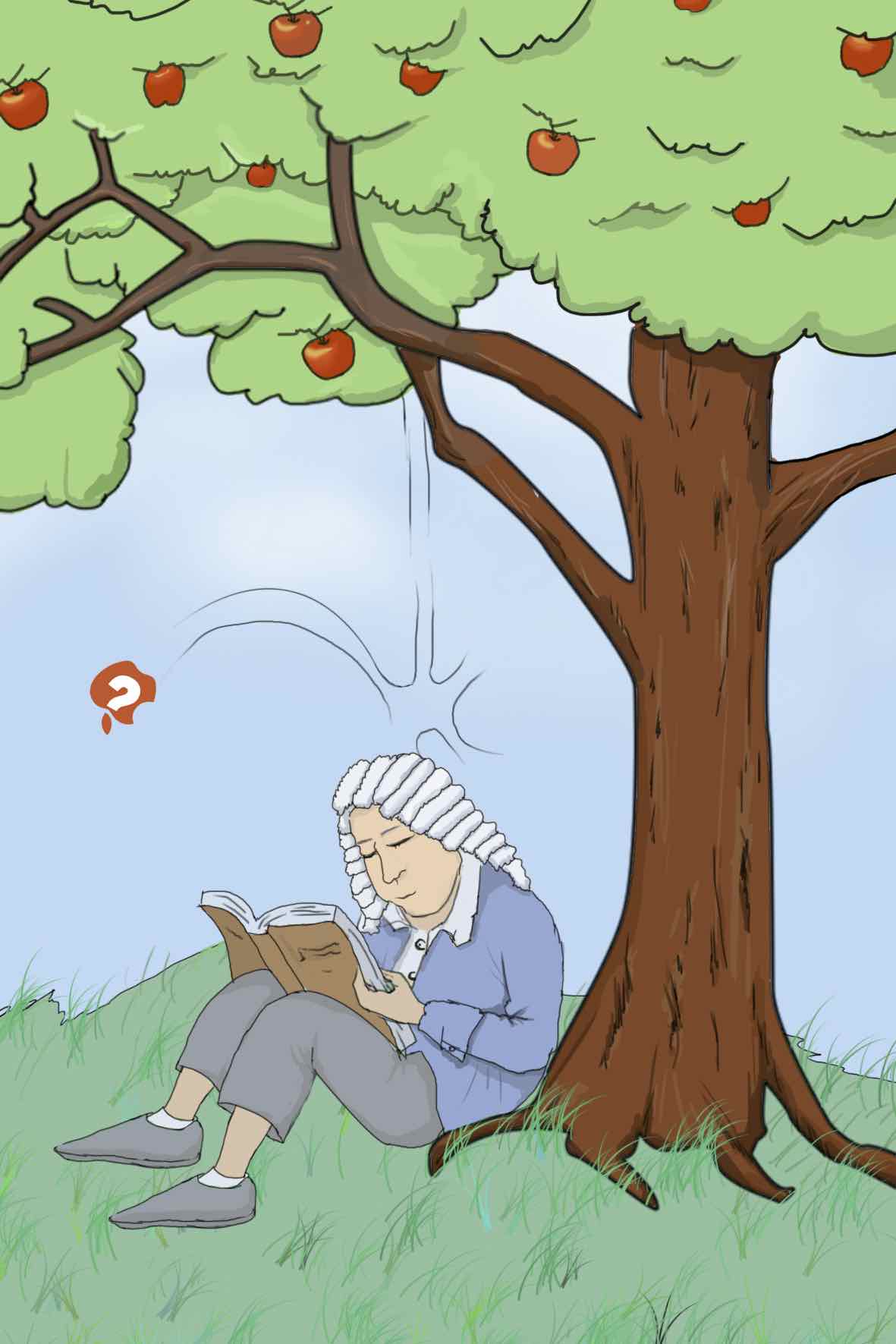}}
\caption{
The perceived trajectories of planets (Mar and Saturn) in the sky.
The falling of apple on earth and the motion of planet
in the sky are unified by Newton theory.
}
\label{mars-path-dec-2015-oct-2016}
\end{figure}

Although the down pull by the earth was realized even before human
civilization, such a phenomena did not arose any curiosity.  On the other hand
the planet motion in the sky has arose a lot of curiosity and led to many
imaginary fantasies.  However, only after Kepler found that planets move in a
certain particular way described by a mathematical formula (see Fig.
\ref{Kepler2Diagram}), people started to wonder: Why are planets so rational?
Why do they move in such a peculiar and precise way.  This motivated Newton to
develop his theory of gravity and his laws of mechanical motion (see Fig.
\ref{newtonlaw}).
Newton's theory not only explains the planets motion, it also explains the
down-pull that we feel on earth.  The planets motion in the sky and the apple
falling on earth look very different (see Fig.
\ref{mars-path-dec-2015-oct-2016}), however, Newton's theory unifies the two
seemingly unrelated phenomena.  This is the first revolution in physics -- the
mechanical revolution.

\begin{svgraybox}
\begin{center}
\textbf{Box 11.1 Mechanical revolution}

All matter are formed by particles, which obey Newton's laws.
Interactions are instantaneous over distance.
\end{center}
\end{svgraybox}

After Newton we view all matter as formed by particles, and use Newton's laws
for particles to understand the motion of all matter.  The success and the
completeness of Newton's theory gave us a sense that we understood everything.

\subsection{Electromagnetic revolution}

\begin{figure}[tb]
\centerline{
	\includegraphics[height=1.2in]{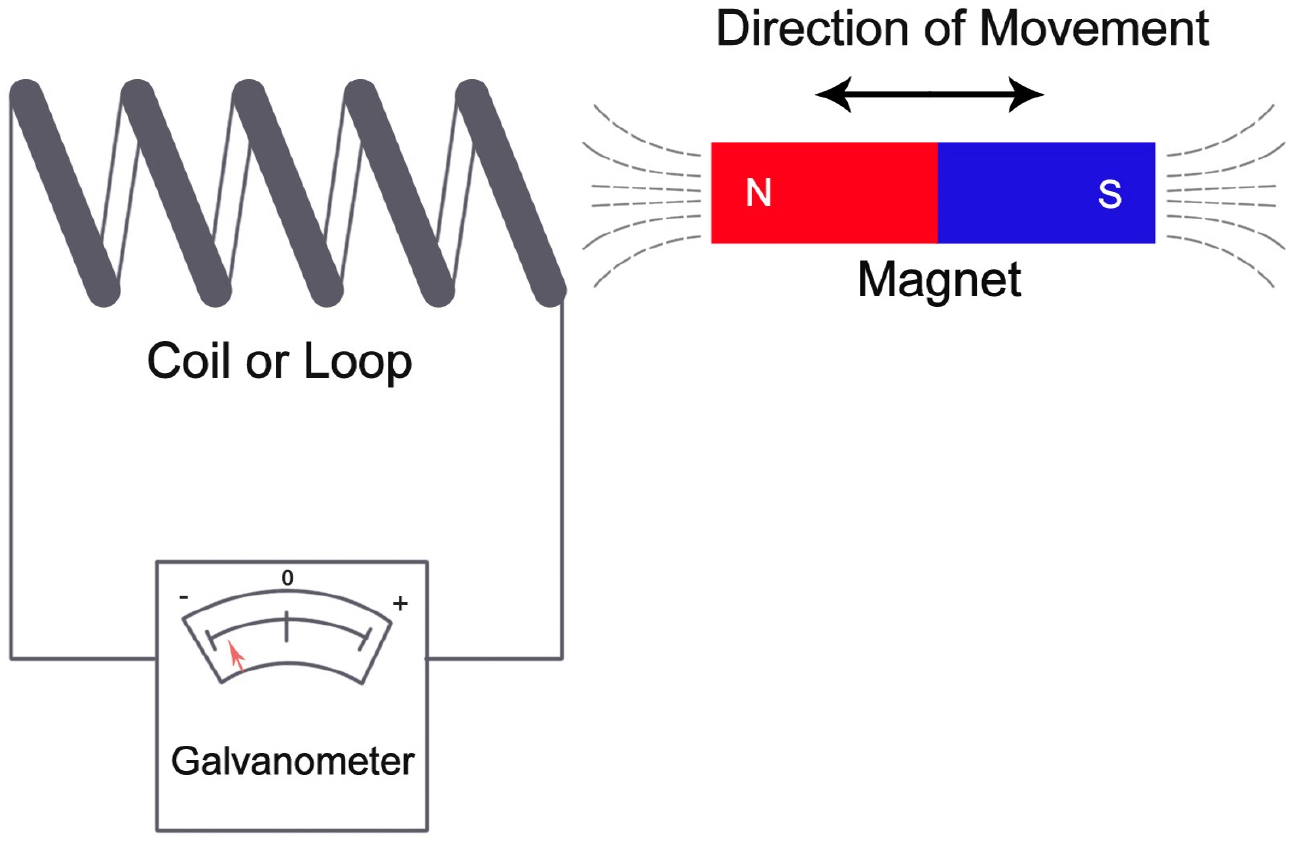}
	\includegraphics[height=1.2in]{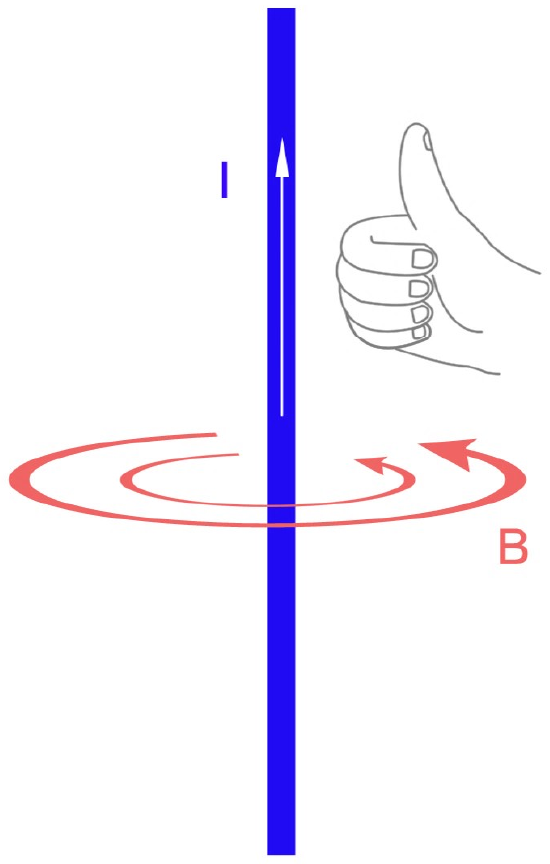} 
	\includegraphics[height=1.2in]{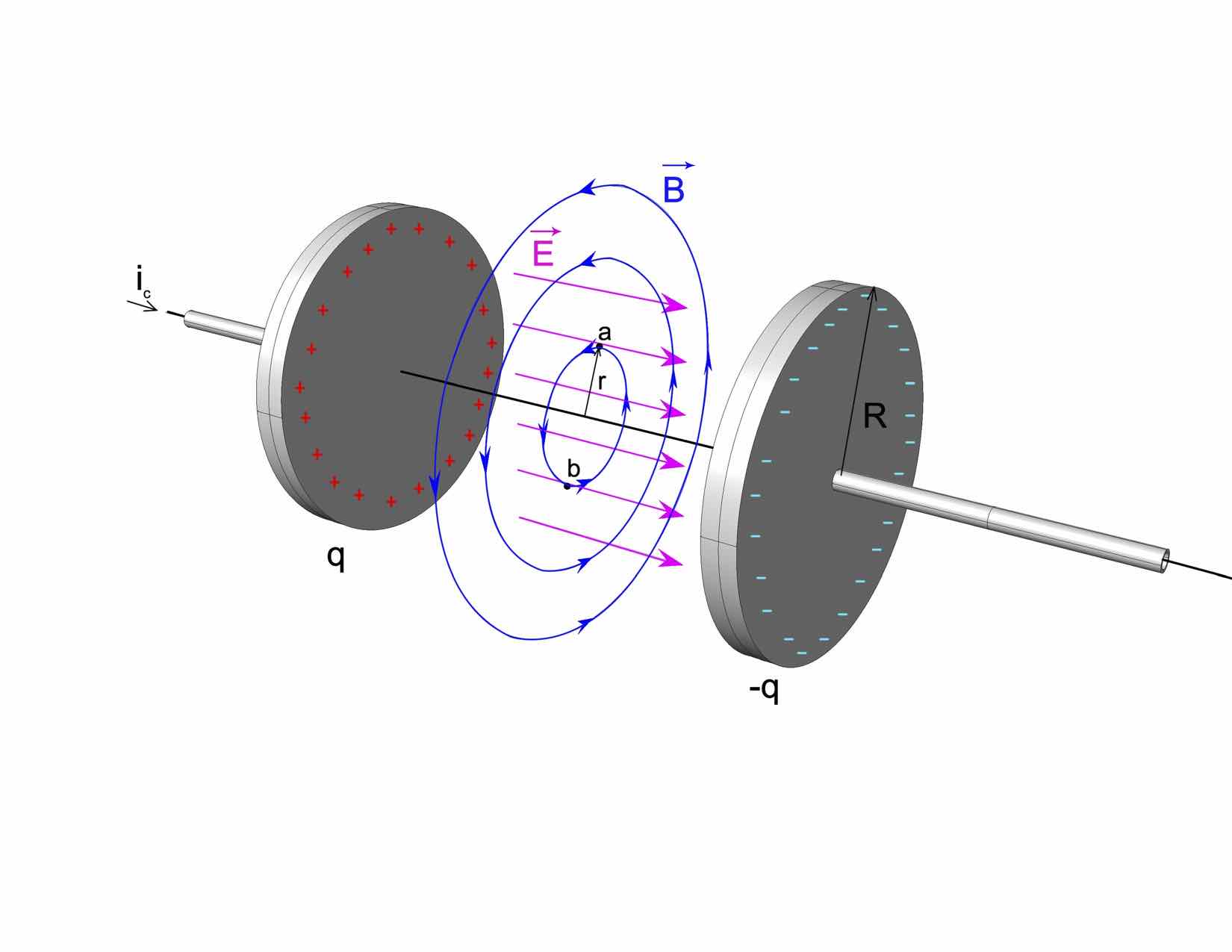}
}
\centerline{
	(a) ~~~~~~~~~~~~~~~~~~~~~~~~~~~~~~~~~~~~~~~~~~~~~~
	(b) ~~~~~~~~~~~~~~~~~~~~~~~~~~~~~~~~~~~~~~~~~~~~~~
	(c)
}
\caption{
	(a) Changing magnetic field can generate an electric field around it, that drives
	an electric current in a coil. (b) Electric current $I$ in a wire can generate a
	magnetic field $B$ around it. (c) A changing electric field $E$ (just like
electric current)  can generate a magnetic field $B$ around it.
}
\label{FAlaw}
\end{figure}

\begin{figure}[tb]
\centerline{
	\includegraphics[height=1.2in]{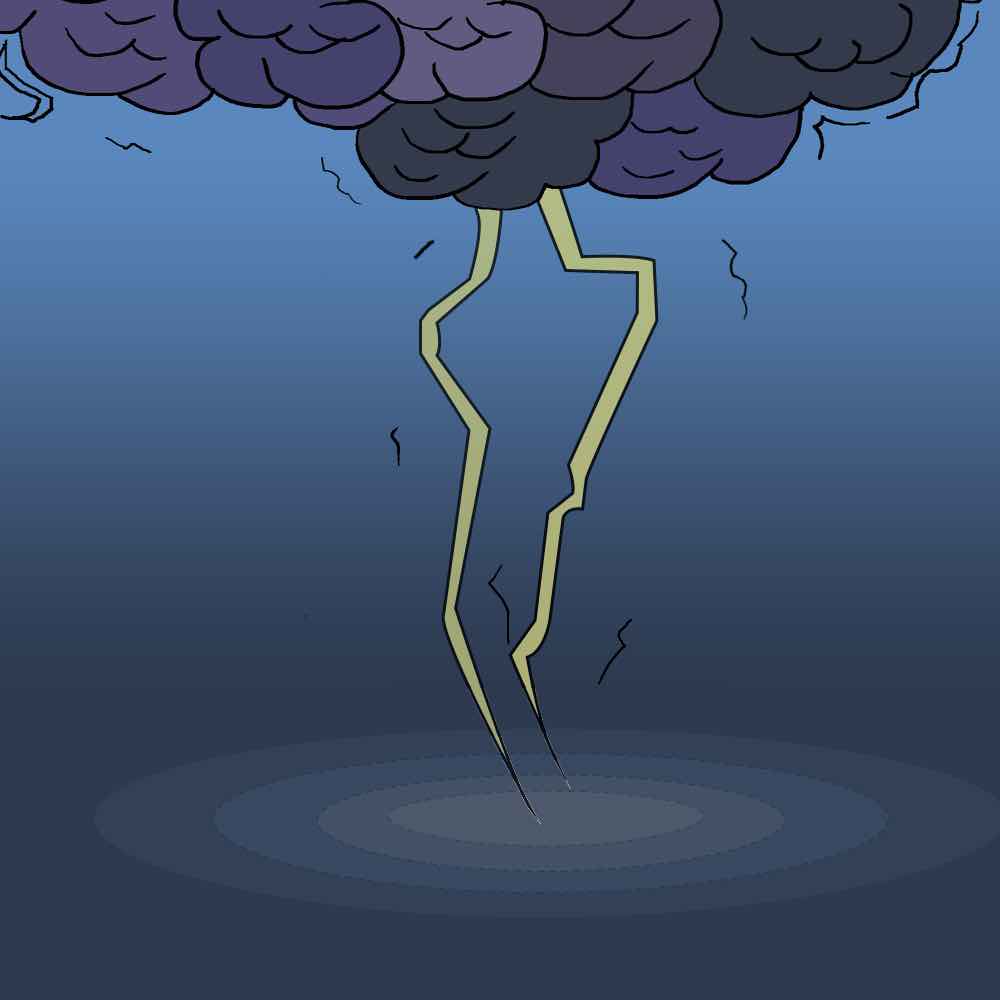}  ~
	\includegraphics[height=1.2in]{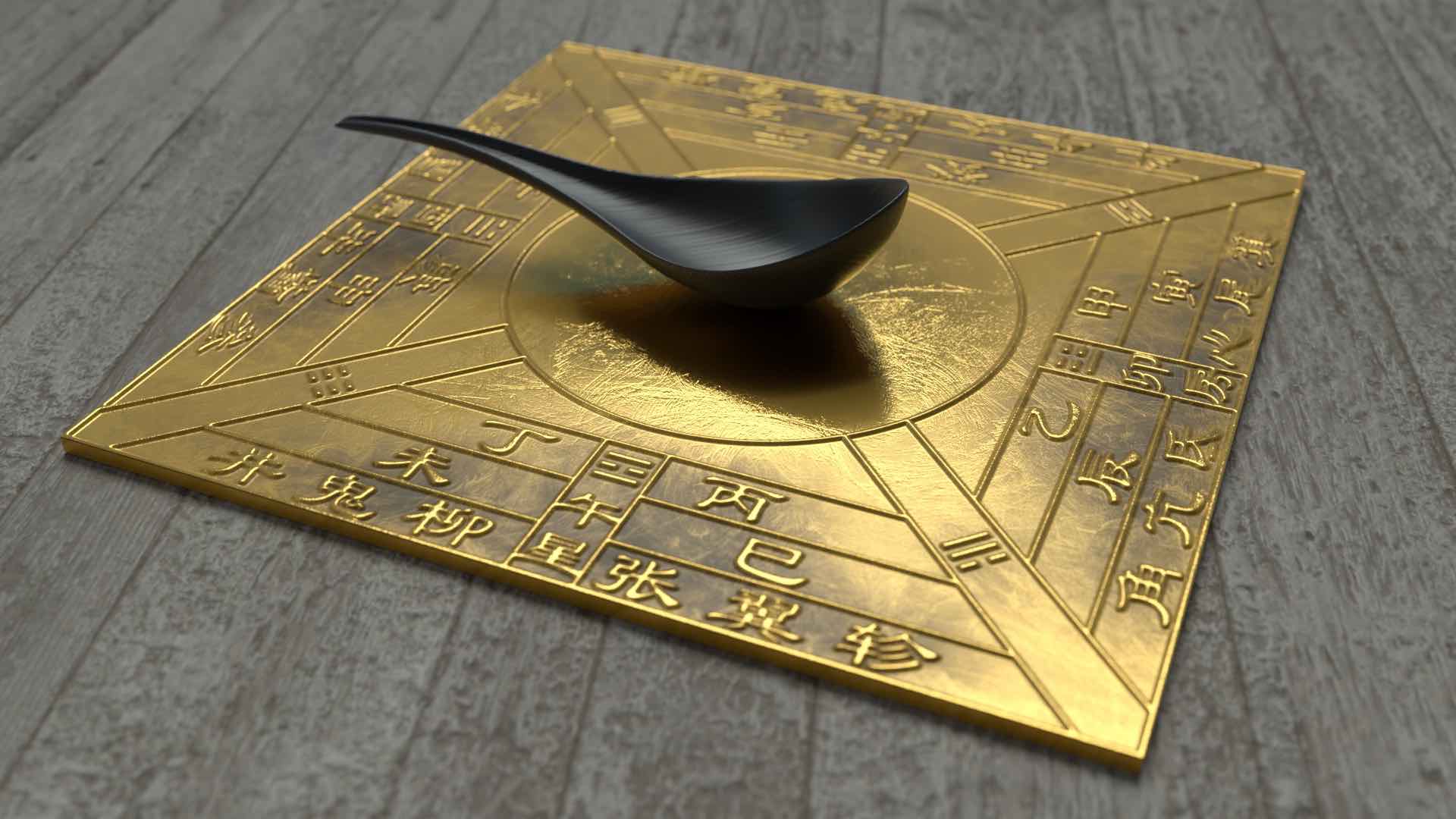}  ~
	}
	
\centerline{		
	\includegraphics[height=1.2in]{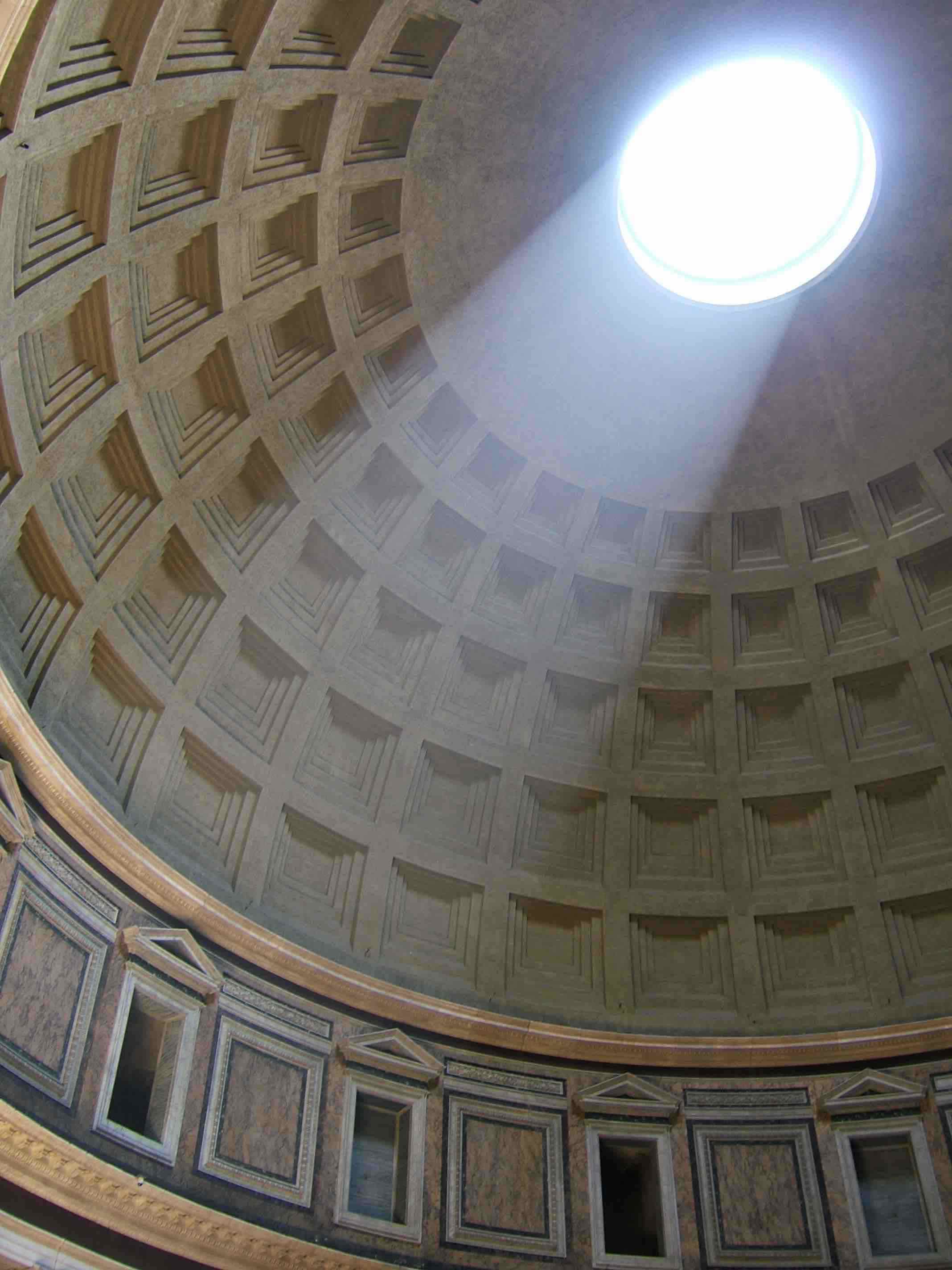} ~~~~~~~
	\includegraphics[height=1.2in]{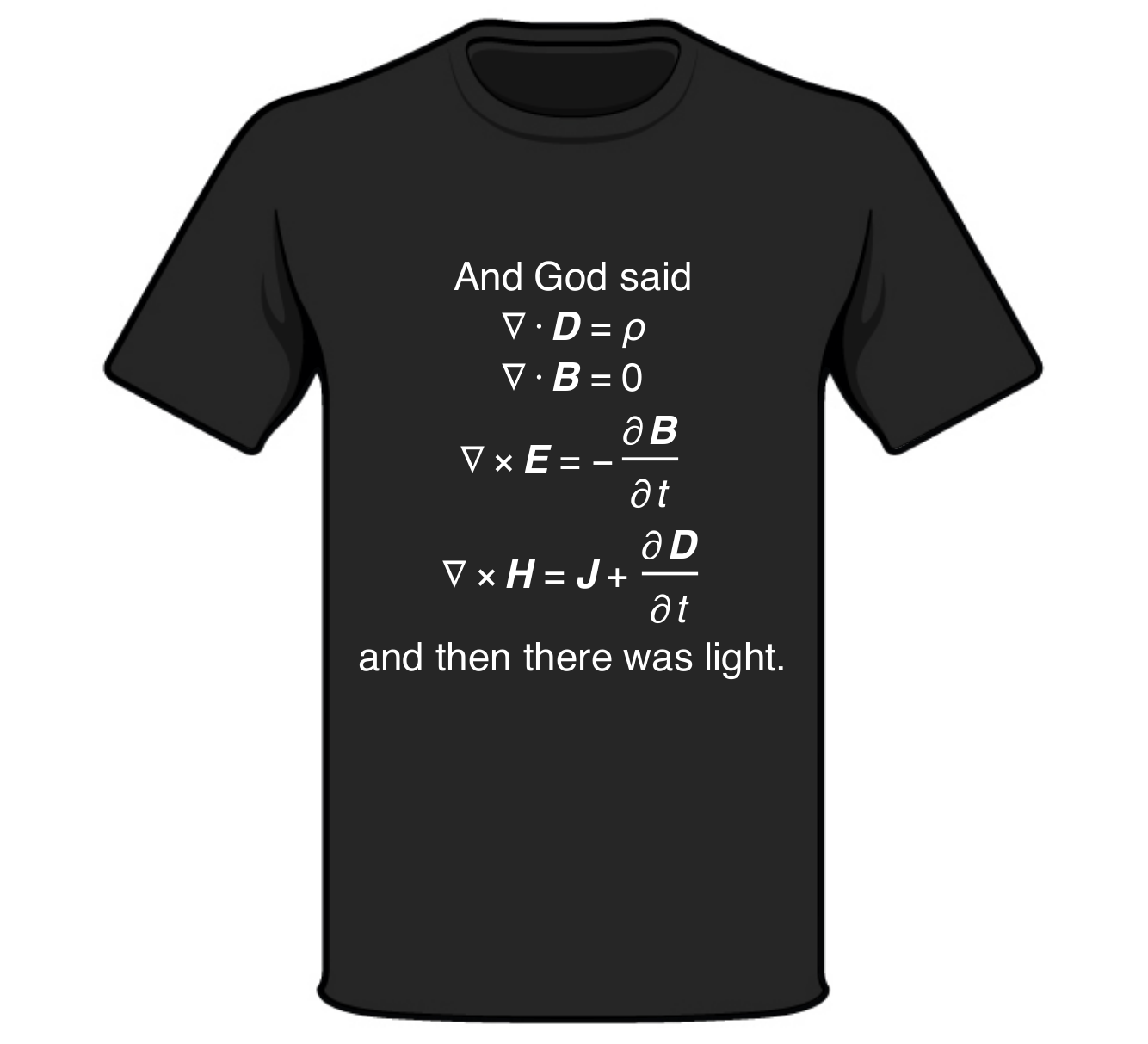}
}
\caption{
Three very different phenomena, electricity, magnetism, and light,
are unified by Maxwell theory.
}
\label{eml}
\end{figure}

But, later we discovered that two other seemingly unrelated phenomena,
electricity and magnetism, can generate each other (see Fig. \ref{FAlaw}).  Our
curiosity about the electricity and magnetism leads to another giant leap in
science, which is summarized by Maxwell equations. Maxwell theory unifies
electricity and magnetism and reveals that light is merely an electromagnetic
wave (see Fig.  \ref{eml}).  We gain a much deeper understanding of light,
which is so familiar and yet so unexpectedly rich and complex in its internal
structure.  This can be viewed as the second revolution -- electromagnetic
revolution.

\begin{svgraybox}
\begin{center}
\textbf{Box 11.2 Electromagnetic revolution}

The discovery of a new form of matter -- wave-like matter: electromagnetic
waves, which obey Maxwell equation.  Wave-like matter causes interaction.
\end{center}
\end{svgraybox}

\begin{figure}[tb]
\centerline{
	\includegraphics[height=1.1in]{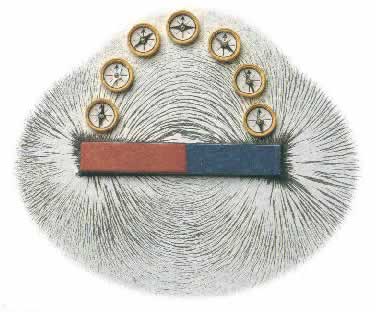}  ~~~~~~~~
	\includegraphics[height=1.1in]{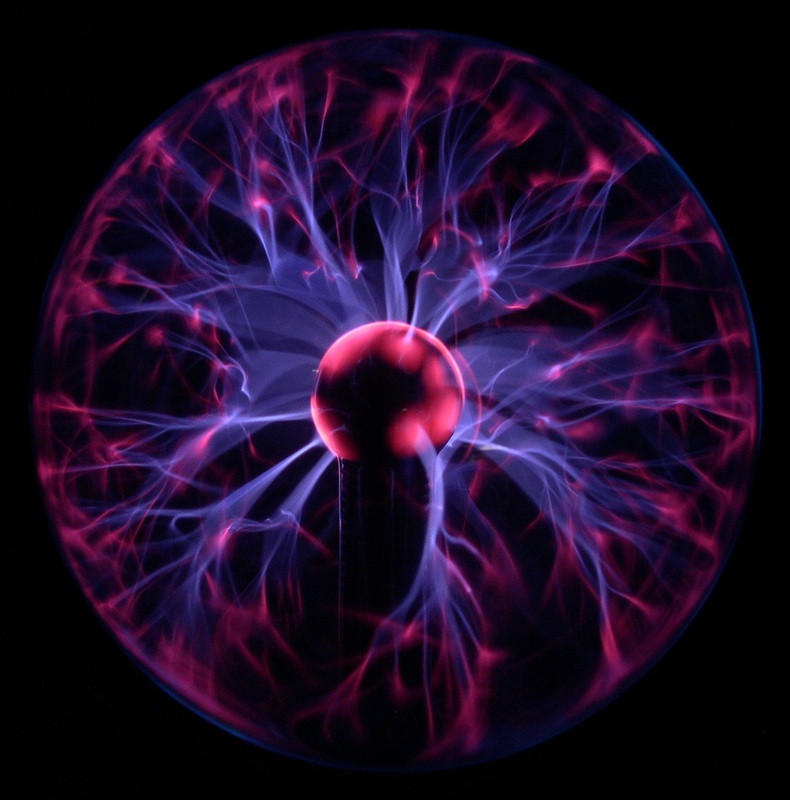}  ~~~~~~~~
	\includegraphics[height=1.3in]{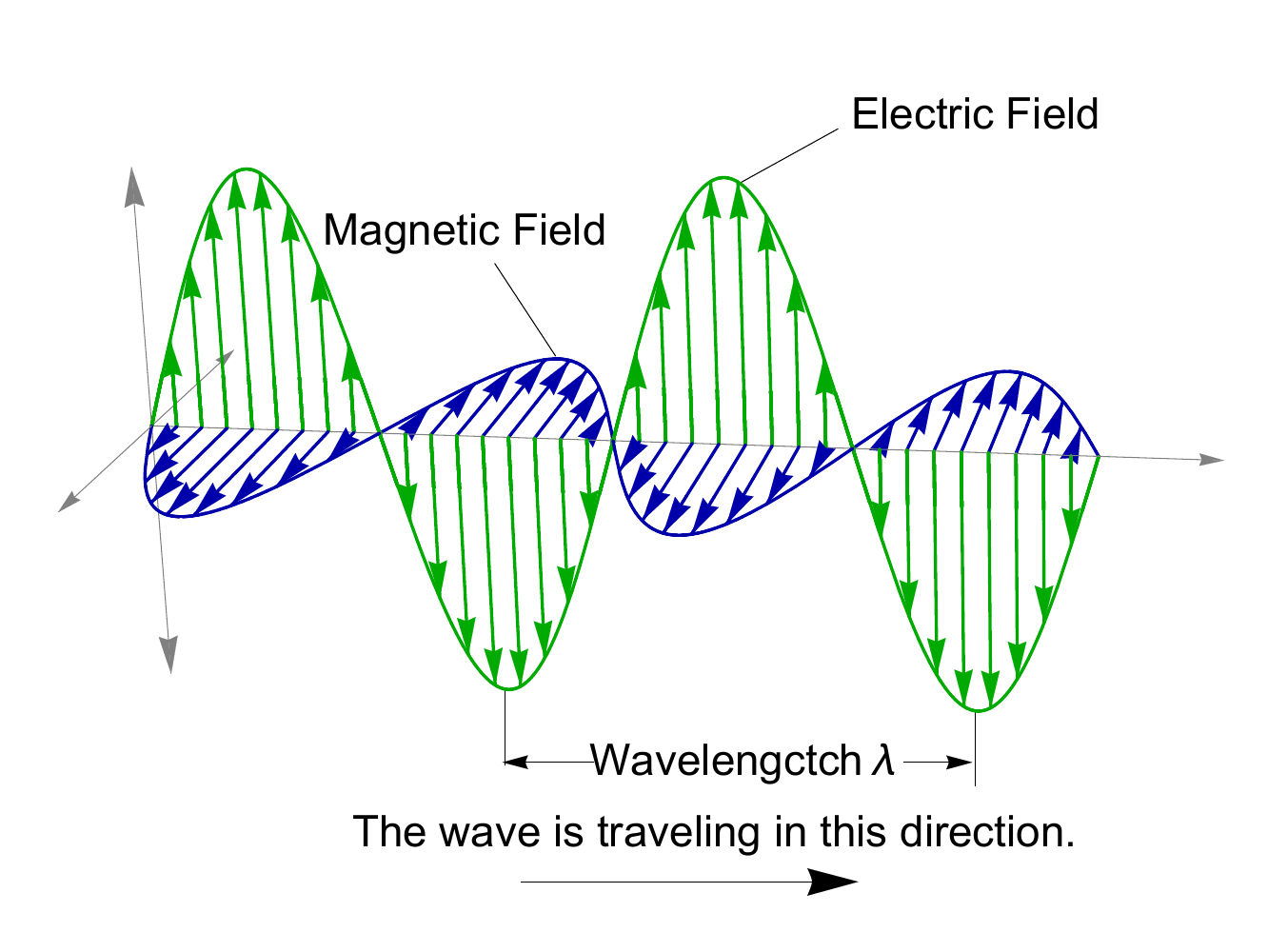}
}
\centerline{
	(a) ~~~~~~~~~~~~~~~~~~~~~~~~~~~~~~~~~~~~~~~~~~
	(b) ~~~~~~~~~~~~~~~~~~~~~~~~~~~~~~~~~~~~~~~~~~~~~~~
	(c) ~~~
}
\caption{
	(a) Magnetic field revealed by iron powder.
	(b) Electric field revealed by glowing plasma.
	(c) They form a new kind of matter: light -- a wave-like matter
}
\label{EMfield}
\end{figure}

However, the true essence of Maxwell theory is the discovery of a new form of
matter -- wave-like (or field-like) matter (see Fig.  \ref{EMfield}), the
electromagnetic wave. The motion of this wave-like matter is governed by
Maxwell equation, which is very different from the particle-like matter
governed by Newton equation $F=ma$.  Thus, the sense that Newton theory
describes everything is incorrect. Newton theory does not apply to wave-like
matter, which requires a new theory -- Maxwell theory.

Unlike the
particle-like matter, the new wave-like matter is closely related to a kind of
interaction -- electromagnetic interaction. In fact, the electromagnetic
interaction can be viewed as an effect of the newly discovered wave-like
matter.

\subsection{Relativity revolution}

After realizing the connection between the interaction and wave-like matter,
one naturally ask: does gravitational interaction also corresponds to a
wave-like matter? The answer is yes.

\begin{figure}[b]
\centerline{
	\includegraphics[height=1.0in]{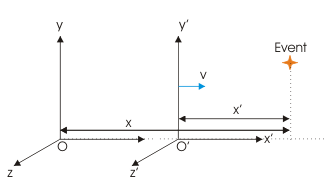}  ~~~~~~~~~~
	\includegraphics[height=0.8in]{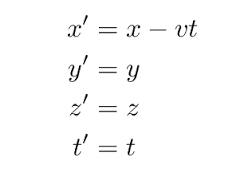}  ~~~~~~~~~~
	\includegraphics[height=1.0in]{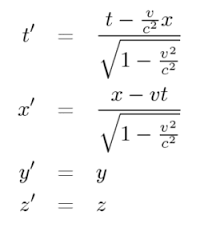}
}
\centerline{
	(a)~~~~~~~~~~~~~~~~~~~~~~~~~~~~~~~~~~~~~~~~~~~~~~~~~~~~~~~~~~~~~~~
	(b) ~~~~~~~~~~~~~~~~~~~~~~~~~~~~~~
	(c)
}
\caption{
	(a) A rest frame and a moving frame with velocity $v$.  An event is
recorded with coordinates $(x,y,z,t)$ in the rest frame and
with $(x',y',z',t')$ in the  moving frame.  There are two opinions on how
$(x,y,z,t)$ and $(x',y',z',t')$ are related: (b) Galilean transformation or (c)
Lorantz transformation where $c$ is the speed of light. In our world, the Lorantz transformation is correct.
}
\label{frame}
\end{figure}

First, people realized that Newton equation and Maxwell equation have different
symmetries under the transformations between two frames moving against each
other.  In other words,  Newton equation $F=ma$ is invariant under Galileo
transformation, while Maxwell equation is invariant under Lorentz
transformation (see Fig. \ref{frame}).  Certainly, only one of the above two
transformation is correct. If one believes that physical law should be the same
in different frames, then the above observation implies that Newton equation
and Maxwell equation are incompatible, and one of them must be wrong.  If
Galileo transformation is correct, then the Maxwell theory is wrong and needs
to be modified.  If Lorentz transformation is correct, then the Newton theory
is wrong and needs to be modified.  Michelson-Morley experiment showed that the
speed of light is the same in all the frames, which implied the Galileo
transformation to be wrong.  So Einstein choose to believe in Maxwell equation.
He modified Newton equation and developed the theory of special relativity.
Thus, Newton theory is not only incomplete, it is also incorrect.

\begin{figure}[tb]
\centerline{
	\includegraphics[height=1.4in]{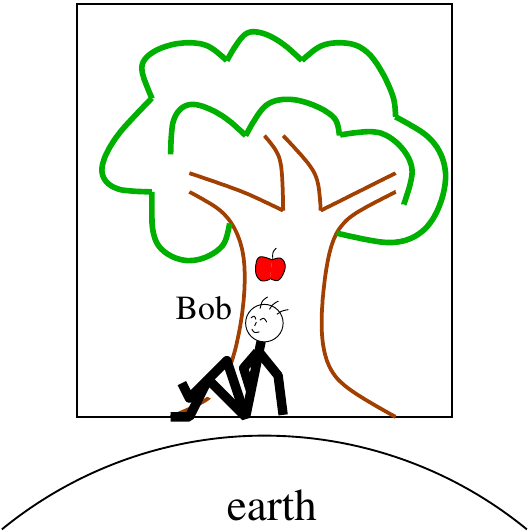}  ~~~
	\includegraphics[height=1.4in]{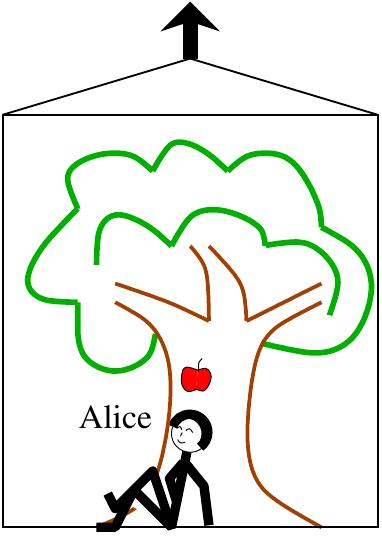}  ~~~~~~~~~~~
	\includegraphics[height=1.2in]{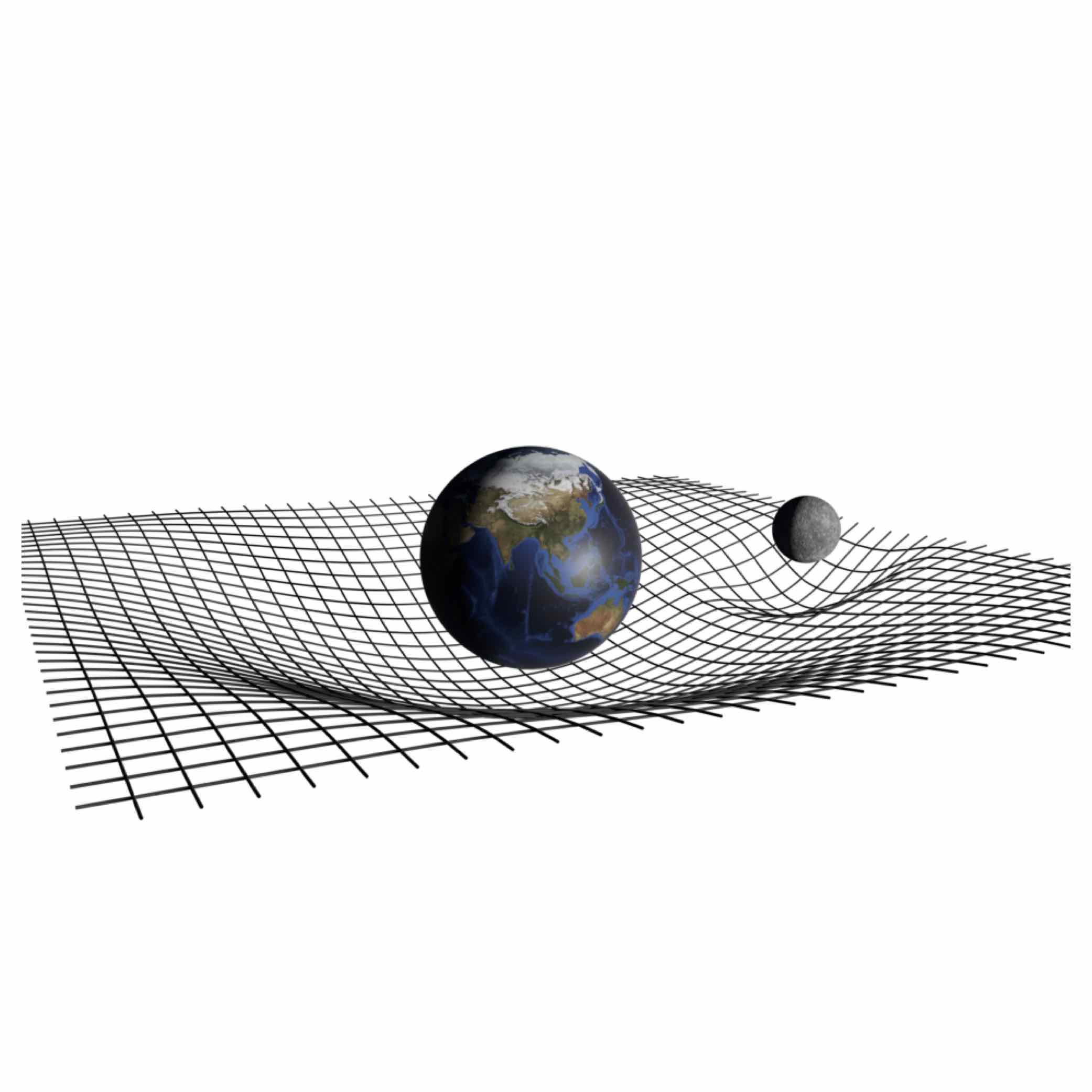}
}
\caption{
The equivalence of the gravitational force of the earth and the force
experienced in an accelerating elevator, leads to an geometric way to
understand gravity: gravity = distortion in space. In other words, the ``gravitational force'' in an
accelerating elevator is related to a geometric feature:
the transformation between the
coordinates in a still elevator and in the accelerating elevator.
}
\label{grav}
\end{figure}

\begin{figure}[tb]
\centerline{
	\includegraphics[height=2.0in]{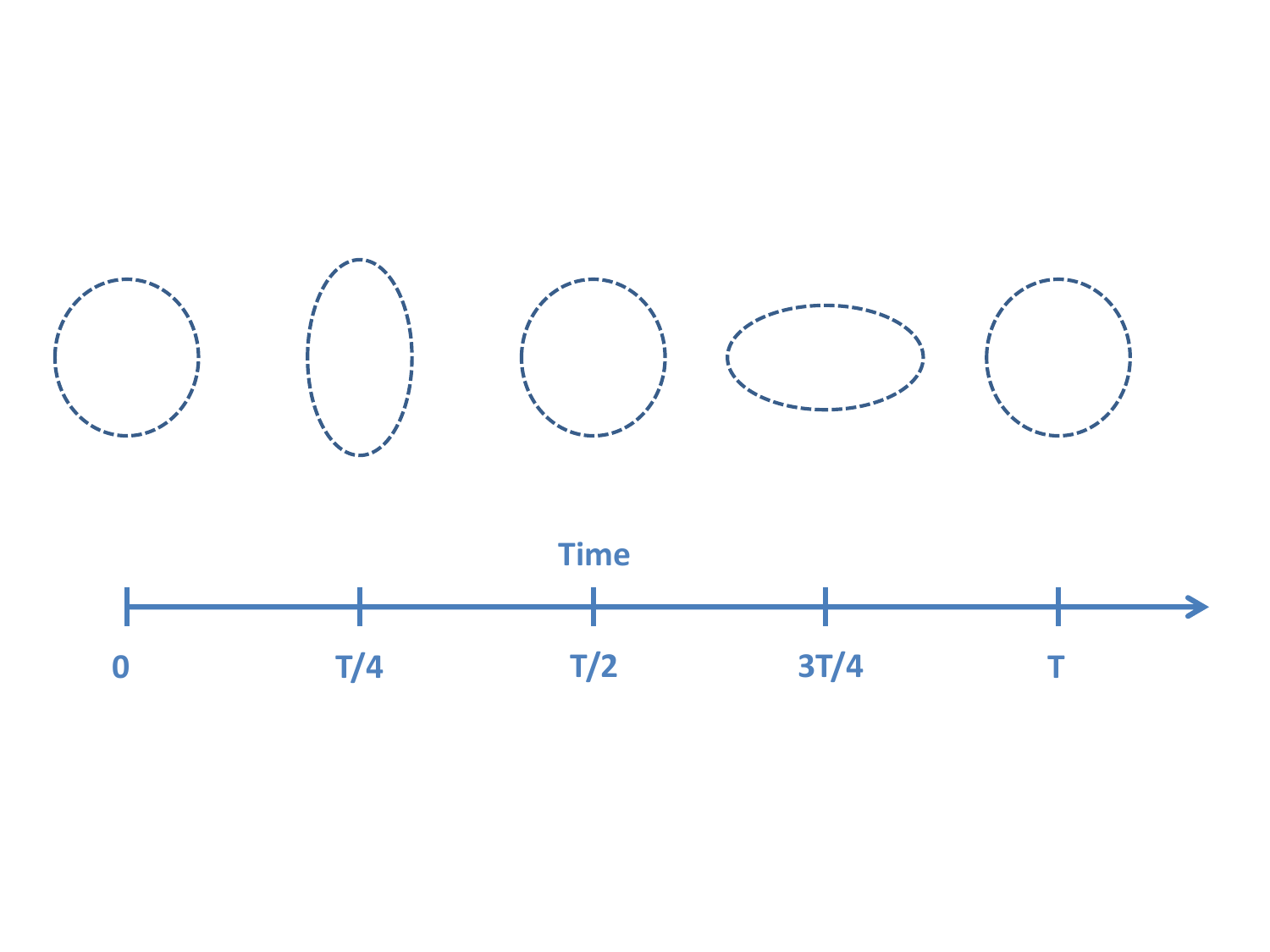}
}
\caption{
Gravitational wave is a propagating distortion of space:
a circle is distorted by a gravitational wave.
}
\label{gravwave}
\end{figure}

Einstein has gone further.  Motivated the equivalence of gravitational force
and the force experienced in an accelerating frame (see Fig. \ref{grav}),
Einstein also developed the theory of general relativity.\cite{E1669}
Einstein theory unifies several seeming unrelated concepts, such as space and
time, as well as interaction and geometry.  Since the gravity is viewed as a
distortion of space and since the distortion can propagate, Einstein discovered
the second wave-like matter -- gravitational wave (see Fig.  \ref{gravwave}).
This is another revolution in physics -- relativity
revolution.

\begin{svgraybox}
\begin{center}
\textbf{Box 11.3 Relativity revolution}

A unification of space and time.
A unification of gravity and space-time distortion.
\end{center}
\end{svgraybox}

\begin{figure}[tb]
\centerline{
	\includegraphics[height=1.5in]{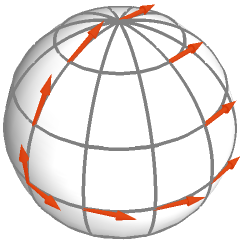}
}
\caption{
A curved space can be viewed as a distortion of
local directions of the space: parallel moving a local direction (represented by an arrow) around a loop in a curved space,
the direction of the arrow does not come back. Such a twist in  local direction
corresponds to a curvature in space.
}
\label{ptrans}
\end{figure}

Motivated by the connection between interaction and geometry in gravity, people
went back to reexamine the electromagnetic interaction, and found that the
electromagnetic interaction is also connected to  geometry.  Einstein's general
relativity views gravity as a distortion of space, which can be viewed as a
distortion of local directions of space (see Fig. \ref{ptrans}).  Motivated by
such a picture, in 1918, Weyl proposed that the unit that we used to measure
physical quantities is relative and is defined only locally.  A distortion of
the unit system can be described by a vector field which is called gauge field.
Weyl proposed that such a vector field (the gauge field) is the vector
potential that describes the electromagnetism.  Although the above particular
proposal turns out to be incorrect, the Weyl's idea is correct.  In 1925, the
complex quantum amplitude was discovered.  If we assume the complex phase is
relative, then a distortion of unit system that measure local complex phase can
also be described by a vector field. Such a vector field is indeed the vector
potential that describes the electromagnetism.  This leads to a unified way to
understand gravity and electromagnetism: gravity arises from the relativity of
spacial directions at different spatial points, while electromagnetism arises
from the relativity of complex quantum phases at different spatial points.
Further more, Nordstr\"om, M\"oglichkeit, Kaluza, and Klein showed that both
gravity and electromagnetism can be understood as a distortion of space-time
provided that we think the space-time as five dimensional with one dimension
compactified into a small circle.\cite{NM1404,K2166,K2695} This can be viewed
as an unification of gravity and electromagnetism.  Those theories are so
beautiful.  Since that time, the geometric way to view our world has dominated
theoretical physics.

\subsection{Quantum revolution}

\begin{figure}[tb]
\centerline{
	\includegraphics[height=1.2in]{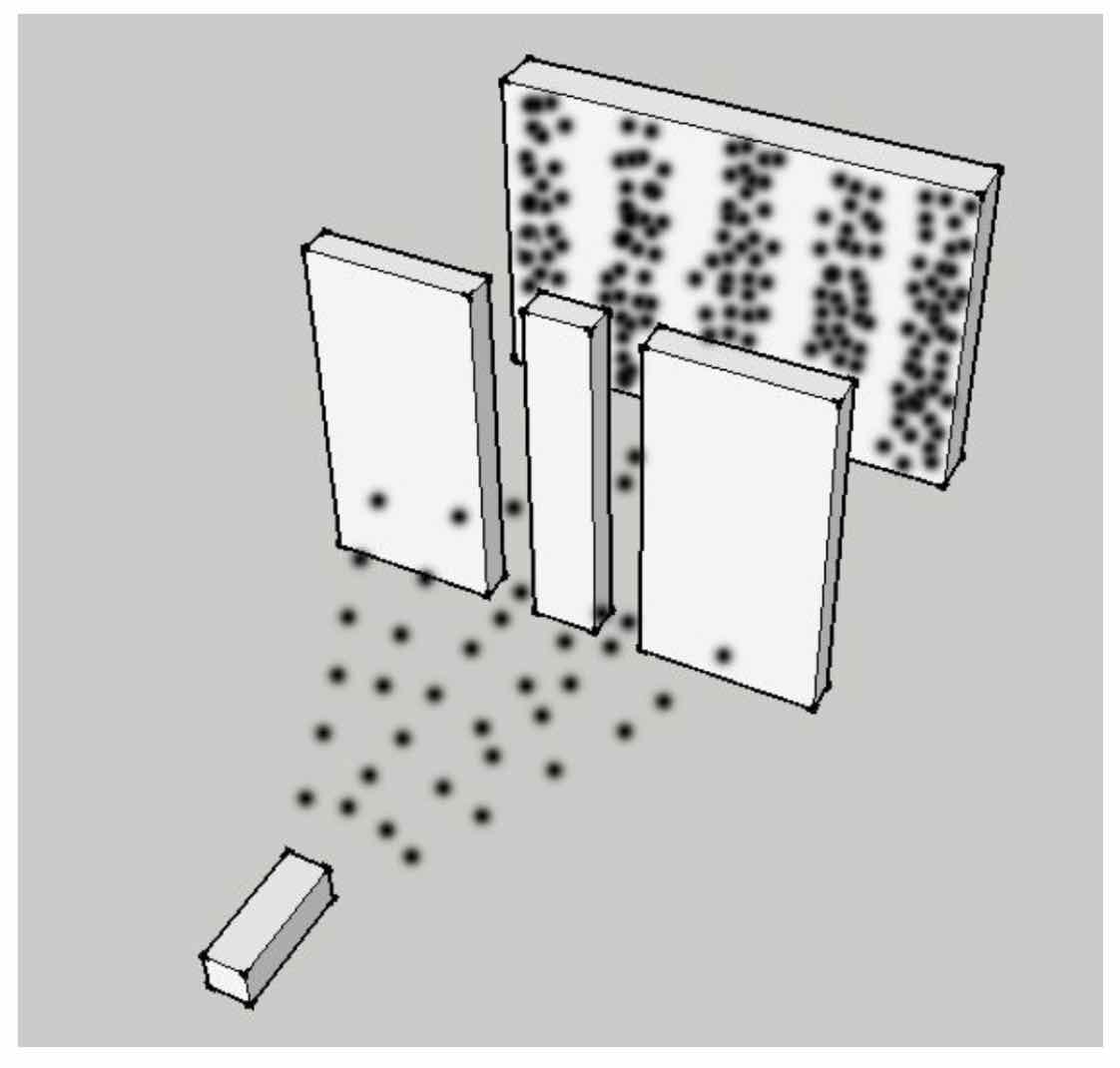} 
	\includegraphics[height=1.2in]{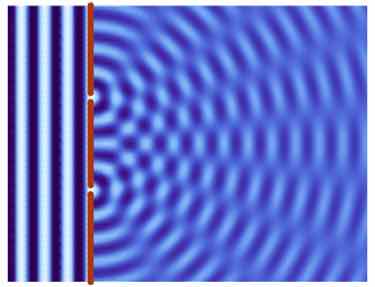} 
	\includegraphics[height=1.2in]{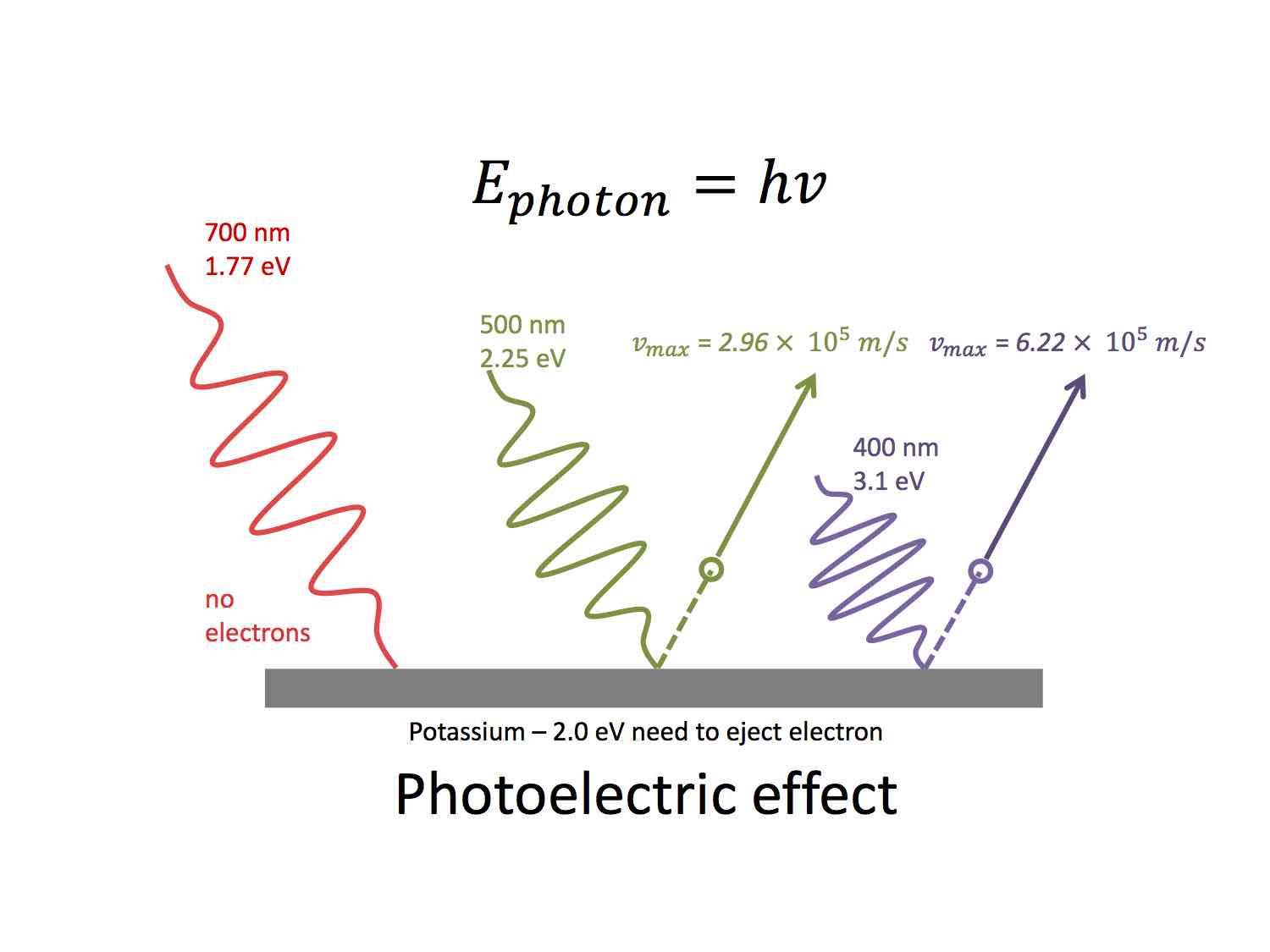}
}
\centerline{
	(a) ~~~~~~~~~~~~~~~~~~~~~~~~~~~~~~~~~~~~~~~~~~~~~~~~~~~~~~~~~~~~~~~~~~~~~~~~~~~
	(b)
}
\caption{
	(a) An electron beam passing through a double-slit can generate an interference
	pattern, indicating that electrons are also waves.  (b) Using light to
eject electrons from a metal (the photoelectric effect) shows that the higher
the light wave frequency (the shorter the wave length), the higher the energy of the ejected electron. This
reveals that a light wave of frequency $f$ can be viewed a beam of particles of
energy $E=h f$, where $h=6.62607004\times 10^{-34}
\frac{\text{m}^2\text{kg}}{\text{s}} $.
}
\label{dslit}
\end{figure}

However, such a geometric view of world was immediately challenged by new
discoveries from microscopic world.\footnote{Many people have ignored such challenges and the geometric view of world becomes the main stream.}
The experiments in  microscopic world tell us that not
only Newton theory is incorrect, even its relativity modification is incorrect.
This is because Newton theory and its relativistic modification are theories
for particle-like matter.  But through experiments on very tiny things, such as
electrons, people found that the particles are not really particles. They
also behave like waves at the same time.  Similarly, experiments also reveal
that the light waves behave like a beam of particles (photons) at the same time
(see Fig. \ref{dslit}).  So the real matter in our world is not what we thought
it was. The matter is neither particle nor wave, and both particle and wave.
So the Newton theory (and its relativistic modification) for particle-like
matter and the Maxwell/Einstein theories for wave-like matter cannot be the
correct theories for matter. We need a new theory for the new form of
existence: particle-wave-like matter.  The new theory is the quantum theory
that explains the  microscopic world. The quantum theory unifies the
particle-like matter and wave-like matter.

\begin{svgraybox}
\begin{center}
\textbf{Box 11.4 Quantum revolution}

There is no particle-like matter nor wave-like matter.
All the matter in our world is particle-wave-like matter.
\end{center}
\end{svgraybox}

From the above, we see that quantum theory reveals the true existence in our
world to be quite different from the classical notion of existence in our mind.
What exist in our world are not particles or waves, but somethings that are
both particle and wave. Such a picture is beyond our wildest imagination, but
reflects the truth about our world and is the essence of quantum theory.  To
understand the new notion of existence more clearly, let us consider another
example. This time it is about a bit (represented by spin-1/2). A bit has two
possible states of classical existence: $|1\>=|\up \rangle$ and
$|0\>=|\down\rangle $.  However,  quantum theory also allows a new kind of
existence $|\up \rangle + |\down \rangle $.  One may say that $|\up \rangle +
|\down \rangle $ is also a classical existence since $|\up \rangle + |\down
\rangle = |\to \rangle $ that describes a spin in $x$-direction.  So let us
consider a third example of two bits. Then there will be four possible states
of classical existence: $|\up\up \rangle $, $|\up\down \rangle $, $|\down\up
\rangle $, and $|\down\down \rangle $.  Quantum theory allows a new kind of
existence $|\up\up \rangle + |\down\down \rangle $.  Such a quantum existence
is entangled and has no classical analogues.

\begin{figure}[tb]
\centerline{
	\includegraphics[height=.9in]{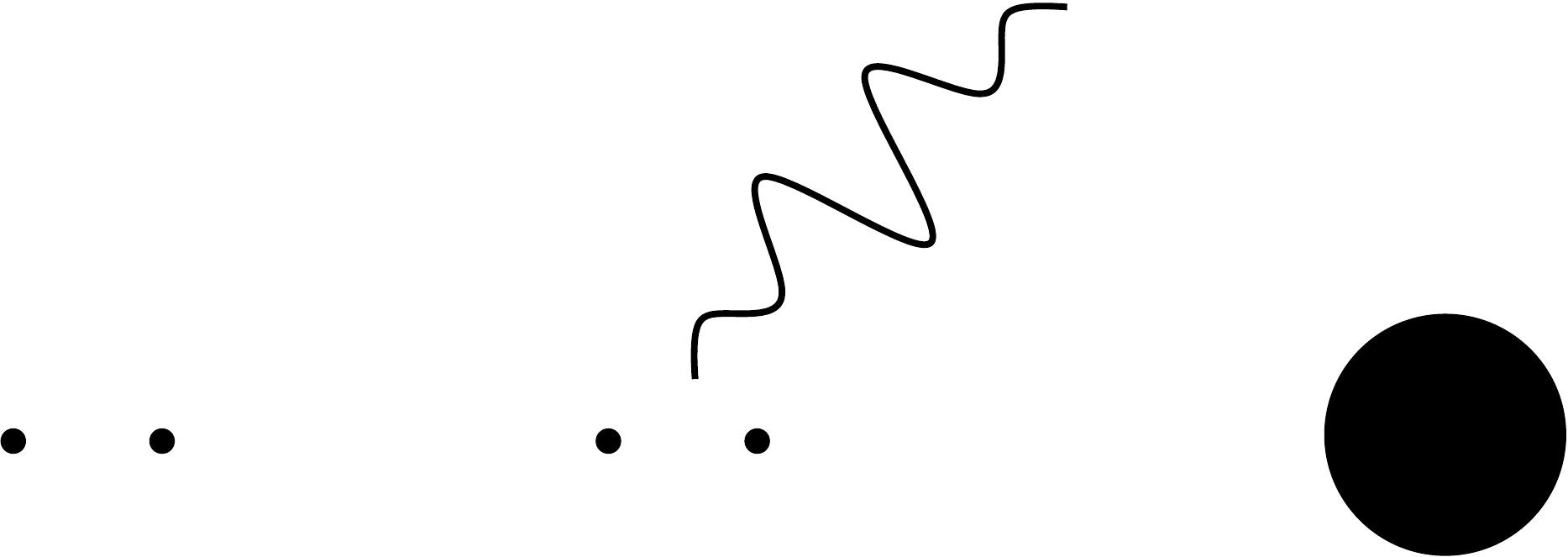}
}
\caption{
To observe two points of distance $l$ apart, we need to send in light of wave
length $\la < l$.  The corresponding photon has an energy $E=hc/\la$. If $l$ is
less than the Planck length $l<l_P$, then the photon will make a back hole of
size larger then $l$.  The black hole will swallow the two points, and we can
never measure  the separation of two points of distance less than $l_P$.  What
cannot be measured cannot exists.  So the notion of ``two points less than $l_P$
apart'' has no physical meaning and does not exist.
}
\label{nomani}
\end{figure}

Although the geometric way to understand our world is a main stream in physics,
here we will take a position that the geometric understanding is not good
enough and will try to advocate a very different non-geometric understanding of
our world.  Why the geometric understanding is not good enough? First the
geometric understanding is not self-consistent. It contradicts with quantum
theory.
The consideration
based quantum mechanics and Einstein gravity indicates that two points
separated by a distance less than the Planck length
\begin{align}
l_P= \sqrt{\frac{\hbar G}{c^3}} = 1.616199\times 10^{35} \text{m}
\end{align}
cannot exist as a physical reality (see Fig. \ref{nomani}). Thus the foundation
of the geometric approach -- manifold -- simply does not exist in our universe,
since manifold contains points with arbitrary small separation. This suggests
that geometry is an emergent phenomenon that appears only at long distances. So
we cannot use geometry and manifold as a foundation to understand fundamental
physical problems.

Second, Maxwell theory of light and Einstein theory of gravity predict light
waves and gravitational waves.  But the theories fail to tell us what is
waving?  Maxwell theory and Einstein theory are built on top of geometry. They
fail to answer what is the origin of the apparent geometry that we see.
In other words,  Maxwell theory and Einstein theory are incomplete, and they should be regarded as effective theories at long distances.

Since geometry does not exist in our world, this is why we say the geometric
view of world is challenged by quantum theory.  The quantum theory tell us such
a point of view to be wrong at length scales of order Planck length. So the
quantum theory represents the most dramatic revolution in physics.


\section{It from qubit, not bit}


After realizing that even the notion of existence is changed by quantum
theory, it is no longer surprising to see that quantum theory also blurs the
distinction between information and matter.  In fact, it implies that
information is matter, and matter is information.  This is because the
frequency is an attribute of information.  Quantum theory tells us that
frequency is energy $E=h f$, and relativity tells us that energy is mass
$m=E/c^2$. Both energy and mass are attributes of matter.  So matter =
information. This represents a new way to view our world.

\begin{svgraybox}
\begin{center}
\textbf{Box 11.5 The essence of quantum theory}

The energy-frequency relation $E=h f$ implies that matter = information.
\end{center}
\end{svgraybox}

The above point of view of ``matter = information'' is similar to Wheeler's ``it
from bit'', which represents a deep desire to unify matter and information. In
fact, such an unification has happened before at a small scale. We introduced
electric and magnetic field to informationally (or pictorially) describe
electric and magnetic interaction.  But later, electric/magnetic field became
real matter with energy and momentum, and even a particle associated with it.

However, in our world, ``it'' are very complicated. (1) Most ``it'' are
fermions, while ``bit'' are bosonic. Can fermionic ``it'' come from bosonic
``bit''? (2) Most ``it'' also carry spin-1/2. Can spin-1/2 arises from ``bit''?
(3) All ``it'' interact via a special kind of interaction -- gauge interaction.
Can ``bit'' produce gauge interaction? Can ``bit'' produce waves that satisfy
Maxwell equation? Can ``bit'' produce photon?

In other words, to understand the concrete meaning of ``matter from
information'' or ``it from bit'', we note that matter are described by Maxwell
equation (photons), Yang-Mills equation (gluons and $W/Z$ bosons), as well as
Dirac and Weyl equations (electrons, quarks, neutrinos).  The statement
``matter = information'' means that those wave equations can all come from
qubits. In other words, we know that elementary particles (\ie matter) are
described by gauge fields and anti-commuting fields in a quantum field theory.
Here we try to say that all those very different quantum fields can arise from
qubits. Is this possible?

All the waves and fields mentioned above are waves and fields in space.  The
discovery of gravitational wave strongly suggested that the space is a
deformable dynamical medium. In fact, the discovery of electromagnetic wave and
the Casimir effect already strongly suggested that the space is a deformable
dynamical medium.  As a dynamical medium, it is not surprising that the
deformation of space give rise to various waves.  But the dynamical medium that
describe our space must be very special, since it should give rise to waves
satisfying Einstein equation (gravitational wave), Maxwell equation
(electromagnetic wave), Dirac equation (electron wave), \etc.  But what is the
microscopic structure of the space?  What kind of microscopic structure can, at
the same time, give rise to waves that satisfy Maxwell equation, Dirac/Weyl
equation, and Einstein equation?

Let us view the above questions from another angle.  Modern science has made
many discoveries and has also unified many seemingly unrelated discoveries
into a few simple structures.  Those simple structures are so beautiful and we
regard them as wonders of our universe.  They are also very myterious since we
do not understand where do they come from and why do they have to be the way
they are.  At moment, the most fundamental mysteries and/or wonders in our
universe can be summarized by the following short list:\\
\begin{svgraybox}
	\centerline{\textbf{Box 11.6 Eight wonders}}
	\hspace*{5ex}(1) Locality.\\
	\hspace*{5ex}(2) Identical particles.\\
	\hspace*{5ex}(3) Gauge interactions.\cite{Wey52,P4103,YM5491}\\
	\hspace*{5ex}(4) Fermi statistics.\cite{F2602,D2661}\\
	\hspace*{5ex}(5) Tiny masses of fermions ($\sim 10^{-20}$ of the Planck
mass).\cite{GW7343,P7346,Wqoem}\\
\hspace*{5ex}(6) Chiral fermions.\cite{LY5654,Wo5713}\\
\hspace*{5ex}(7) Lorentz invariance.\cite{E0591}\\
\hspace*{5ex}(8) Gravity.\cite{E1669}
\end{svgraybox}

In the current physical theory of nature (such as the standard model), we take
the above properties for granted and do not ask where do they come from.  We
put those wonderful properties into our theory by hand, for example, by
introducing one field for each kind of interactions or elementary particles.

However, here we would like to question where do those wonderful and mysterious
properties come from?  Following the trend of science history, we wish to have
a single unified understanding of all of the above mysteries.  Or more
precisely, we wish that we can start from a single structure to obtain all of
the above wonderful properties.

The simplest element in quantum theory is qubit $|0\>$ and $|1\>$ (or
$|\down\>$ and $|\up\>$).  Qubit is also the simplest element in quantum
information.  Since our space is a dynamical medium, the simplest choice is to
assume the space to be an ocean of qubits.  We will give such an ocean a
formal name ``qubit ether''.  Then the matter, \ie the elementary particles,
are simply the waves, ``bubbles'' and other defects in the qubit ocean (or
quibt ether).  This is how ``it from qubit'' or ``matter = information''.


Qubit, having only two states $|\down  \rangle $ and $| \up \rangle $, is very
simple.  We may view the many-qubit state with all qubits in $|\down \rangle $
as the quantum state that correspond to the empty space (the vacuum).  Then the
many-qubit state with a few qubits in  $|\up  \rangle $ correspond to a space
with a few spin-0 particles described by a scaler field.  Thus, it is easy to
see that a scaler field can emerge from  qubit ether as a density wave of
up-qubits.  Such a wave satisfy the Eular eqution, but not Maxwell equation or
Yang-Mills equation. So the above particular qubit ether is not the one that
correspond to our space. It has a wrong microscopic structure and cannot carry
waves satisfying Maxwell equation and Yang-Mills equation.  But this line of
thinking may be correct. We just need to find a qubit ether with a different
microscopic structure.

However, for a long time, we do not know how waves satisfying Maxwell equation
or Yang-Mills equation can emerge from any qubit ether.  The anti-commuting
wave that satisfy Dirac/Weyl equation seems even more impossible.  So, even
though quantum theory strongly suggests ``matter = information'', trying to
obtain all elementary particles from an ocean of simple qubits is regarded as
impossible by many and has never become an active research effort.

So the key to understand ``matter = information'' is to identify the microscopic
structure of the qubit ether (which can be viewed as space).  The microscopic
structure of our space must be very rich, since our space
not only can carry gravitational wave and electromagnetic wave, it can also
carry electron wave, quark wave, gluon wave, and the waves that correspond to
all elementary particles. Is such a qubit ether possible?


\begin{figure}[tb]
\centerline{
	\includegraphics[height=1.5in]{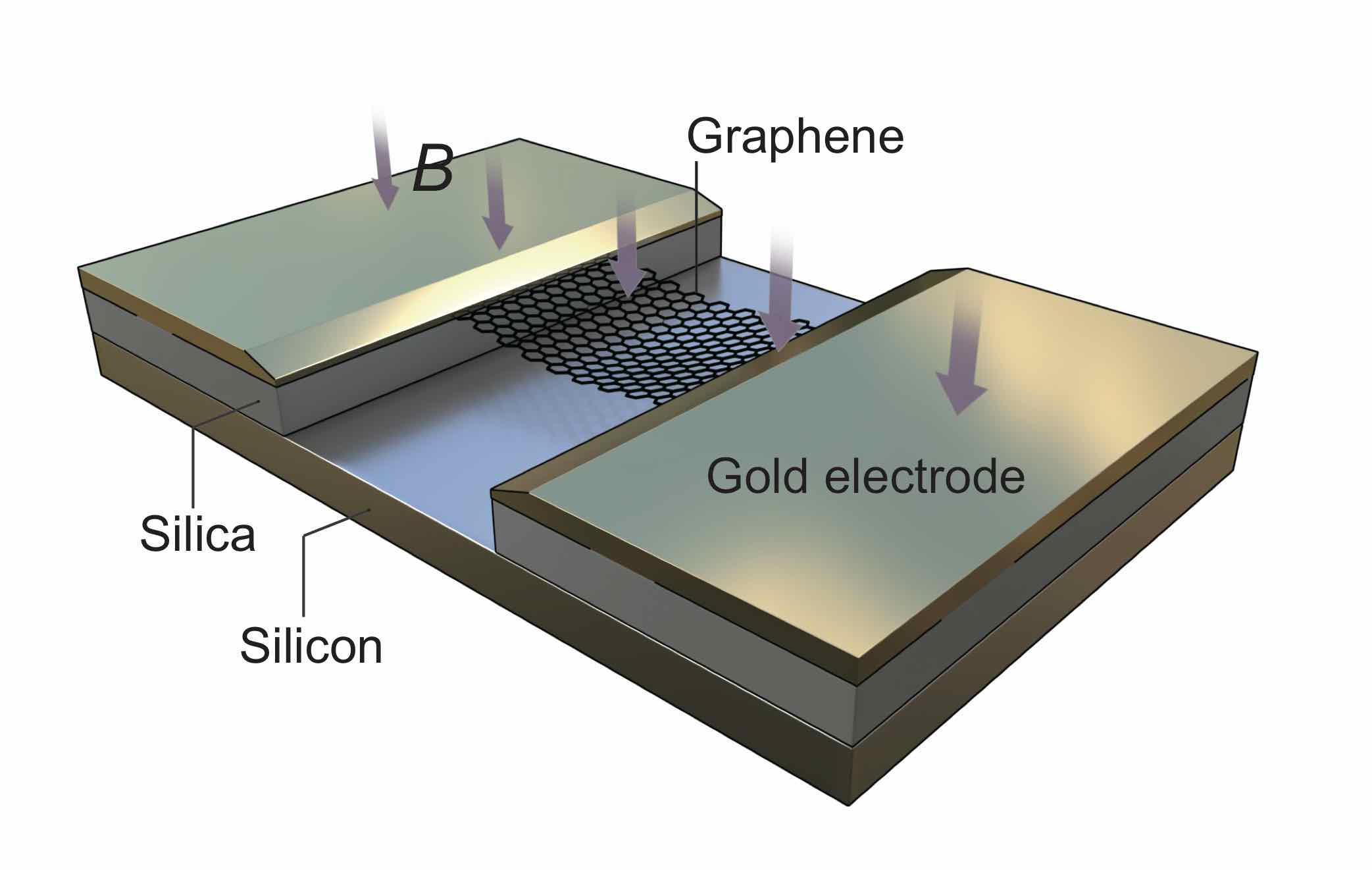}
}
\caption{
Fractional quantum Hall states are new states of quantum matter formed by
electrons traped at the interface of two semiconductors, or by electrons on a
sheet of graphene, under a strong magnetic field $B$.
}
\label{graphene}
\end{figure}

In condensed matter physics, the discovery of fractional quantum Hall
states\cite{TSG8259} (see Fig. \ref{graphene}) bring us into a new world of highly entangled many-body
systems.  When the strong entanglement becomes long range
entanglement\cite{CGW1038}, the systems will possess a new kind of order --
topological order\cite{Wtop,Wrig}, and represent new states of matter.  We find
that the waves (the excitations) in topologically ordered states can be very
strange: they can be waves that satisfy Maxwell equation, Yang-Mills equation,
or Dirac/Weyl equation.  So the impossible become possible: all elementary
particles can emerge from long range entangled qubit ether.

We would like to stress that the above picture is ``it from qubit'', which is very
different from Wheeler's ``it from bit''. As we have explained, our observed
elementary particles can only emerge from  long range entangled qubit ether.
The requirement of quantum entanglement implies that ``it cannot from bit''.
In fact ``it from entangled qubits''.

\section{Emergence approach}

\subsection{Two approaches}


\begin{figure}[tb]
\centerline{
	\includegraphics[height=1.2in]{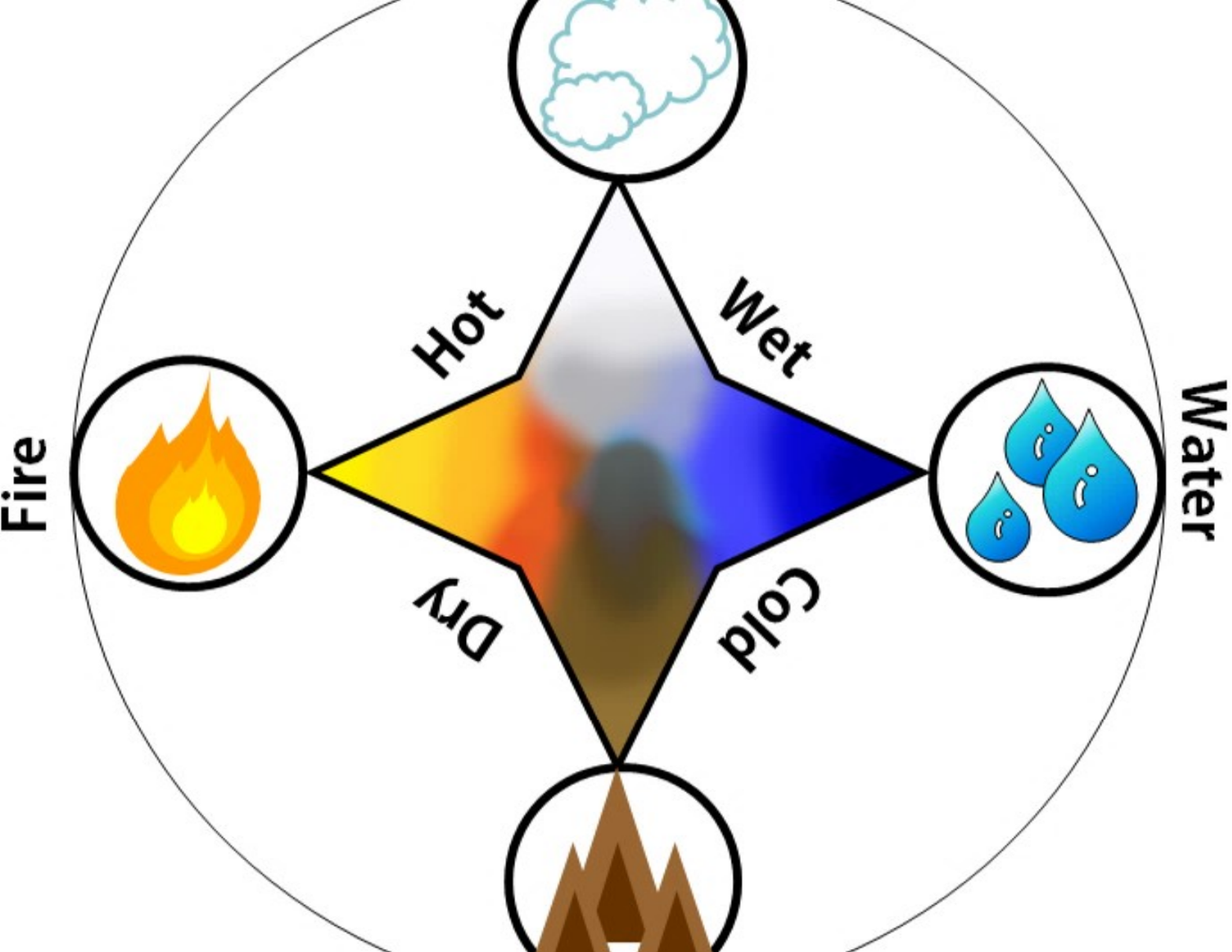}~~~
	\includegraphics[height=1.2in]{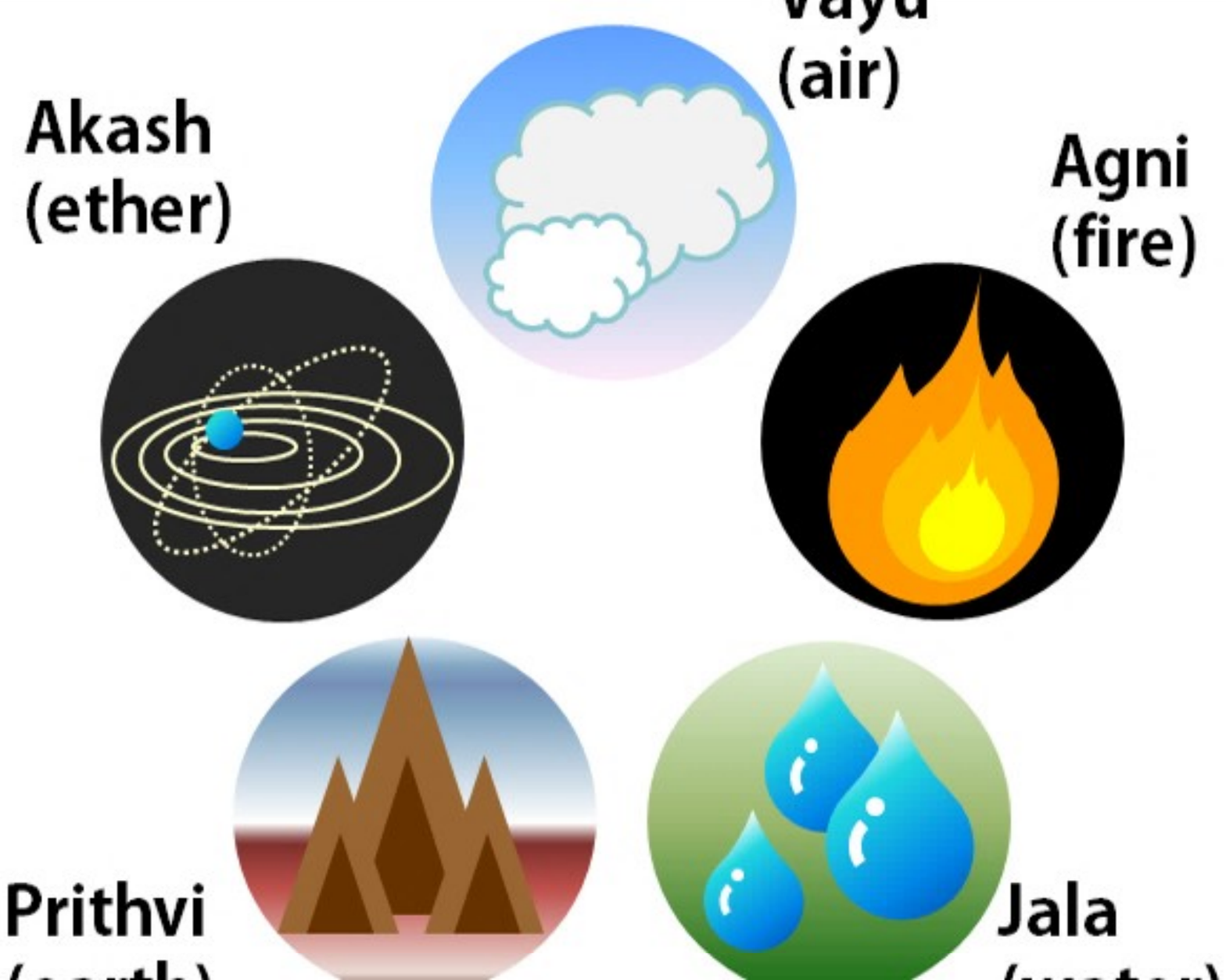}~~~
}

\centerline{
(a) ~~~~~~~~~~~~~~~~~~~~~~~~~~~~~~~~~
(b) 
}

\centerline{
\includegraphics[height=1.0in]{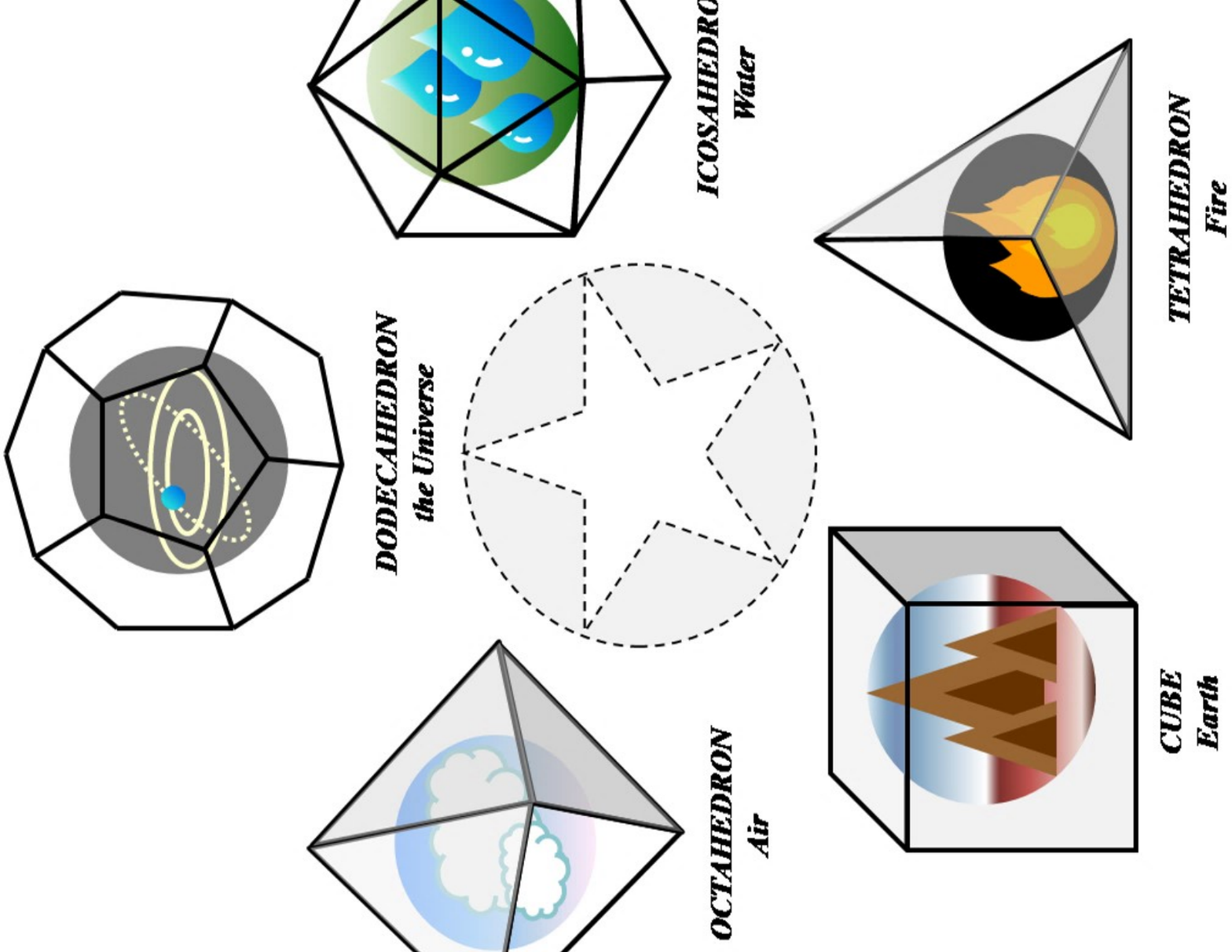}~~~
	\includegraphics[height=1.0in]{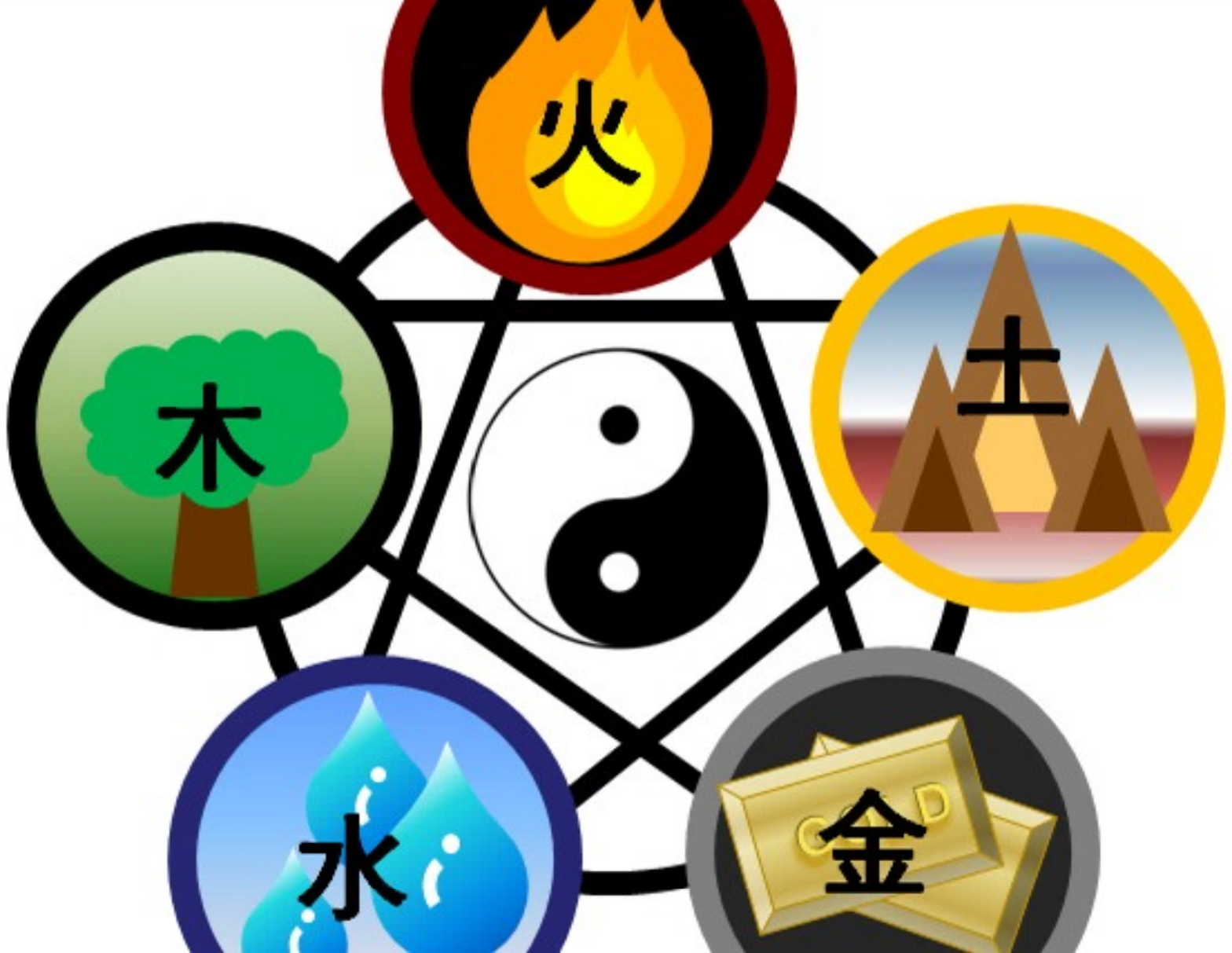}
}

\centerline{
(c) ~~~~~~~~~~~~~~~~~~~~~~~~~~~~~~~~~
(d) 
}

\caption{
Acient ``atomic'' theory of matter:
(a) Aristotle's four-elements theory: all matter is formed by air, water, earth, and fire.
(b) Ayurveda's five-elements theory: air, fire, water, earth, and ether (space).
(c) Plato's five-elements theory:  water, fire, earth, air, and universe.
(d) Chinese five-elements theory: all matter is formed by five ``features''
gold, wood, water, fire, earth, which come from two more basic features ying
(negtive) and yang (positive). Ying and yang come from the grand unifier Taiji
(represented by the circle in the center).
}
\label{elements}
\end{figure}

In the reductionism approach, we try to understand various things by dividing
them into smaller and smaller parts.  If we assume the division has to end at a
certain level, then we conclude that all things are formed by the parts that
cannot be divided further.  The indivisible parts are called ``atoms'' or
elementary particles (see Fig. \ref{elements}).  So in the reductionism
approach, we view all matter in our world as made of some simple beautiful
building blocks, the elementary particles.  A deeper understanding is gained if
we find some elementary particles are not actually elementary and are formed by
even smaller objects.  A large part of science is devoted in finding those
smaller and smaller objects, as represented by the discoveries of atoms,
electrons and protons, and then quarks.

However, the reductionism approach that we followed in last 200 years may not
represent a right direction.  For example, phonons (the quanta of sound waves)
in a solid is as particle-like as any other elementary particles at low
energies.  But if we look at phonons closely, we do not see smaller parts that
form a phonon.  We see the atoms that fill the entire space.  The phonons are
not formed by those atoms, the phonons are simply collective motions of those
atoms.

This leads us to wonder that maybe photons, electrons, gravitons, \etc, are
also collective motions of a certain underlying structure that fill the entire
space. They may not have smaller parts. Looking for the smaller parts of
photons, electrons, and gravitons to gain a deeper understanding of those
elementary particles may not be a right approach.

\begin{figure}[tb]
\centerline{
	\includegraphics[height=1.2in]{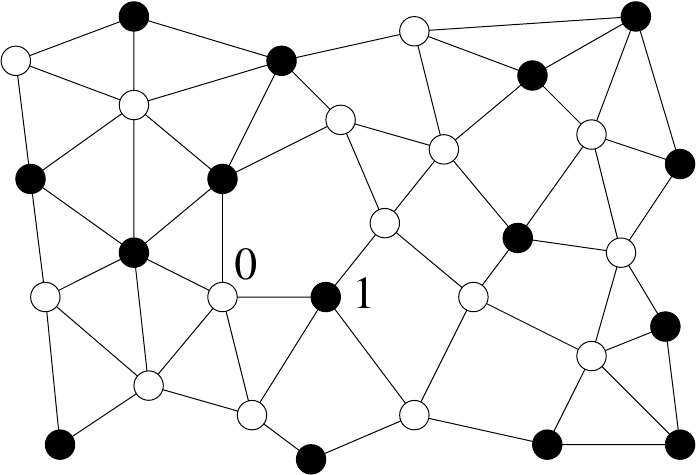}
}
\caption{
In the emergence approach, there is only one form of ``matter'' --
the space (the vaccum) itself, which is formed by qubits.
What we regarded as matter are distortions and defects in this ``space-matter''.
}
\label{qubitether}
\end{figure}

Here, we will use a different approach, emergence approach, to gain a deeper
understanding of elementary particles.  In the emergence approach, we view
space as an ocean of qubits, \ie a qubit ether (see Fig. \ref{qubitether}). The
empty space (the vacuum) corresponds to the ground state of the qubit ether,
and the elementary particles (that form the matter) correspond to the
excitations of the qubit ether.

As we have pointed out that the elementary particles in our world have very
rich and strange properties.  Can excitations of simple qubits have those rich
strange properties?  How to answer such an question?  Here is our plan: due to
the particle-wave duality in quantum theory, particles and waves are the same
thing. So we can try to understand the rich strange properties of elementary
particles by trying to understand the rich strange properties of waves.

\subsection{Principle of emergence} \label{pem}

One might think the properties of a material should be determined by the
components that form the material.  However, this simple intuition is
incorrect, since all the materials are made of the same components: electrons,
protons and neutrons, with about the same numerical density.  So we cannot use
the richness of components to understand the richness of the materials. In fact,
the various properties of different materials originate from various ways in
which the particles are organized.  Different orders (the organizations of
particles) give rise to different physical properties of a material. It is the
richness of the orders that gives rise to the richness of material world.

\begin{svgraybox}
\begin{center}
\textbf{Box 11.7 Principle of emergence}

The physical properties of a many-body state mainly come from the organization
(\ie the order) of the degrees of freedom in the state.
\end{center}
\end{svgraybox}

\begin{figure}[t]
\centerline{
	\includegraphics[scale=0.7]{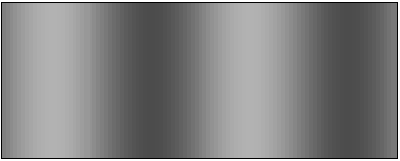}
}
\caption{
Liquids only have a compression wave -- a wave of density fluctuations.
}
\label{liquidC}
\end{figure}
\begin{figure}[t]
\centerline{
	\includegraphics[scale=0.38]{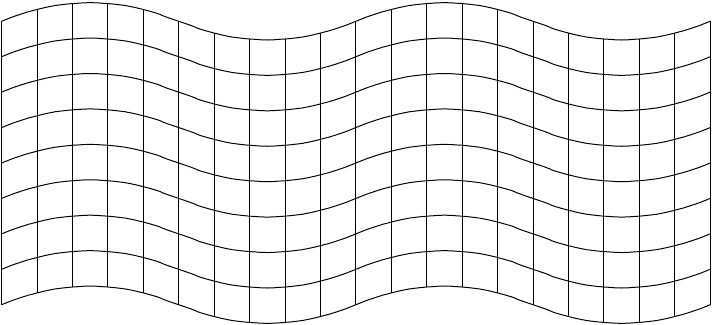}
}
\caption{
Drawing a grid on a sold helps us to see the deformation of the solid.  The
vector $u^i$ in \eqn{NavEq} is the displacement of a vertex in the grid.  In
addition to the compression wave (\ie the density wave), a solid also supports
transverse wave (wave of shear deformation) as shown in the above figure.
}
\label{crystalC}
\end{figure}

We know that a deformation in a material can propagate just like
the ripple on the surface of water.  The propagating deformation corresponds to
a wave traveling through the material.  Since liquids can resist only
compression deformation, so liquids can only support a single kind of wave --
compression wave (see Fig.  \ref{liquidC}). (Compression wave is also
called longitudinal wave.) Mathematically the motion of the compression wave is
governed by the Euler equation
\begin{equation}
\label{EulEq}
 \frac{\prt^2 \rho}{\prt t^2}-v^2 \frac{\prt^2 \rho}{\prt x^2}=0,
\end{equation}
where $\rho$ is the density of the liquid.

Solid can resist both compression and shear deformations.  As a result, solids
can support both compression wave and transverse wave. The transverse wave
correspond to the propagation of shear deformations. In fact there are two
transverse waves corresponding to two directions of shear deformations.  The
propagation of the compression wave and the two transverse waves in solids are
described by the elasticity equation
\begin{equation}
\label{NavEq}
 \frac{\prt^2 u^i}{\prt t^2}- T^{ikl}_j \frac{\prt^2 u^j}{\prt x^k\prt x^l}
=0
\end{equation}
where the vector field $u^i(\v x, t)$  describes the local displacement of the
solid.

We would like to point out that the elasticity equation and the Euler equations
not only describe the propagation of waves, they actually describe all small
deformations in solids and liquids.  Thus, the two equations represent a
complete mathematical description of the mechanical properties of solids and
liquids.

But why do solids and liquids behave so differently? What makes a solid to have
a shape and a liquid to have no shape?  What are the origins of
elasticity  equation and Euler equations?

\begin{figure}[t]
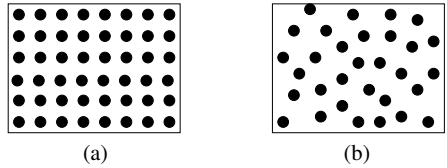

\centerline{
	\includegraphics[scale=0.45]{Chapters/Chap11/crystal}~~~~~~~~~~~~~~~
	\includegraphics[scale=0.45]{Chapters/Chap11/liquid}
}
\centerline{
(a)  ~~~~~~~~~~~~~~~~~~~~~~~~~~~~~~~~~~~~~~~~
(b)
}
\caption{
(a) In a crystal, particles form a regular array (and break the continuous
translation symmetry down to a discrete translation symmetry).
(b) In a liquid, particles have a random fluctuating distribution
(which do not break any symmetry).
}
\label{cryliq1}
\end{figure}

To answer the above questions, we have to use the microscopic structure of
liquids and solids: they are all formed by atoms.  In liquids, the positions of
atoms fluctuate randomly (see Fig.  \ref{cryliq1}a),  while in solids, atoms
organize into a regular fixed array (see Fig.  \ref{cryliq1}b).\footnote{The
solids here should be more accurately referred as crystals.}  It is the
different organizations of atoms that lead to the different mechanical
properties of liquids and solids.  In other words, it is the different
organizations of atoms that make liquids to be able to flow freely and solids
to be able to retain its shape.

\begin{figure}[t]
\centerline{
	\includegraphics[scale=0.45]{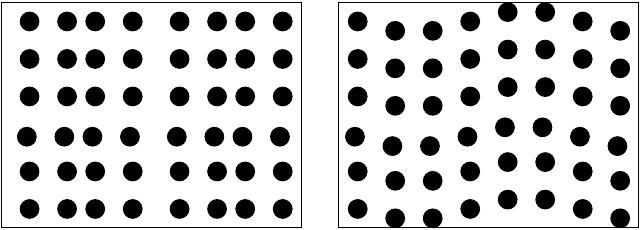}
}
\centerline{
\hfil
(a) ~~~~~~~~~~~~~~~~~~~~~~~~~~
(b)
\hfil
}
\caption{
The atomic picture of
(a) the compression wave and
(b) the transverse wave in a crystal.
}
\label{sndwavs1}
\end{figure}

\begin{figure}[t]
\centerline{
	\includegraphics[scale=0.45]{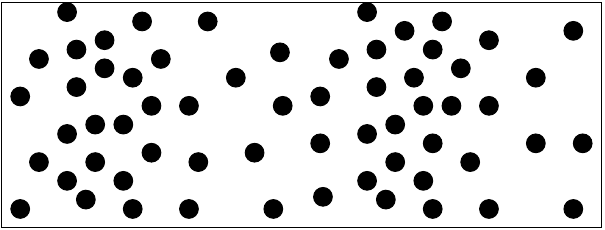}
}
\caption{
The atomic picture of the compression wave in liquids.
}
\label{liquidwav1}
\end{figure}

How can different organizations of atoms affect mechanical properties of
materials?  In solids, both the compression deformation (see Fig.
\ref{sndwavs1}a) and the shear deformation (see Fig. \ref{sndwavs1}b) lead to
real physical changes of the atomic configurations. Such changes cost energies.
As a result, solids can resist both kinds of deformations and can retain their
shapes.  This is why we have both the compression wave and the transverse wave
in solids.

In contrast, a shear deformation of atoms in liquids does not result in a new
configuration since the atoms still have the same uniformly random
distribution. So the shear deformation is a do-nothing operation for liquids.
Only the compression deformation which changes the density of the atoms results
in a new atomic configuration and costs energies. As a result, liquids can only
resist compression and have only compression wave.  Since shear deformations do
not cost any energy for liquids, liquids can flow freely.

We see that the properties of the propagating wave are entirely determined by
how the atoms are organized in the materials.  Different organizations lead to
different kinds of waves and different kinds of mechanical laws.  This point of
view is called the principle of emergence.

In the above, we see that the  Euler equation and elasticity equation
originated from the different organizations of atoms.  Elementary particles are
described by Maxwell/Yang-Mills equations and Dirac/Weyl equations, or in
other words they are described by quantum field theory (such as the standard
model). But quantum field theory (\ie the  Maxwell/Yang-Mills equations and
Dirac/Weyl equations) are effective theories like Euler/elasticity equations.
They are not a complete description of physical systems, since they lack of
description of the microscopic structure.  We know that Euler/elasticity
equations come from atoms. But what is the origin (\ie the microscopic
structure) of  quantum field theory? The motion of what give rise to
Maxwell/Yang-Mills equations and Dirac/Weyl equations? In the following, we
like to show that the motion of particles or qubits can give rise both to
Maxwell/Yang-Mills equations and Dirac/Weyl equations, as long as the
particles/qubits have a proper organization.  We will concentrate on how
Maxwell and Dirac equations arise from the motion of particles/qubits.

\subsection{String-net liquid of qubits unifies light and electrons}

When Maxwell equation was first introduced, people firmly believed that any
wave must corresponds to motion of something.  So people want to find out
what is the origin of the Maxwell equation?  The motion of what gives rise
electromagnetic wave?

First, one may wonder: can Maxwell equation comes from a certain symmetry
breaking order?  Based on Landau symmetry-breaking theory, the different
symmetry breaking orders can indeed lead to different waves satisfying
different wave equations.  So maybe a certain symmetry breaking order can give
rise to a wave that satisfy Maxwell equation.  But people have been searching
for ether -- a medium that supports light wave -- for over 100 years, and could
not find any symmetry breaking states that can give rise to waves satisfying
the Maxwell equation.  This is one of the reasons why people give up the idea
of ether as the origin of light and Maxwell equation.

However, the discovery of topological order~\cite{Wtop,Wrig} suggests that
Landau symmetry-breaking theory does not describe all possible organizations of
particles/qubits.  This gives us a new hope: Maxwell equation may arise from a
new kind of organizations of particles/qubits that have non-trivial topological
orders (or their gapless generalization, quantum orders).

In addition to the Maxwell equation, there is an even stranger equation, Dirac
equation, that describes wave of electrons (and other fermions).  Electrons
have Fermi statistics. They are fundamentally different from the quanta of
other familiar waves, such as photons and phonons, since those quanta all have
Bose statistics.  To describe the electron wave, the amplitude of the wave must
be anti-commuting Grassmann numbers, so that the wave quanta will have Fermi
statistics. Since electrons are so strange, few people regard electrons and the
electron waves as collective motions of something. People accept without
questioning that electrons are fundamental particles, one of the building
blocks of all that exist.

\begin{figure}[tb]
\centerline{
	\includegraphics[height=1.3in]{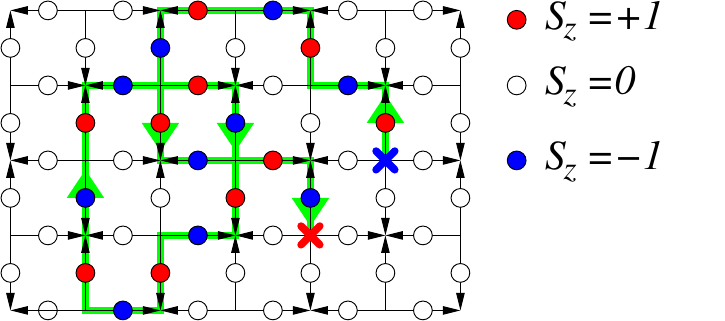}
}
\caption{
Oriented strings from spin-1-qubits living on the links of a cubic lattice
(only one slice is shown).
A spin-1-qubit has three states
$S_z=0$ (open circle),
$S_z=+1$ (red circle),
$S_z=-1$ (blue circle).
$S_z=0$ corresponds to no string on the link,
$S_z=+1$ a string with the same orientation of the link,
$S_z=-1$ a string with the opposite orientation.
(Oriented strings can also arise from atoms as oriented polymers.)
}
\label{SqStrnet}
\end{figure}

\begin{figure}[tb]
\centerline{
	\includegraphics[height=1.in]{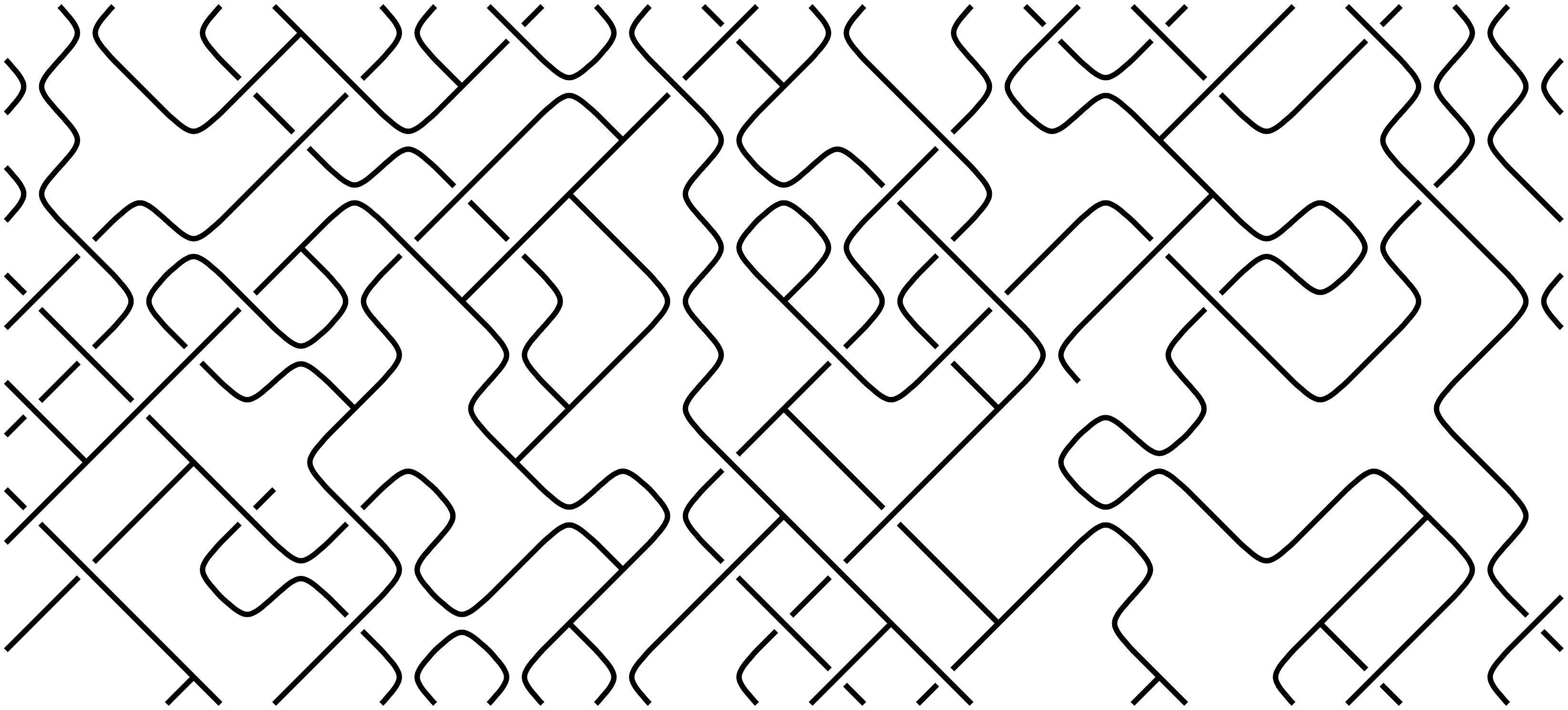}
}
\caption{
A quantum ether: The fluctuation of oriented strings give rise to
electromagnetic waves (or light). The ends of strings give rise to electrons.
Note that oriented strings have directions which should be described by curves
with arrow. For ease of drawing, the arrows on the curves are omitted in the
above plot.
}
\label{stringnetS}
\end{figure}

However, in a recent study~\cite{LWstrnet,LWuni,LWqed}, we find that if
particles/qubits form large oriented strings (see Fig. \ref{SqStrnet})  and if those strings form a quantum
liquid state, then the collective motion of the such organized particles/qubits
will correspond to waves described by Maxwell equation and Dirac equation.  The
strings in the string liquid are free to join and cross each other. As a
result, the strings look more like a network (see Fig.  \ref{stringnetS}).  For
this reason, the string liquid is actually a liquid of string-nets, which is
called string-net condensed state.

\begin{figure}[tb]
\centerline{
	\includegraphics[width=1.7in]{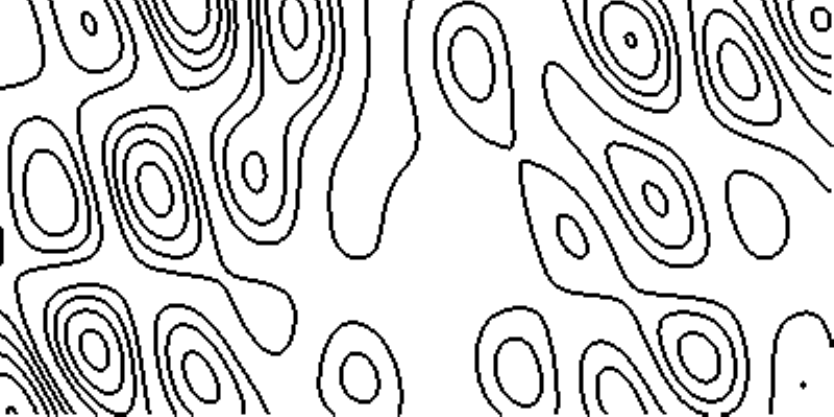}
}
\caption{
The fluctuating strings in a string liquid.
}
\label{vacBW}
\end{figure}

\begin{figure}[tb]
\centerline{
	\includegraphics[width=1.7in]{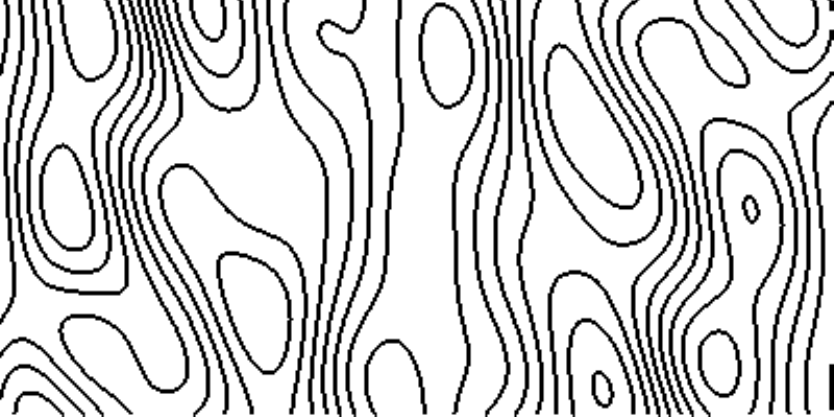}
}
\caption{
A `density' wave of oriented strings in a string liquid.  The wave propagates
in $\v x$-direction.  The `density' vector $\v E$ points in $\v y$-direction.
For ease of drawing, the arrows on the oriented strings are omitted in the
above plot.
}
\label{vacwvBW}
\end{figure}

But why the waving of strings produces waves described by the Maxwell equation?
We know that the particles in a liquid have a random but uniform distribution.
A deformation of such a distribution corresponds a density fluctuation, which
can be described by a scaler field $\rho(\v x,t)$.  Thus the waves in a liquid
is described by the scaler field $\rho(\v x,t)$ which satisfy the Euler
equation \eq{EulEq}.  Similarly, the strings in a string-net liquid also have
a random but uniform distribution (see Fig.  \ref{vacBW}). A deformation of
string-net liquid corresponds to a change of the density of the strings (see
Fig. \ref{vacwvBW}).  However, since strings have an orientation, the
`density' fluctuations are described by a vector field $\v E(\v x, t)$, which
indicates there are more strings in the $\v E$ direction on average.  The
oriented strings can be regarded as flux lines. The vector field $\v E(\v x,
t)$ describes the smeared average flux.
Since strings are continuous (\ie they cannot end), the
flux is conserved: $\v\prt\cdot\v E(\v x, t)=0$.  The vector density $\v E(\v
x, t)$ of strings cannot change in the direction along the strings (\ie along
the $\v E(\v x, t)$ direction). $\v E(\v x, t)$ can change only in the
direction perpendicular to $\v E(\v x, t)$.  Since the direction of the
propagation is the same as the direction in which $\v E(\v x, t)$ varies, thus
the waves described by $\v E(\v x, t)$ must be transverse waves: $\v E(\v x,
t)$ is always perpendicular to the direction of the propagation.  Therefore,
the waves in the string liquid have a very special property: the waves have
only transverse modes and no longitudinal mode.  This is exactly the property
of the light waves described by the Maxwell equation.  We see that `density'
fluctuations of strings (which are described be a transverse vector field)
naturally give rise to the light (or electromagnetic) waves and the Maxwell
equation~\cite{Walight,Wqoem,MS0312,HFB0404,LWuni,LWqed}.

To understand how electrons appear from string-nets, we would like to point out
that if we only want photons and no other particles, the strings must be closed
strings with no ends.  The fluctuations of closed strings produce only photons.
If strings have open ends, those open ends can move around and just behave like
independent particles.  Those particles are not photons. In fact, the ends of
strings are nothing but electrons (the blue $\times$ corresponds to
an electron and the red $\times$ corresponds to
a positron in Fig. \ref{SqStrnet}).

\begin{svgraybox}
\begin{center}
\textbf{Box 11.8 String density wave}

String density wave in a quantum liquid of oriented strings is a
divergence-free vector field, which give rise to a wave with only two transverse
modes -- an electromagnetic wave.
\end{center}
\end{svgraybox}

How do we know that ends of strings behave like electrons?  First, since the
waving of string-nets is an electromagnetic wave, a deformation of string-nets
correspond to an electromagnetic field.  So we can study how an end of a string
interacts with a deformation of string-nets.  We find that such an interaction
is just like the interaction between a charged electron and an electromagnetic
field. Also electrons have a subtle but very important property -- Fermi
statistics, which is a property that exists only in quantum theory.  Amazingly,
the ends of strings can reproduce this subtle quantum property of Fermi
statistics~\cite{LWsta,LWstrnet}:
For string liquid state described by wave function
\begin{align}
|\Phi\> =\sum_\text{all conf.}
\left  |\bmm \includegraphics[height=0.30in]{Chapters/Chap11/stringnetS}\emm
\right \>,
\end{align}
then the end of strings are bosons (\ie Higgs bosons).  For string liquid state
\begin{align}
\label{Fwave}
|\Phi\> =\sum_\text{all conf.}
(-)^\text{\# of crossings} \left  |\bmm
\includegraphics[height=0.30in]{Chapters/Chap11/stringnetS}\emm\right \>
,
\end{align}
then the end of strings are fermions.  Here ``\# of crossings'' is obtained by
first project the 3D string configuration to a fixed 2D plan, then ``\# of
crossings'' is the number of string crossings $\bmm
\includegraphics[height=0.20in]{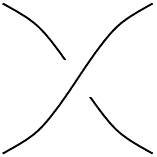} \emm$ (for details, see
Chapter \ref{chap7}).  Actually, string-net liquids explain why Fermi
statistics should exist.

We see that qubits that organize into string-net liquid naturally explain both
light and electrons (gauge interactions and Fermi statistics).  In other words,
string-net theory provides a way to unify light and
electrons~\cite{LWuni,LWqed}. So, the fact that our vacuum contains both light
and electrons may not be a mere accident. It may actually suggest that the
vacuum is indeed a long-range entangled qubit state, whose order is described
by a string-net liquid.

\begin{svgraybox}
\begin{center}
\textbf{Box 11.9 A qubit unification of light and electrons}
\end{center}

\noindent
Q: Where do light and electrons come from?\\
A: Light and electrons come from the qubits that form the space.\\
Q: Why do light and electrons exist?\\
A: Light and fermions exist because the space-forming qubits form a string-net condensed state.\\
Q: What are light and electrons?\\
A: Light waves are collective motions of strings and electrons are ends of open strings in the string-net condensed state.\\
\end{svgraybox}

%

We would like to stress that the string-nets are formed by qubits.  So in the
string-net picture, both the Maxwell equation and Dirac equation, emerge from
\emph{local} qubit model, as long as the qubits form a long-range entangled
state (\ie a string-net liquid).  In other words, light and electrons are
unified by the long-range entanglement of qubits.  Information unifies matter!

The electric field and the magnetic field in the Maxwell equation are called
gauge fields.  The field in the Dirac equation are Grassman-number valued
field.\footnote{Grassmann numbers are anti-commuting numbers.} For a long time,
we thought that we have to use gauge fields to describe light waves that have
only two transverse modes, and we thought that we have to use Grassmann-number
valued fields to describe electrons and quarks that have Fermi statistics. So
gauge fields and Grassmann-number valued fields become the fundamental build
blocks of quantum field theory that describe our world.  The string-net liquids
demonstrate that we do not have to introduce  gauge fields and
Grassmann-number valued fields to describe photons, gluons, electrons, and
quarks. It demonstrates how gauge fields and Grassmann fields emerge from local
qubit models that contain only complex scaler fields at the cut-off scale.

\subsection{Evolving views for light and gauge theories}

\begin{figure}[tb]
\centerline{
	\includegraphics[height=0.9in]{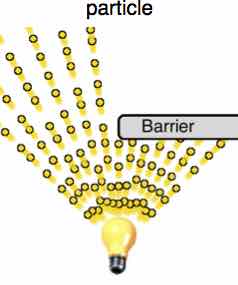}~~~~~
\includegraphics[height=0.9in]{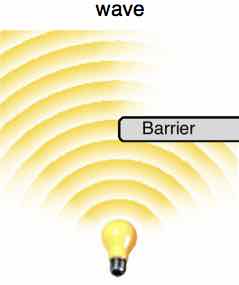}~~~~~
	\includegraphics[height=1.1in]{Chapters/Chap11/EMwave}~~~~~
	\includegraphics[height=0.9in]{Chapters/Chap11/ptrans}
}
\centerline{
(a) ~~~~~~~~~~~~~~~~~~~~~~~~~
(b) ~~~~~~~~~~~~~~~~~~~~~~~~~~~~~~~~~~~
(c) ~~~~~~~~~~~~~~~~~~~~~~~~~~~~~~~
(d) ~~
}
\centerline{
	\includegraphics[height=1.4in]{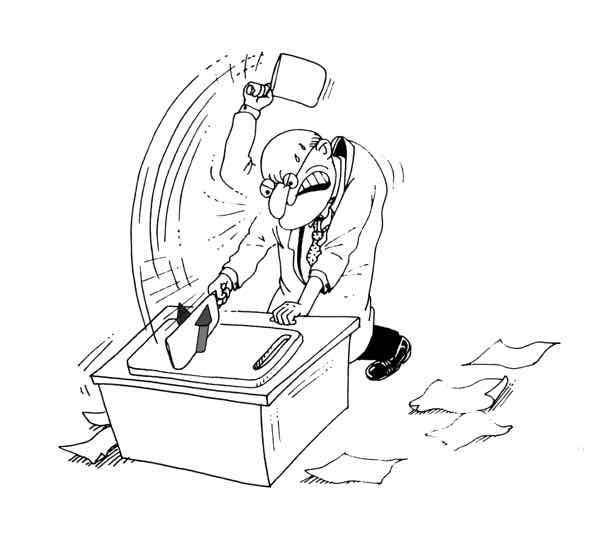}~~~~~
	\includegraphics[height=0.75in]{Chapters/Chap11/vacwvBW}~~~~~~~~~~~~
	\includegraphics[height=1.0in]{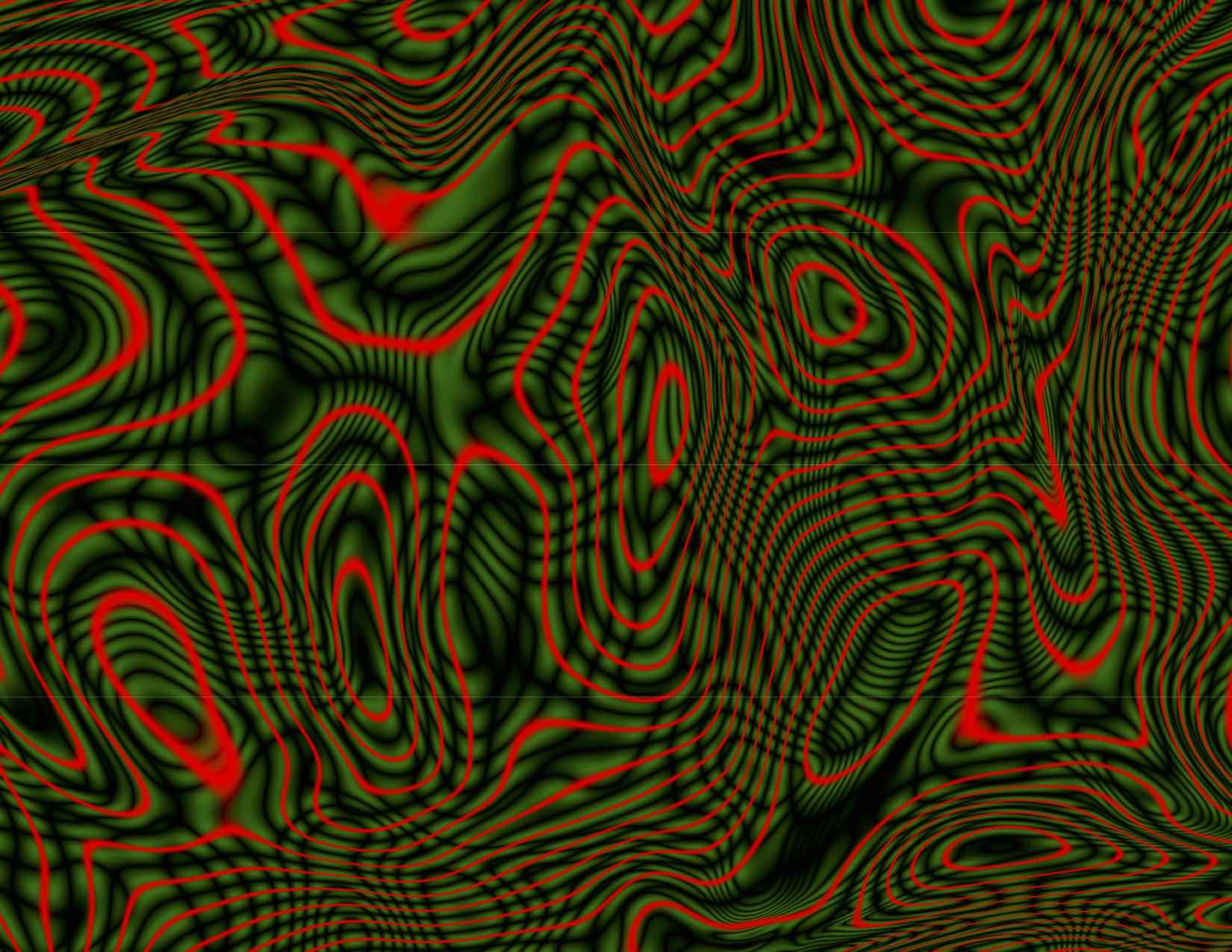}
}
\centerline{
(e) ~~~~~~~~~~~~~~~~~~~~~~~~~~~~~~~~~~~~~~~~~~~~~~~~~~~
(f) ~~~~~~~~~~~~~~~~~~~~~~~~~~~~~~~~~~~~~~~~~~~~~~
(g)
}
\caption{
The evolution of our understanding of light (and gauge interaction):
(a) particle beam, (b) wave, (c) electromagnetic wave, (d)
curvature in fiber bundle, (e) glue of partons, (f) wave in string-net liquid,
(g)  wave in long-range entanglement of many qubits.
}
\label{gauge}
\end{figure}

Our attempt to understand light has a long and evolving history.  We first
thought light to be a beam of particles (see Fig. \ref{gauge}a).  After Maxwell, we understand light as
electromagnetic waves (see Fig. \ref{gauge}b,c).  After Einstein's theory of general relativity, where
gravity is viewed as curvature in space-time, Weyl and others try to view
electromagnetic field as curvatures in the `unit system' that we used to
measure complex phases.  It leads to the notion of gauge theory.  The general
relativity and the gauge theory are two corner stones of modern physics. They
provide a unified understanding of all four interactions in terms of a
beautiful mathematical frame work: all interactions can be understood
geometrically as curvatures in space-time and in `unit systems' (or more
precisely, as curvatures in the tangent bundle and other vector bundles in
space-time, see Fig. \ref{gauge}d).

Later, people in high-energy physics and in condensed matter physics have found
another way in which gauge field can emerge~\cite{DDL7863,W7985,BA8880,AM8874}:
one first cut a particle (such as an electron) into two partons (see Fig.
\ref{gauge}e) by writing the field of the particle as the product of the two
fields of the two partons.  Then one introduces a gauge field to glue the  two
partons back to the original particle.  Such a `glue-picture' of gauge fields
(instead of the fiber bundle picture  of gauge fields) allow us to understand
the emergence of gauge fields in models that originally contain no gauge field
at the cut-off scale.

For long time, people think, by definition,  gauge theories are theories with
gauge symmetries (a kind of local symmetries).  Since all interactions in our
world are described by gauge theories (the abelian ones and non-abelian ones),
gauge symmetry is regarded as a founding principle in our understanding of the
world.  The geometric fiber bundle picture of the gauge theory has stressed the
gauge symmetry.  However, some people are unhappy with the gauge-symmetry point
of view for gauge theory, since it involves many unphysical quantities.  An
attempt to describe gauge theory only in terms physical quantities leads to a
string-net picture of gauge theory~\cite{Walight,LWstrnet}, which represent the
third way to understand gauge theory (see
Fig.  \ref{gauge}f).  Before the string-net theory of gauge
interactions, string operators has appeared in the Wilson-loop
characterization~\cite{W7445} of gauge theory. The Hamiltonian and the duality
description of lattice gauge theory also reveal string
structures~\cite{KS7595,BMK7793,K7959,S8053}, which lead to the string-net
theory for all gauge interactions.

Lattice gauge theories are not local bosonic models since the strings are
unbreakable in lattice gauge theories.  String-net theory points out that we do
not really need strings and qubits themselves are capable of generating gauge
fields and the associated Maxwell/Yang-Mills equation.  This is because even
breakable strings can give rise to gauge fields~\cite{HWcnt}.  This phenomenon
was discovered in several qubit
models~\cite{FNN8035,BA8880,Wlight,MS0204,HFB0404} before realizing their
connection to the string-net liquids~\cite{Walight}.  In other words, opposite
to our opinion that gauge symmetry is a founding principle of our world, in
fact gauge symmetry is not important for gauge theory.  A lattice gauge theory
will always produce gauge interaction at low energies even if we break the
gauge symmetry (by not too big amount) at lattice scale~\cite{FNN8035,HWcnt}.
So gauge theory does not need gauge symmetry!  Since gauge field can emerge
from local qubit models without gauge symmetry, the string picture evolves into
the entanglement picture -- the fourth way to understand gauge field: gauge
fields are fluctuations of long-range entanglement (see Fig.  \ref{gauge}g).
String-net is only a description of the patterns of long-range entanglement.

We feel that the  entanglement picture capture the essence of
gauge theory.  Despite the beauty of the geometric picture, the essence of
gauge theory is not the curved fiber bundles.  In fact, we can view gauge
theory as a theory for long-range entanglement, despite the gauge theory is
discovered long before the notion of long-range entanglement. The evolution of
our understanding of light and gauge interaction: particle beam $\to$ wave
$\to$ electromagnetic wave $\to$ curvature in fiber bundle $\to$ glue of
partons $\to$ string-net density wave $\to$  wave in long-range entanglement
(see Fig. \ref{gauge}), represents 200 year's effort of human race to unveil
the mystery of universe.

\begin{figure}[tb]
\centerline{
	\includegraphics[height=0.8in]{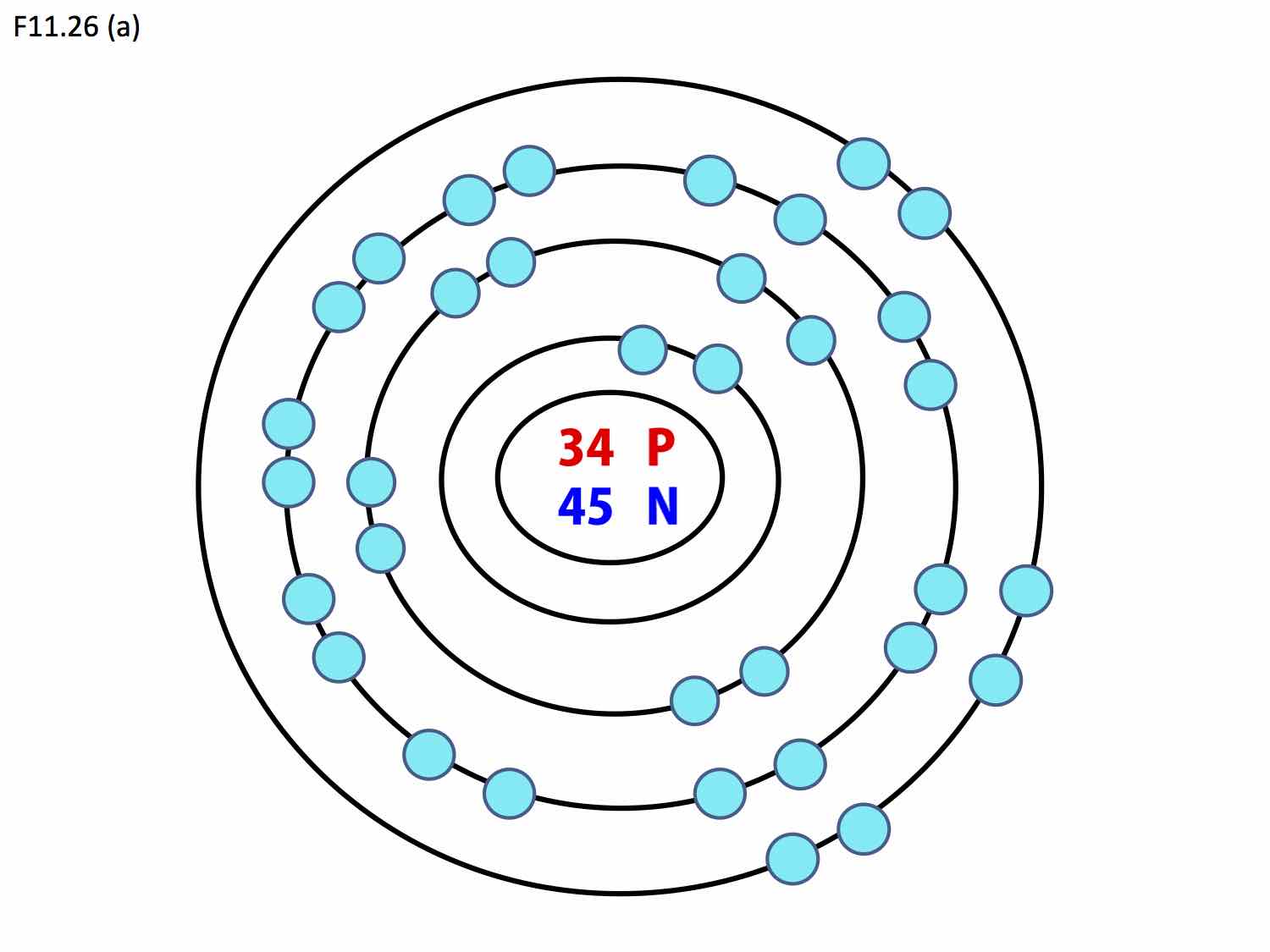}~~~~~
	\includegraphics[height=0.7in]{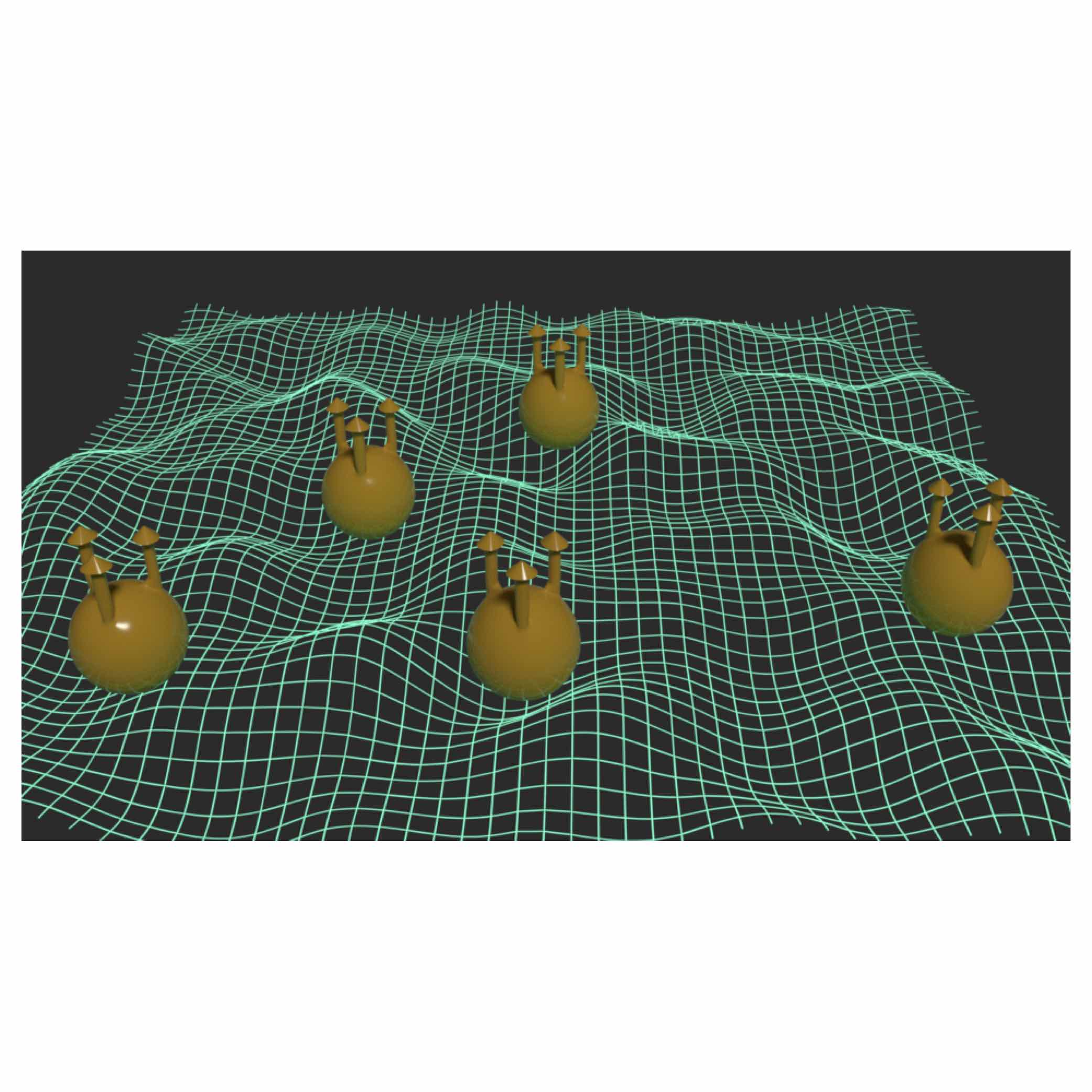}~~~~~
	\includegraphics[height=0.8in]{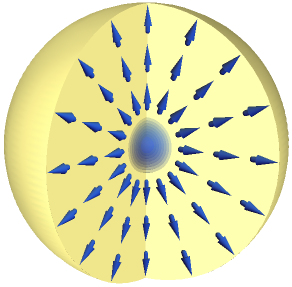}~~~~~
	\includegraphics[height=0.5in]{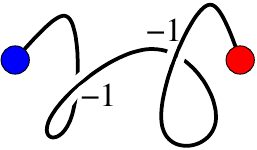}
}
\centerline{
(a) ~~~~~~~~~~~~~~~~~~~~~~~~~~~~~~~~~~
(b) ~~~~~~~~~~~~~~~~~~~~~~~~~~~~~~~~~~~
(c) ~~~~~~~~~~~~~~~~~~~~~~~~~~~~~~~
(d) ~~
}
\caption{
The evolution of our understanding of fermions: (a) elementary particles,
(b) charge-flux bound state in 2D, (c) charge-monople bound state in 3D, (d)
ends of string in quantum string liquids, with $-1$ factor for each string crossing (see \eqn{Fwave}).
}
\label{fermion}
\end{figure}

Viewing gauge field (and the associated gauge bosons) as fluctuations of
long-range entanglement has an added bonus: we can understand the origin of
Fermi statistics in the same way: fermions emerge as defects of long-range
entanglement, even though the original model is purely bosonic.  Previously,
there are two ways to obtain emergent fermions from purely bosonic model: by
binding gauge charge and gauge flux in (2+1)D (see Fig.
\ref{fermion}b~\cite{LM7701,W8257}), and by binding the charge and the monopole
in a $U(1)$ gauge theory in (3+1)D (see Fig.
\ref{fermion}c~\cite{T3141,JR7616,W8246,G8205,LM0012}). But those approaches
only work in (2+1)D or only for $U(1)$ gauge field.  Using long-range
entanglement and their string-net realization, we can obtain the simultaneous
emergence of both gauge bosons (as string density waves) and fermions (as
string ends) in \emph{any} dimensions and for any gauge group (see Fig.
\ref{fermion}d~\cite{LWsta,LWstrnet,LWuni,Wqoem}). This result gives us hope
that maybe all elementary particles are emergent and can be unified using
local qubit models.  Thus, long-range entanglement offer us a new option to
view our world: maybe our vacuum is a long-range entangled state.  It is the
pattern of the long-range entanglement in the vacuum that determines the
content and the structures of observed elementary particles.

We would like to point out that the string-net unification of gauge
bosons and fermions is very different from the superstring theory for gauge
bosons and fermions.  In the string-net theory, gauge bosons and fermions come
from the qubits that form the space, and `string-net' is simply the name that
describe how qubits are organized in the ground state.  So string-net is not a
thing, but a pattern of qubits.  In the string-net theory, the gauge bosons are
waves of collective fluctuations of the string-nets, and a fermion corresponds
to one end of string.  In contrast, gauge bosons and fermions come from strings
in the superstring theory. Both gauge bosons and fermions correspond to small
pieces of strings.  Different vibrations of the small pieces of strings give
rise to different kind of particles. The fermions in the superstring theory are
put in by hand through the introduction of Grassmann fields.

%
%
%
%

\subsection{Where to find long-range entangled quantum matter?}

In this book, we described the world of quantum phases.  We pointed out that
there are symmetry breaking quantum phases, and there are topologically ordered
quantum phases.  The topologically ordered quantum phases are a totally new
kind of phases which cannot be understood using the conventional concepts (such
as symmetry breaking, long-range order, and order parameter) and conventional
mathematical frame work (such as group theory and Ginzburg-Landau theory).  The
main goal of this book is to introduce new concepts and pictures to describe
the new topologically ordered quantum phases.

In particular, we described how to use global dancing pattern to gain an
intuitive picture of topological order (which is a pattern of long-range
entanglement). We further point out that we can use local dancing rules to
\emph{quantitatively} describe the  global dancing pattern (or topological
order). Such an approach leads to a systematic description of topological
order in terms of string-net (or unitary fusion category
theory)~\cite{LWstrnet,H0904,CGW1038,GWW1017} and systematic description of 2D
chiral topological order in terms of pattern of
zeros~\cite{WW0808,WW0809,R0634,SRC0706,BH0802,BH0802a,BW0932,BW1001a,LWW1024}
(which is a generalization of `charge-density-wave' description of FQH
states~\cite{SL0604,BKW0608,SL0701,S0802,SY0802,ABK0816,S1002,FS1115}).

The local-dancing-rule approach also leads to concrete and explicit
Hamiltonians, that allow us to realize each string-net state and each FQH state
described by pattern of zeros.  However, those Hamiltonians usually contain
three-body or more complicated interactions, and are hard to realize in real
materials.  So here we would like to ask: can topological order be realized by
some simple Hamiltonians and real materials?

Of cause, non-trivial topological orders -- FQH states -- can be realized by 2D
electron gas under very strong magnetic fields and very low
temperatures~\cite{TSG8259,L8395}. Recently, it was proposed that FQH states
might appear even at room temperatures with no magnetic field in flat-band
materials with spin-orbital coupling and spin
polarization~\cite{TMW1106,SGK1103,NSC1104,SGS1189,GNC1297}. Finding such
materials and realizing FQH states at high temperatures will be an amazing
discovery. Using flat-band materials, we may even realize non-Abelian
fractional quantum Hall states~\cite{MR9162,Wnab,WES8776,RMM0899} at high
temperatures.

Apart from the FQH effects, non-trivial topological order may also appear in
quantum spin systems.  In fact, the concept of topological order was first
introduced~\cite{Wtop} to describe a chiral spin liquid~\cite{KL8795,WWZ8913},
which breaks time reversal and parity symmetry.  Soon after, time reversal and
parity symmetric topological order was proposed in
1991~\cite{RS9173,Wsrvb,MLB9964,MS0181}, which has spin-charge separation and
emergent fermions.  The new topological spin liquid is called $\mathbb{Z}_2$
spin liquid or $\mathbb{Z}_2$ topological order since the low energy effective
theory is a $\mathbb{Z}_2$ gauge theory.  In 1997, an exactly soluble
model~\cite{K032} (that breaks the spin rotation symmetry) was obtained that
realizes the  $\mathbb{Z}_2$ topological order.  Since then, the
$\mathbb{Z}_2$ topological order become widely accepted.

More recently, extensive new numerical calculations indicated that the
$J_1$-$J_2$-$J_3$
Heisenberg model on Kagome lattice~\cite{HC13123461,ZS14104883,GS14121571}
\begin{align}
 H=
\sum_\text{1sf} J_1 \v S_i\cdot \v S_j
+\sum_\text{2nd} J_2 \v S_i\cdot \v S_j
+\sum_\text{3rd} J_3 \v S_i\cdot \v S_j, \ \
\ \ \ J_2/J_1\sim J_3/J_1 \sim 0.5 ,
\end{align}
has gapped spin liquid ground state. Such spin liquid is the chiral spin liquid.\cite{KL8795,WWZ8913}

The nearest neighbor Heisenberg model on Kagome lattice can be realized in
Herbertsmithite $ZnCu_3(OH)_6Cl_2$~\cite{HMS0704,IFH1111}. Although $J_1$ is as
large as $150$K, no spin ordering and other finite temperature phase
transitions are found down to 50mK.  So Herbertsmithite may realize a 2D spin
liquid state.  However, experimentally, it is not clear if the spin liquid is a
gapped spin liquid or a gapless spin liquid.  Theoretically, both a gapped
$\mathbb{Z}_2$ spin liquid~\cite{JWS0803,YHW1173,LR1120,LRL1113} and a gapless
$U(1)$ spin liquid~\cite{H0013,RHL0705,HRL0813} are proposed for the Heisenberg
model on Kagome lattice.  The theoretical study suggests that the spin liquid
state in Herbertsmithite may have some very interesting characteristic
properties: A magnetic field in $z$-direction may induce a spin order in
$xy$-plane~\cite{RKL0774}, and an electron (or hole) doping may induce a charge
$4e$ topological superconductor~\cite{KLW0902}.

To summarize, topological order and long-range entanglement give rise to new
states of quantum matter.  Topological order, or more generally, quantum order
have many new emergent phenomena, such as emergent gauge theory, fractional
charge, fractional statistics, non-Abelian statistics, perfect conducting
boundary, etc.  In particular, if we can realize a quantum liquid of oriented
strings in certain materials, it will allow us to make artificial elementary
particles (such as artificial photons and artificial electrons).  So we can
actually create an artificial vacuum, and an artificial world for that matter,
by making an oriented string-net liquid.  This would be a fun experiment to do!

%
%
\bibliographystyle{plain}
\bibliography{all}






\end{document}